\documentclass[%
 reprint,
superscriptaddress,
 amsmath,amssymb,
 aps,
prb,
]{revtex4-1}

\usepackage{graphicx}
\usepackage{dcolumn}
\usepackage{bm}
\usepackage{mathtools}
\usepackage{bbold}
\usepackage{physics}

\begin{document}

\preprint{}

\title{Self-Energies and Quasiparticle Scattering Interference}

\author{Miguel Antonio Sulangi}
\author{Jan Zaanen}%
\affiliation{%
	Instituut-Lorentz for Theoretical Physics, Leiden University, Leiden, Netherlands 2333 CA
}

\date{\today}

\begin{abstract}
The cuprate high-temperature superconductors are known to host a wide array of effects due to interactions and disorder. In this work, we look at some of the consequences of these effects which can be visualized by scanning tunneling spectroscopy in order to provide a guide for future experiments. These interaction and disorder effects can be incorporated into a mean-field description by means of a self-energy appearing in the Green's function. We first examine the quasiparticle scattering interference (QPI) spectra in the superconducting state at optimal doping as temperature is increased. Assuming agreement with angle-resolved photoemission experiments which suggest that the scattering rate depends on temperature, resulting in the filling of the $d$-wave gap, we find that the peaks predicted by the octet model become progressively smeared as temperature is increased. When the scattering rate is of the same order of magnitude as the superconducting gap, the spectral function shows Fermi-arc-like patterns, while the power spectrum of the local density of states shows the destruction of the octet-model peaks. We next consider the normal state properties of the optimally-doped cuprates. We model this by adding a marginal Fermi liquid (MFL) self-energy to the normal-state propagator, and consider the dependence of the QPI spectra on frequency, temperature, and doping. We demonstrate that the MFL self-energy leads to a smearing of the caustics appearing in the normal-state QPI power spectrum as either temperature or frequency is increased at fixed doping. The smearing is found to be more prominent in the MFL case than in an ordinary Fermi liquid. We also consider the case of a marginal Fermi liquid with a strongly momentum-dependent self-energy which gives rise to a visible ``nodal-antinodal'' dichotomy at the normal state, and discuss how the spectra as seen in ARPES and STS differ from both an isotropic metal and a broadened $d$-wave superconductor. Finally, we discuss how these results become modified in the presence of weak distributed disorder and finite-temperature smearing.
\end{abstract}

\maketitle

\section{\label{sec:level1}Introduction}

The copper-oxide superconductors are well-known to be strongly correlated materials. Many phenomena exhibited by the cuprates evade explanations based on weakly interacting quasiparticles. Perhaps the most notorious example of this is the ``strange metal,'' the normal state of these materials near optimal doping. This shows behavior that, as probed by transport, is very different from that seen in conventional metals, which are described well by Fermi-liquid theory.\cite{keimer2015quantum} Another similarly perplexing phase of these cuprates is the pseudogap, in which the density of states is prominently suppressed near the Fermi energy, exhibiting numerous exotic phenomena such as various ordered phases, gap inhomogeneities, and  ``Fermi arcs''--- disconnected segments in momentum space hosting gapless excitations---as seen in experiments such as scanning tunneling spectroscopy (STS) and angle-resolved photoemission spectroscopy (ARPES).\cite{timusk1999pseudogap, norman2005pseudogap} Even the superconducting state, which is comparatively well-understood among the various phases of these materials, is highly unusual: it has $d$-wave pairing, leading to gapless, Dirac-like nodal quasiparticles.\cite{shen1993anomalously, ding1996angle} It appears to be much more stable against disorder than $d$-wave mean-field BCS theory predicts, while unusual interaction effects, as probed by ARPES, are seen to emerge as the temperature approaches $T_c$,\cite{reber2012origin, reber2013prepairing, reber2015pairing, reber2015coordination} which in turn is much higher than in conventional superconductors. The $T = 0$ states at low and high doping are rather firmly established as an antiferromagnetic Mott insulator and a conventional Fermi liquid, respectively, but the intermediate-doping states remain to be fully understood. A full microscopic theory of these materials consistent with all of these phenomena has yet to be developed.

Much understanding can nevertheless be gained by adopting a phenomenological approach towards modeling these various phases of the cuprates. Starting from a weakly-interacting picture,  interaction or disorder effects can be included by putting in the appropriate self-energy, which ``dresses'' up the mean-field description one starts out with. For instance, many of the unusual transport properties of the strange metal, such as linear-in-$T$ resistivity, can be captured by the marginal Fermi liquid self-energy first introduced by Varma \emph{et al.}\cite{varma1989phenomenology} While this self-energy enters the Fermi-liquid propagator in what appears to be an innocuous manner,  it results in the complete absence of quasiparticles at $T = 0$: the quasiparticle weight of a marginal Fermi liquid vanishes at zero temperature. This MFL self-energy has been shown to account for much of the transport anomalies seen in the cuprates, although its microscopic origins remain largely unknown. Similarly, much insight can be derived by treating the $d$-wave superconducting state as a mean-field, albeit unconventional, BCS superconductor. This starting point is largely justified by experiment. In general, ARPES finds that the Bogoliubov quasiparticles inside the superconducting state are well-defined excitations,\cite{kaminski2000quasiparticles, feng2000signature, tanaka2006distinct, lee2007abrupt, vishik2009momentum} while STS similarly finds behavior consistent with \emph{coherent} quasiparticles scattering off of disorder, leading to quasiparticle scattering interference (QPI).\cite{hoffman2002imaging, mcelroy2003relating,hanaguri2007quasiparticle, kohsaka2008cooper, lee2009spectroscopic, fujita2014simultaneous} From a purely phenomenological standpoint, the $d$-wave superconducting state can be reasonably studied starting with this mean-field description, with self-energies included to model phenomena that deviate strikingly from the mean-field expectation.

Recently, a number of ARPES experiments on both normal and superconducting Bi$_2$Sr$_2$CaCu$_2$O$_{8+\delta}$ (Bi-2212) across a wide doping range have provided a more complete picture of the various phenomena in these materials, with the self-energy playing a crucial role in both phases. In the superconducting state, it is observed that the superconducting gap is not the sole factor determining $T_c$---contrary to expectations from BCS theory. Instead, the quasiparticle scattering rate exhibits a pronounced uptick near $T_c$. It appears that $T_c$ is set by the scale at which the gap and the scattering rate cross over into each other, and the temperature at which the gap closes is \emph{larger} than $T_c$.\cite{reber2012origin, reber2013prepairing, reber2015pairing, reber2015coordination} Meanwhile, in the normal state, experiments affirm the validity of the marginal Fermi liquid description at optimal doping, but in addition find that the ARPES data are well-described by a self-energy that interpolates smoothly between a Fermi-liquid one at extreme overdoping and a marginal-Fermi-liquid one at optimal doping. Such a doping-dependent self-energy has been central to the proposed ``power-law liquid'' phenomenology first proposed by Reber \emph{et al.}\cite{reber2015power}

Our goal in this chapter is to provide a detailed theoretical exploration of the effects of these self-energies, in both the normal and superconducting phases, on the real-space local density of states as probed by experiments. We will focus on QPI, which has not been looked at in related high-temperature STS studies on Bi-2212. Very few experimental studies on the temperature-dependent behavior in the superconducting state have been performed thus far \cite{gomes2007visualizing, pasupathy2008electronic, parker2010fluctuating}, and the effects of self-energies on STS spectra have been largely unexplored save for a small number of studies \cite{pasupathy2008electronic, alldredge2008evolution,dahm2014quasi}. In particular, Dahm and Scalapino have studied how to extract the form of the self-energy from the QPI response of a normal metal with a cuprate-like Fermi surface.\cite{dahm2014quasi} Given this situation, we will provide a template demonstrating how STS results might look like, providing a guide for future experiments. QPI can be used as a real-space method of probing the momentum-space structure of the excitations: one takes the power spectrum of the differential conductance maps, and the most prominent wavevectors appearing can be used to map out the underlying band structure of these materials.\cite{hoffman2002imaging, mcelroy2003relating, wang2003quasiparticle, capriotti2003wave, hanaguri2007quasiparticle,  kohsaka2008cooper, lee2009spectroscopic, fujita2014simultaneous} In addition, STS experiments can, in principle, demonstrate whether the excitation spectra are coherent or not. The presence of sharp peaks in the power spectrum of the differential conductance maps taken from the $d$-wave superconducting state at low temperatures is a clear-cut demonstration of the existence of the Bogoliubov quasiparticles as sharp, phase-coherent excitations.\cite{zaanen2003superconductivity} This fact is corroborated by evidence from ARPES suggesting that the excitations in the superconducting state at optimal doping are long-lived, unlike those in the normal state at the same doping.\cite{kaminski2000quasiparticles, feng2000signature, tanaka2006distinct, lee2007abrupt, vishik2009momentum} These peaks in the power spectrum behave exactly as the heuristic ``octet model'' suggests. If these Bogoliubov quasiparticles are no longer long-lived, there is no reason to suspect that these peaks will continue to be present. These will be broadened and, if the scattering rate is large enough, will be rendered diffuse enough that these no longer exist as well-defined peaks. Throughout this chapter we will examine closely in several case studies the effect of the quasiparticle scattering rate on the power spectrum of the LDOS in the superconducting and normal state.

We first study the superconducting state as temperature is varied, and consider three different possible scenarios and how these can be seen in ARPES and STS. The first, which we call the ``gap-closing'' scenario, is well-known from BCS theory. Here the gap shrinks continuously as temperature is increased until $T_c$ is reached, at which point it vanishes. The scattering rate is constant as a function of temperature. The second scenario is ``gap-filling/closing''  and is argued to be seen in ARPES experiments. Here the gap shrinks with increasing $T$, vanishing at some temperature $T_p$, but, importantly, $T_c \neq T_p$. In addition, additional spectral weight fills in at low energies as $T$ is increased. This can be accounted for by a temperature-dependent imaginary part of the self-energy which takes on a value comparable to that of the gap at temperatures near $T_c$. The third scenario is ``gap-filling,'' wherein the superconducting gap is temperature-independent, while the scattering rate is strongly temperature-dependent, as in the second scenario. We observe the gradual disappearance of the octet-model peaks as the scattering rate becomes very large. In the two scenarios where the gap closes, we observe that the octet-model peaks can be seen to disperse when the energy is fixed and temperature is varied, but that these peaks lose coherence if the scattering rate becomes very large. 

As for the normal state, three scenarios are also considered. The first is the ordinary Fermi liquid, the second is the marginal Fermi liquid, and the third is a realistic marginal Fermi liquid whose spectral function exhibits considerable momentum-space anisotropy, with the nodal regions being much more coherent than the antinodal ones. We see that the power spectrum of the LDOS in the first two cases appears superficially similar to each other---the main feature for both is a set of caustics which correspond to the scattering wavevectors between points along the Fermi surface. The difference between the two sets of spectra is quite subtle: the caustics in the marginal Fermi liquid power spectrum are much more broadened than those in the ordinary Fermi liquid power spectrum, owing to the smaller self-energies present in the ordinary Fermi liquid compared to those in the marginal Fermi liquid. Finally, the momentum-dependence of the self-energy of the anisotropic marginal Fermi liquid results in a highly anisotropic LDOS power spectrum as well---scattering between incoherent portions of the Fermi surface results in very broadened segments of the caustics in the power spectrum, while the wavevectors corresponding to scattering between coherent quasiparticles give rise to sharp caustic segments.

We note that STS as a probe is particularly vulnerable to finite-temperature smearing, which can obscure the features described in our numerics. We thus augment our single-impurity results without thermal smearing with macroscopically disordered and thermally smeared simulations to provide guides to experimentalists. It is in principle possible to deconvolute the thermally smeared STS data to obtain differential conductances that feature only \emph{intrinsic} broadening; this has been performed in a number of STS studies.\cite{pasupathy2008electronic} However it is nevertheless worthwhile to examine the extent to which the features described in the single-impurity, thermally unsmeared case survive when multiple impurities and thermal smearing are included. We find that the thermally smeared case obscures many of the features seen in the superconducting state, with the octet model peaks disappearing even when the thermally unsmeared simulations suggest they are present. In the normal state cases we study, however, the general features of the thermally unsmeared results---the caustics---survive even with thermal smearing included.

\section{Self-Energies and Broadening} \label{selfenergies}

When considered as phenomenological inputs and in the limit of weak disorder, self-energies do not fundamentally alter any of the fundamental physics of quasiparticle scattering interference in both the normal and the superconducting state. Their main nontrivial effect is to broaden the density of states relative to the clean, non-interacting limit. In what follows we will illustrate these effects in the normal and superconducting states.

Consider a normal metallic system described by a Hamiltonian $H$ without any self-energies. The density of states $\rho$ at energy $E$ is 
\begin{equation}
\rho(E) = \sum_{n} \delta(E - \epsilon_n),
\end{equation}
where $\epsilon_n$ is an eigenvalue of $H$ and $n$ is some quantum number. (In a translationally-invariant system, this quantum number could be the momentum $\mathbf{k}$, for example, and the sum amounts to integrating over $\mathbf{k}$.) If one has a finite-sized system, the spectrum of $H$ is discrete, and the DOS consists of spikes at energies equal to $\epsilon_n$. Now we include the effect of self-energies. In this system, the self-energy is defined as
\begin{equation}
\Sigma^N(n, \omega) = G_0^{-1}(n, \omega) - G^{-1}(n, \omega),
\end{equation}
where $G$ is the Green's function for the full system (with interactions, disorder, or both) and $G_0$ is the noninteracting/clean Green's function, written in the basis which diagonalizes $H$ (\emph{i.e.}, the set of eigenstates $\ket{n}$).\cite{mahan2013many} The self-energy is assumed to incorporate all the effects of interactions or disorder, so the Green's function for the full system has the same symmetries as that of the non-interacting/clean one. The retarded full Green's function can be written as
\begin{equation}
G(n, \omega) = \frac{1}{\omega - \epsilon_n -\Sigma^N(n, \omega)}.
\end{equation}
We can define a spectral function $A(n, \omega) = -\frac{1}{\pi}\text{Im}G(n, \omega)$. It has the following form:
\begin{equation}
A(n, \omega) = -\frac{1}{\pi}\frac{\Sigma^N_2(n, \omega)}{[\omega- \Sigma^N_1(n, \omega) - \epsilon_n]^2 + [\Sigma^N_2(n, \omega)]^2}.
\label{eq:lorentzian}
\end{equation}
Here $\Sigma^N_1$ and $\Sigma^N_2$ are the real and imaginary parts, respectively, of the self-energy. In the limit $\Sigma^N \to 0$, this reduces to a delta function describing the noninteracting system:
\begin{equation}
\begin{aligned}
\lim\limits_{\Sigma \to 0}A(n, \omega) &=-\lim\limits_{\Sigma^N \to 0} \frac{1}{\pi}\frac{\Sigma^N_2(n, \omega)}{[\omega- \Sigma^N_1(n, \omega) - \epsilon_n]^2 + [\Sigma^N_2(n, \omega)]^2}\\
 &= \delta(\omega - E_n).
\end{aligned}
\end{equation}
The DOS for the full system is
\begin{equation}
\rho(E) = \sum_{n} A(n, \omega \to E).
\end{equation}
This means that, in the presence of $\Sigma^N$, the density of states at an energy $E$ does not consist merely of states which statisfy $\epsilon_n = E$. For one, $\Sigma^N_1(n, \omega)$ shifts the real parts of the poles of the Green's function from $\omega = \epsilon_n$ to $\omega - \Sigma^N_1(n, \omega) = \epsilon_n$. More importantly, the spectrum is broadened and $\rho(E)$ now incorporates nonlocal contributions from states located away from $E$ in energy space. This will be reflected in the local density of states (LDOS) as well: a map of the LDOS taken at energy $E$ will include contributions from states at other energies, weighted by Eq.~\ref{eq:lorentzian}.

None of our discussion fundamentally changes when one considers the superconducting state. The full Green's function in Nambu space, including self-energies, is
\begin{equation}
\tilde{G}^{-1} = 
\begin{pmatrix*}[c]
\omega - \epsilon_n - \Sigma^N(n, \omega)   & -\Sigma^A(n, \omega) \\
-\Sigma^A(n, \omega) & \omega + \epsilon_n + \Sigma^N(n, -\omega)^*
\end{pmatrix*}.
\label{eq:sc_gf}
\end{equation}
$\epsilon_n$ is the normal-state energy, and $\Sigma^N(n, \omega)$ and $\Sigma^A(n, \omega)$ are the normal and anomalous self-energies, respectively.\cite{civelli2008nodal} Under this definition, in the superconducting state the real part of the anomalous self-energy is equal to the pairing gap. In the cases involving $d$-wave superconductors that we will discuss, we will focus only on normal-state self-energies, and we will take the anomalous self-energy to be frequency-independent, so that in the translationally-invariant case the gap has the usual noninteracting $d$-wave form given by $\Sigma^A(\mathbf{k}) = \Delta(\mathbf{k})= 2\Delta_0(\cos k_x - \cos k_y)$. 

We start with a normal-state self-energy of the form $\Sigma^N(\omega) = \Sigma^N_1(\omega) + i\Sigma^N_2(\omega)$. We assume that the self-energy depends only on $\omega$, and that $\Sigma^N_1(\omega) = -\Sigma^N_1(-\omega)$ and $\Sigma^N_2(\omega) = \Sigma^N_2(-\omega)$. It can be shown that the the spectral functions corresponding to the particle and hole parts of the Green's functions, $A_1(n, \omega) = -\frac{1}{\pi}\text{Im}G_{11}(n, \omega)$ and $A_2(n, \omega) = -\frac{1}{\pi}\text{Im}G_{22}(n, \omega)$, are
\begin{equation}
A_1(n, \omega) = -\frac{1}{\pi}\frac{\Sigma^N_2(\omega)}{[\omega- \Sigma^N_1(\omega) - E_n]^2 + [\Sigma^N_2(\omega)]^2} 
\label{eq:sclorentzian1}
\end{equation}
and
\begin{equation}
A_2(n, \omega) = -\frac{1}{\pi}\frac{\Sigma^N_2(\omega)}{[\omega- \Sigma^N_1(\omega) + E_n]^2 + [\Sigma^N_2(\omega)]^2},
\label{eq:sclorentzian2}
\end{equation}
where $E_n = \sqrt{\epsilon_n + \Delta_n}$ are the energies of the Bogoliubov quasiparticles. Without self-energies these spectral functions consist of delta functions at energies $E_n$. As in the normal case, the spectral functions are broadened by $\Sigma^N_2(\omega)$, and the presence of $\Sigma^N_1(\omega)$ shifts the real parts of the poles by $\Sigma^N_1(\omega)$. The full spectral function $A(n, \omega)$ is 
\begin{equation}
A(n, \omega) = u^2_nA_1(n, \omega) + v^2_nA_2(n, \omega),
\label{eq:sccf}
\end{equation}
where $u^2_n$ and $v^2_n$ are coherence factors, given by $u^2_n = \frac{1}{2}(1 + \frac{\epsilon_n}{E_n})$ and $v^2_n = \frac{1}{2}(1 - \frac{\epsilon_n}{E_n})$.\cite{campuzano2004photoemission} Consequently the density of states at energy $E$ takes the following form:
\begin{equation}
\rho(E) = \sum_{n} \big[u^2_nA_1(n, \omega \to E) +v^2_n A_2(n, \omega \to E)\big].
\end{equation}

\section{Methods}

Here we will briefly sketch the methods we utilize in the paper. Both real- and momentum-space methods are used to ensure that our numerical results can be compared well with STS and ARPES.  We first focus on real-space methods. To obtain quantities such as the local density of states, we start with the Bogoliubov-de Gennes Hamiltonian, written in a site basis:
\begin{equation}
H = -\sum_{ij\sigma} t_{ij}c_{i\sigma}^{\dagger}c_{j\sigma} + \sum_{ij}(\Delta_{ij}^{\ast}c_{i \uparrow}c_{j \downarrow} + \text{h.c.}).
\label{eq:hamiltonian}
\end{equation}
We will parametrize the normal-state Fermi surface with a minimal single-band model capturing most of the salient features of the normal state of optimally-doped BSCCO. We set the nearest-neighbor and next-nearest neighbor hopping amplitudes to be $t = 1$ and $t' = -0.3$, respectively. The chemical potential $\mu$ is tuned to ensure that the hole doping concentration is $p \approx 16\%$. In the superconducting state the pairing amplitude is of a $d$-wave nature; this is ensured by taking $\Delta_{ij} = \Delta_0$ and $\Delta_{ij} = -\Delta_0$ whenever $i$ and $j$ are nearest-neighbor sites in the $x$- and $y$-directions, respectively.

All real-space information about the spectrum of Eq.~\ref{eq:hamiltonian} can be extracted from the Green's function $G$. The bare Green's function $G_0$---without disorder or interactions---can be written in terms of the lattice Hamiltonian H as follows:
\begin{equation}
G_0^{-1}(\omega) = \omega\mathbb{1} - H.
\label{eq:green}
\end{equation}
As we have defined them, $G_0$ and $H$ are $2N_xN_y \times 2N_xN_y$ matrices living in Nambu space, as in Eq.~\ref{eq:sc_gf}.  We will incorporate disorder or interactions into this mean-field formalism by means of a self-energy $\Sigma(\omega)$, which is another $2N_xN_y \times 2N_xN_y$ matrix with a similar Nambu-space substructure as $G_0$. Most generally, $\Sigma(\omega)  = \Sigma^N(\omega)  + \Sigma^A(\omega)$ in the $d$-wave state; however we will assume that $d$-wave pairing has already been incorporated into the bare Green's function, so only the normal part of the self-energy enters into consideration. The full Green's function becomes
\begin{equation}
G^{-1}(\omega) = G_0^{-1}(\omega) - \Sigma(\omega),
\label{eq:greendressed}
\end{equation}
and before proceeding we need to input first the needed form of $\Sigma(\omega)$. Note that in principle, $\Sigma(\omega)$ could be momentum-dependent; this can be incorporated into a lattice description by putting the appropriate off-diagonal couplings into Eq.~\ref{eq:greendressed}. 

By judiciously choosing the indexing of the sites, $G^{-1}$ can be written in block tridiagonal form. We then invert $G^{-1}$ using an efficient algorithm for block tridiagonal matrices.\cite{godfrin1991method, hod2006first, drouvelis2006parallel, petersen2008block, wimmer2009optimal, li2012extension, reuter2012efficient, li2013fast, kuzmin2013fast} The details of this method have been worked out in detail in prior work, so we will not repeat them here.\cite{sulangi2017revisiting, sulangi2018quasiparticle} The advantage of this method is that it is extremely fast compared to exact diagonalization; allows general forms of disorder to be included, unlike the $T$-matrix method, which is exact only for pointlike impurities; and enables self-energies to be included explicitly in the Green's function, allowing the study of the unusual effects of self-energies on measurable real-space quantities. We take $N_x = 1000$ and $N_y = 120$.

The local density of states $\rho(\mathbf{r}, E)$ can be obtained from $G$ using the following equation:
\begin{equation}
\rho(\mathbf{r}, E) = -\frac{1}{\pi}\text{Im}G_{11}(\mathbf{r}, \omega \to E).
\label{eq:ldos}
\end{equation}
To study quasiparticle scattering interference, we first introduce a single weak ($V = 0.5$) pointlike scatterer in the middle of the sample. The choice of a weak pointlike scatterer has been shown to reproduce, on a phenomenological level, the octet-model peaks and the real-space modulations in the LDOS indicative of quasiparticle scattering interference. While experiments do show sharper octet-model peaks than simulations do, it is very likely that the inclusion of microscopic details relating to the tunneling of electrons from the STM tip to the copper-oxide planes plays an important role in resolving this apparent discrepancy.\cite{sulangi2017revisiting, kreisel2015interpretation} We obtain the LDOS map of the central $100 \times 100$ region from Eq.~\ref{eq:ldos} and take the \emph{absolute value} of its Fourier transform to obtain the QPI power spectrum $P(\mathbf{q}, \omega)$. The general case of a macroscopically disordered sample can be modeled by randomly distributing a number of these weak scatterers across the sample. To provide a guide for experimentalists, we also include results in which thermal broadening is present. It is known that the differential conductance as measured by STS at temperature $T$ is broadened by a factor $\Gamma_t \approx 3.5k_B T$---this is simply the width of the first derivative of the Fermi-Dirac distribution function which enters into the expression for the density of states at temperature $T$.\cite{pasupathy2008electronic} This can be incorporated into our model by adding this temperature-dependent thermal smearing factor---note here that $\Gamma_t$ is the full width at half maximum---to the intrinsic broadening due to disorder and interactions.

Another major quantity we are interested in is the spectral function $A(\mathbf{k}, \omega)$. Assuming that we have only normal self-energies entering the Green's function, the spectral function can be computed directly from the dispersion of the Bogoliubov quasiparticles using Eqs.~\ref{eq:sclorentzian1},~\ref{eq:sclorentzian2} and~\ref{eq:sccf}, with $n \to \mathbf{k}$. Here, $E_{\mathbf{k}} = \sqrt{\epsilon_{\mathbf{k}} + \Delta_{\mathbf{k}}}$, and $\epsilon_{\mathbf{k}} = -2t(\cos k_x + \cos k_y ) - 4t' \cos k_x \cos k_y - \mu$ and $\Delta_{\mathbf{k}} = 2\Delta_0(\cos k_x - \cos k_y)$. To numerically calculate this, the first Brillouin zone is divided into a discrete $1000 \times 1000$ grid, which is large enough to render finite-size effects insignificant.

\section{Self-energies in the superconducting state}

\begin{figure*}
	\centering
	\includegraphics[width=.3\textwidth]{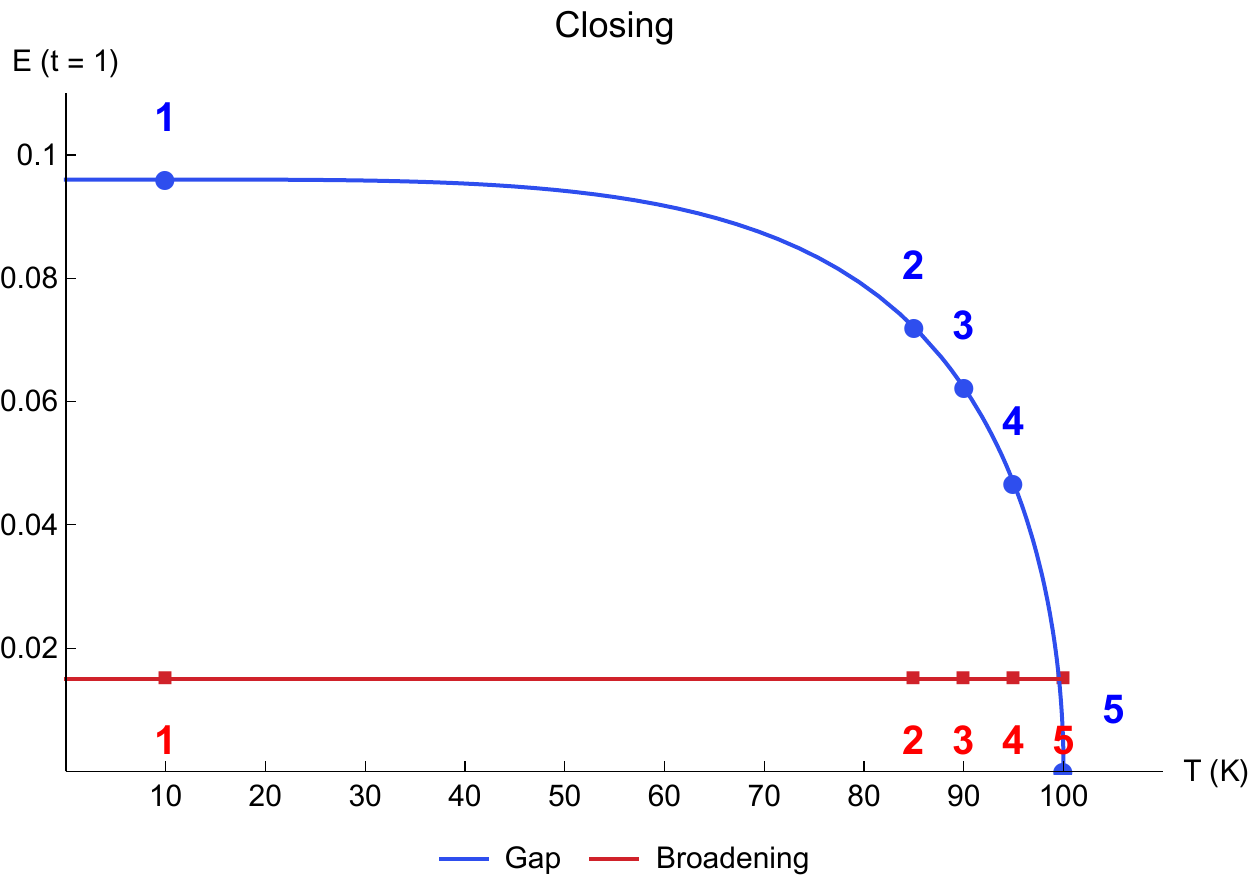} 
	\includegraphics[width=.3\textwidth]{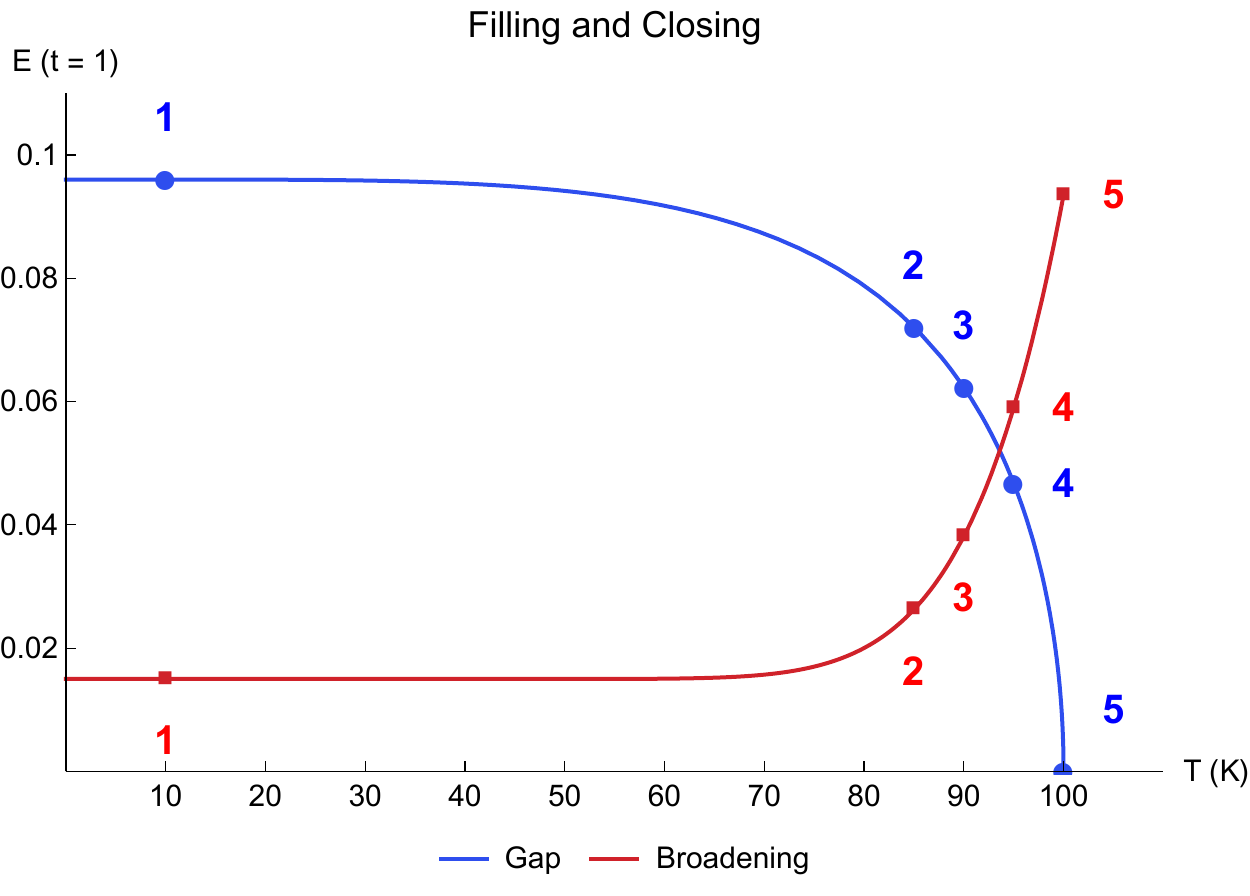}
		\includegraphics[width=.3\textwidth]{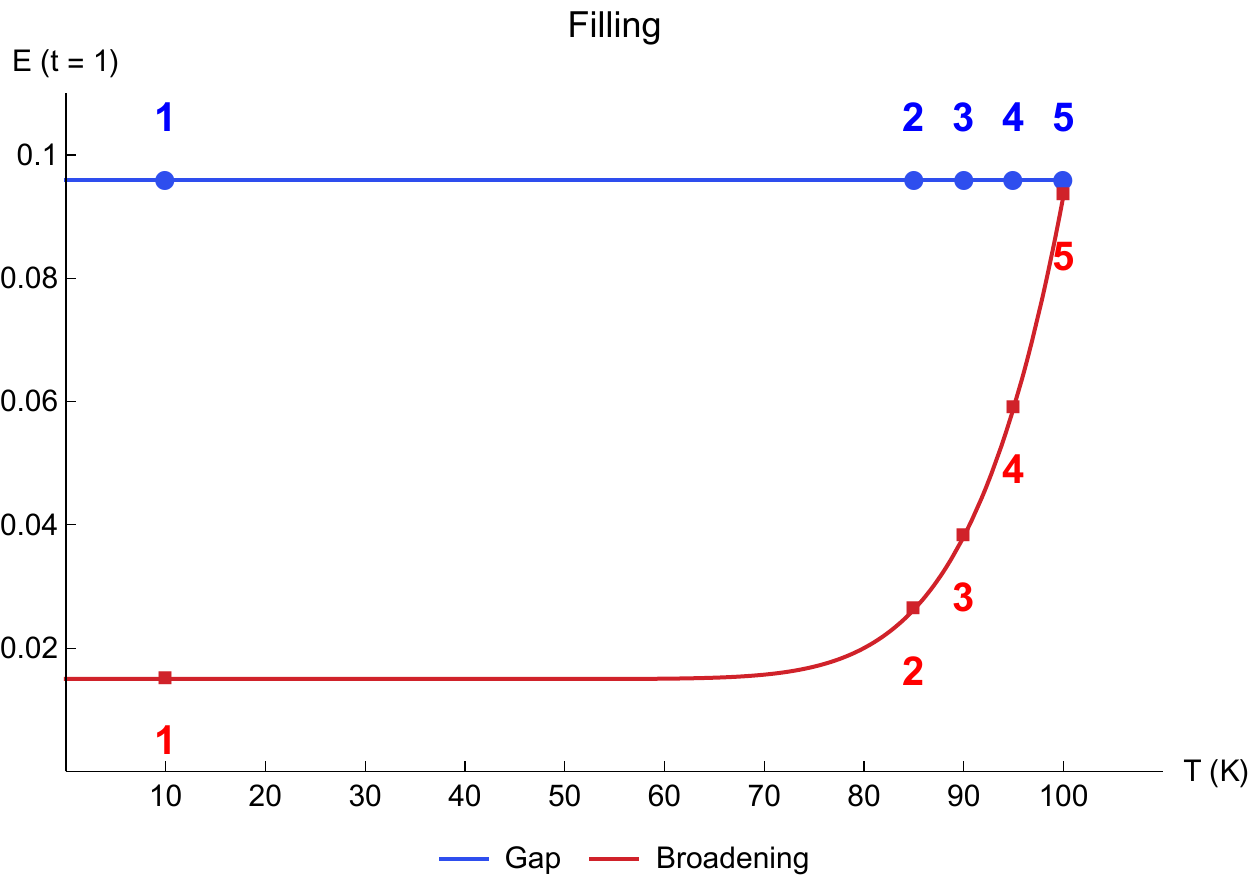}
	\caption{Plot of the gap and the quasiparticle scattering rate as a function of temperature in the gap-closing (left), gap-filling and -closing (middle), and gap-filling (right) scenarios. The behavior seen in the middle plot---corresponding to the gap-filling/closing scenario---is seen in ARPES measurements by Reber \emph{et al.} on optimally-doped BSCCO. The markers label the values of the gap and scattering rate at selected temperatures which are used in plots throughout this section.}
	\label{fig:gbtemp}
\end{figure*}

\begin{figure*}
	\centering
	
	\includegraphics[height=0.18\textwidth]{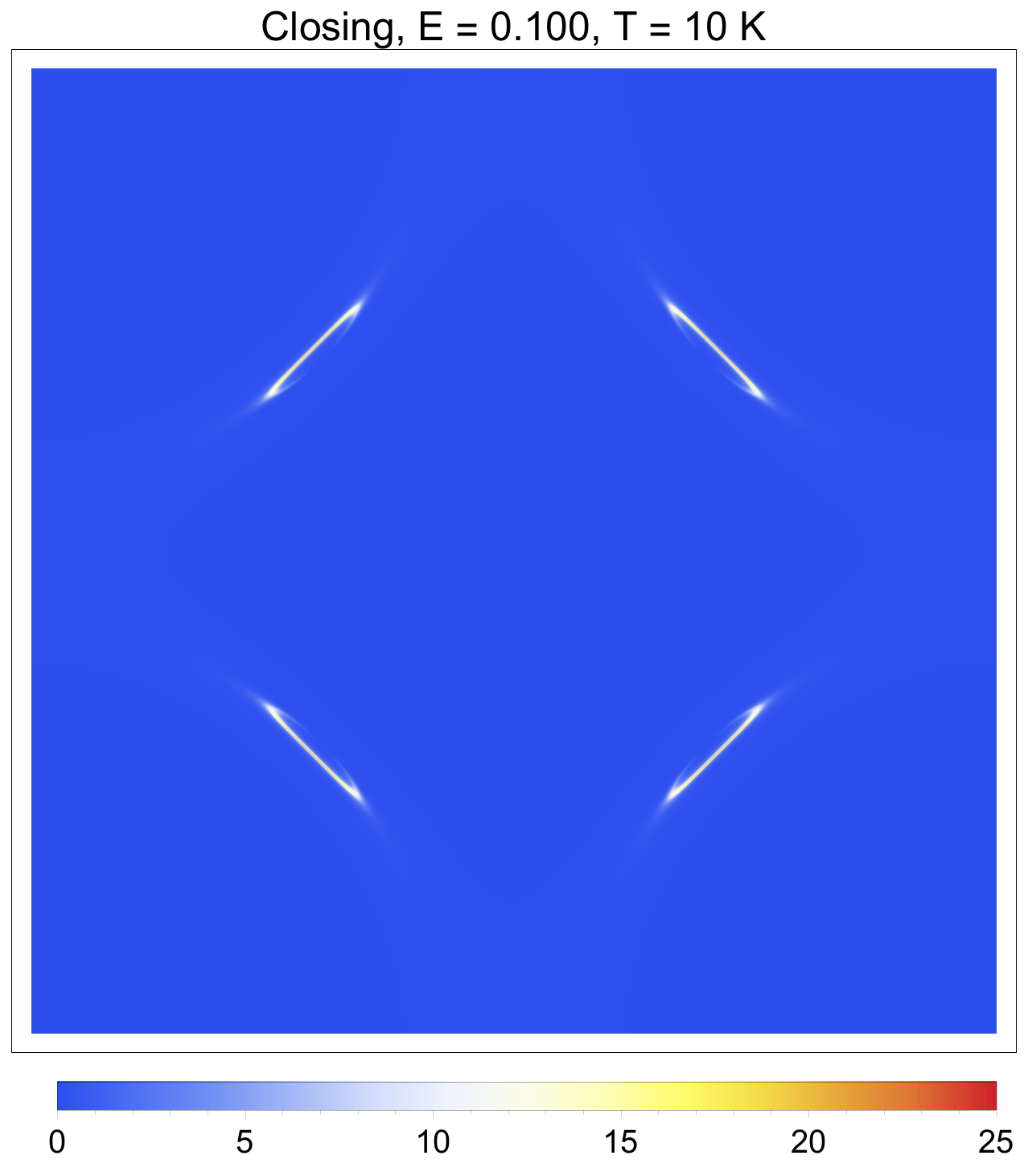}
	\includegraphics[height=0.18\textwidth]{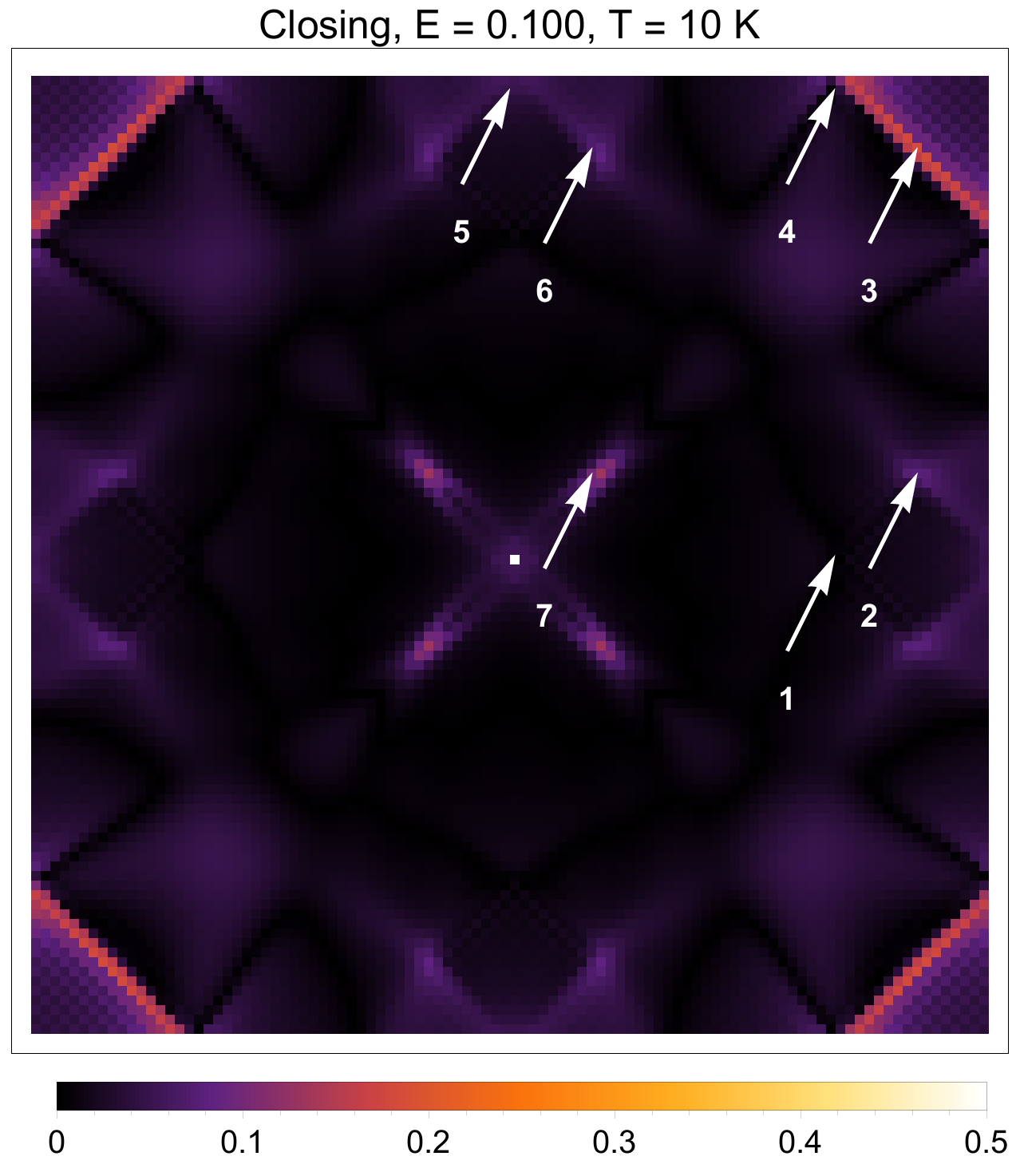}
	\includegraphics[height=0.18\textwidth]{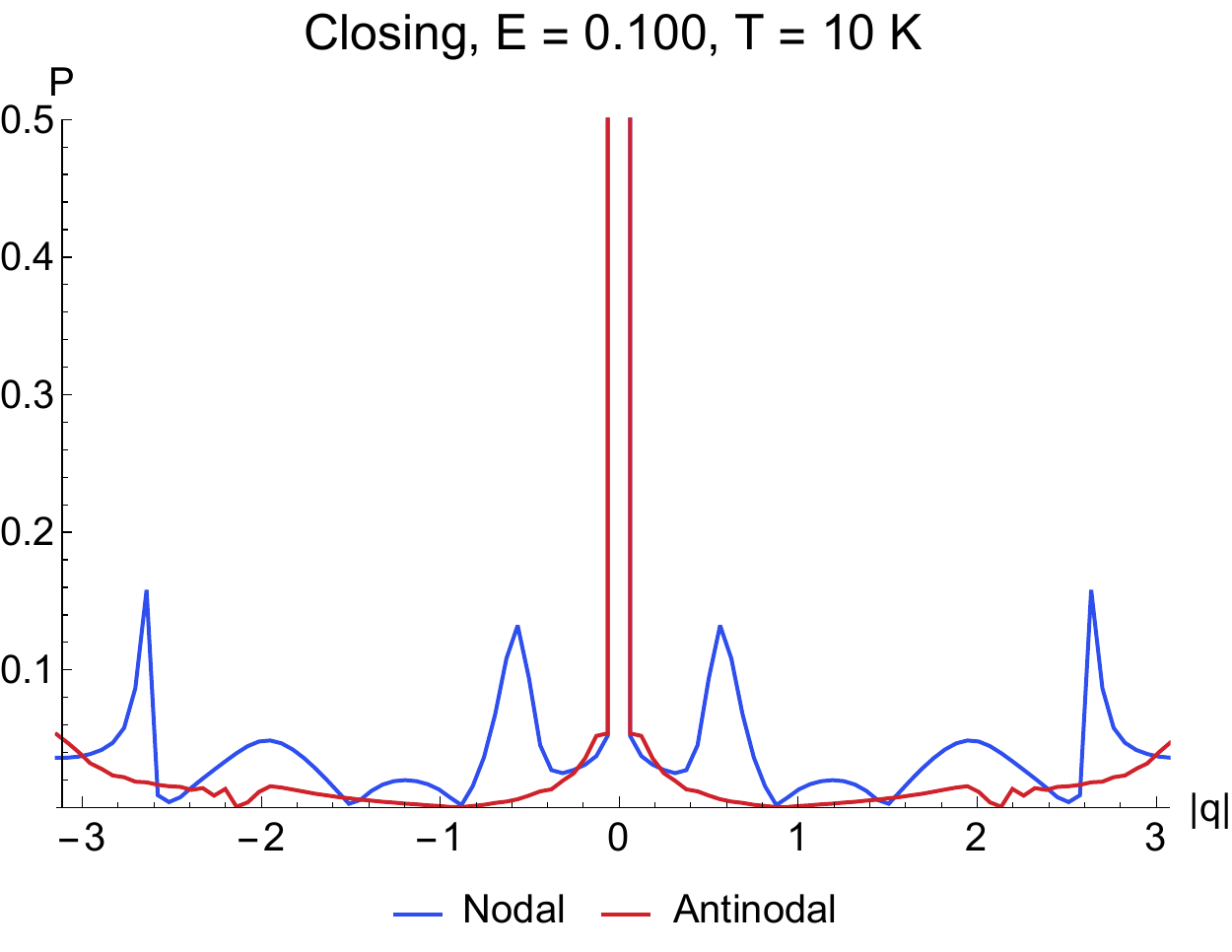} 
	\includegraphics[height=0.18\textwidth]{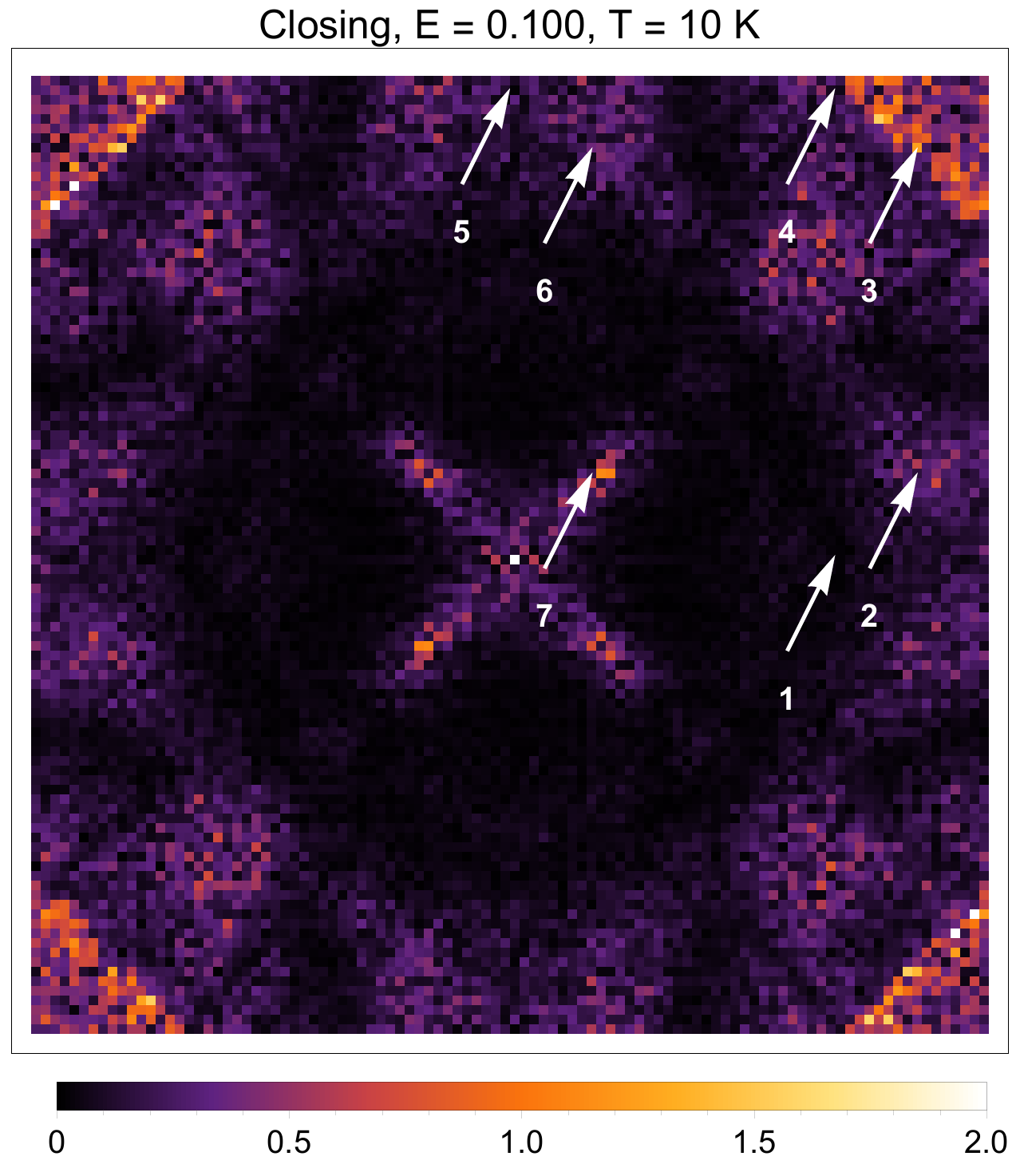}
	\includegraphics[height=0.18\textwidth]{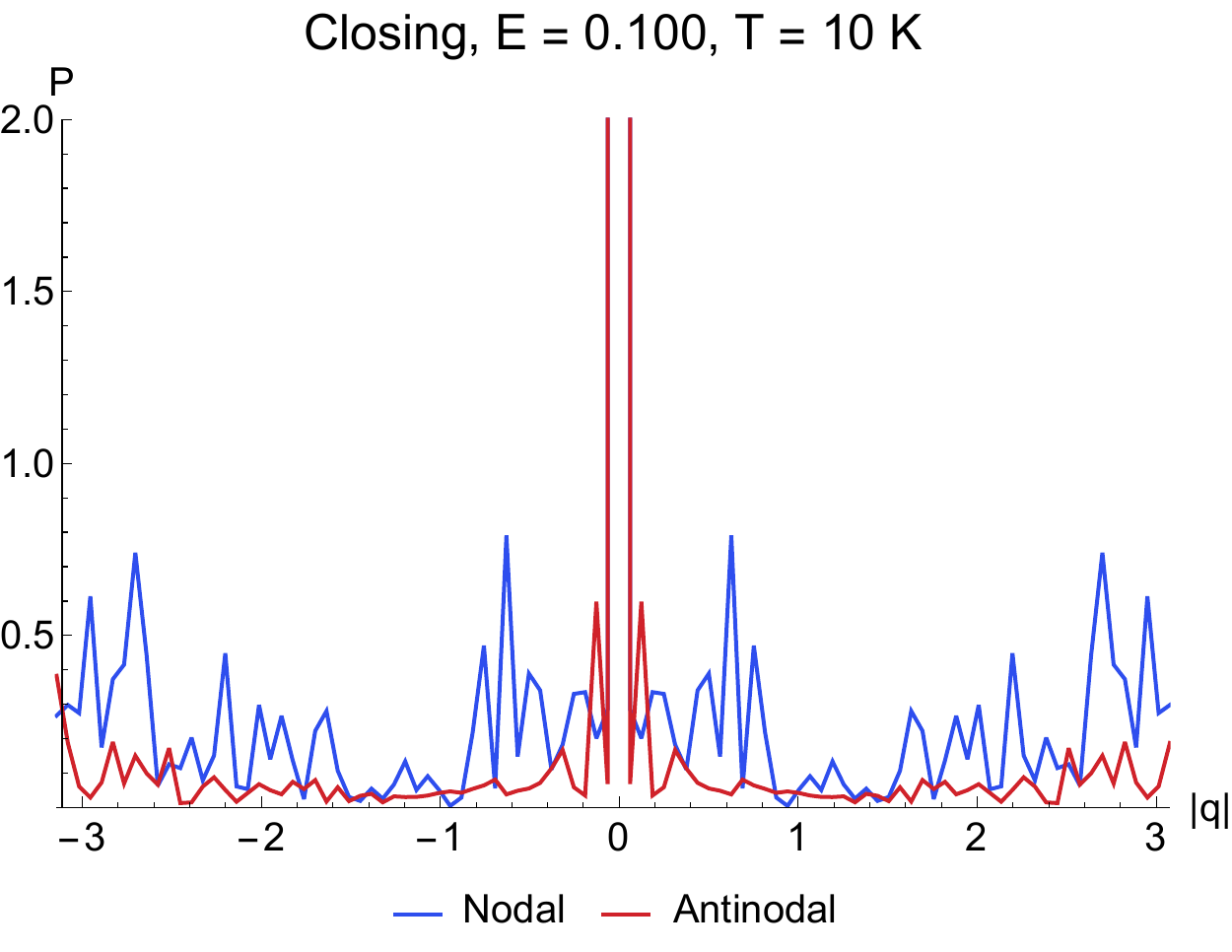} \\
	\includegraphics[height=0.18\textwidth]{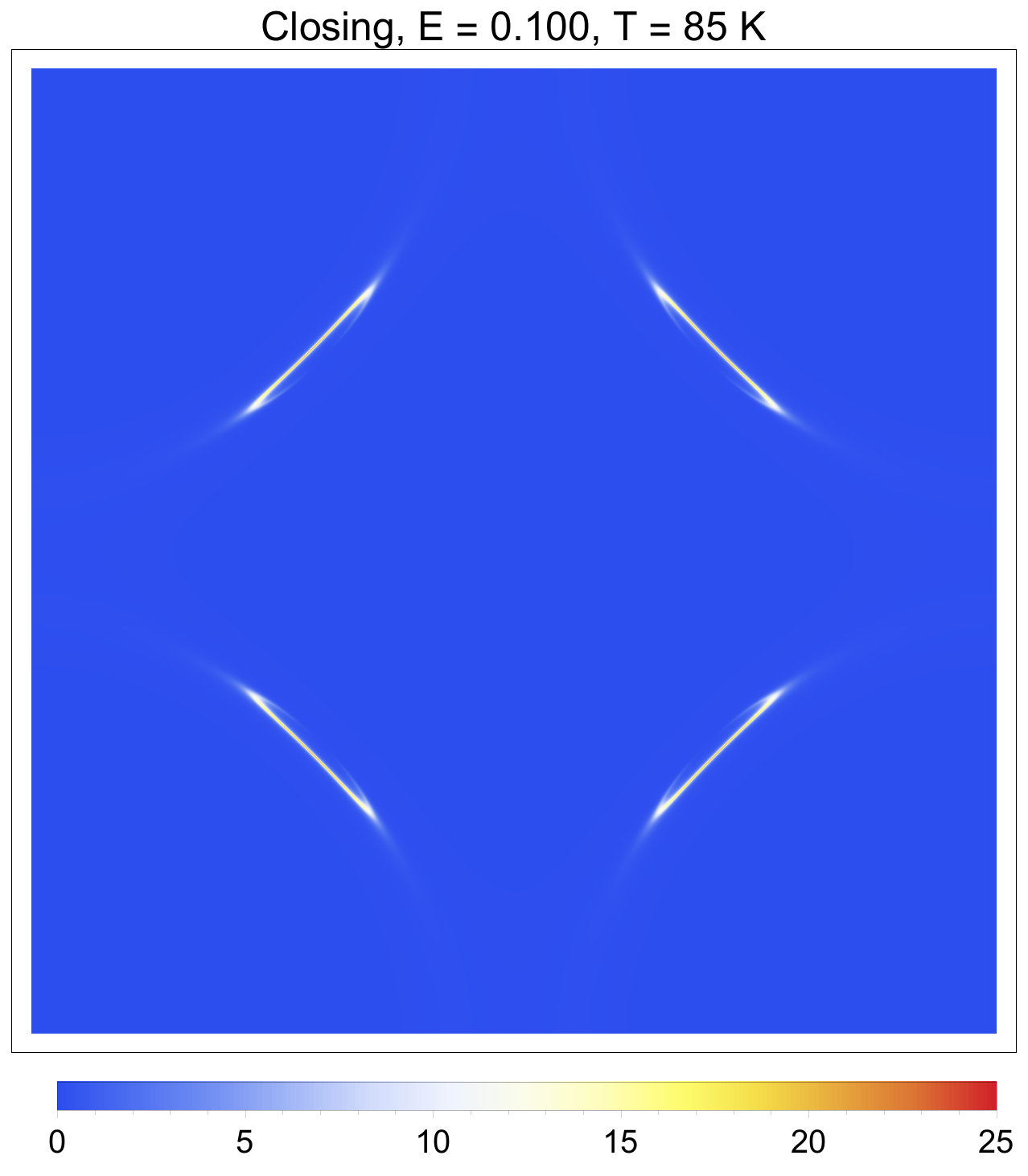}
	\includegraphics[height=0.18\textwidth]{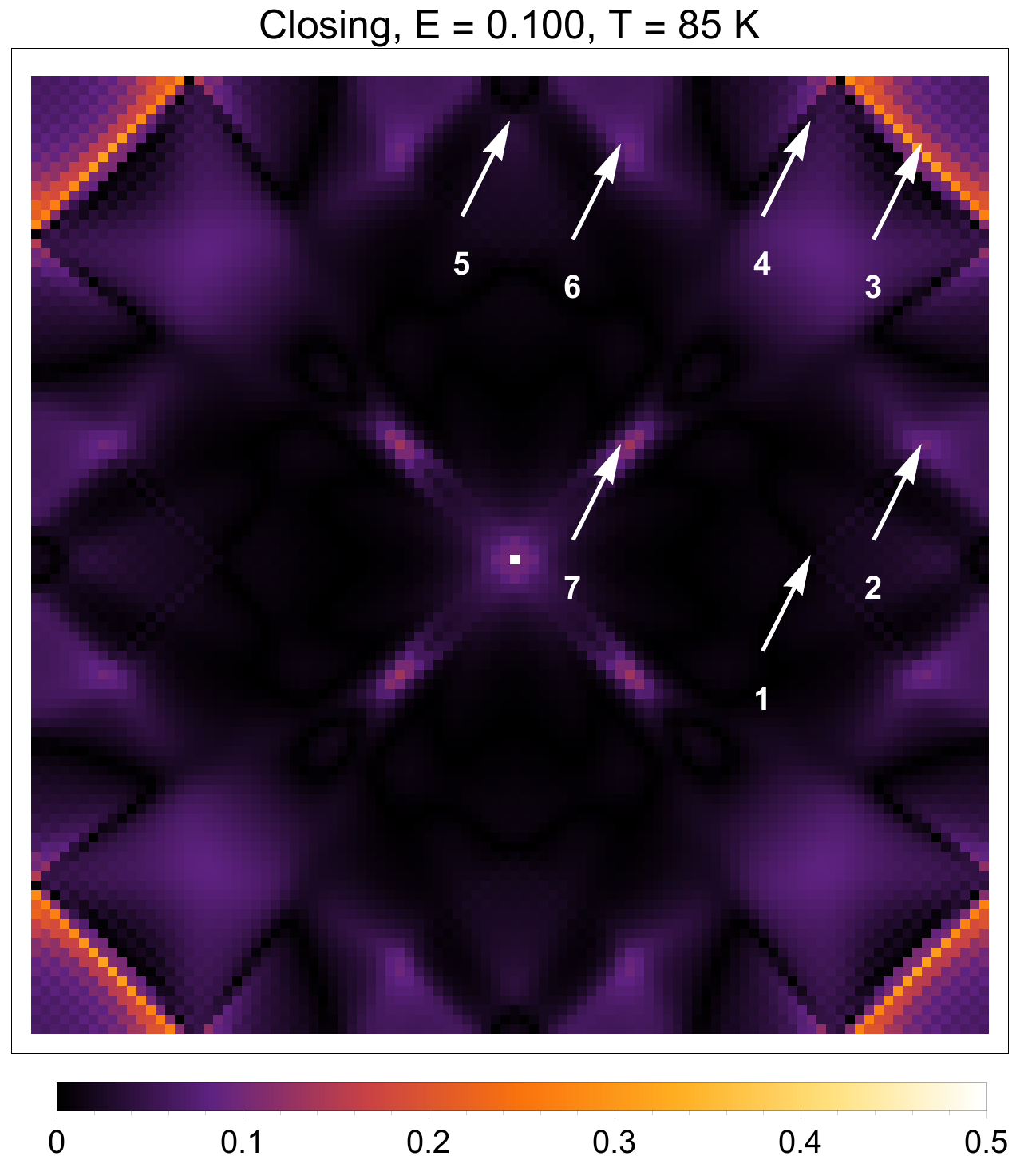}
	\includegraphics[height=0.18\textwidth]{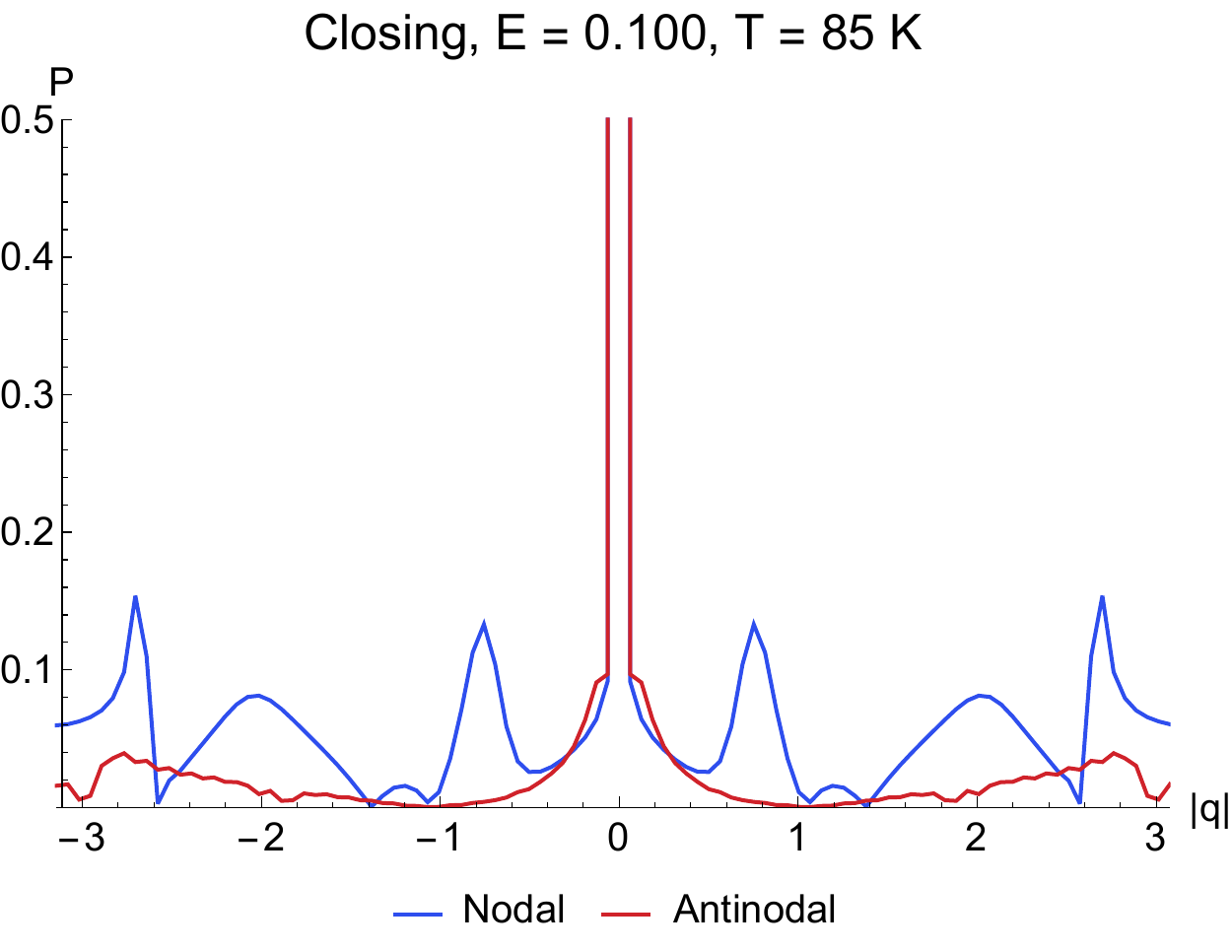}
	\includegraphics[height=0.18\textwidth]{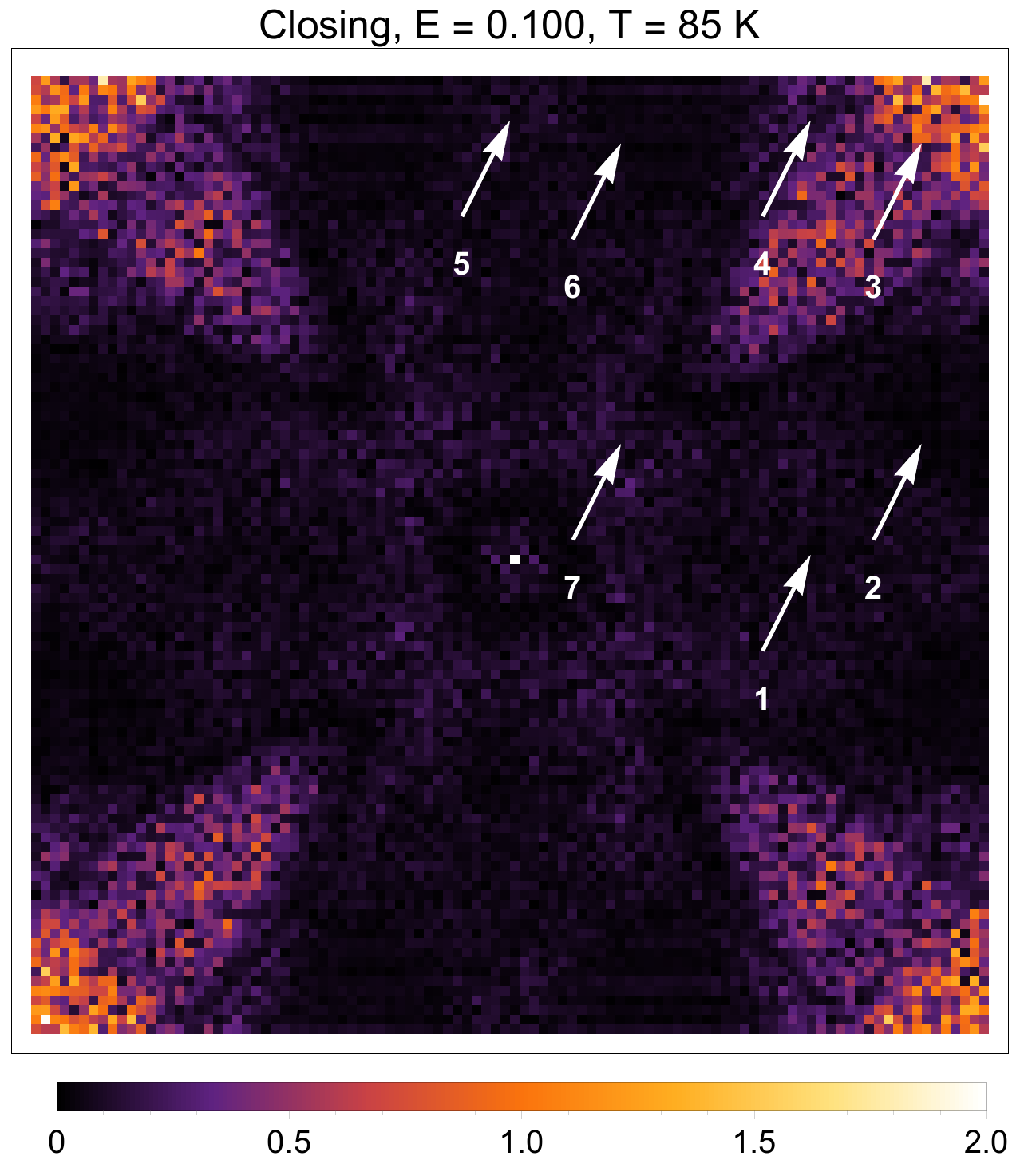}
	\includegraphics[height=0.18\textwidth]{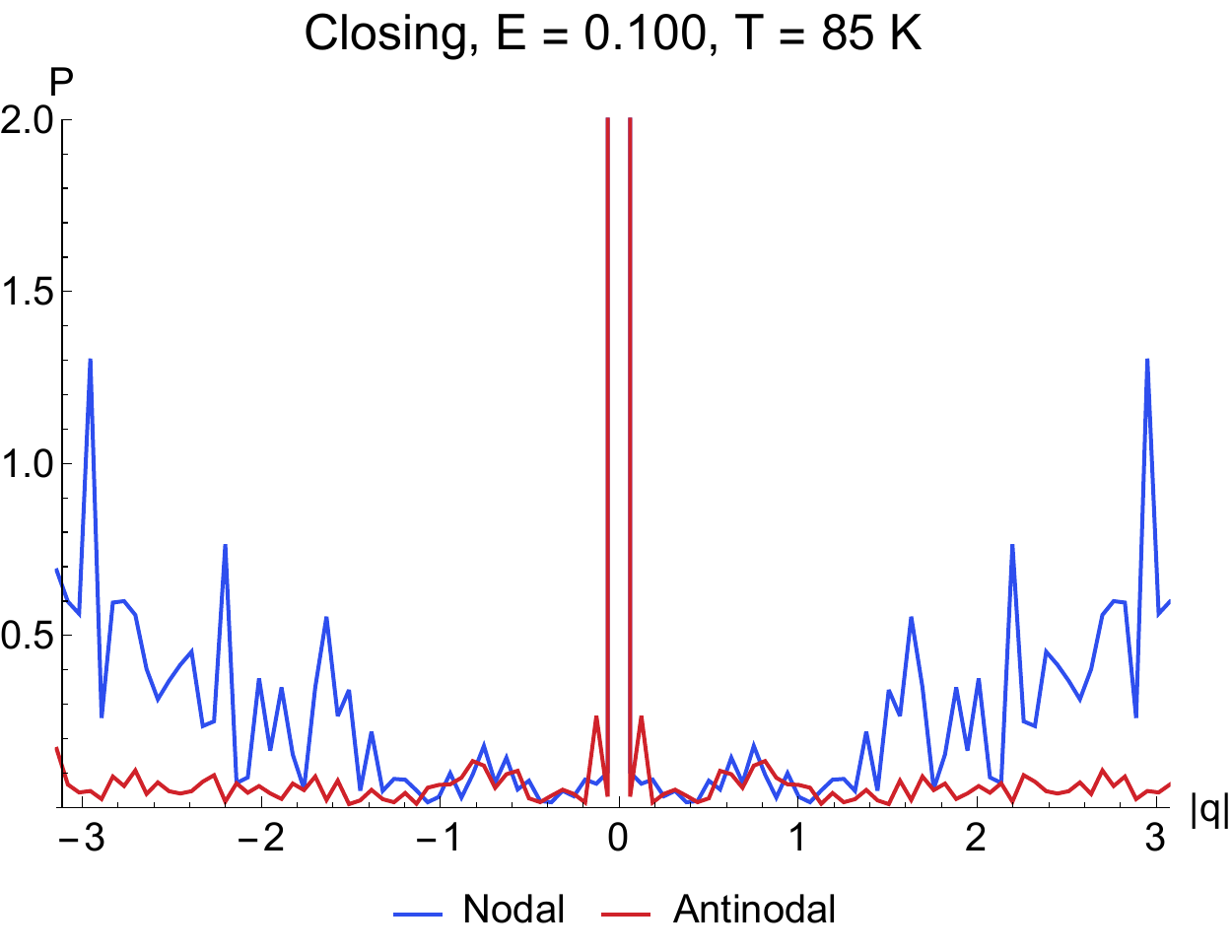} \\
	\includegraphics[height=0.18\textwidth]{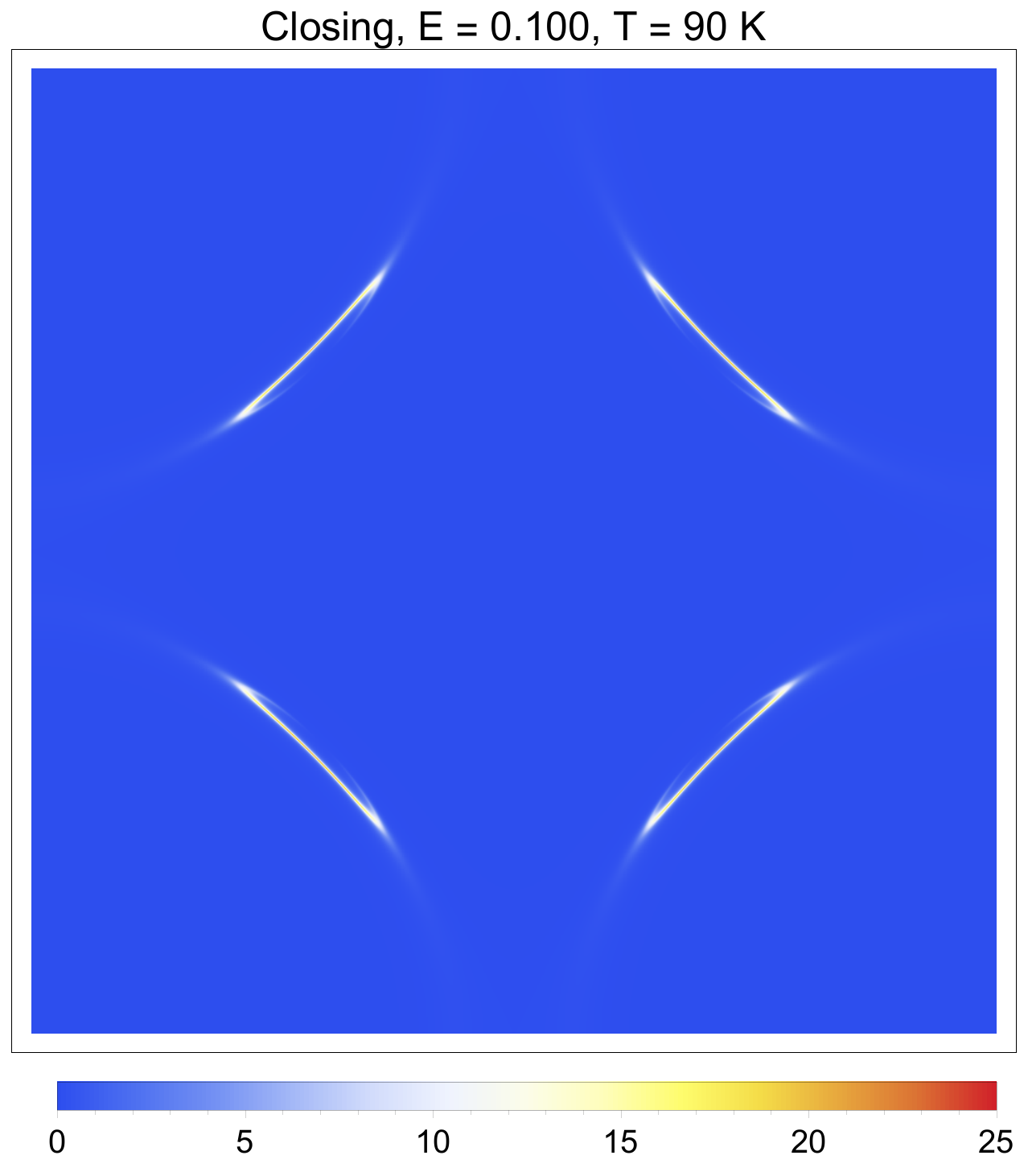}
	\includegraphics[height=0.18\textwidth]{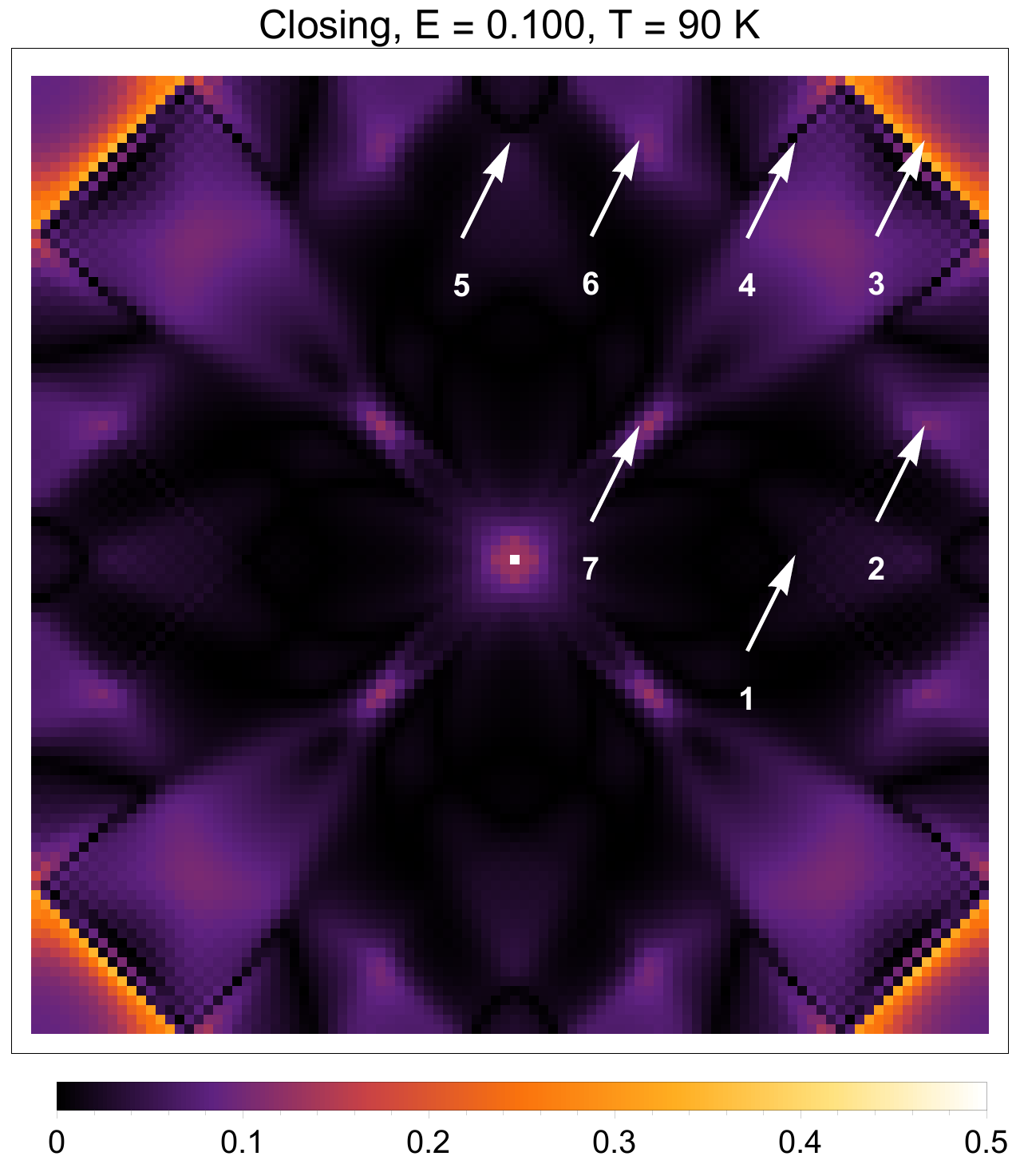}
	\includegraphics[height=0.18\textwidth]{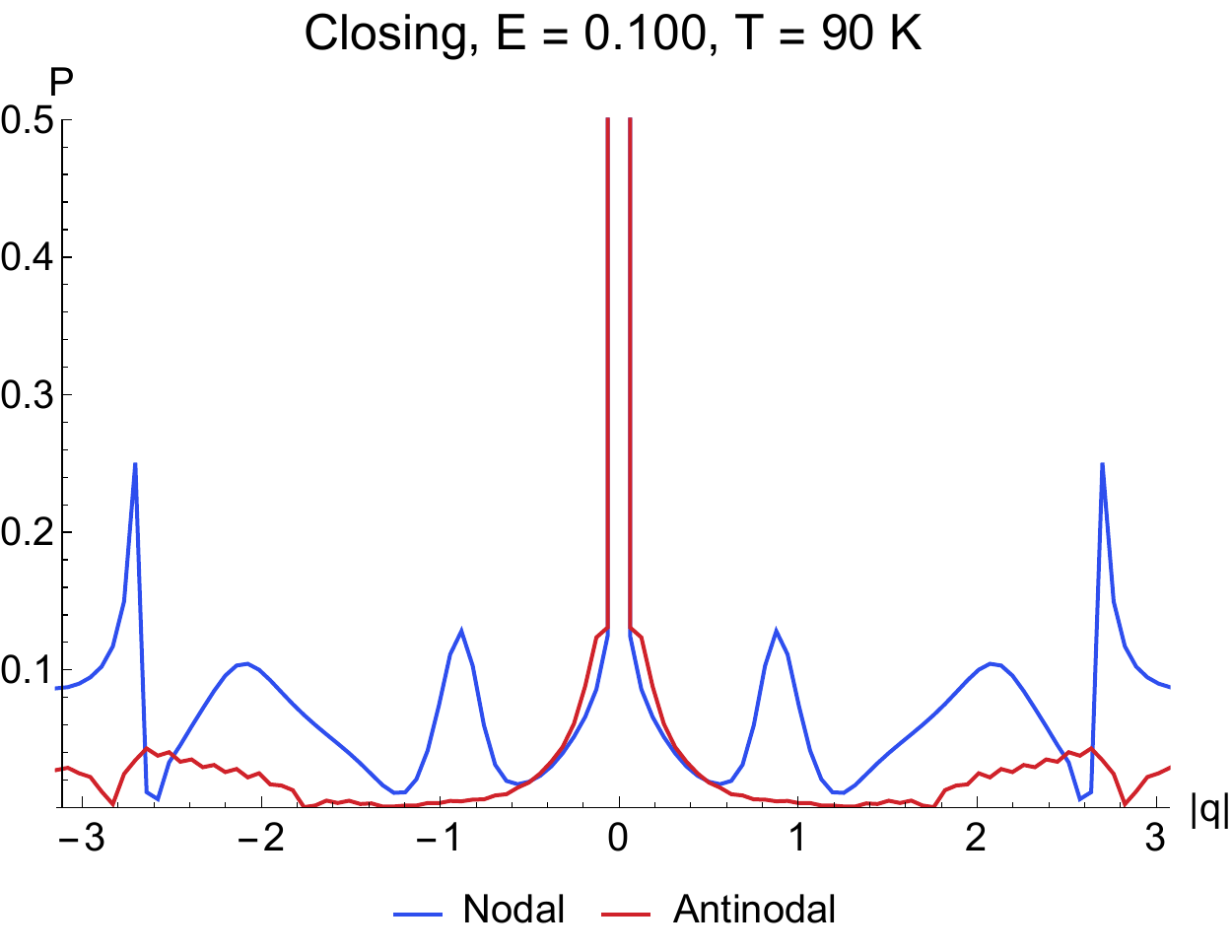}
	\includegraphics[height=0.18\textwidth]{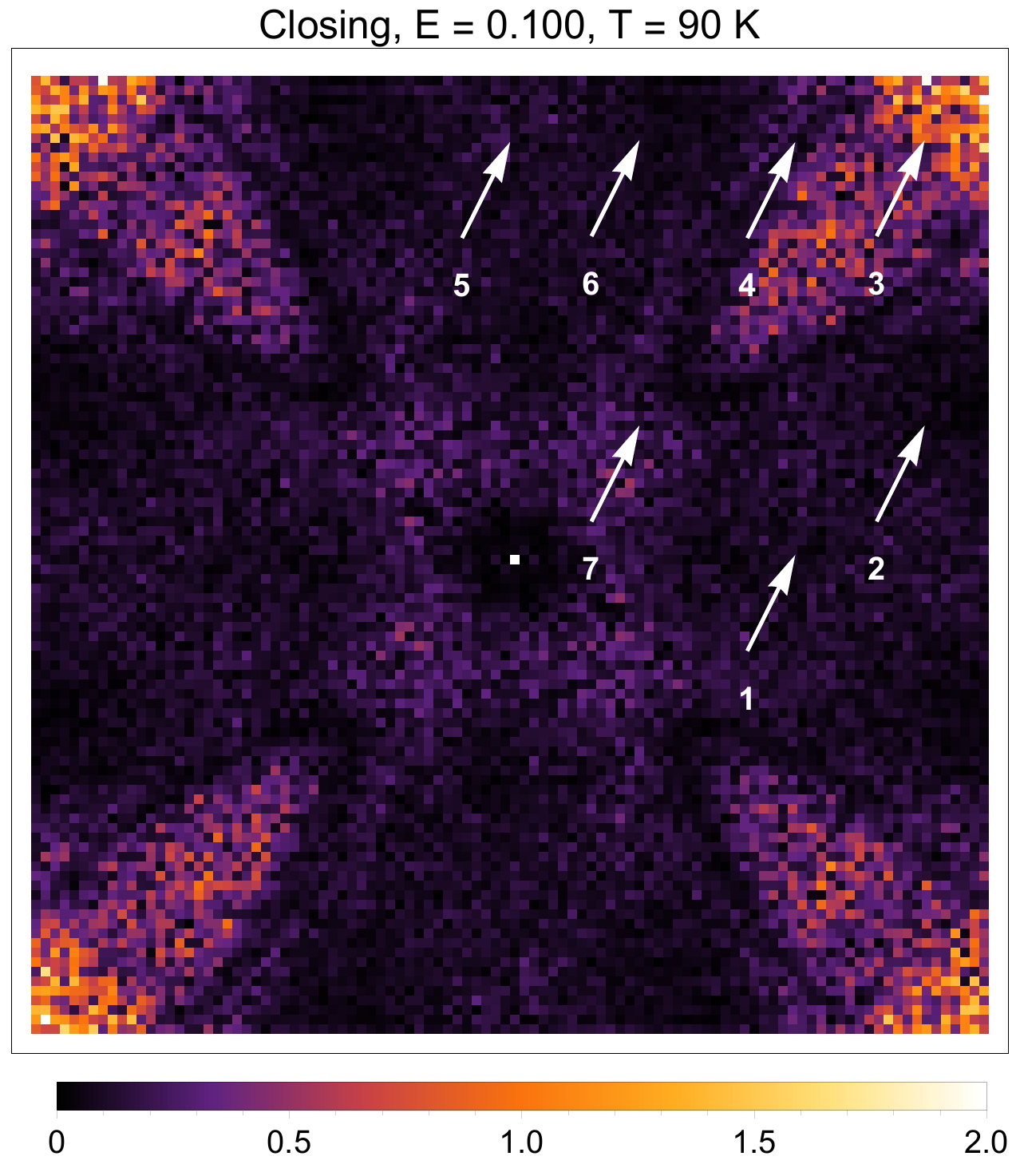}
	\includegraphics[height=0.18\textwidth]{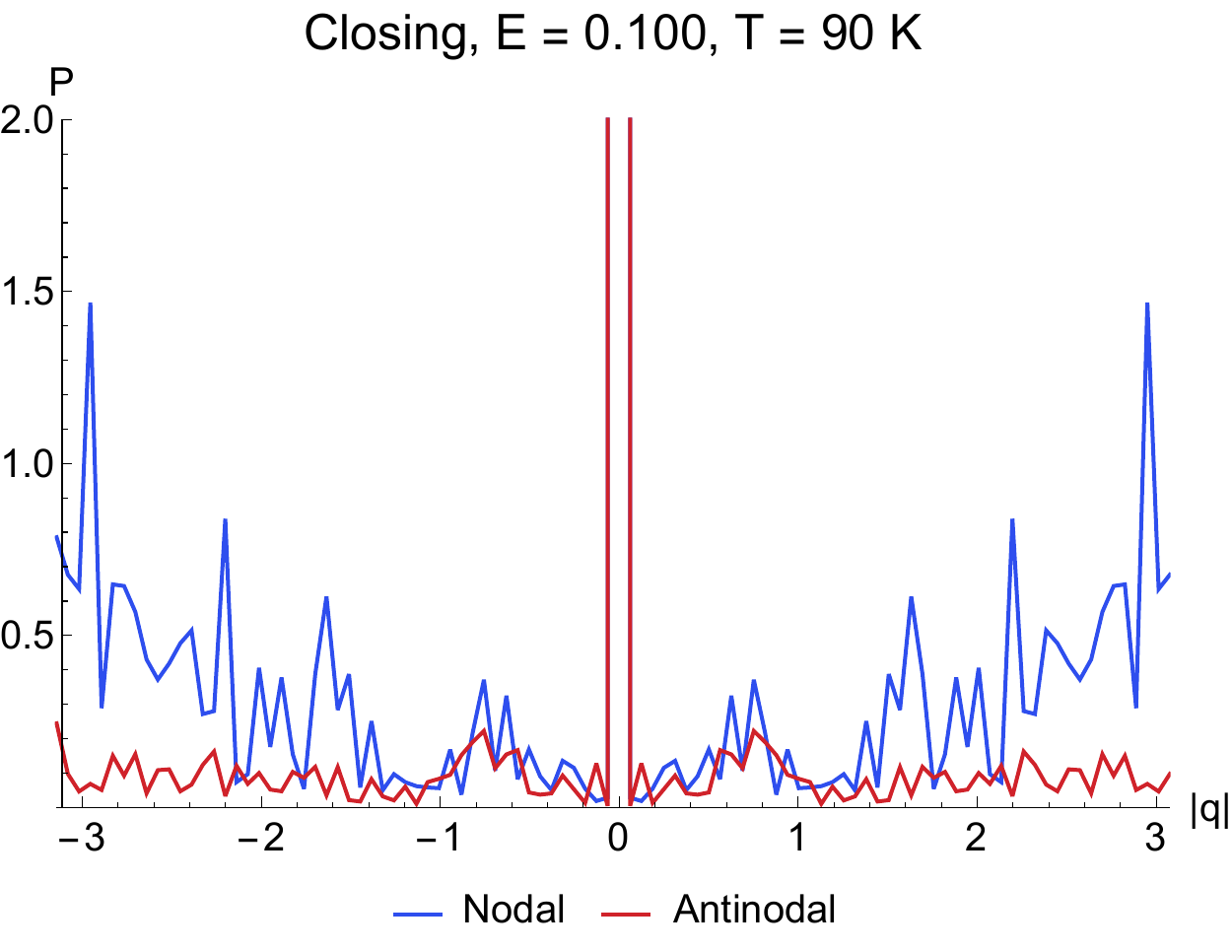} \\
	\includegraphics[height=0.18\textwidth]{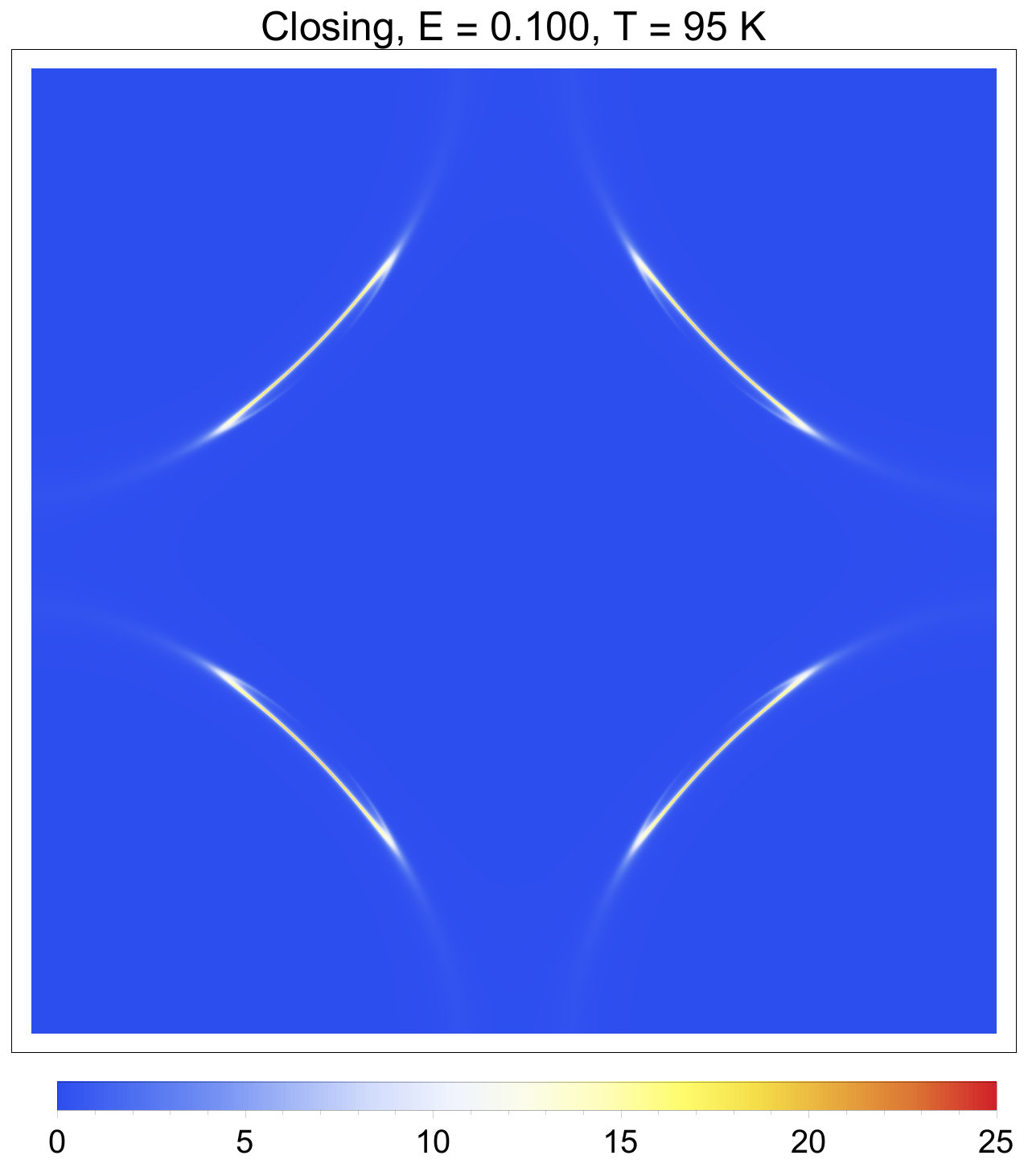}
	\includegraphics[height=0.18\textwidth]{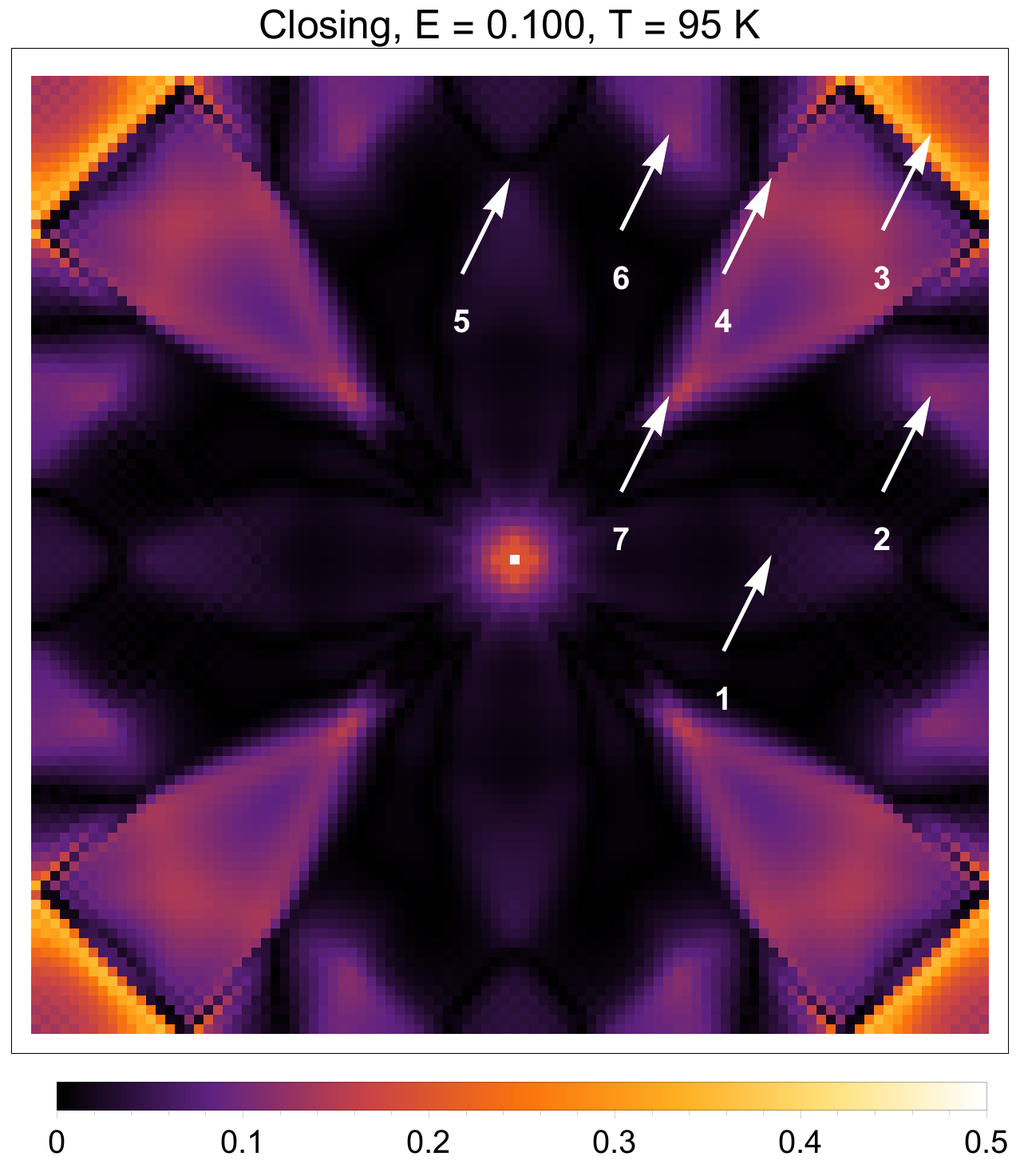}
	\includegraphics[height=0.18\textwidth]{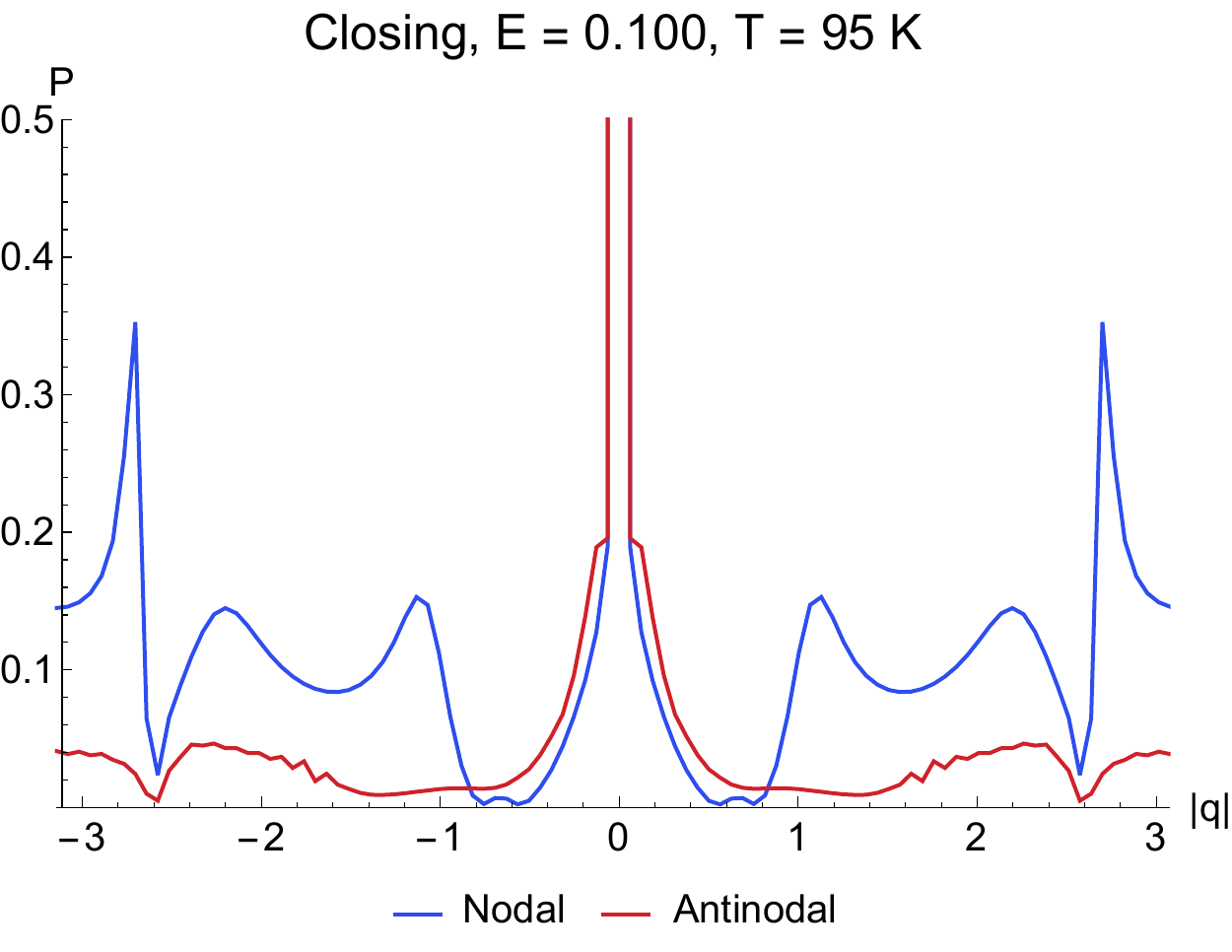}
	\includegraphics[height=0.18\textwidth]{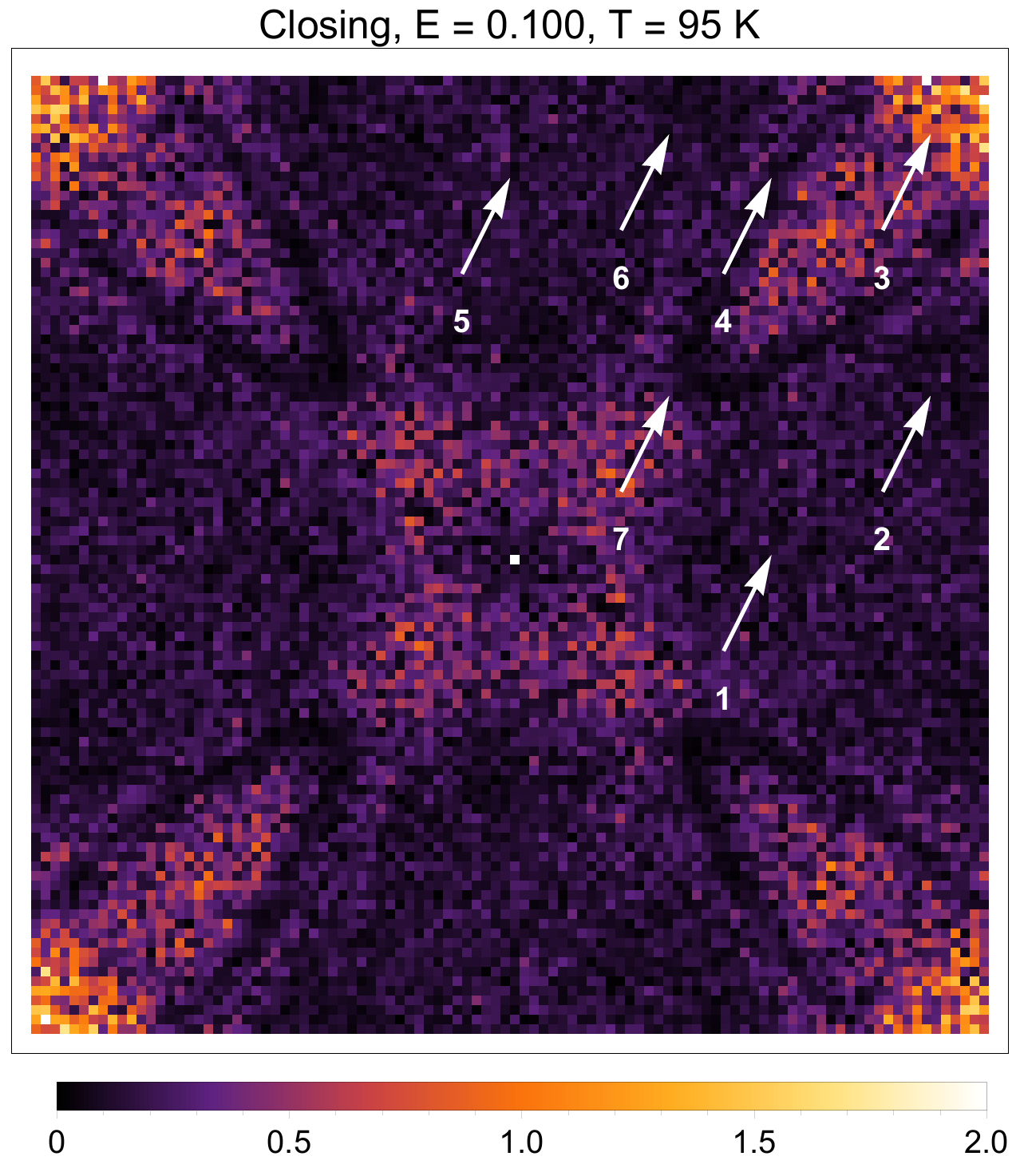}
	\includegraphics[height=0.18\textwidth]{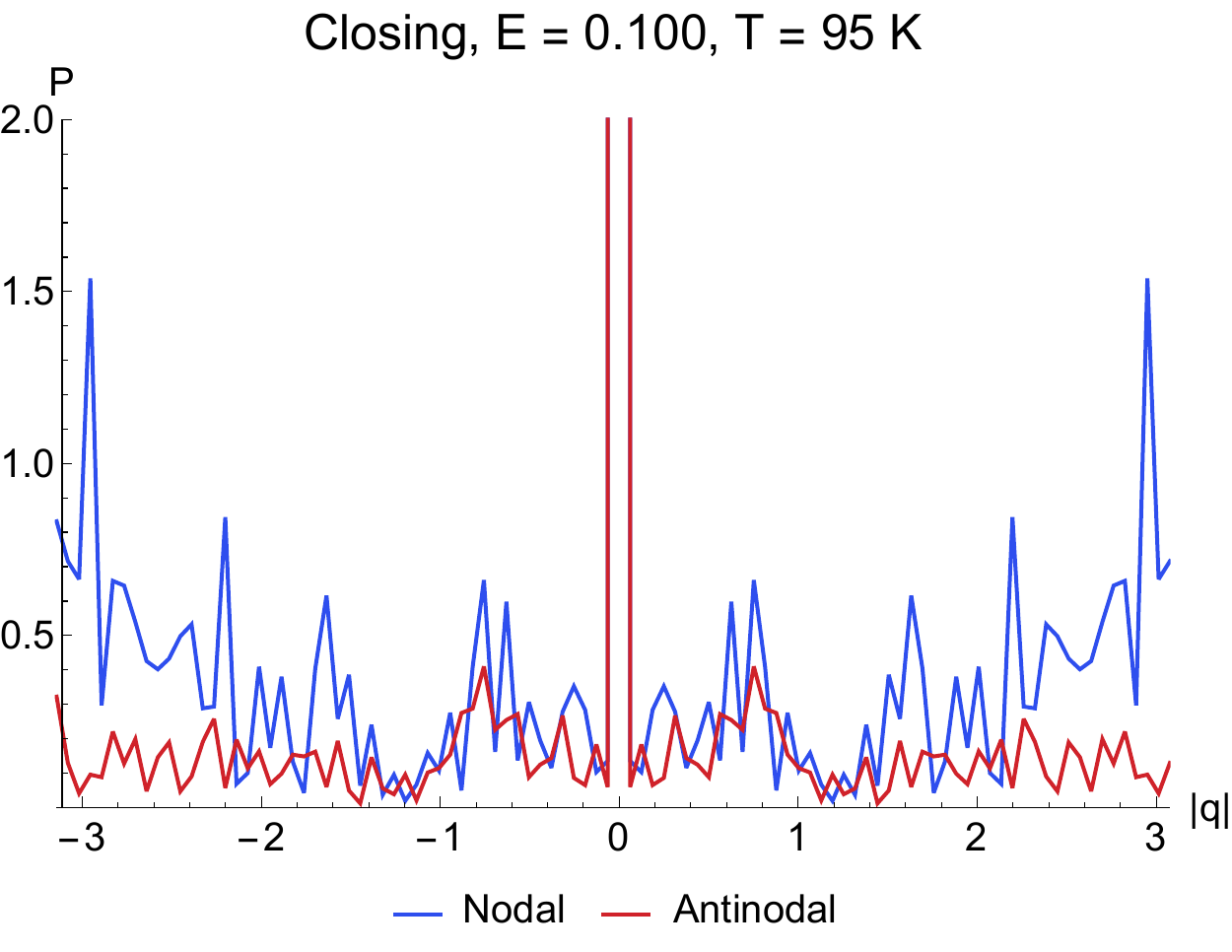} \\
	\includegraphics[height=0.18\textwidth]{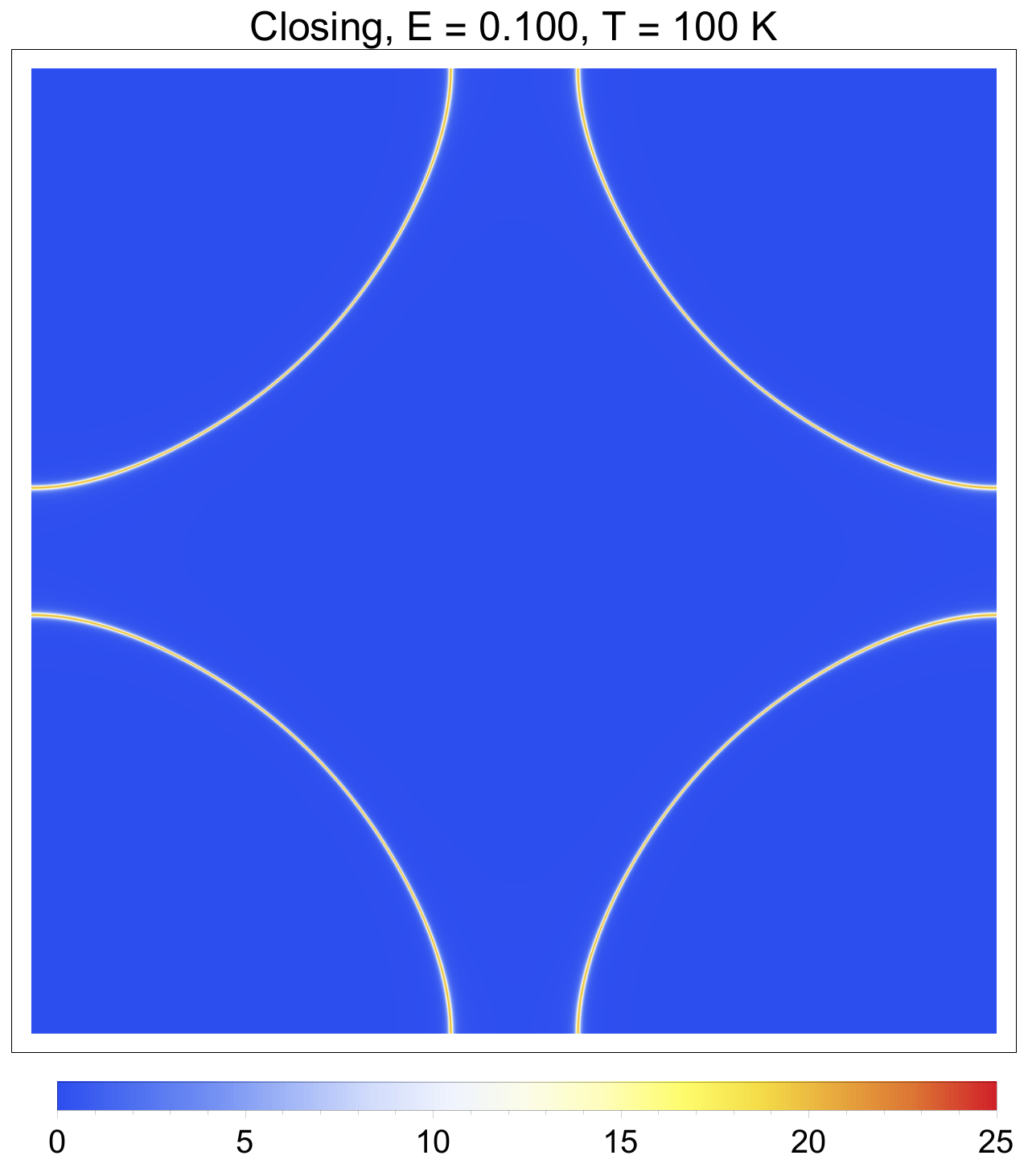}
	\includegraphics[height=0.18\textwidth]{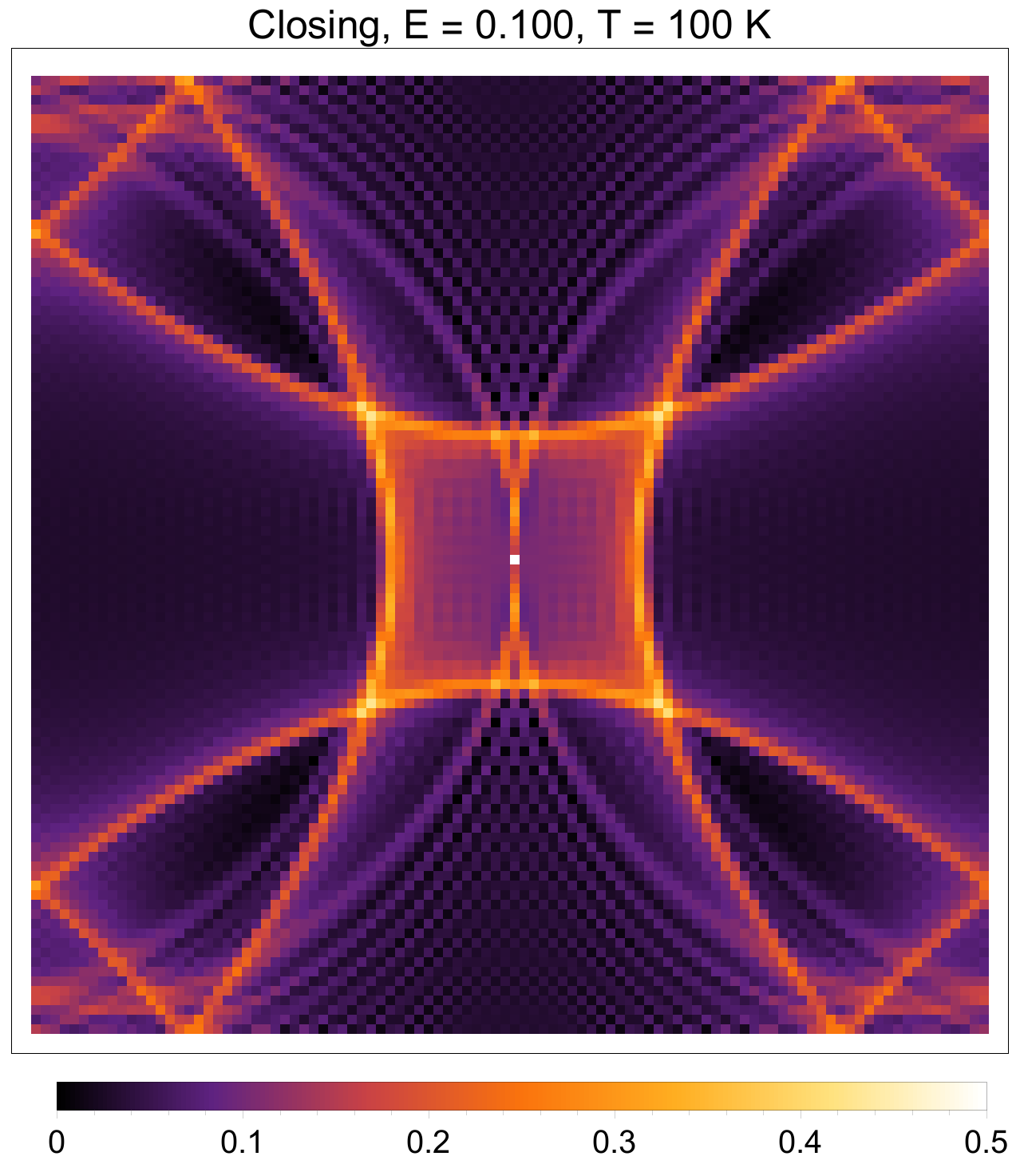}
	\includegraphics[height=0.18\textwidth]{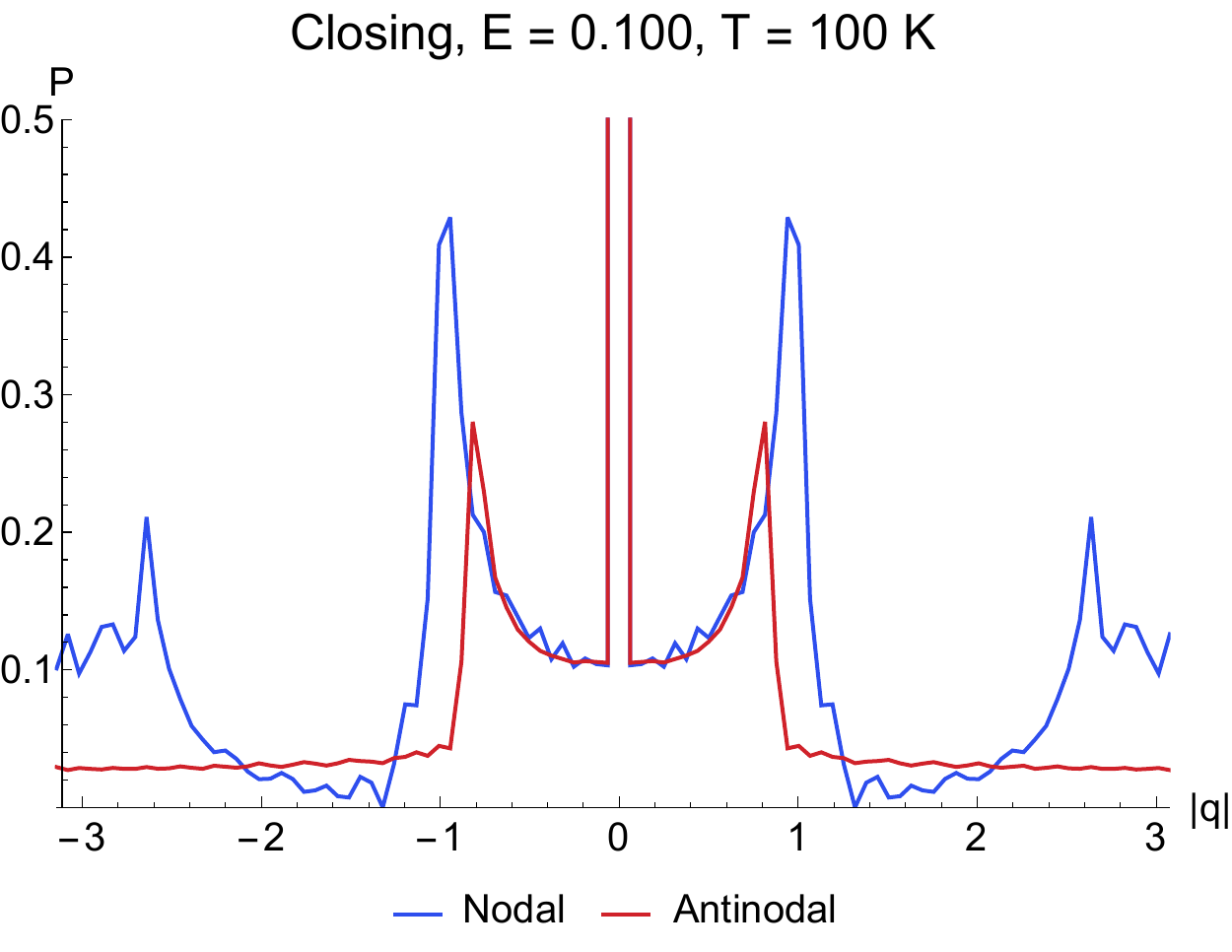}
	\includegraphics[height=0.18\textwidth]{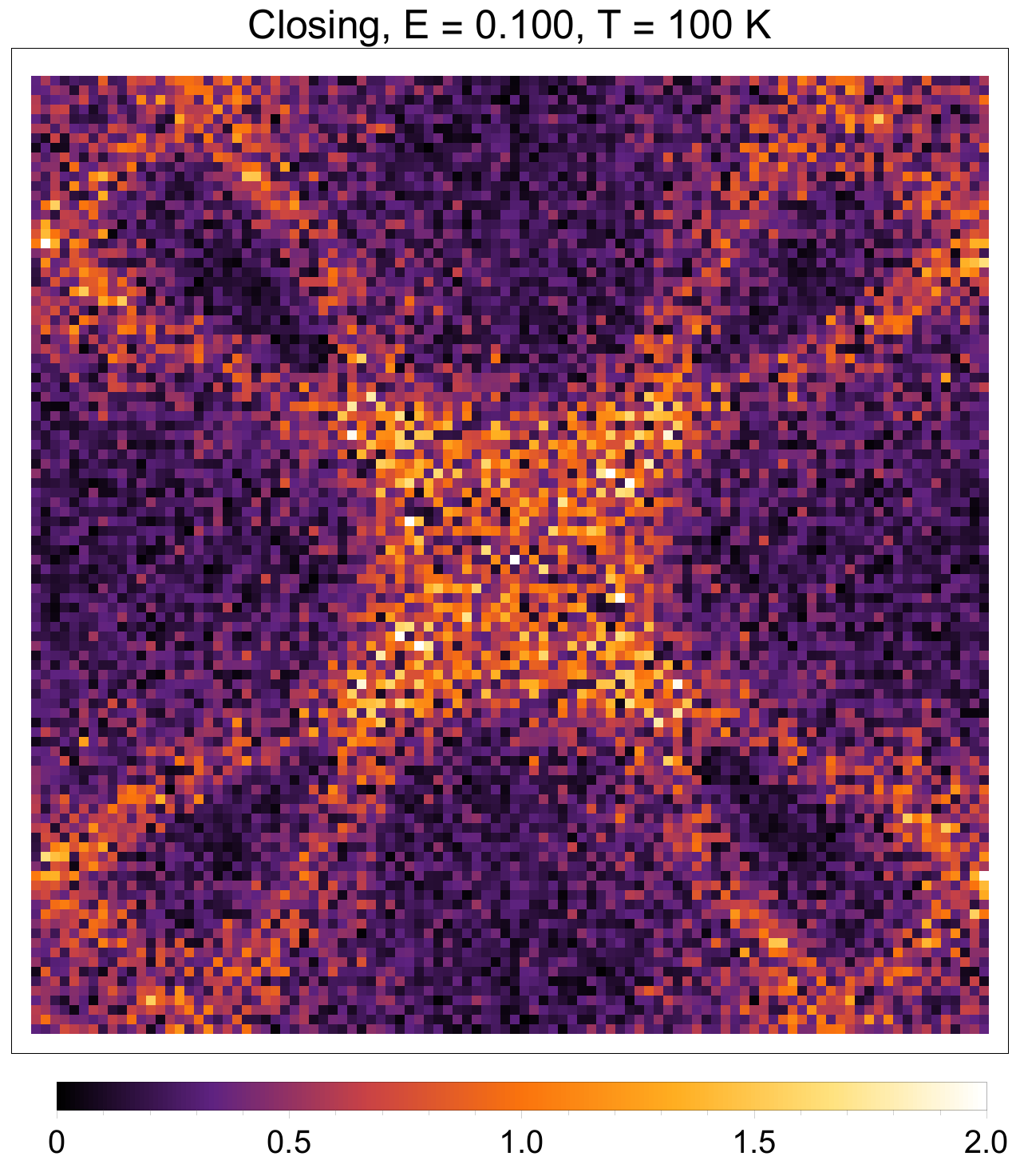}
	\includegraphics[height=0.18\textwidth]{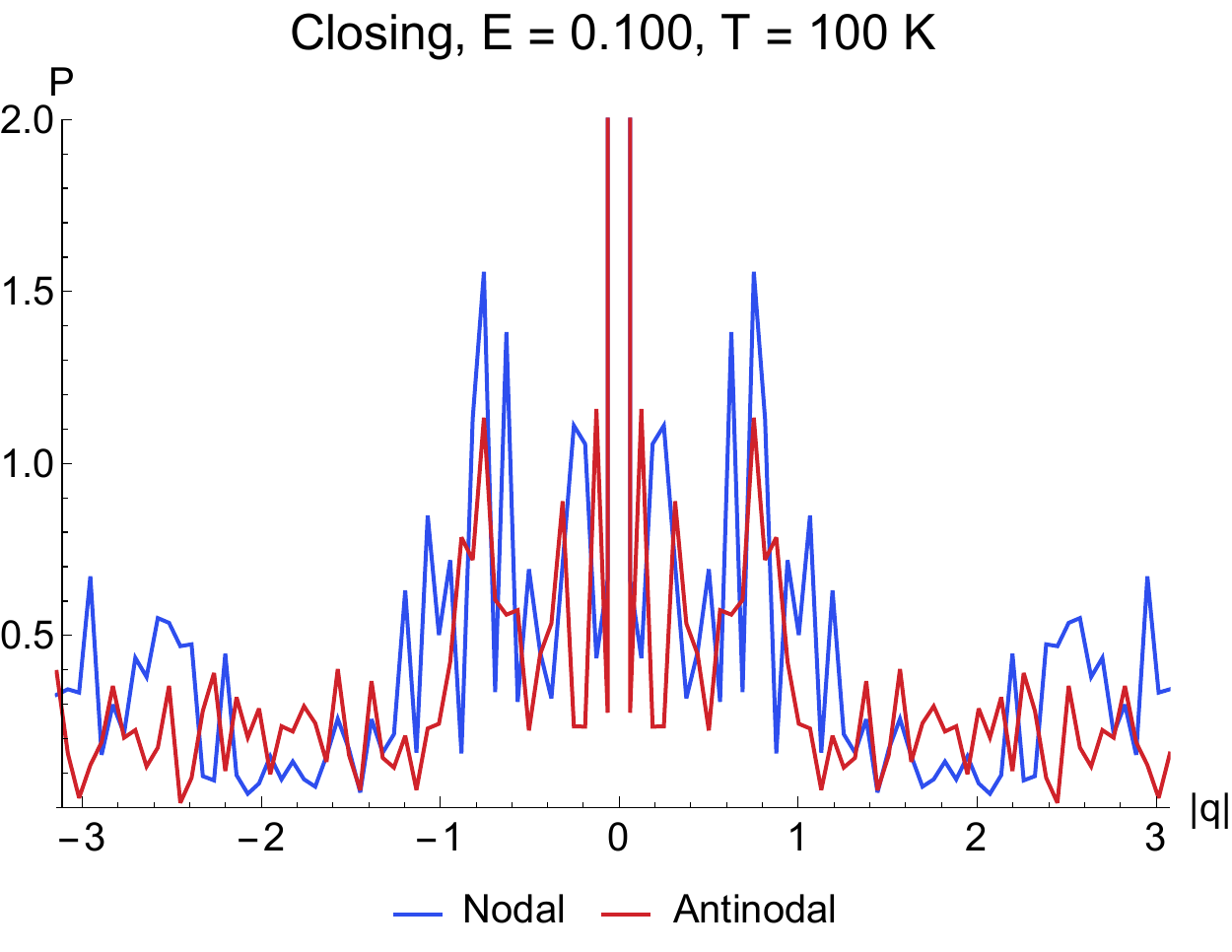} \\
	
	\caption{Gap-closing phenomenology at various temperatures. $T_c$ here is 100 K. Left to right: The spectral function $A(\mathbf{k}, \omega)$; the Fourier transform of the LDOS $P(\mathbf{q}, \omega)$; linecuts of $P(\mathbf{q}, \omega)$ in the nodal and antinodal directions; $P(\mathbf{q}, \omega)$ in the presence of multiple weak impurities and finite-temperature smearing; and linecuts of $P(\mathbf{q}, \omega)$ in the presence of multiple weak impurities and finite-temperature smearing. Arrows indicate the locations of the peaks predicted by the octet model. All plots are taken at $E = 0.100$.}
	\label{fig:temperature_bcs}
\end{figure*}

\begin{figure*}
	\centering
	
	\includegraphics[height=0.18\textwidth]{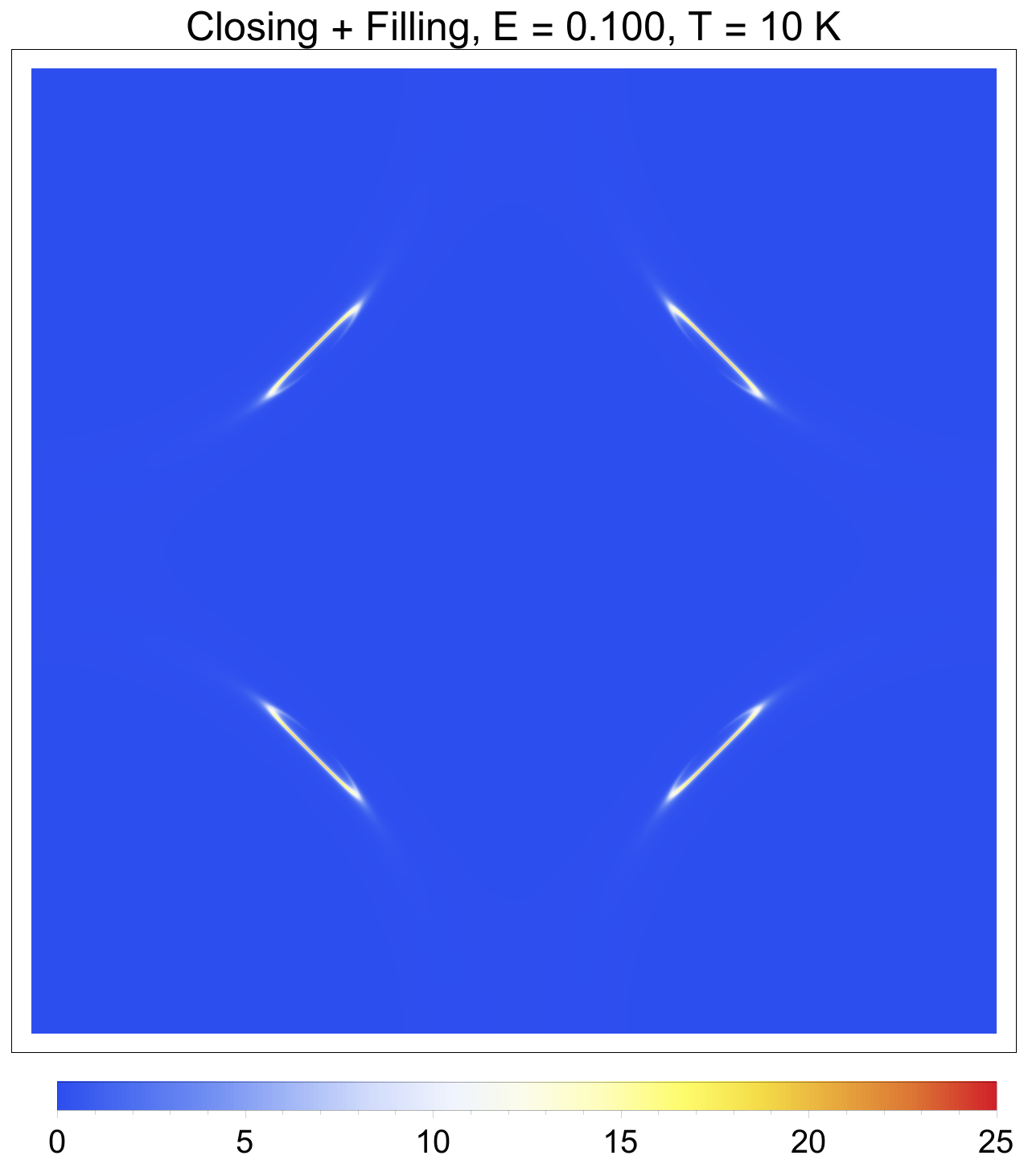}
	\includegraphics[height=0.18\textwidth]{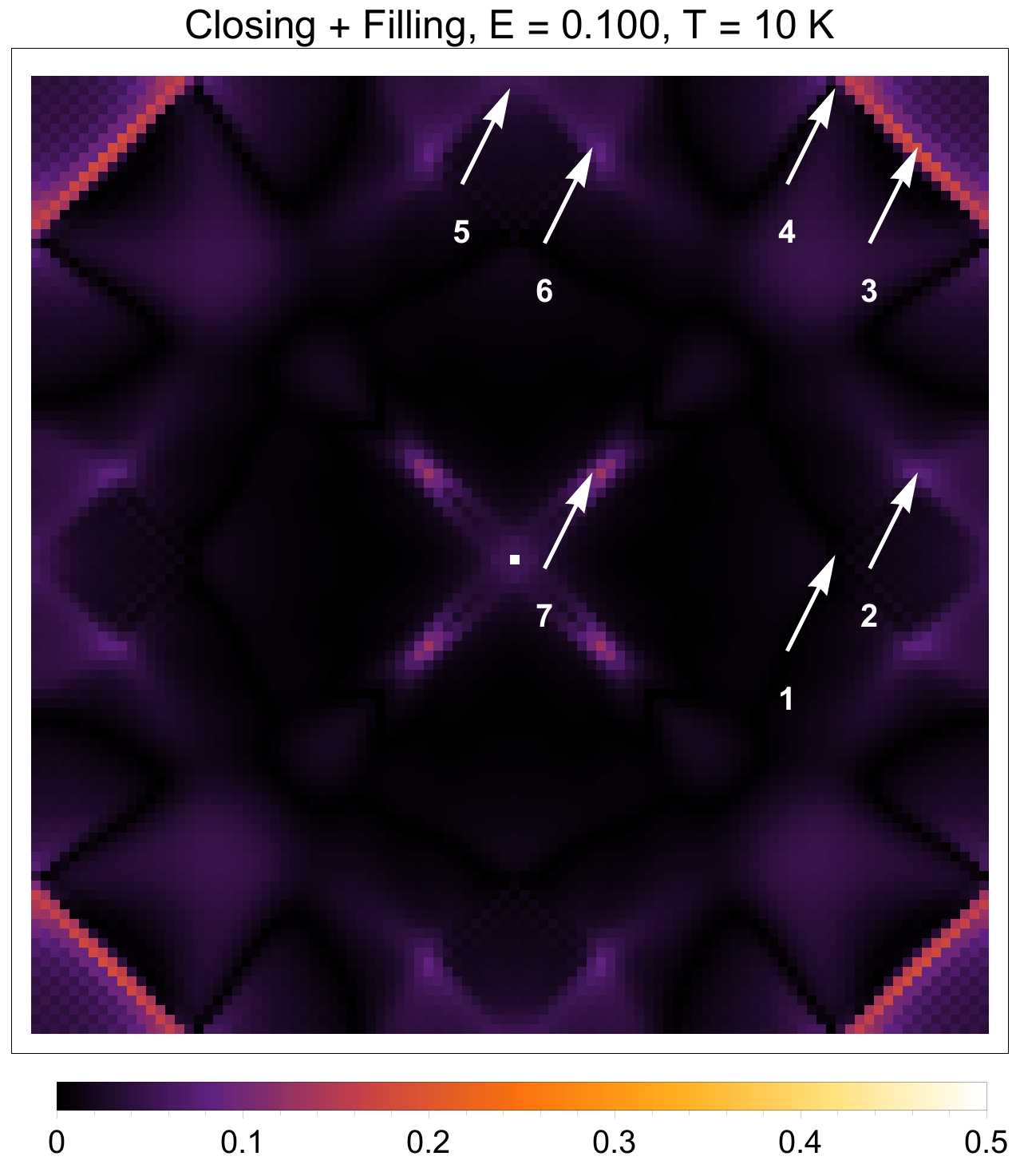}
	\includegraphics[height=0.18\textwidth]{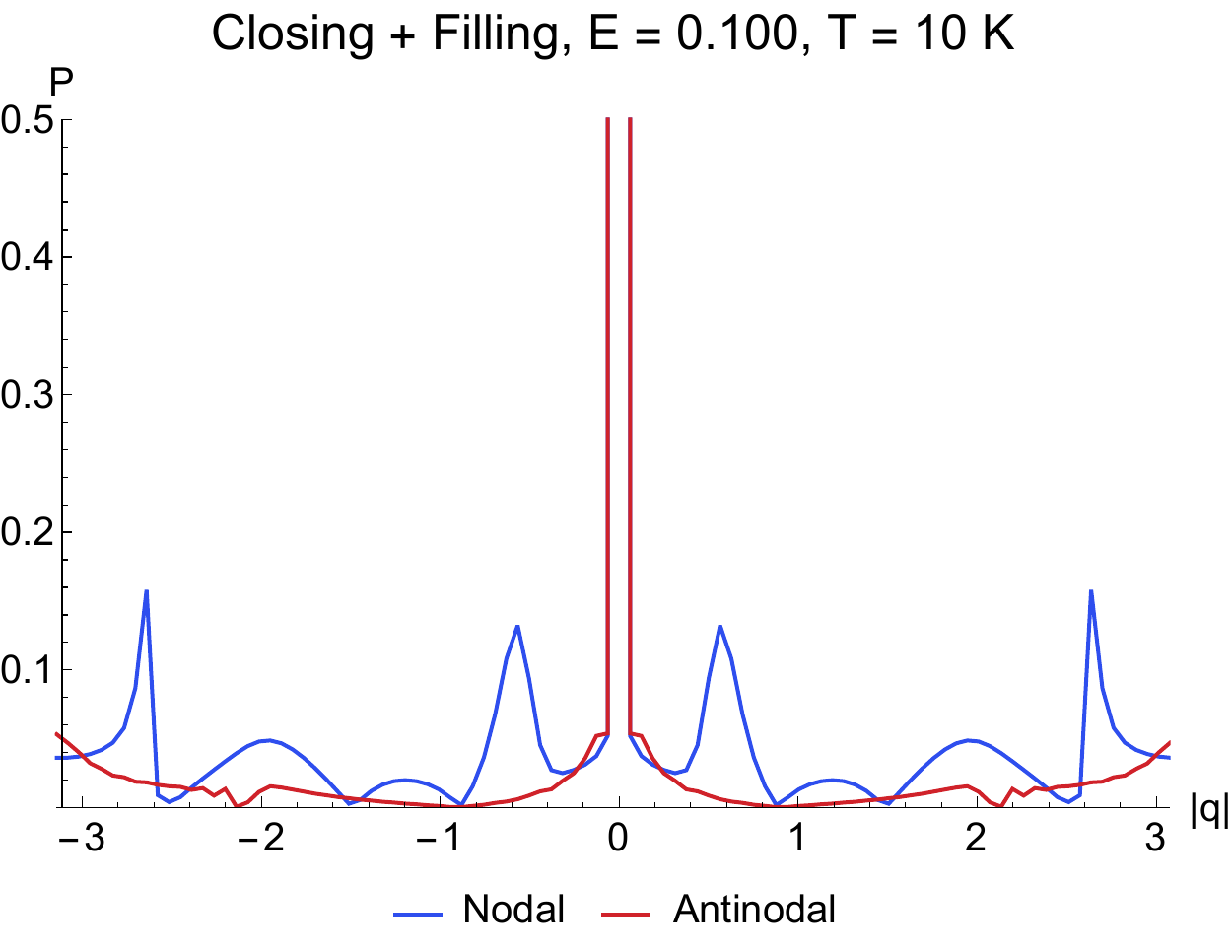} 
	\includegraphics[height=0.18\textwidth]{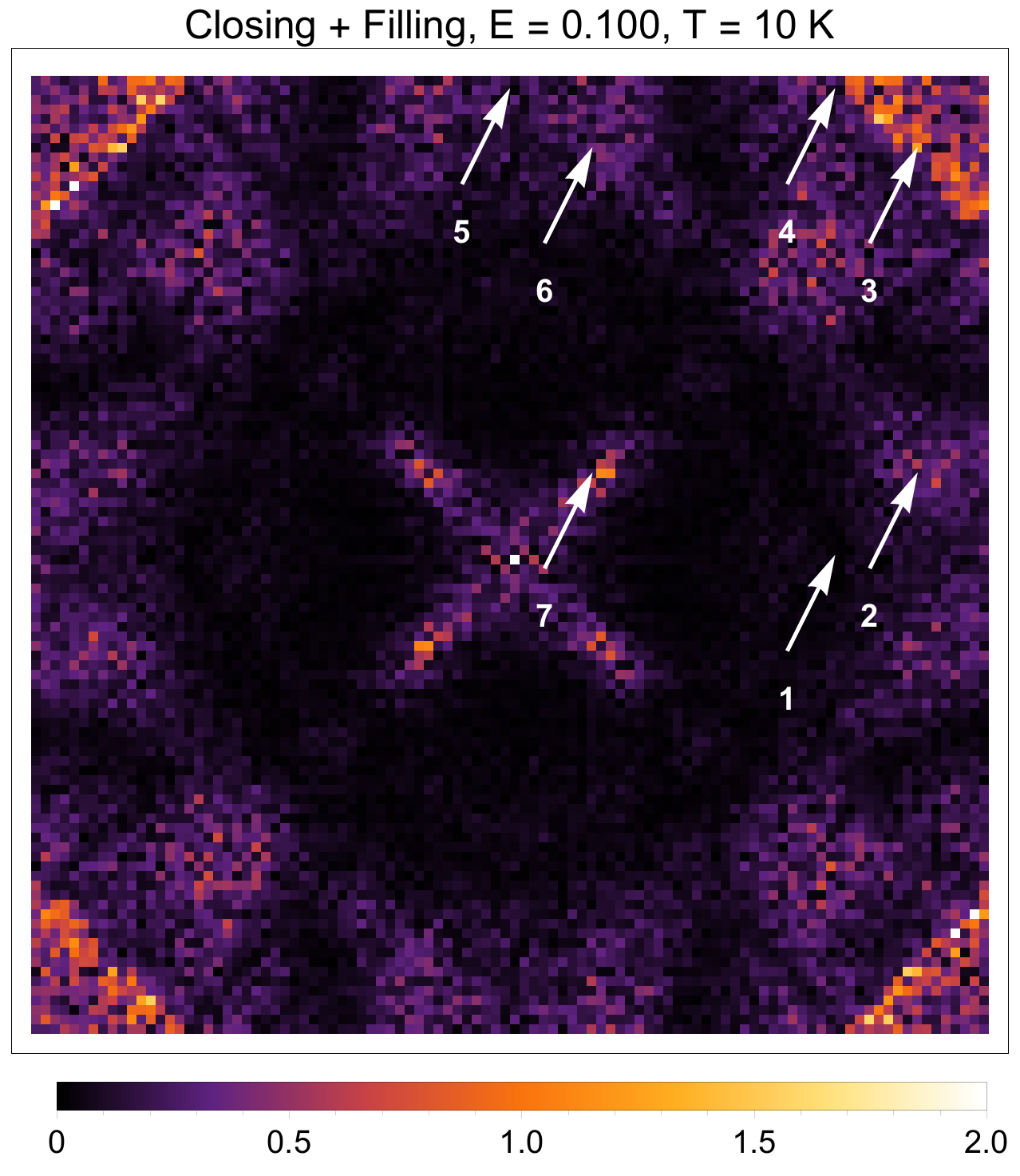}
	\includegraphics[height=0.18\textwidth]{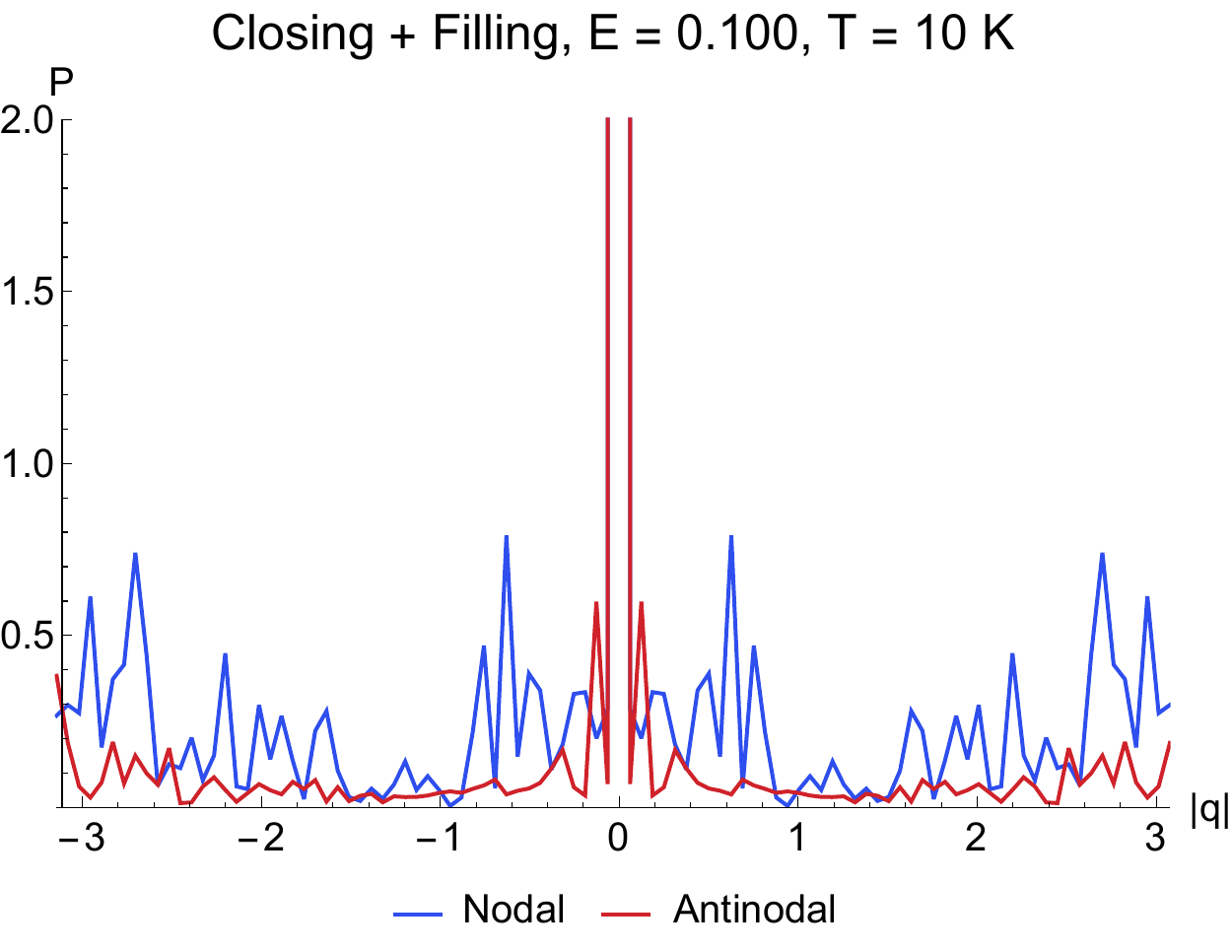} \\
	\includegraphics[height=0.18\textwidth]{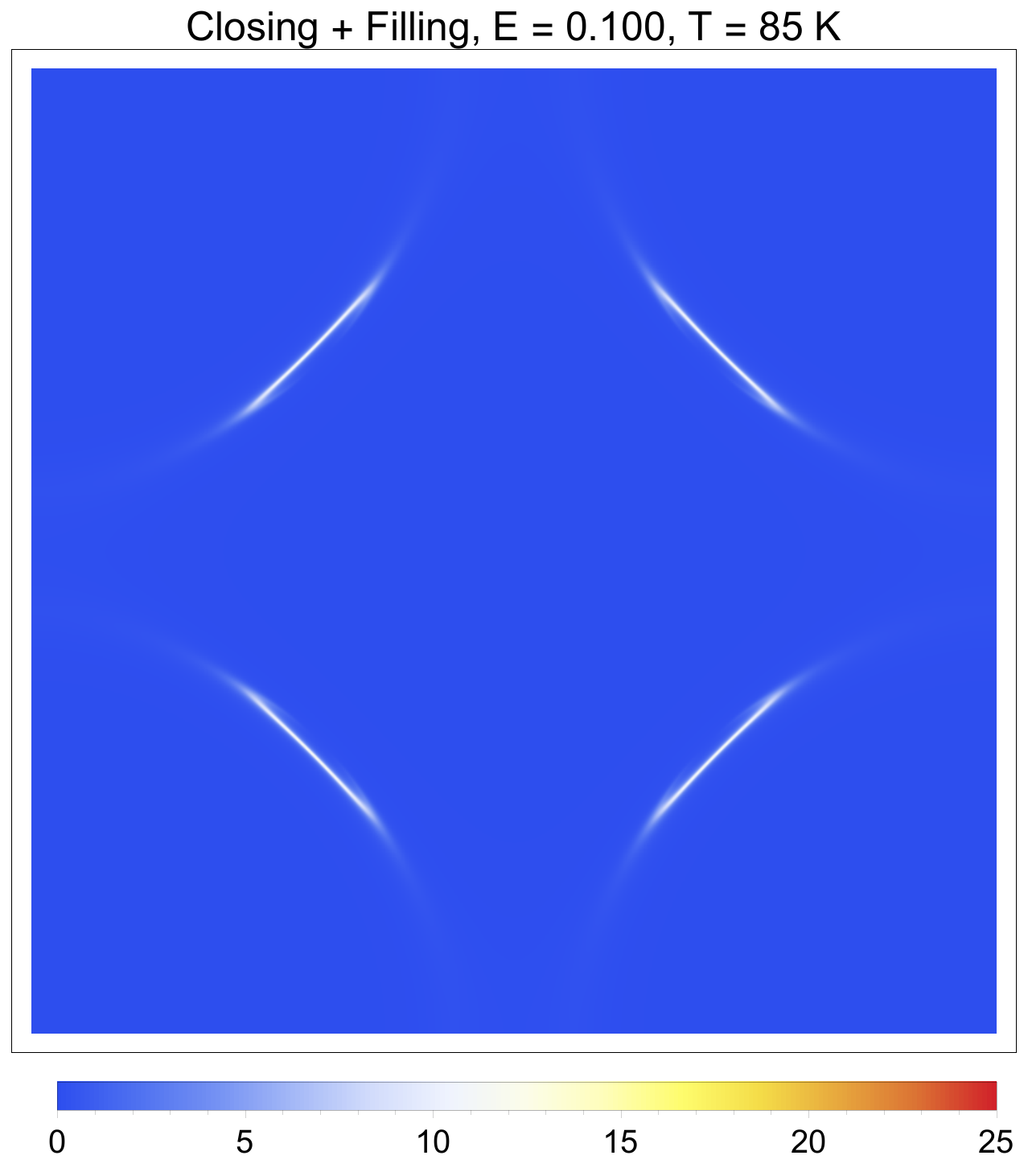}
	\includegraphics[height=0.18\textwidth]{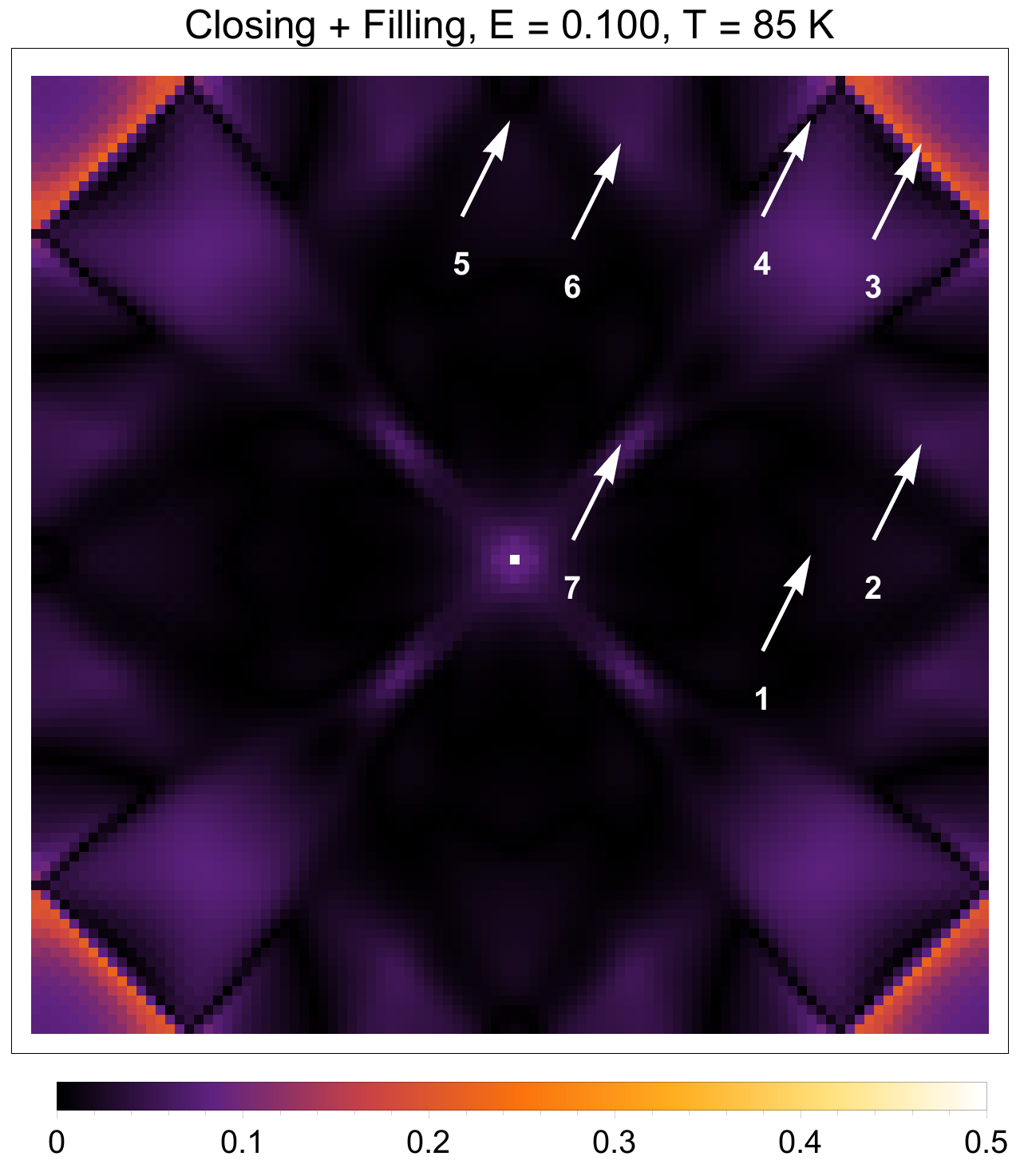}
	\includegraphics[height=0.18\textwidth]{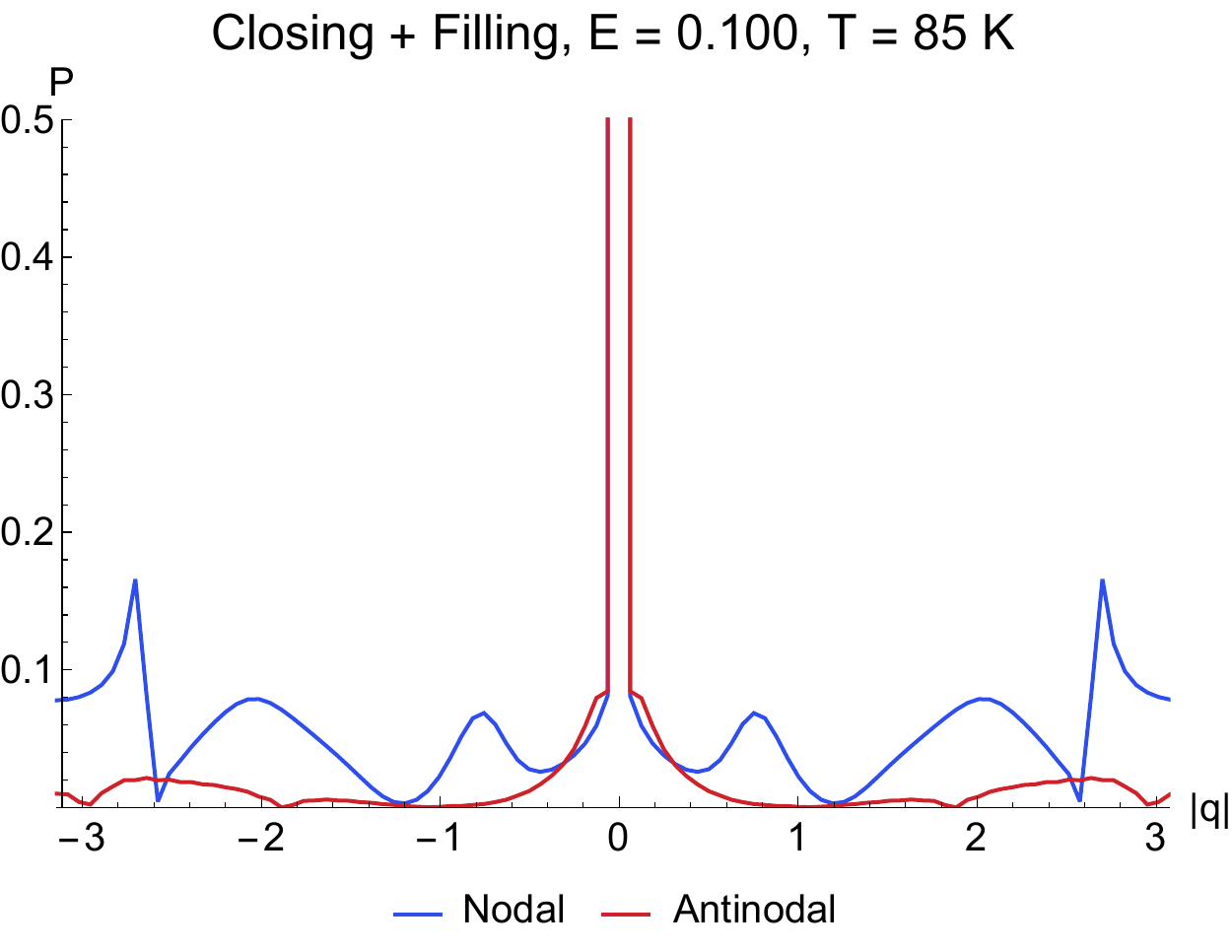}
	\includegraphics[height=0.18\textwidth]{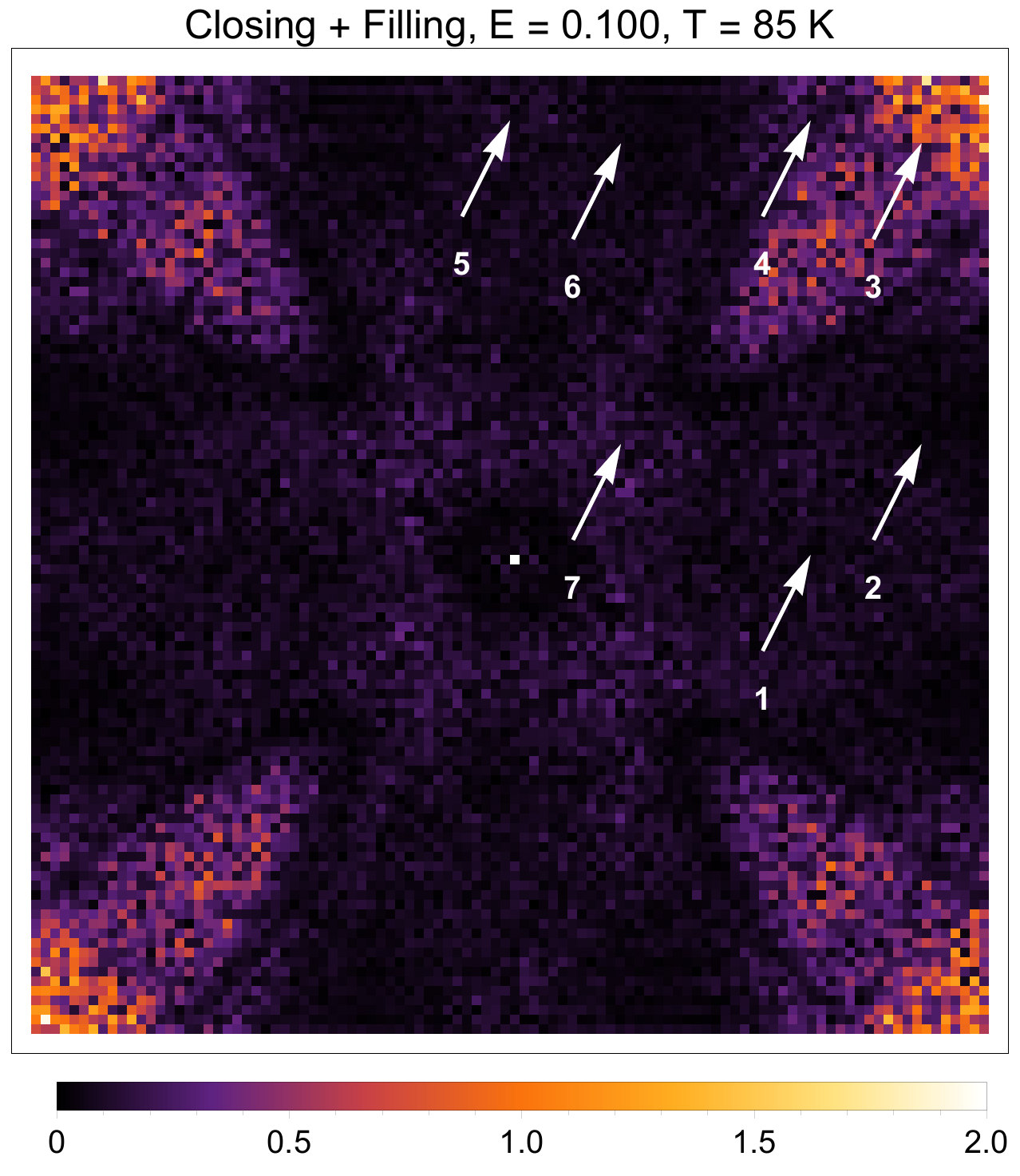}
	\includegraphics[height=0.18\textwidth]{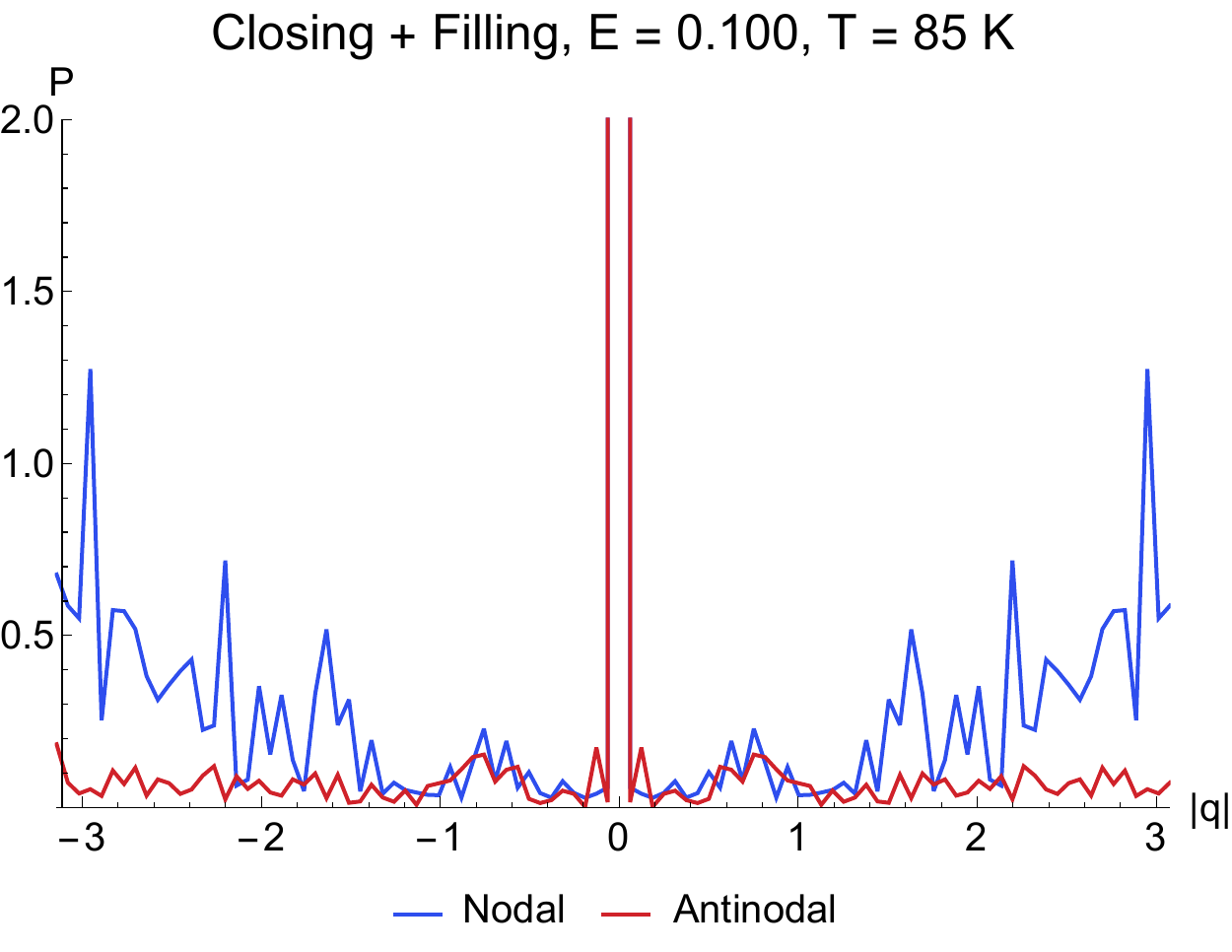} \\
	\includegraphics[height=0.18\textwidth]{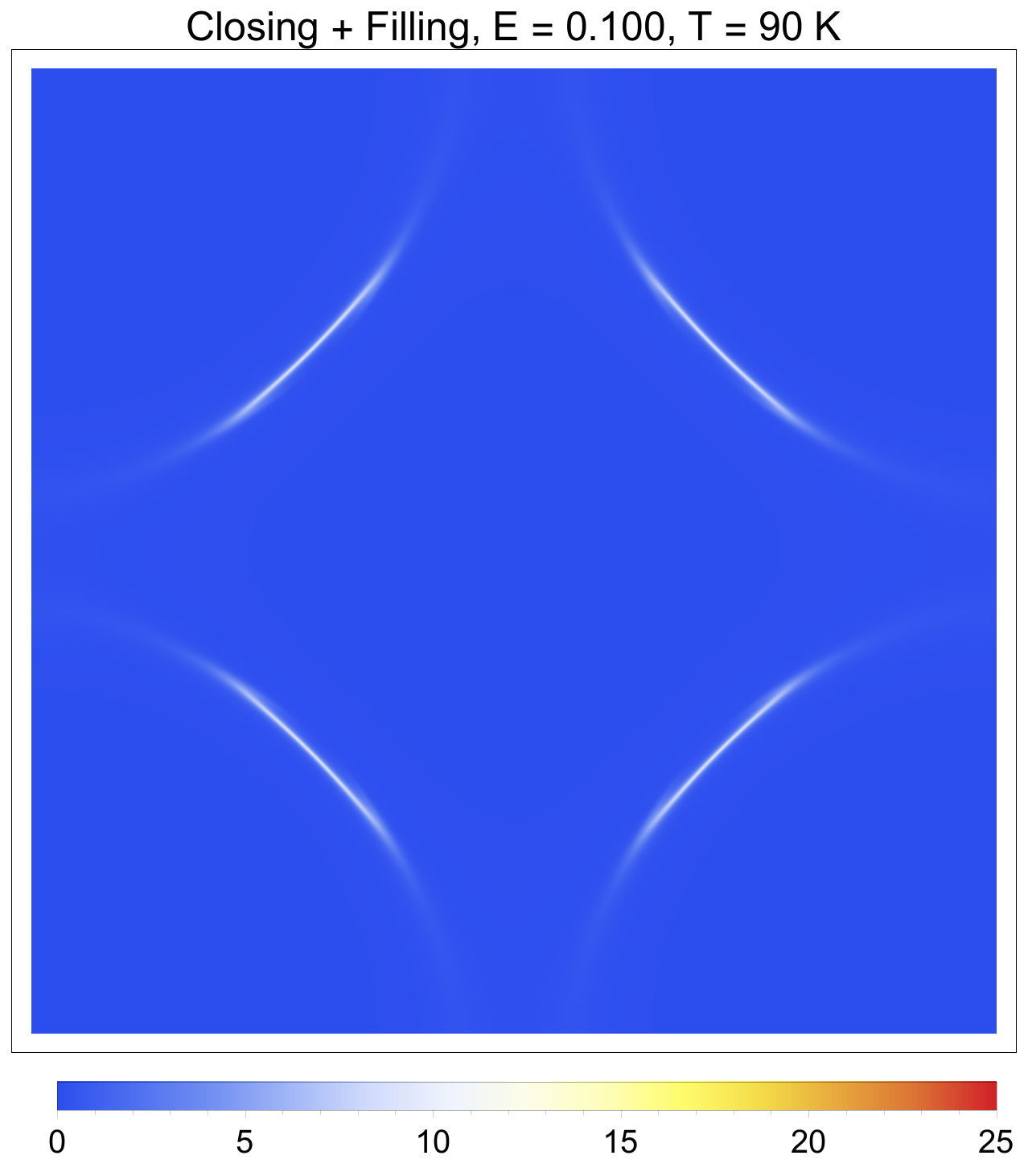}
	\includegraphics[height=0.18\textwidth]{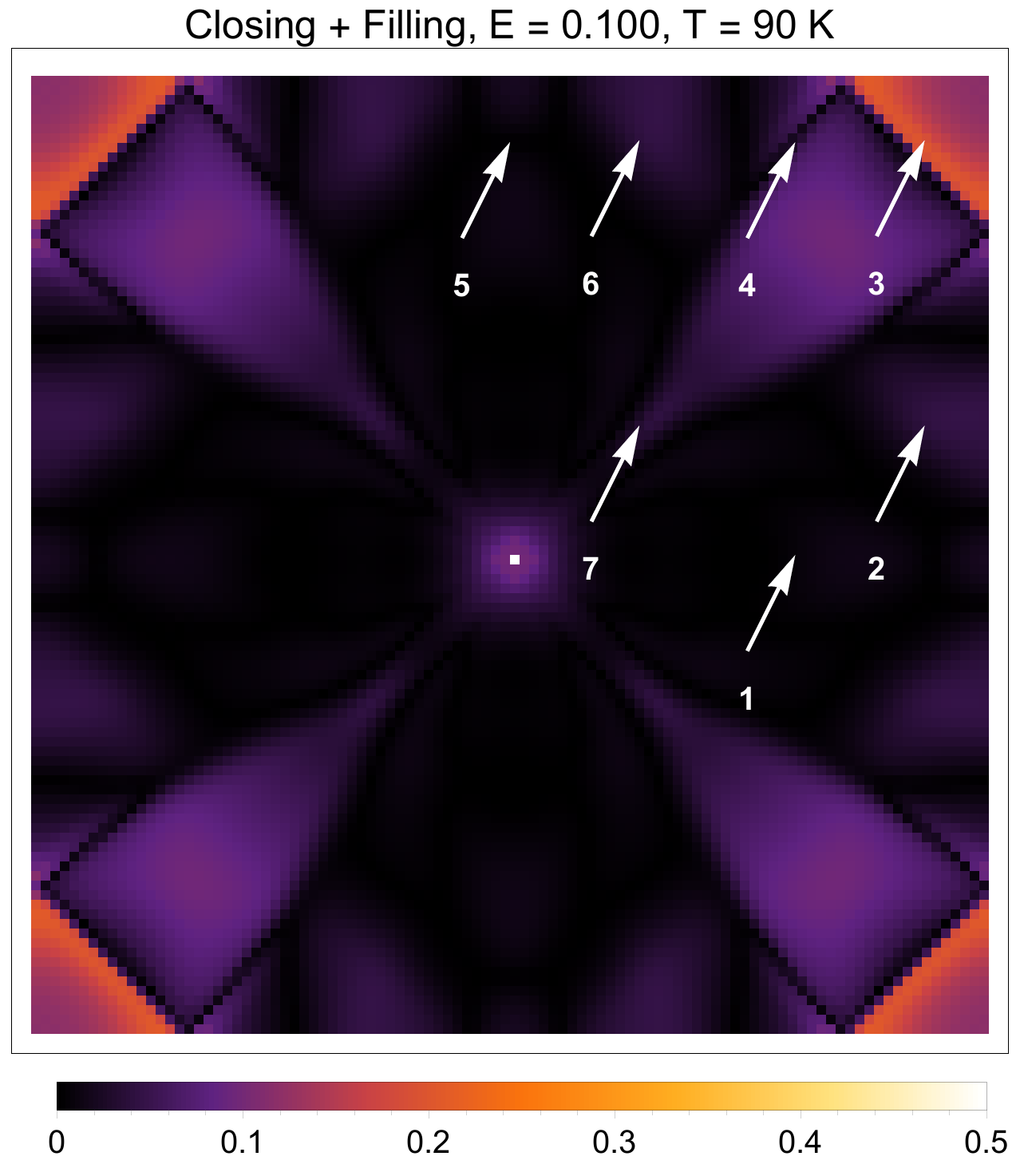}
	\includegraphics[height=0.18\textwidth]{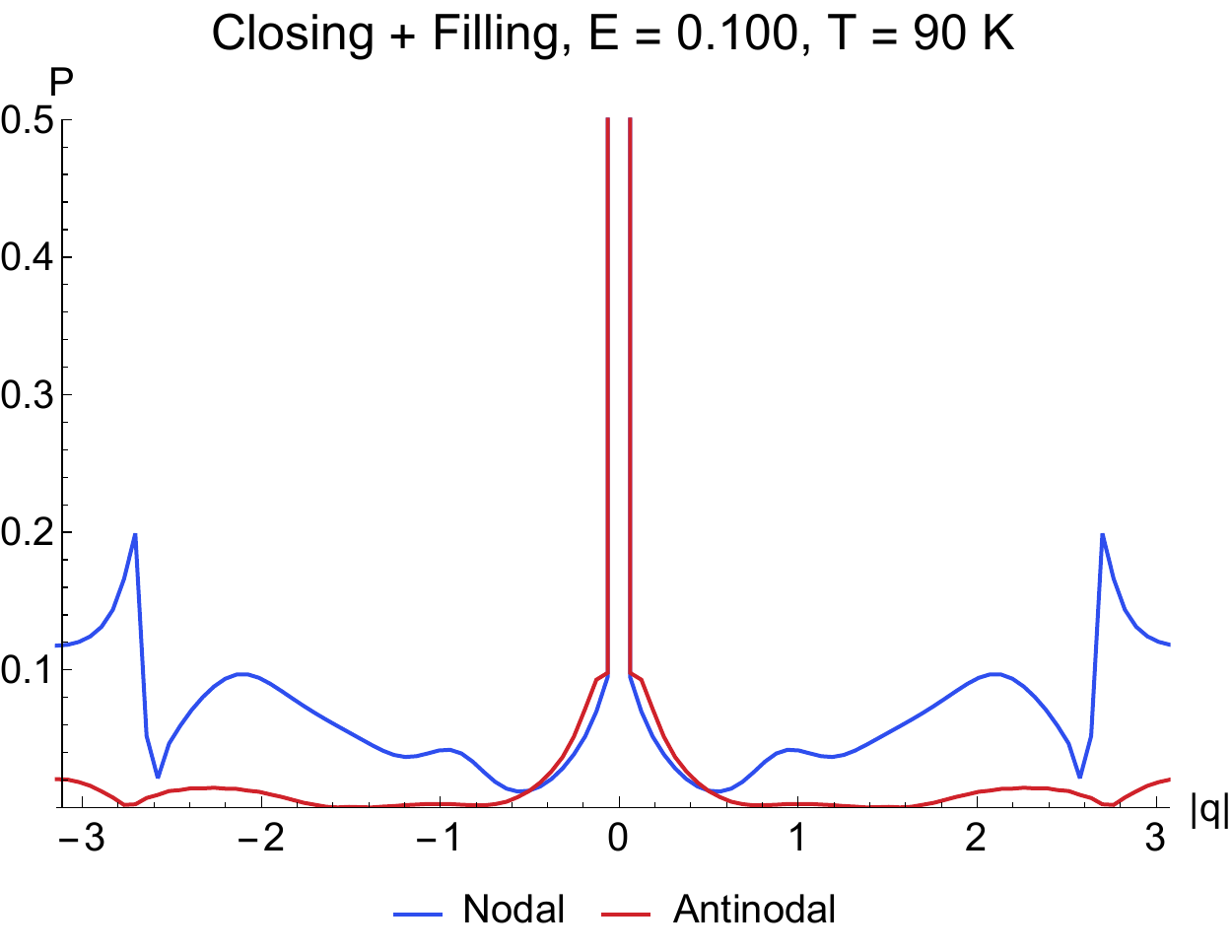}
	\includegraphics[height=0.18\textwidth]{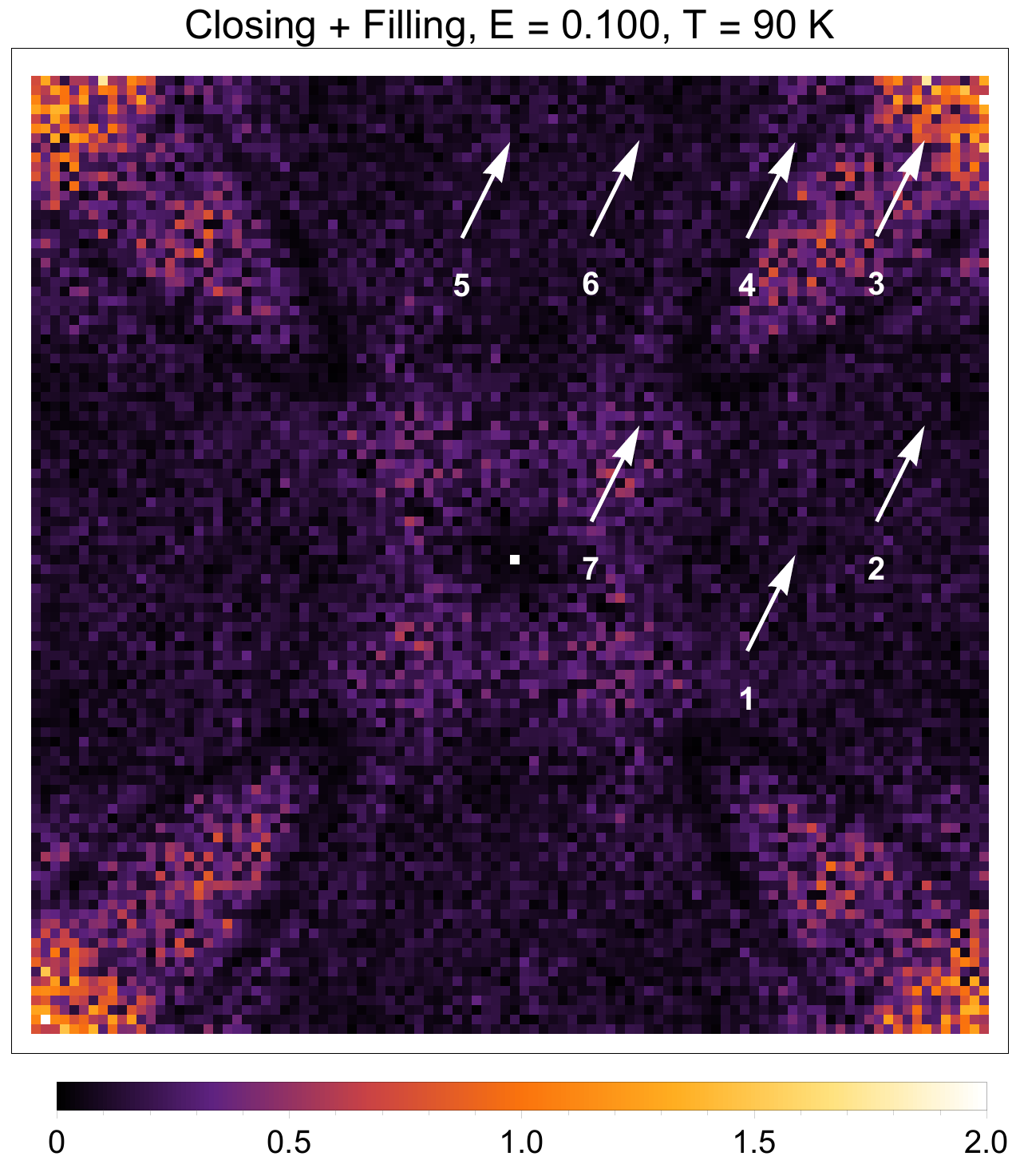}
	\includegraphics[height=0.18\textwidth]{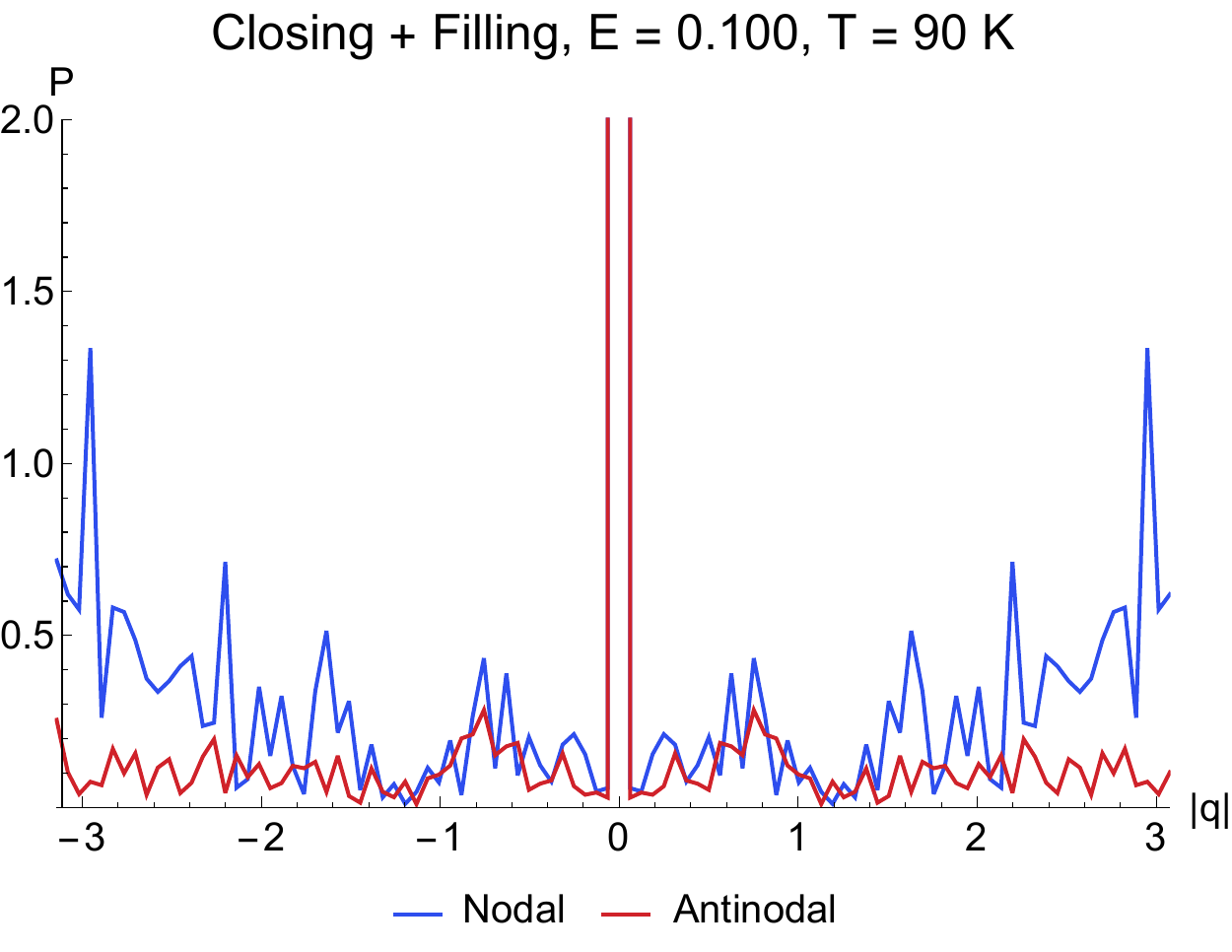} \\
	\includegraphics[height=0.18\textwidth]{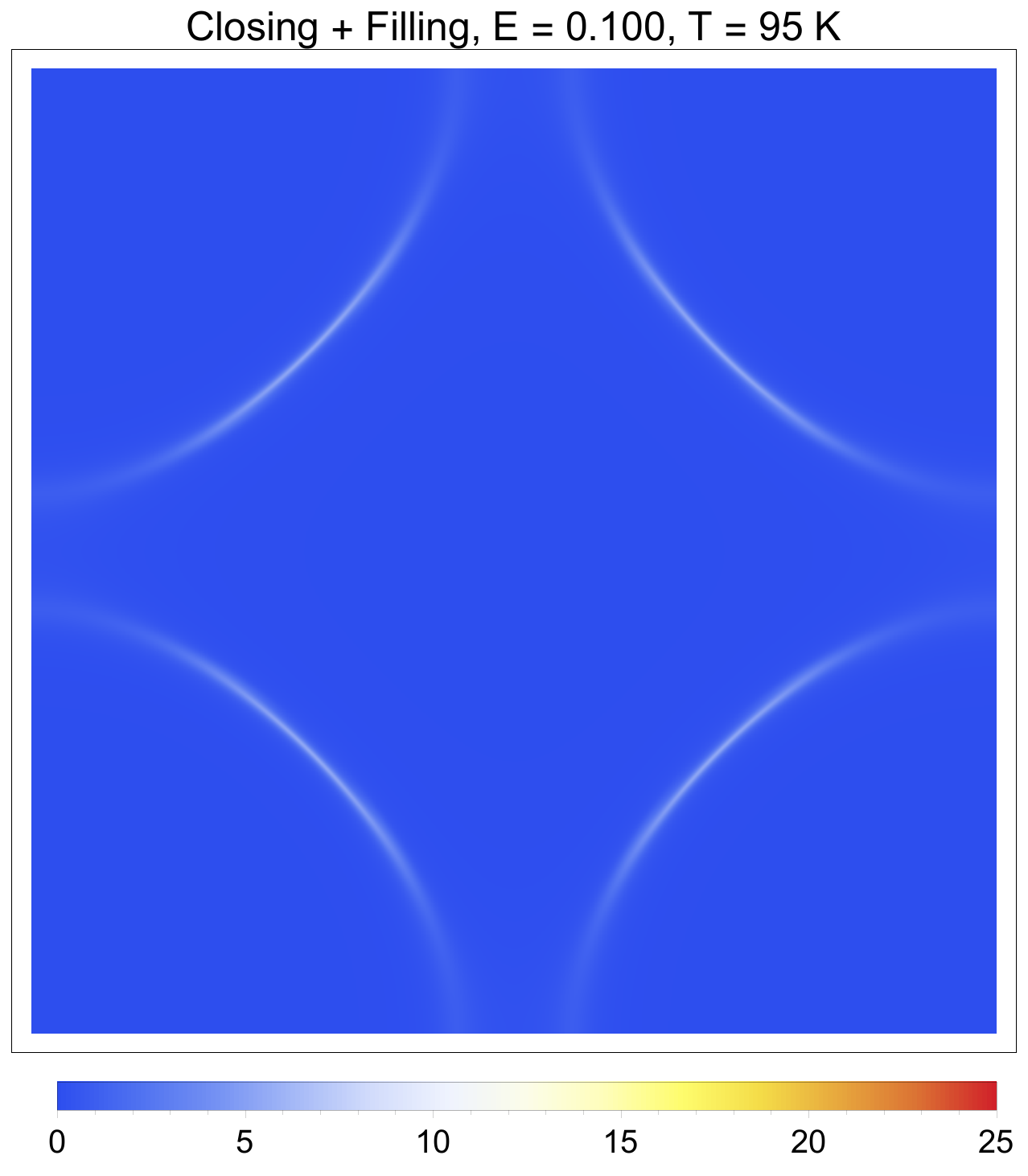}
	\includegraphics[height=0.18\textwidth]{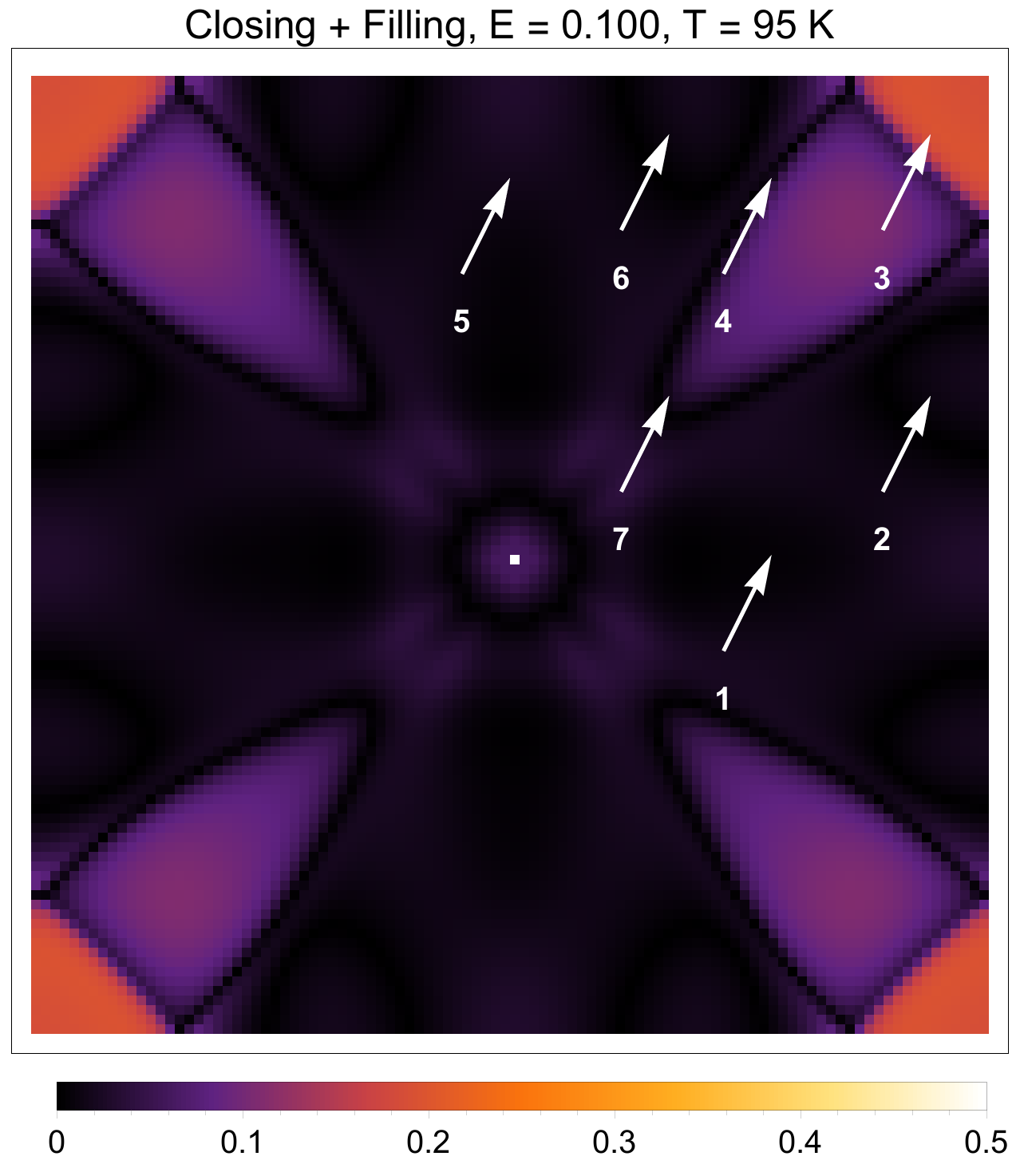}
	\includegraphics[height=0.18\textwidth]{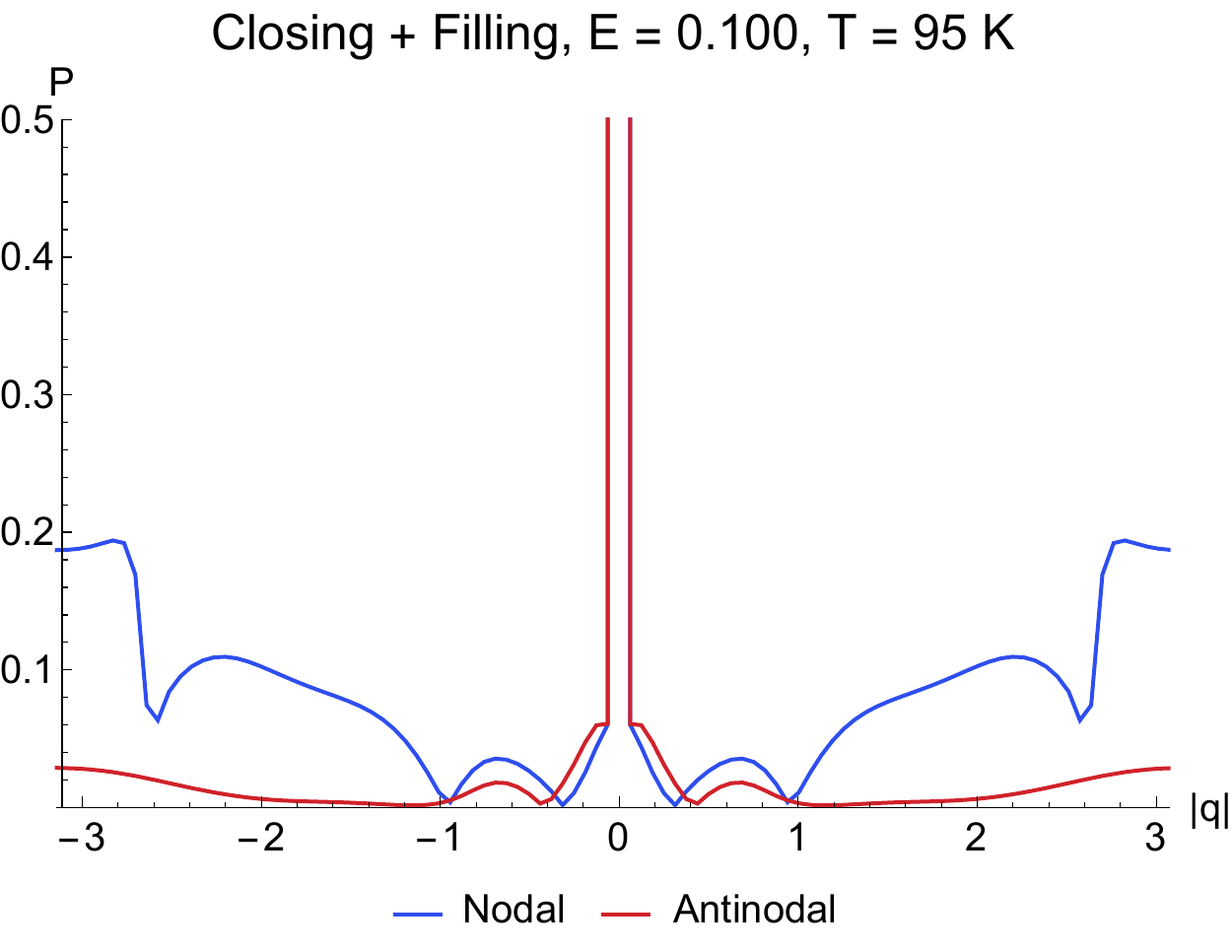}
	\includegraphics[height=0.18\textwidth]{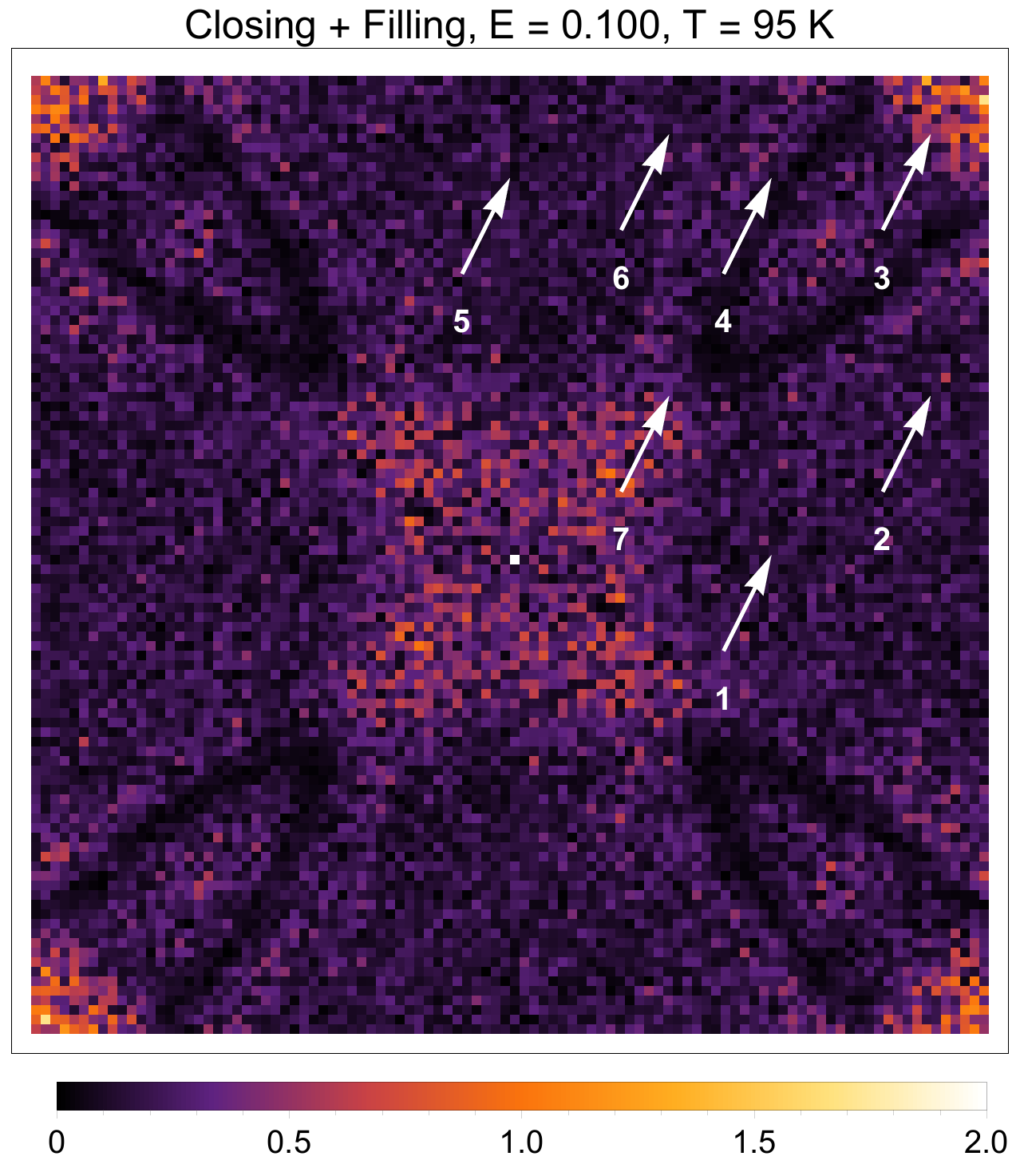}
	\includegraphics[height=0.18\textwidth]{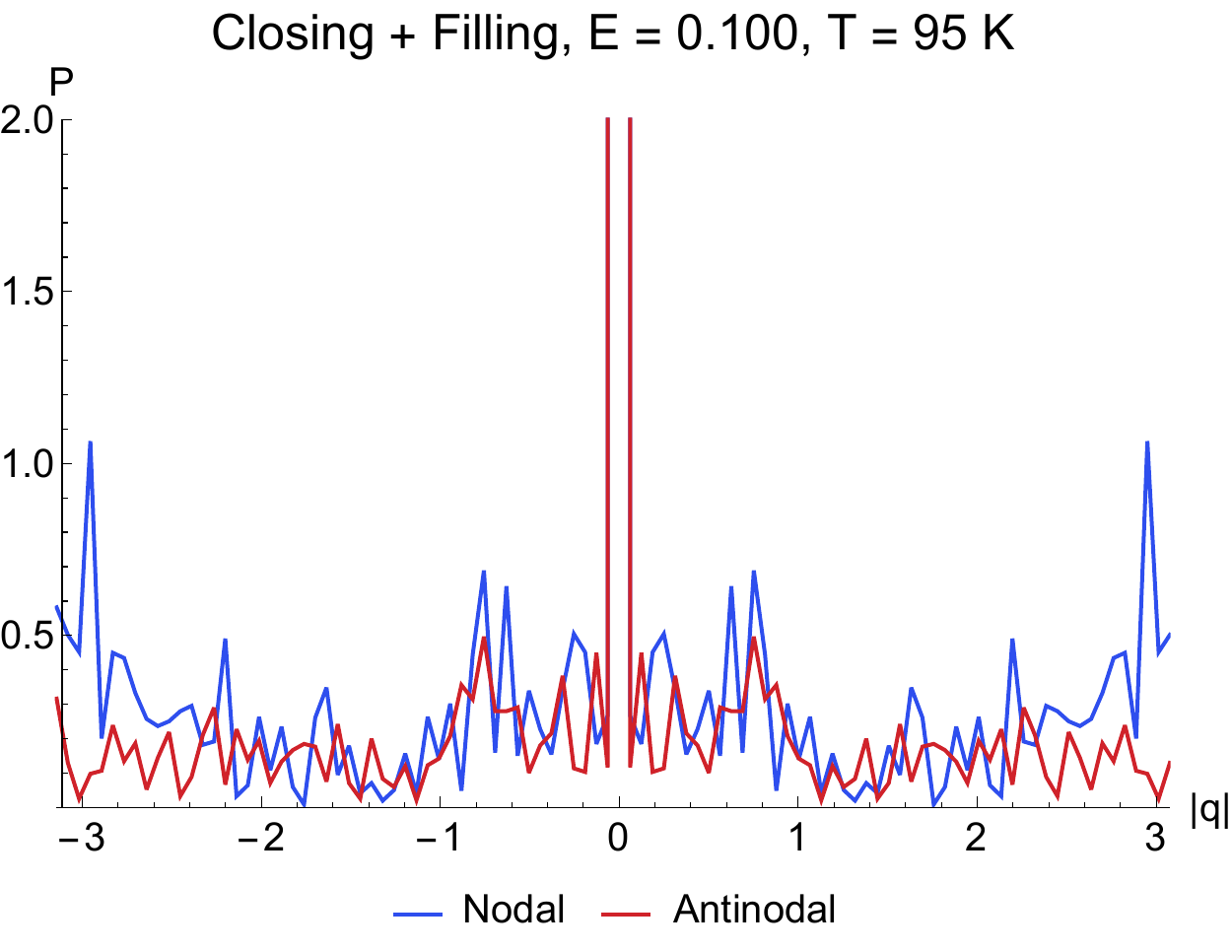} \\
	\includegraphics[height=0.18\textwidth]{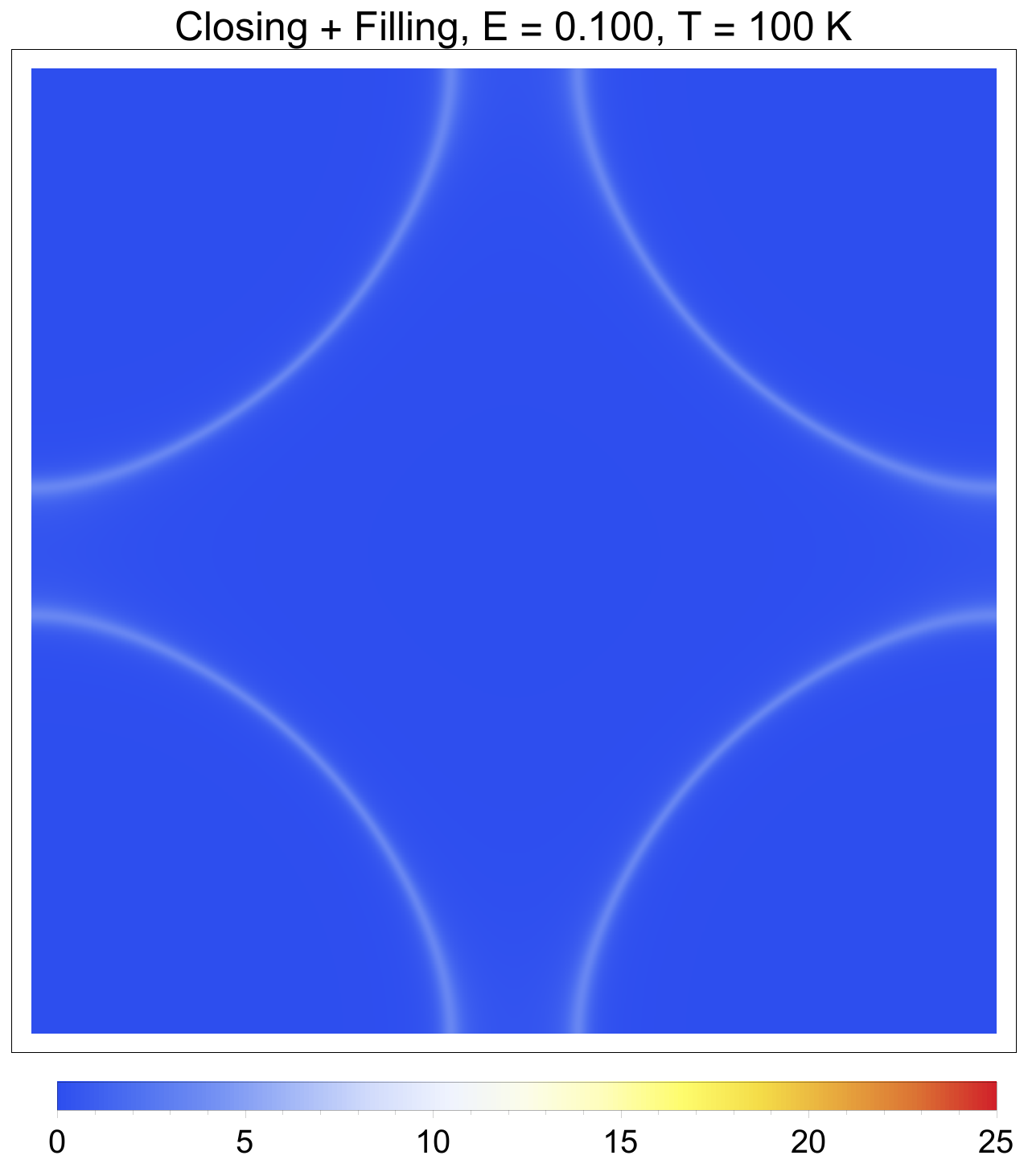}
	\includegraphics[height=0.18\textwidth]{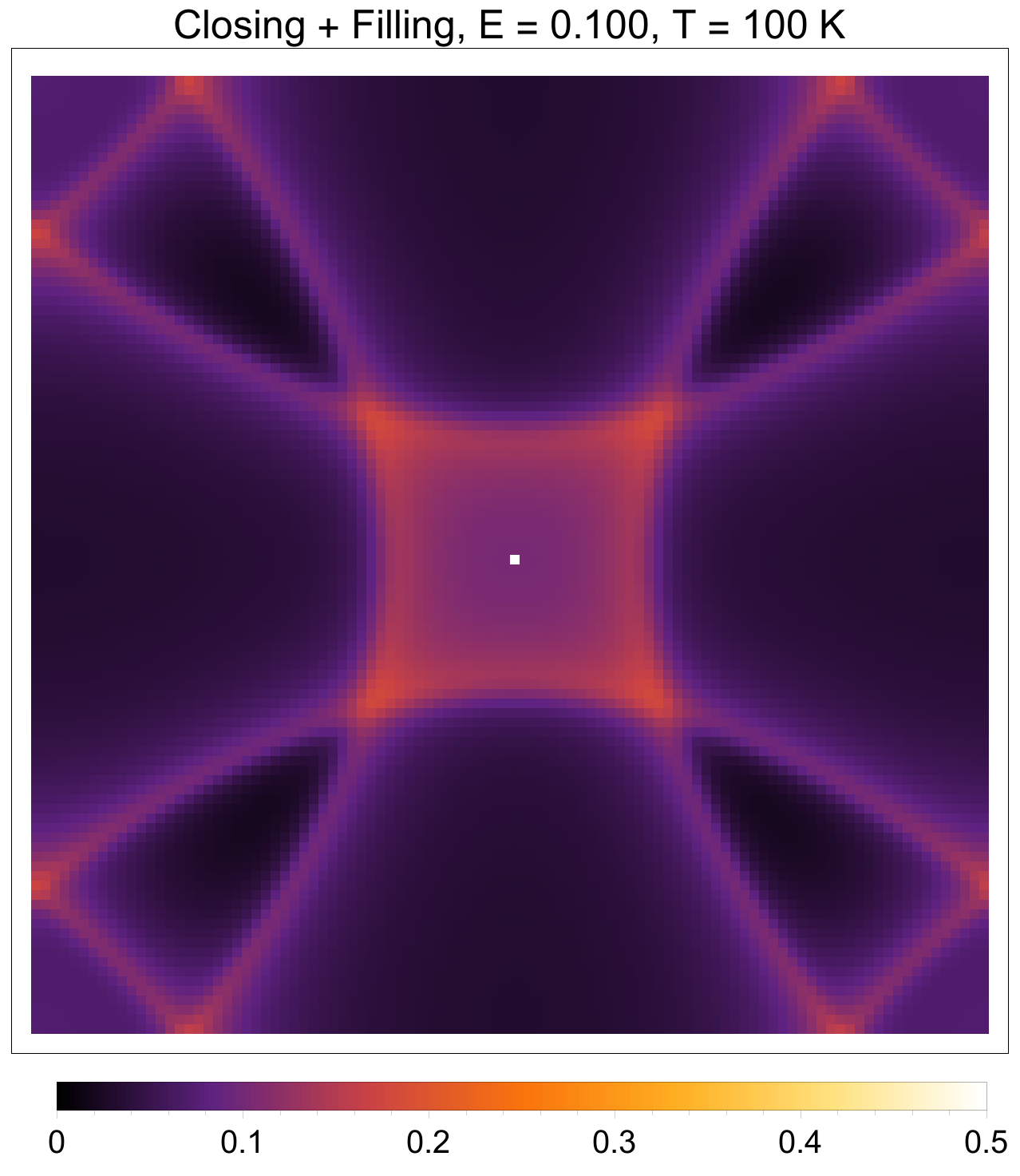}
	\includegraphics[height=0.18\textwidth]{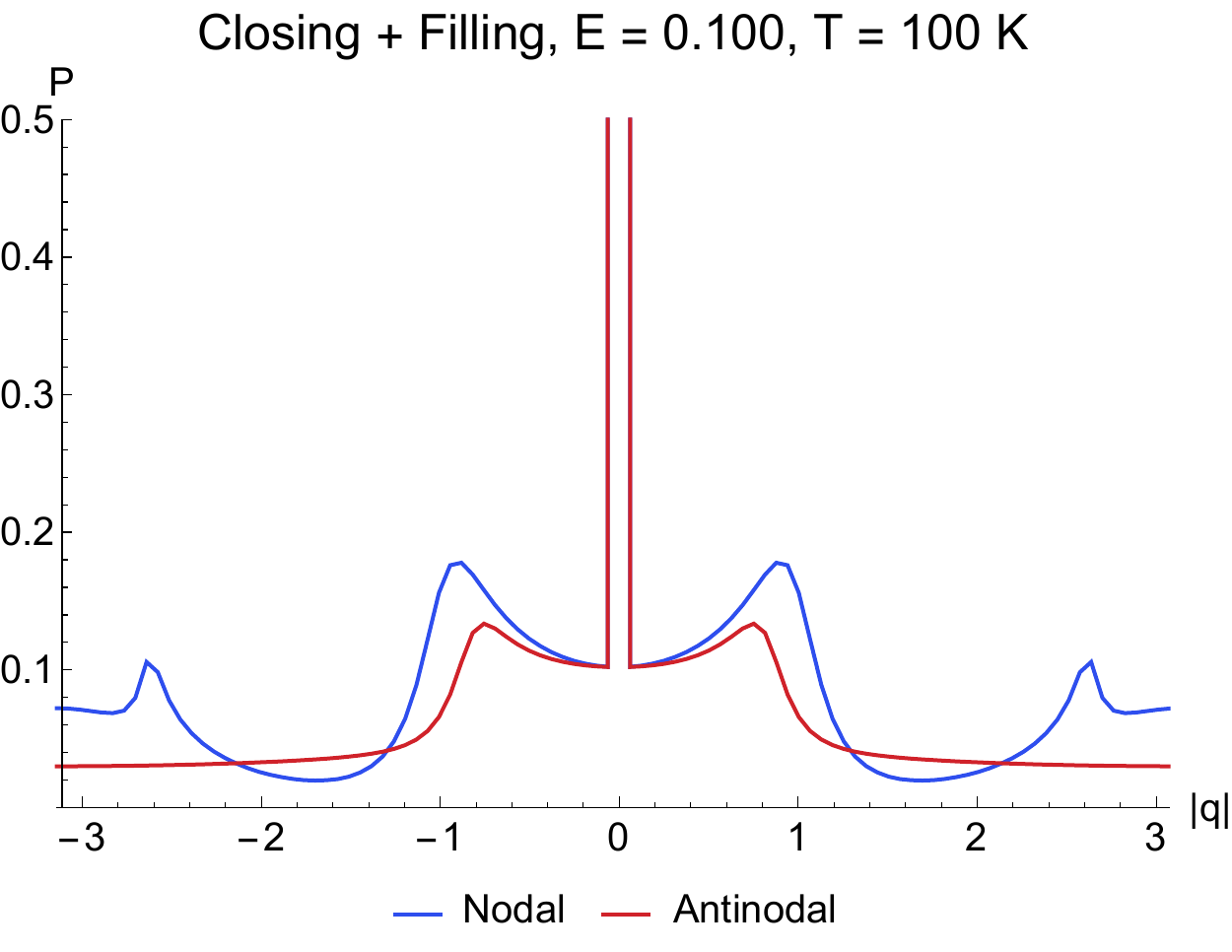}
	\includegraphics[height=0.18\textwidth]{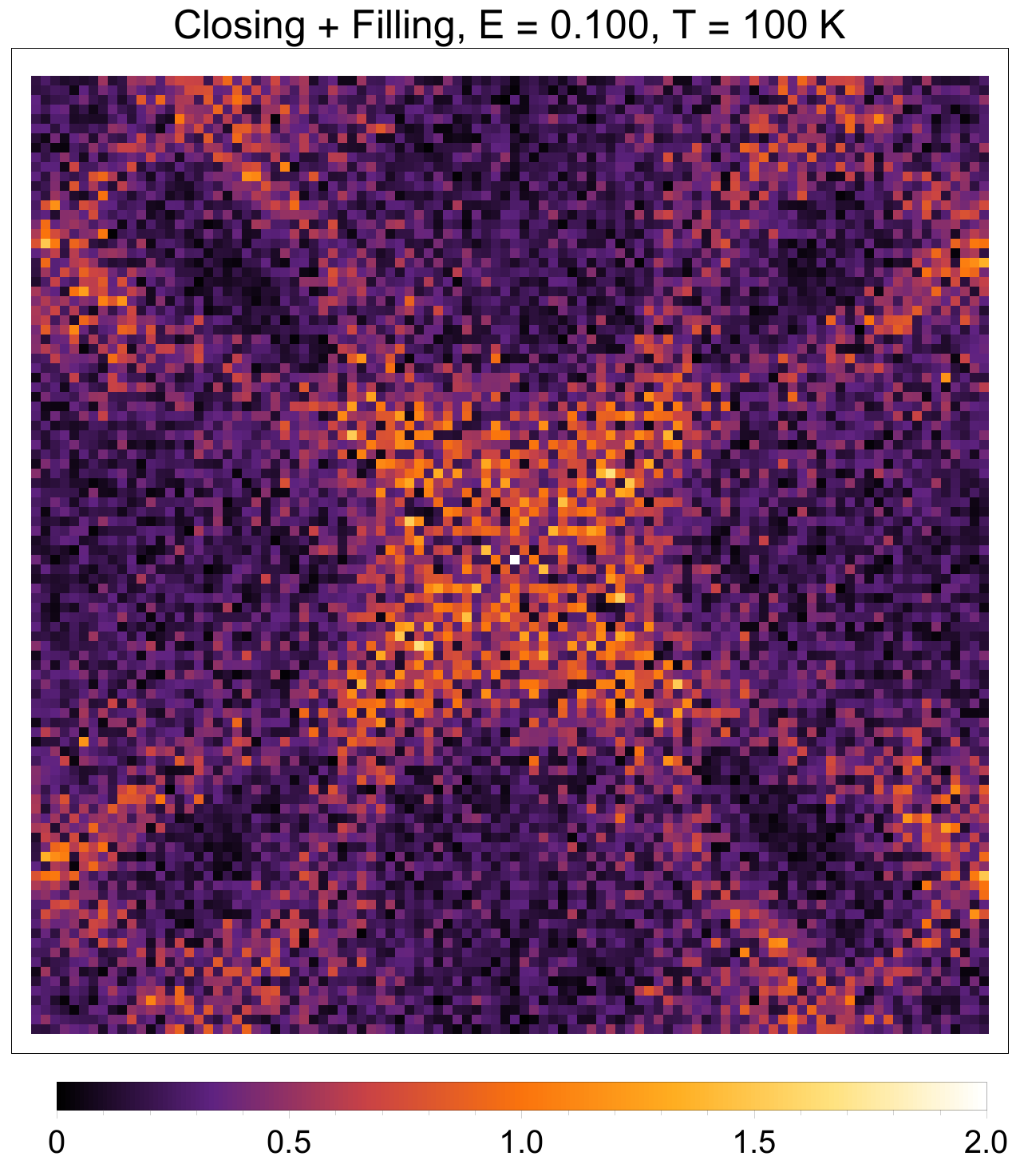}
	\includegraphics[height=0.18\textwidth]{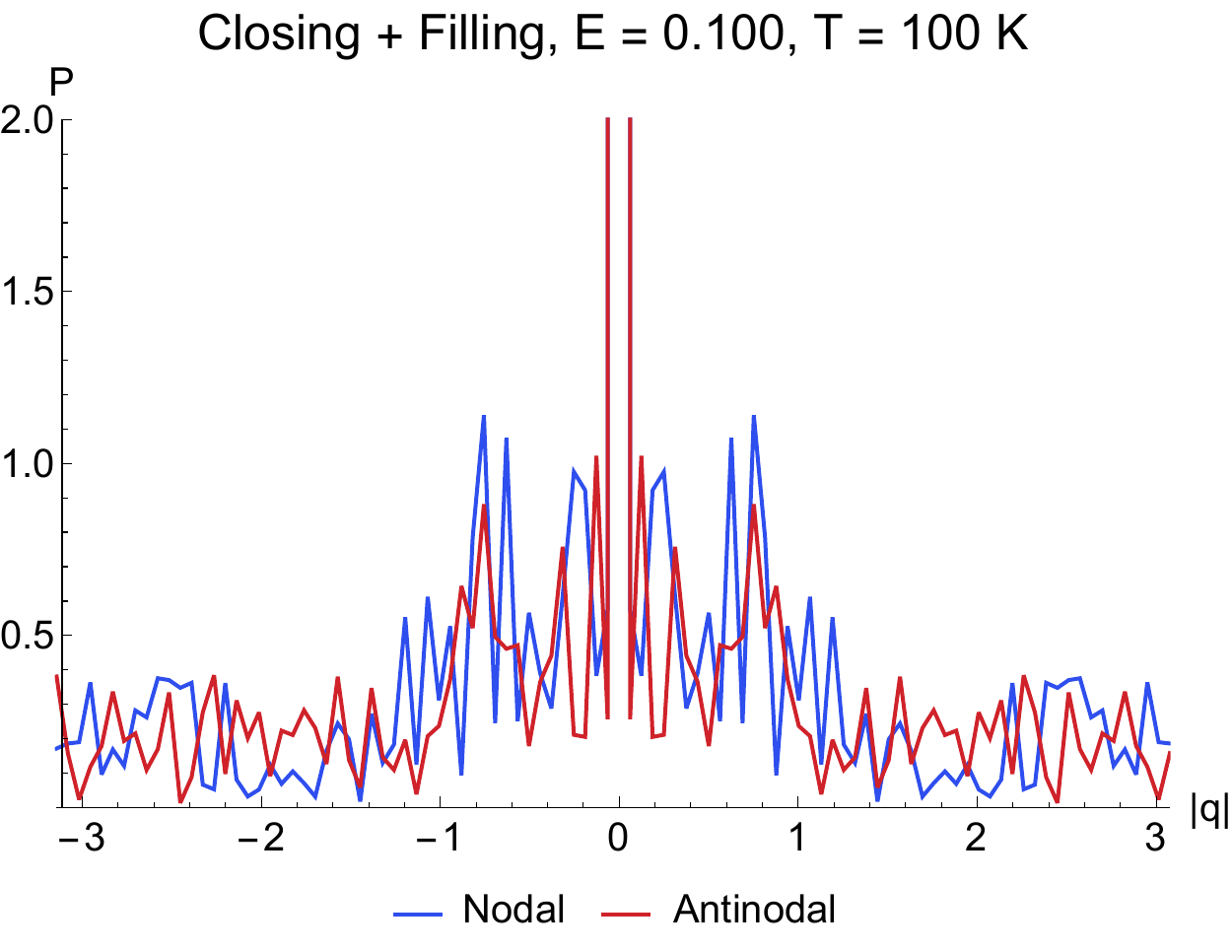} \\
	
	\caption{Gap-filling and -closing phenomenology at various temperatures. $T_c$ here is 90 K. Left to right: The spectral function $A(\mathbf{k}, \omega)$; the Fourier transform of the LDOS $P(\mathbf{q}, \omega)$; linecuts of $P(\mathbf{q}, \omega)$ in the nodal and antinodal directions; $P(\mathbf{q}, \omega)$ in the presence of multiple weak impurities and finite-temperature smearing; and linecuts of $P(\mathbf{q}, \omega)$ in the presence of multiple weak impurities and finite-temperature smearing. Arrows indicate the locations of the peaks predicted by the octet model. All plots are taken at $E = 0.100$.}
	\label{fig:temperature_gf}
\end{figure*}

\begin{figure*}
	\centering
	
	\includegraphics[height=0.18\textwidth]{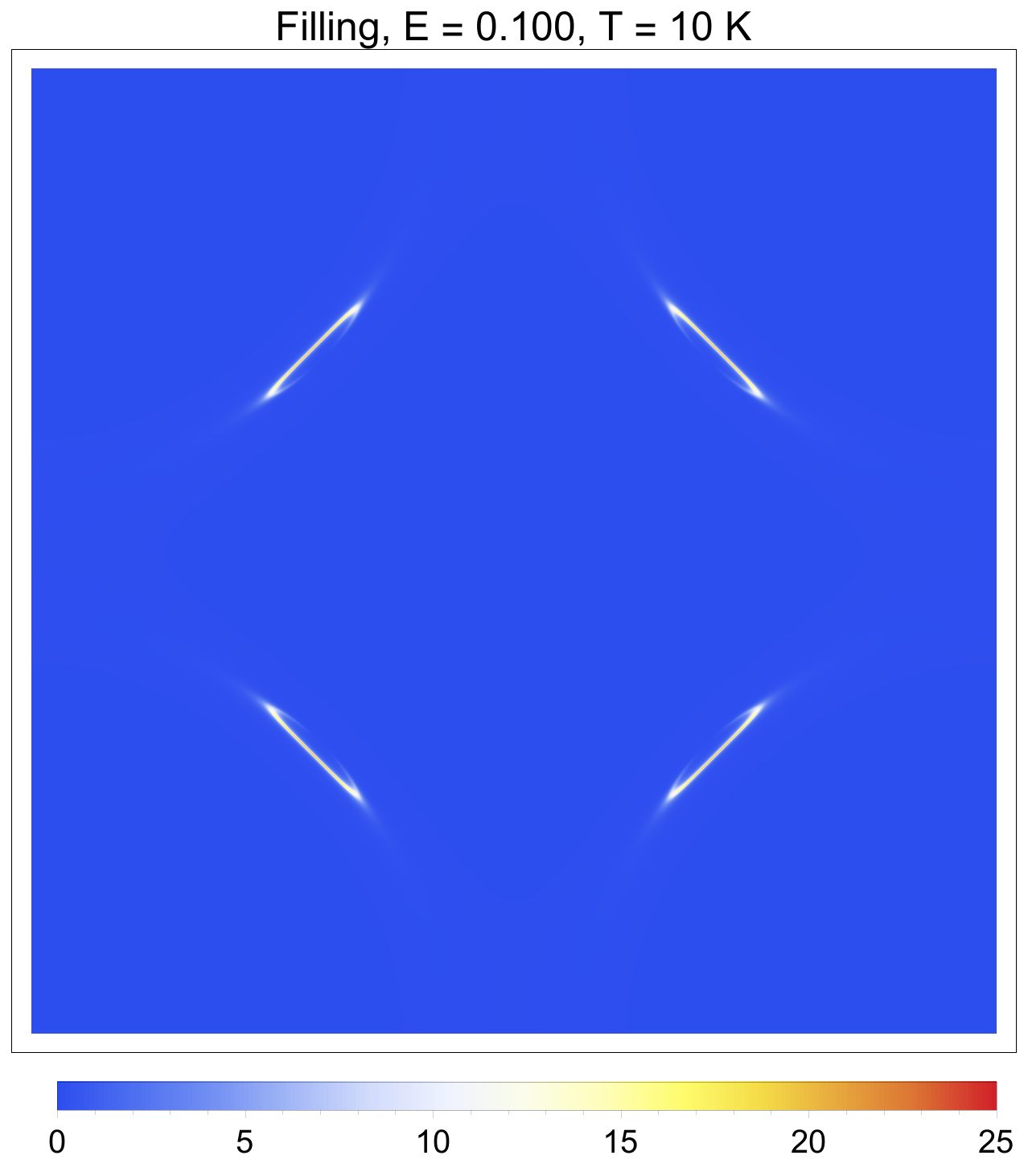}
	\includegraphics[height=0.18\textwidth]{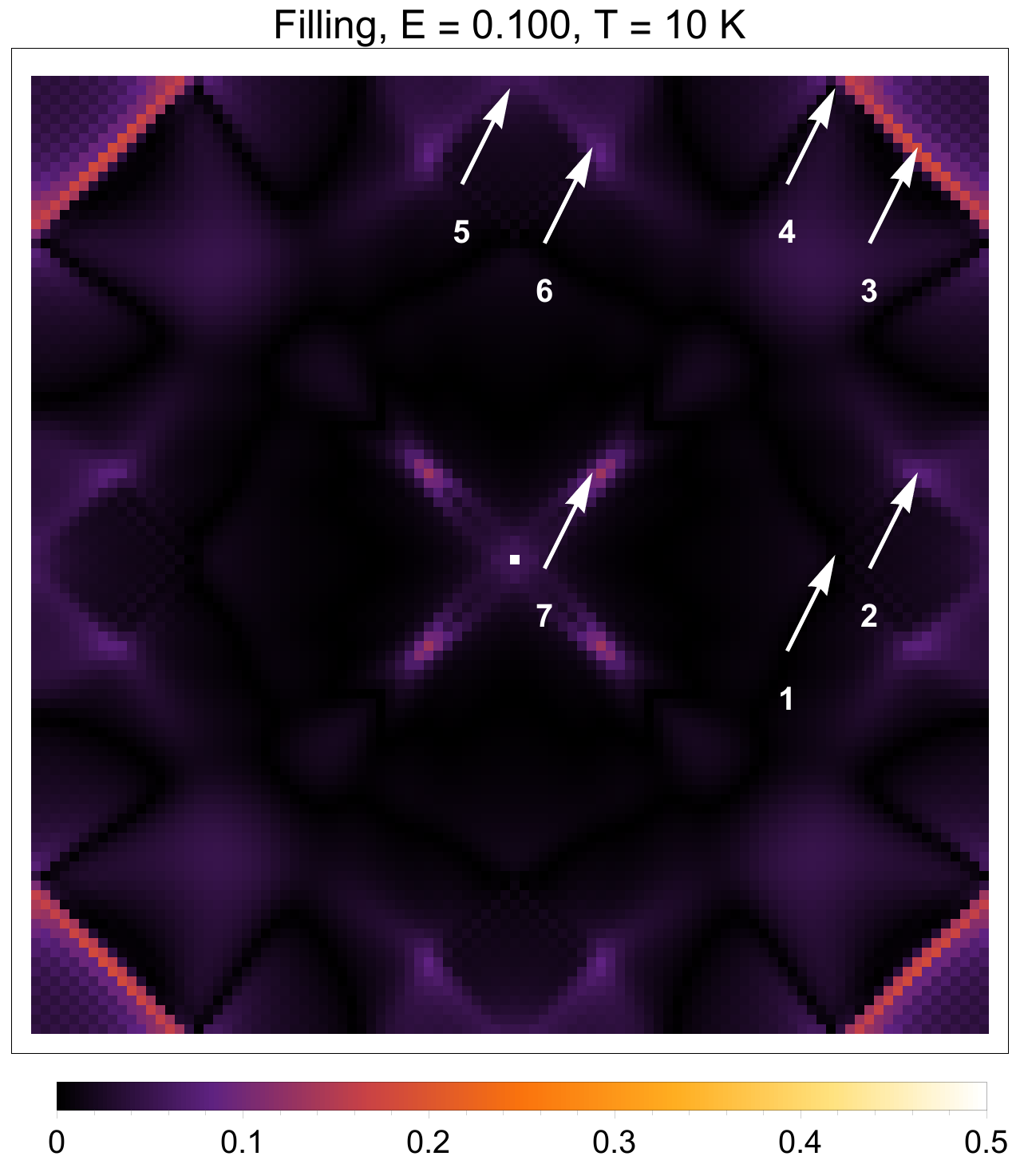}
	\includegraphics[height=0.18\textwidth]{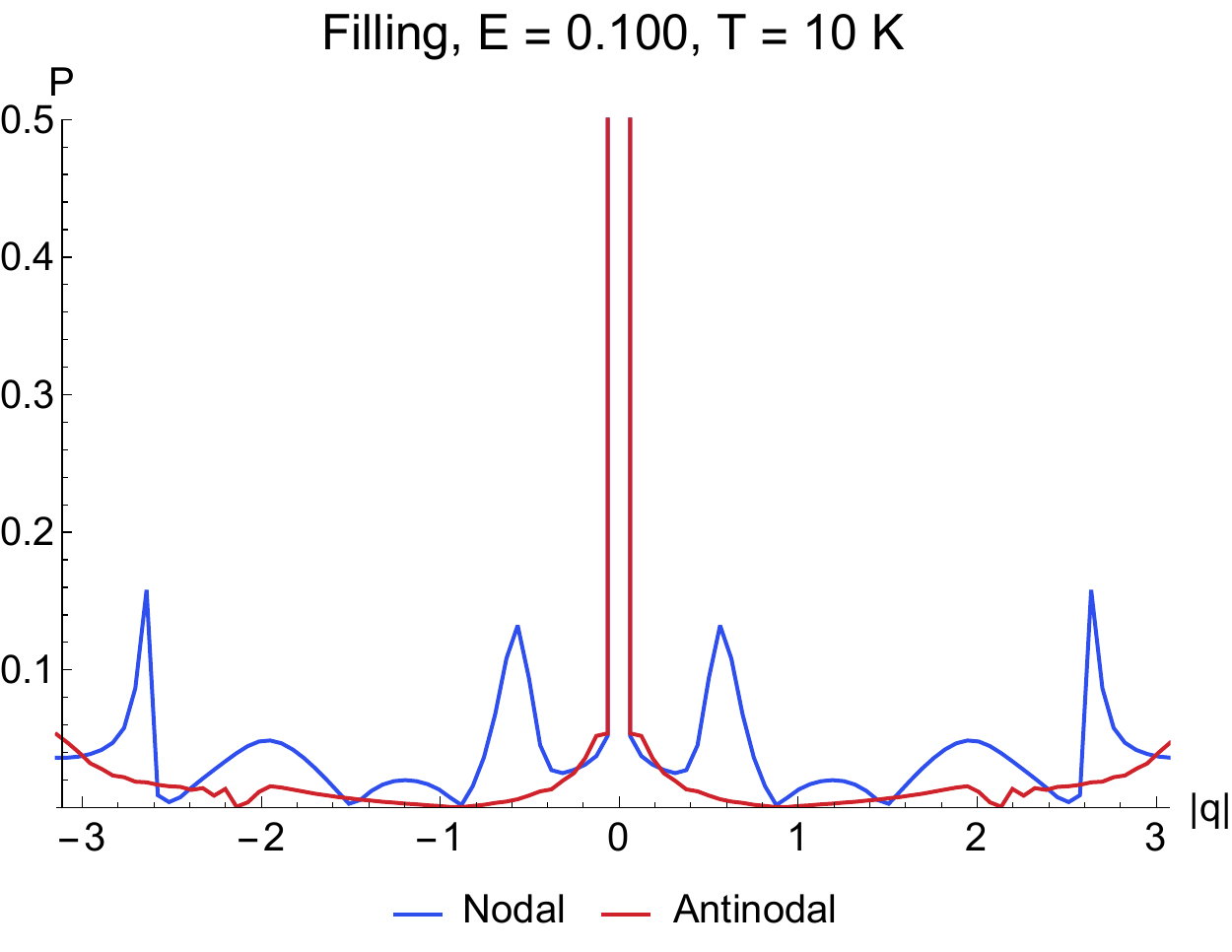} 
	\includegraphics[height=0.18\textwidth]{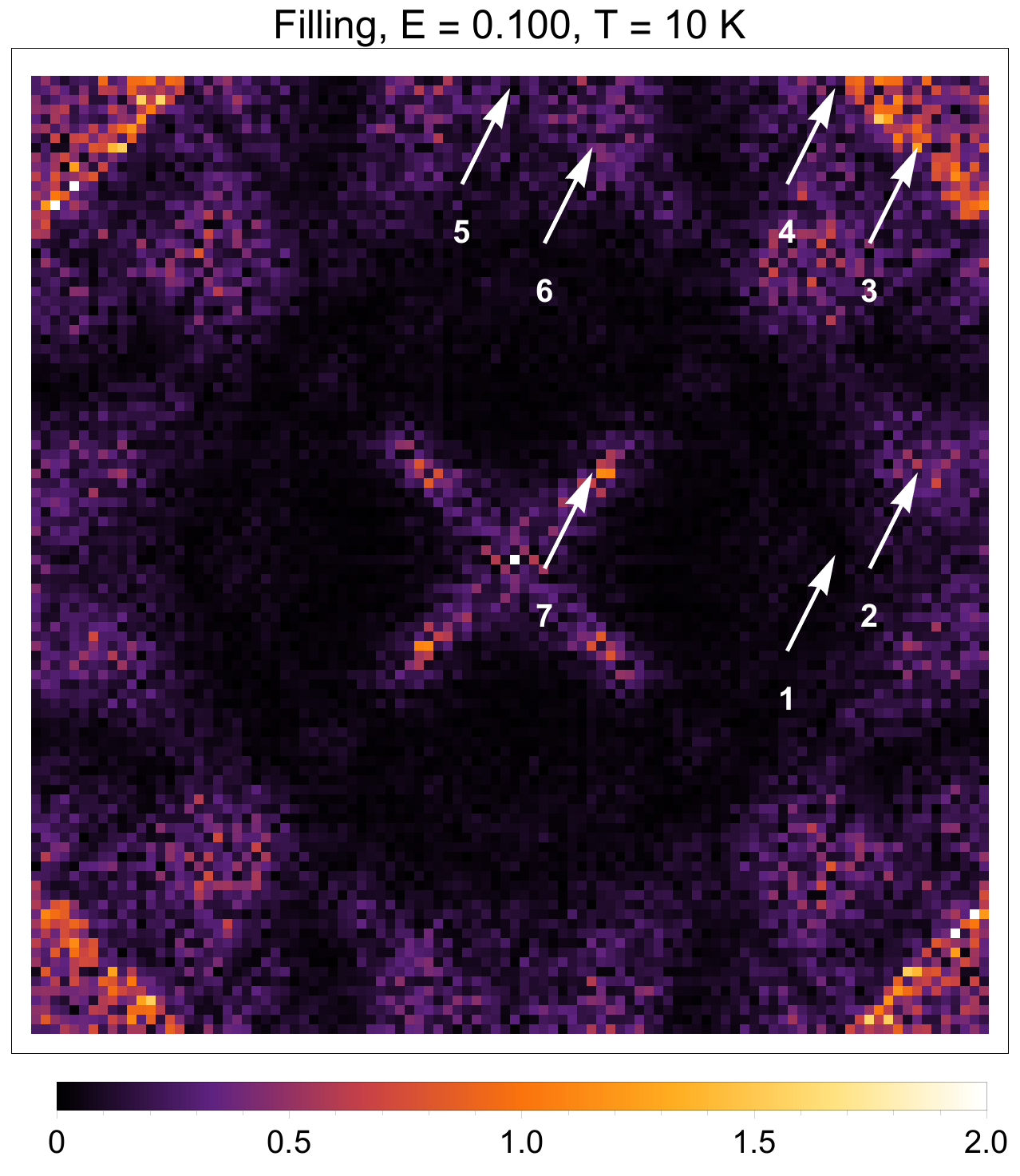}
	\includegraphics[height=0.18\textwidth]{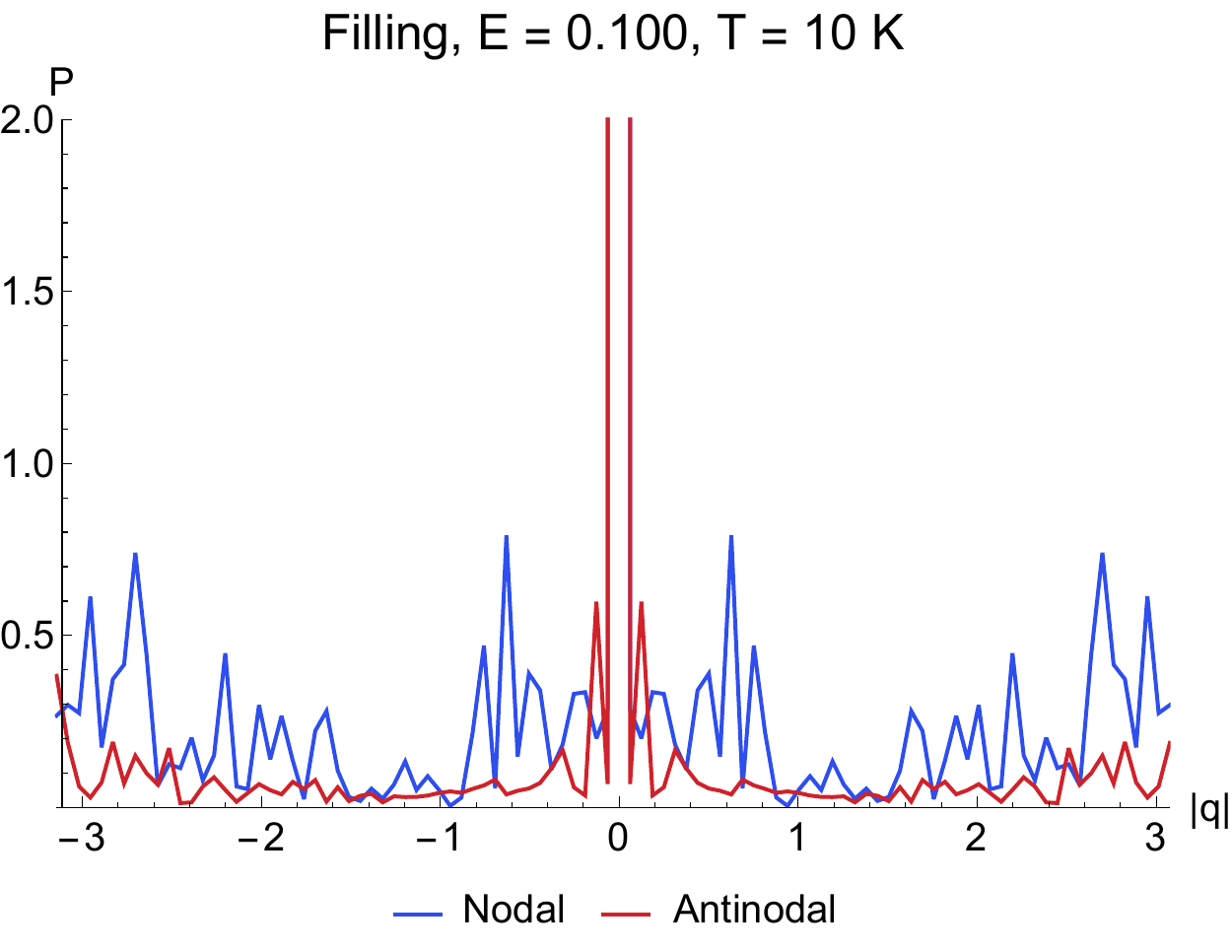} \\
	\includegraphics[height=0.18\textwidth]{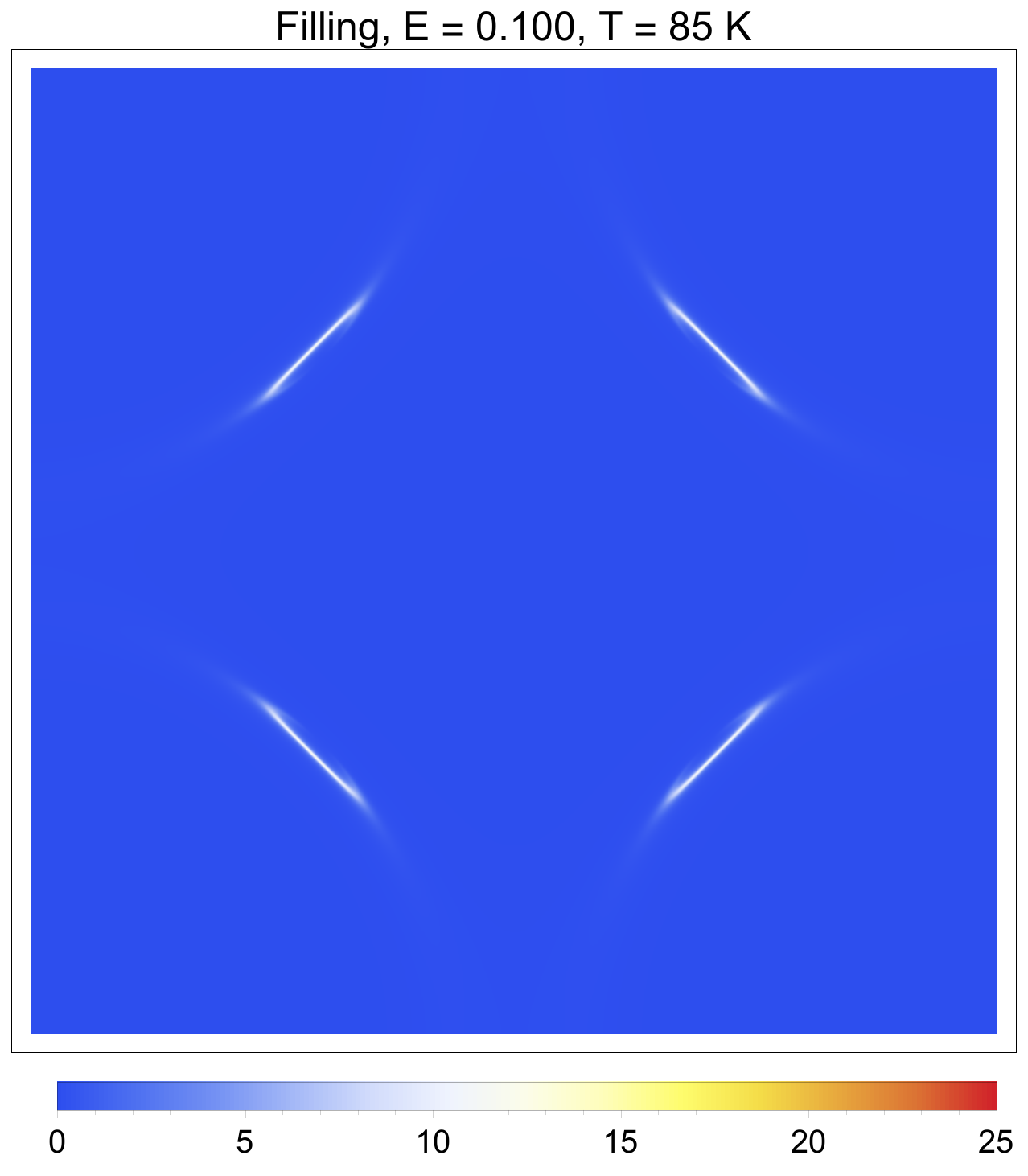}
	\includegraphics[height=0.18\textwidth]{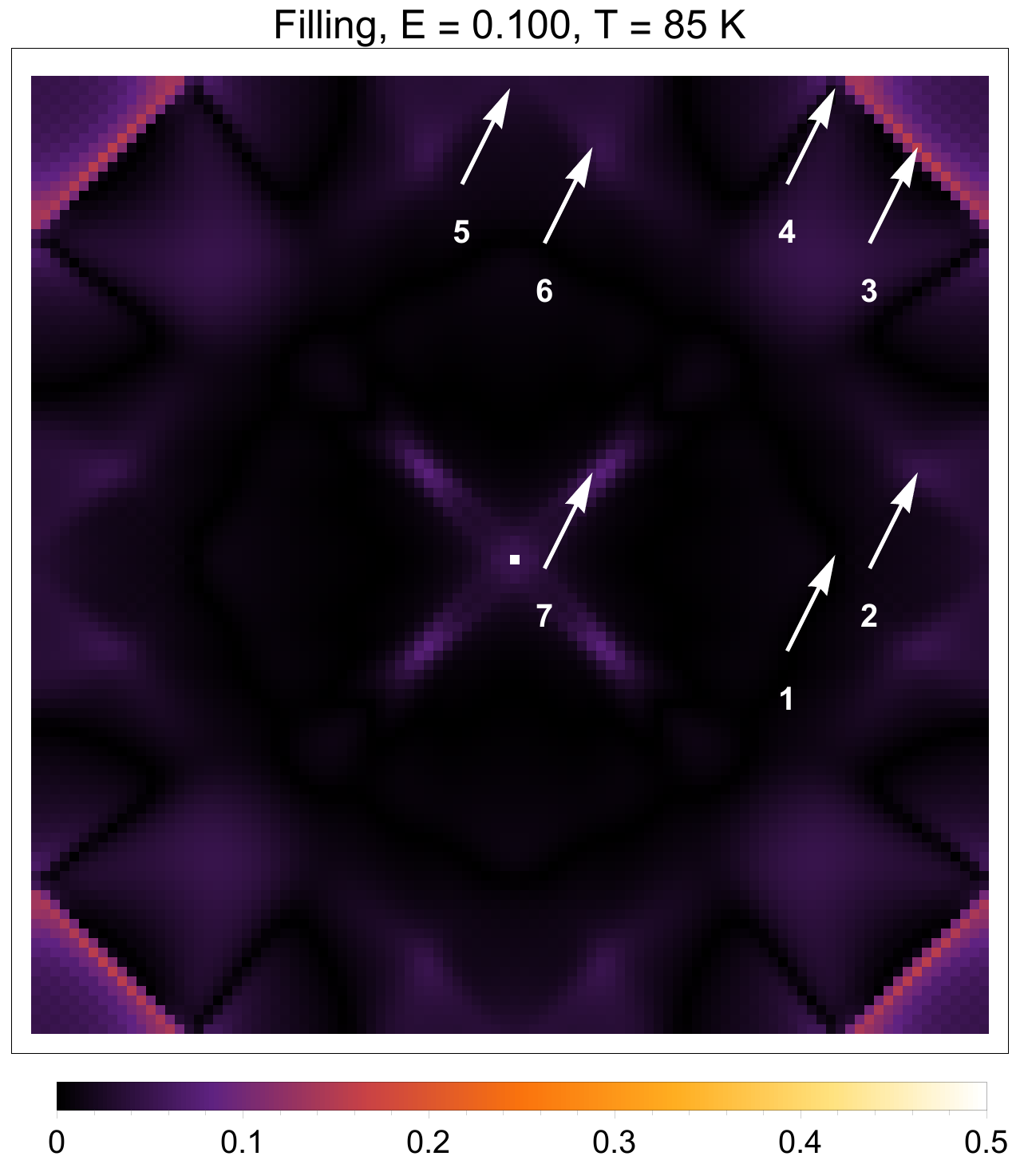}
	\includegraphics[height=0.18\textwidth]{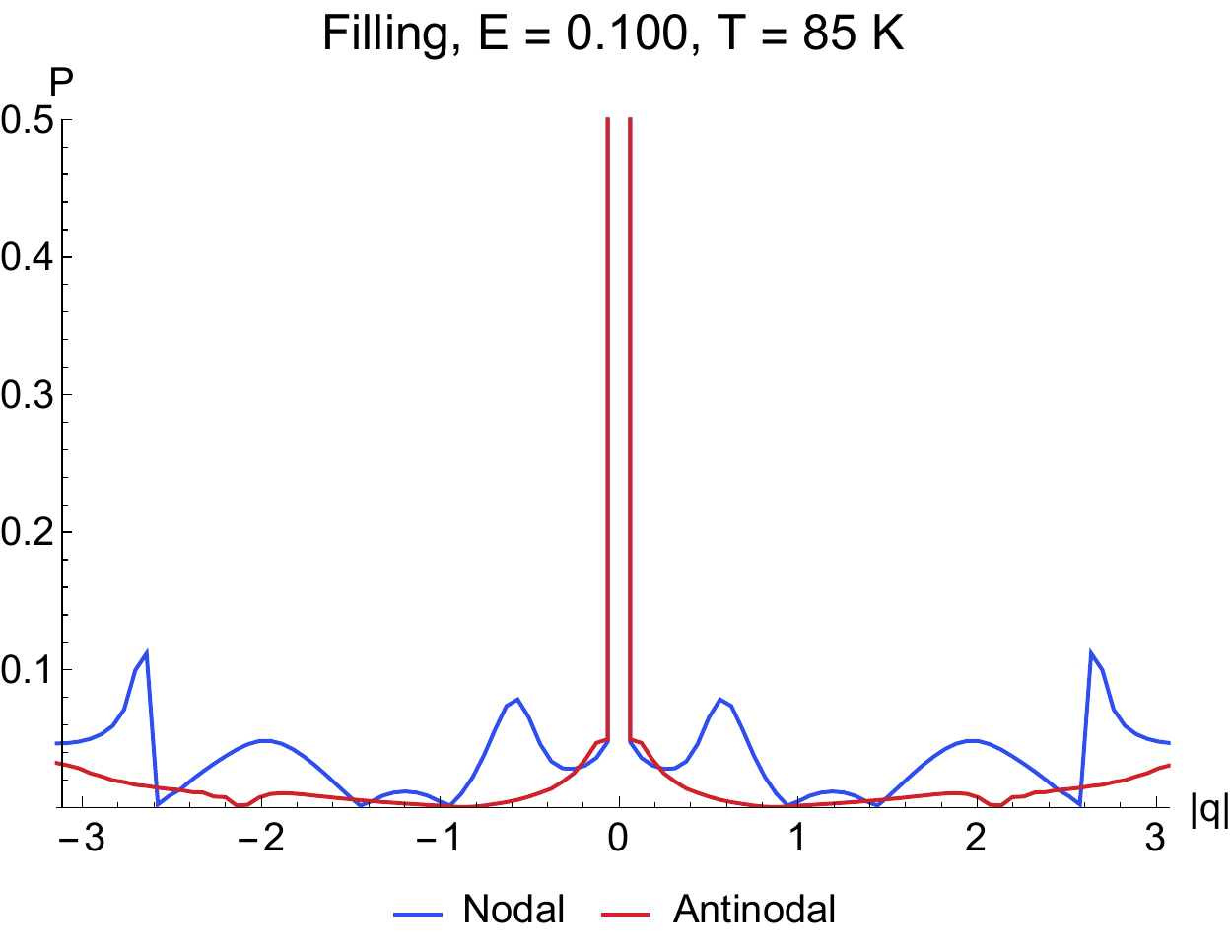}
	\includegraphics[height=0.18\textwidth]{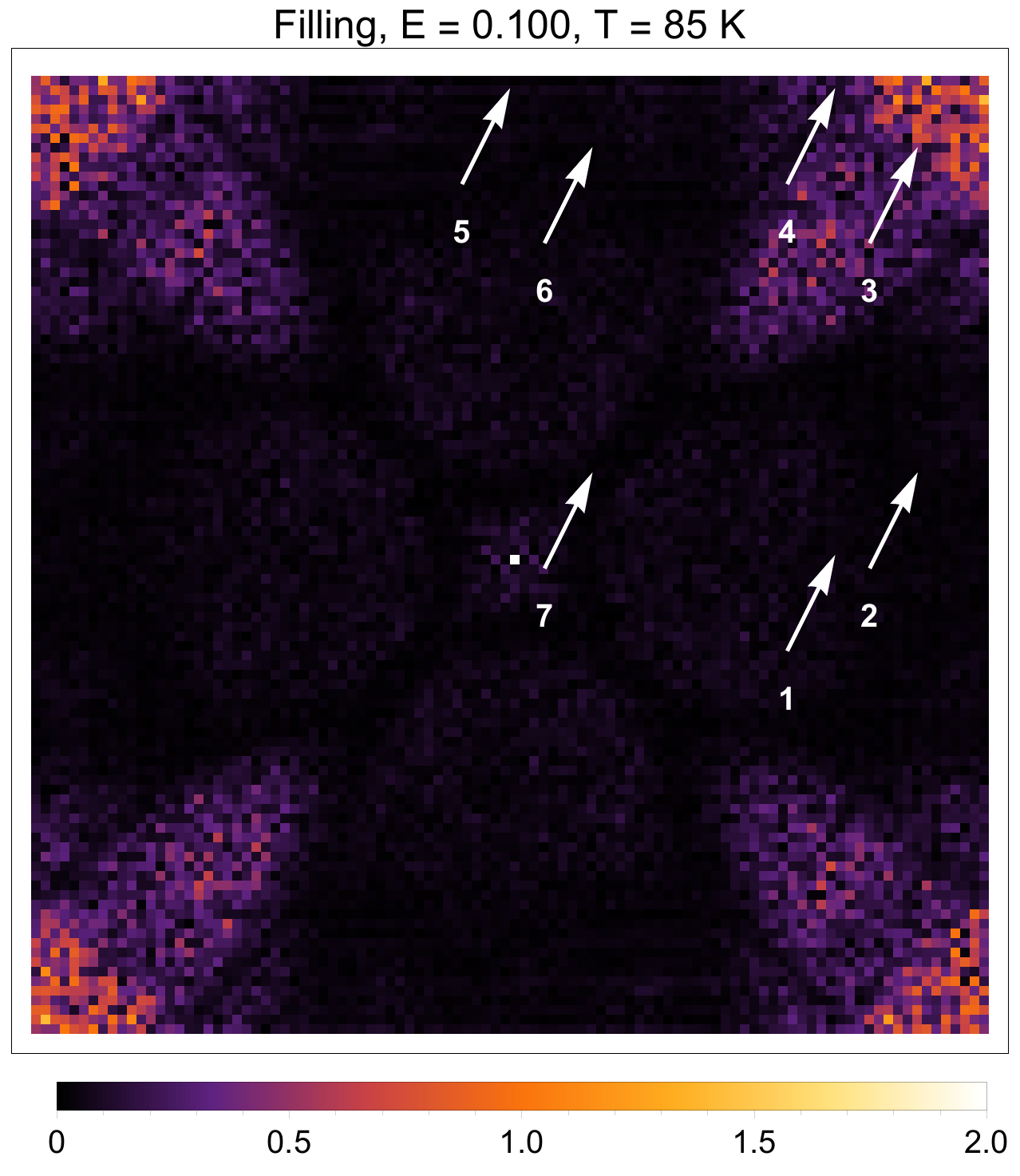}
	\includegraphics[height=0.18\textwidth]{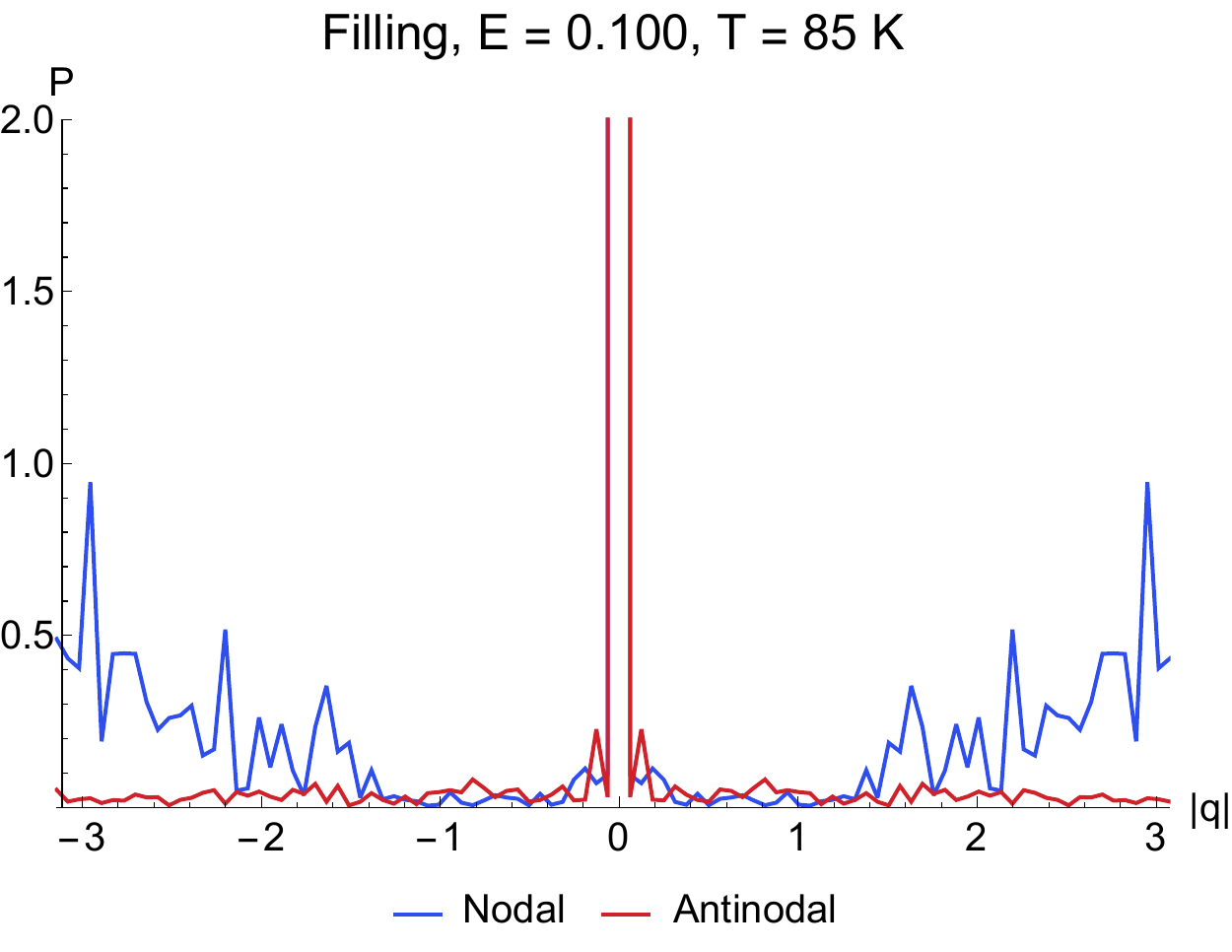} \\
	\includegraphics[height=0.18\textwidth]{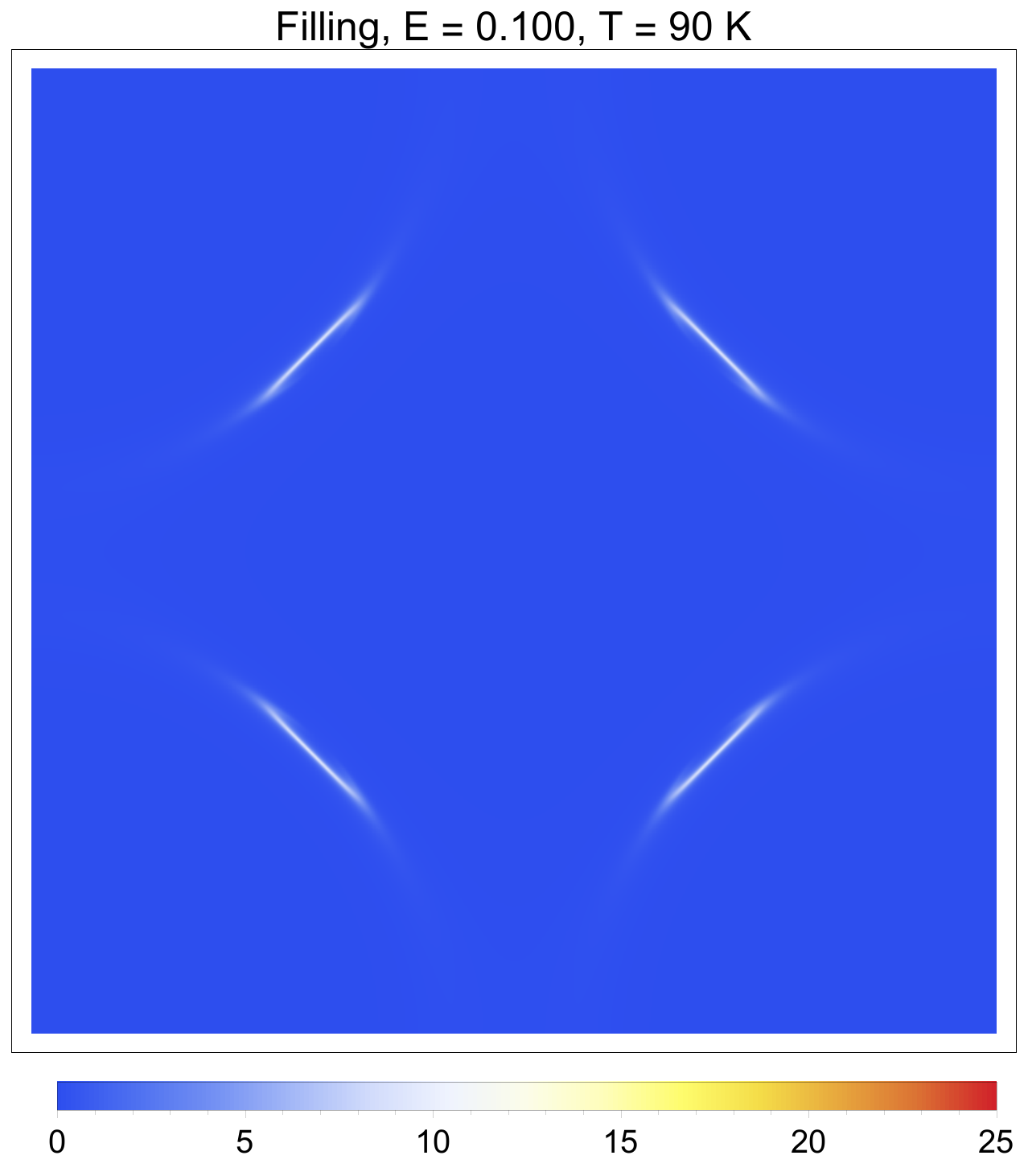}
	\includegraphics[height=0.18\textwidth]{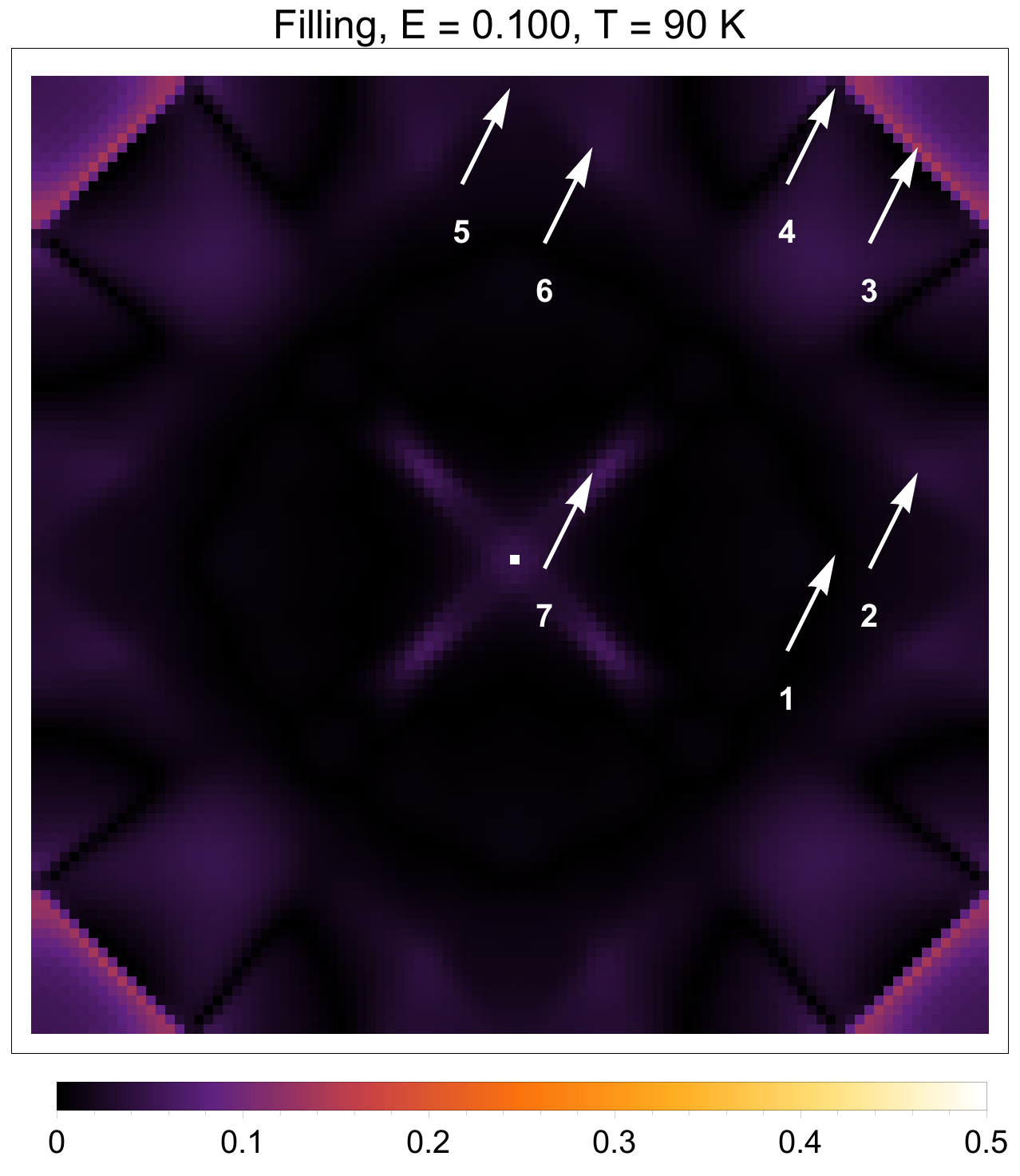}
	\includegraphics[height=0.18\textwidth]{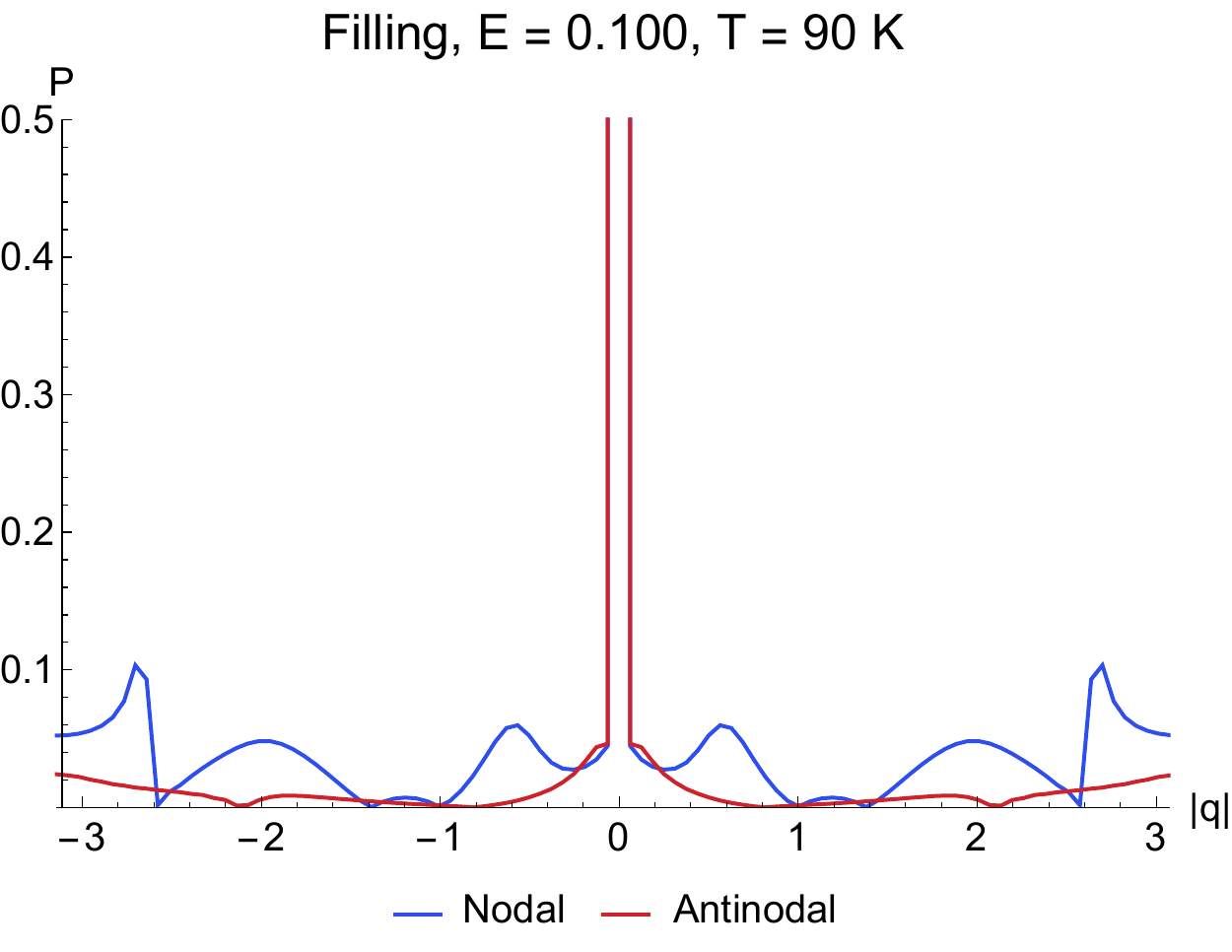}
	\includegraphics[height=0.18\textwidth]{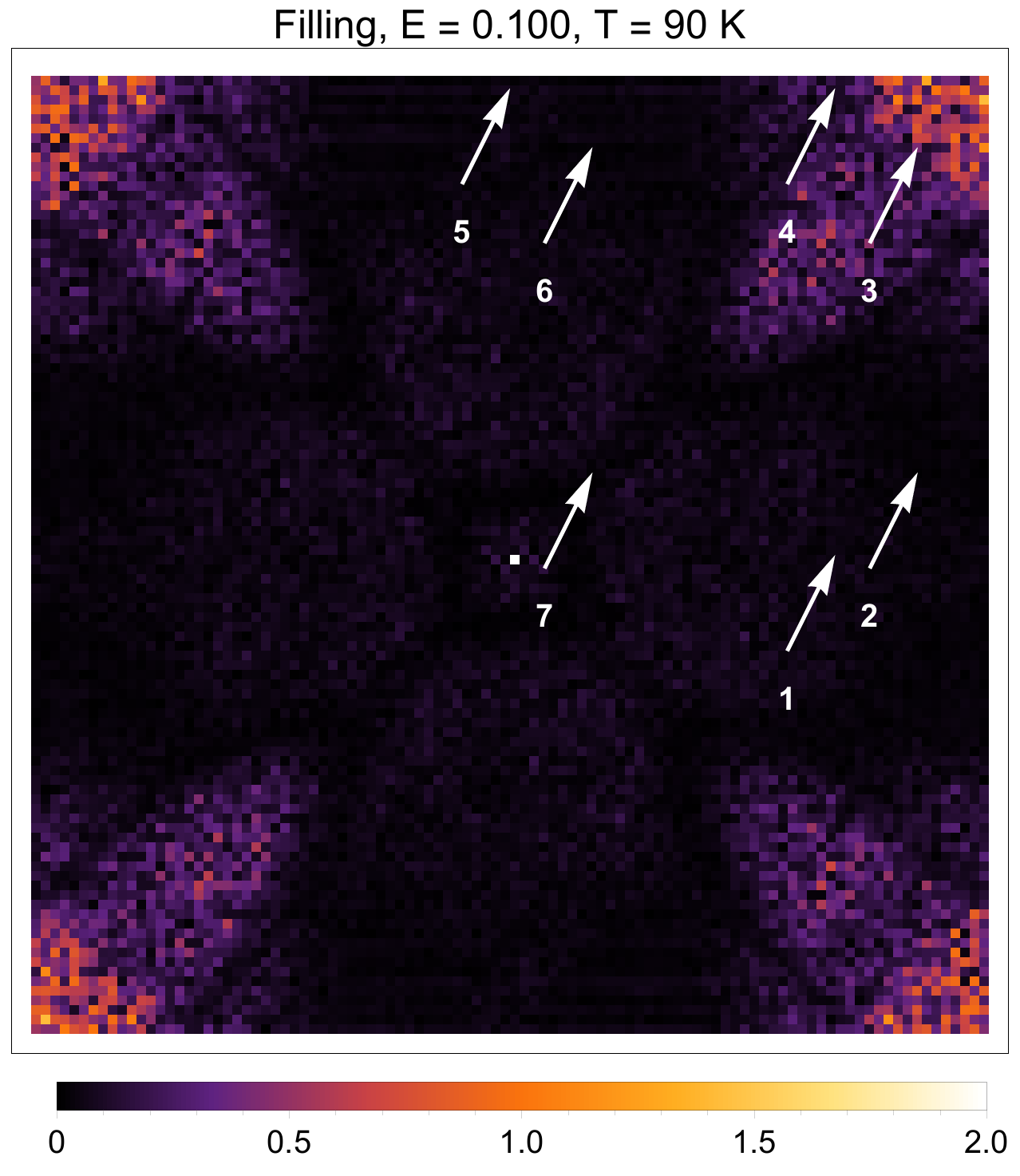}
	\includegraphics[height=0.18\textwidth]{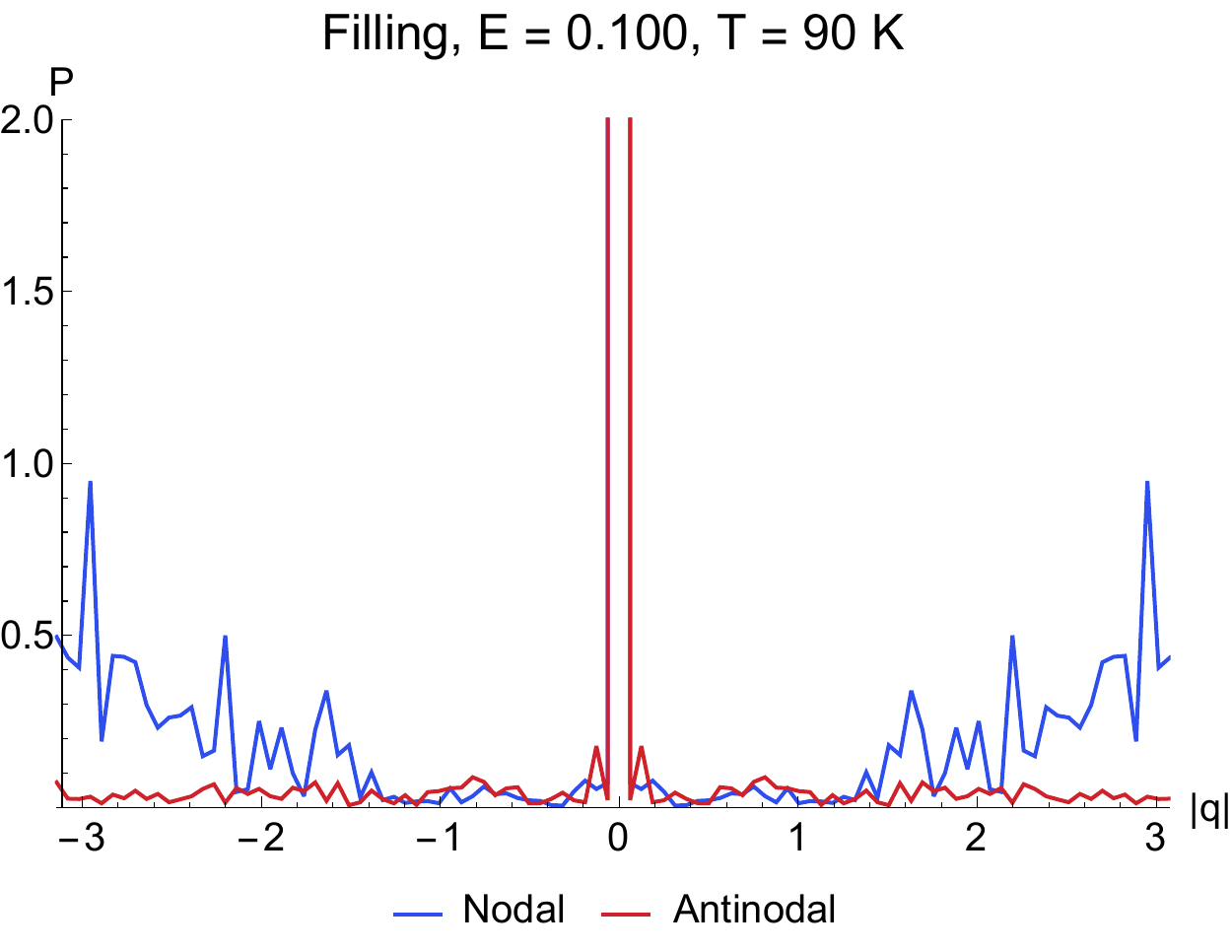} \\
	\includegraphics[height=0.18\textwidth]{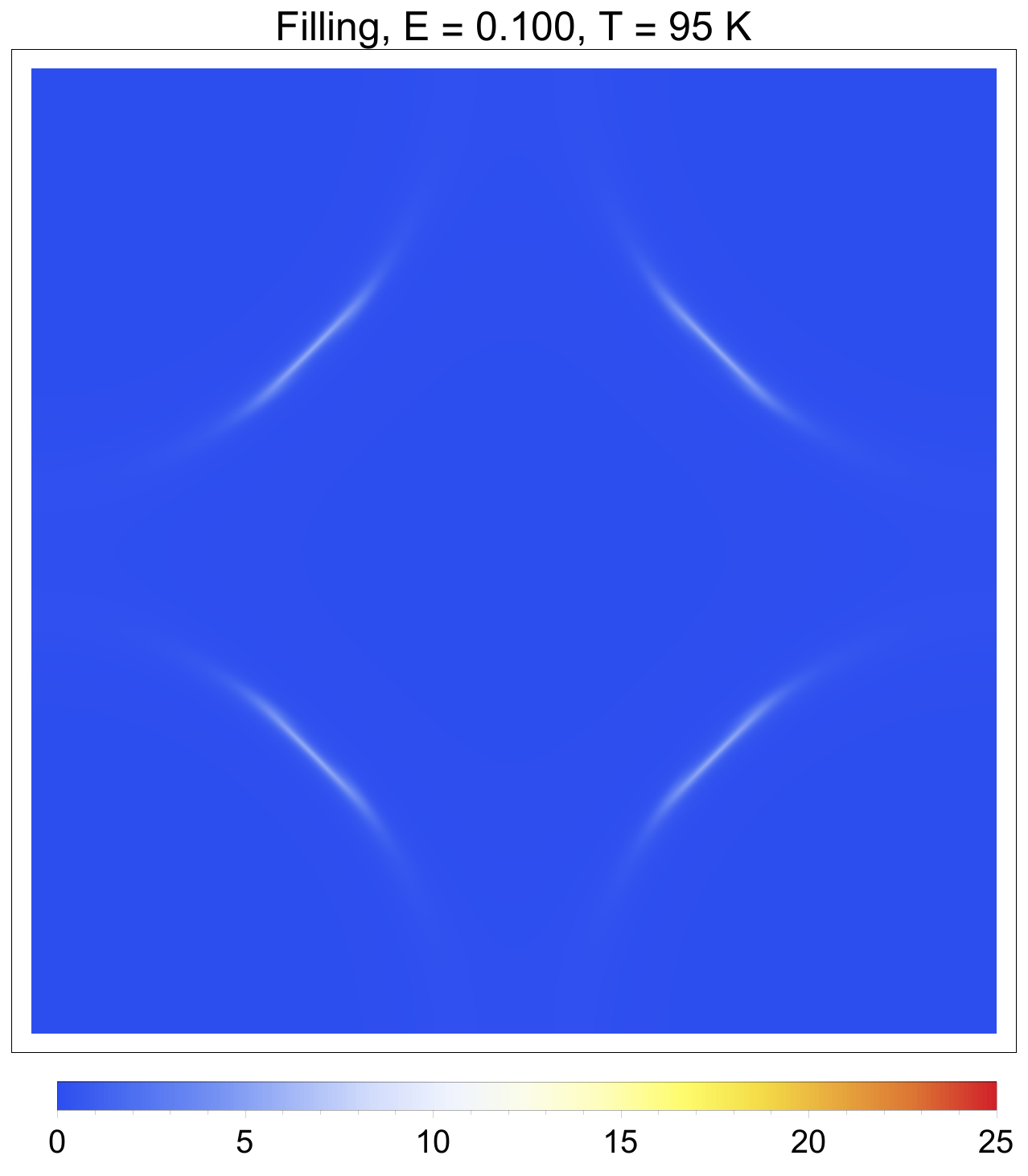}
	\includegraphics[height=0.18\textwidth]{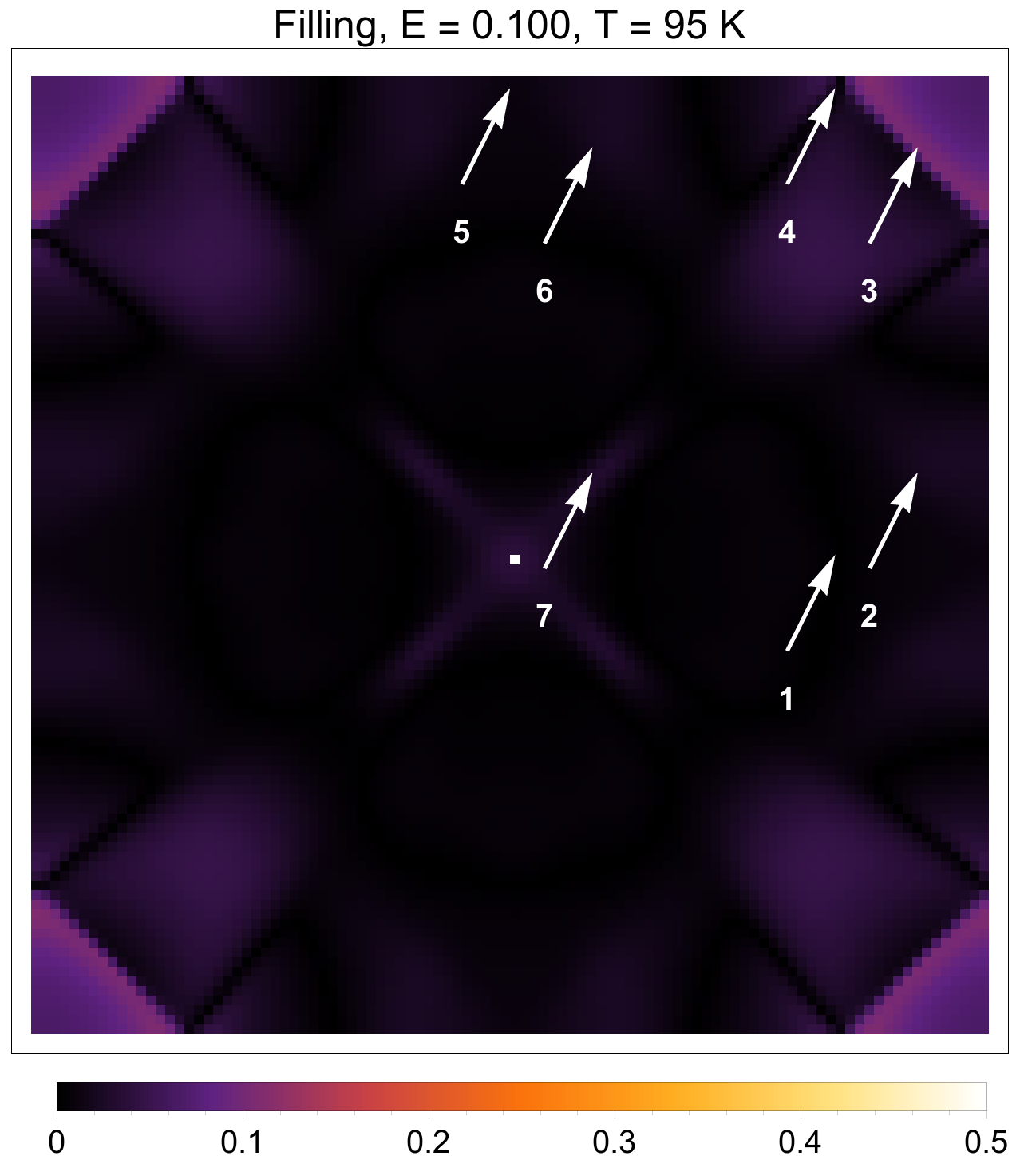}
	\includegraphics[height=0.18\textwidth]{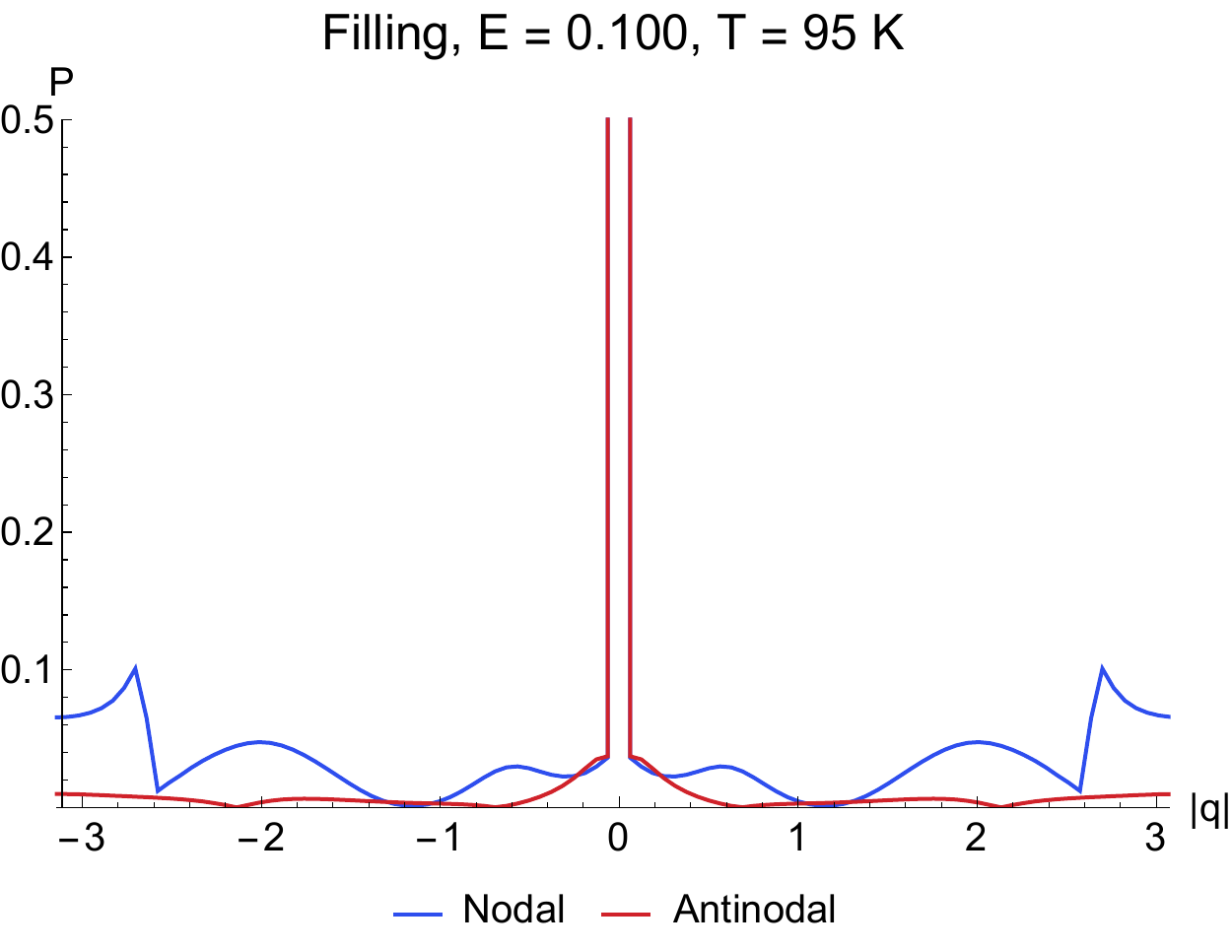}
	\includegraphics[height=0.18\textwidth]{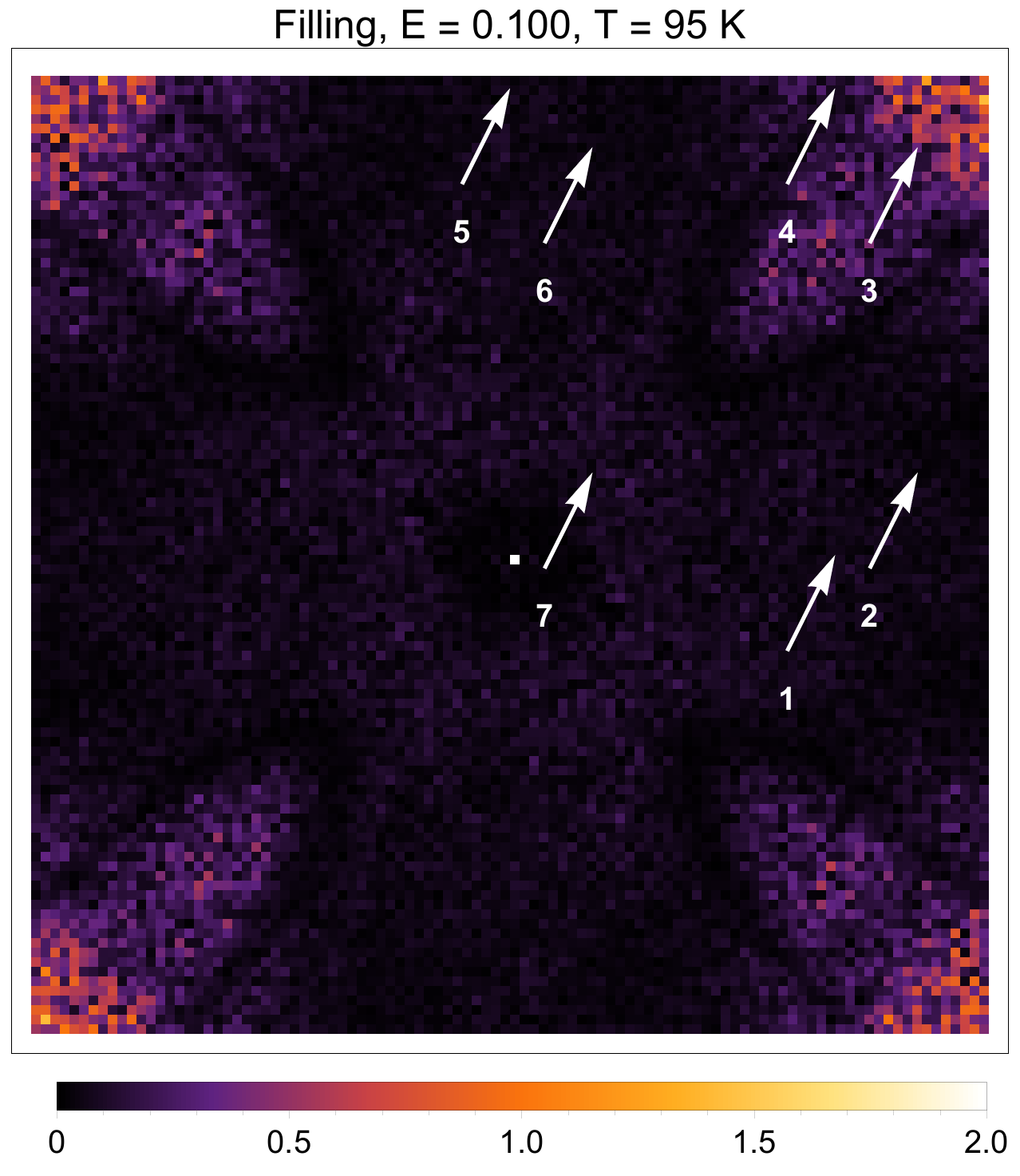}
	\includegraphics[height=0.18\textwidth]{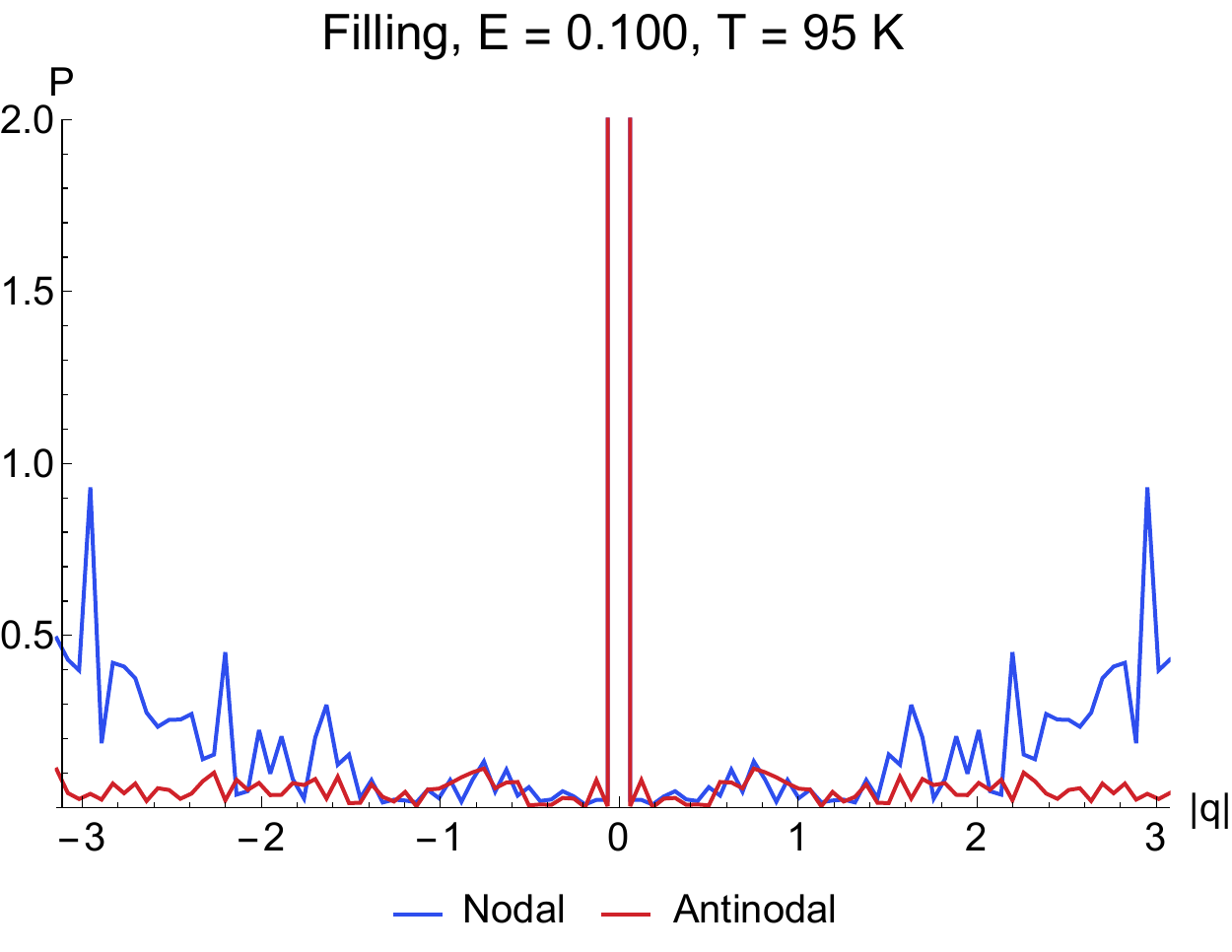} \\
	\includegraphics[height=0.18\textwidth]{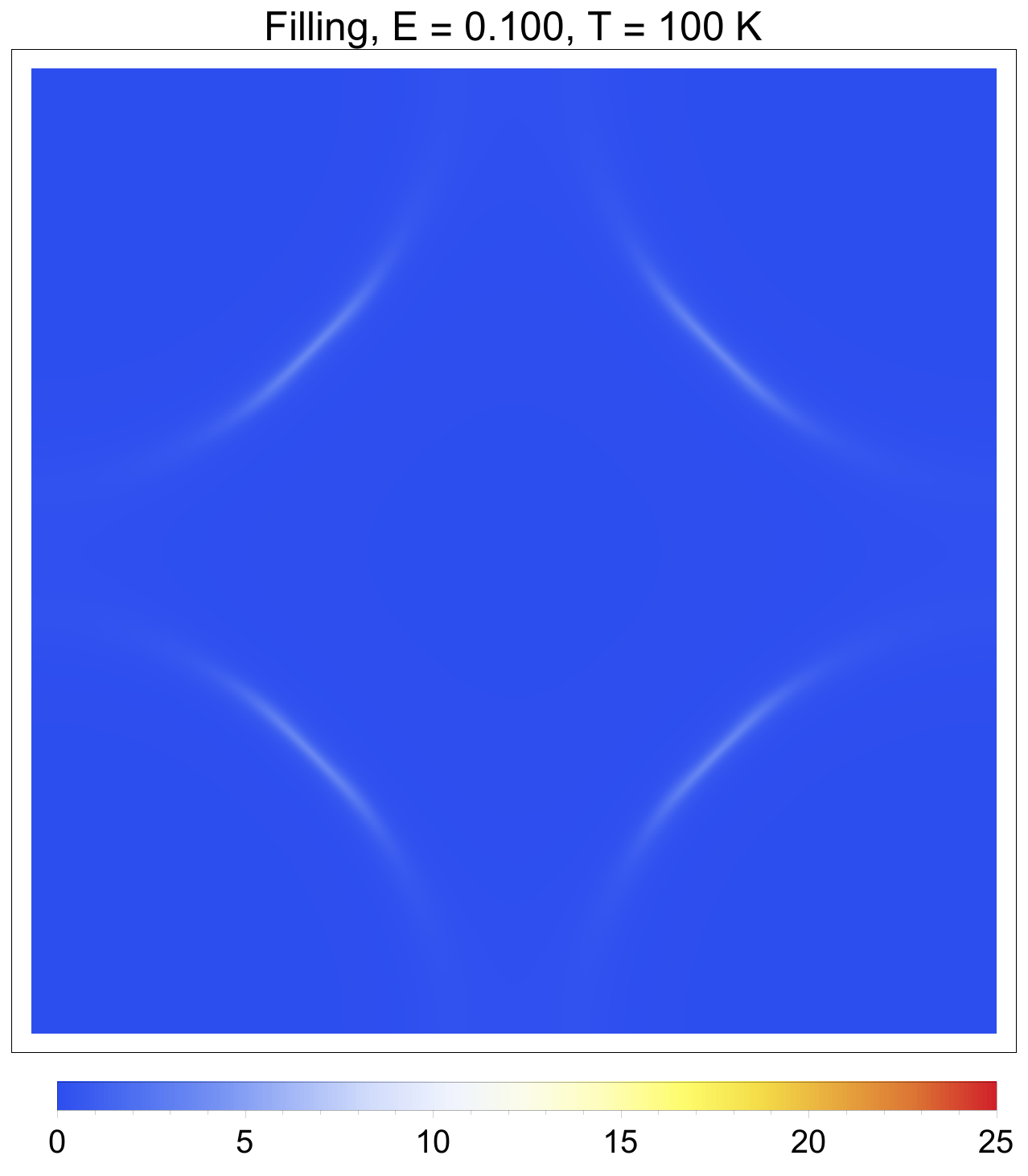}
	\includegraphics[height=0.18\textwidth]{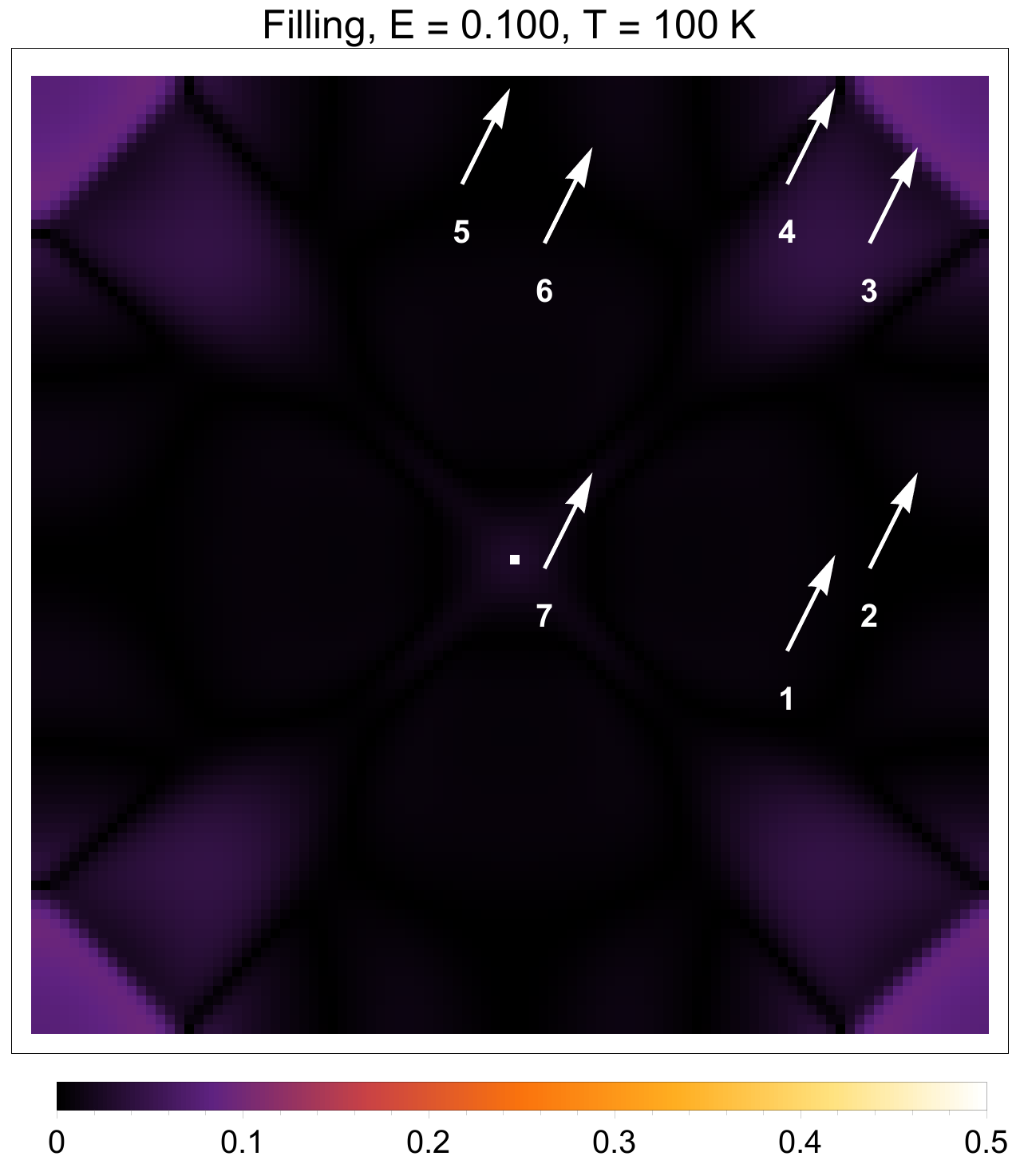}
	\includegraphics[height=0.18\textwidth]{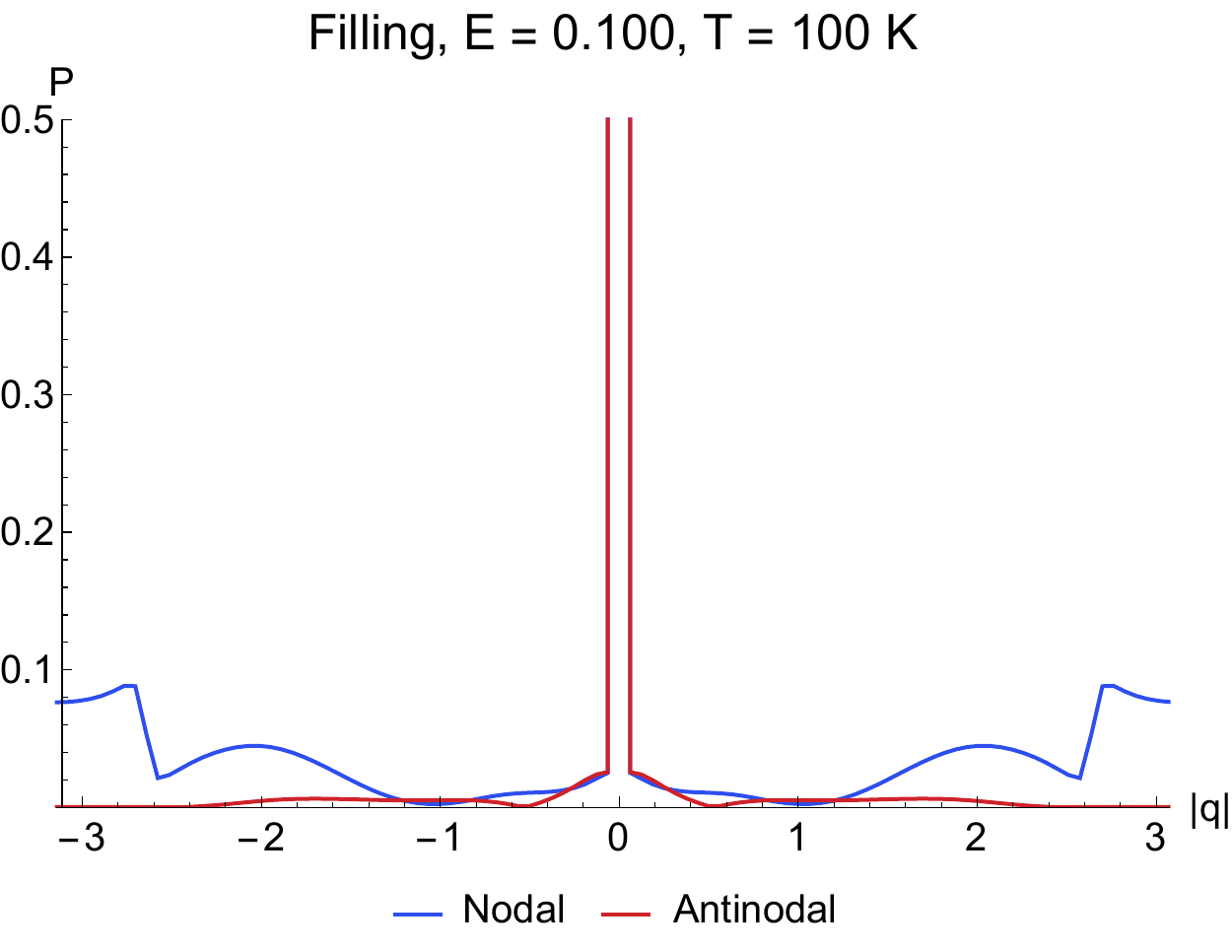}
	\includegraphics[height=0.18\textwidth]{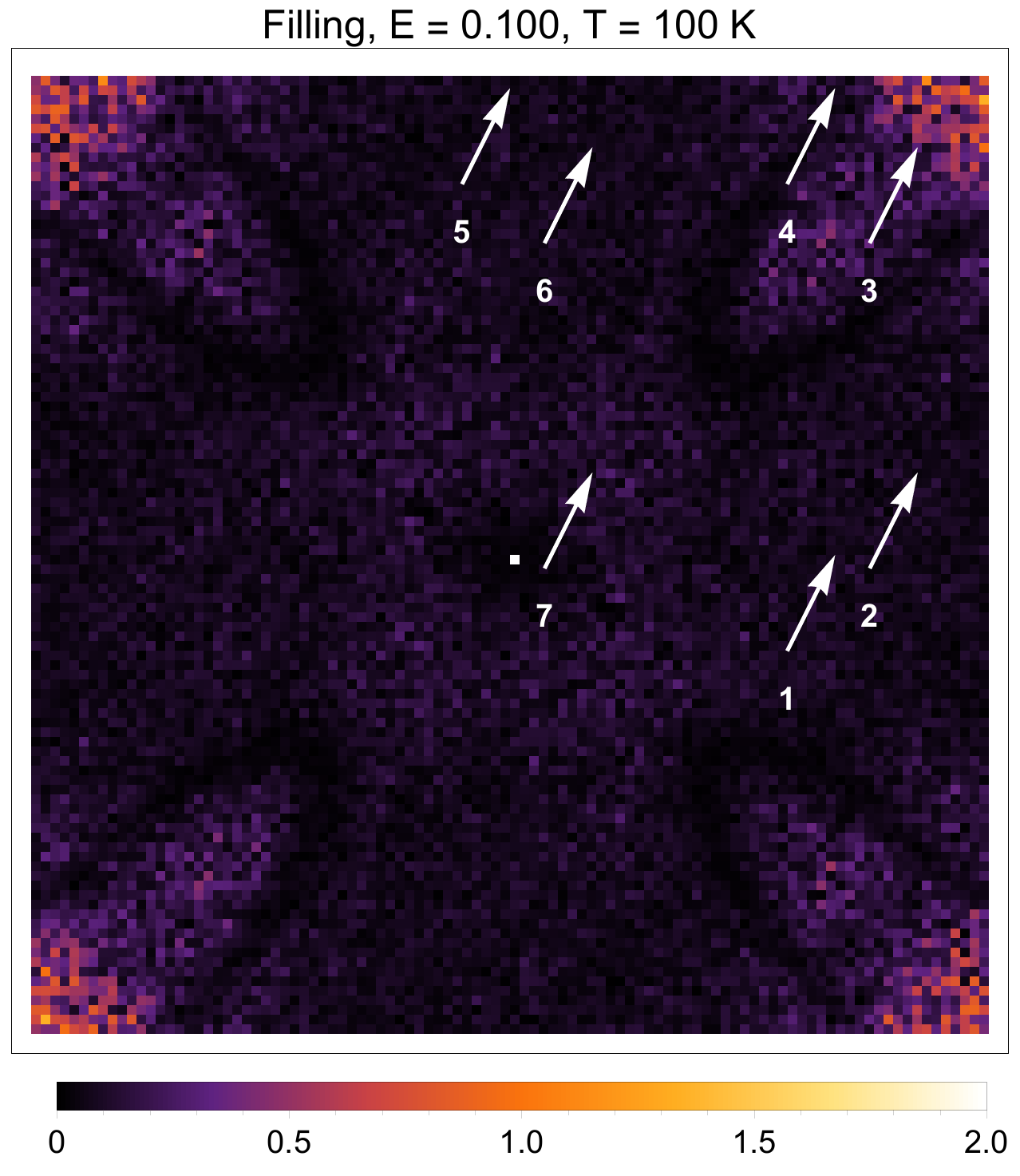}
	\includegraphics[height=0.18\textwidth]{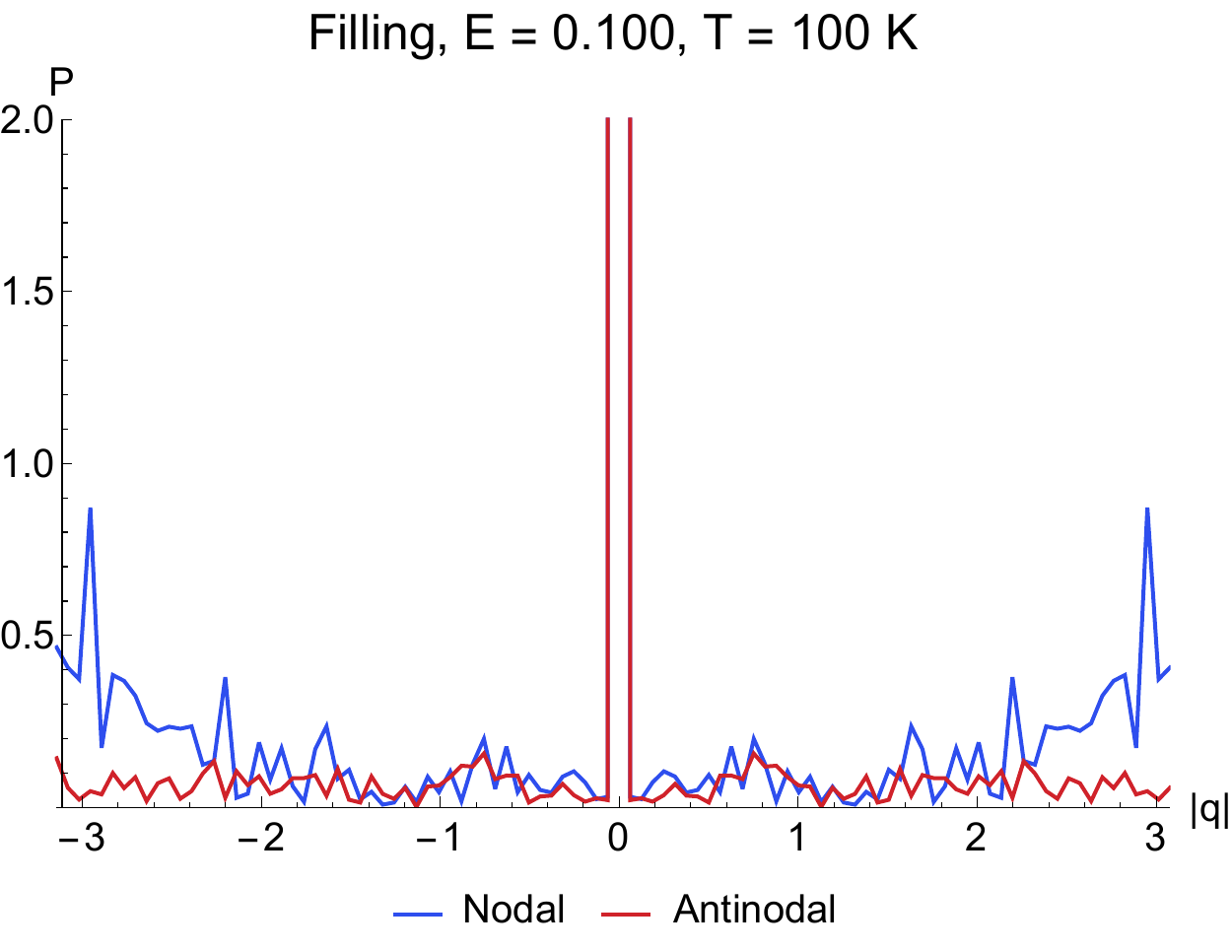} \\
	
	\caption{Gap-filling phenomenology at various temperatures. Left to right: The spectral function $A(\mathbf{k}, \omega)$; the Fourier transform of the LDOS $P(\mathbf{q}, \omega)$; linecuts of $P(\mathbf{q}, \omega)$ in the nodal and antinodal directions; $P(\mathbf{q}, \omega)$ in the presence of multiple weak impurities and finite-temperature smearing; and linecuts of $P(\mathbf{q}, \omega)$ in the presence of multiple weak impurities and finite-temperature smearing. Arrows indicate the locations of the peaks predicted by the octet model. All plots are taken at $E = 0.100$.}
	\label{fig:temperature_cg}
\end{figure*}

\begin{figure*}
	\centering
	
	\includegraphics[width=0.16\textwidth]{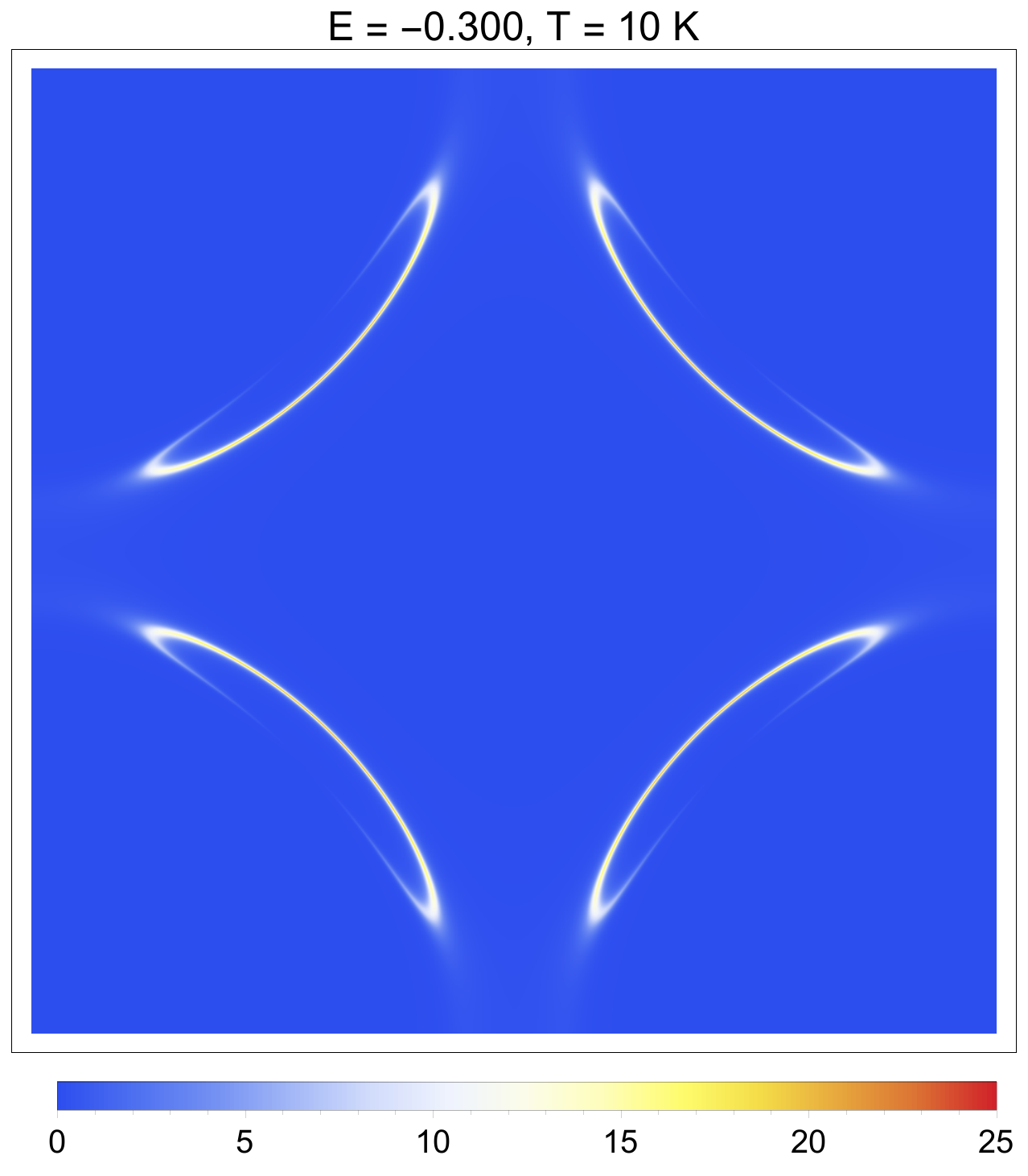}
	\includegraphics[width=0.16\textwidth]{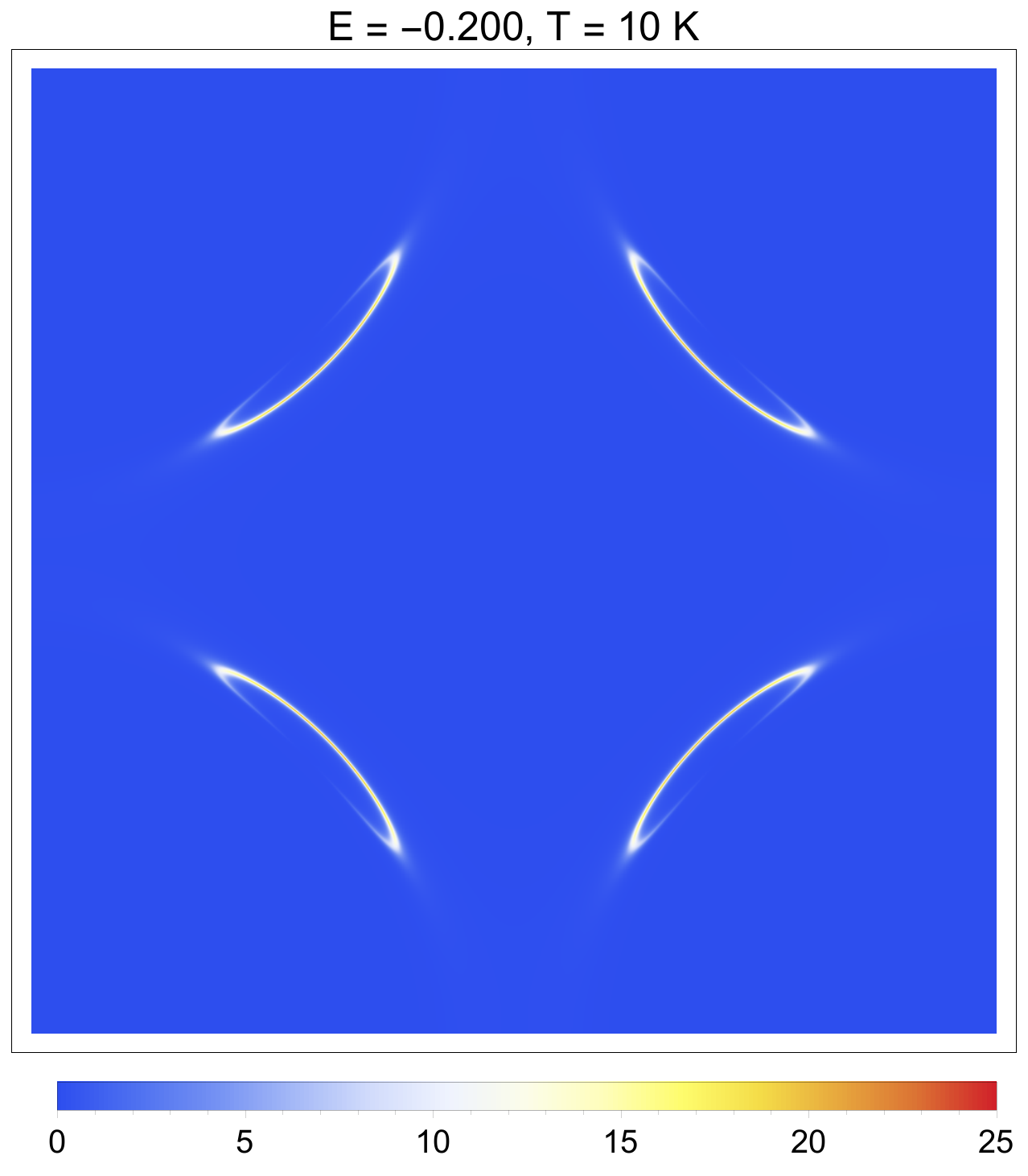}
	\includegraphics[width=0.16\textwidth]{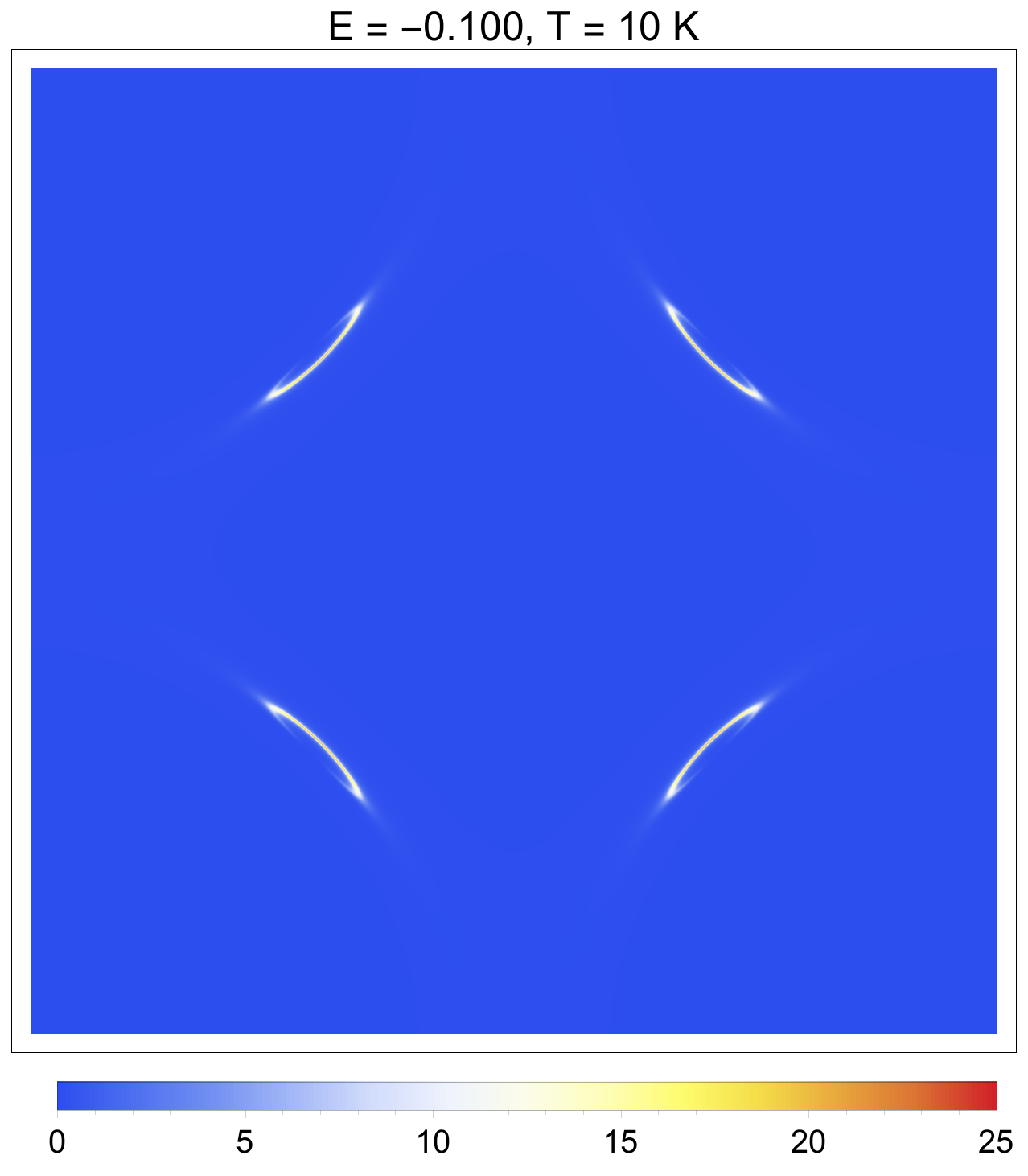}
	\includegraphics[width=0.16\textwidth]{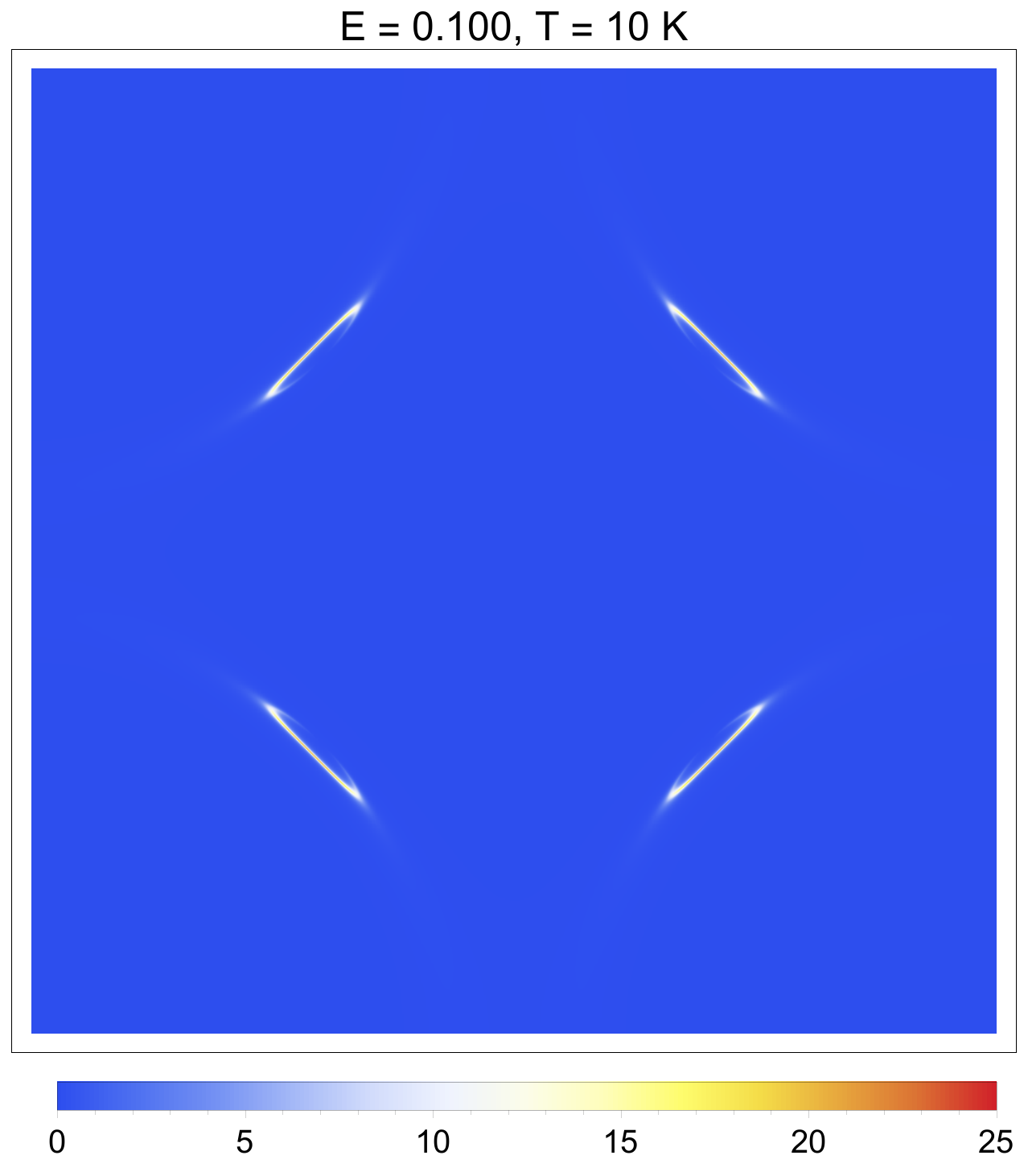}
	\includegraphics[width=0.16\textwidth]{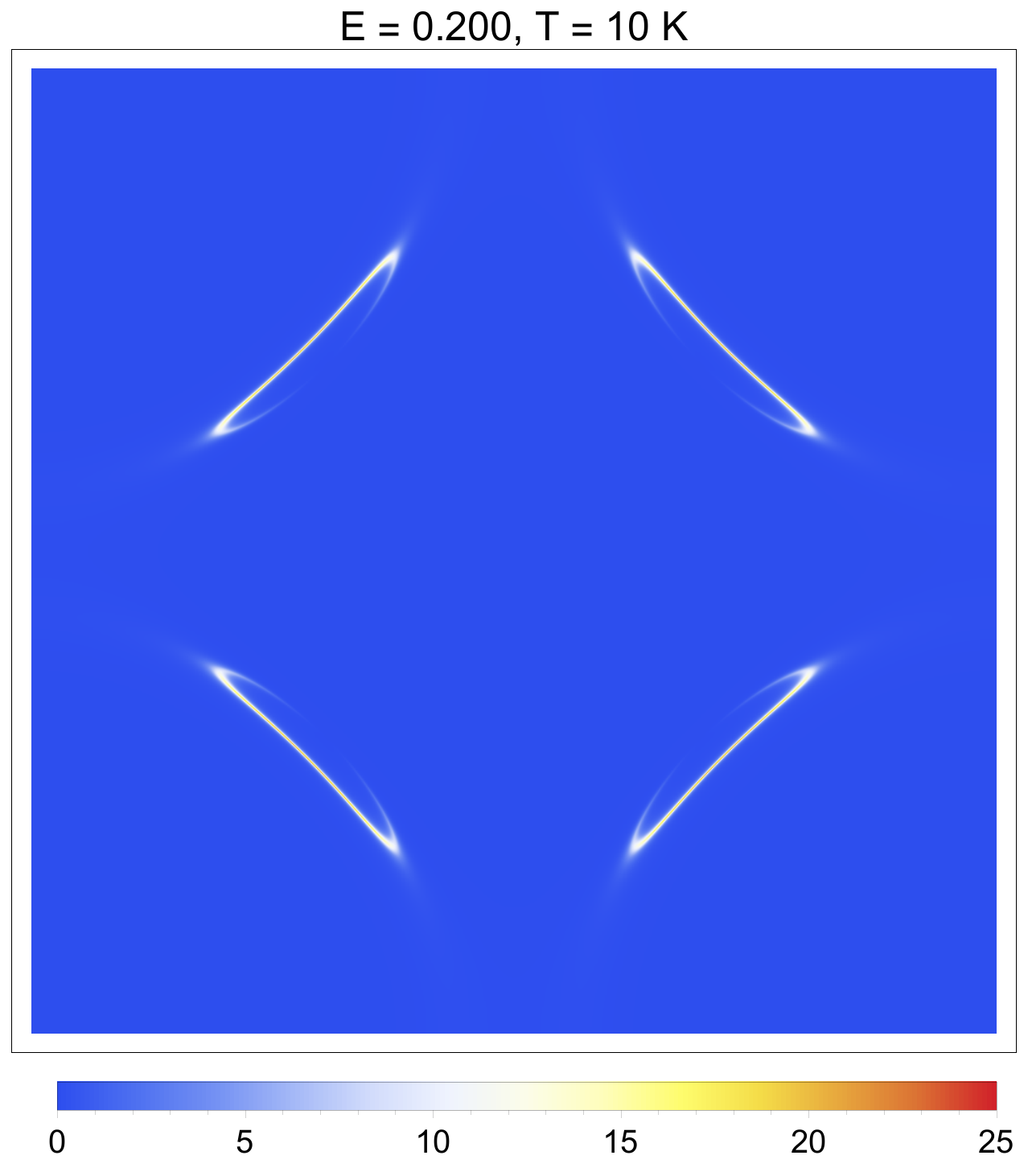}
	\includegraphics[width=0.16\textwidth]{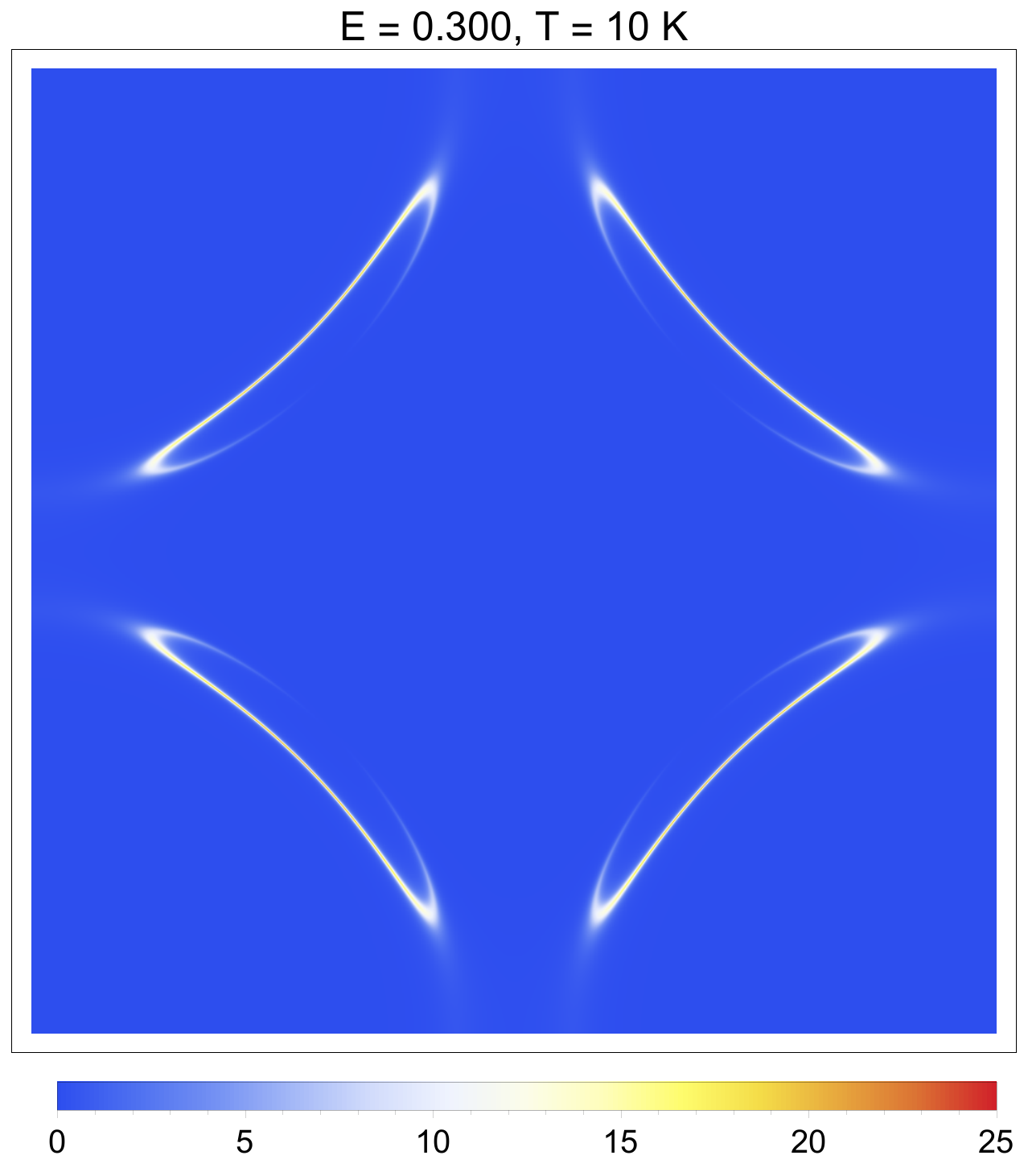} \\
	\includegraphics[width=0.16\textwidth]{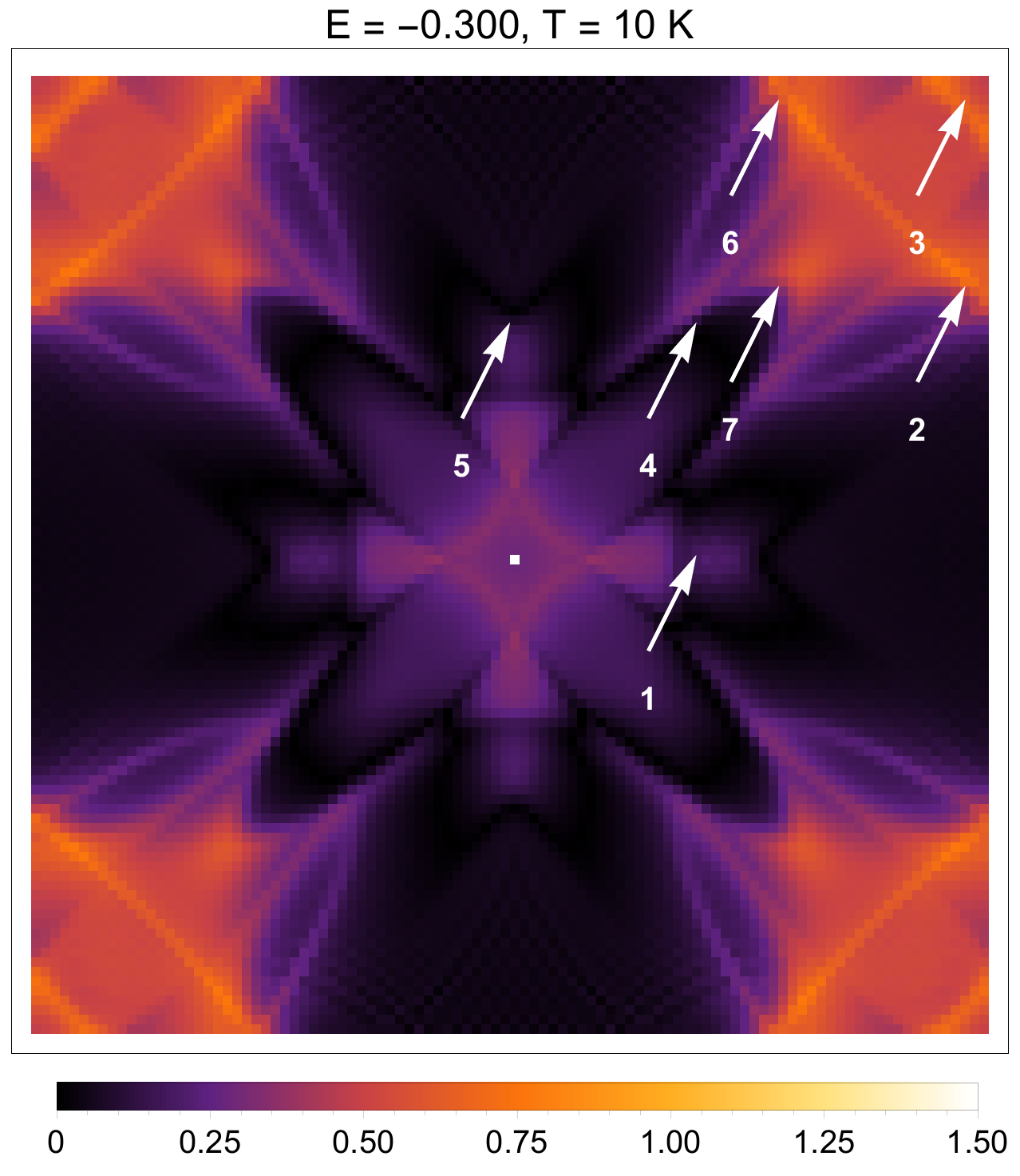}
	\includegraphics[width=0.16\textwidth]{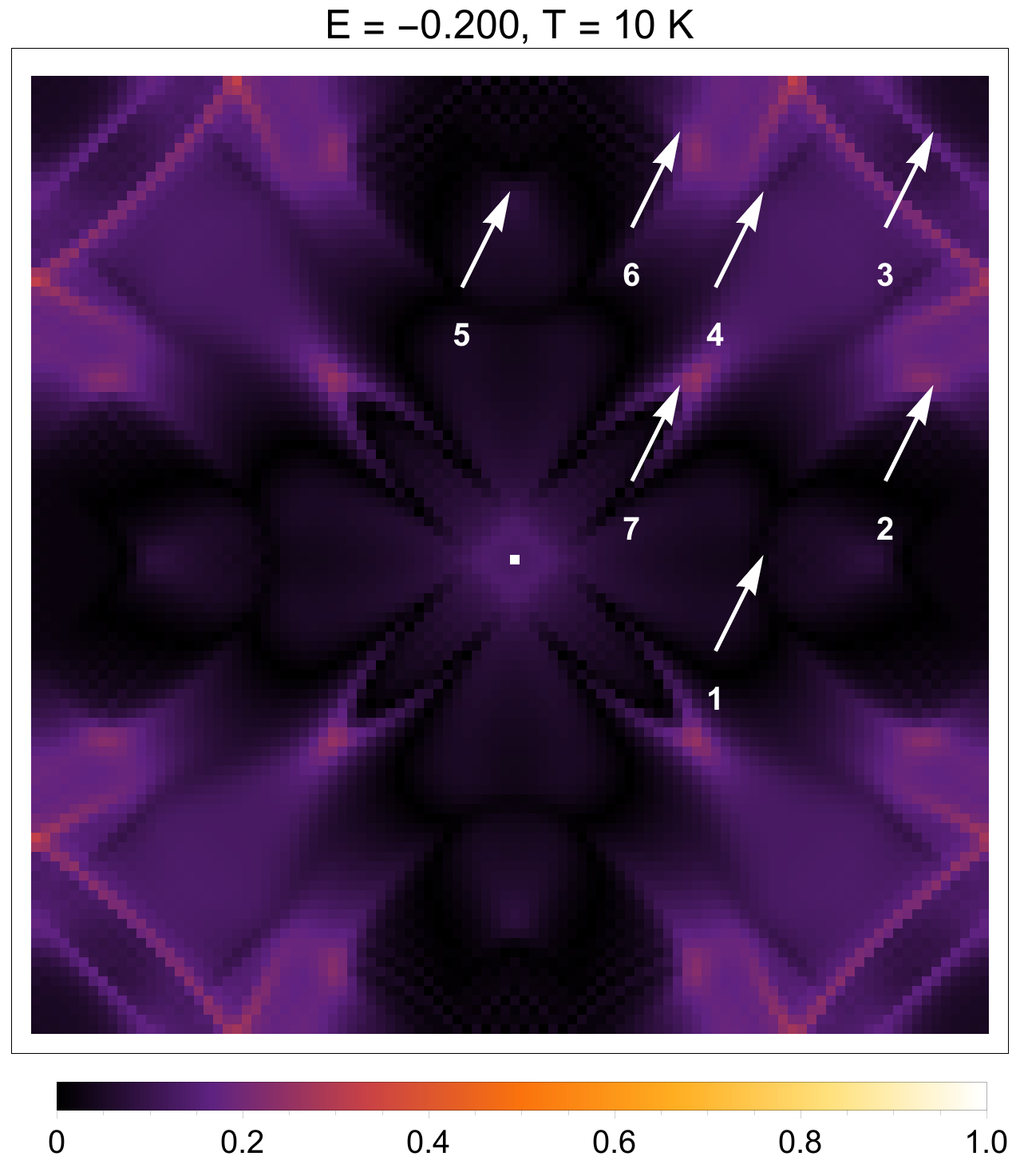}
	\includegraphics[width=0.16\textwidth]{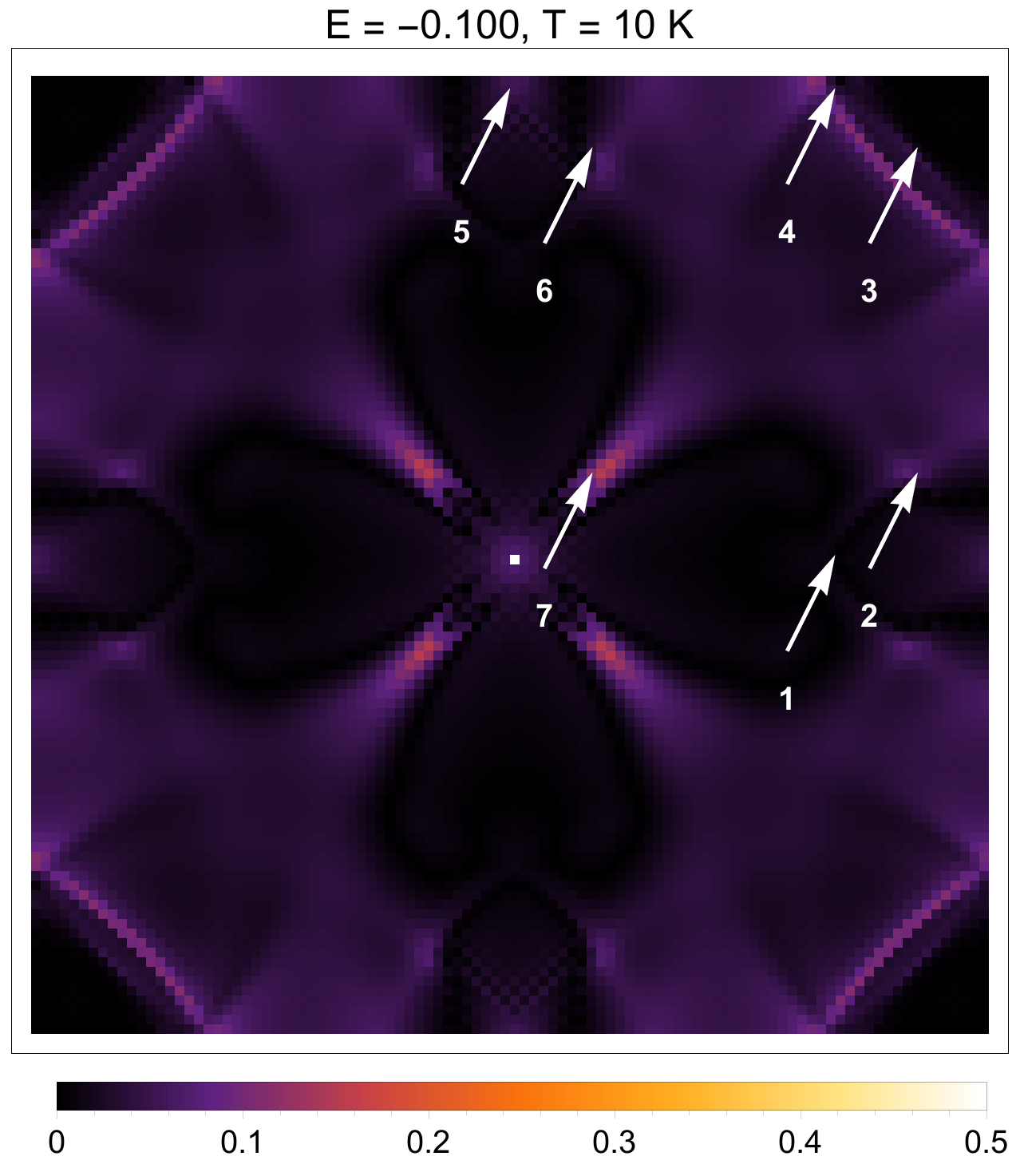}
	\includegraphics[width=0.16\textwidth]{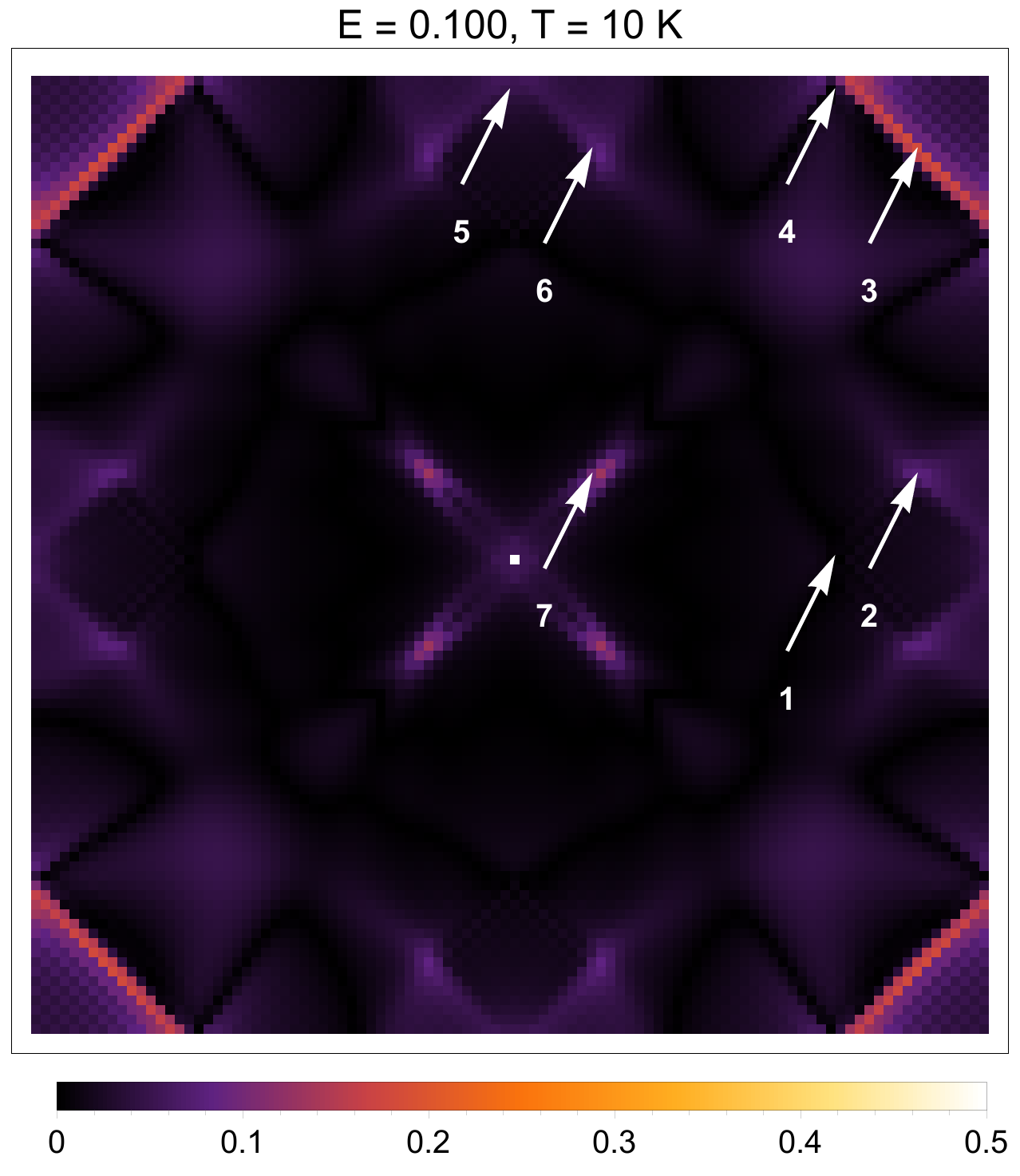}
	\includegraphics[width=0.16\textwidth]{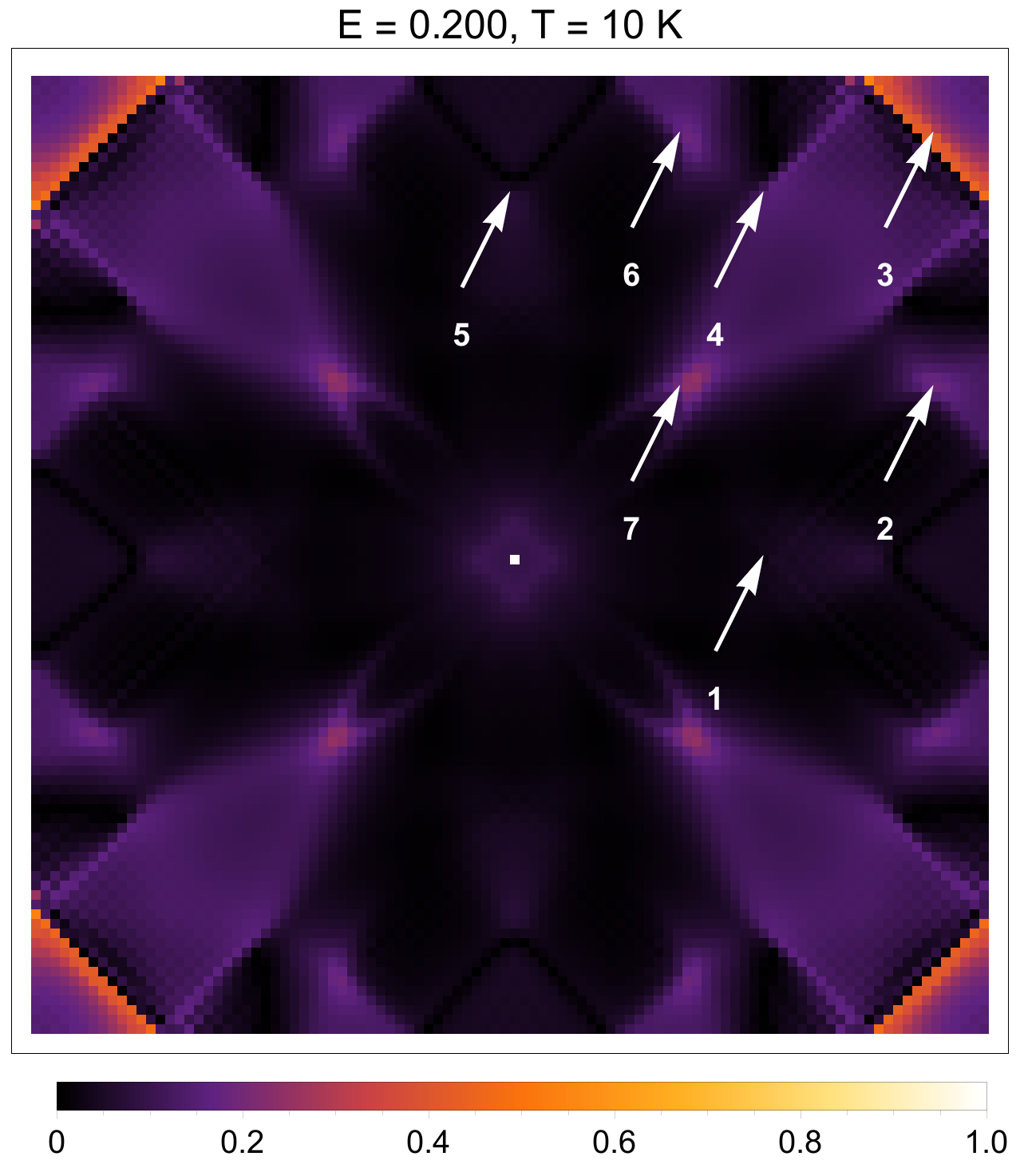}
	\includegraphics[width=0.16\textwidth]{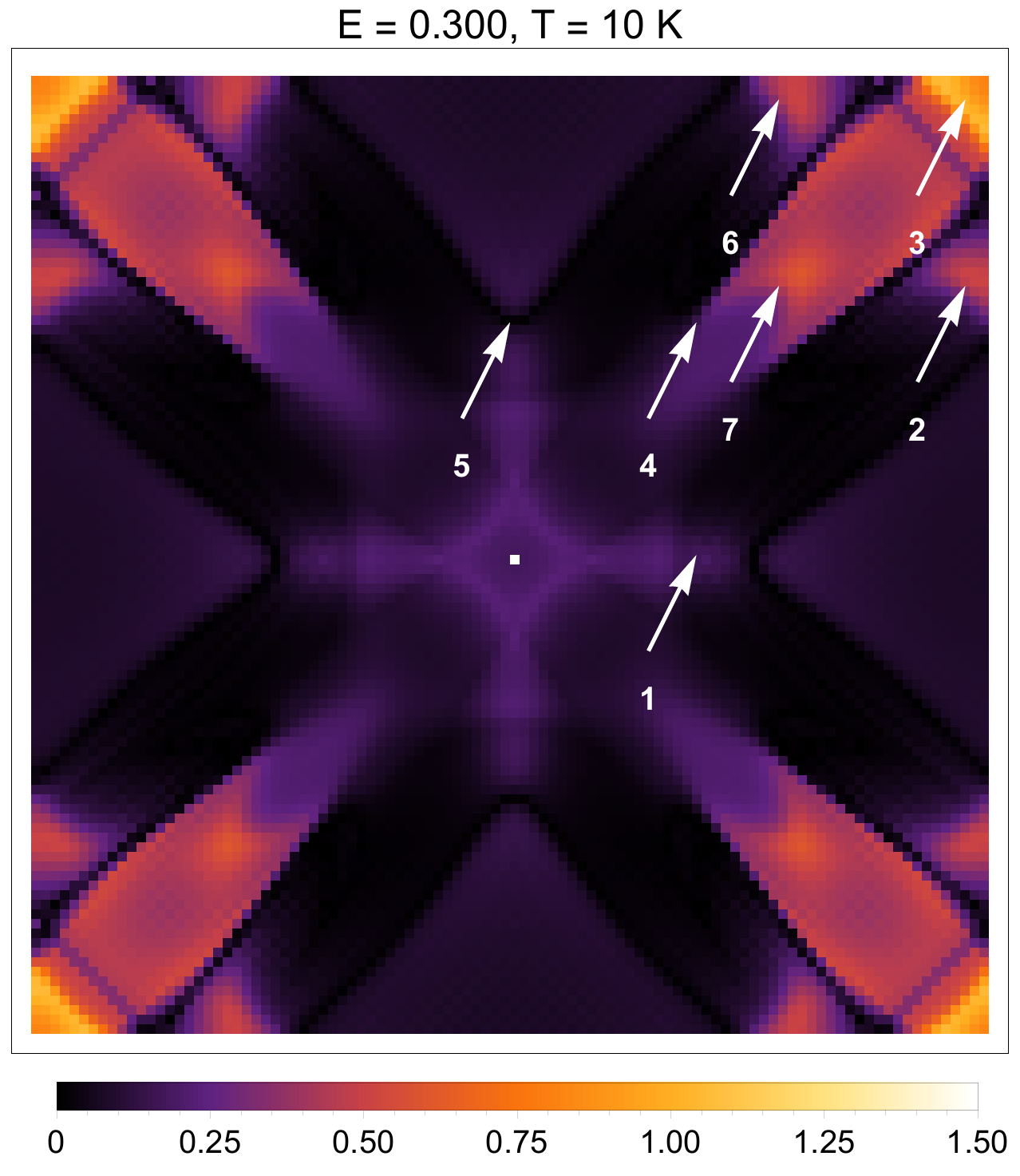} \\
	\includegraphics[width=0.16\textwidth]{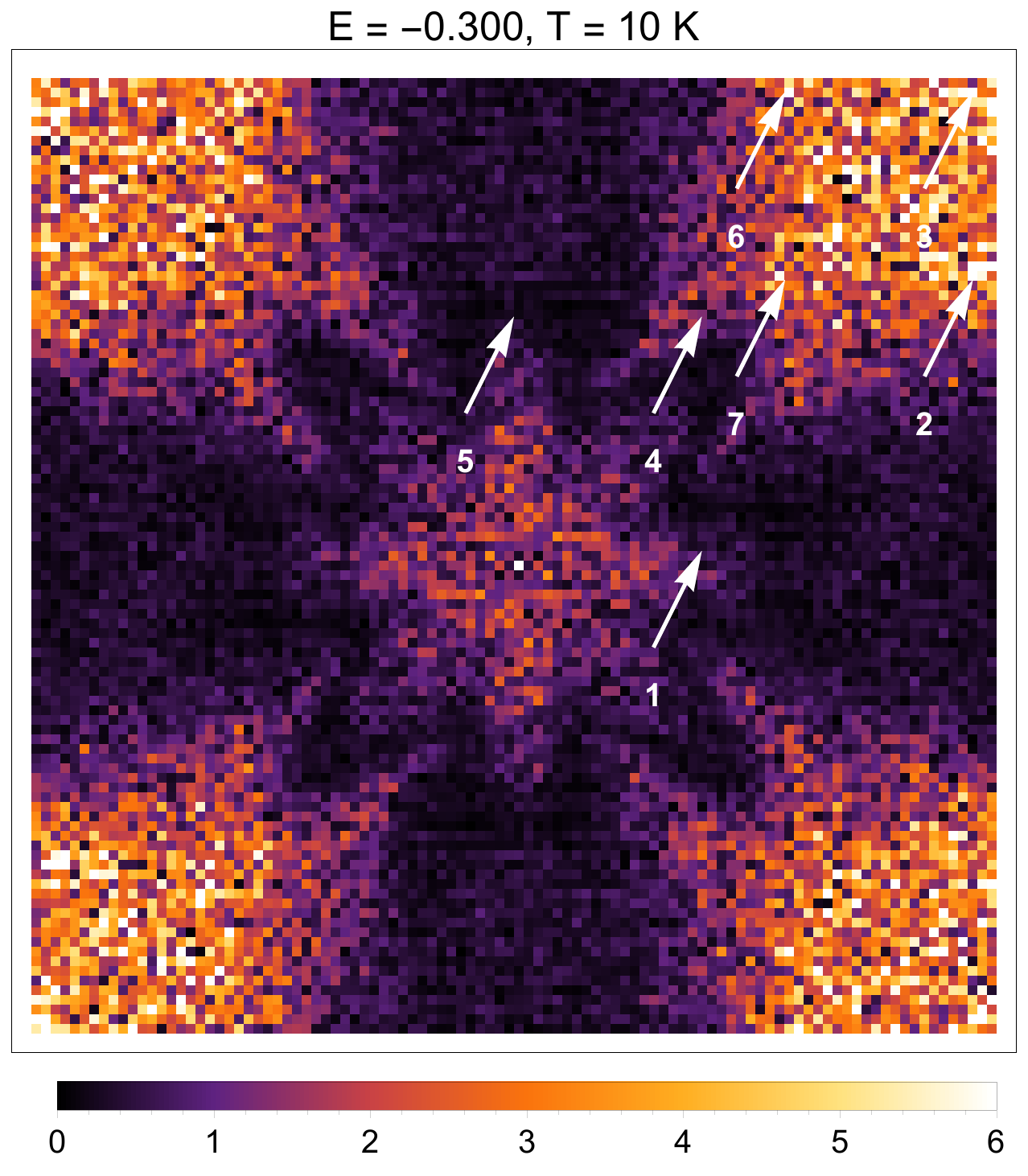}
	\includegraphics[width=0.16\textwidth]{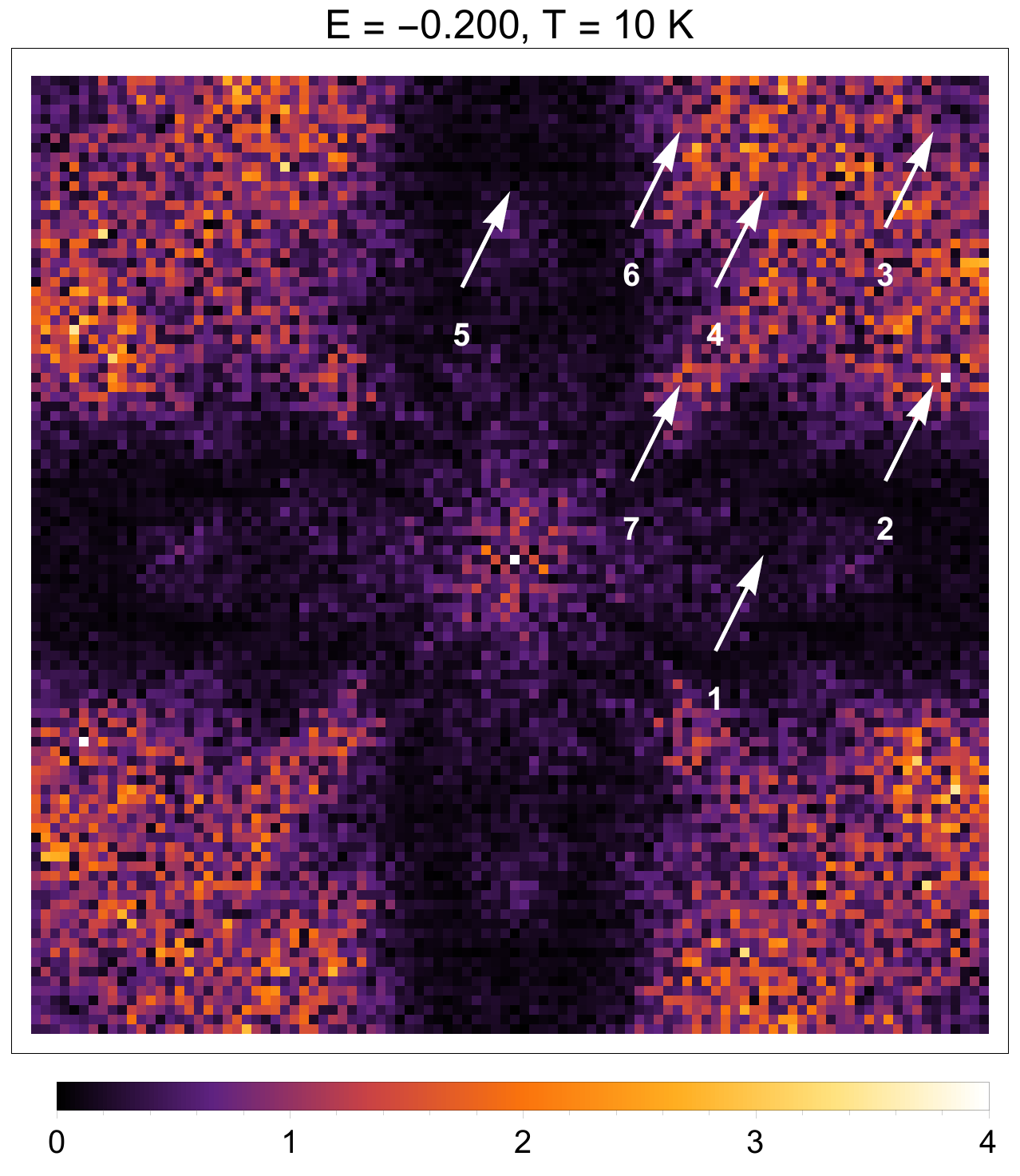}
	\includegraphics[width=0.16\textwidth]{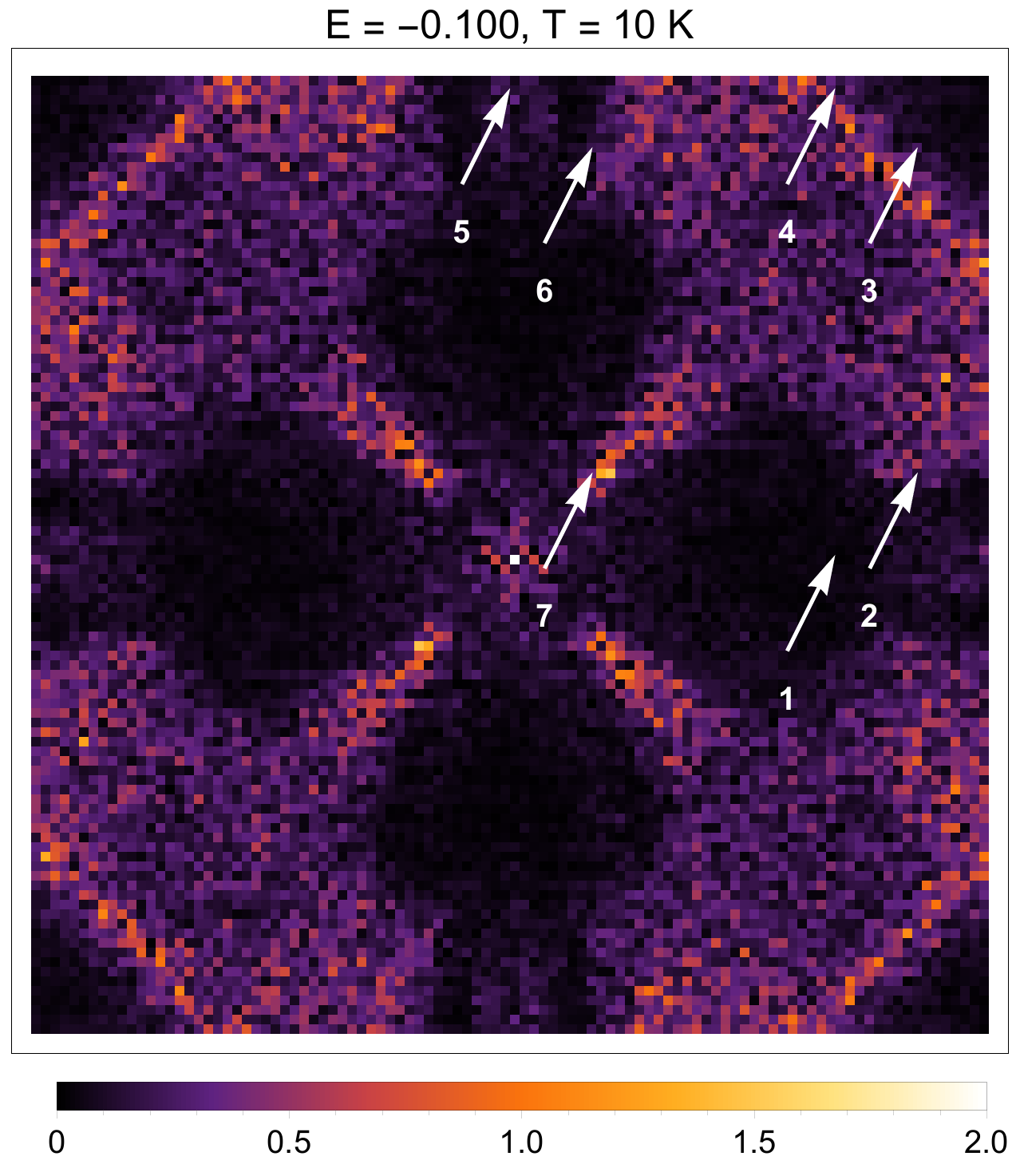}
	\includegraphics[width=0.16\textwidth]{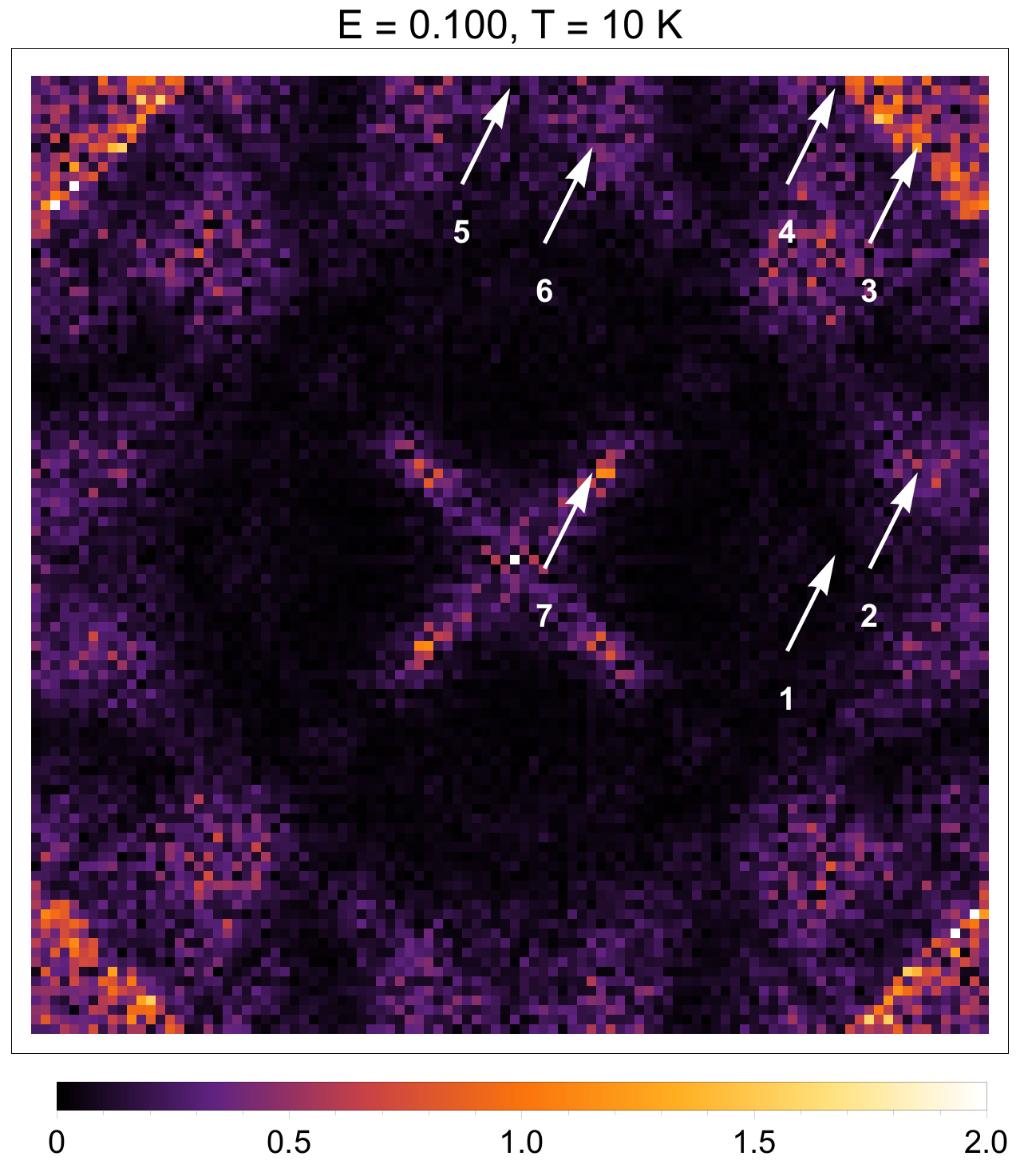}
	\includegraphics[width=0.16\textwidth]{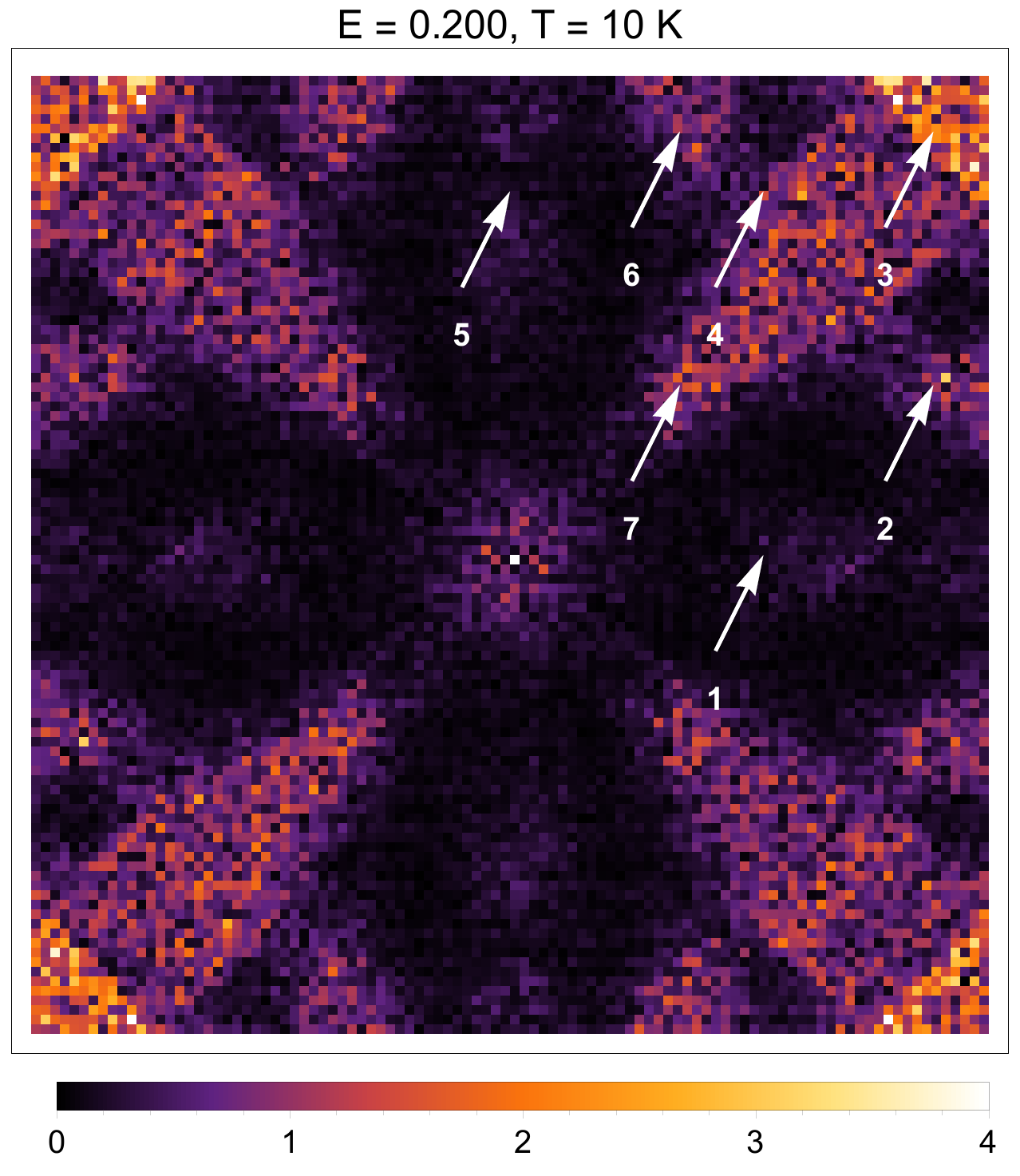}
	\includegraphics[width=0.16\textwidth]{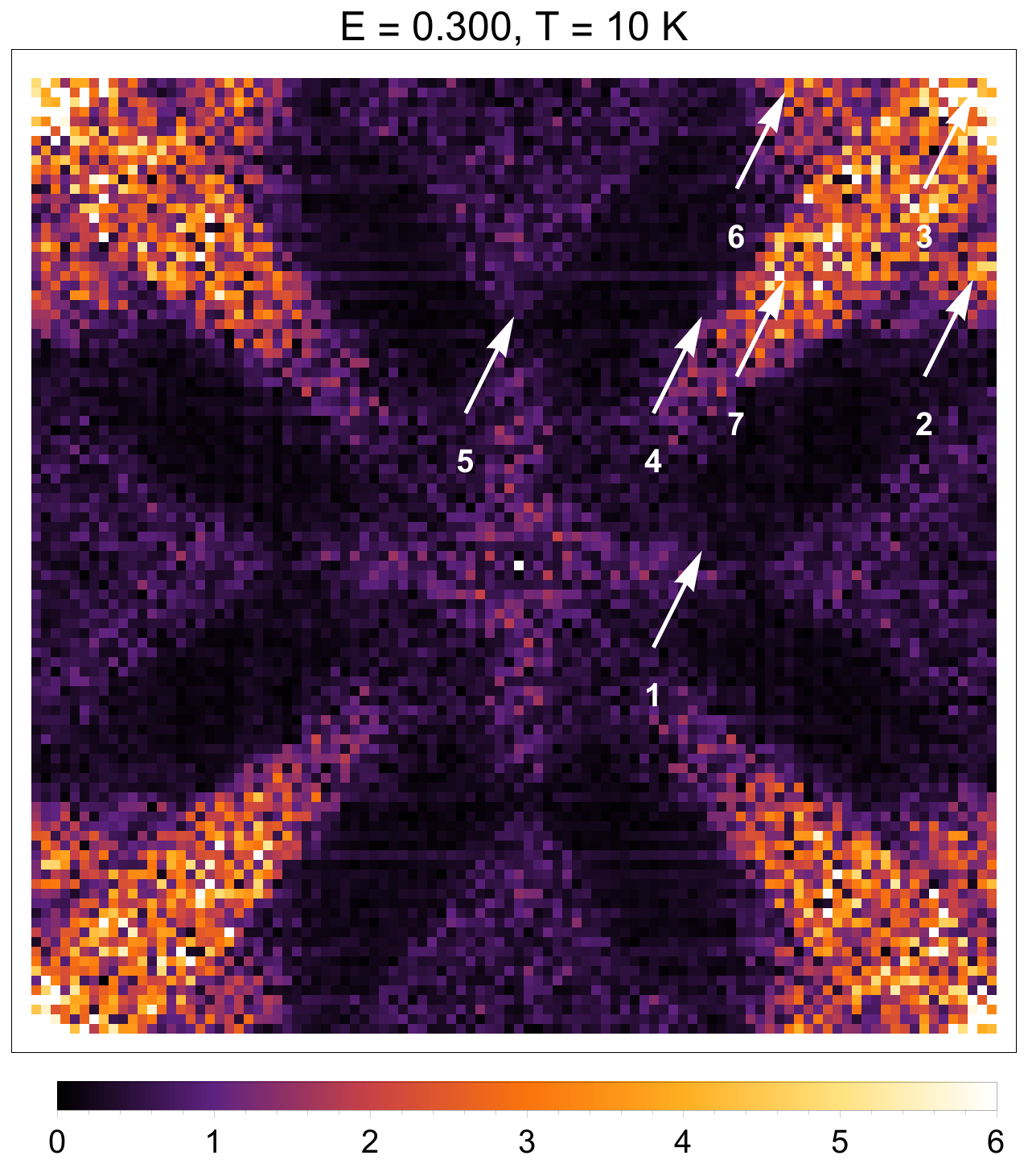} 
	
	\caption{Frequency-dependence at $T = 10$ K---a temperature at which all three scenarios are essentially identical---of the spectral function $A(\mathbf{k}, \omega)$ (upper row); the LDOS power spectrum with a single pointlike scatterer without thermal smearing (middle row); and the LDOS power spectrum with both a 0.5\% concentration of pointlike scatterers and thermal smearing (bottom row). Arrows indicate the locations of the peaks predicted by the octet model. Note that the scales used for plotting the LDOS power spectra change with frequency.}
	\label{fig:frequency_gfbcs_10k}
\end{figure*}

\begin{figure*}
	\centering
	
	\includegraphics[width=0.16\textwidth]{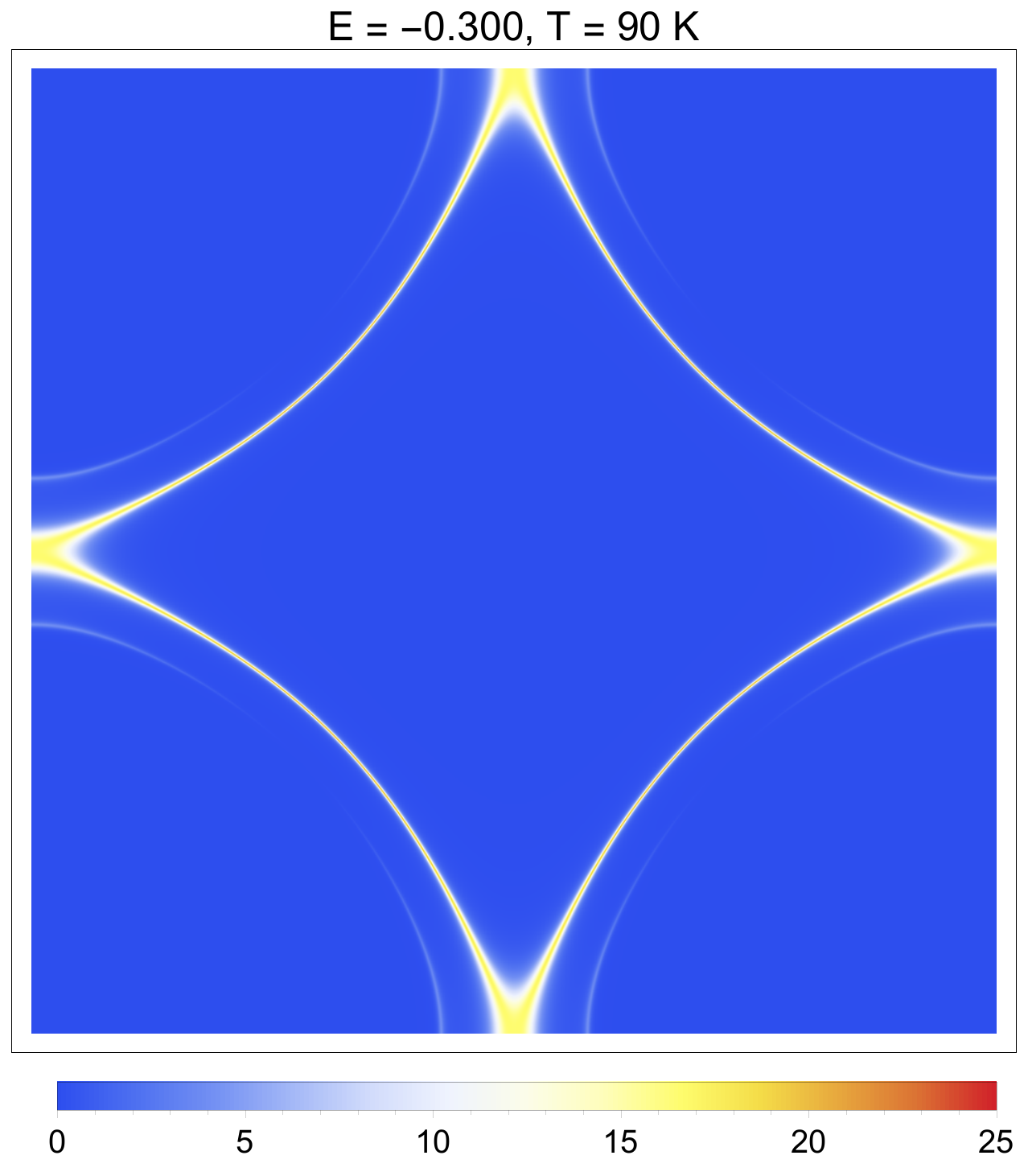}
	\includegraphics[width=0.16\textwidth]{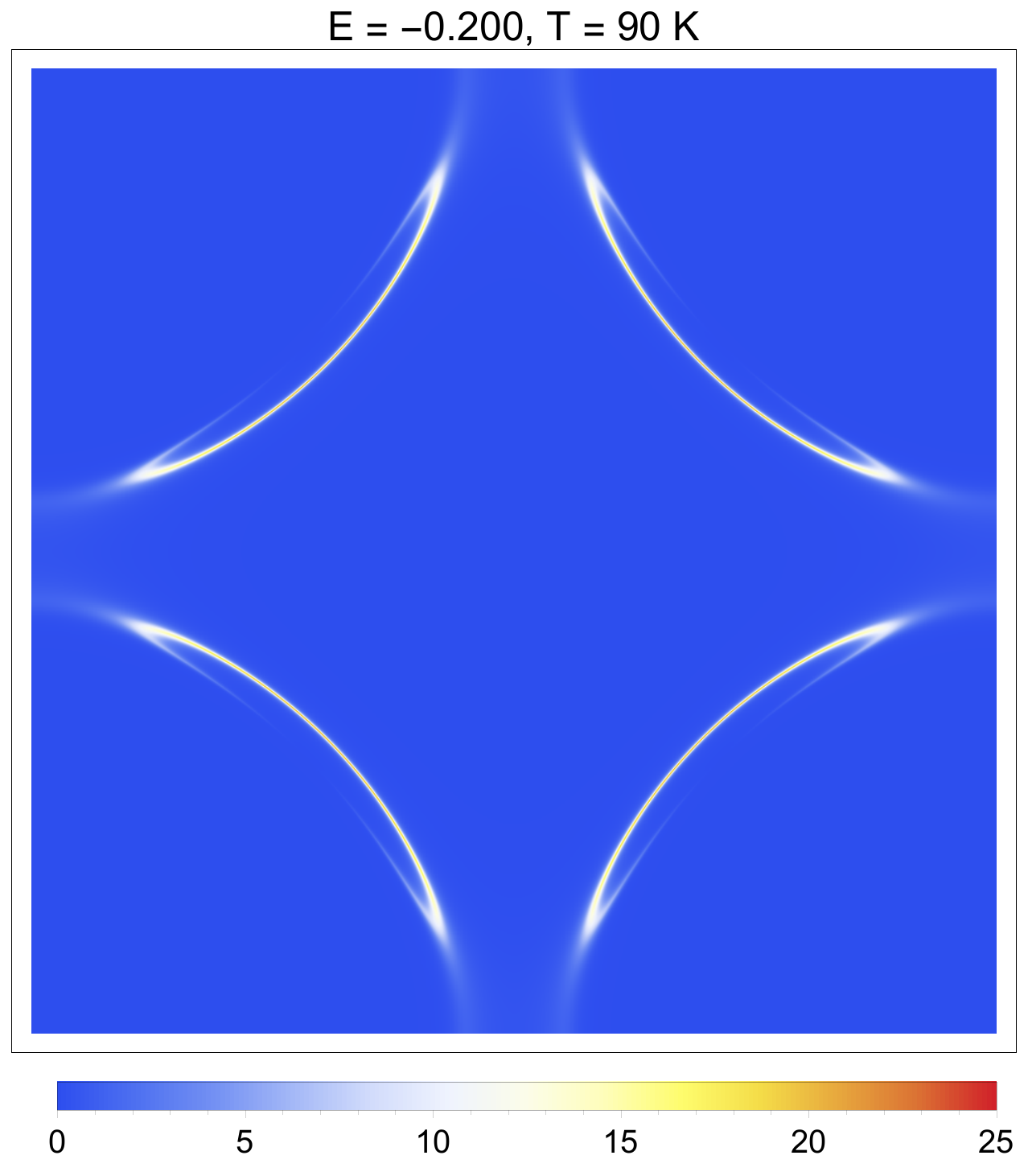}
	\includegraphics[width=0.16\textwidth]{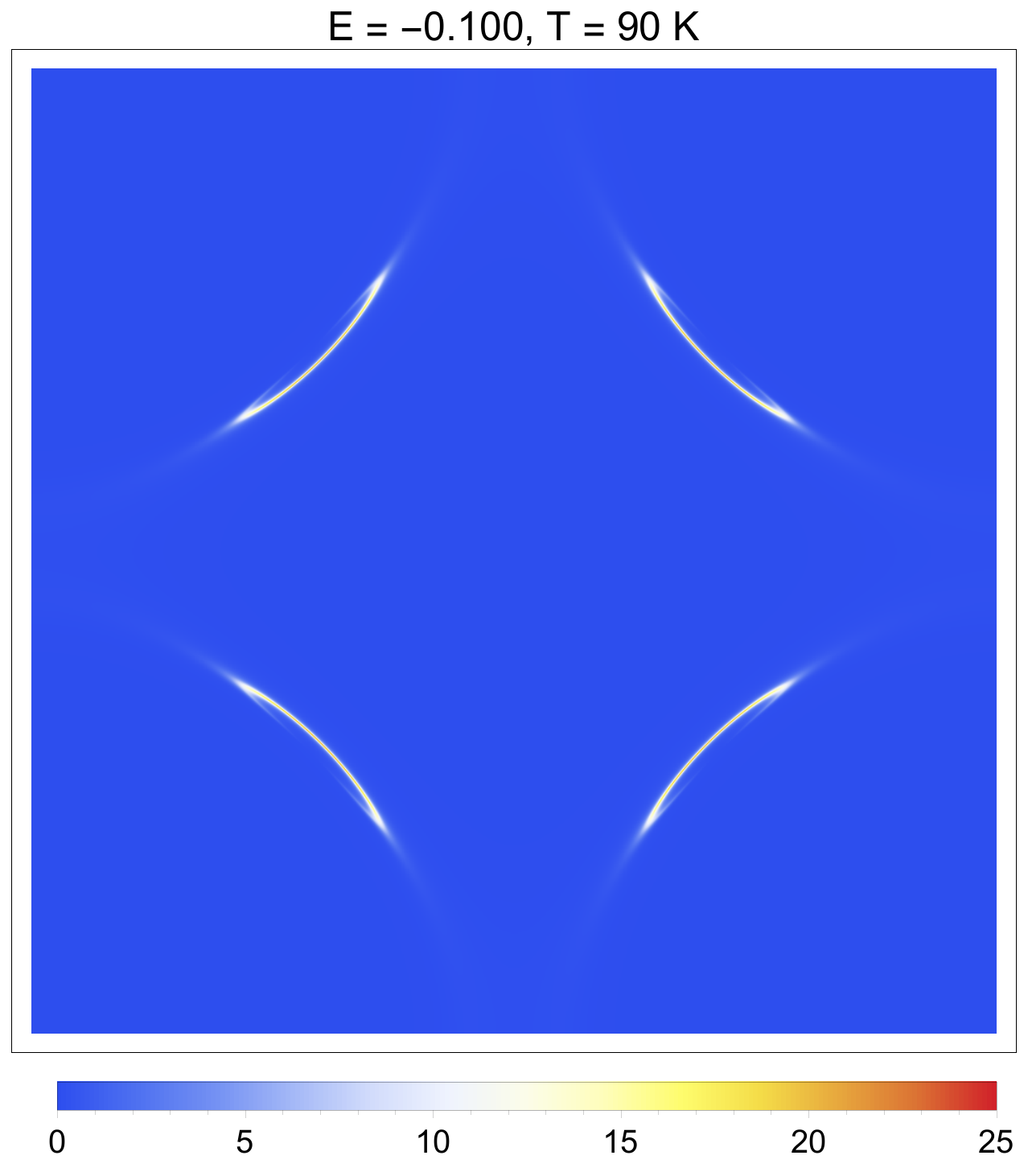}
	\includegraphics[width=0.16\textwidth]{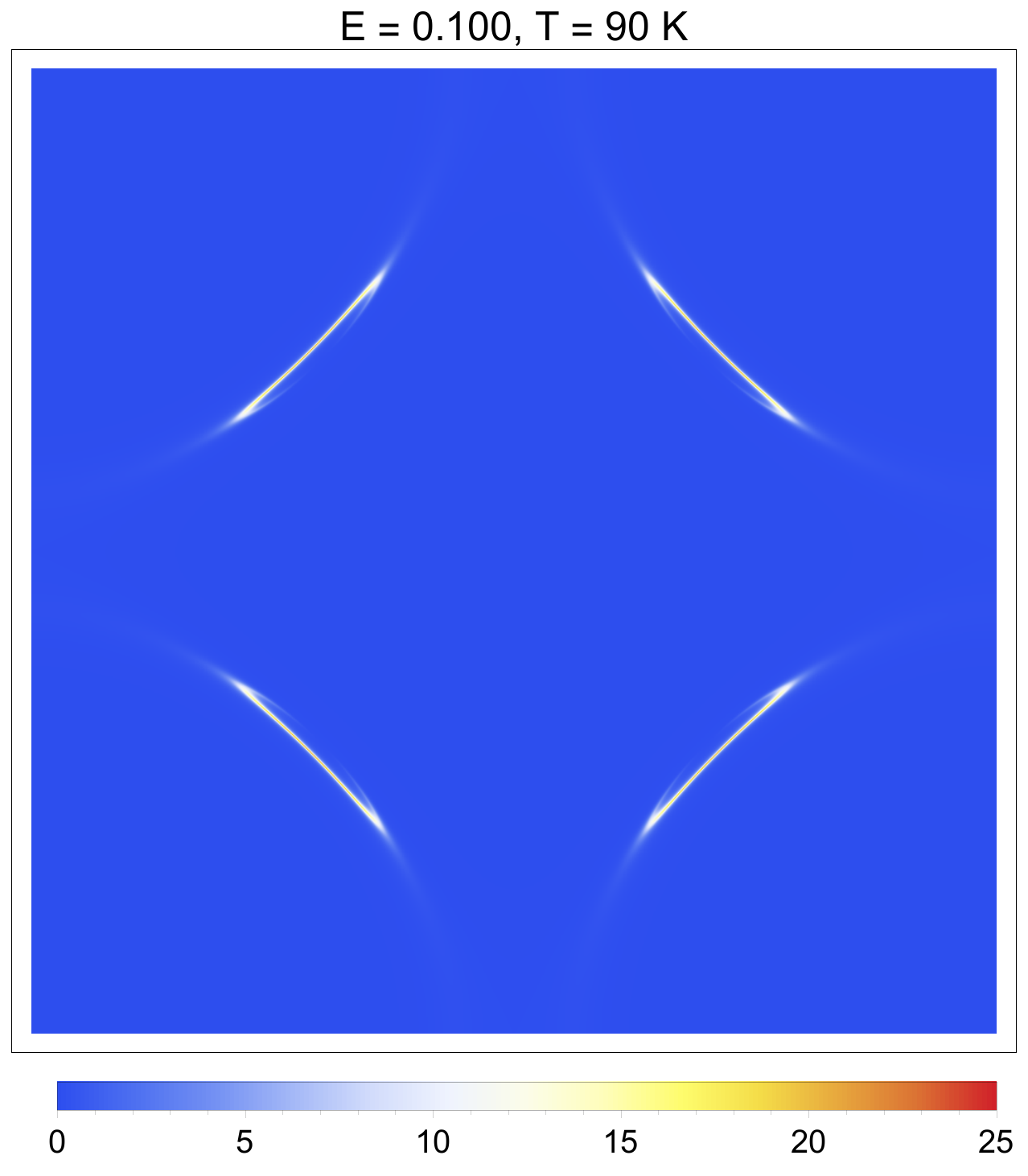}
	\includegraphics[width=0.16\textwidth]{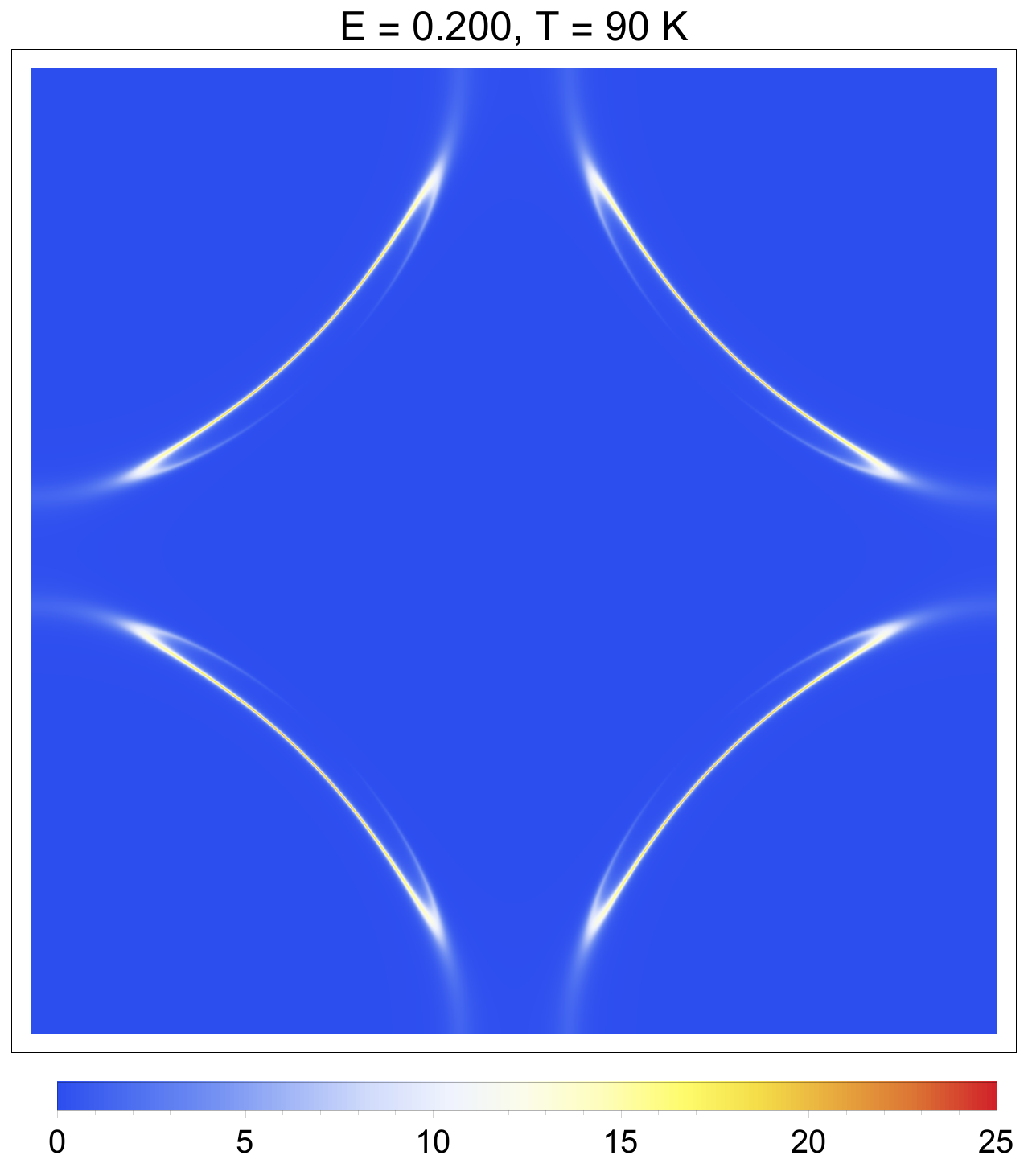}
	\includegraphics[width=0.16\textwidth]{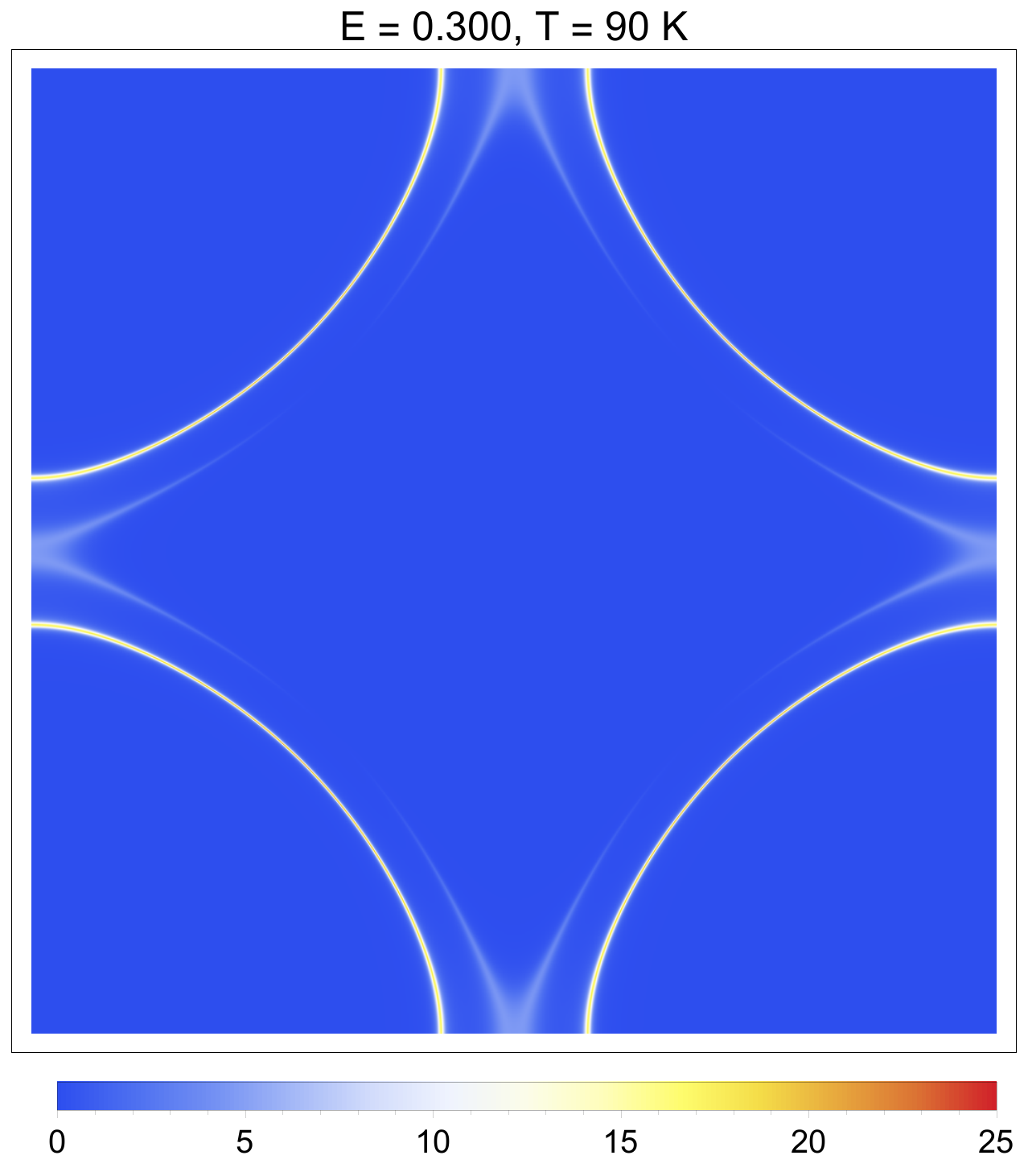} \\
	\includegraphics[width=0.16\textwidth]{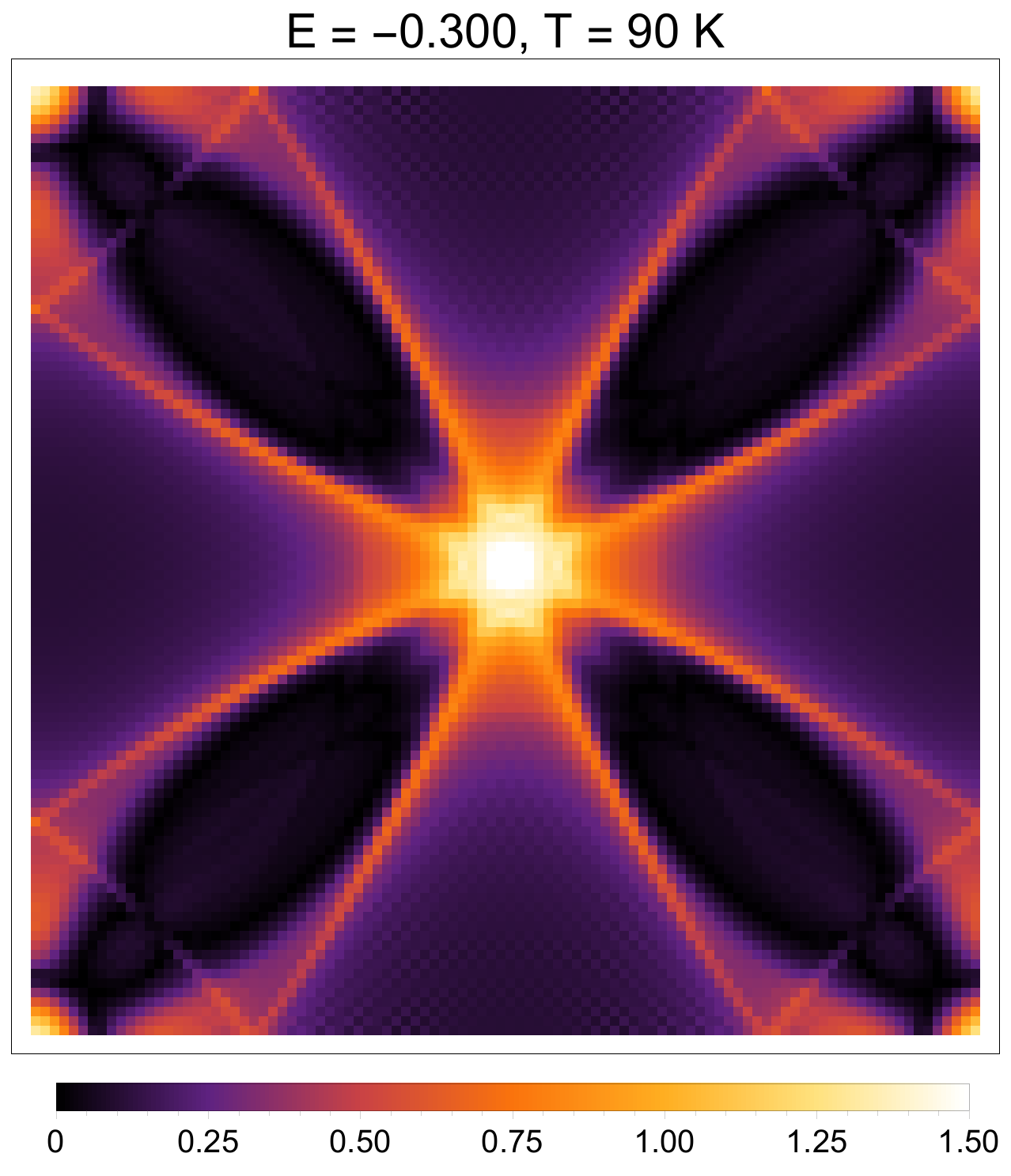}
	\includegraphics[width=0.16\textwidth]{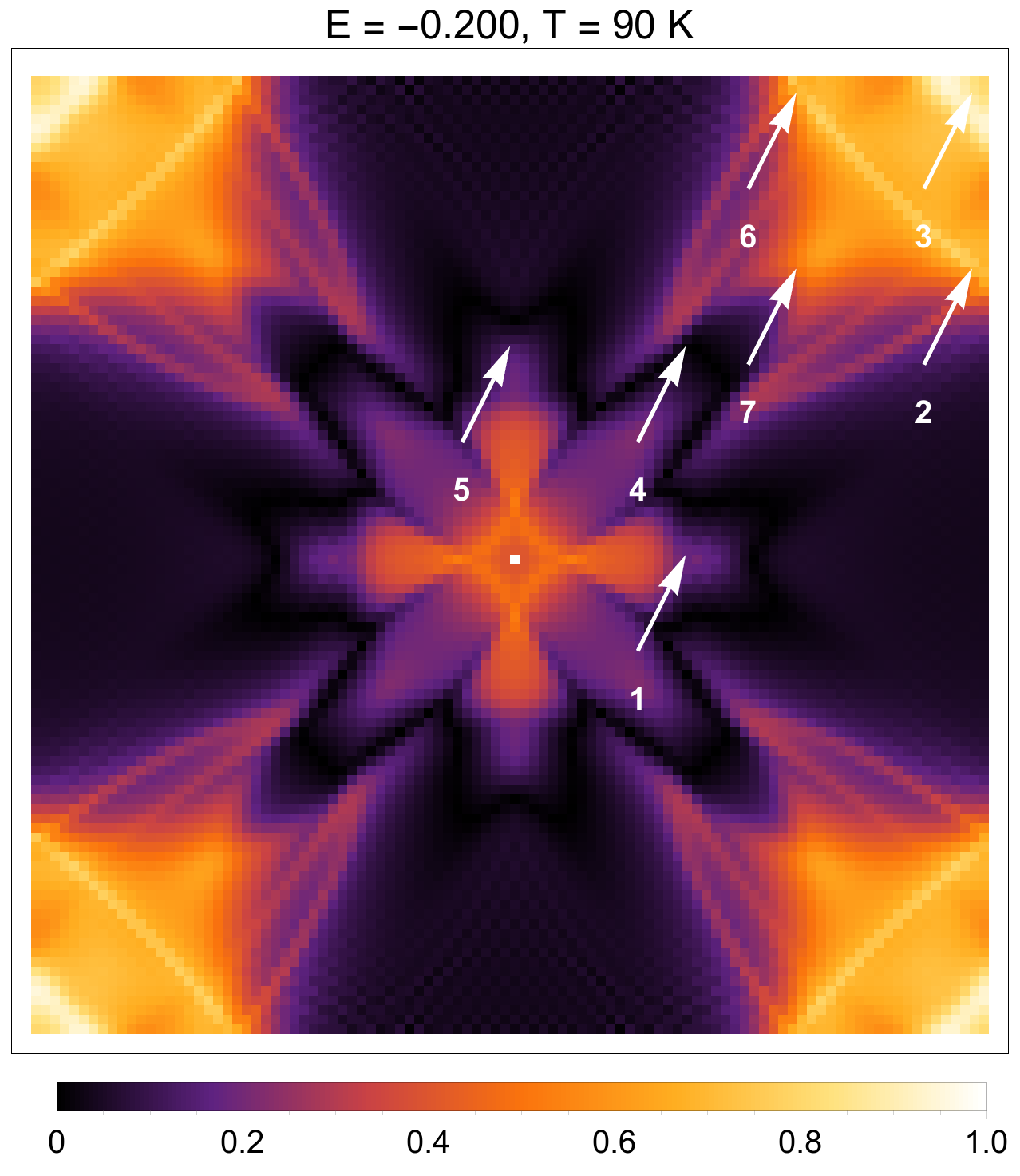}
	\includegraphics[width=0.16\textwidth]{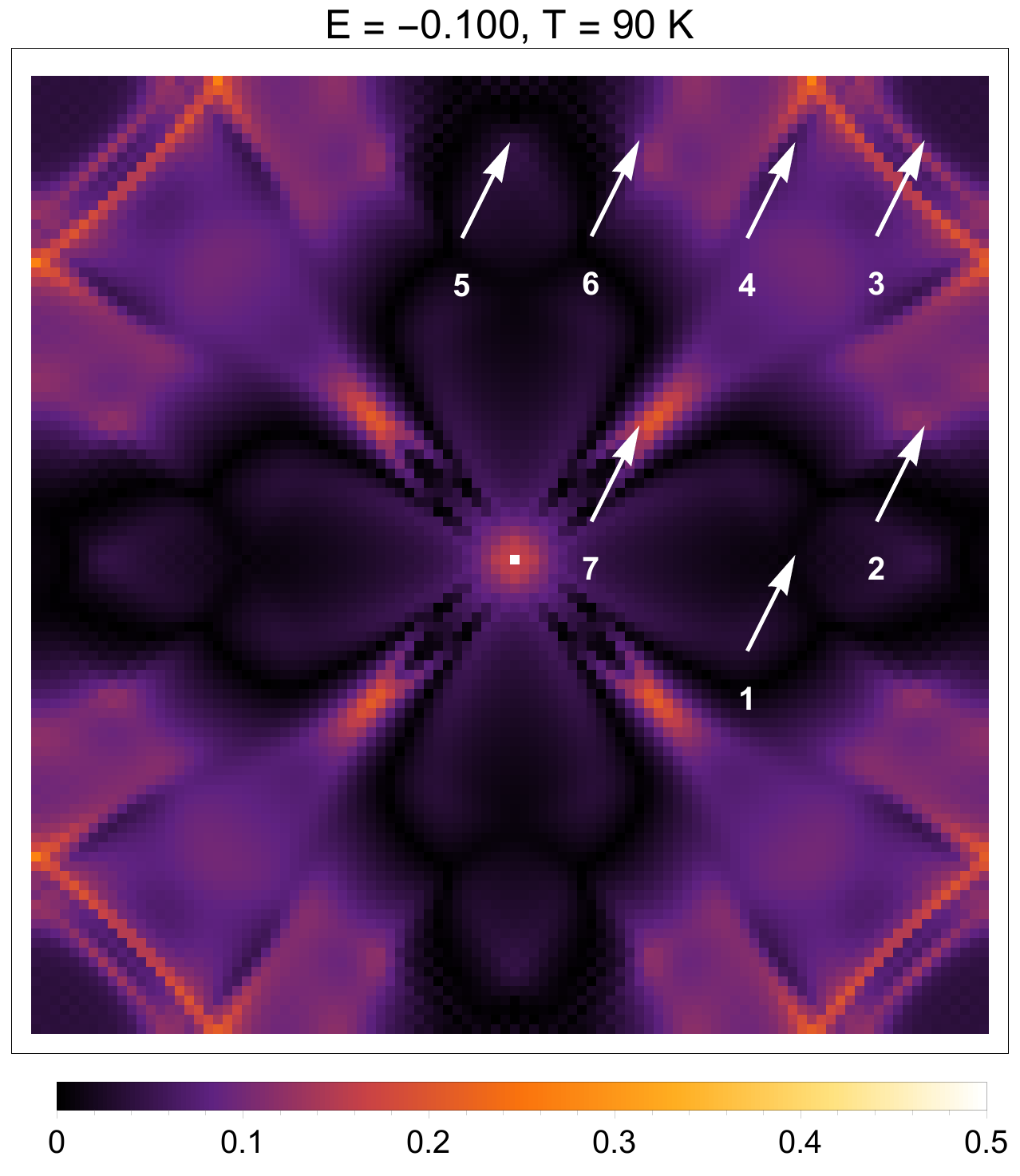}
	\includegraphics[width=0.16\textwidth]{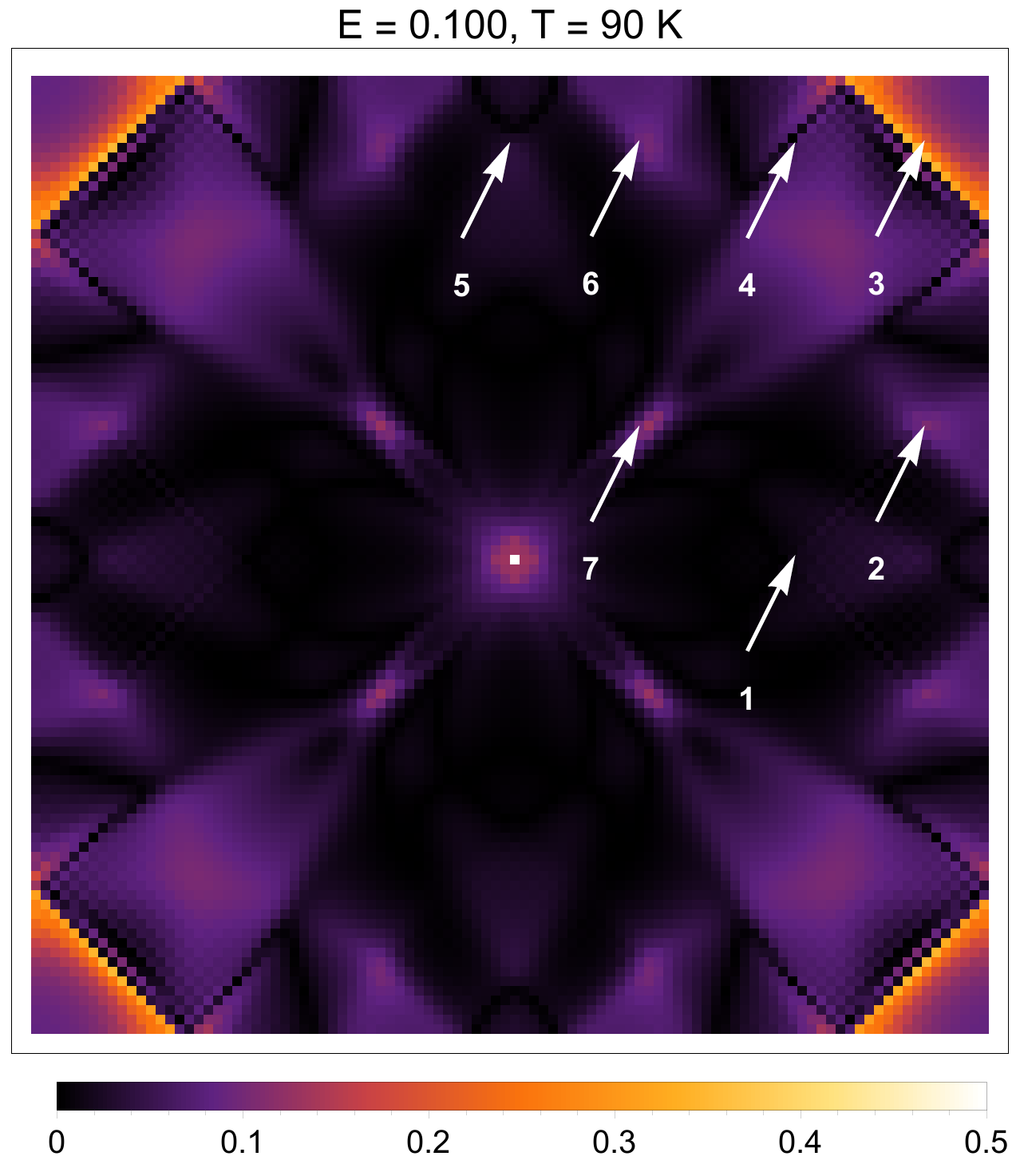}
	\includegraphics[width=0.16\textwidth]{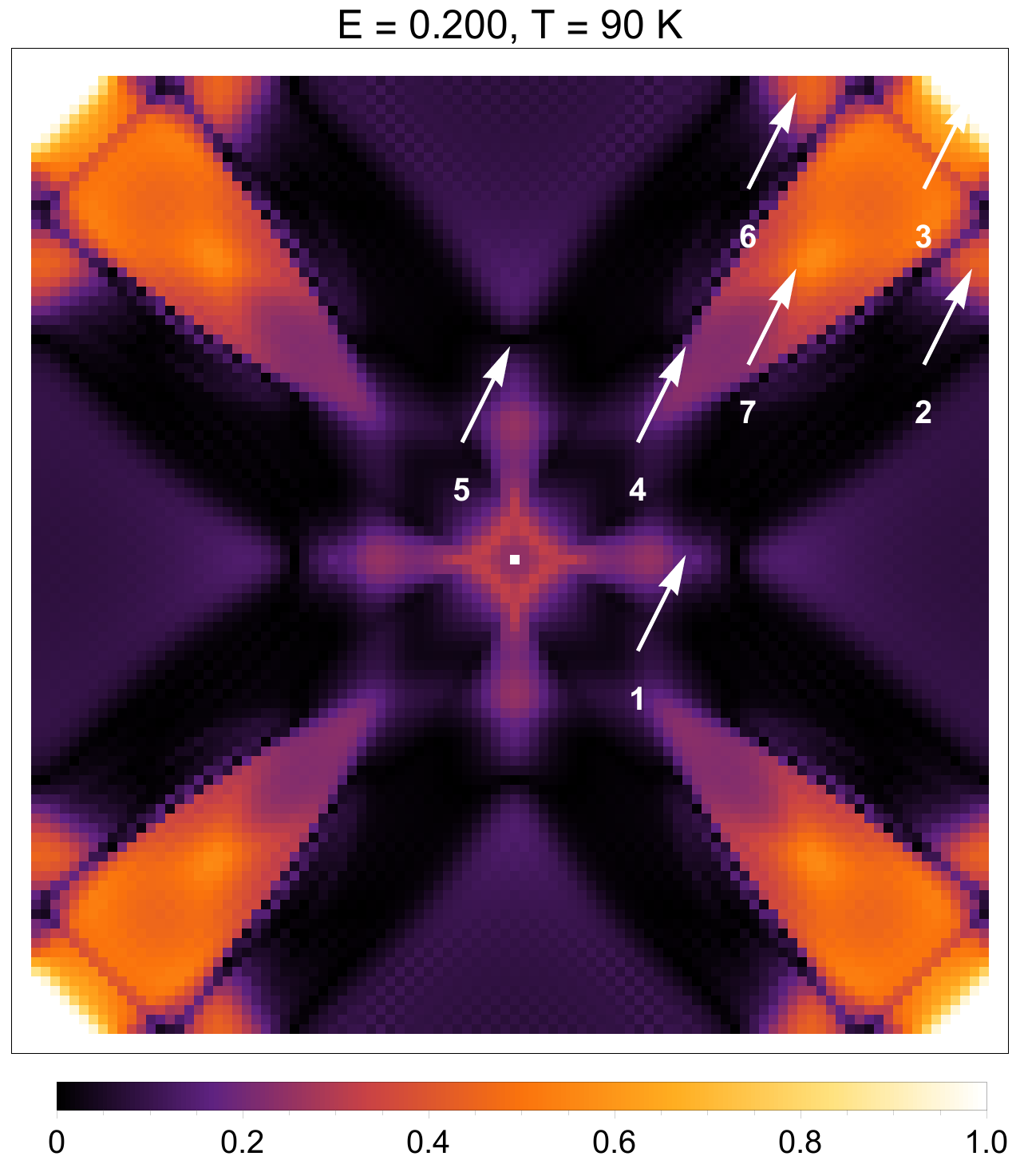}
	\includegraphics[width=0.16\textwidth]{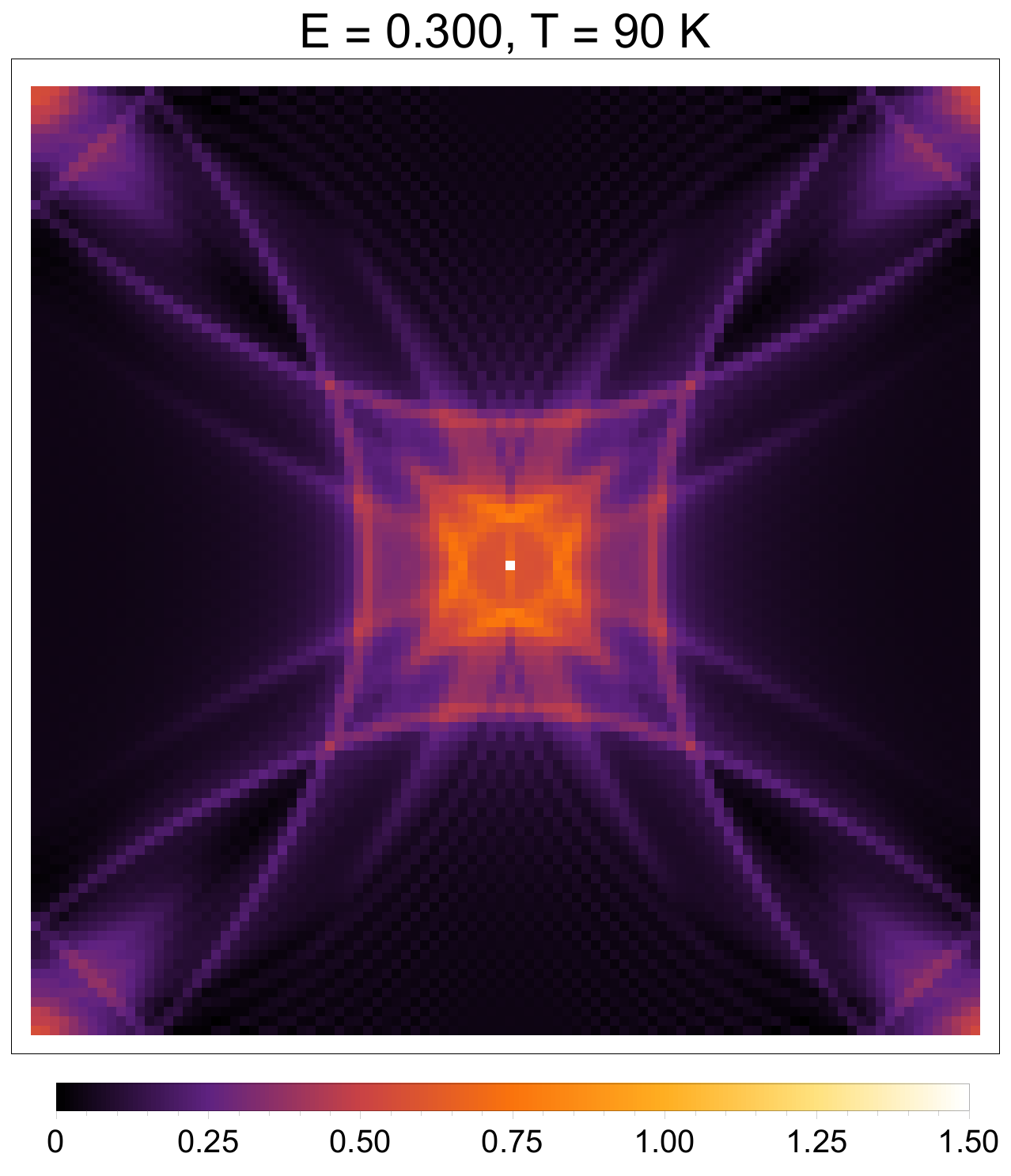} \\
	\includegraphics[width=0.16\textwidth]{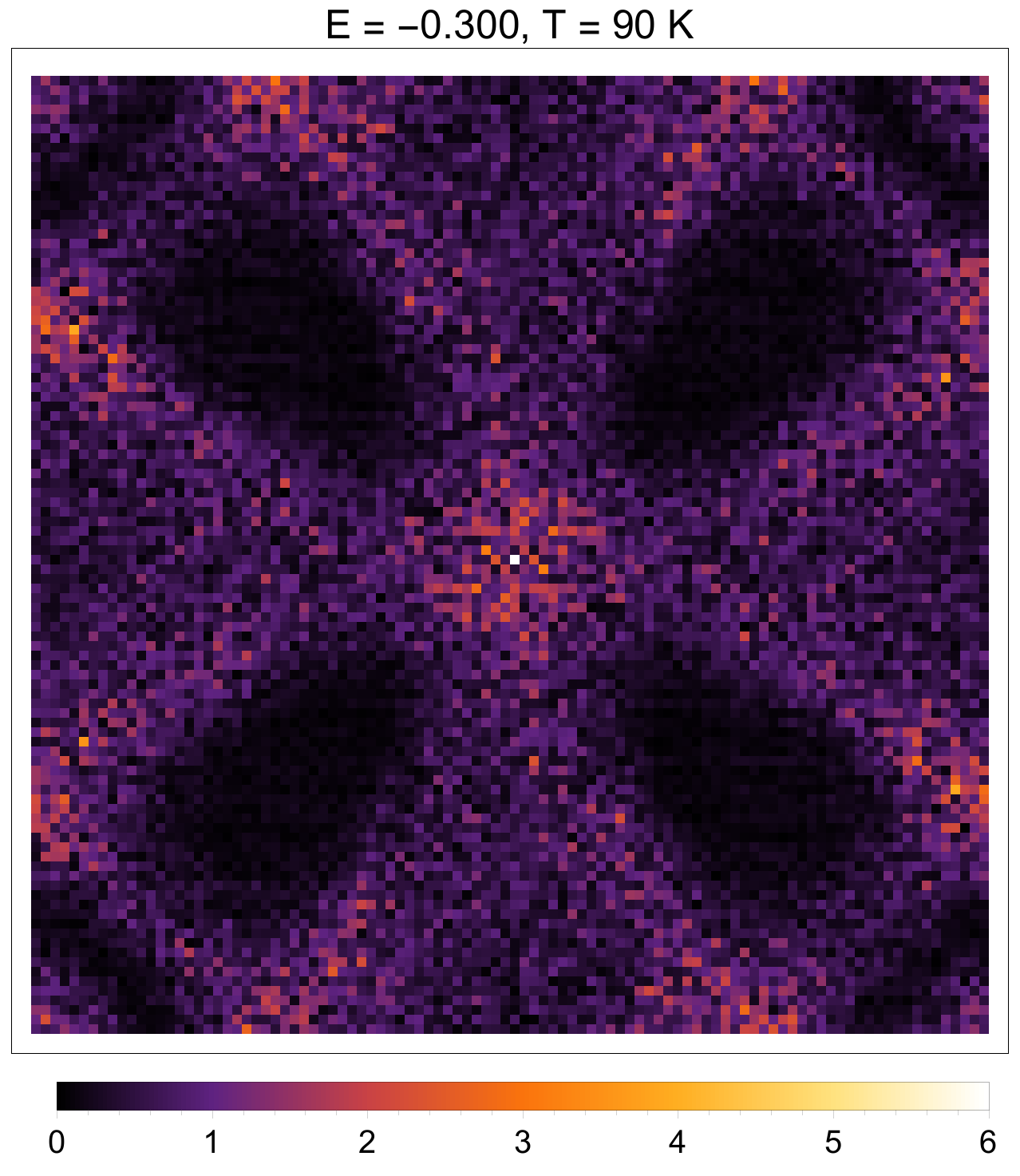}
	\includegraphics[width=0.16\textwidth]{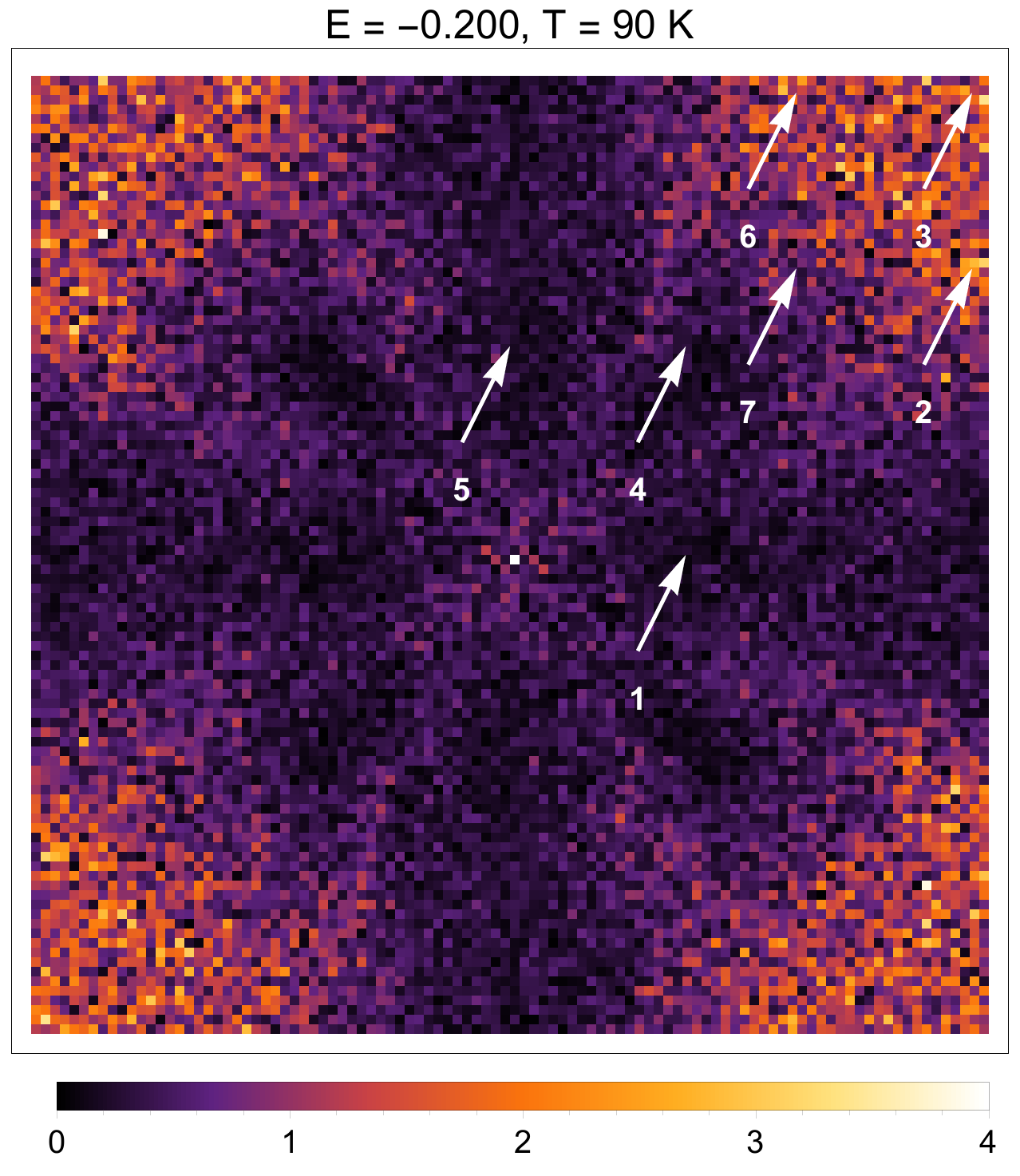}
	\includegraphics[width=0.16\textwidth]{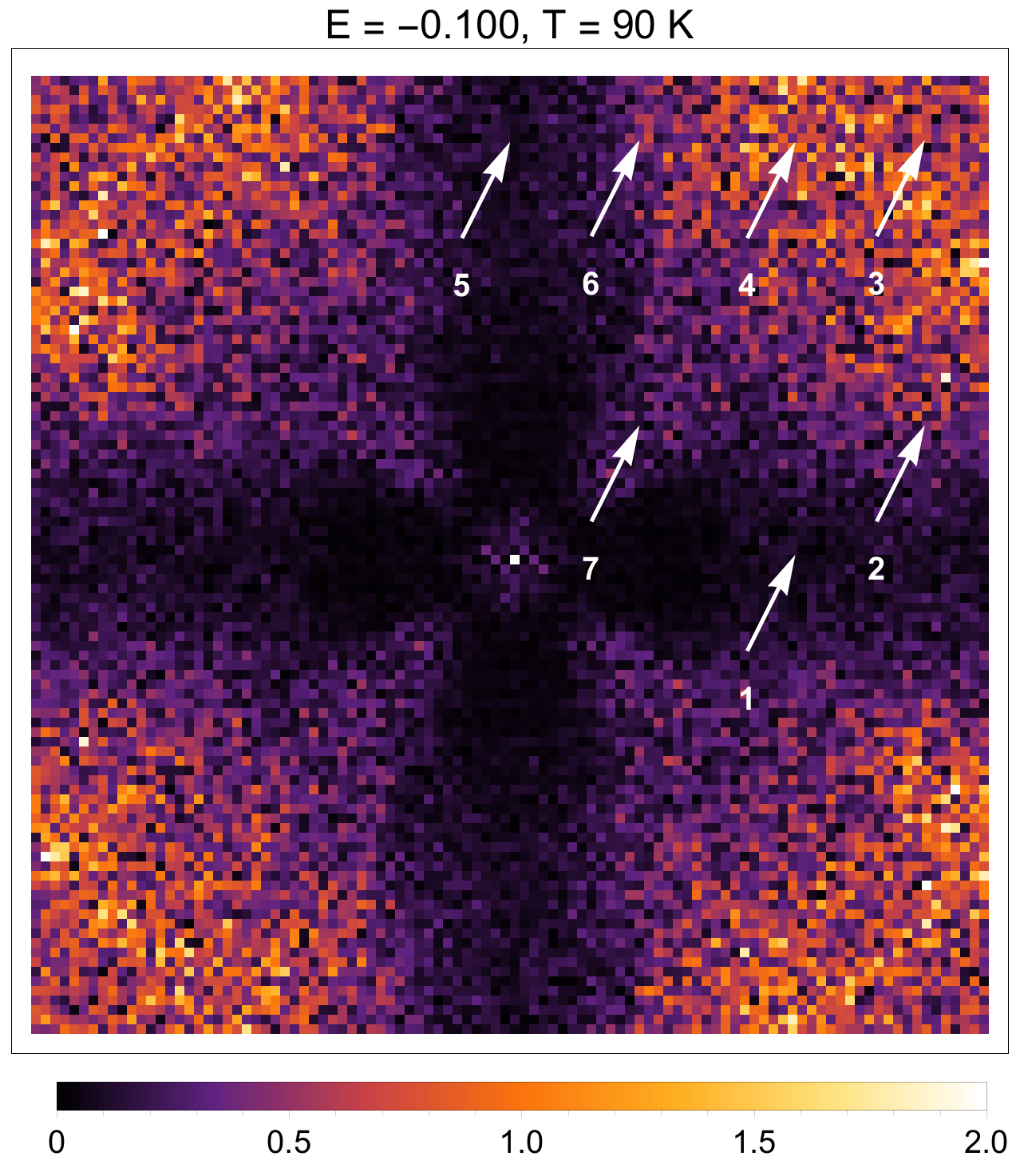}
	\includegraphics[width=0.16\textwidth]{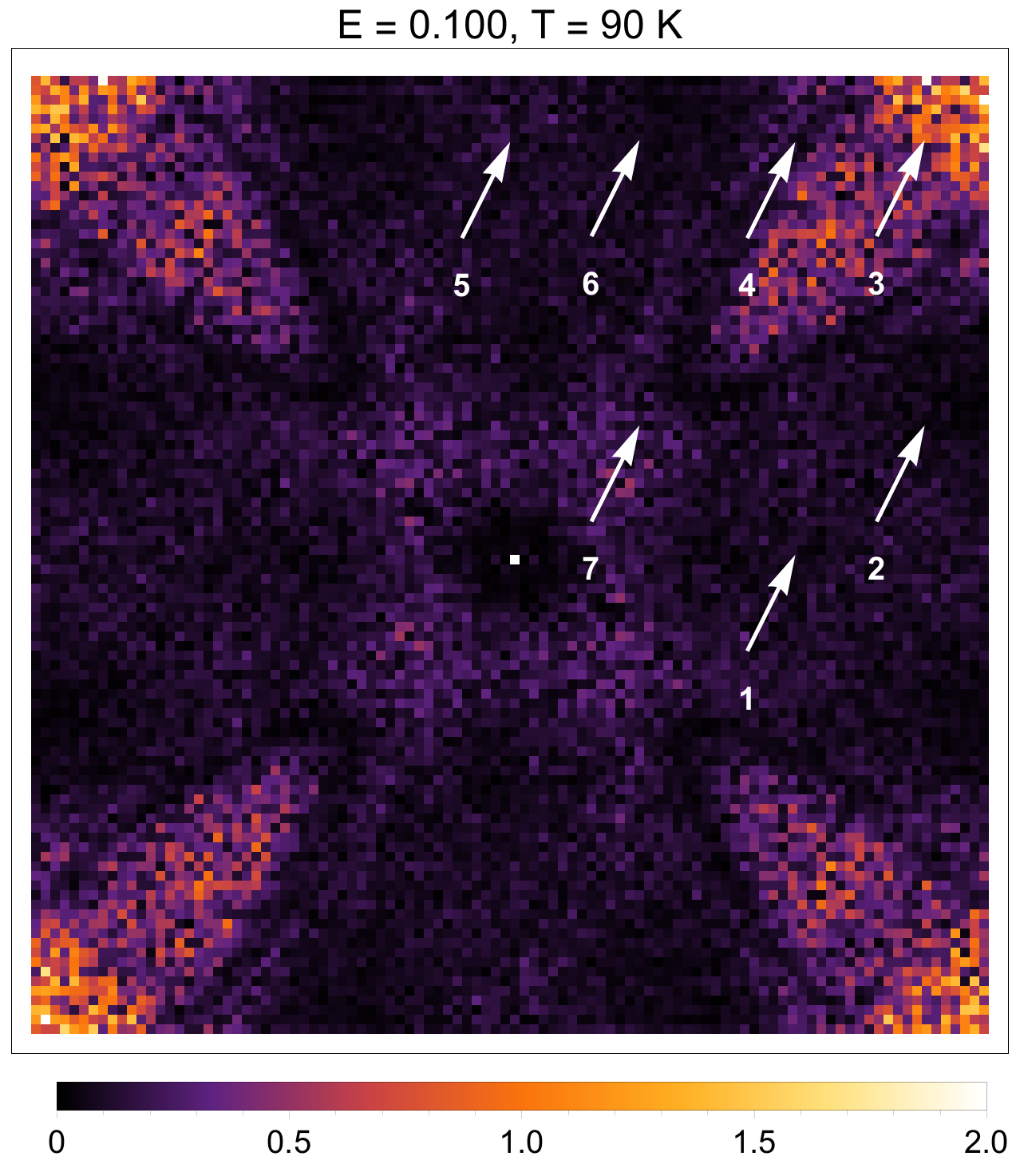}
	\includegraphics[width=0.16\textwidth]{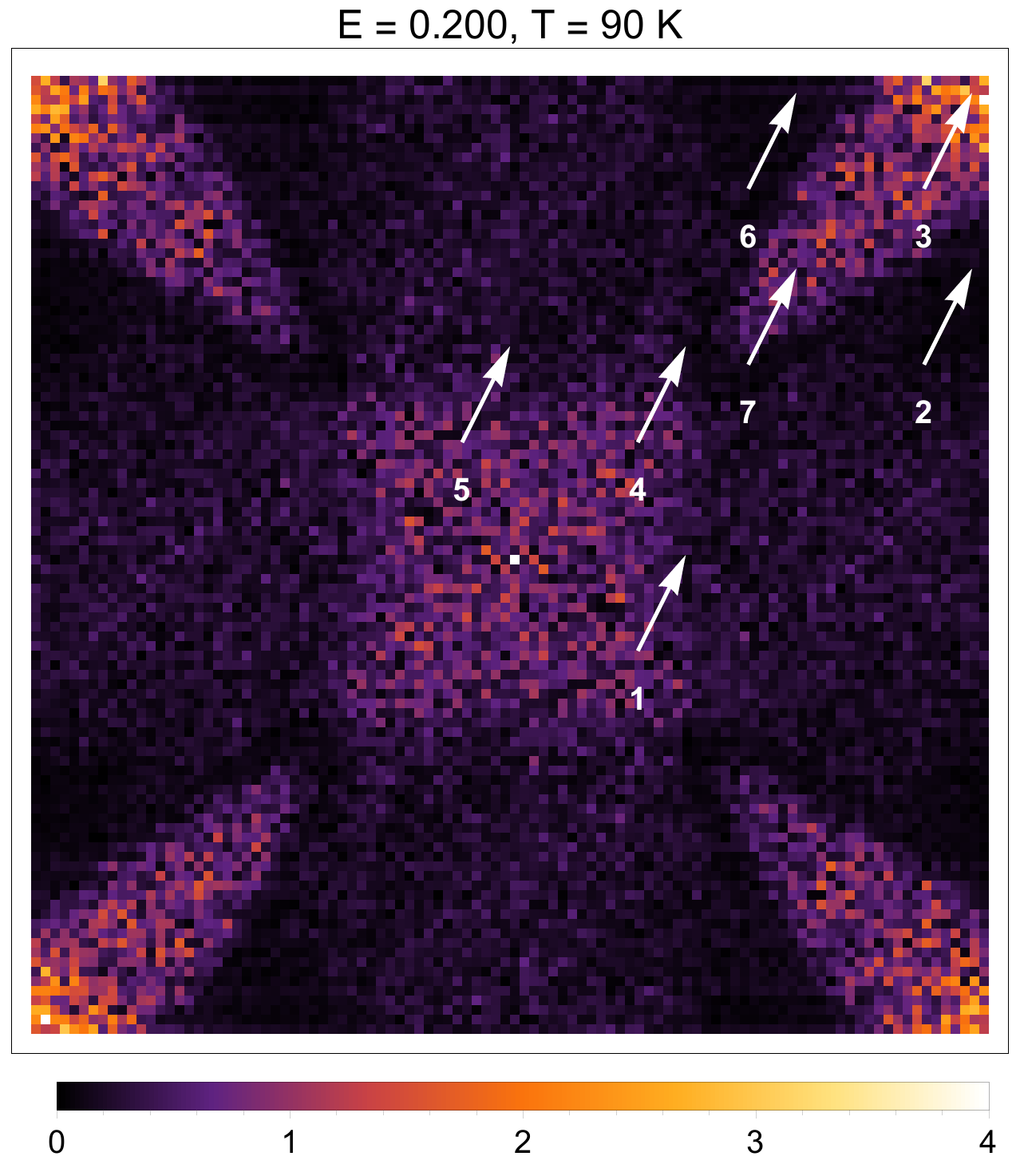}
	\includegraphics[width=0.16\textwidth]{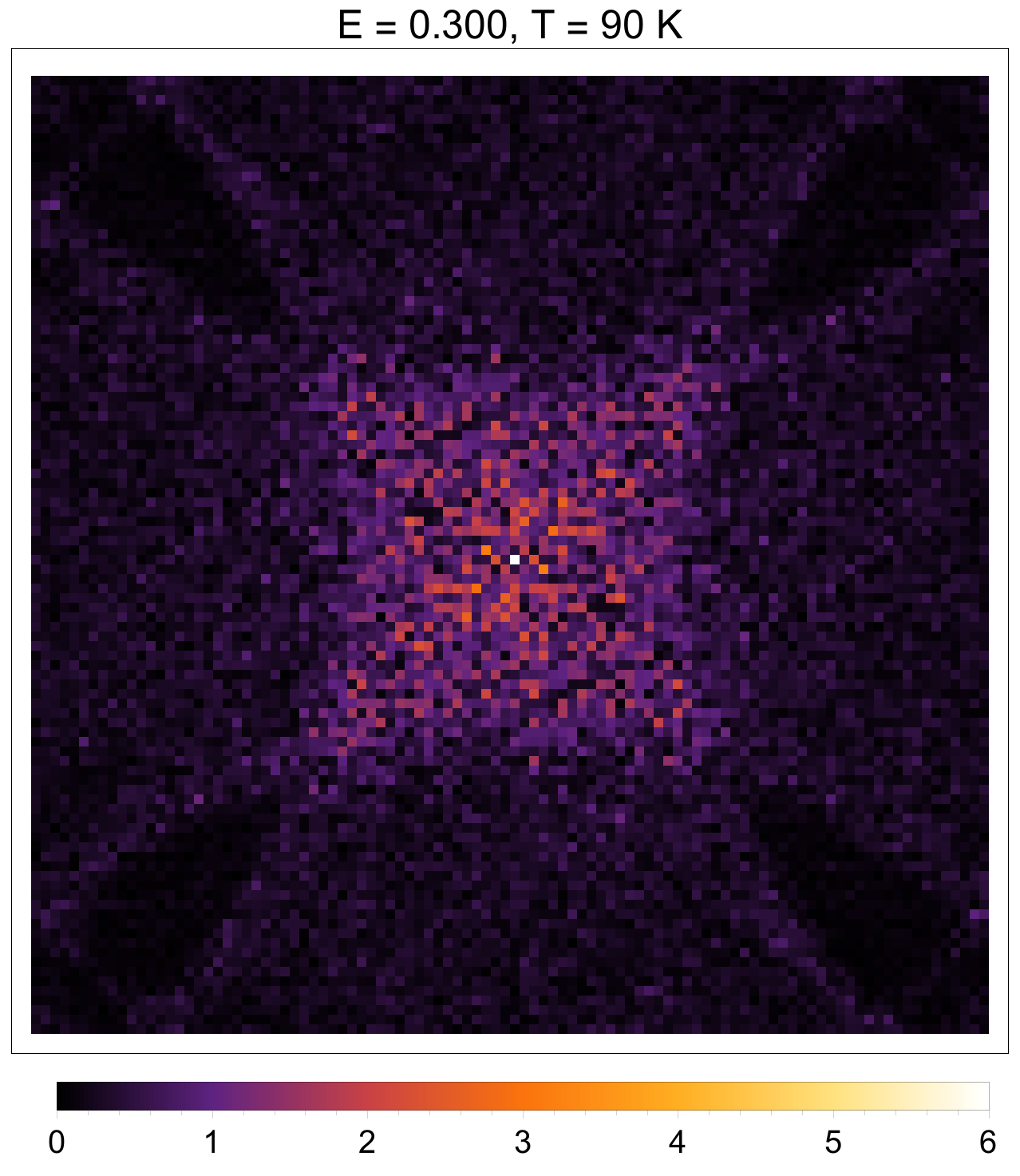} 
	
	\caption{Frequency-dependence at $T = 90$ K in the gap-closing scenario of the spectral function $A(\mathbf{k}, \omega)$ (upper row); the LDOS power spectrum with a single pointlike scatterer without thermal smearing (middle row); and the LDOS power spectrum with both a 0.5\% concentration of pointlike scatterers and thermal smearing (bottom row). Arrows indicate the locations of the peaks predicted by the octet model. Note that the scales used for plotting the LDOS power spectra change with frequency. In this scenario, this temperature is \emph{less} than $T_c$.}
	\label{fig:frequency_bcs_90k}
\end{figure*}

\begin{figure*}
	\centering
	
	\includegraphics[width=0.16\textwidth]{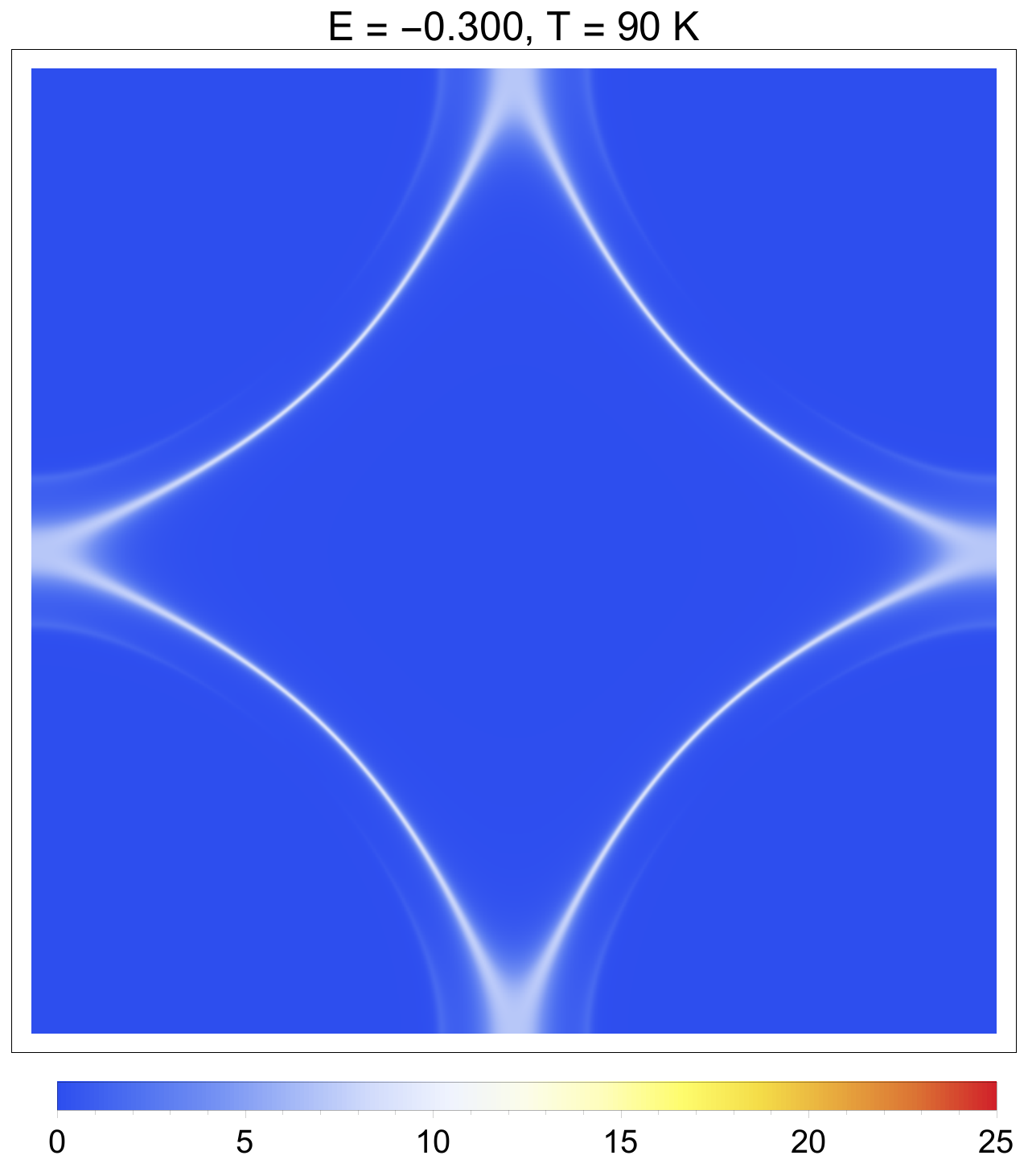}
	\includegraphics[width=0.16\textwidth]{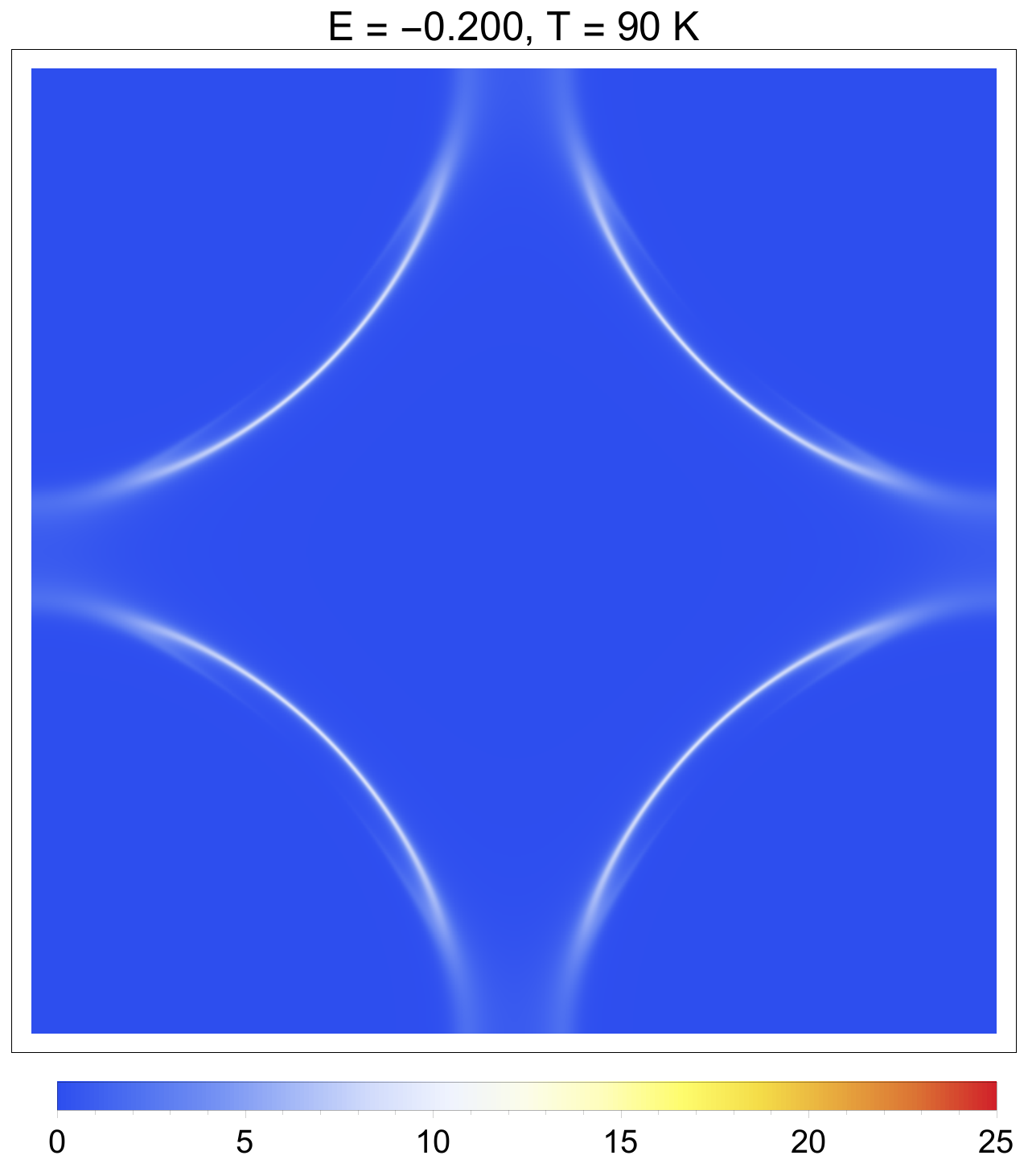}
	\includegraphics[width=0.16\textwidth]{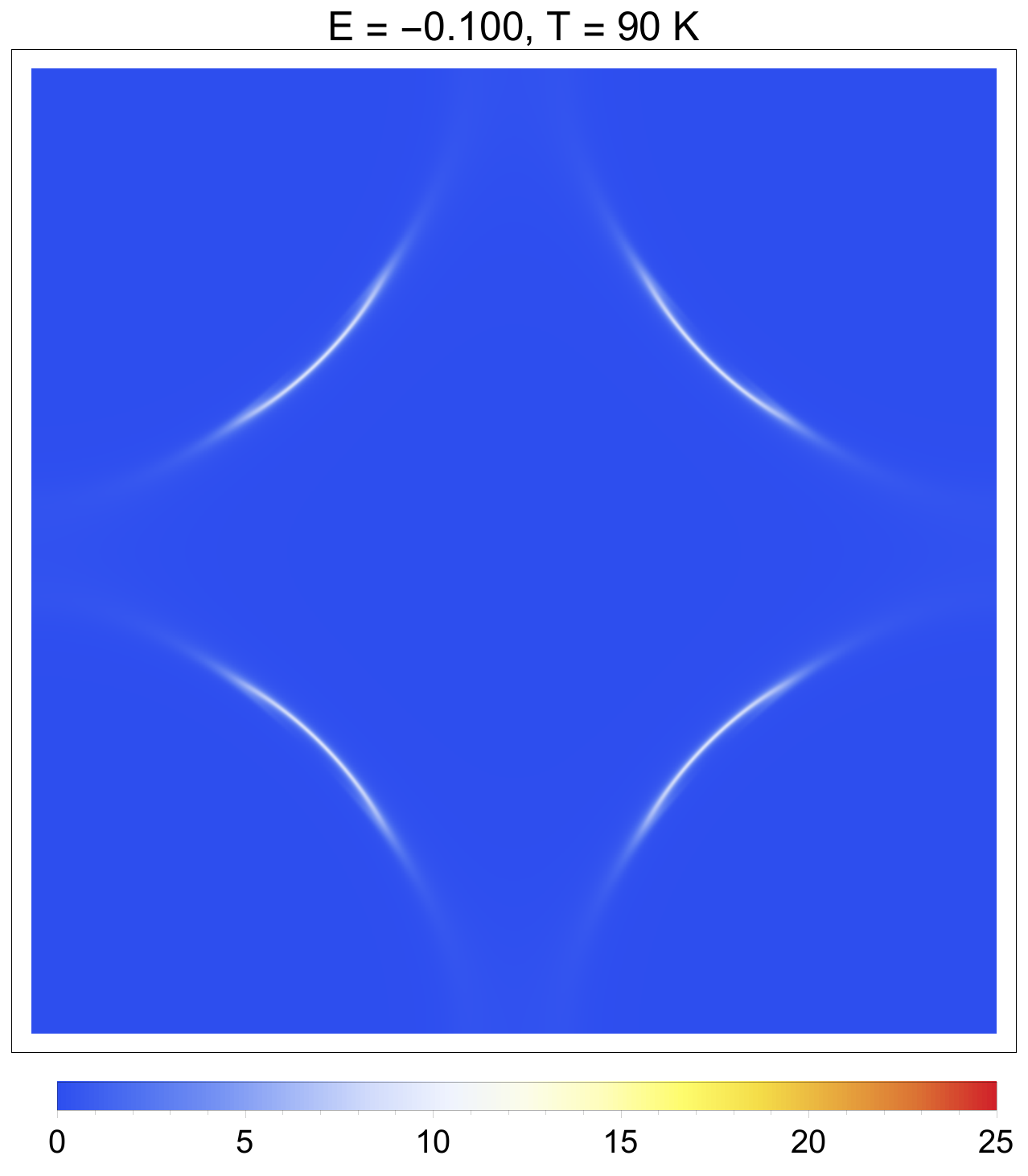}
	\includegraphics[width=0.16\textwidth]{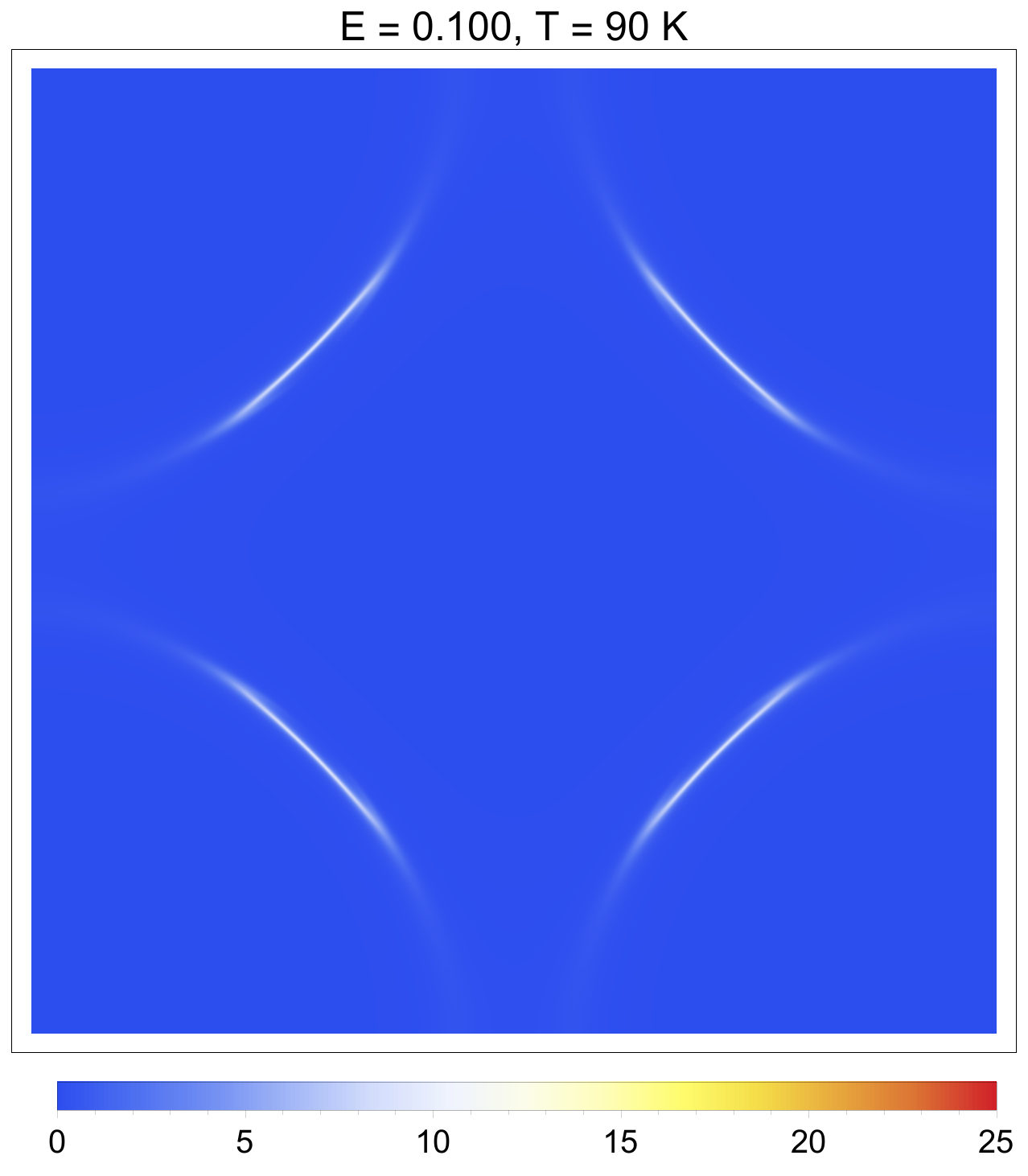}
	\includegraphics[width=0.16\textwidth]{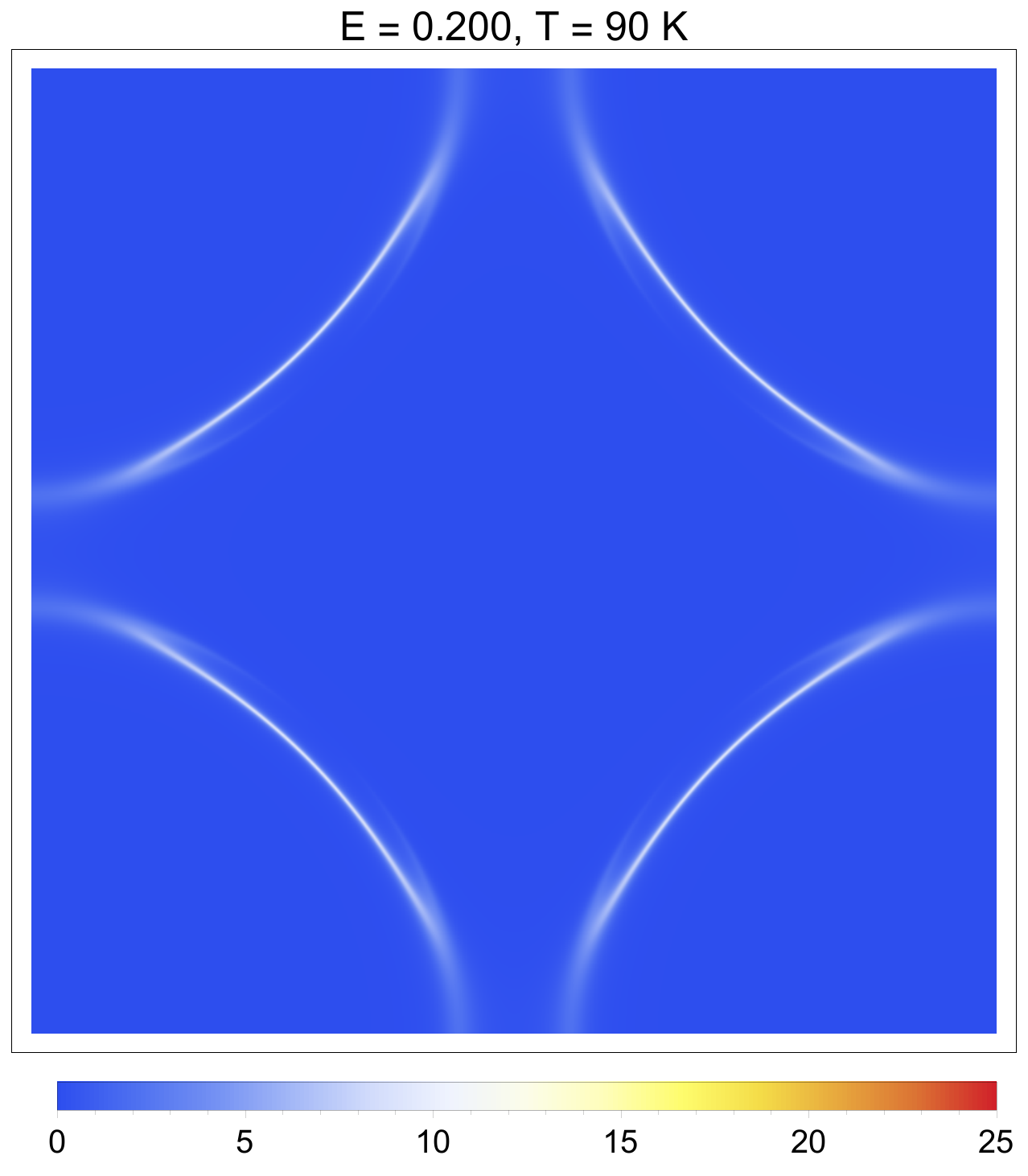}
	\includegraphics[width=0.16\textwidth]{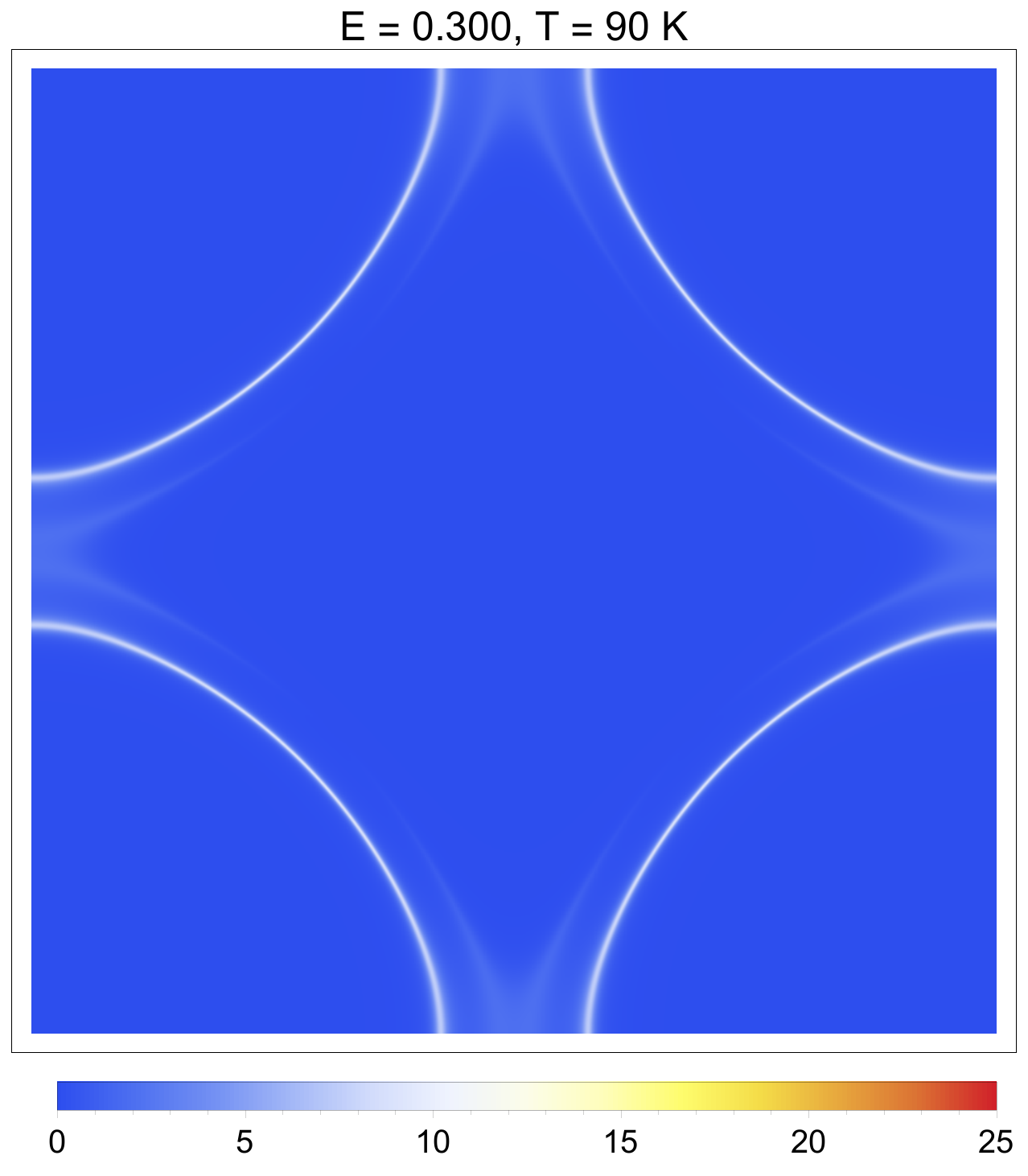} \\
	\includegraphics[width=0.16\textwidth]{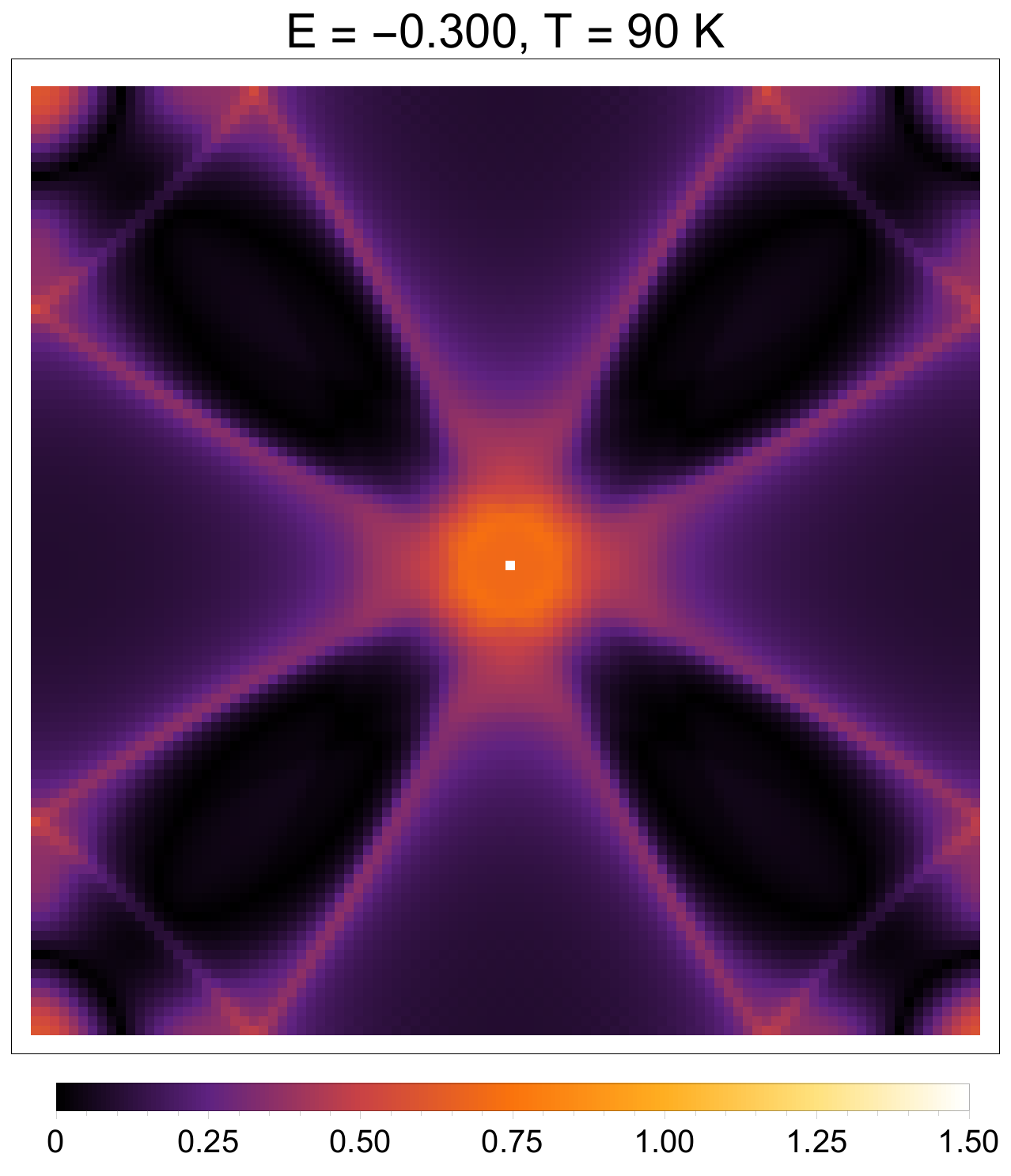}
	\includegraphics[width=0.16\textwidth]{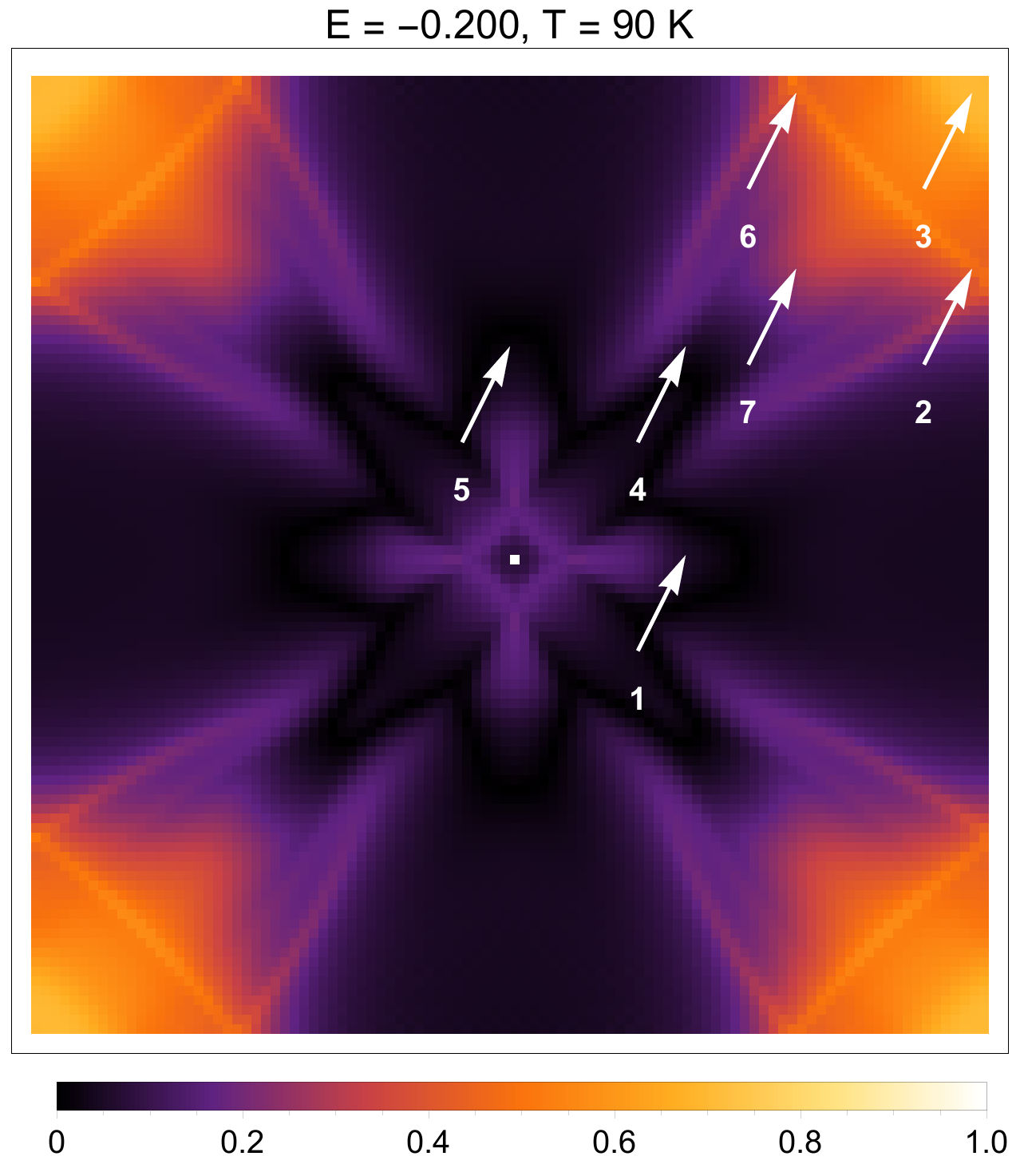}
	\includegraphics[width=0.16\textwidth]{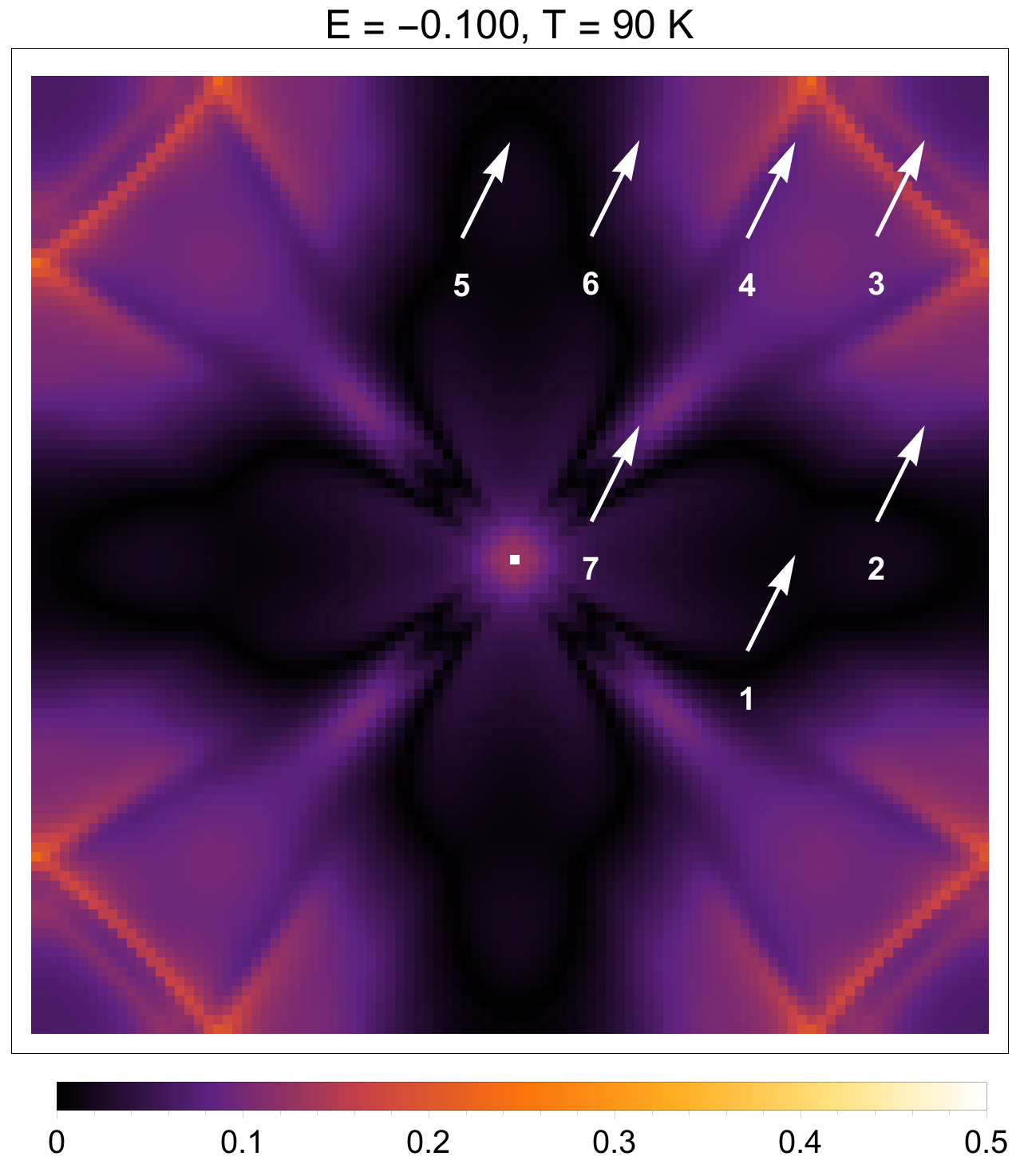}
	\includegraphics[width=0.16\textwidth]{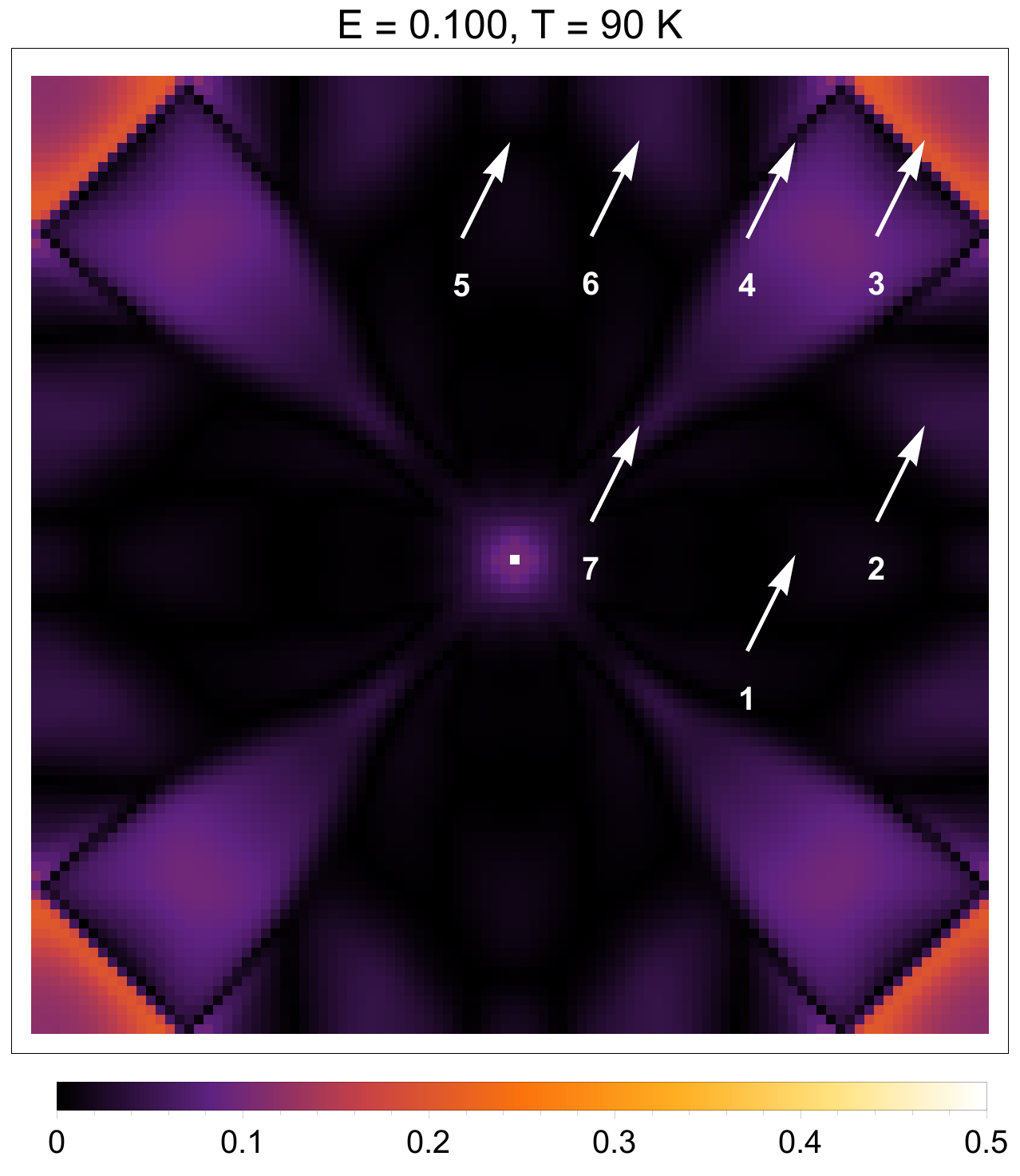}
	\includegraphics[width=0.16\textwidth]{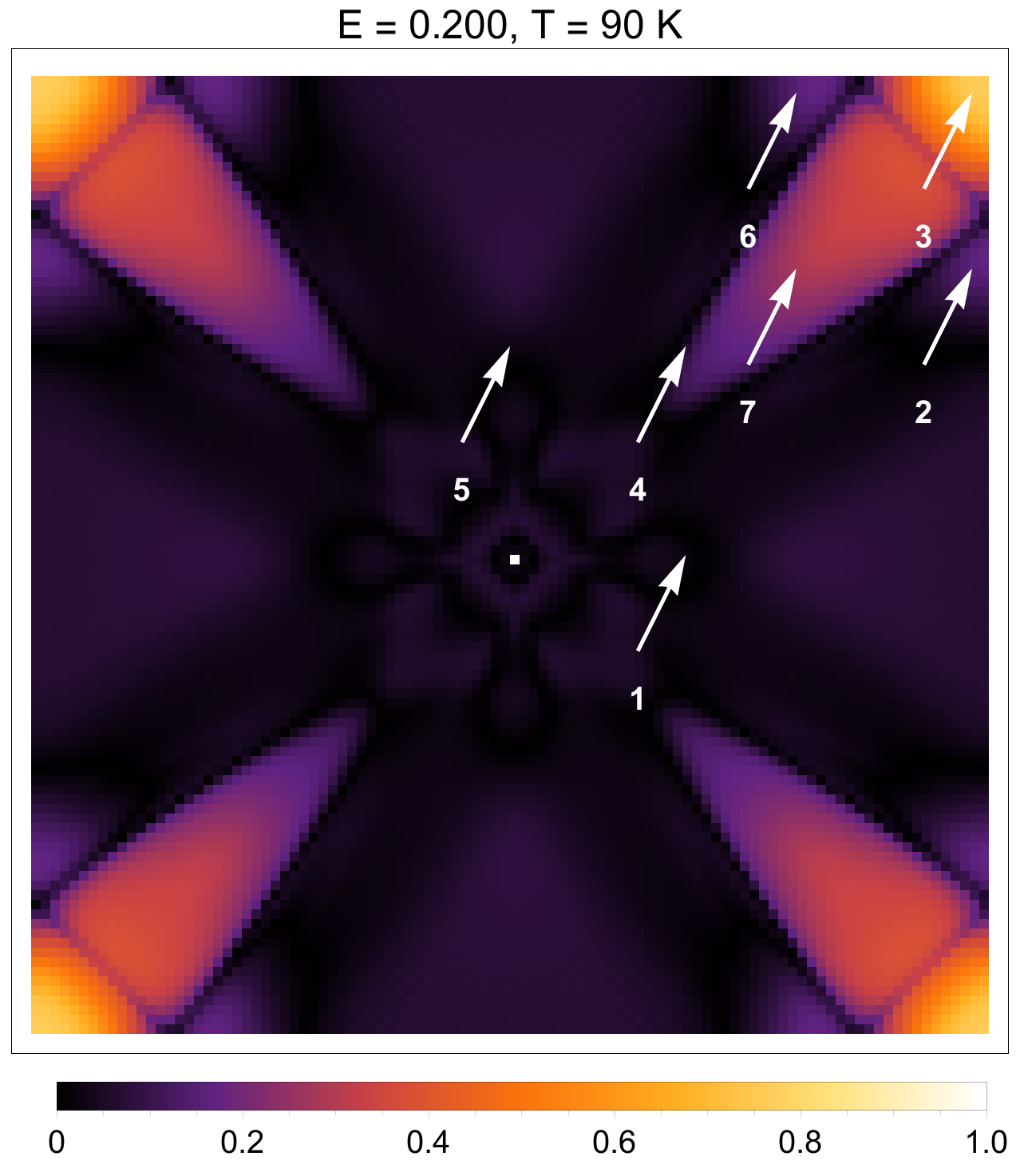}
	\includegraphics[width=0.16\textwidth]{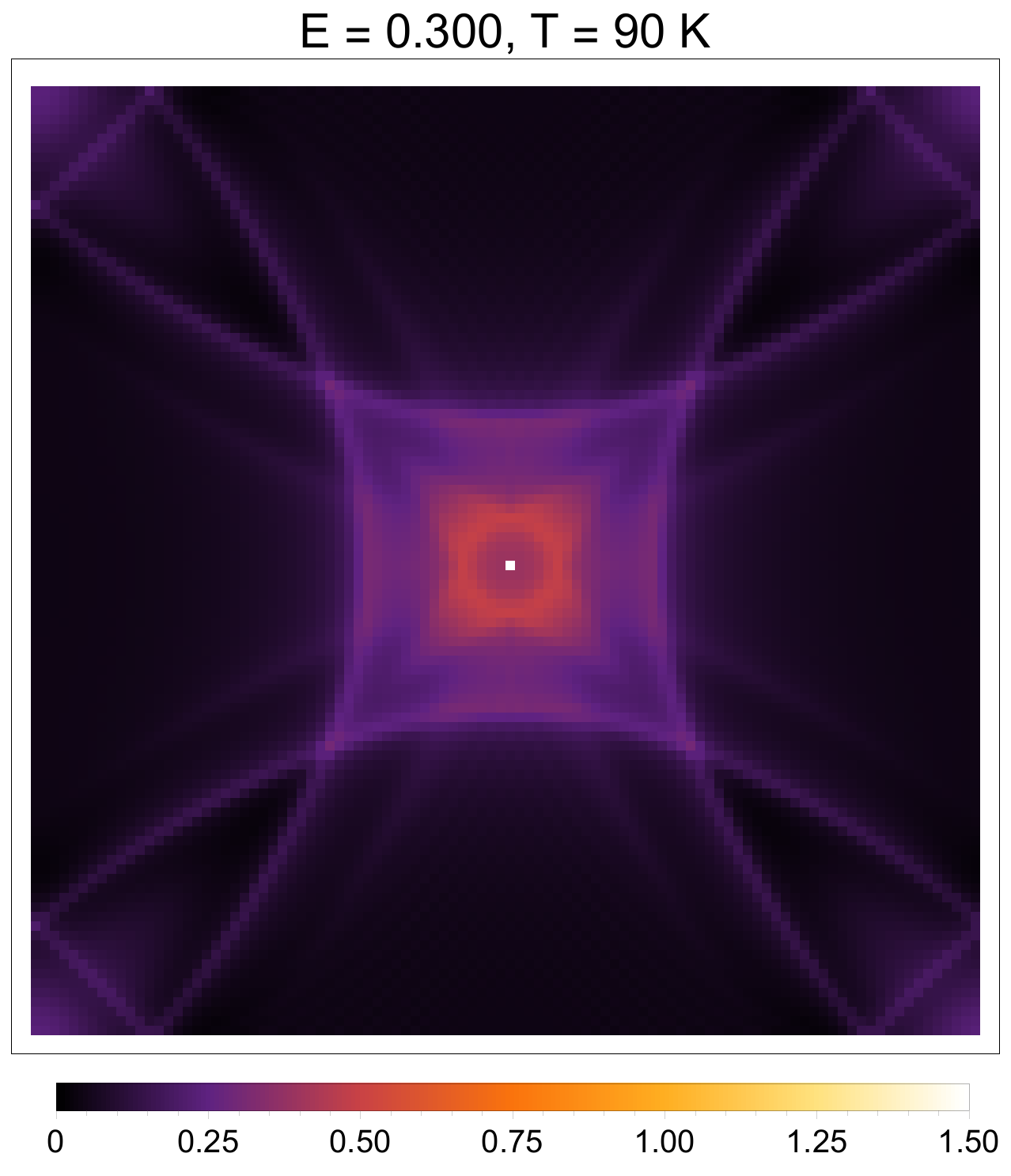} \\
	\includegraphics[width=0.16\textwidth]{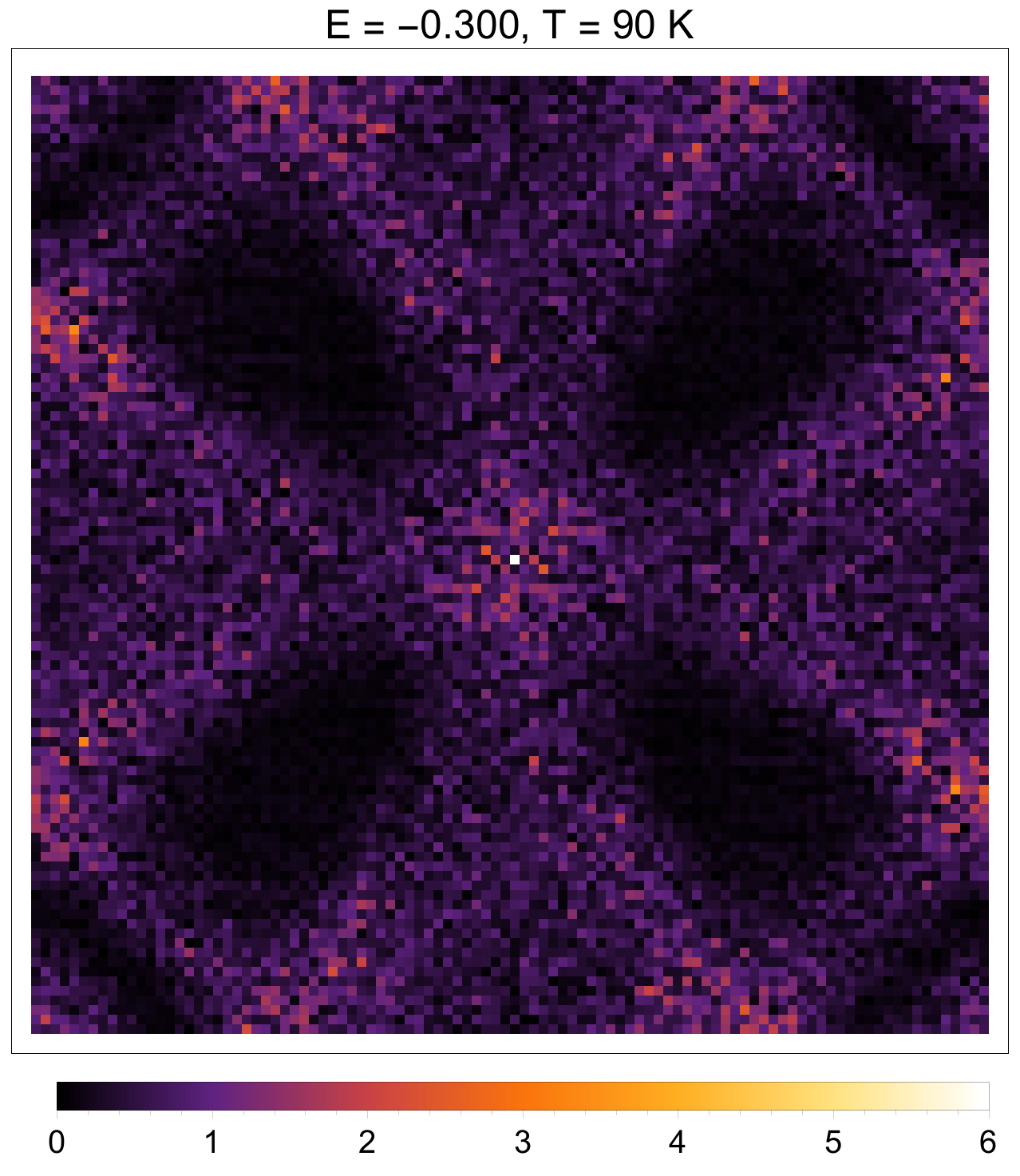}
	\includegraphics[width=0.16\textwidth]{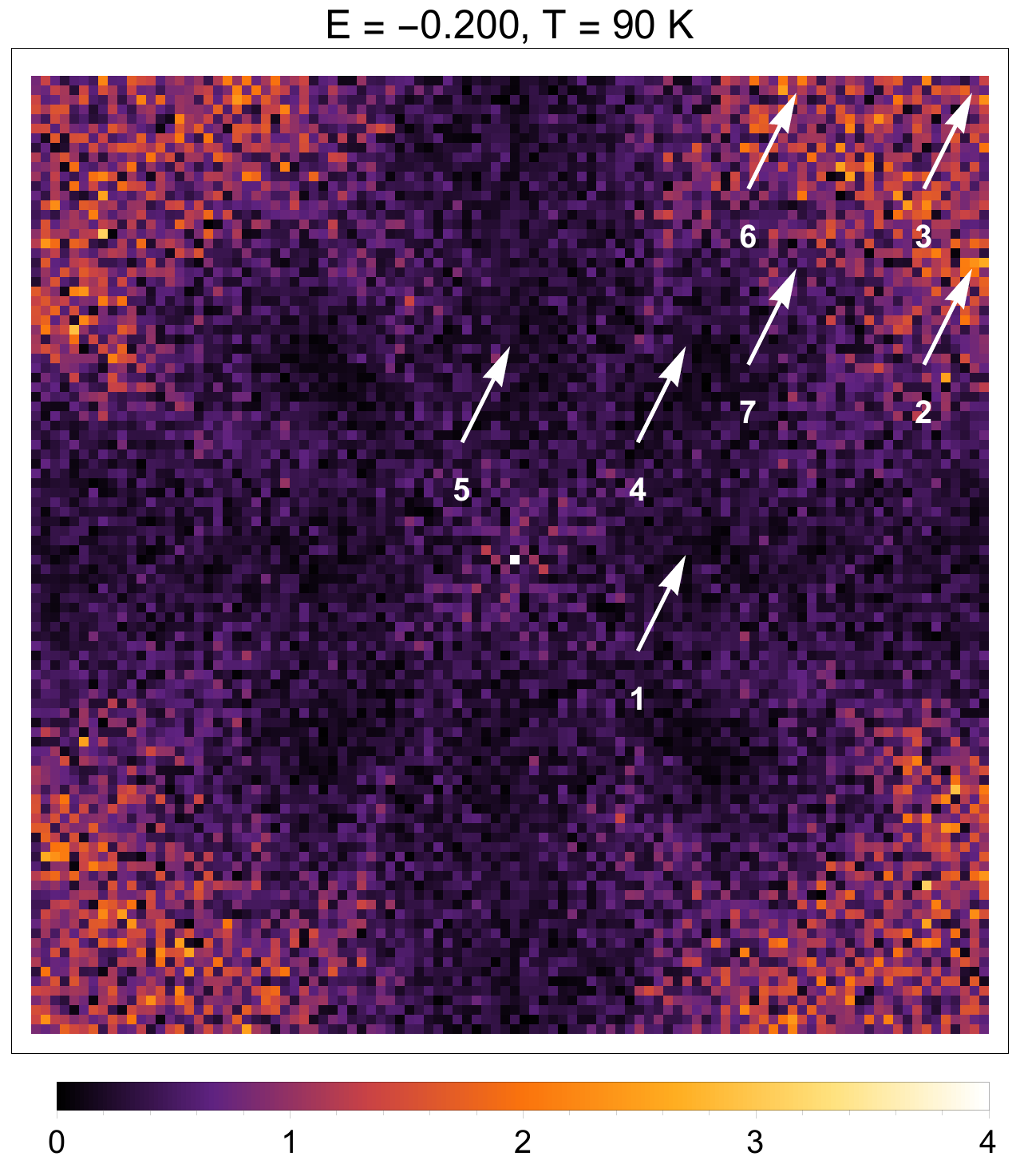}
	\includegraphics[width=0.16\textwidth]{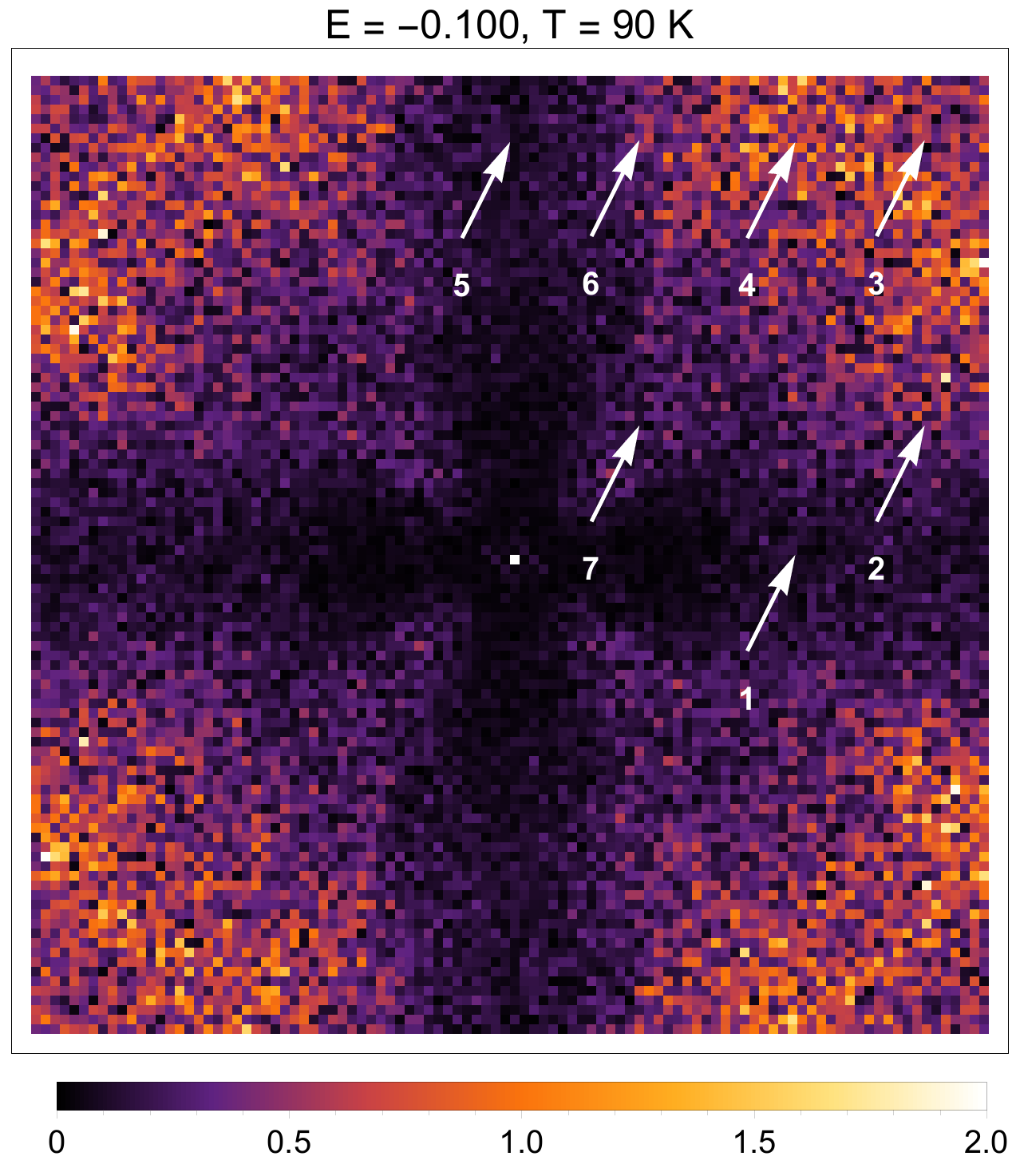}
	\includegraphics[width=0.16\textwidth]{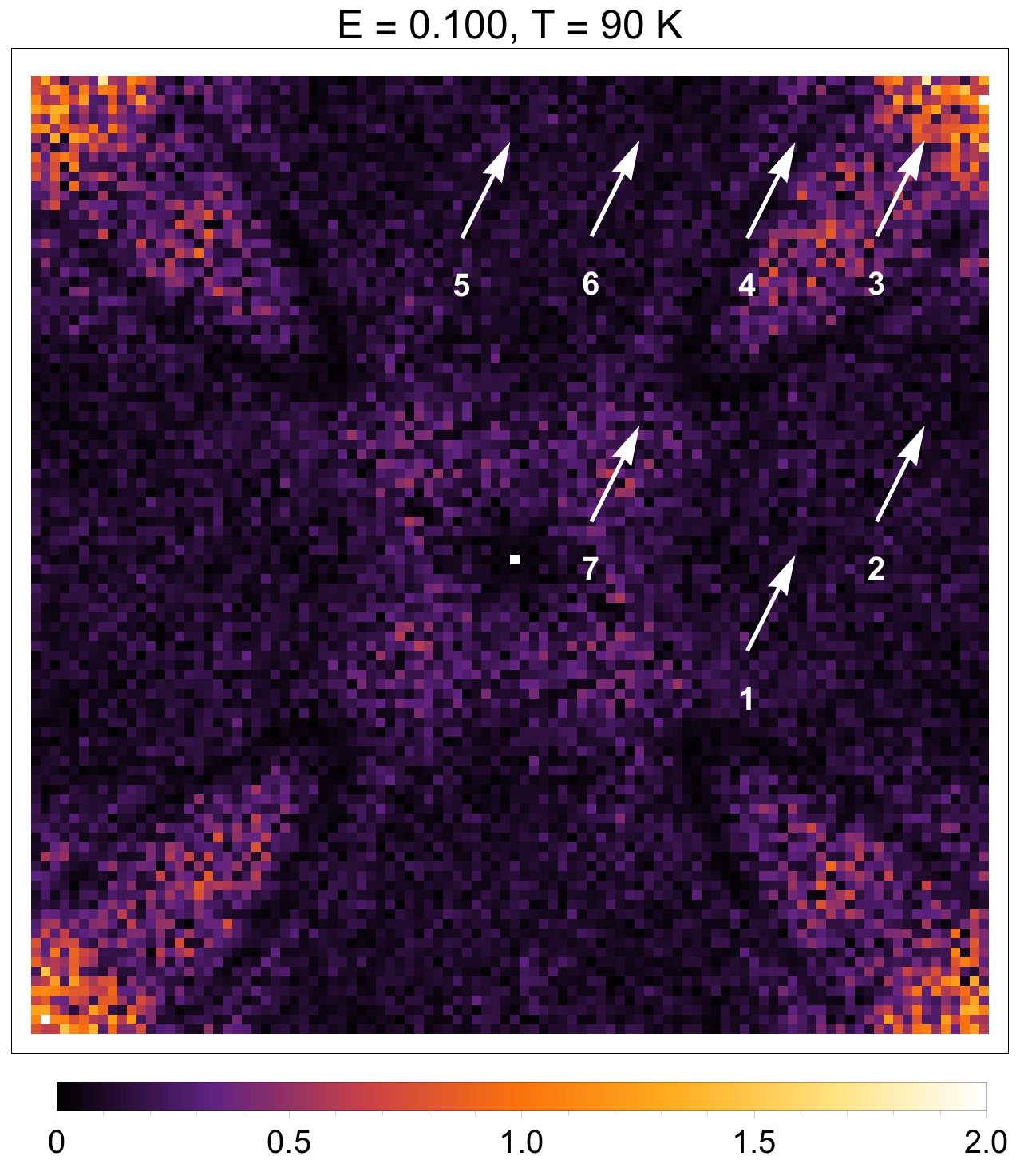}
	\includegraphics[width=0.16\textwidth]{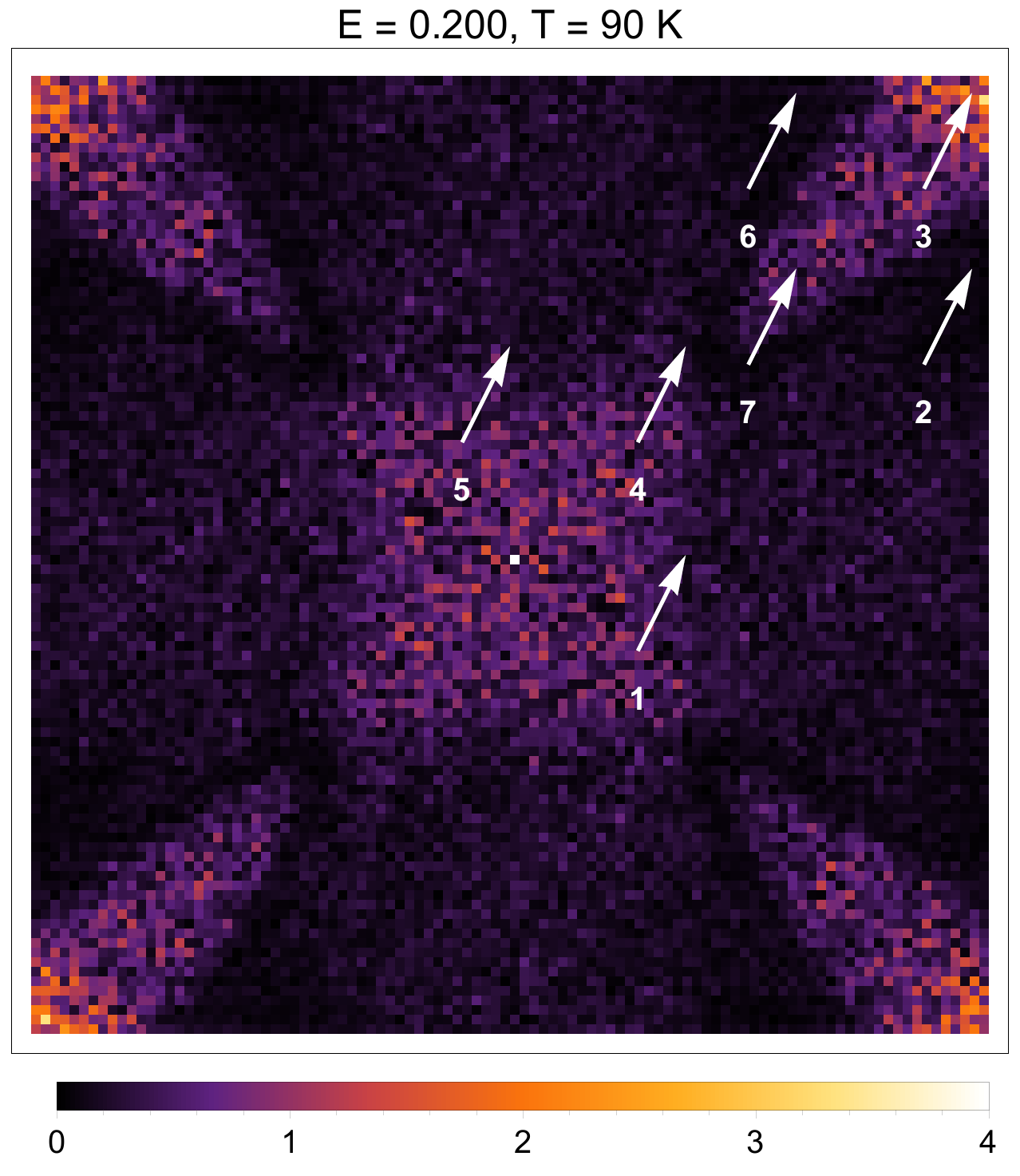}
	\includegraphics[width=0.16\textwidth]{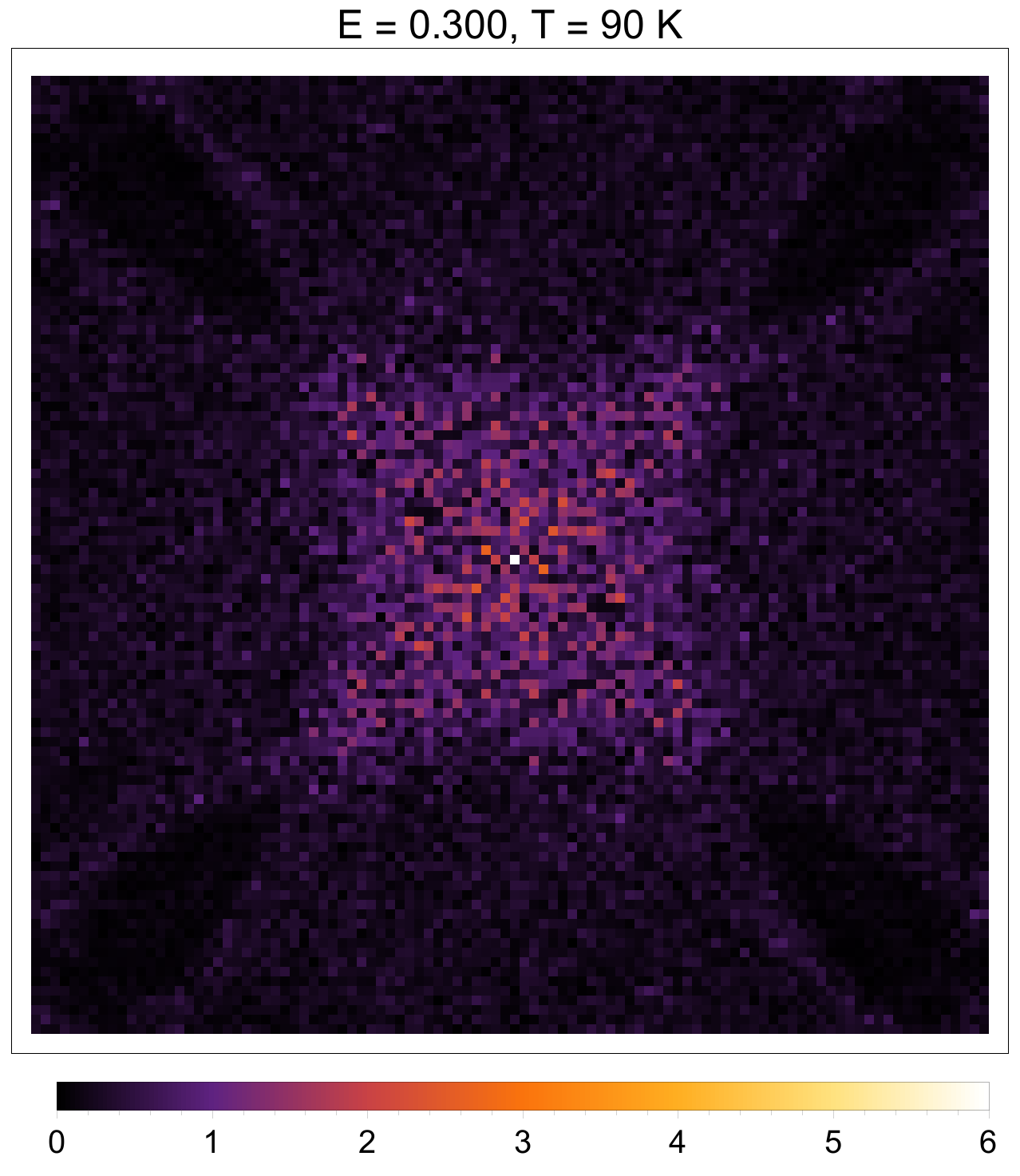} 
	
	\caption{Frequency-dependence at $T = 90$ K in the gap-closing/filling scenario of the spectral function $A(\mathbf{k}, \omega)$ (upper row); the LDOS power spectrum with a single pointlike scatterer without thermal smearing (middle row); and the LDOS power spectrum with both a 0.5\% concentration of pointlike scatterers and thermal smearing (bottom row). Arrows indicate the locations of the peaks predicted by the octet model. Note that the scales used for plotting the LDOS power spectra change with frequency. In this scenario, this temperature is the same as $T_c$.}
	\label{fig:frequency_gf_90k}
\end{figure*}

\begin{figure*}
	\centering
	
	\includegraphics[width=0.16\textwidth]{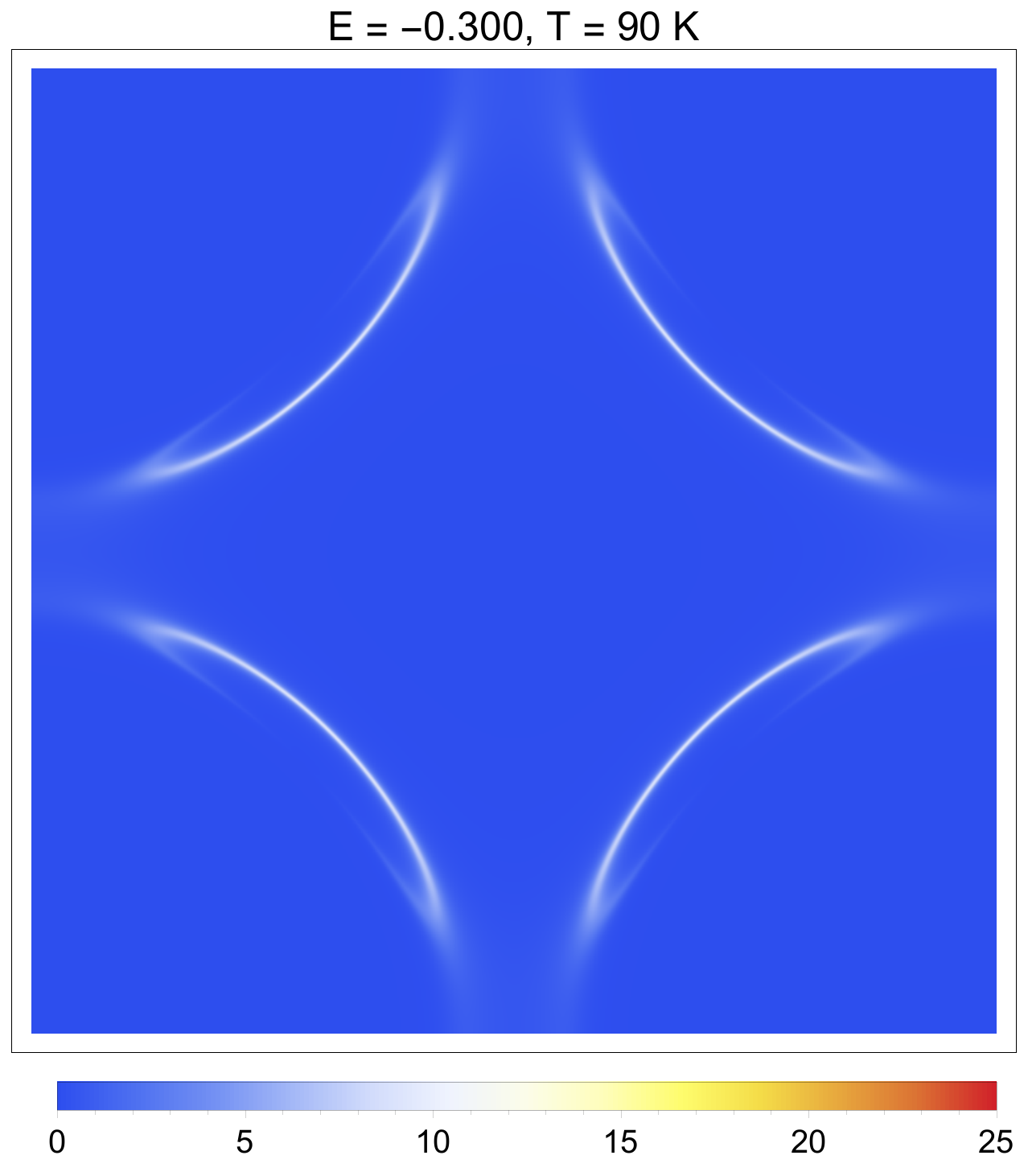}
	\includegraphics[width=0.16\textwidth]{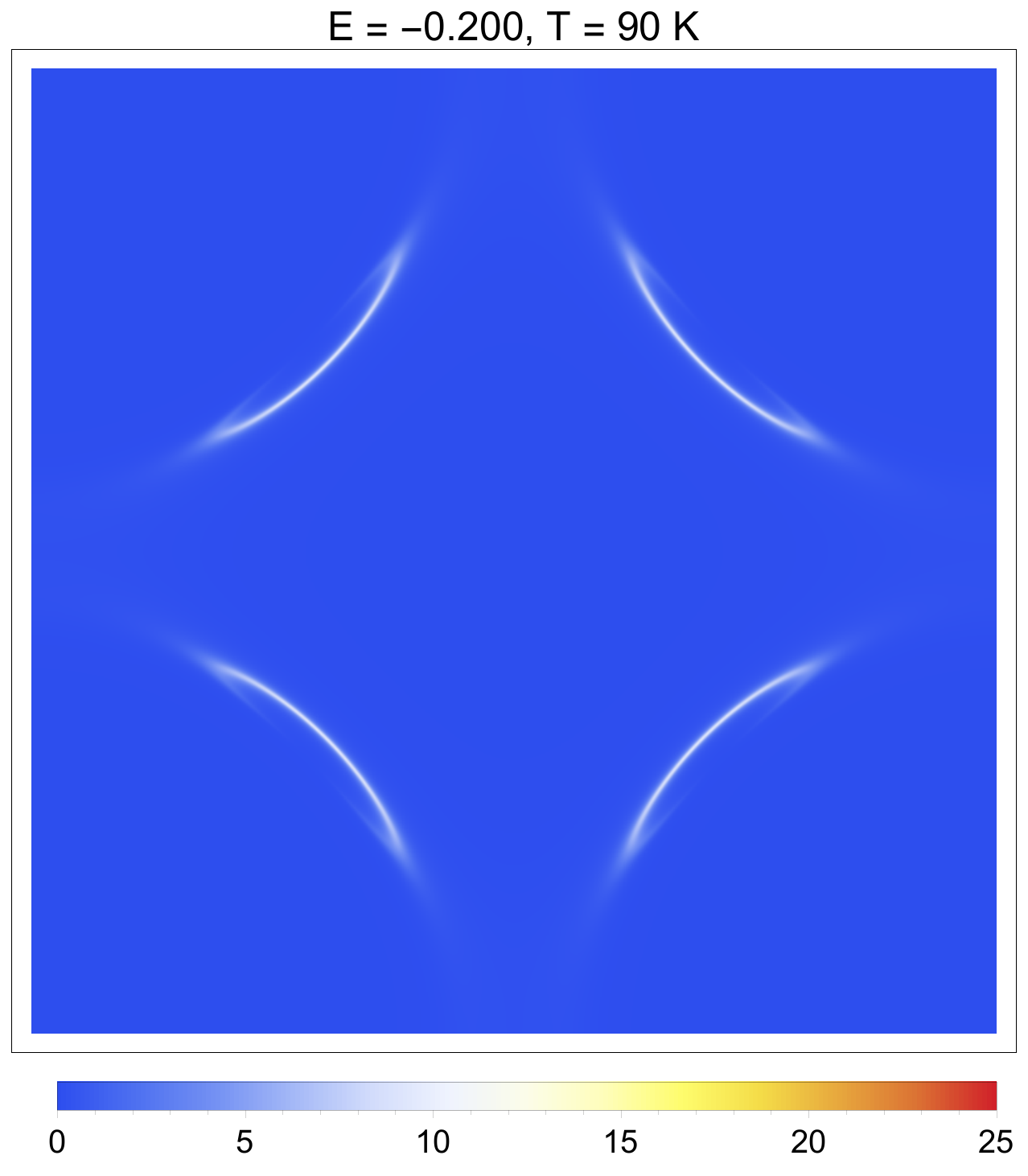}
	\includegraphics[width=0.16\textwidth]{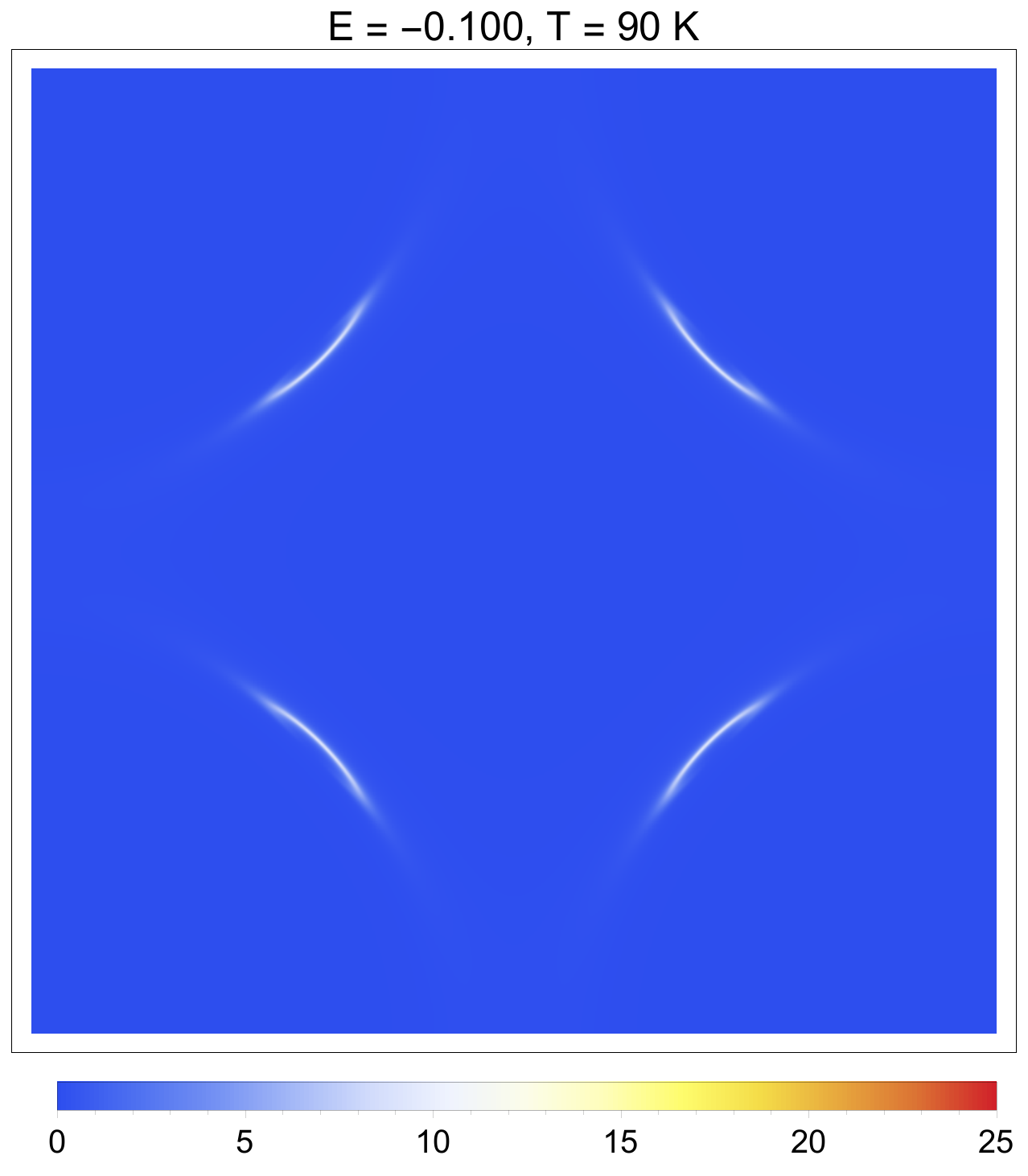}
	\includegraphics[width=0.16\textwidth]{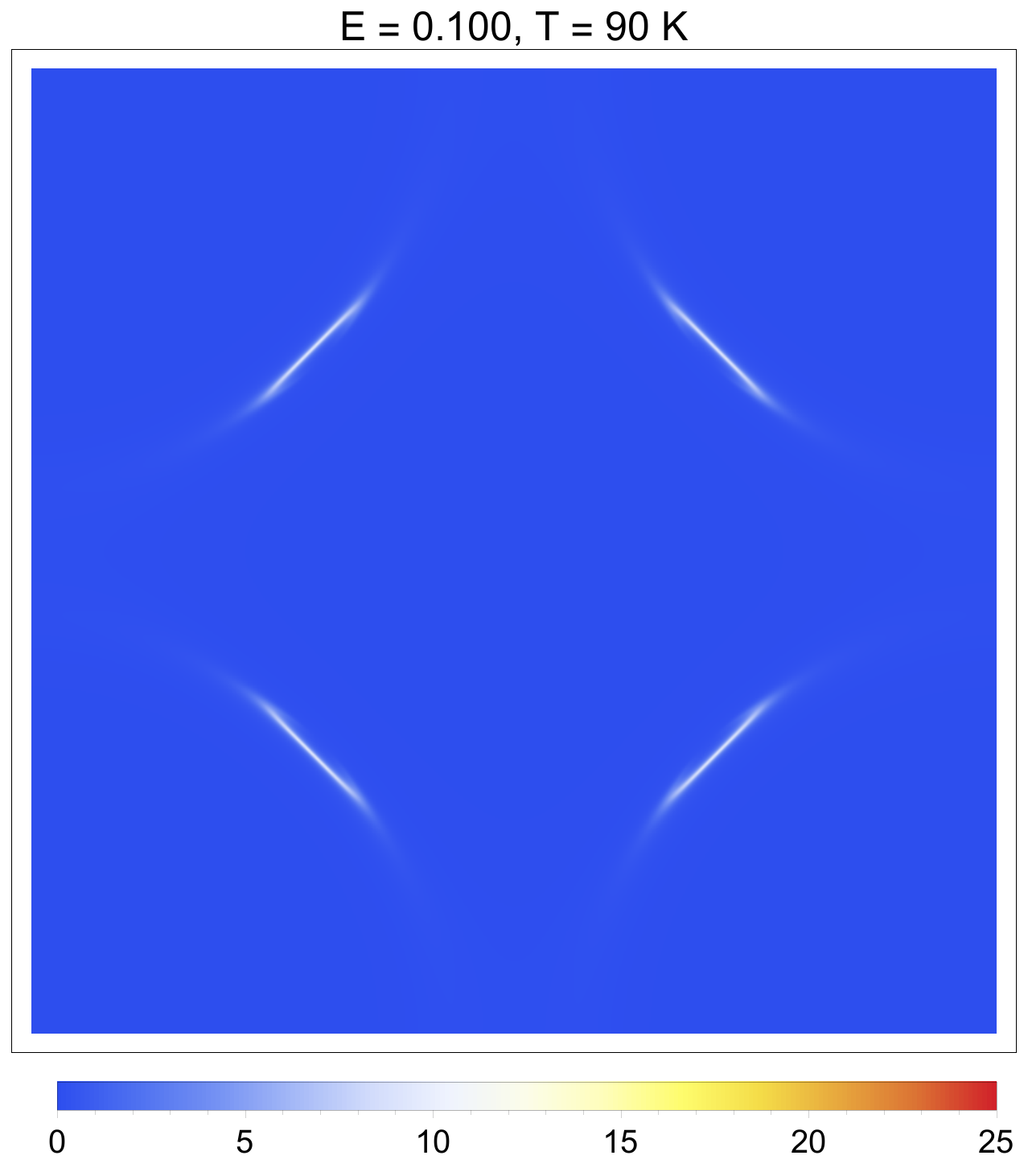}
	\includegraphics[width=0.16\textwidth]{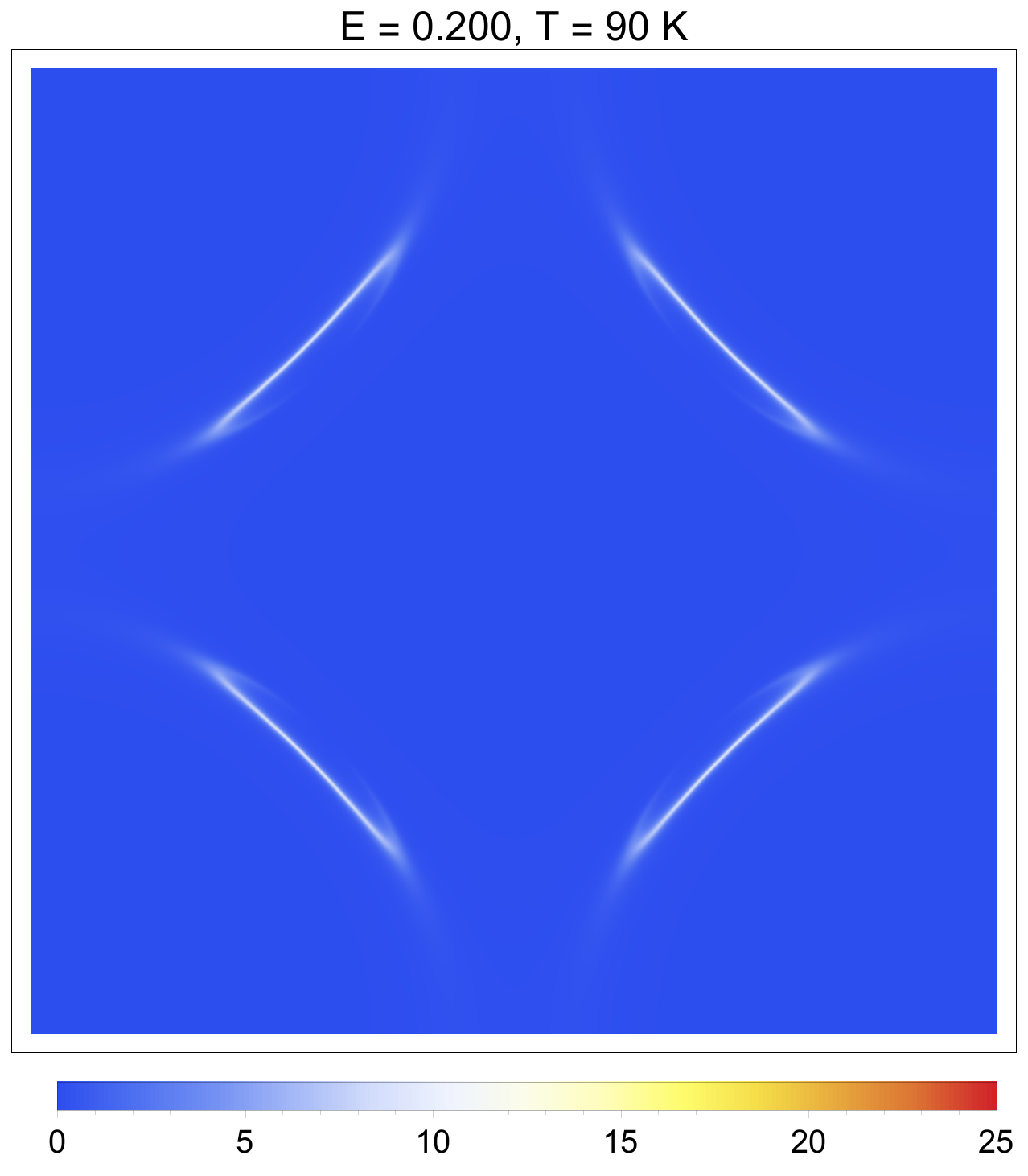}
	\includegraphics[width=0.16\textwidth]{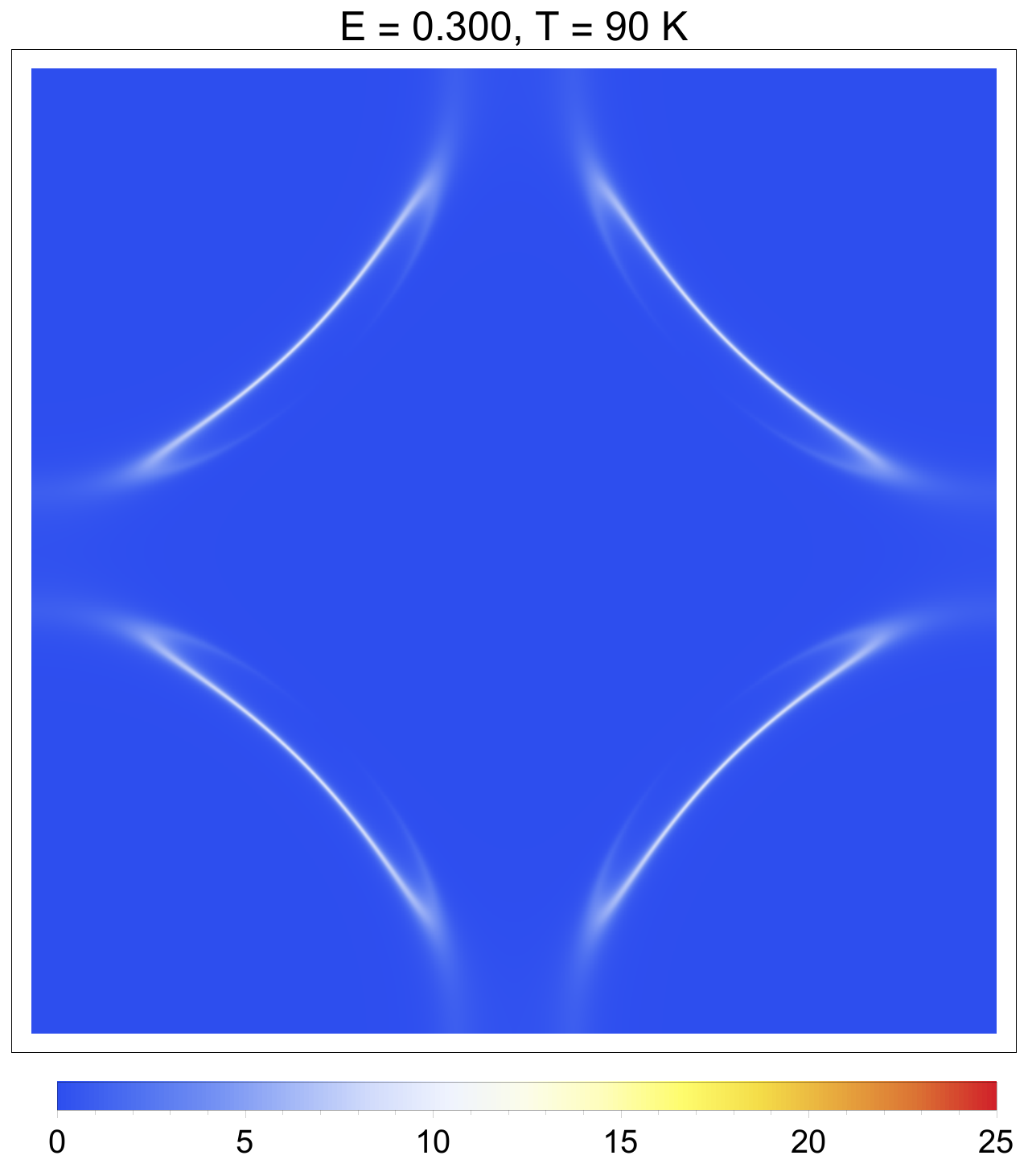} \\
	\includegraphics[width=0.16\textwidth]{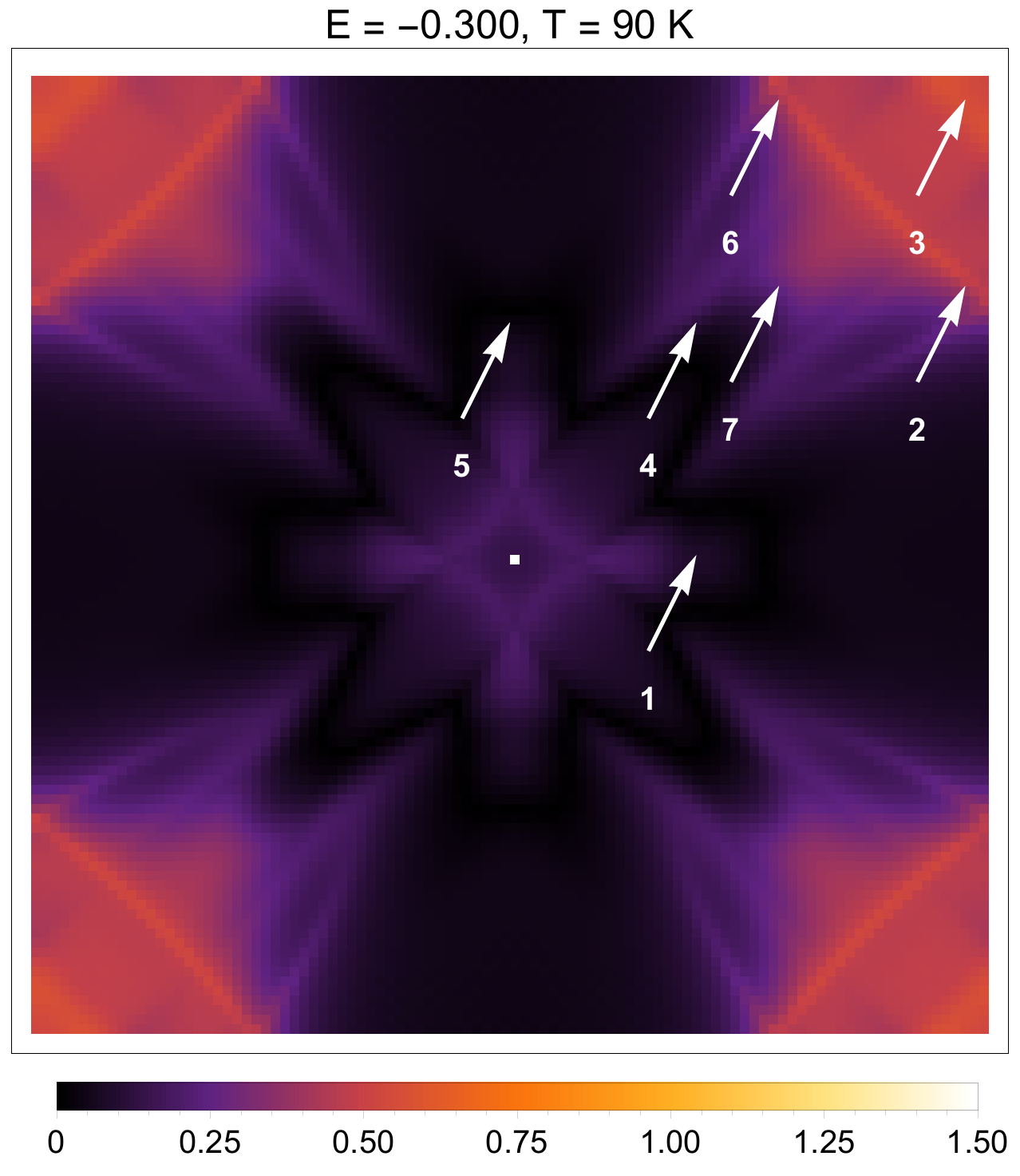}
	\includegraphics[width=0.16\textwidth]{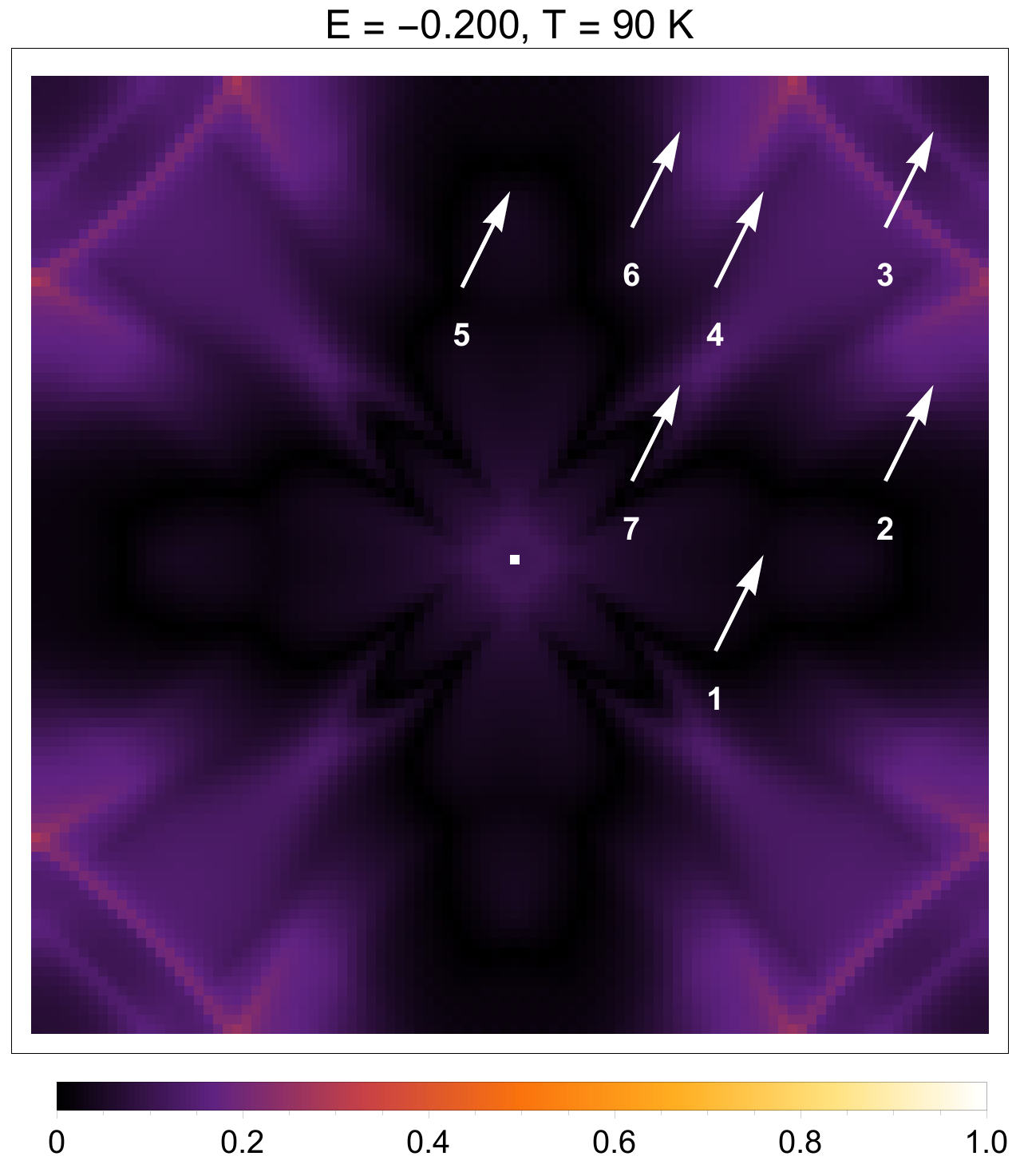}
	\includegraphics[width=0.16\textwidth]{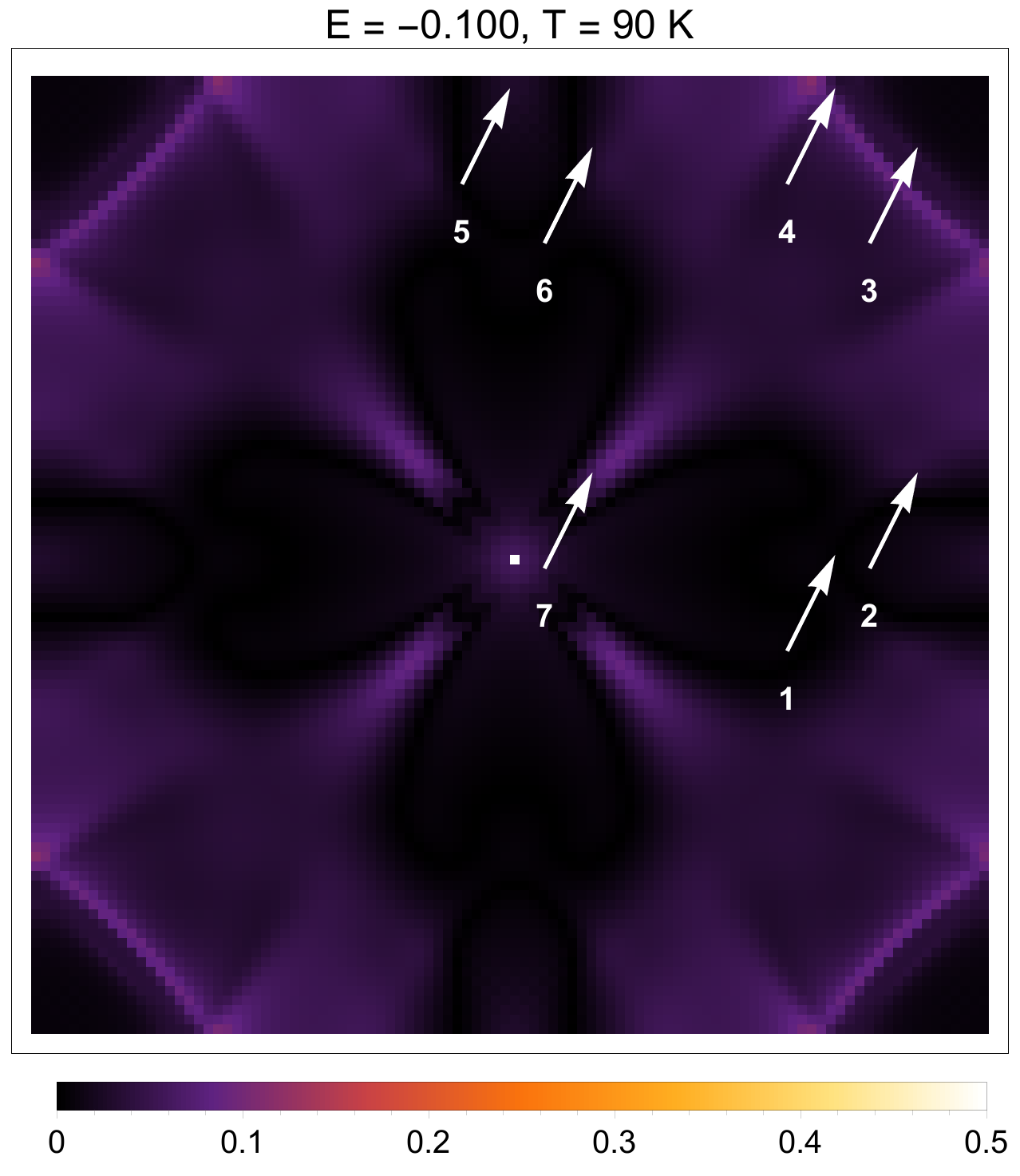}
	\includegraphics[width=0.16\textwidth]{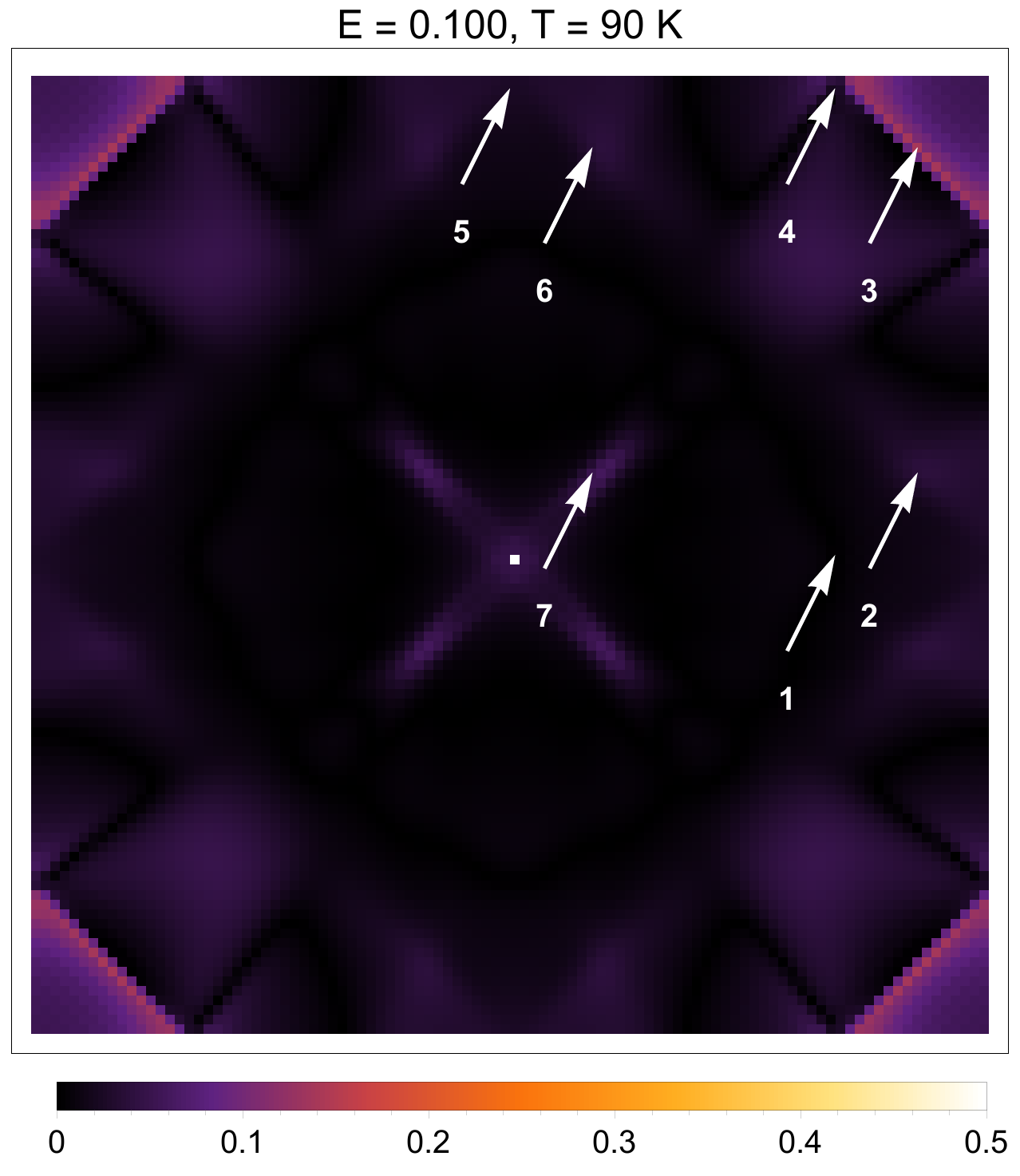}
	\includegraphics[width=0.16\textwidth]{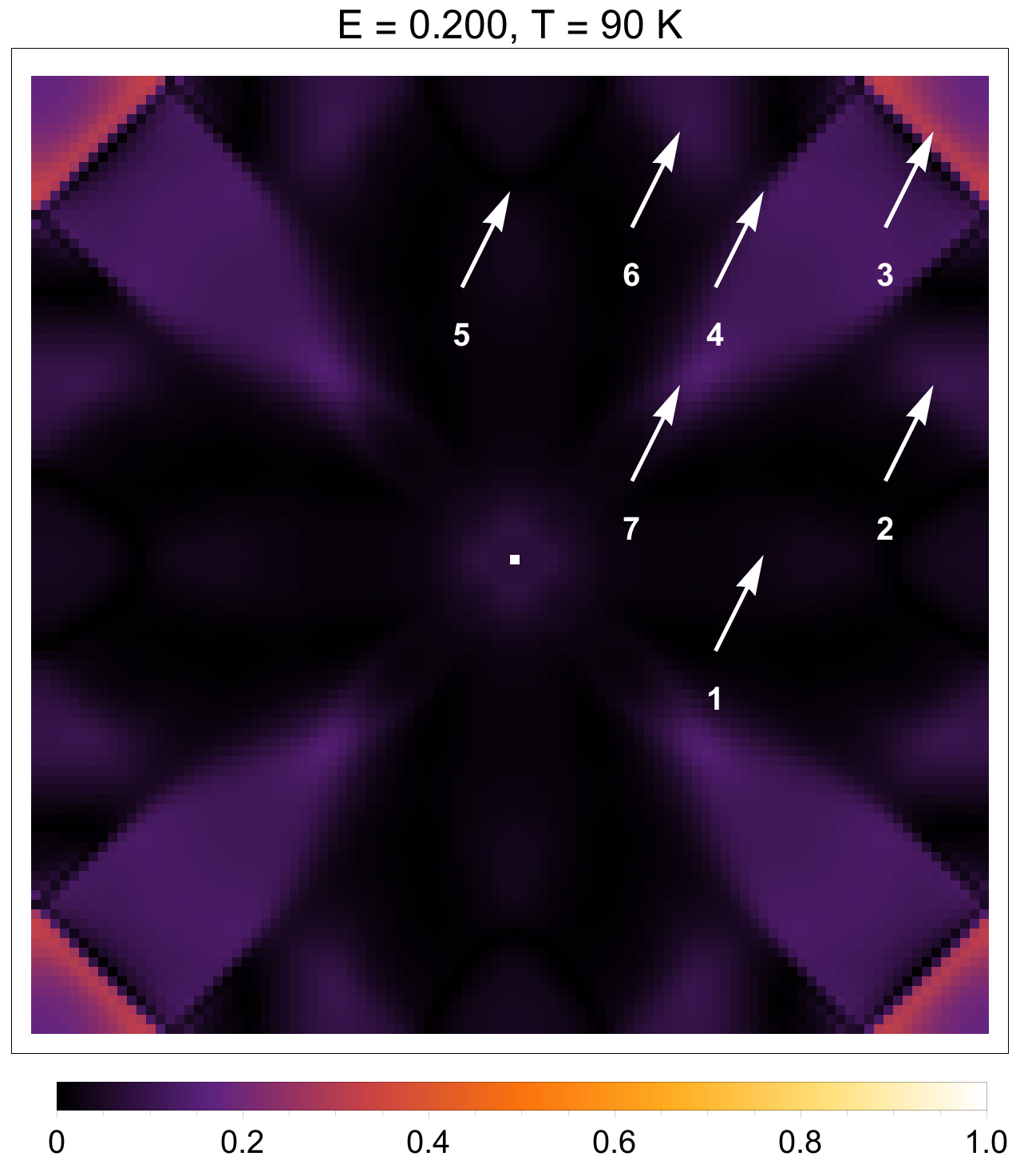}
	\includegraphics[width=0.16\textwidth]{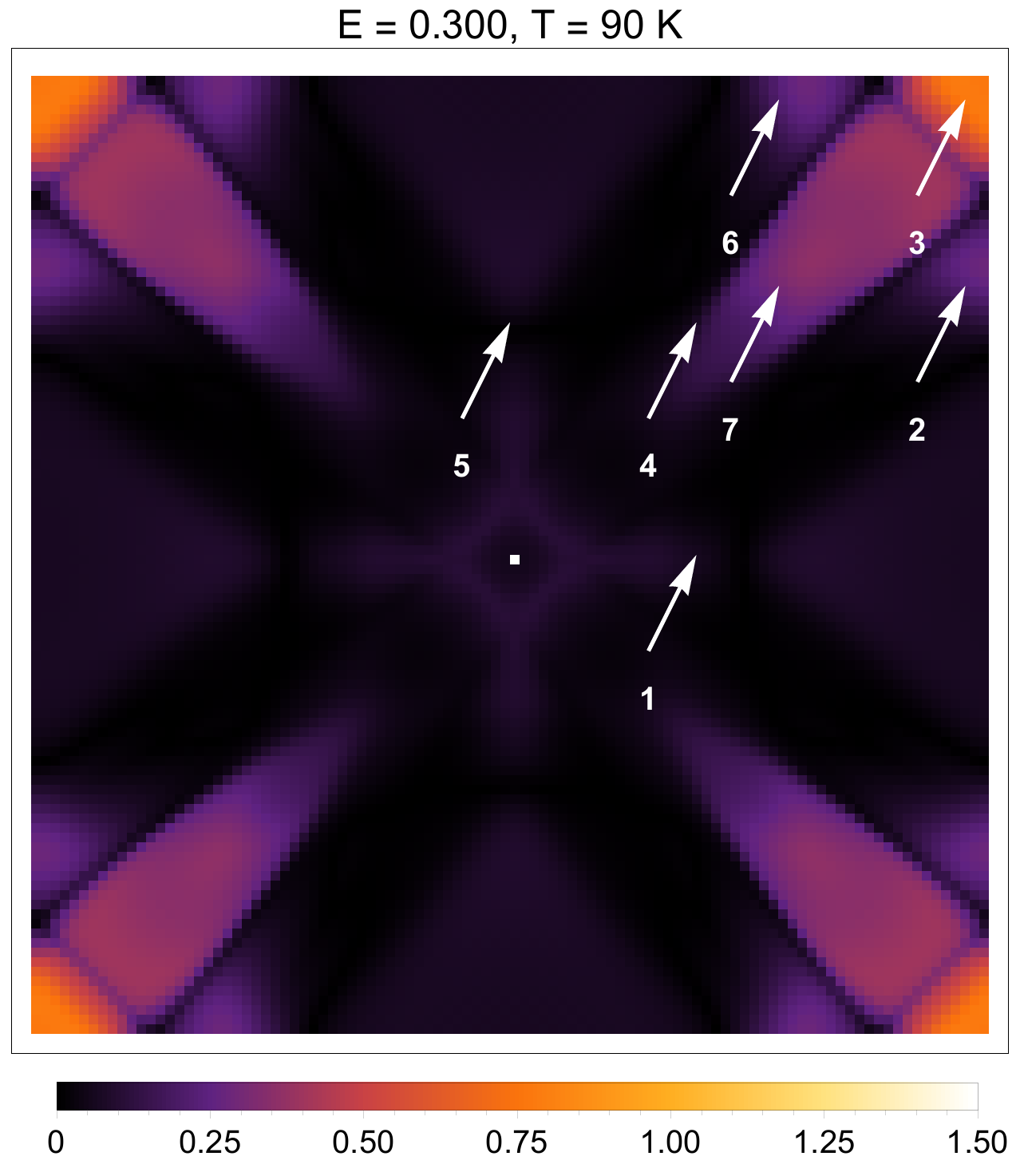}\\
	\includegraphics[width=0.16\textwidth]{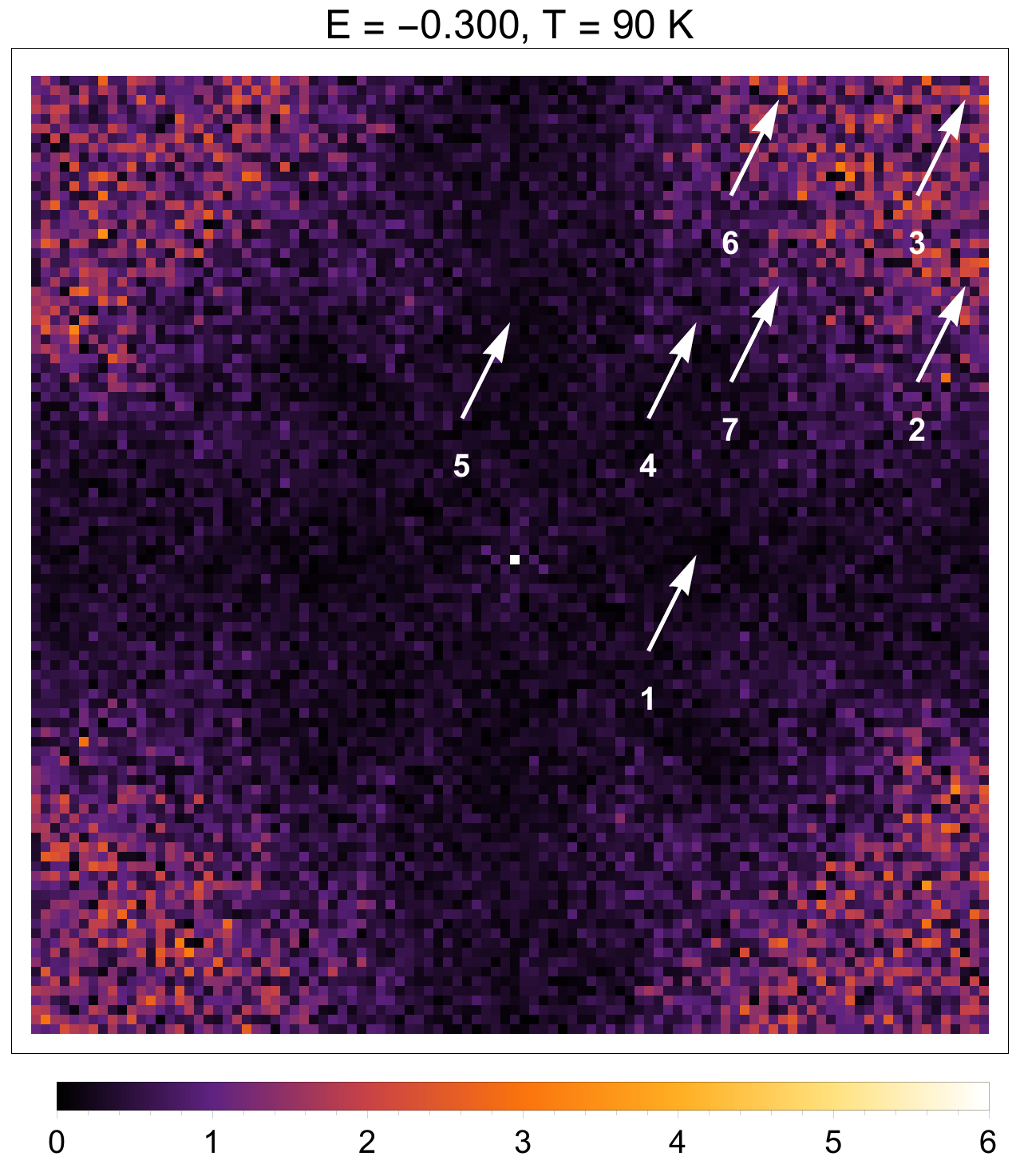}
	\includegraphics[width=0.16\textwidth]{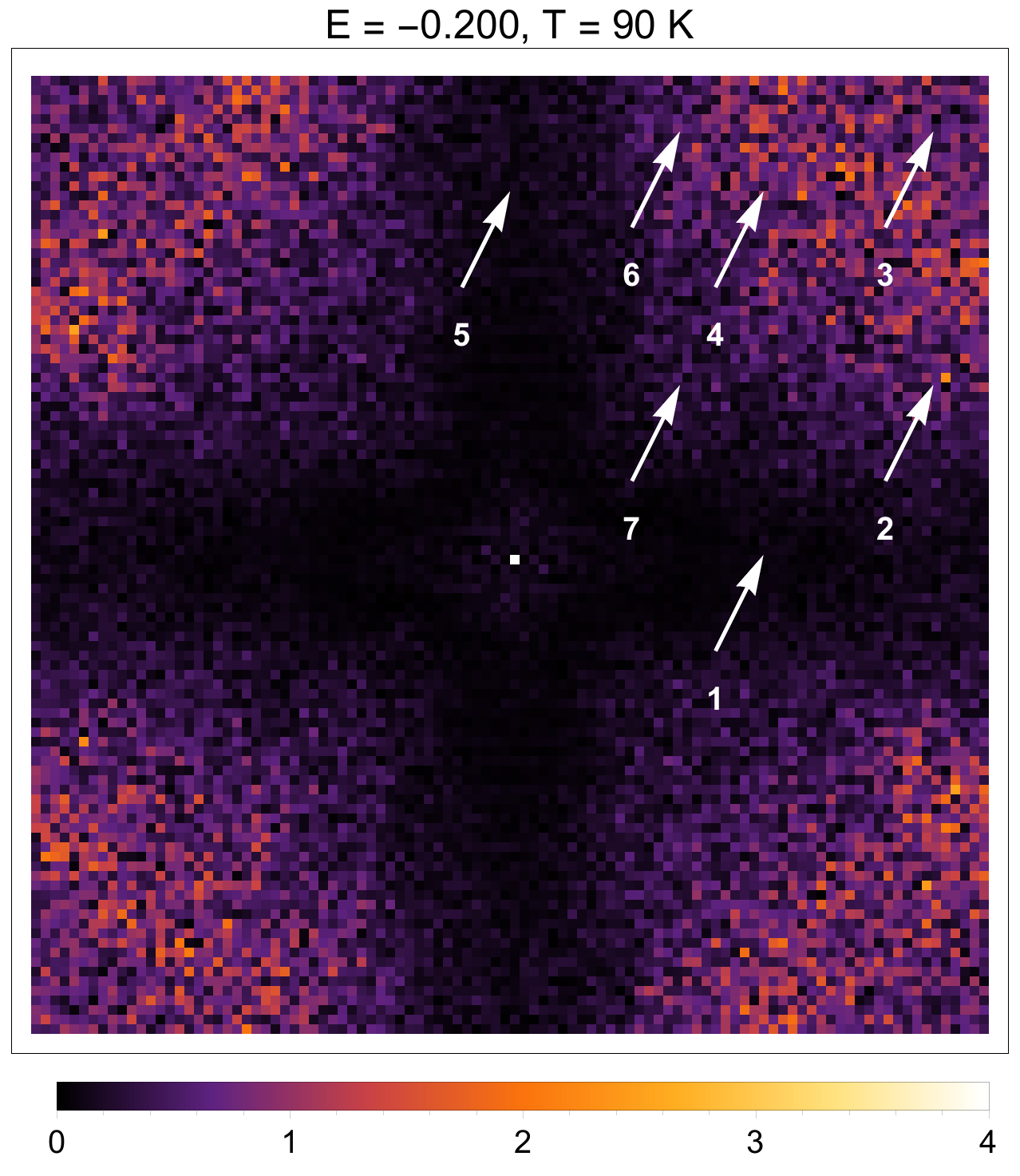}
	\includegraphics[width=0.16\textwidth]{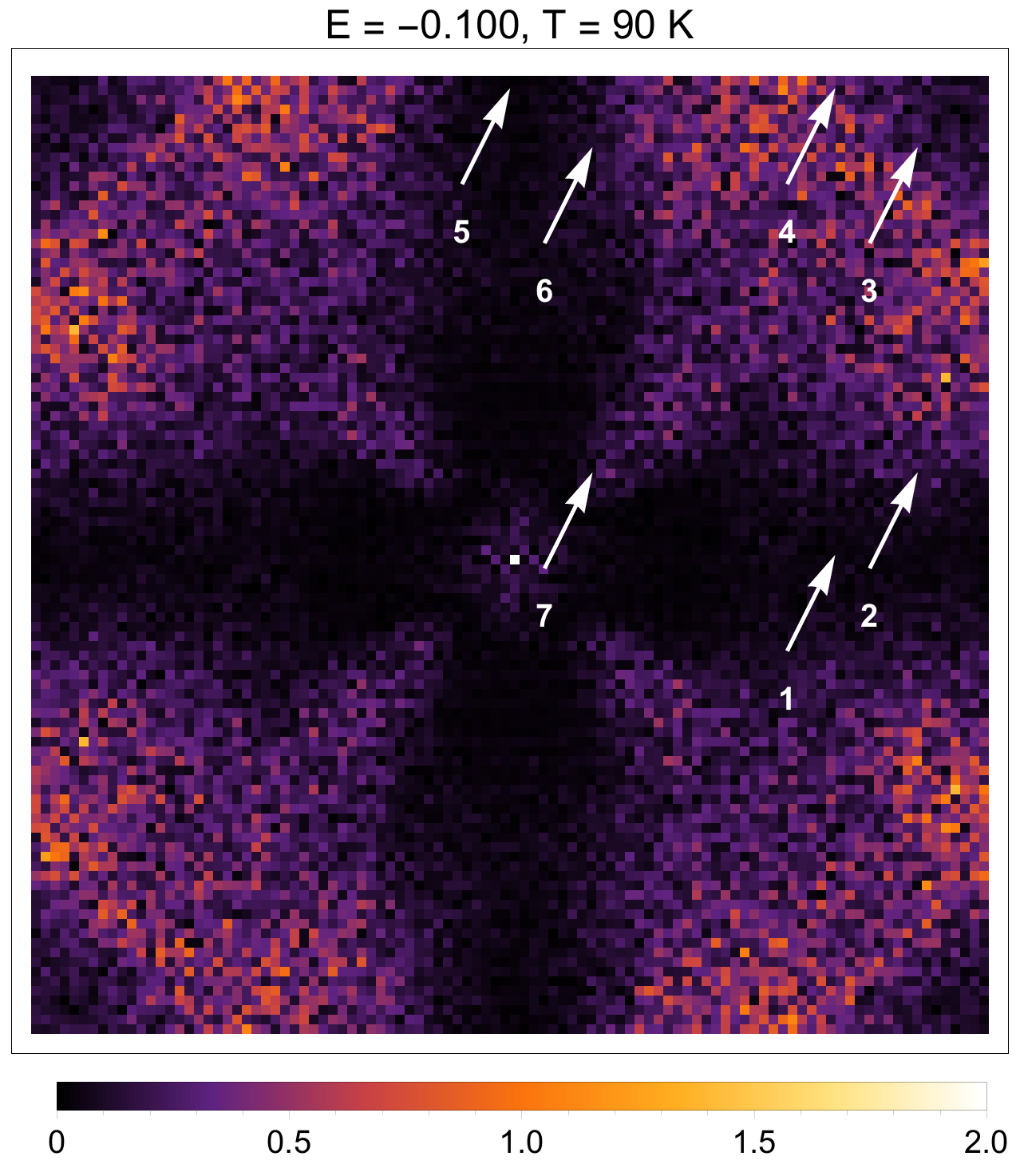}
	\includegraphics[width=0.16\textwidth]{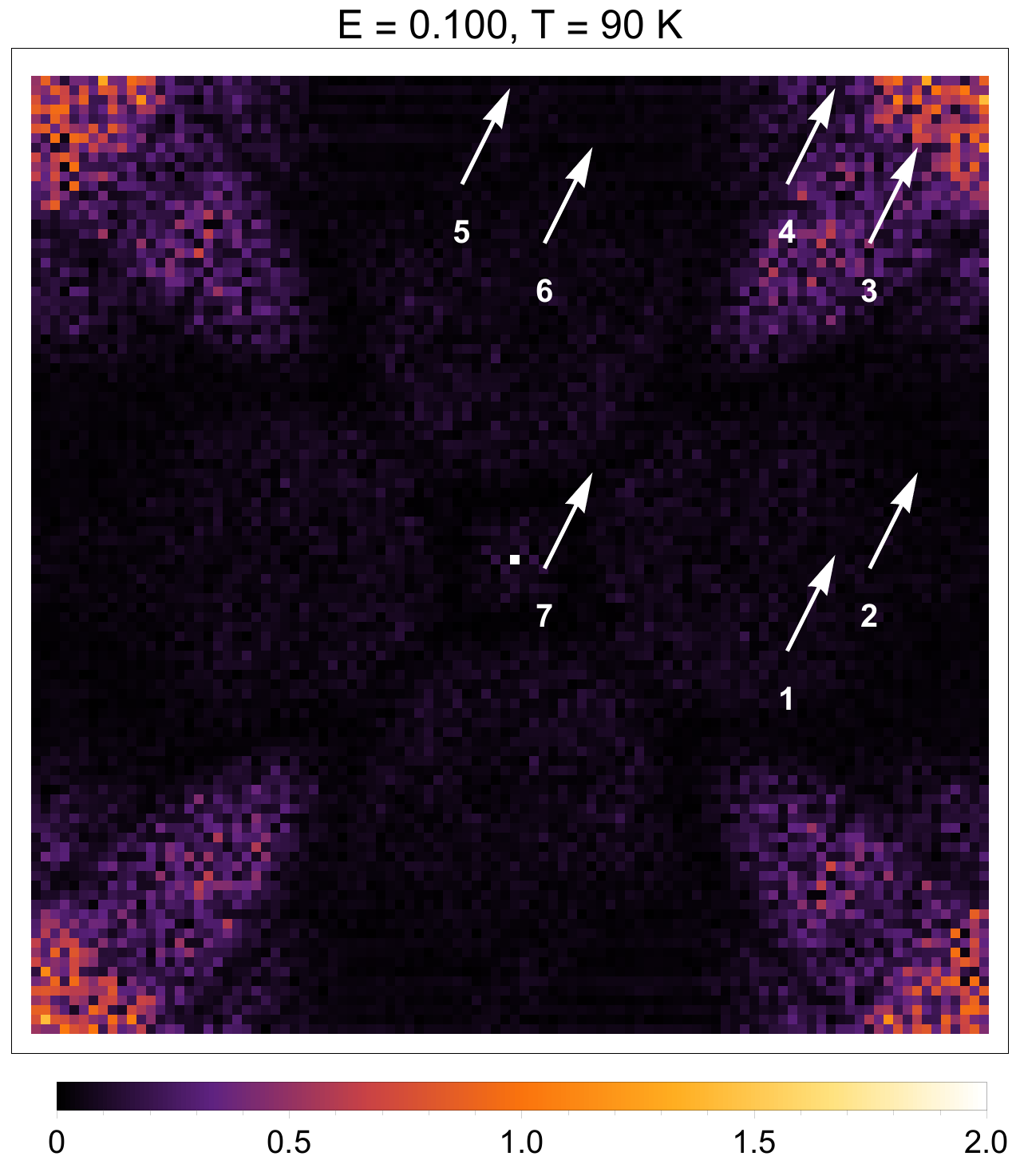}
	\includegraphics[width=0.16\textwidth]{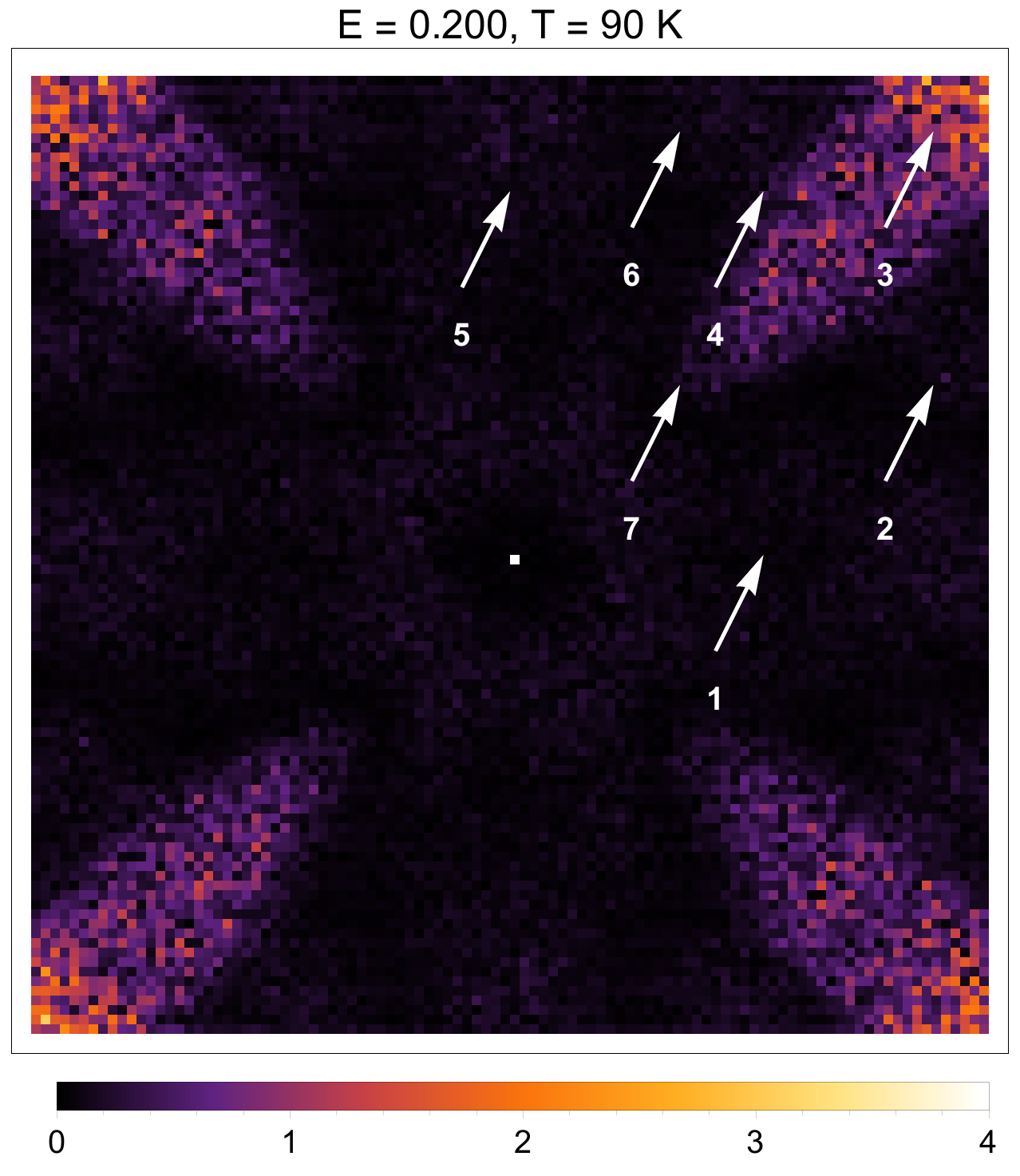}
	\includegraphics[width=0.16\textwidth]{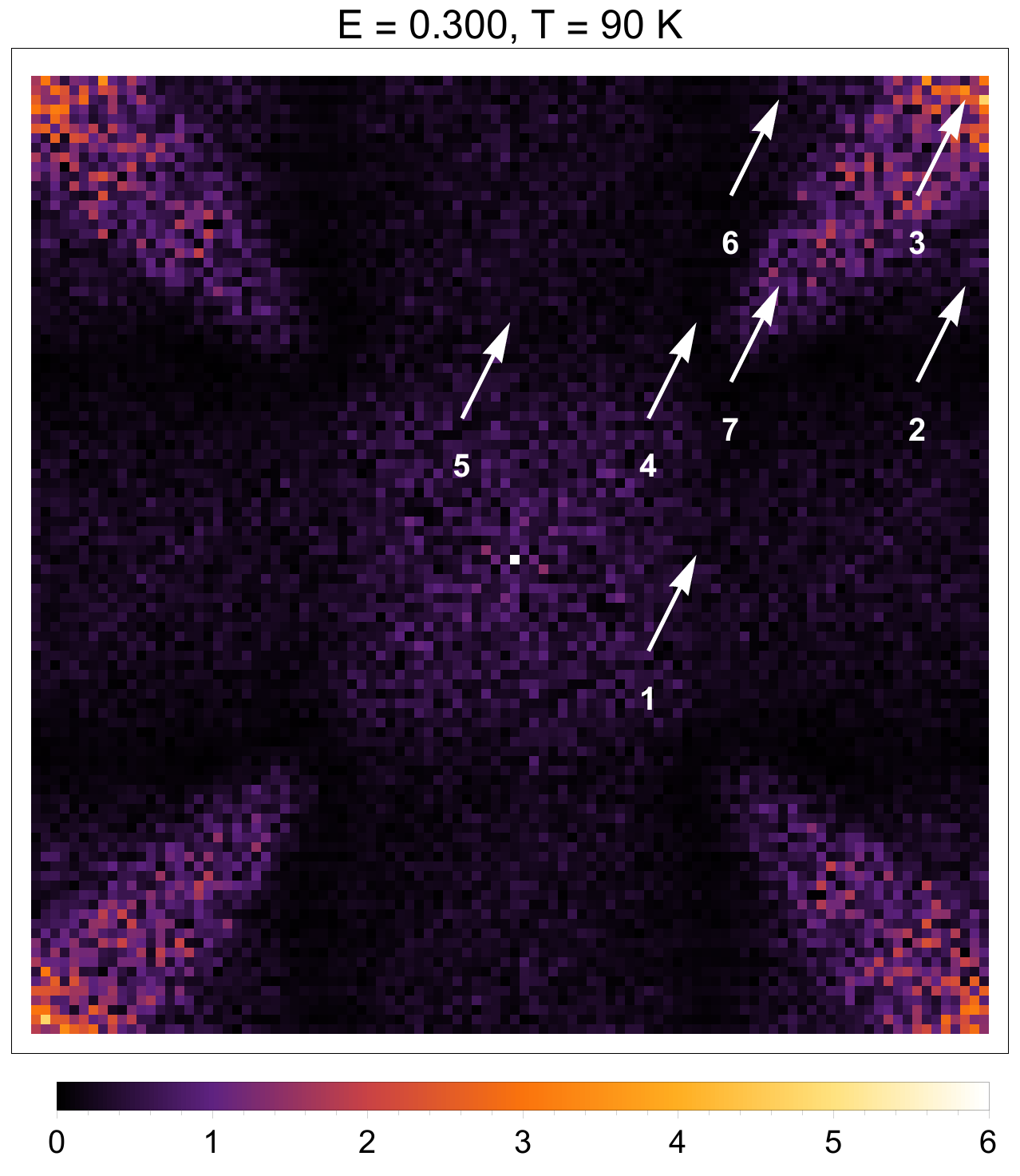} 
	
	\caption{Frequency-dependence at $T = 90$ K in the gap-filling scenario of the spectral function $A(\mathbf{k}, \omega)$ (upper row); the LDOS power spectrum with a single pointlike scatterer without thermal smearing (middle row); and the LDOS power spectrum with both a 0.5\% concentration of pointlike scatterers and thermal smearing (bottom row). Arrows indicate the locations of the peaks predicted by the octet model. Note that the scales used for plotting the LDOS power spectra change with frequency.}
	\label{fig:frequency_cg_90k}
\end{figure*}

\begin{figure}
	\centering
	
	\includegraphics[height=0.18\textwidth]{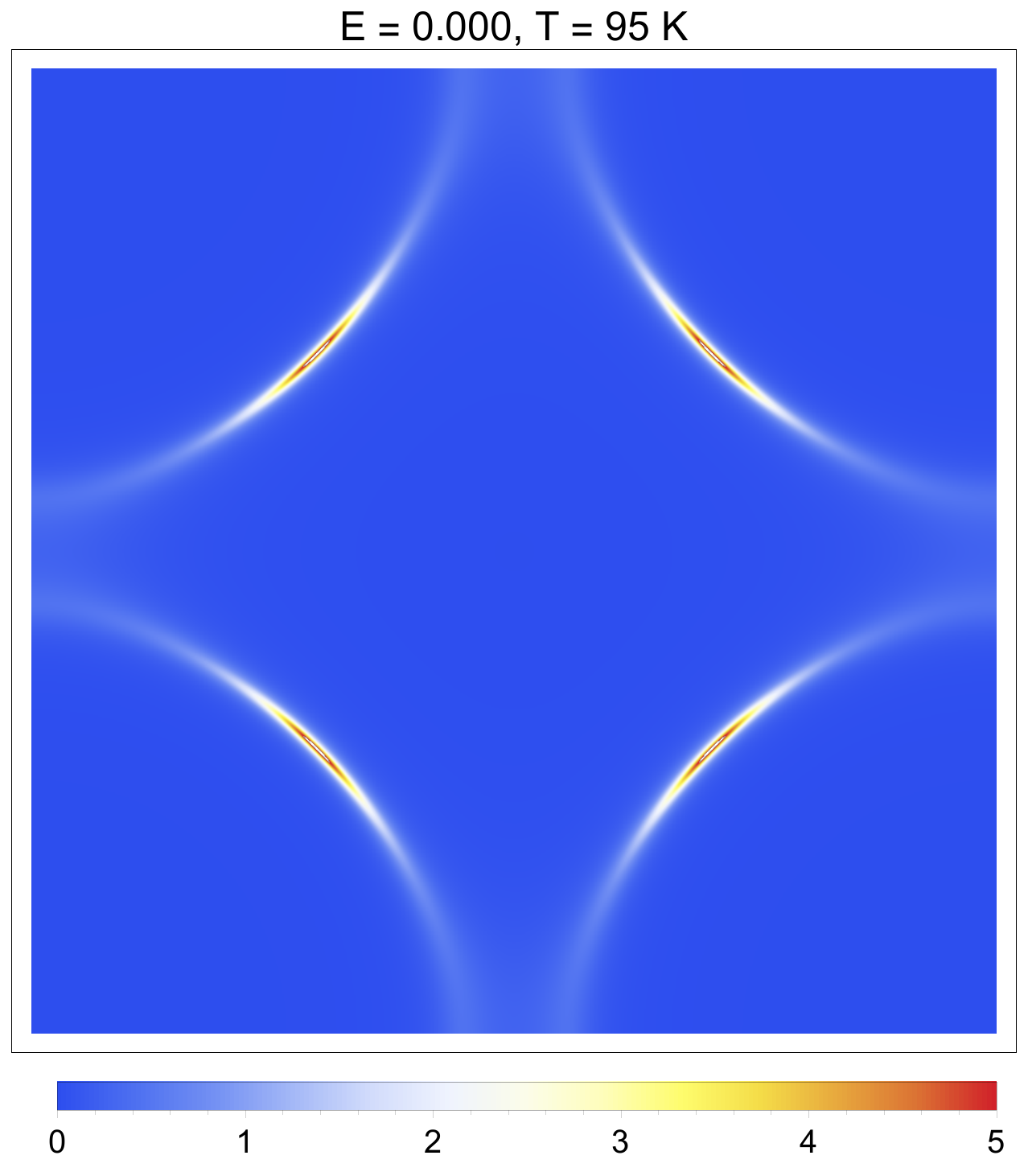}
	\includegraphics[height=0.18\textwidth]{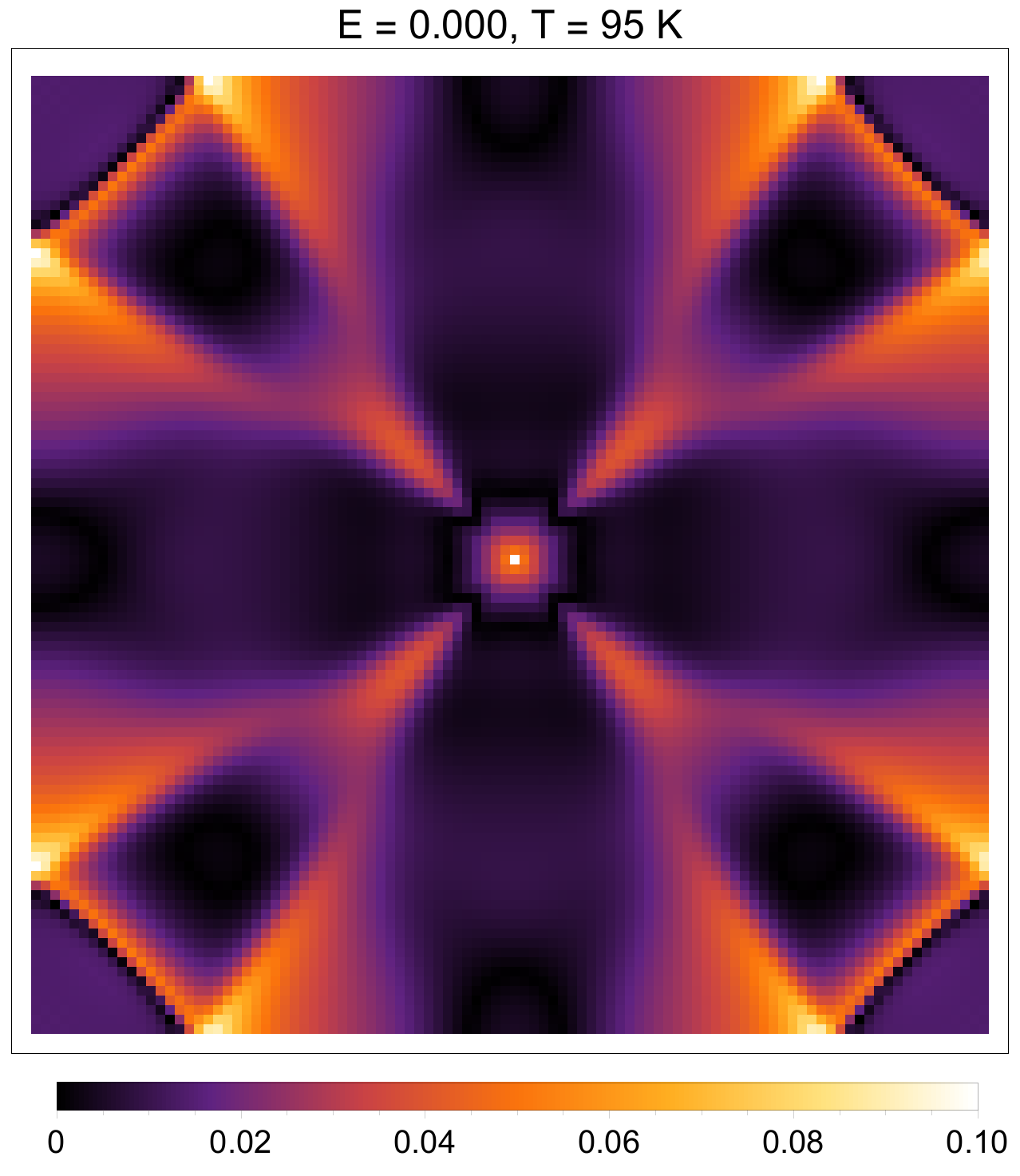} 
	\includegraphics[height=0.18\textwidth]{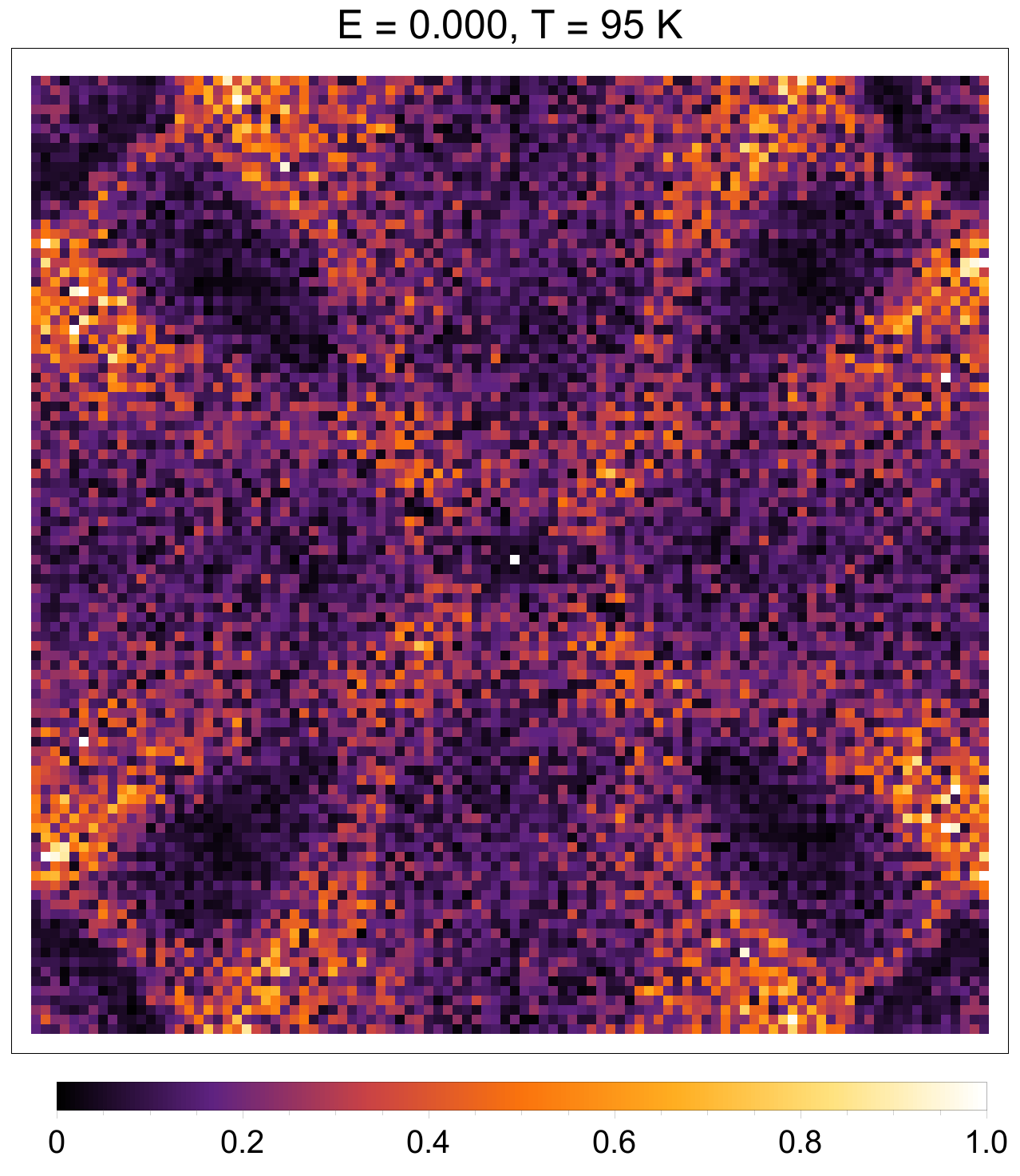} 
	
	\caption{Plots of the spectral function $A(\mathbf{k}, \omega)$ (left), the power spectrum of the single-impurity LDOS without thermal smearing (middle), and the power spectrum of the multiple-impurity LDOS with thermal smearing (right) at $T = $ 95 K in the gap-closing/filling scenario, taken at the Fermi energy ($E = 0$). The spectral function at this regime bears a marked resemblance to the ``Fermi arcs'' found in the pseudogap regime of the underdoped cuprates.}
	\label{fig:fermi_arc_95k}
\end{figure}

In this section we will focus our attention on the various effects of self-energies in the superconducting state which can be seen in STS and ARPES. The main phenomenon of interest is ``gap filling,'' which is seen across a wide range of dopings via ARPES and STS.\cite{reber2012origin, reber2013prepairing, reber2015pairing, reber2015coordination, pasupathy2008electronic} We will examine the phenomenological consequences of a nontrivial temperature-dependence of the scattering rate $\Gamma$ on the observed spectral function, $A(\mathbf{k}, \omega)$, and the power spectrum of the LDOS, $P(\mathbf{q}, \omega)$, both for the single-impurity case without thermal smearing (the idealized case) and the case with an dilute array of weak impurities with thermal smearing (to simulate actual tunneling data). 

Recall that STS experiments on the superconducting cuprates show weak and energy-dependent modulations in the LDOS due to QPI. QPI results from scattering off of weak impurities, which generate Friedel oscillations around impurities. Because of the unusual, banana-like shape of the contours of constant energy (CCEs) of $d$-wave superconductors, the most dominant scattering processes are those from states on one tip of a ``banana'' to those on another, and these dominant wavevectors appear as peaks in the power spectrum of the LDOS---this in a nutshell is the so-called ``octet model.'' Indeed, the peaks seen in experimentally-obtained power spectra behave entirely in accordance with the predictions of this simple model of $d$-wave Bogoliubov quasiparticles scattering off of weak impurities. That said, the vast majority of STS experiments on the superconducting state of the cuprates have been performed at temperatures well below $T_c$, where the quasiparticle scattering rate $\Gamma$ is fairly small and is only weakly dependent on temperature. However, various experiments have shown that $\Gamma$ is not temperature-independent, as one would expect from elastic scattering off of impurities---it instead exhibits a very pronounced dependence on $T$ as temperatures are increased. In fact, recent ARPES results suggest that $\Gamma(T)$ is roughly of the same size as the superconducting gap $\Delta_0(T)$ itself as $T \to T_c $.\cite{reber2015pairing} Furthermore, the same ARPES results show that $\Delta_0(T)$ does not go to zero at $T_c$, as one would expect from BCS theory. Instead, $d$-wave pairing correlations are seen to exist beyond $T_c$, and persist up to a higher temperature scale which appears to decrease as doping is increased. We show in Fig.~\ref{fig:gbtemp} plots of the superconducting gap and the quasiparticle scattering rate as a function of temperature for three different scenarios: the BCS scenario, in which the gap closes as $T$ is increased, becoming zero at $T_c$; the gap-filling and -closing scenario, seen in ARPES experiments by Reber \emph{et al.} on optimally-doped BSCCO ($T_c \approx 90$ K), in which the gap shrinks and the quasiparticle scattering rate increases as $T$ is increased, but the gap remains finite past $T_c$;\cite{reber2015pairing} and the gap-filling scenario, in which the gap remains roughly temperature-independent while the scattering rate increases at $T$ near $T_c$. We will carry out the exercise of obtaining results measurable by STS experiments in the cuprates as temperature is increased, assuming consistency with ARPES results. It is an interesting experiental question to see if the peaks suggested by the ``octet model'' still appear when the quasiparticle scattering rate is very large, as appears to be the case when $T \approx T_c$.

The temperature-dependence of the superconducting gap and the scattering rate can be parametrized simply as follows.  As argued by Reber \emph{et al.}, the experimentally-measured gap amplitude at optimal doping can be fit to the following BCS-like functional form,
\begin{equation}
\Delta_0(T) = \Delta_0(0)\times\text{tanh}\left({\alpha\sqrt{\frac{T_p}{T} - 1}}\right),
\label{eq:gap}
 \end{equation}
 where $T_p$ is the temperature at which the gap fully closes, $\Delta_0(0)$ is the value of the gap at $T = 0$, and $\alpha$ is a dimensionless number of order unity.\cite{reber2015coordination} In our numerics we will take $\Delta_0(0) = 0.096$, $T_p = 100$ K, and $\alpha = 2.32$. We remind the reader that $\Delta_0(T)$ enters the momentum-space gap function as $\Delta(\mathbf{k}, T) = 2\Delta_0(T) \times (\cos k_x - \cos k_y)$. As for the scattering rate, we use the form obtained by Chubukov \emph{et al.},\cite{chubukov2007gapless} which we write in the following manner:
 \begin{equation}
\Gamma(T) = \Gamma_0 + \Omega\text{sinh}\left(\frac{T_b}{T} \right).
\label{eq:scatteringrate}
 \end{equation}
 Here $T_b$ is some very large temperature scale included phenomenologically in order to provide a good fit with the experimental results and $\Gamma_0$ is the elastic scattering rate. Reasonably good fits can be obtained by using $\Gamma_0 = 0.015$, $\Omega = 2350$, and $T_b = 1100$ K. We will neglect any momentum-dependence of the scattering rate. These two functional forms in tandem with each other explain very well the phenomenology of the closing and the filling of the gap as seen in experiments.
 
The BCS case features only the closing of the gap, and only elastic scattering is present; as such we will take $\Gamma(T) = 0.015$ in that case. To allow us to compare the results of the first case with the BCS case, we take the same functional form for the BCS case as in Eq.~\ref{eq:gap}. This ensures that the values of the superconducting gap are the same at each temperature, and that the effects of the self-energies in the first case can be isolated very clearly and contrasted with the trivial effects seen in the BCS case. It is very important to note that in the case with both the filling and closing of the gap, $T_p$ is \emph{not} equal to $T_c$, whereas in the BCS case $T_p = T_c$. Finally, for the gap-filling case, we will freeze $\Delta_0(T)$ at its $T = 0$ value, and let the scattering rate vary with temperature as in Eq.~\ref{eq:scatteringrate}.

To illustrate clearly the differences between BCS and gap-filling phenomenology, we first discuss the BCS case with only the closing of the gap and show in Fig.~\ref{fig:temperature_bcs} plots of $A(\mathbf{k}, \omega \to E = 0.100)$ and $P(\mathbf{q}, \omega \to E = 0.100)$ for various temperatures. The main changes one can observe with increasing temperature at fixed frequency are due to way the CCEs---as seen directly in $A(\mathbf{k}, \omega)$---are altered by the decreasing size of $\Delta_0$ as $T$ is increased. At $T = 10$ K, the superconducting gap is large, implying that at the low frequencies ($E = 0.100 \approx 15$ meV) at which these plots were taken the banana-shaped contours only cover a small part of the underlying normal-state Fermi surface. As $\Delta_0$ shrinks with increasing temperature, more and more of the underlying Fermi surface becomes covered by the ``bananas.'' However, because $\Gamma$ is constant as a function of temperature, the spectral functions taken at various temperatures remain similarly sharp---the CCEs maintain their shape without much visible smearing. These imply that for the power spectrum of the LDOS, the peaks corresponding to the ``octet model'' remain very much visible. Because no change in intrinsic broadening occurs as temperature is increased, the octet-model peaks retain their sharpness throughout the temperature ranges we consider, and even the caustics corresponding to scattering between the off-tip segments of the ``bananas'' are still visible and do not get broadened. The only change that occurs as temperature is changed is in the positions of the characteristic peaks of the power spectrum. Because $\Delta_0$ decreases in size as $T$ increases at fixed frequency, the CCEs all increase in size with increasing $T$, and consequently the seven octet-model peaks disperse as $T$ is changed at fixed frequency. For instance, $\mathbf{q}_7$---the smallest diagonal scattering wavevector, corresponding to tip-to-tip scattering within one ``banana''---is seen to increase in magnitude as $T$ is increased. When the gap finally fully closes, the QPI power spectrum consists of sharp, well-defined caustics characteristic of a normal metal. With realistic disorder (\emph{i.e.}, a 0.5\% concentration of weak pointlike scatterers) and finite-temperature smearing, the expected (unconvoluted) LDOS power spectra is seen to feature the loss of the octet-model peaks as temperature is increased. In particular, only at $10$ K does the disordered and thermally-smeared power spectrum show these peaks. However, a sharp transition in the features of the power spectrum once the gap fully closes is still visible even at the high temperatures at which these occur---there is a change from a highly anisotropic power spectrum in the superconducting state, with pronounced spectral weight near the corners and suppressed antinodal scattering wavevectors, to the caustics seen in the zero-gap case. If deconvolution is carefully applied to the real-space differential conductance data, only the intrinsic (that is, non-thermal) broadening will affect the LDOS and the octet-model peaks should be recovered.

Dramatically different behavior is seen once a strongly temperature-dependent but momentum- and frequency-independent quasiparticle scattering rate is included, as is the case in the gap-filling/closing scenario, the results for which we plot in Fig.~\ref{fig:temperature_gf}. We used the same superconducting gap for each temperature as in the BCS case, so all differences between the two sets of plots can be attributed to the presence of a $T$-dependent $\Gamma$. At low temperatures both $A(\mathbf{k}, \omega \to E = 0.100)$ and $P(\mathbf{q}, \omega \to E = 0.100)$ are identical, as in that particular regime there is little difference between the two scenarios. However, when $T \approx T_c$, $\Gamma$ is no longer parametrically smaller than $\Delta_0$ but is instead almost of similar size, and thus the effects of the intrinsic broadening are no longer trivial. Consider first the behavior of $A(\mathbf{k}, \omega)$. At 85 K, the CCEs are still well-defined, albeit broadened considerably compared to the BCS case, with more spectral weight found in the streaks emanating from the ends of the contours which follow the underlying Fermi surface. At $T = T_c = 90$ K, even more broadening is present, and yet more spectral weight moves towards the streaks. At 95 K, $\Gamma \approx \Delta_0$, and consequently the spectral function resembles neither that of a $d$-wave superconductor nor that of a normal metal. Instead, it shows a quasiparticle excitation spectrum which resembles Fermi arcs. That is, there is considerable spectral weight present near the nodes, and one sees less spectral weight as one moves along the underlying Fermi surface towards the antinodes. Once the gap fully closes, what is seen is the expected isotropic normal-state spectrum in which the spectral weight along the CCE is uniform. In our plots where the gap is fully zero, we have assumed the value of the scattering rate to be equal to that given by the marginal Fermi liquid self-energy at $T = 100$ K.

The strong temperature-dependence of $\Gamma$ has an even more pronounced effect on the single-impurity $P(\mathbf{q}, \omega)$ without thermal smearing. At 85 K, the patterns are the same as in the BCS case, with the difference that the octet-model peaks that were visible and sharp in the BCS case are now muted---the intensities of the peaks are quite weak in the gap-filling scenario. At 90 K even more smearing of the QPI patterns becomes apparent, and some peaks, such as $\mathbf{q}_7$, almost completely disappear. The points corresponding to certain other octet-model peaks such as $\mathbf{q}_2$ and $\mathbf{q}_6$, while still discernible, are so broadened as to be almost undefined at this point, and only streaks corresponding to diagonal internodal scattering remain as the prominent signal. At 95 K and beyond, all octet-model peaks cease to be well-defined signals. Instead what remains are caustics which track scattering along the underlying normal-state Fermi surface, but with variable weights depending on the location of the initial and final states on the Fermi surface, resulting in a nonuniform distribution of spectral weight along the caustics. This is considerably different from the QPI power spectrum of a normal metal with a momentum-independent scattering rate, wherein the magnitude of $P$ along  the caustics is uniform. Finally, at the point where the gap has fully closed, we see a return to a metallic QPI power spectrum, with uniform weight along all the caustics, but which is considerably smeared compared to that seen in the BCS scenario. The addition of thermal broadening and distributed disorder however results in power spectra which are very similar to those of the BCS case. This makes it difficult to distinguish the gap-filling/closing scenario from the BCS one from unconvoluted data, and one needs to perform a deconvolution of the data to recover the power spectrum with only intrinsic broadening present. 

The change in the behavior of $P(\mathbf{q}, \omega)$ from a small-gap $d$-wave superconductor with very large $\Gamma$ to a normal-state metal with zero gap is quite stark---in contrast to what is seen in $A(\mathbf{k}, \omega)$, where the change appears to occur smoothly. Compared to the QPI power spectrum for the normal metal, the spectrum at energies below the gap for a broadened $d$-wave superconductor is much more suppressed in the antinodal directions (\emph{i.e.}, the $(0,0) \to (0, \pm\pi)$ and $(0,0) \to (\pm\pi, 0)$ directions in $\mathbf{q}$-space). It also features much more spectral weight near the corners. All of this can be attributed to the fact that, because the gap is finite in this regime, the scattering matrix element is affected by coherence factors. In the presence of a weak perturbation in the chemical potential (as in the case of our numerics), scattering between two states where the $d$-wave gap has the same sign is suppressed compared to that between states where the gap has opposite signs.\cite{wang2003quasiparticle,pereg2003theory} With a finite gap, this would explain why the intensities of wavevectors corresponding to scattering in the antinodal directions (which would be between states where the gap has the same sign) are weaker compared to those of internodal scattering wavevectors, resulting in the comparatively strong signal near $(\pm\pi, \pm\pi)$. This coherence-factor effect completely disappears upon the closing of the gap. It should be emphasized that this dramatic change in the spectrum as $\Delta_0 \to 0$ is still visible even in the unconvoluted thermally-smeared spectrum.

The third scenario we consider is one in which the superconducting gap remains finite and temperature-independent, while the quasiparticle scattering rate increases monotonically as temperature is raised. We plot results for this case in Fig.~\ref{fig:temperature_cg}. We used the same scattering rate as in the gap-filling/closing case; note that at higher temperatures the scattering rate becomes of the same size as the gap. Unlike in the second case, because the gap remains a constant as $T$ is increased, there is no change in the position and size of the CCEs as seen in $A(\mathbf{k}, \omega \to E = 0.100)$. What differs is the sharpness of these contours in momentum space. At $T = 85$ K, the contours remain sharp, with only a small amount of spectral weight found beyond the ends of the ``banana.''  When temperature is increased, the scattering rate increases, and the contours become less sharp, with more and more spectral weight found in the tails. At the highest temperatures, what had once been well-defined banana-shaped contours resemble more and more the underlying Fermi surface, but with anisotropy in the spectral weight along the Fermi surface. While most of the spectral weight remains near the nodes, considerably more weight has shifted towards the tails, which track the Fermi surface and which now extend all the way to the antinodes. However, unlike the scenario in which the gap both closes and fills, here the shape of the contours is largely preserved even with increasing broadening.

Because no change in the gap occurs with increasing temperature, the peaks seen in the single-impurity $P(\mathbf{q}, \omega)$ without thermal smearing do not disperse when frequency is fixed and temperature is varied. The main change that occurs is in the sharpness of the peaks, which is affected by how large the quasiparticle scattering rate is. At $T = 85$ K, the peaks can still be seen, but with more bluriness than at lower temperatures due to the large $\Gamma$ at this temperature scale. Increasing $T$ from this point onwards results in these peaks becoming progressively more broadened and less visible, turning into blurry patches with nonzero spectral weight. At the highest value of the scattering rate we considered, no isolated peaks are visible. With distributed disorder and thermal smearing, the plots show similar behavior as the thermally unsmeared single-impurity results, insofar as no shifts in the spectral weight as $T$ increases appear in the spectra due to the constancy of the gap, but no peaks can be discerned at these high temperatures, and in experiment one has to deconvolute the $dI/dV$ data to disentangle the intrisinc broadening from finite-temperature effects.

Further differences between the BCS and the two gap-filling scenarios can be seen by plotting both $A(\mathbf{k}, \omega)$ and $P(\mathbf{q}, \omega)$ for various frequencies. At the lowest temperatures, all three scenarios result in the same behavior, as seen in Fig.~\ref{fig:frequency_gfbcs_10k}: the small scattering rate results in both sharp features in the spectral function, and well-defined peaks in the LDOS power spectrum whose position in $\mathbf{q}$-space changes as $\omega$ is varied, in agreement with the octet model. Because at low temperatures thermal smearing only has a weak effect, the disordered and thermally-smeared power spectra show octet-model peaks that are clearly discernable. 

At higher temperatures, how these CCEs and QPI peaks appear in measurements depends on the degree of broadening present. In the BCS scenario, the CCEs in the spectral function remain sharp as frequency and temperature are changed. In comparison, in the two gap-filling scenarios the CCEs feature much more smearing, which in turn affects how prominent the QPI peaks appear in the LDOS power spectrum. Fig.~\ref{fig:frequency_bcs_90k} shows $A(\mathbf{k}, \omega)$ and $P(\mathbf{q}, \omega)$ taken for the gap-closing scenario at $T = 90$ K, while Fig.~\ref{fig:frequency_gf_90k} shows similar quantities with gap-closing/filling assumed, and Fig.~\ref{fig:frequency_cg_90k} shows the case with only the filling of the gap. At this temperature it is already apparent that in the gap-closing/filling case the QPI peaks broaden so much that it is difficult to see them clearly.  What had been very visible QPI peaks in the gap-closing scenario have turned into barely-discernible patches in the gap-closing/filling scenario, while at higher energies no trace of the QPI peaks remain. Similarly, in the gap-filling scenario, one can see that because the gap is temperature-independent, the LDOS power spectrum resembles that of the low-temperature case, but with so much more smearing that the octet-model peaks become far less discernable.  In all three cases, the thermal smearing at $T = 90$ K is so large that the fine features seen in the single-impurity unsmeared data are lost in the smeared data, and the plots appear qualitatively similar to each other.

We note further that in the two scenarios in which the gap closes (shown in Figs.~\ref{fig:frequency_bcs_90k} and~\ref{fig:frequency_gf_90k}), the shrinking of the gap with increasing $T$ alters the shape of the CCEs as seen in $A(\mathbf{k}, \omega)$, and consequently the positions of the QPI peaks in $P(\mathbf{q}, \omega)$ change as well. The smallness of the gap ensures that the superconducting coherence peaks, located at $E_c \approx \pm 4\Delta_0$,  are shifted closer to the Fermi level. At energies which satisfy $|E| > |E_c|$ the spectral function and QPI power spectrum in the superconducting state are largely similar to those of the normal state at $E$, except for additional features which arise from the presence of shadow-like streaks in the spectral function, which in turn are an effect of the coherence factors which enter Eq.~\ref{eq:sccf}. The similarity to the normal-state LDOS power spectrum here is such that even in the BCS case, which has minimal broadening, no traces of the octet-model peaks appear at these high energies.

We end this section by revisiting our earlier observation that the combination of small but nonzero $d$-wave pairing correlations and a large scattering rate at $T > T_c$ can give rise to Fermi arc-like patterns in the spectral function. It is interesting to note that this can be seen right \emph{at} the Fermi energy itself. In Fig.~\ref{fig:fermi_arc_95k} we plot the spectral function and the LDOS power spectrum at the Fermi energy at $T = 95$ K for the gap-filling and -closing scenario. In the absence of broadening, the $d$-wave superconducting state would result in zero-energy states being localized only at the nodes---the four points on the Fermi surface at which the superconducting gap is zero. With broadening, however, there is now a finite density of zero-energy states in the neighborhood of the nodes. When the scattering rate is small, the effect is minor, and apart from a small arc centered near the nodes the zero-energy states disappear a short distance away from the nodes. However, once $\Gamma \approx \Delta_0$ the regions about the nodes which support low-energy states become large: the ``arc'' along the Fermi surface which supports zero-energy states becomes longer and broader, and a comparable lack of spectral weight is found at the antinodes. The QPI power spectrum is quite pronounced even at the Fermi energy, and is completely different from that of a $d$-wave superconductor or a normal metal. Instead it shows streaks near the corners due to strong internodal scattering, and large low-$\mathbf{q}$ patches showing strong intranodal scattering. We note that this particularly simple set of ingredients (nonzero $d$-wave pairing past $T_c$ and a large quasiparticle scattering rate) has already been proposed as an explanation for the Fermi arcs found via ARPES in the underdoped cuprates.\cite{norman1998phenomenology, norman2007modeling, chubukov2007gapless, vishik2018photoemission} It is an interesting experimental challenge to see if these Fermi arc-like patterns can be seen by STS in the optimally-doped cuprates above $T_c$. 

\section{Self-energies in the normal state}

\begin{figure}
	\centering
	\includegraphics[width=0.4\textwidth]{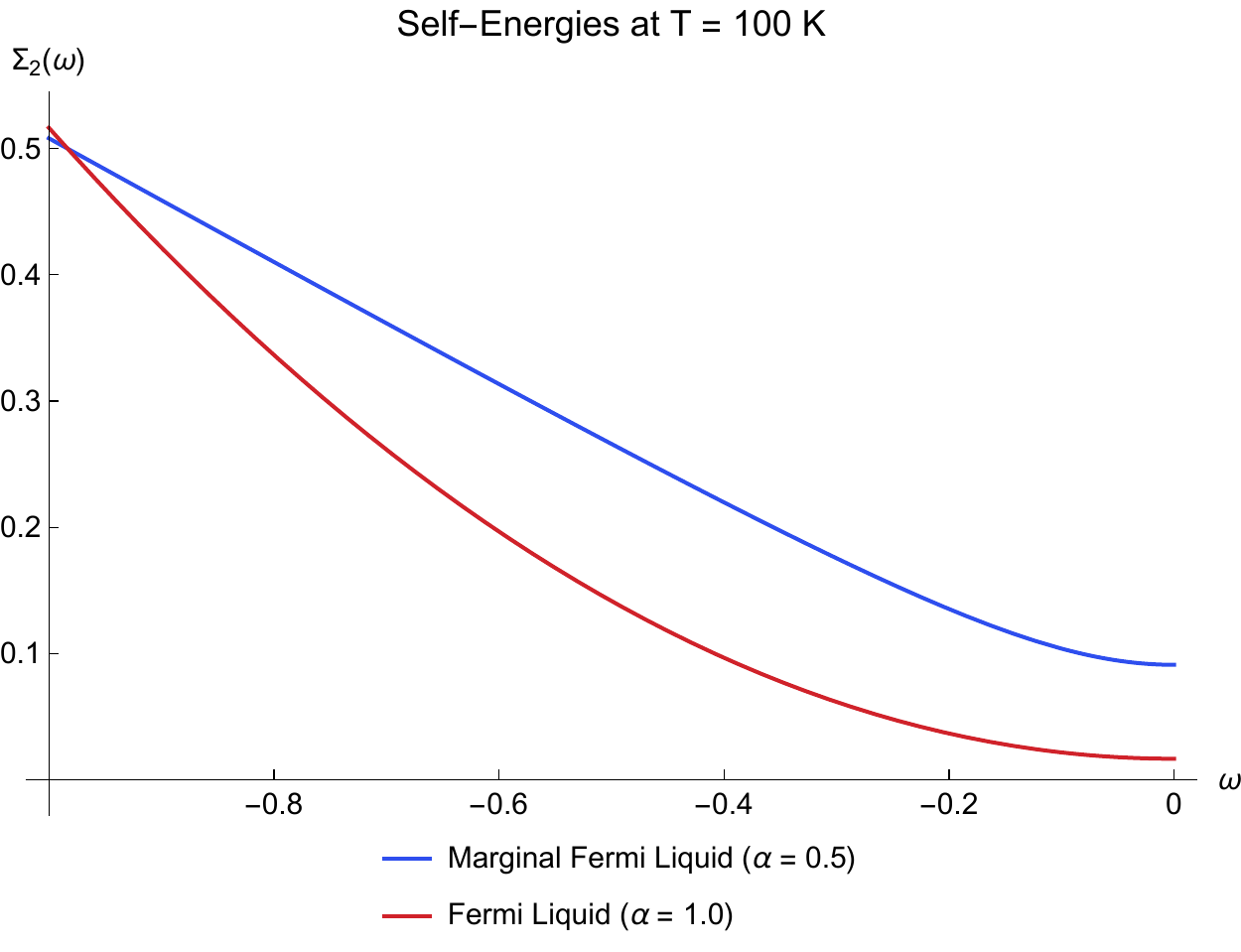}
	\caption{Plots of the self-energies for the Fermi liquid (red line) and marginal Fermi liquid (blue line) at $T = 100$ K. Here $\lambda = 0.5$, $\Gamma_0 = 0$, and $\omega_c = 1$.} 
	\label{fig:flmflse_100K}
\end{figure}

\begin{figure*}
	\centering
	
	\includegraphics[height=0.18\textwidth]{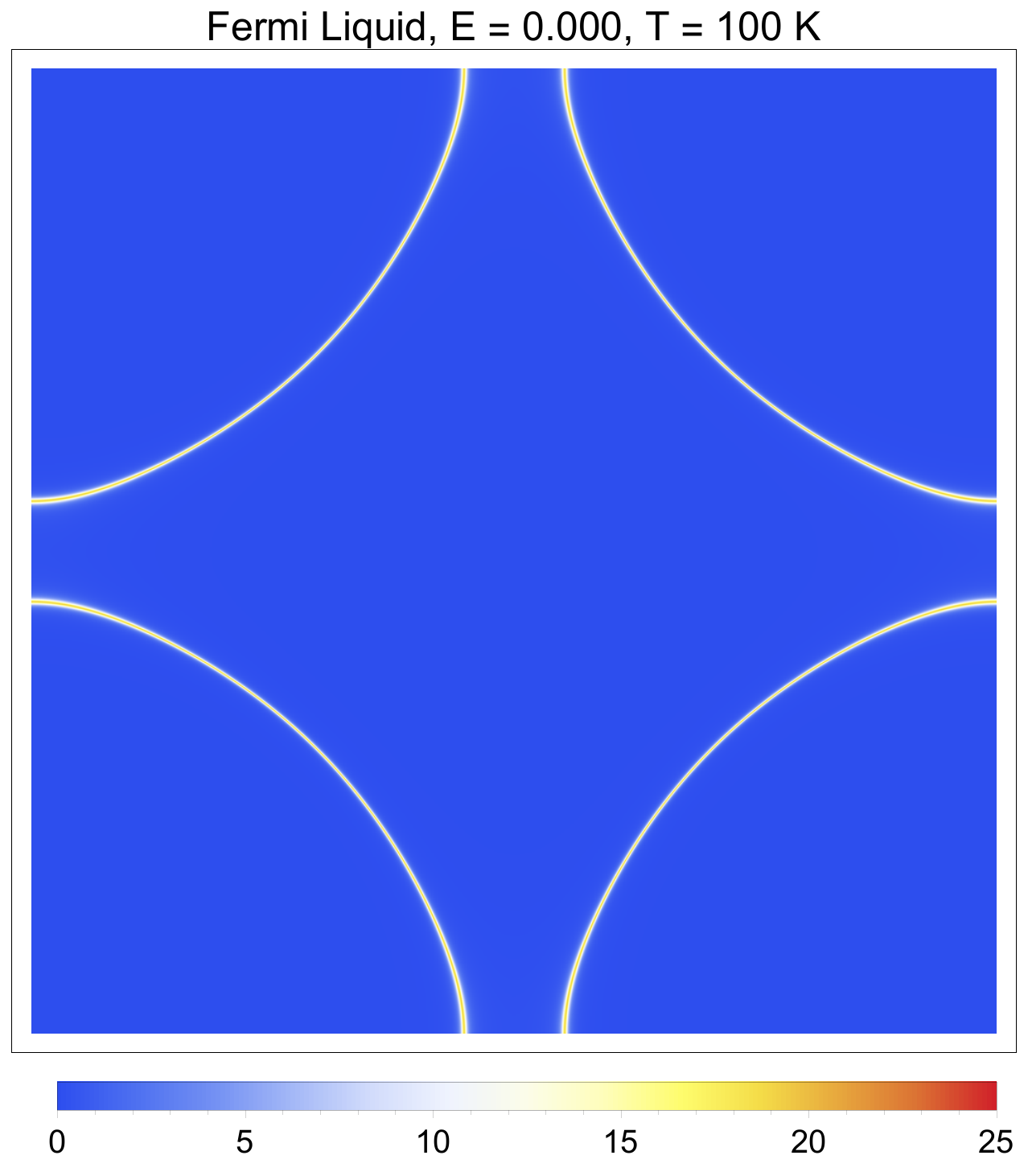}
	\includegraphics[height=0.18\textwidth]{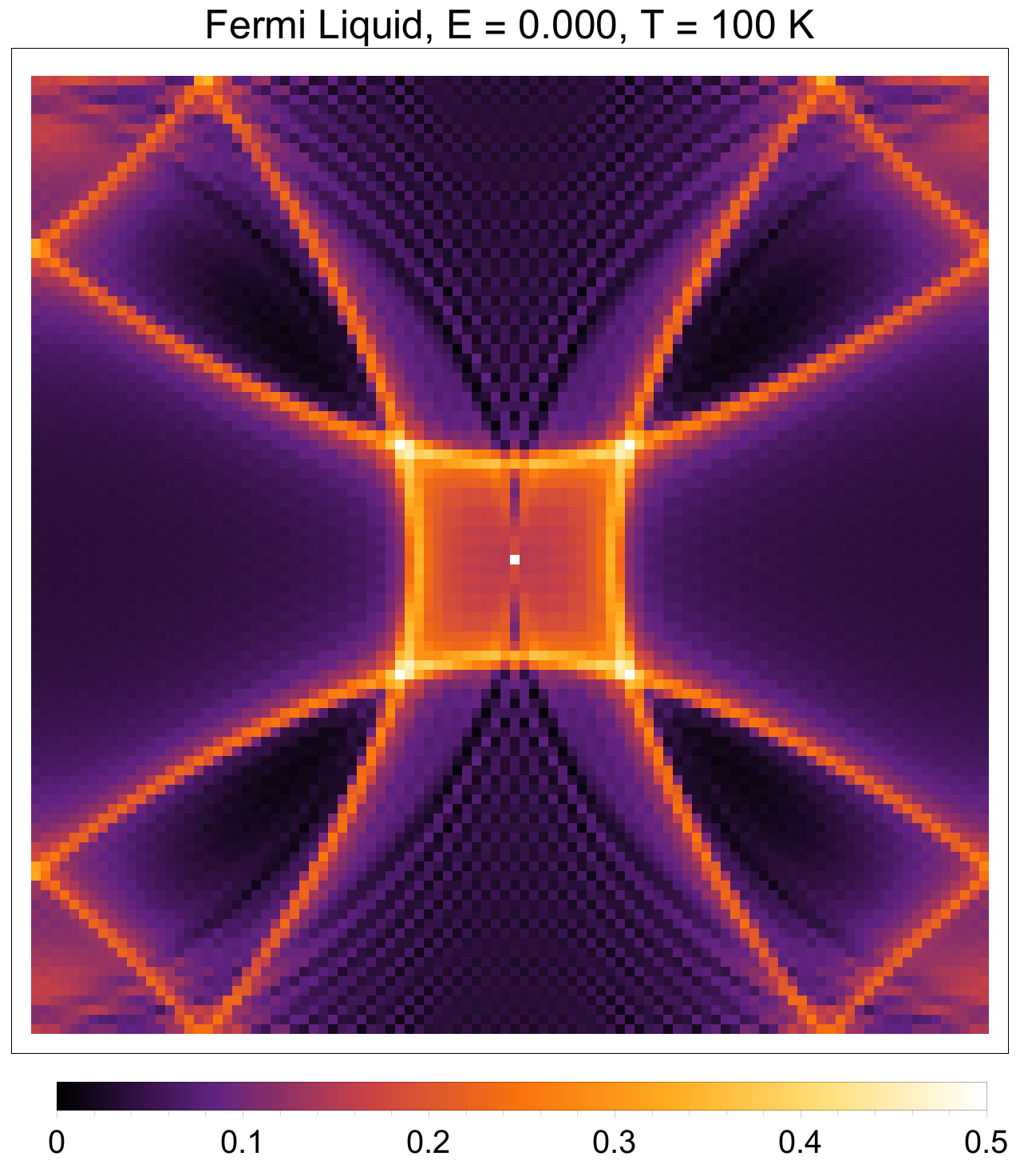}
	\includegraphics[height=0.18\textwidth]{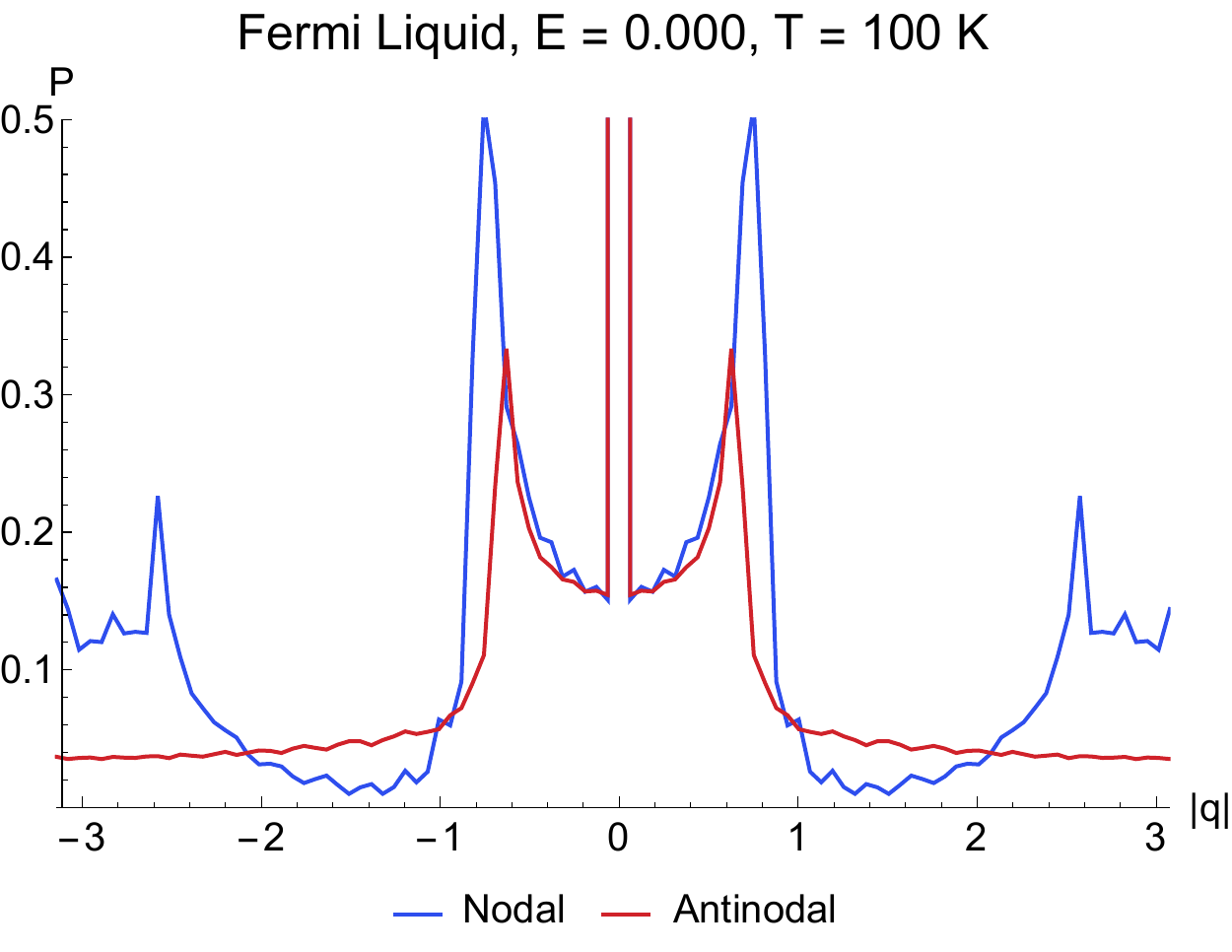}
	\includegraphics[height=0.18\textwidth]{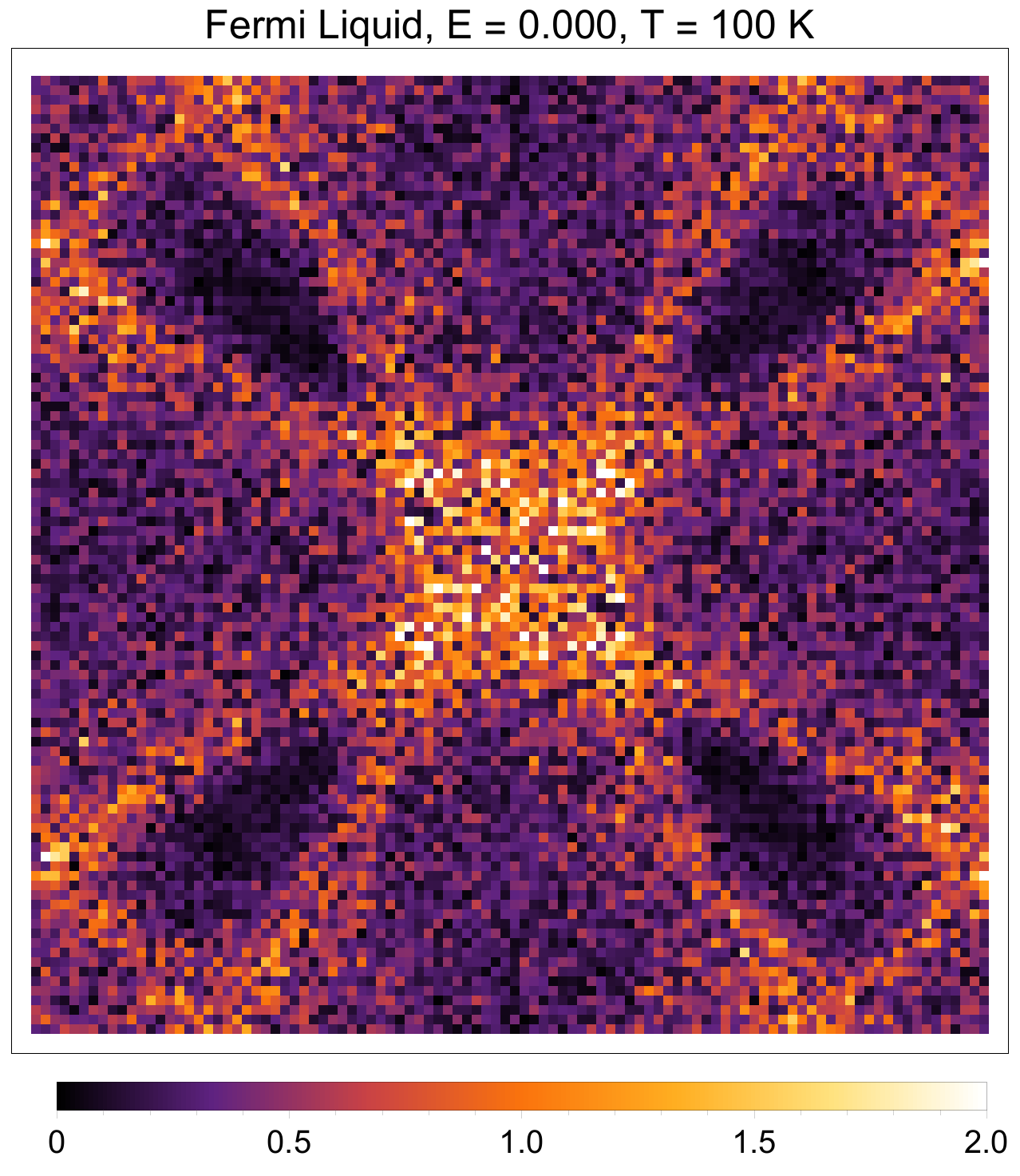}
	\includegraphics[height=0.18\textwidth]{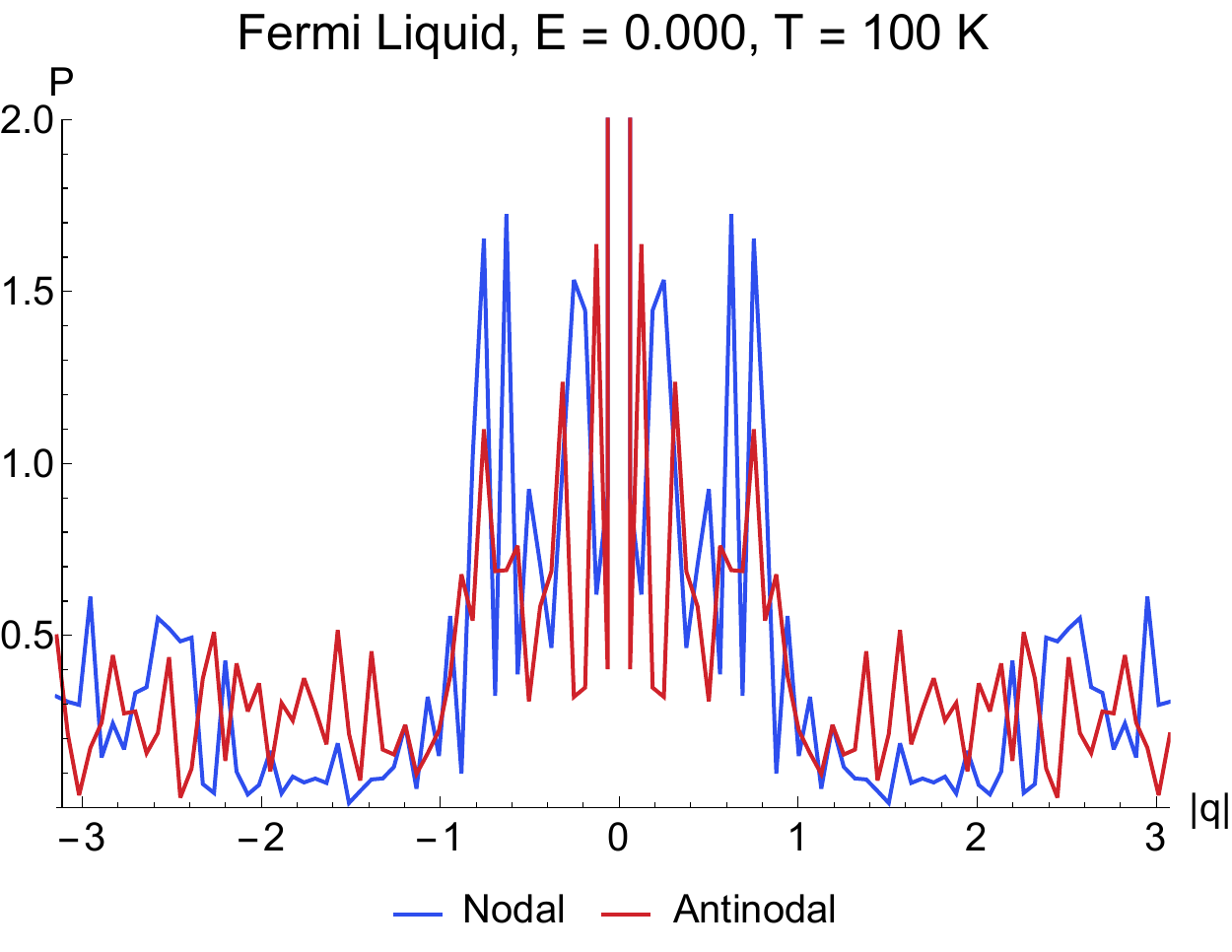} \\
	\includegraphics[height=0.18\textwidth]{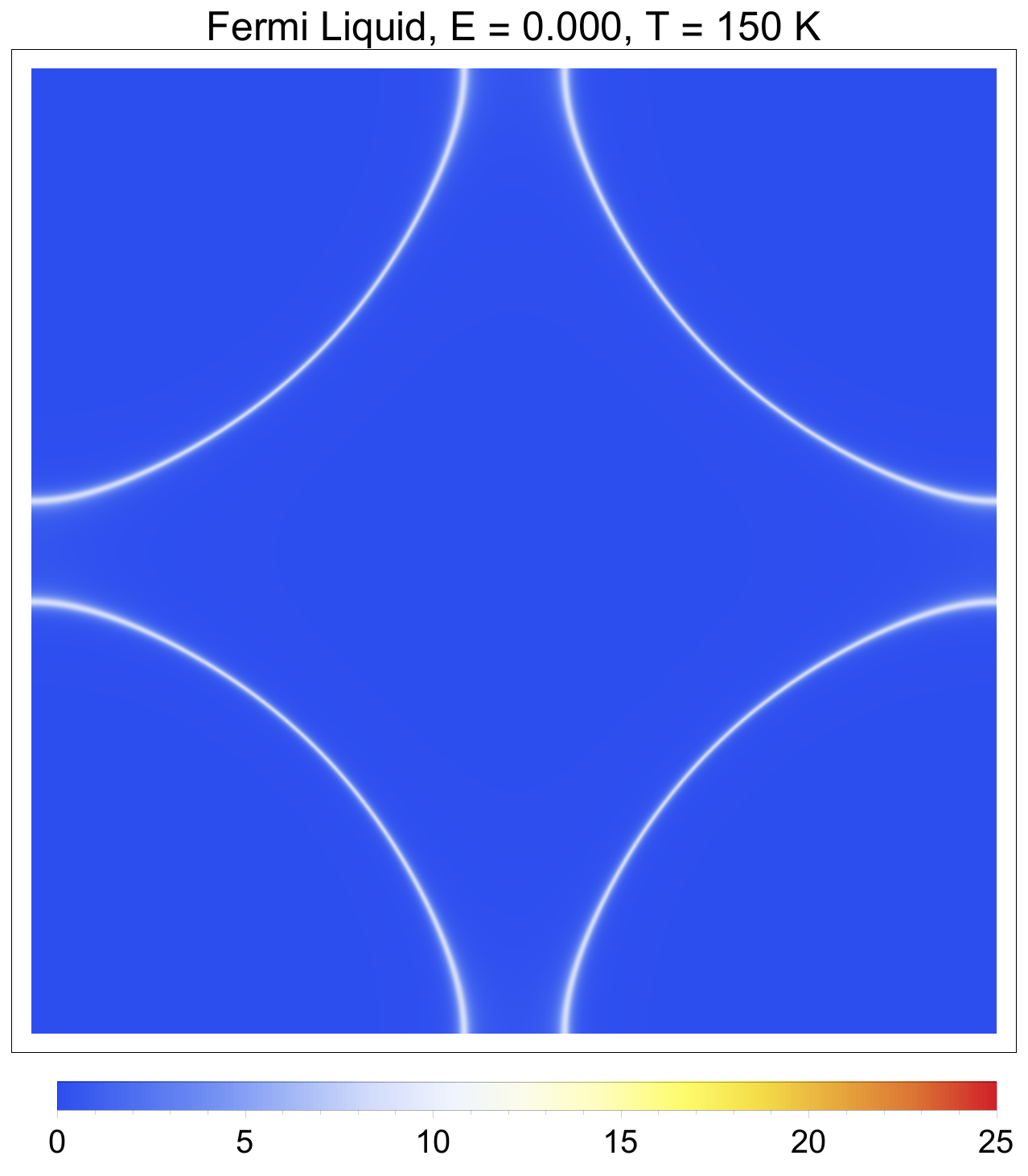}
	\includegraphics[height=0.18\textwidth]{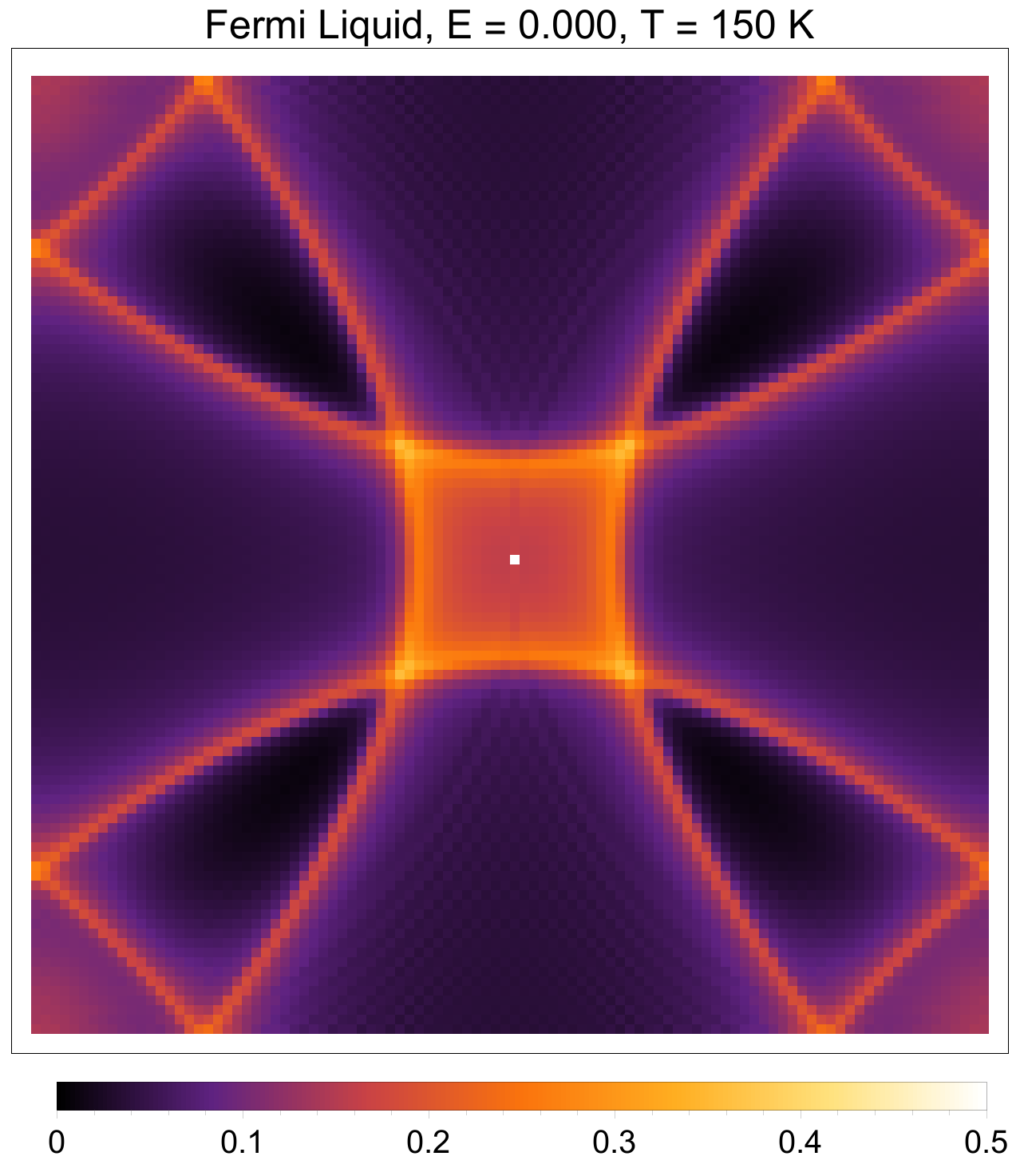}
	\includegraphics[height=0.18\textwidth]{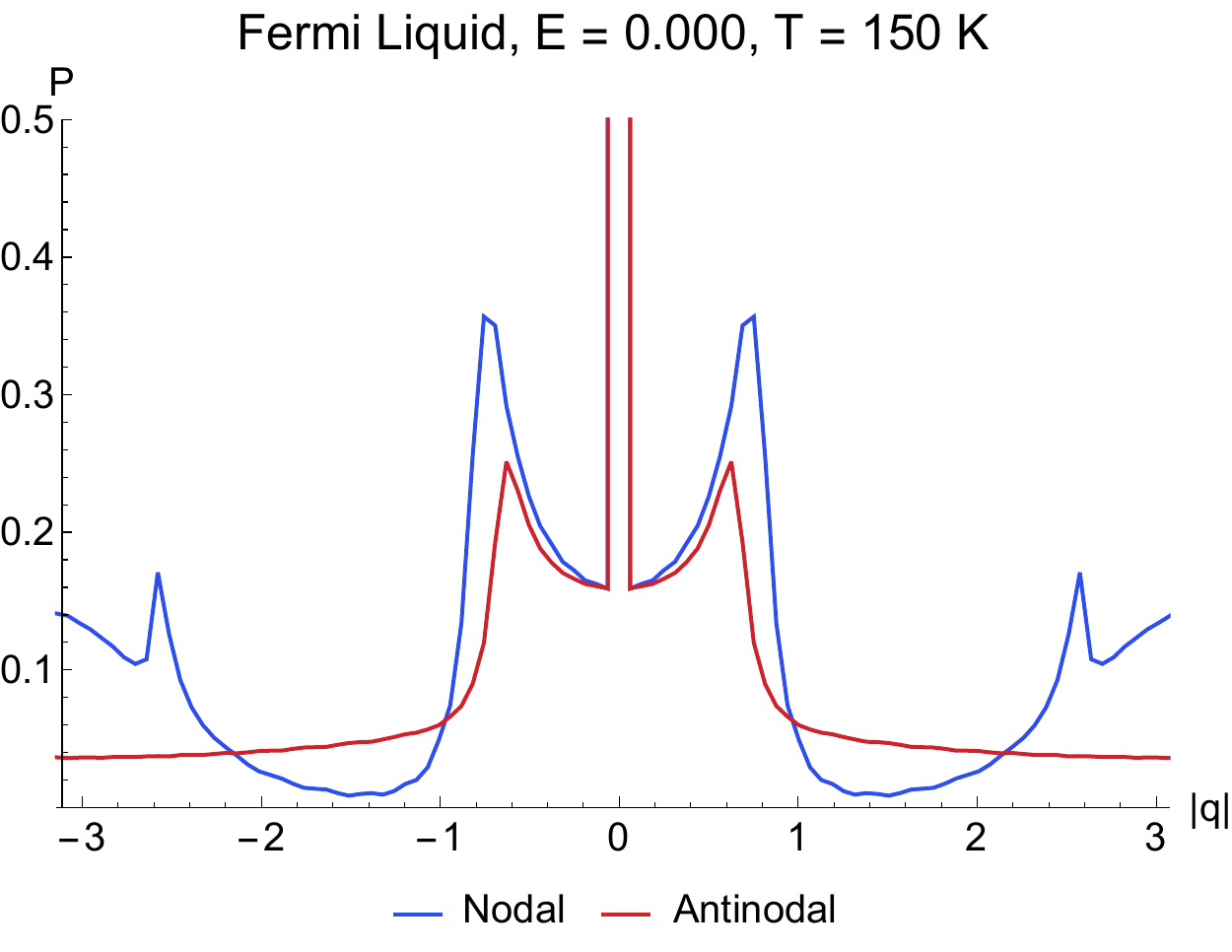}
	\includegraphics[height=0.18\textwidth]{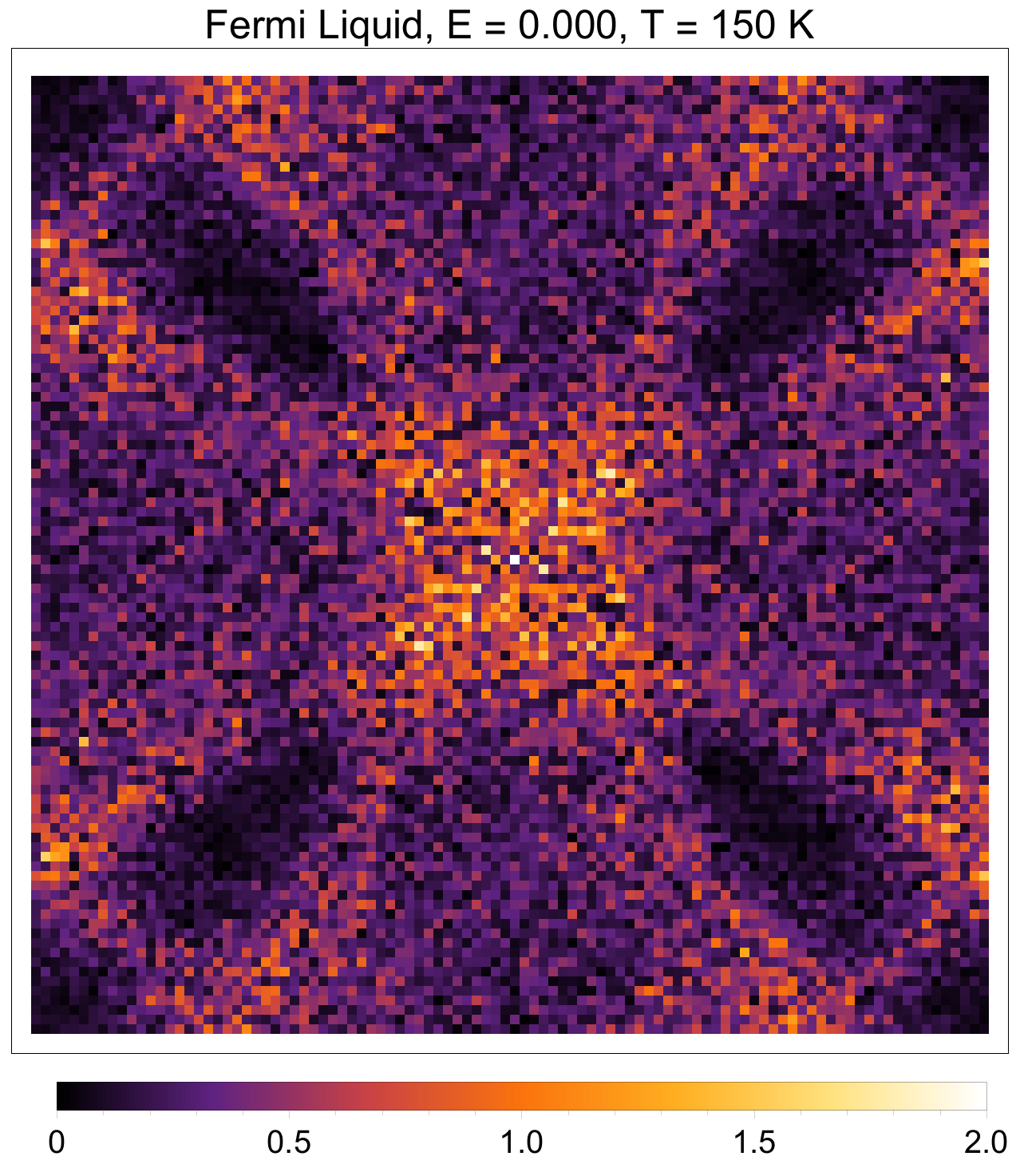}
	\includegraphics[height=0.18\textwidth]{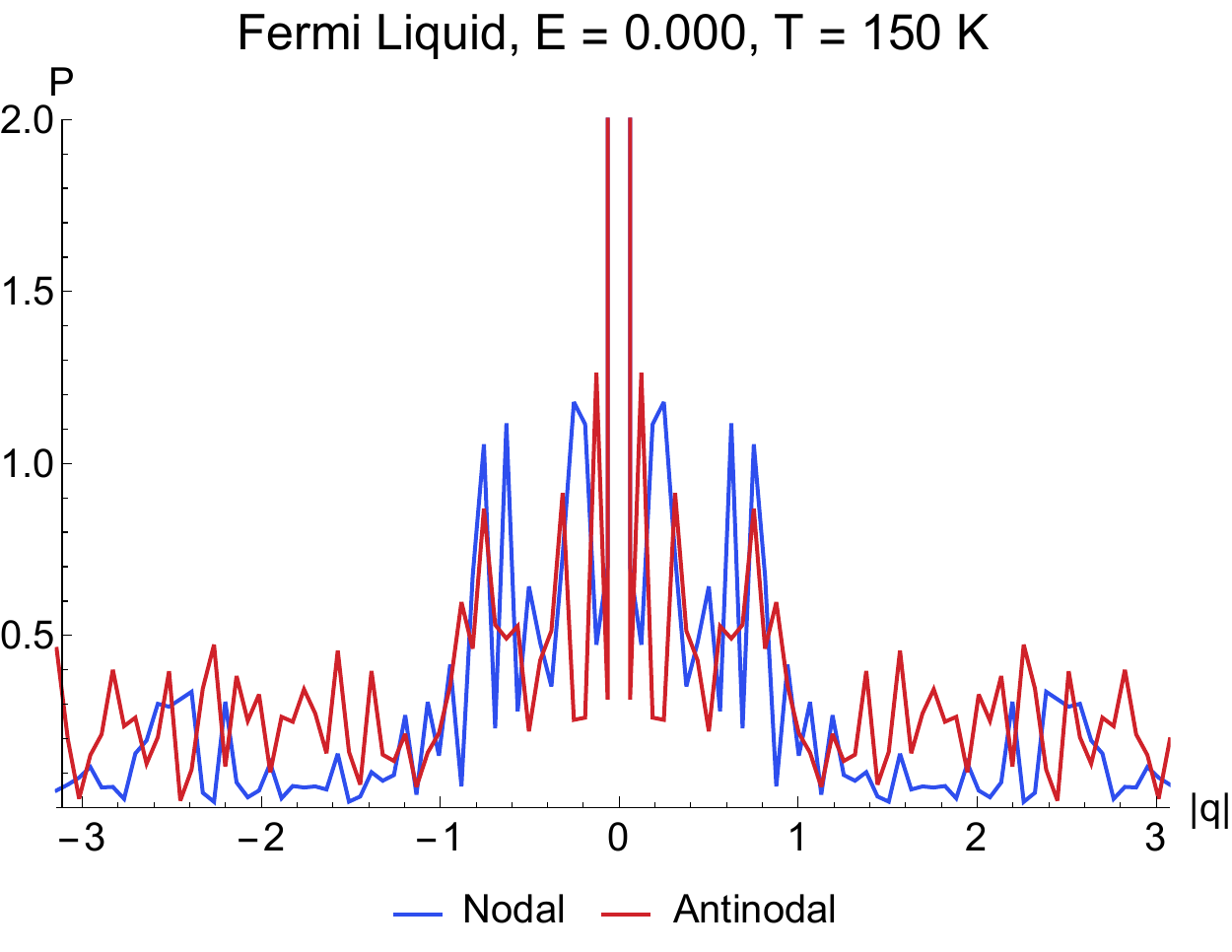} \\
	\includegraphics[height=0.18\textwidth]{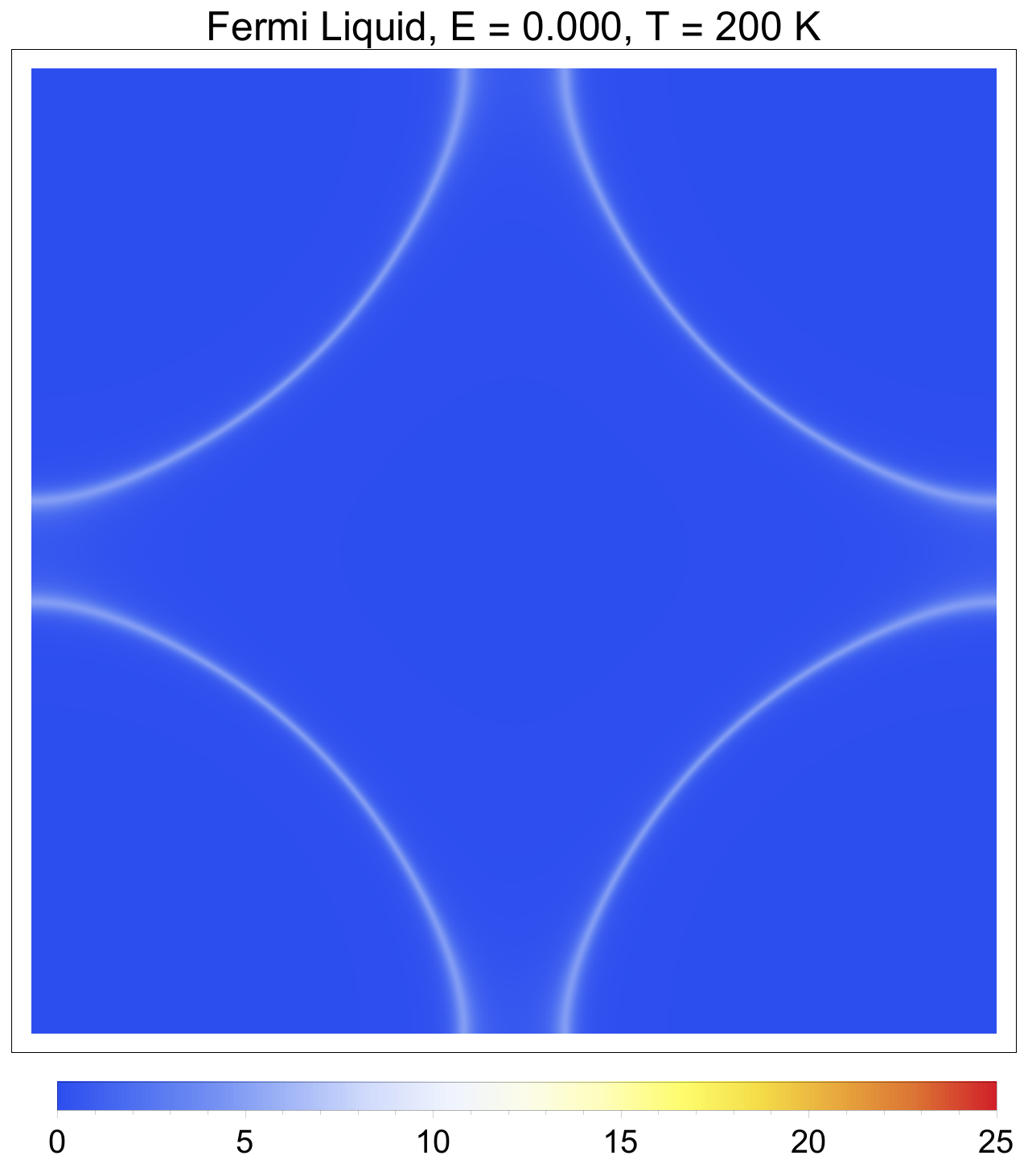}
	\includegraphics[height=0.18\textwidth]{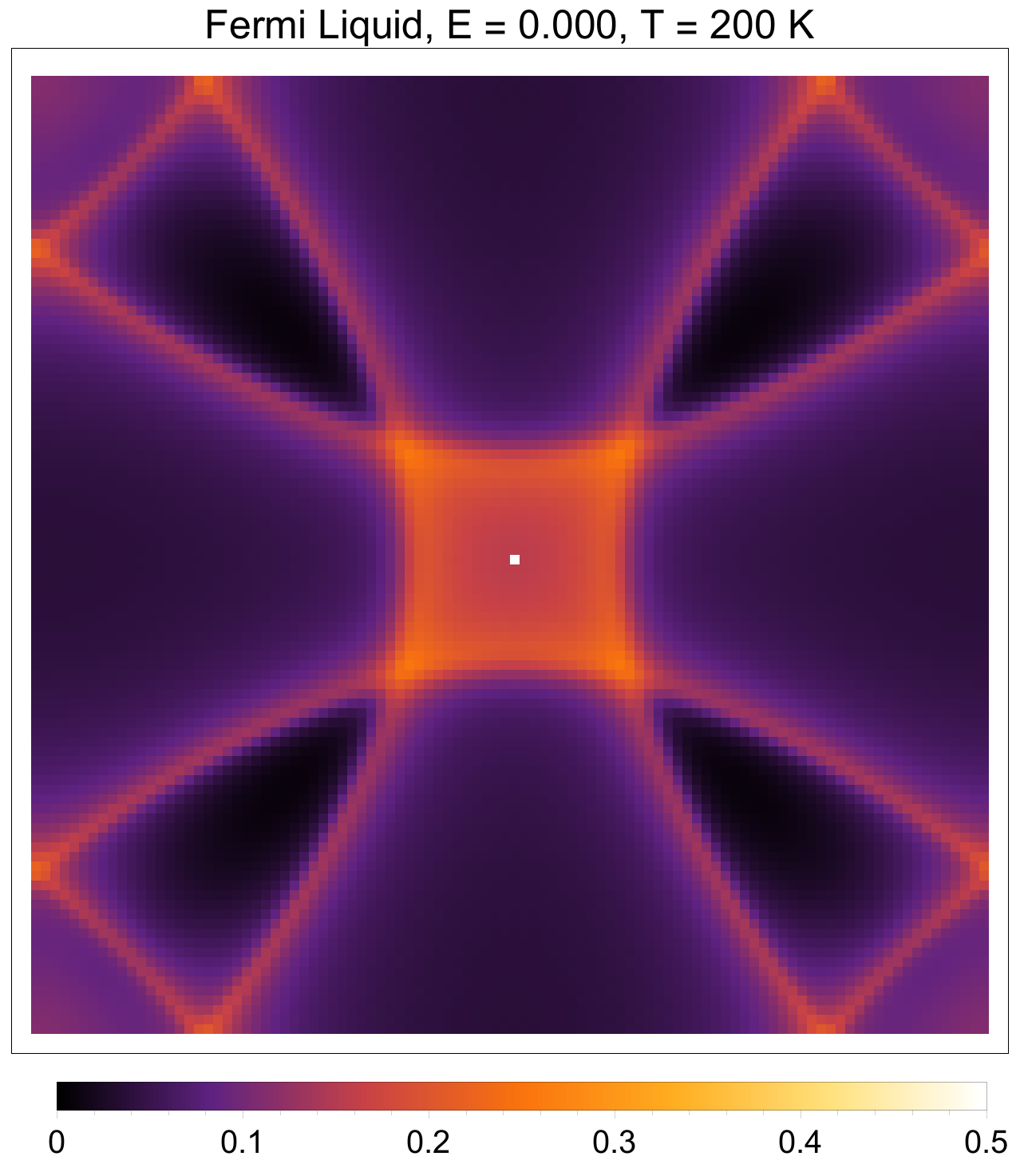}
	\includegraphics[height=0.18\textwidth]{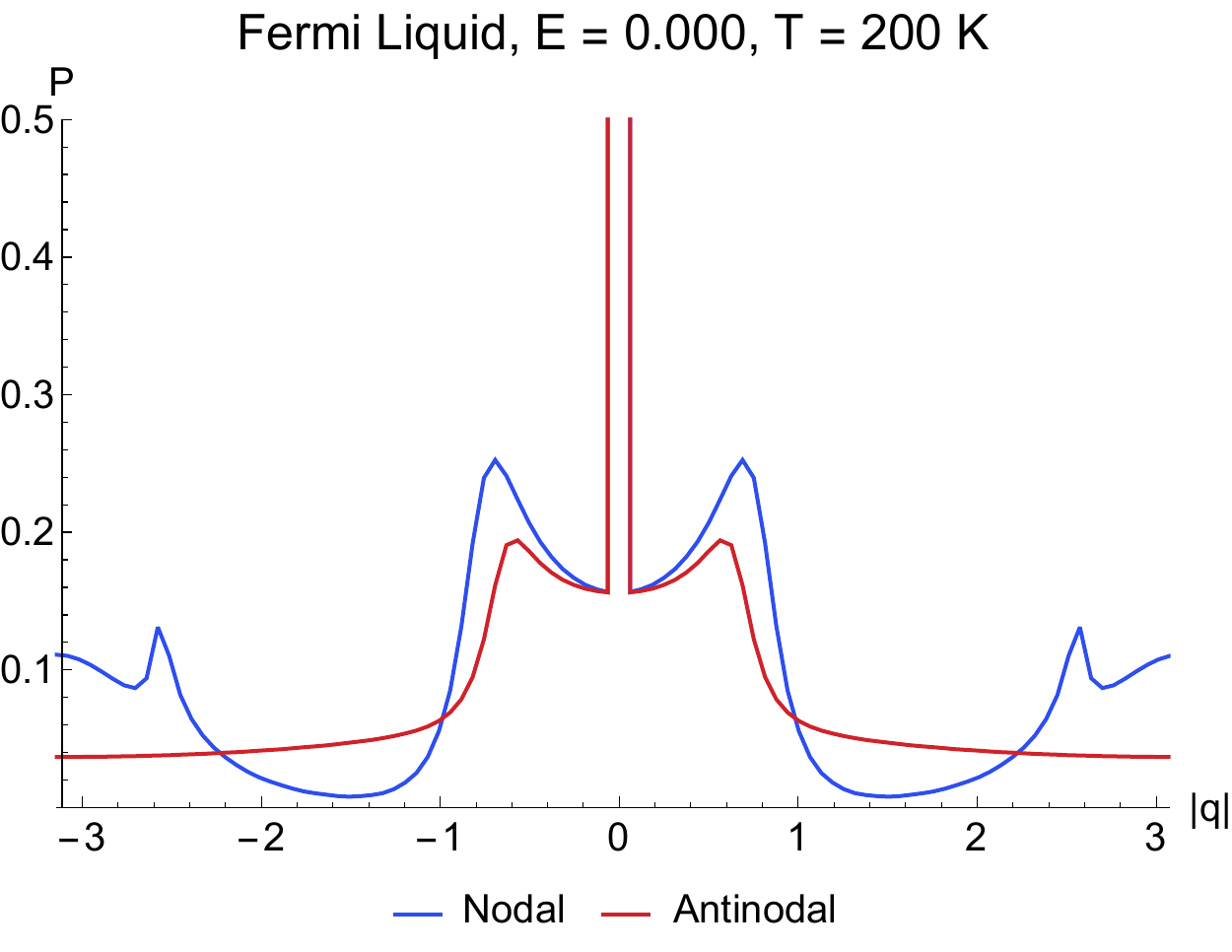}
	\includegraphics[height=0.18\textwidth]{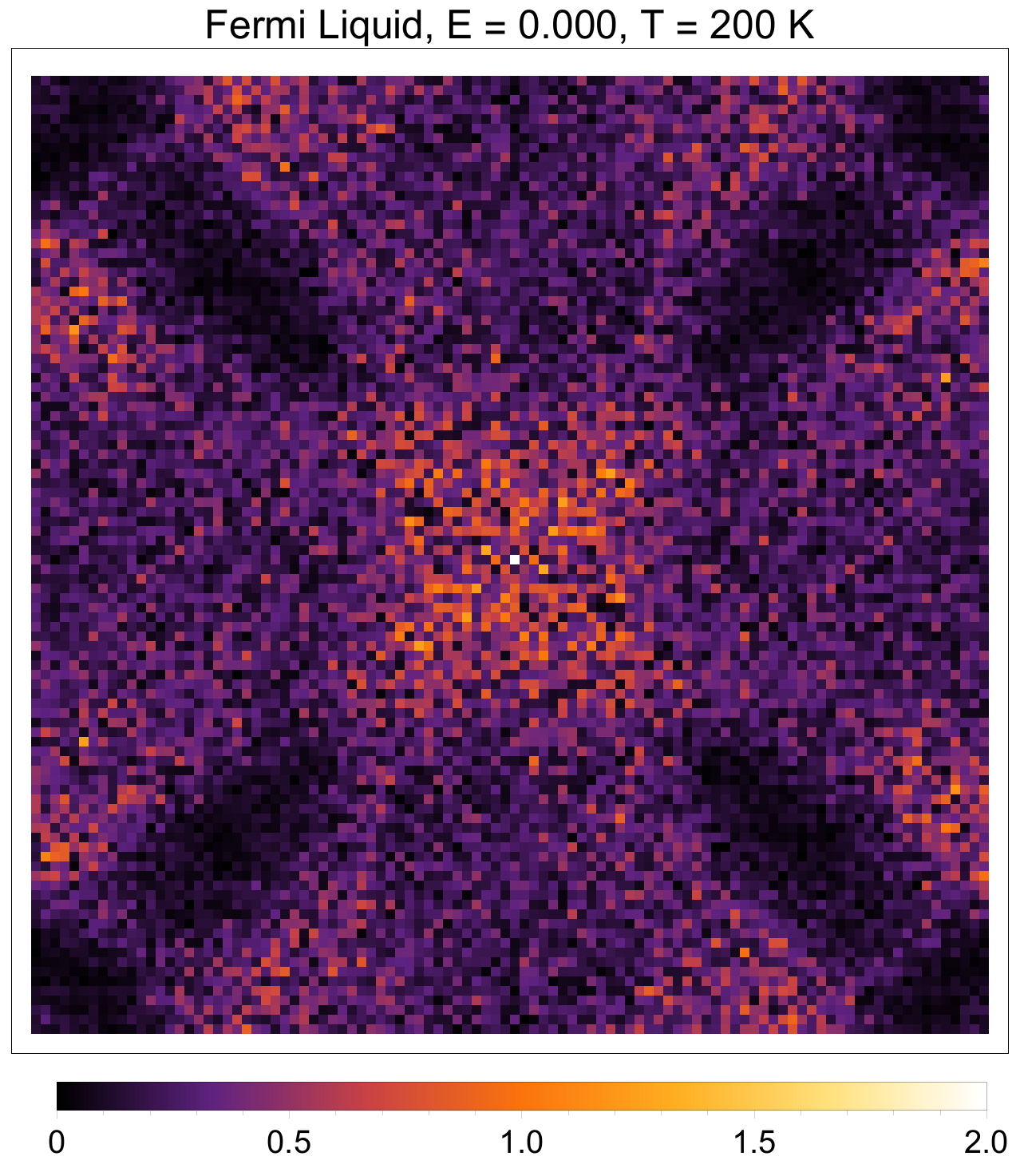}
	\includegraphics[height=0.18\textwidth]{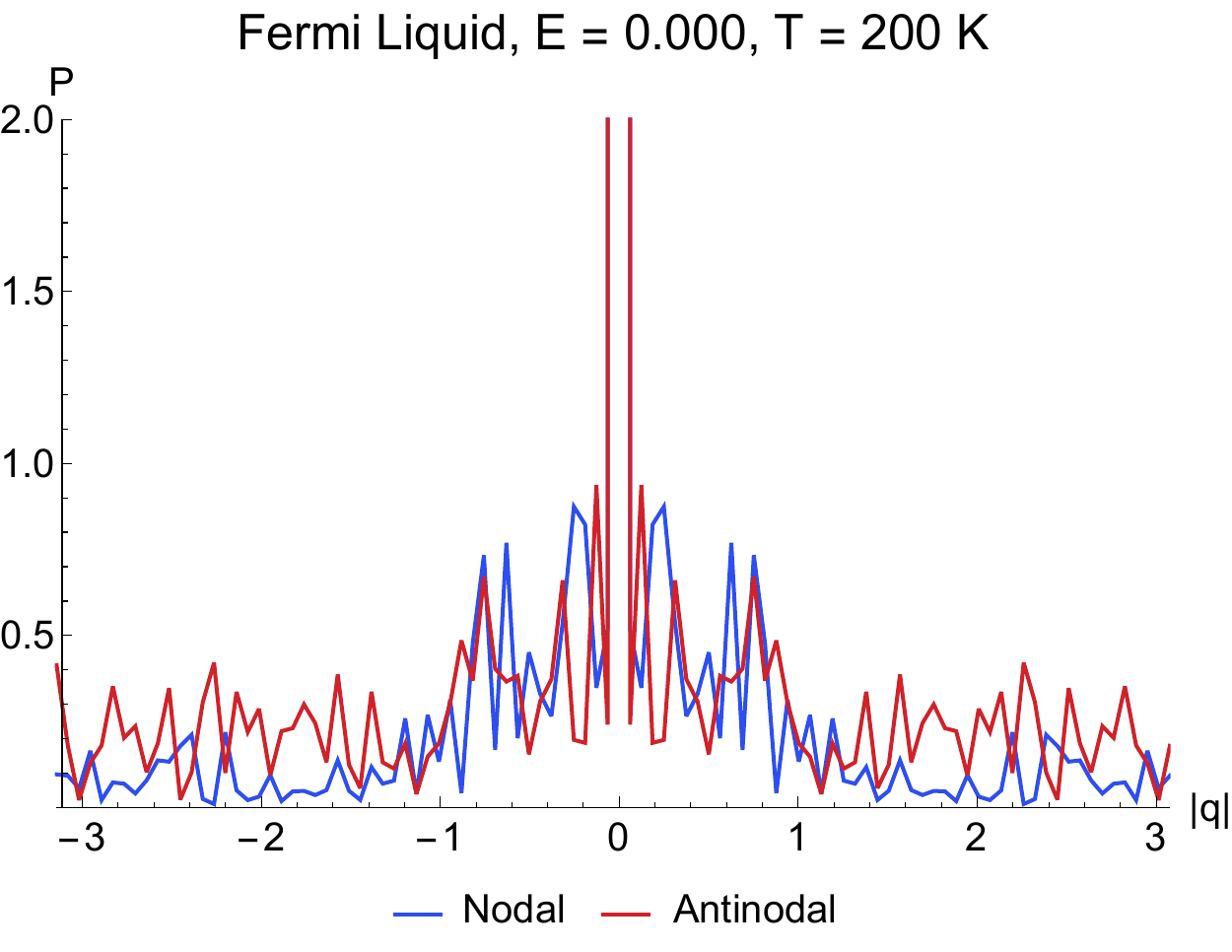} \\
	\includegraphics[height=0.18\textwidth]{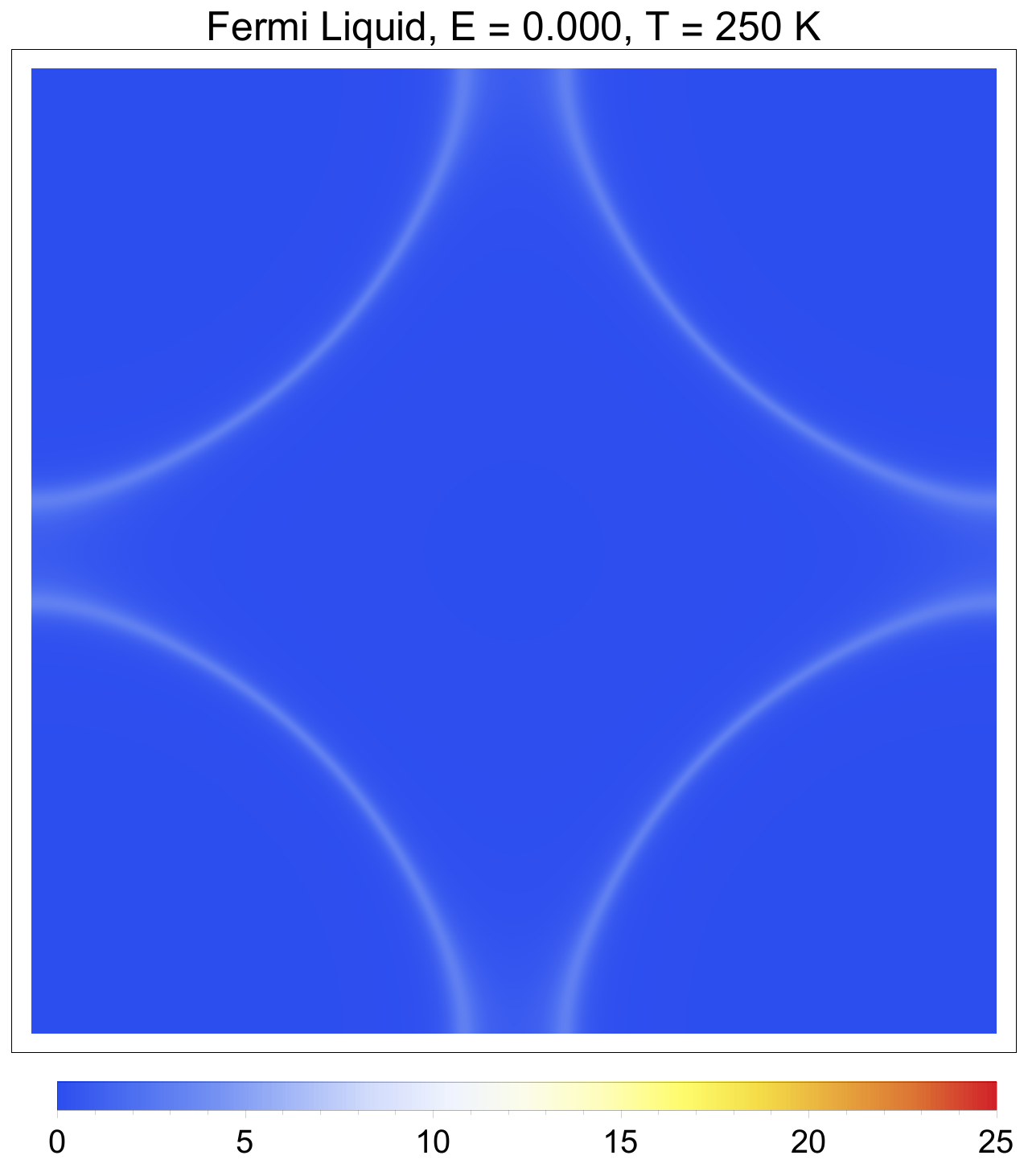}
	\includegraphics[height=0.18\textwidth]{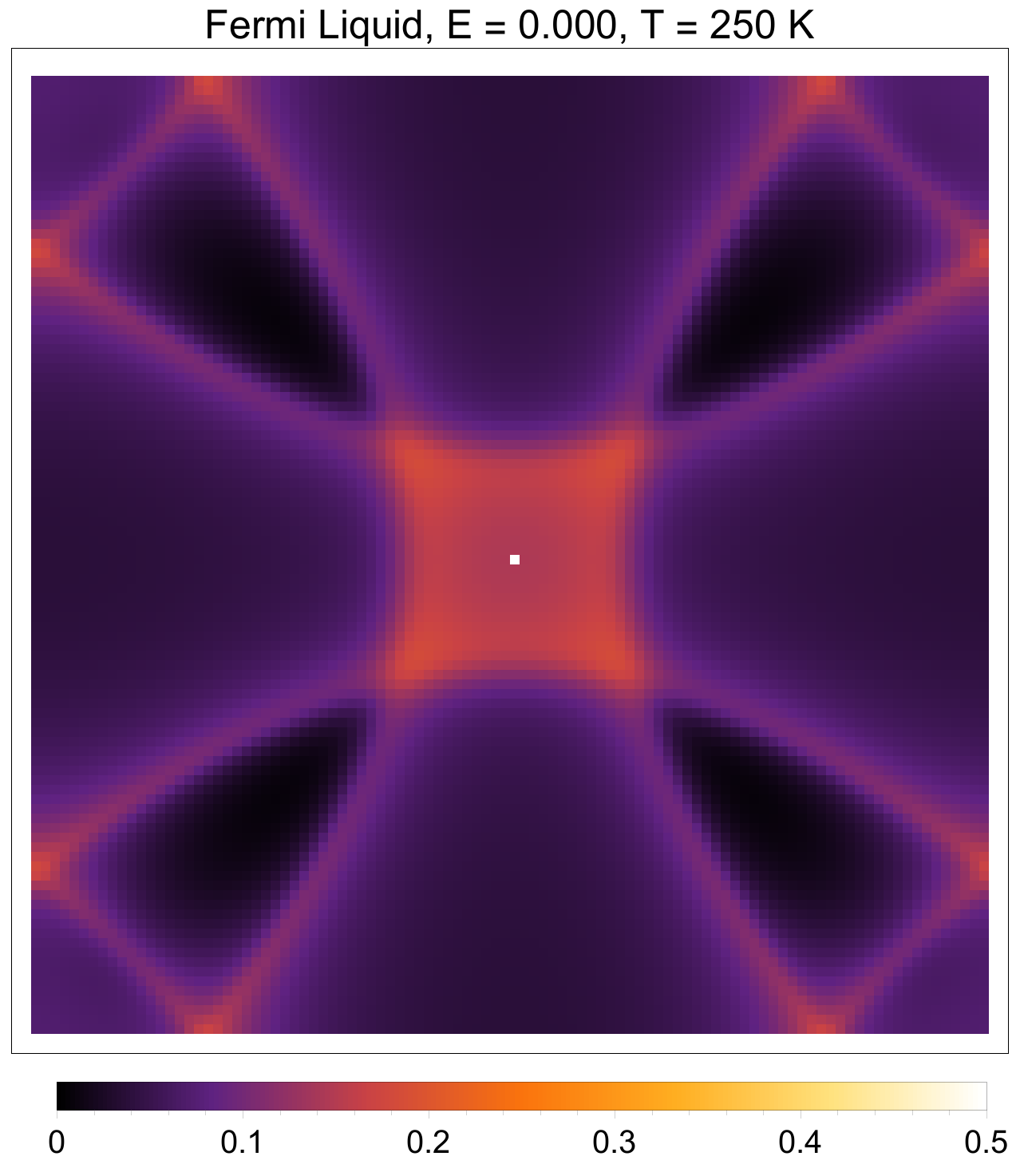}
	\includegraphics[height=0.18\textwidth]{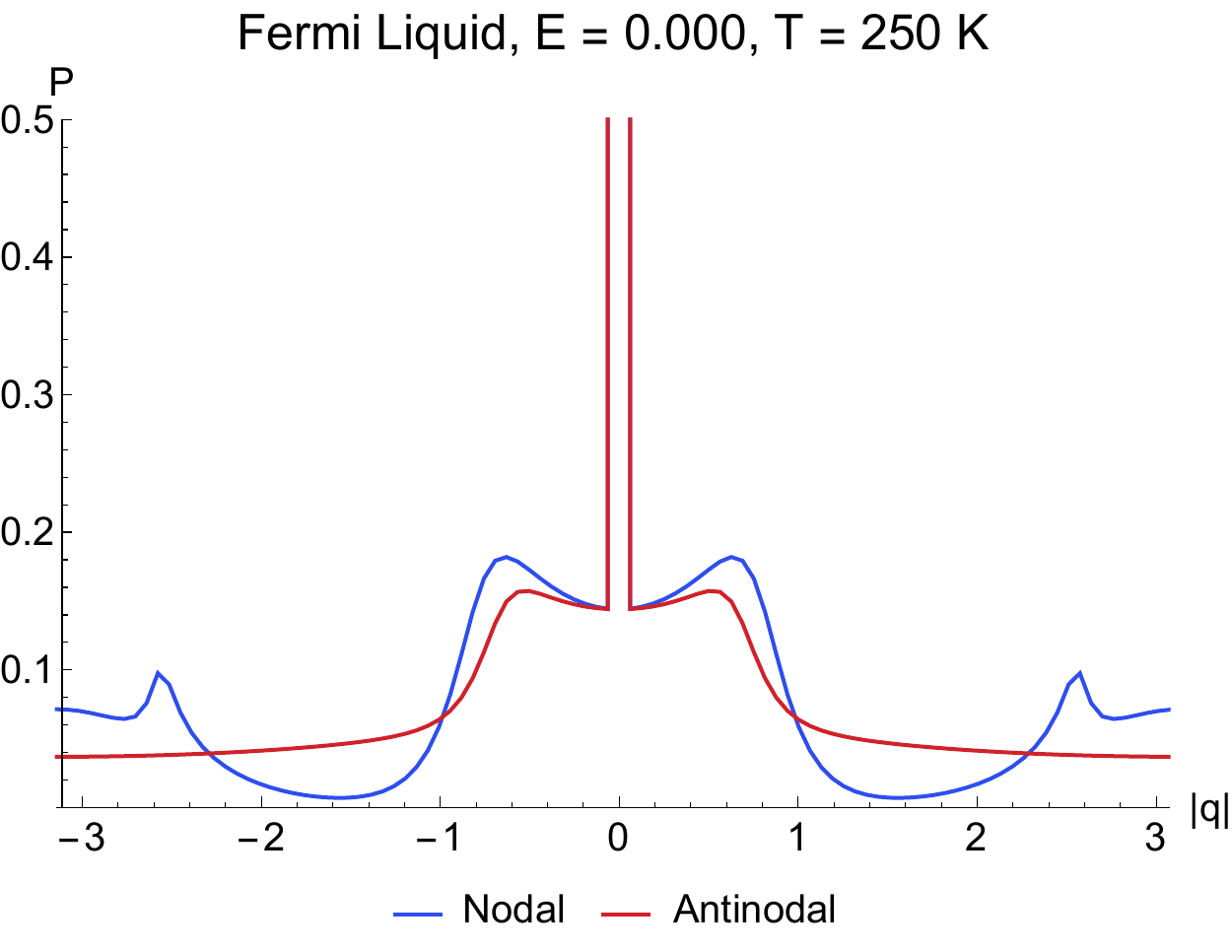}
	\includegraphics[height=0.18\textwidth]{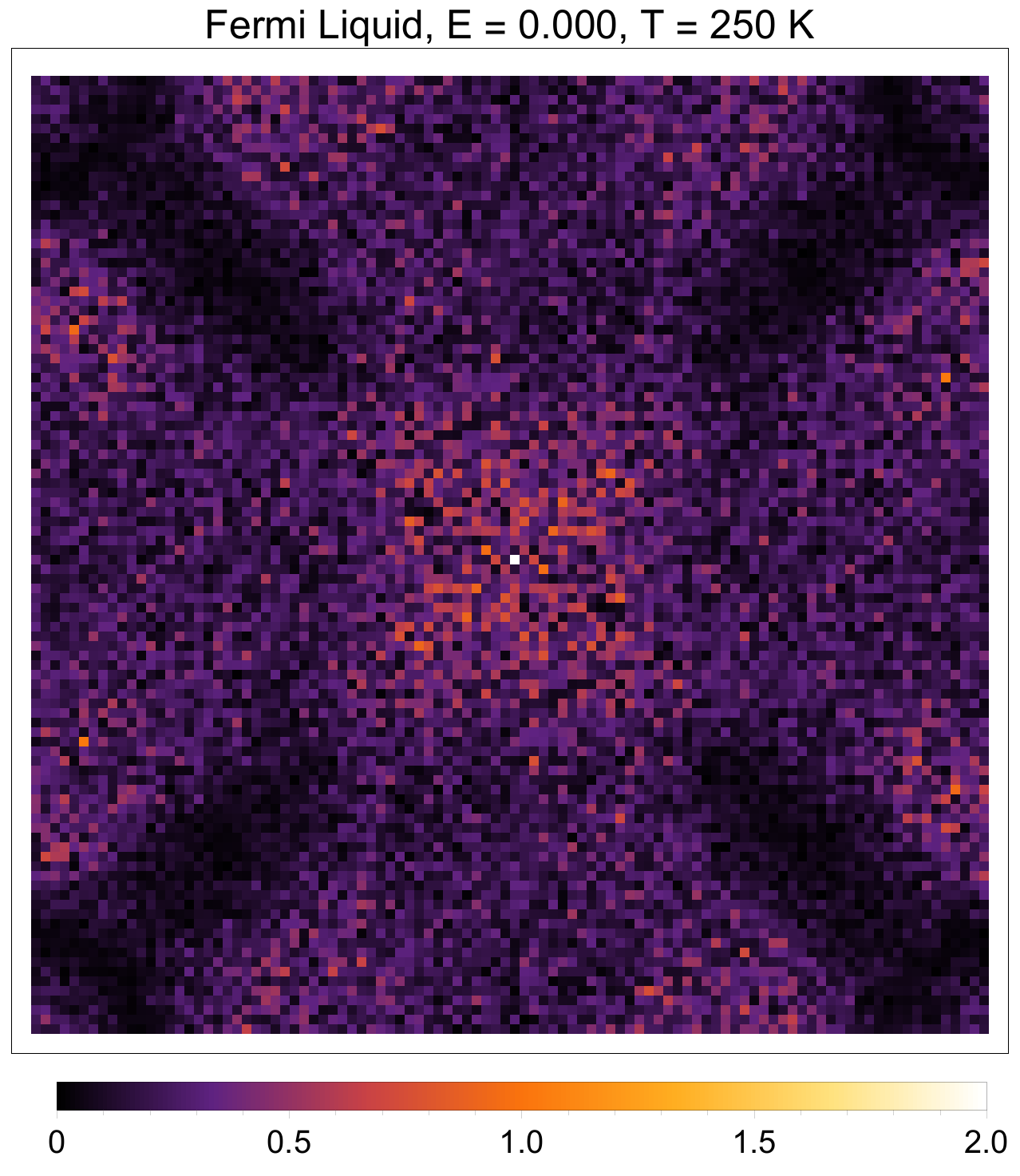}
	\includegraphics[height=0.18\textwidth]{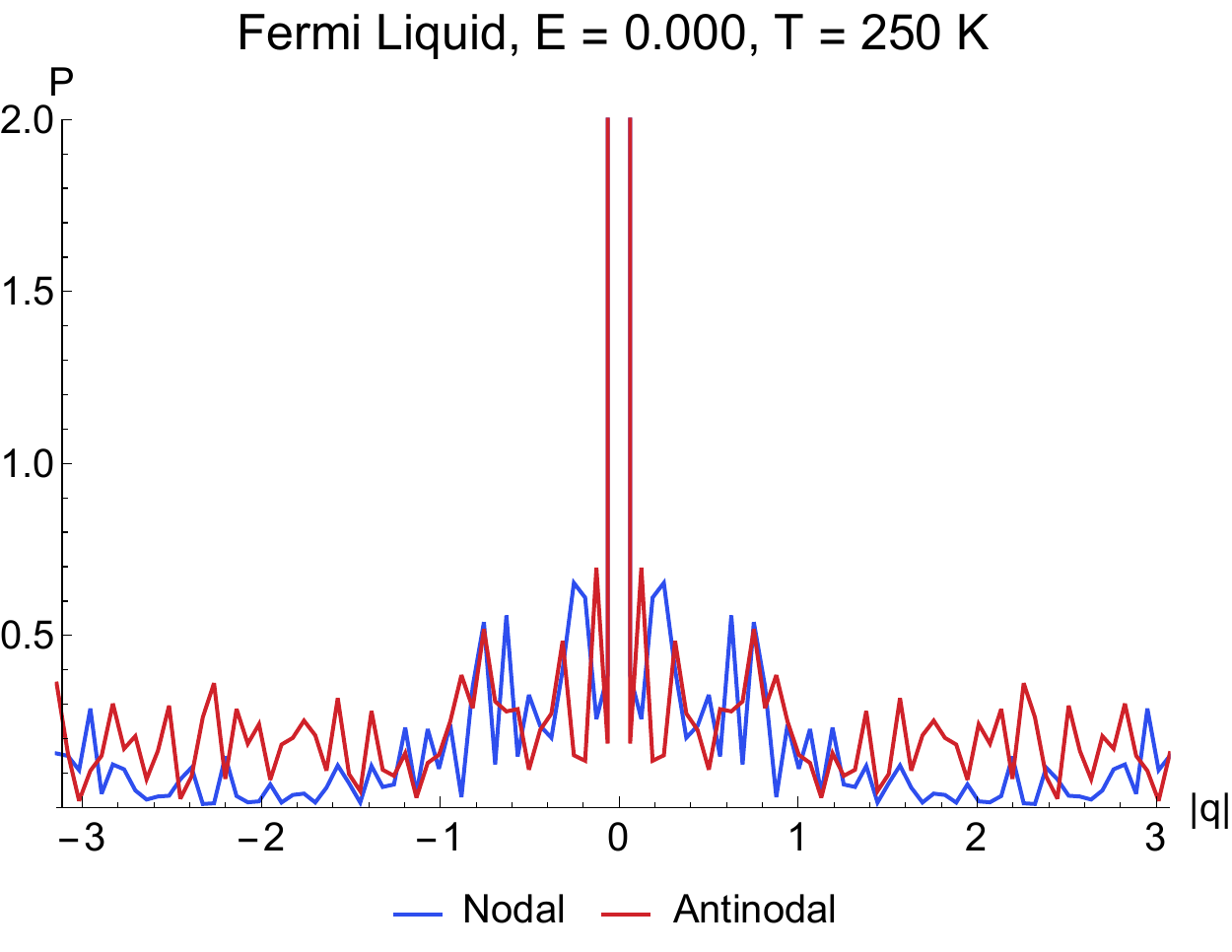} \\
	\includegraphics[height=0.18\textwidth]{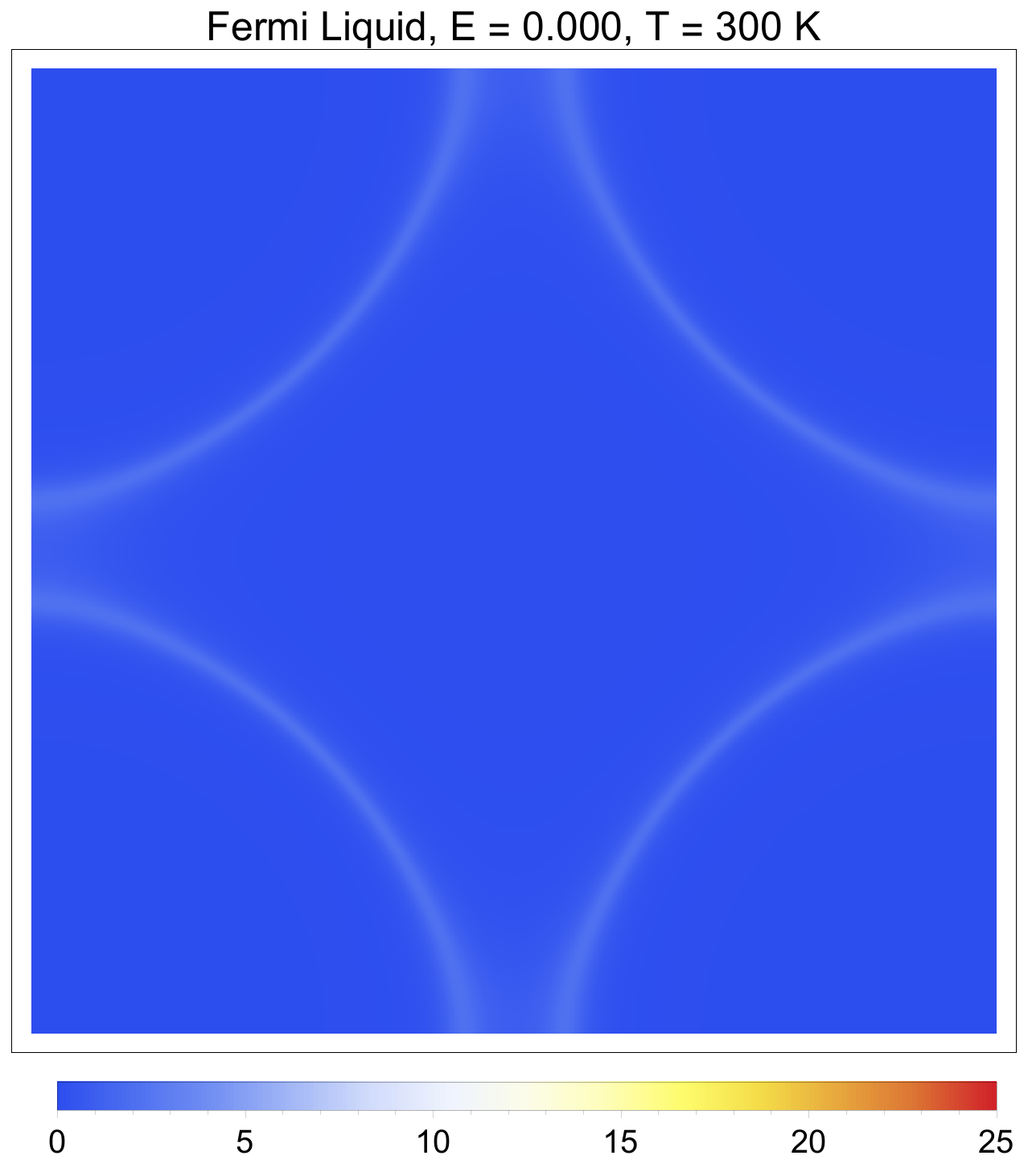}
	\includegraphics[height=0.18\textwidth]{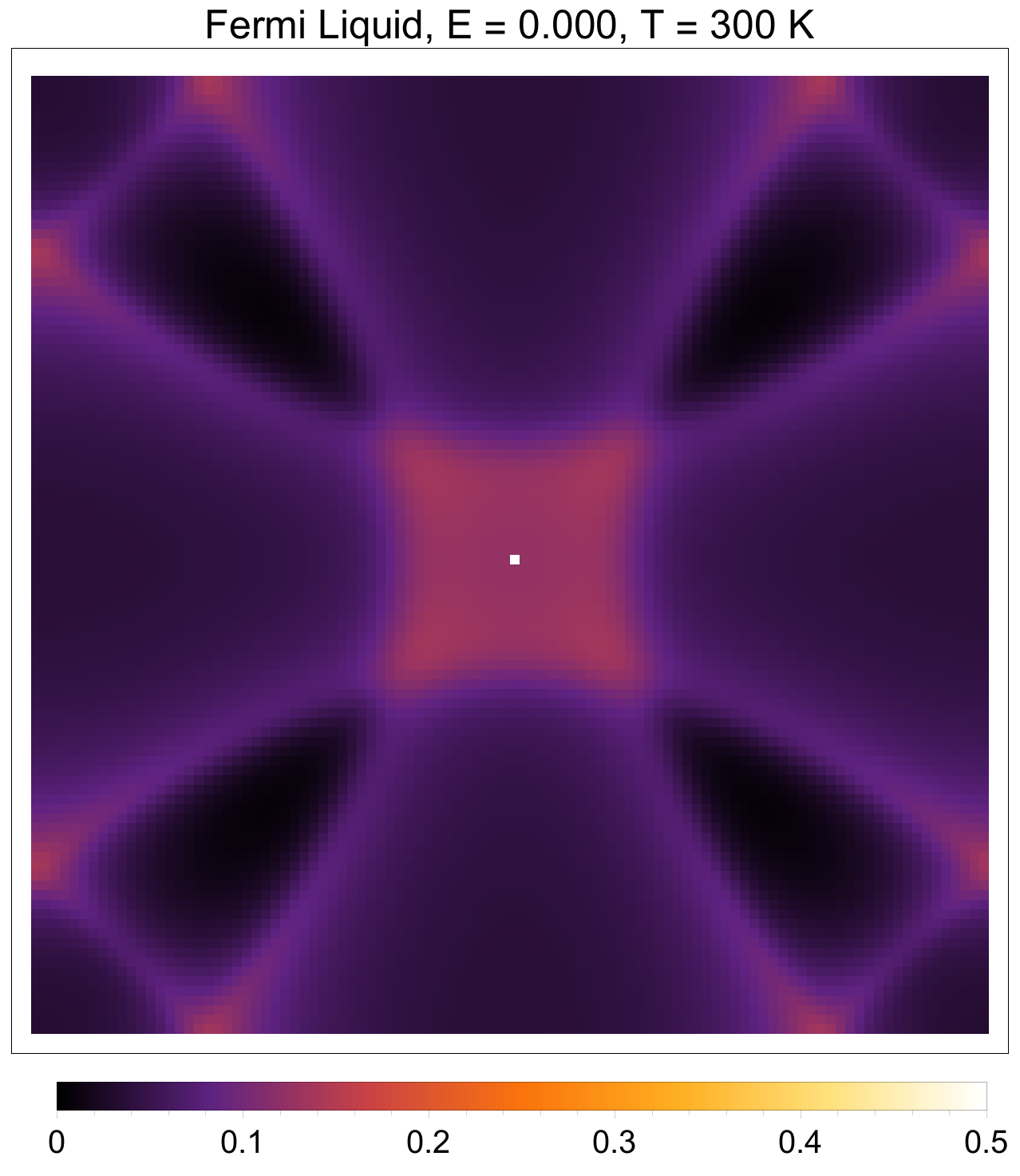}
	\includegraphics[height=0.18\textwidth]{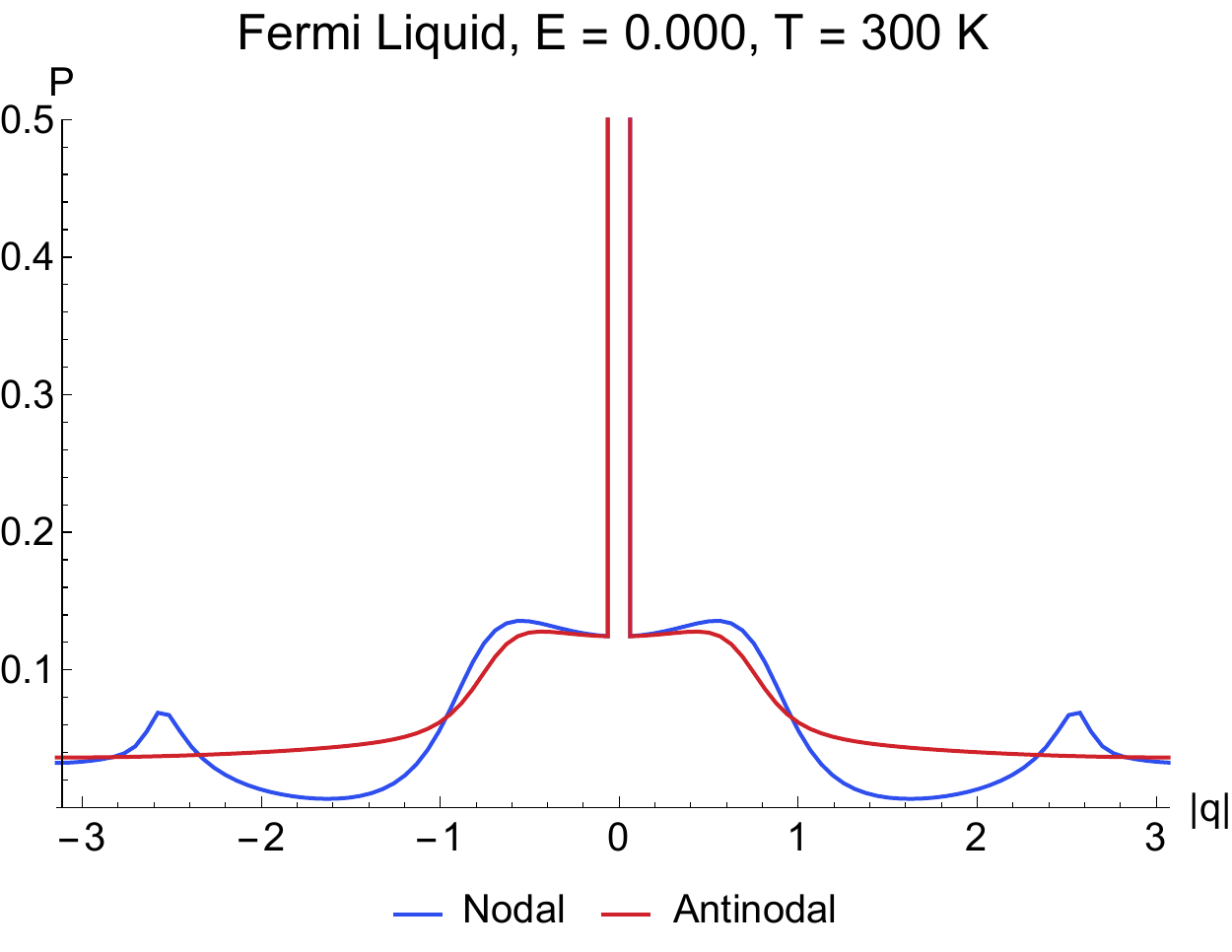}
	\includegraphics[height=0.18\textwidth]{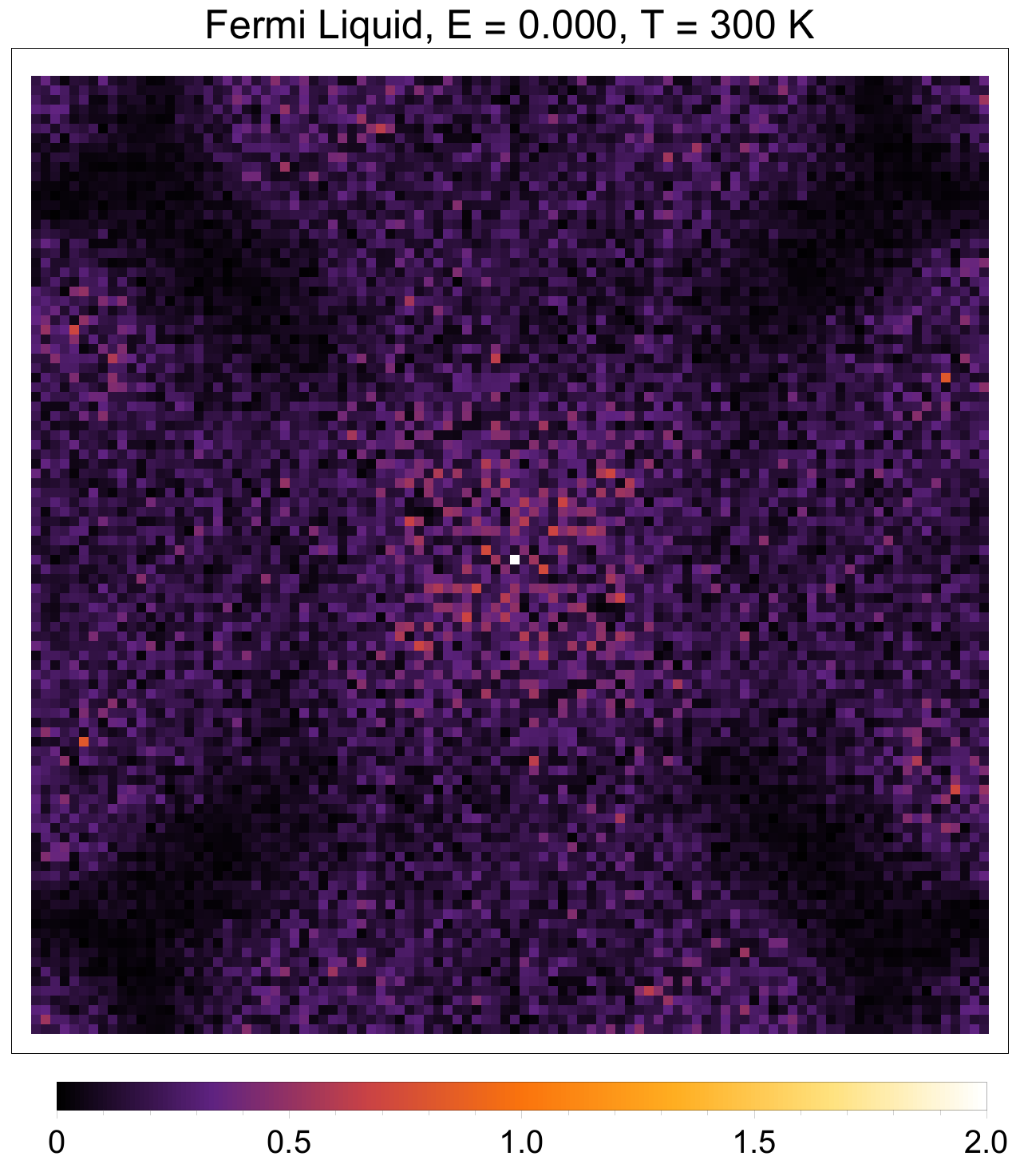}
	\includegraphics[height=0.18\textwidth]{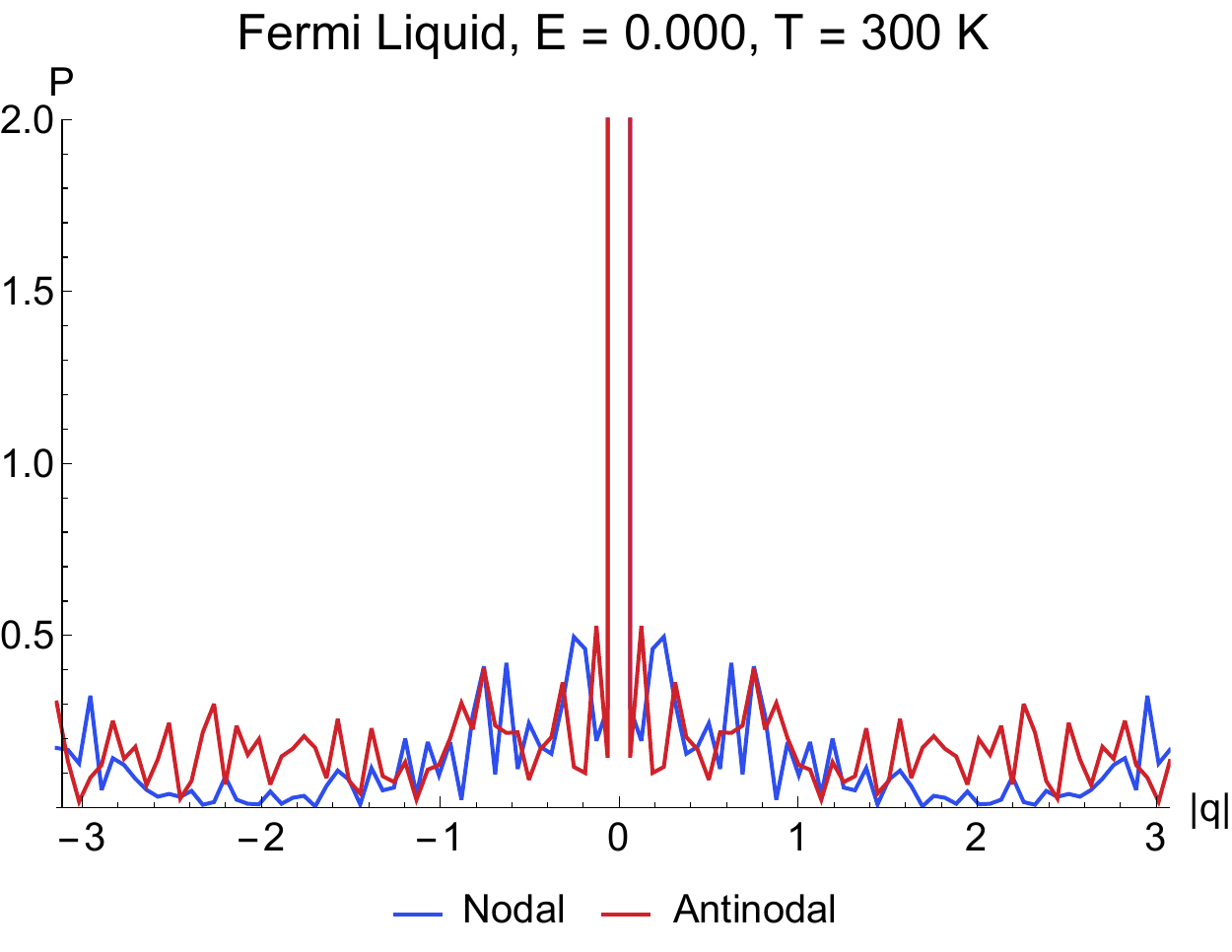} \\
	
	\caption{Ordinary Fermi liquid phenomenology at various temperatures.  Left to right: The spectral function $A(\mathbf{k}, \omega)$; the Fourier transform of the LDOS $P(\mathbf{q}, \omega)$; linecuts of $P(\mathbf{q}, \omega)$ in the nodal and antinodal directions; $P(\mathbf{q}, \omega)$ in the presence of multiple weak impurities and finite-temperature smearing; and linecuts of $P(\mathbf{q}, \omega)$ in the presence of multiple weak impurities and finite-temperature smearing. All plots are taken at $E = 0.000$.}
	
	\label{fig:temperature_fl}
\end{figure*}

\begin{figure*}
	\centering
	
	\includegraphics[height=0.18\textwidth]{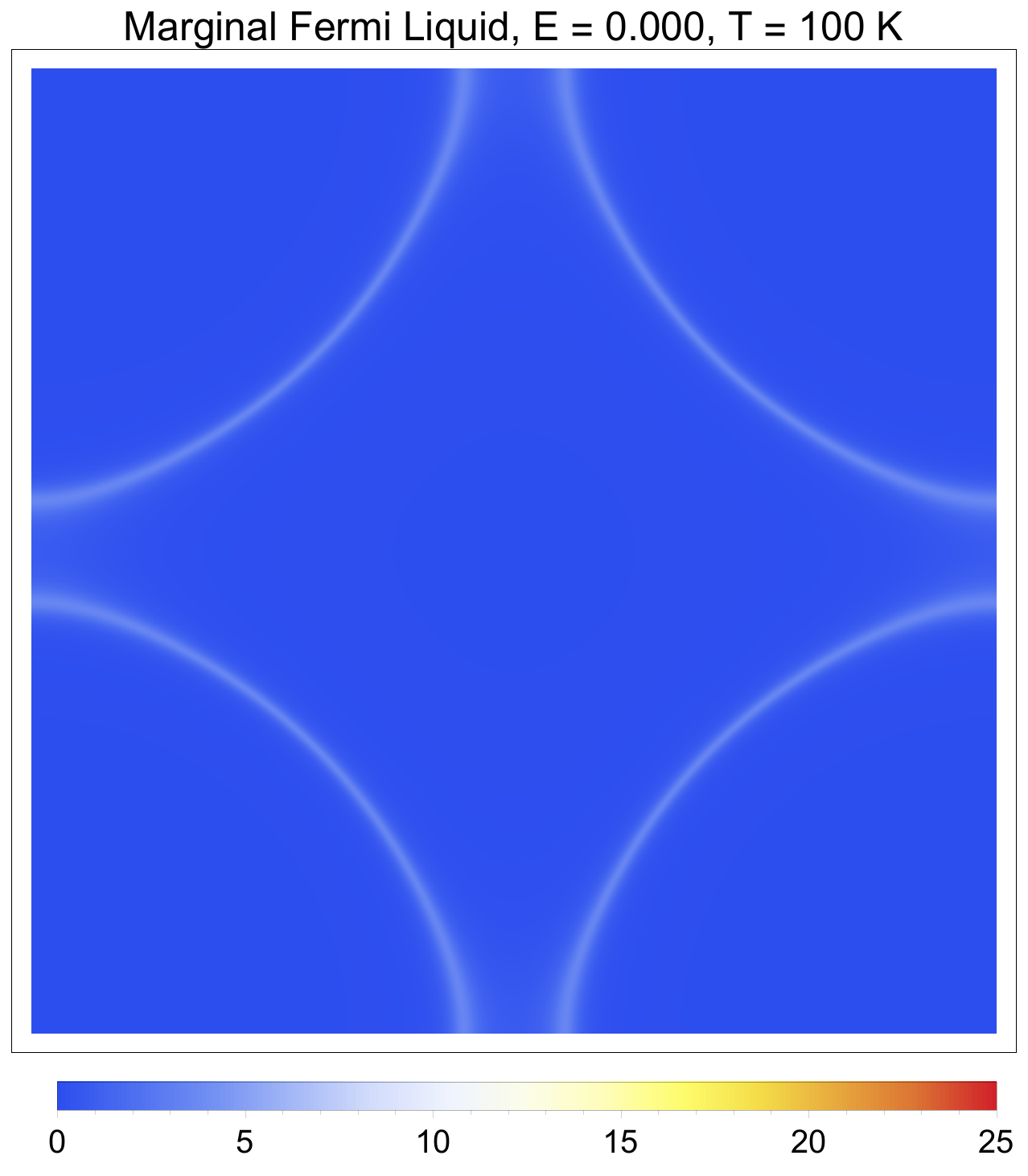}
	\includegraphics[height=0.18\textwidth]{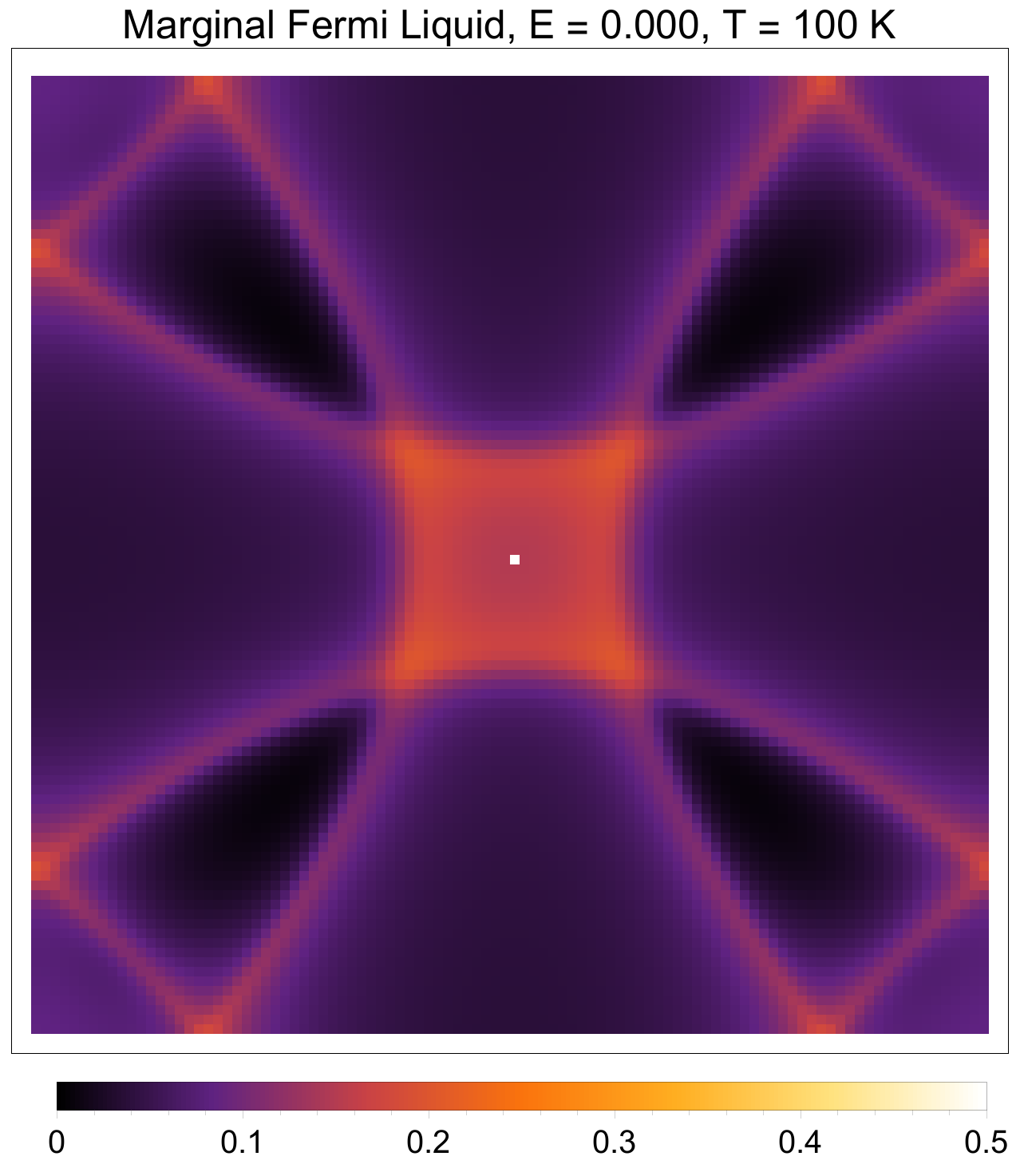}
	\includegraphics[height=0.18\textwidth]{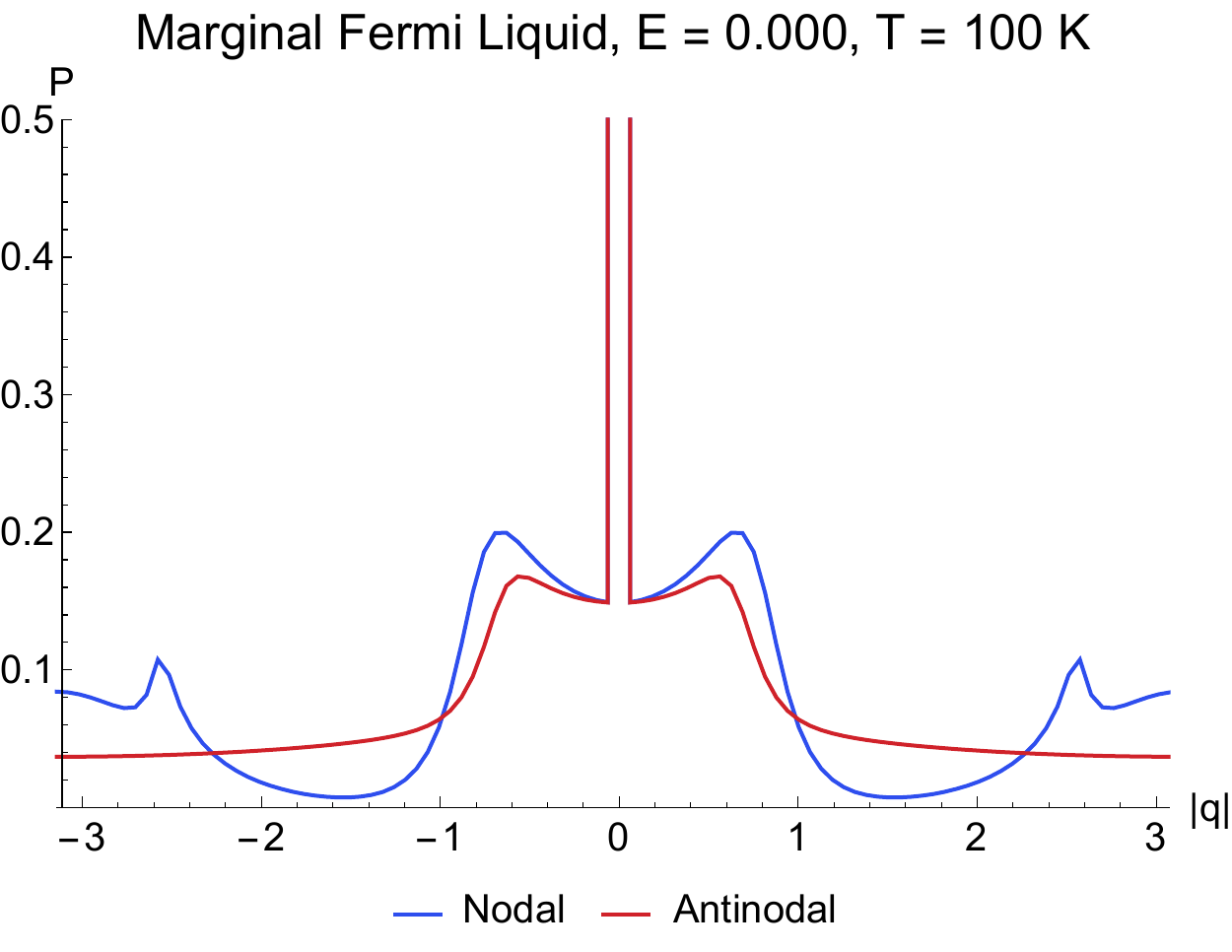}
	\includegraphics[height=0.18\textwidth]{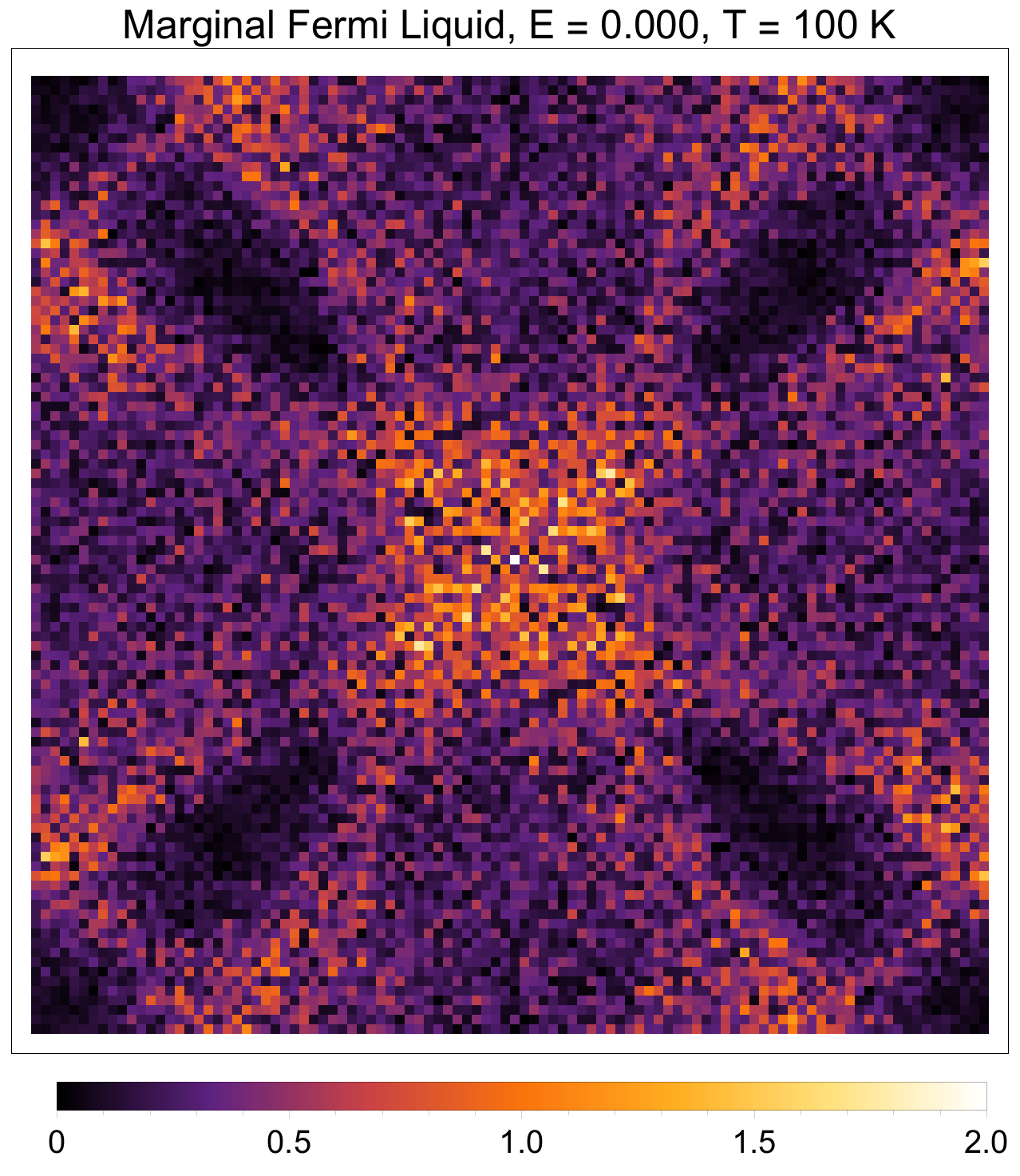}
	\includegraphics[height=0.18\textwidth]{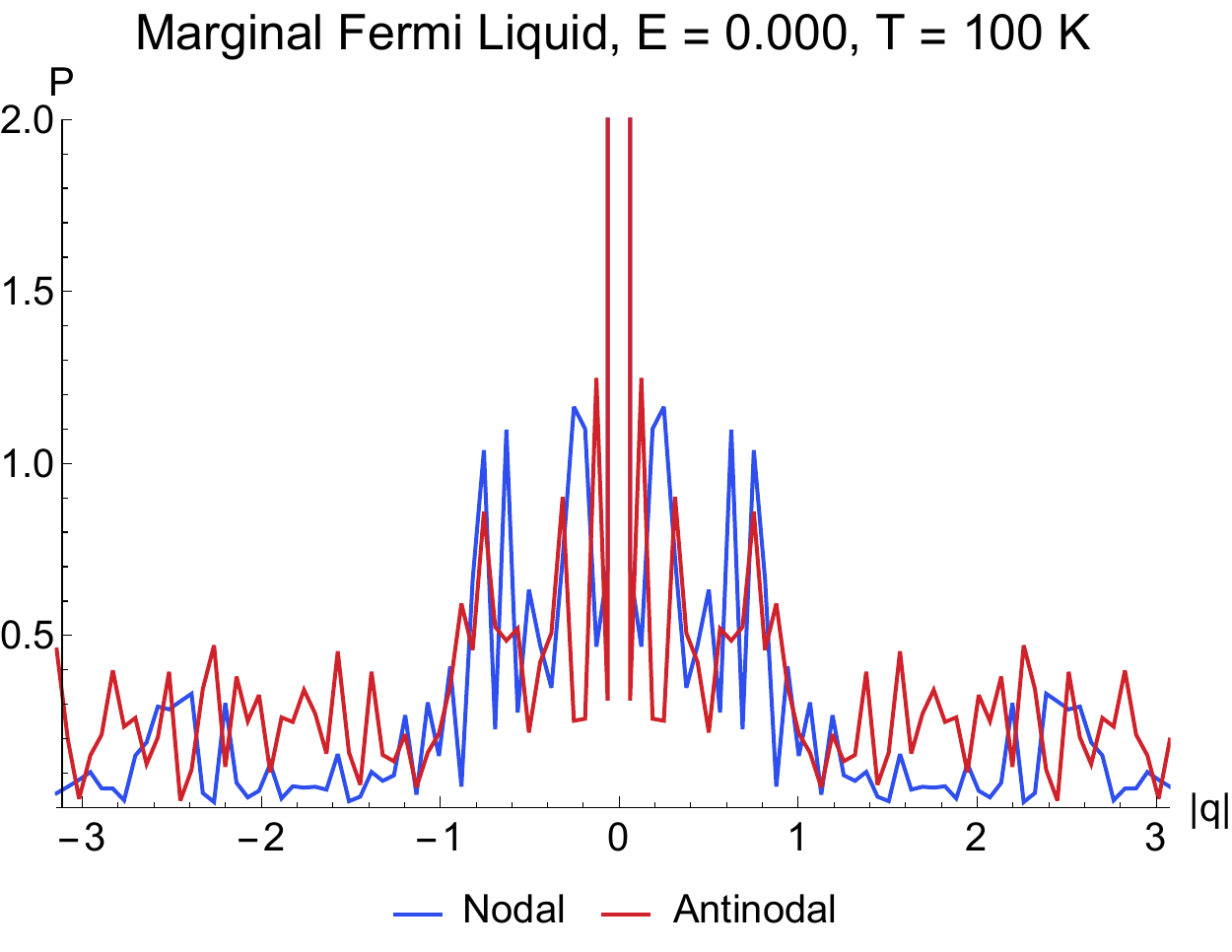} \\
	\includegraphics[height=0.18\textwidth]{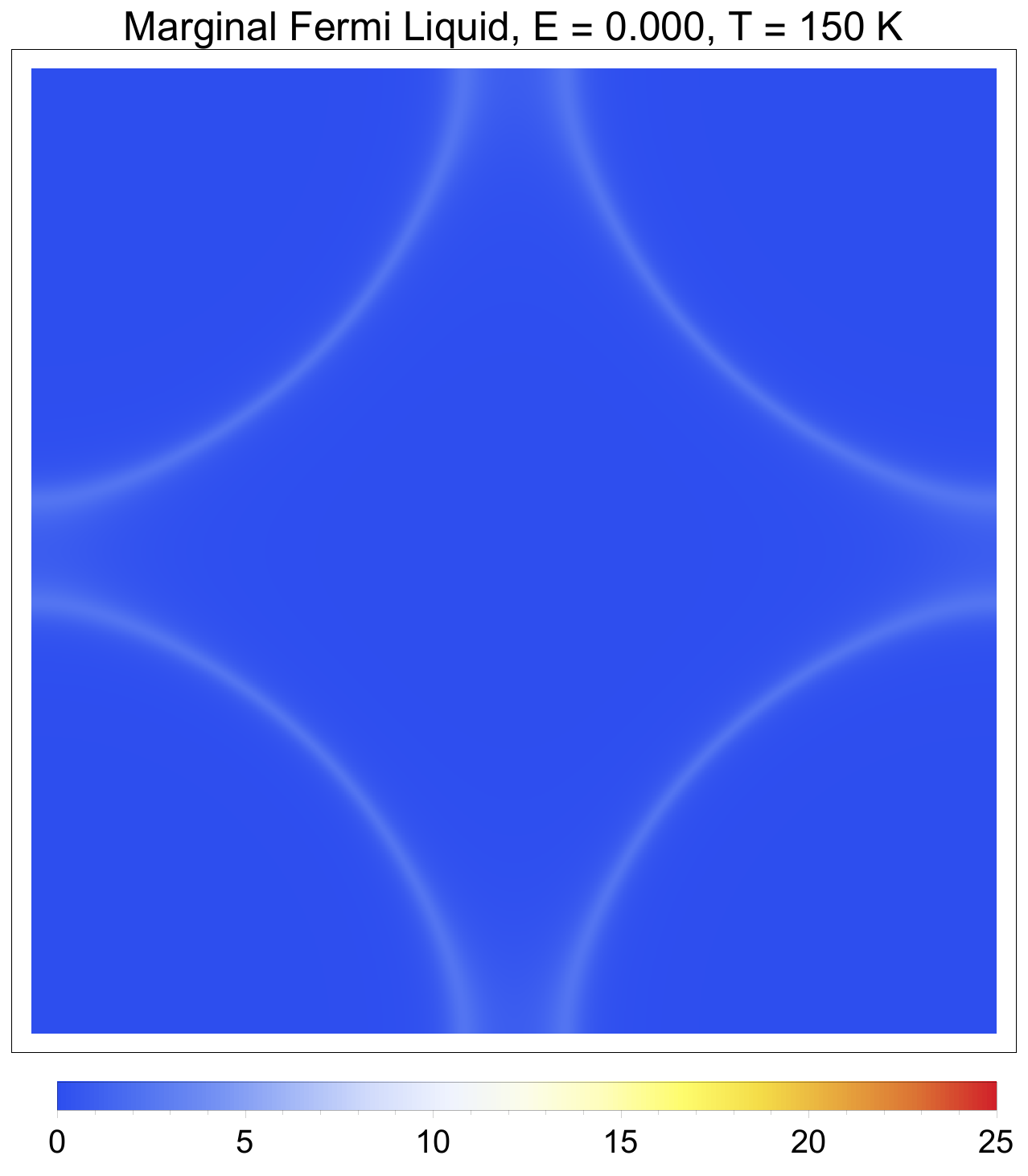}
	\includegraphics[height=0.18\textwidth]{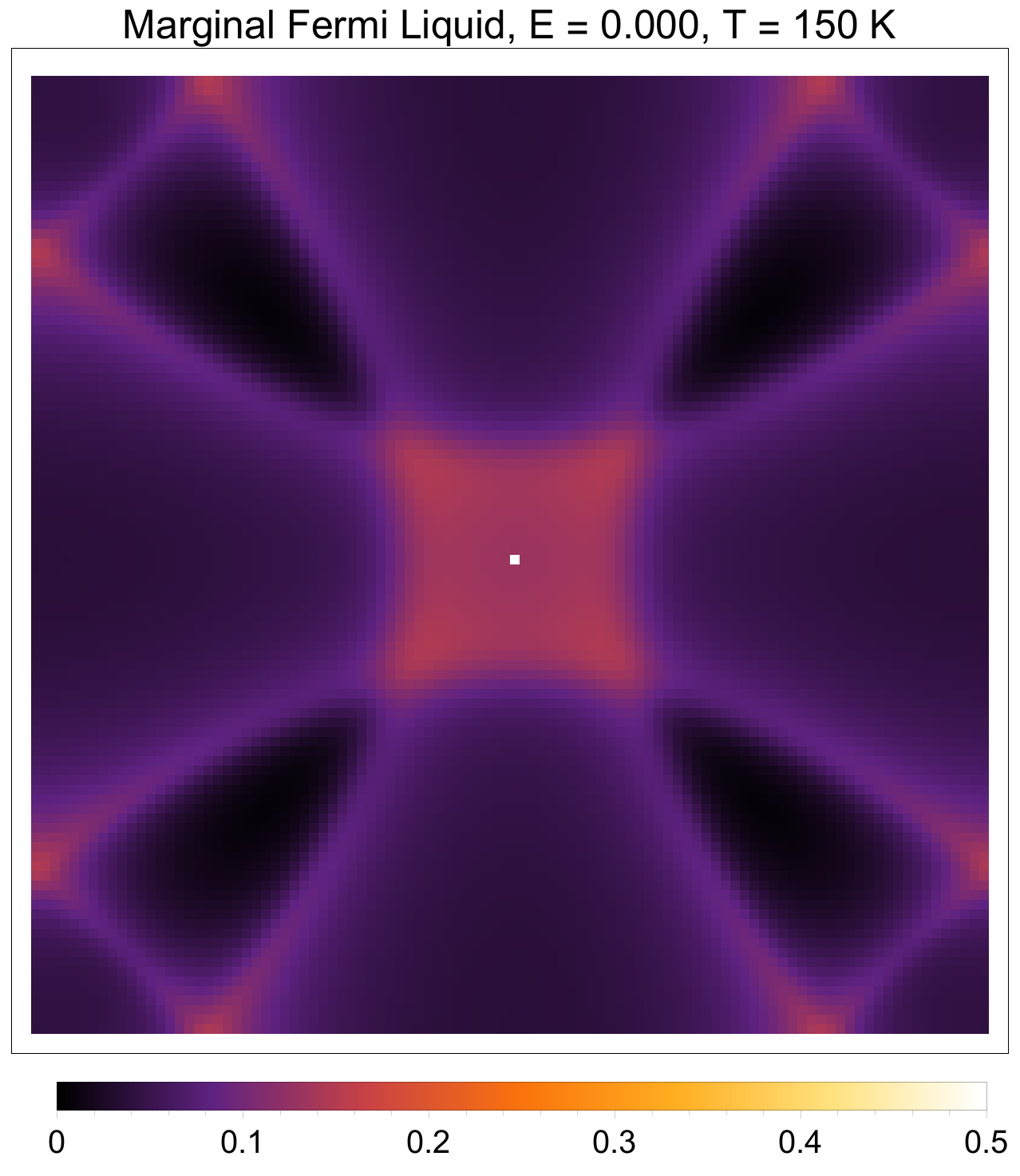}
	\includegraphics[height=0.18\textwidth]{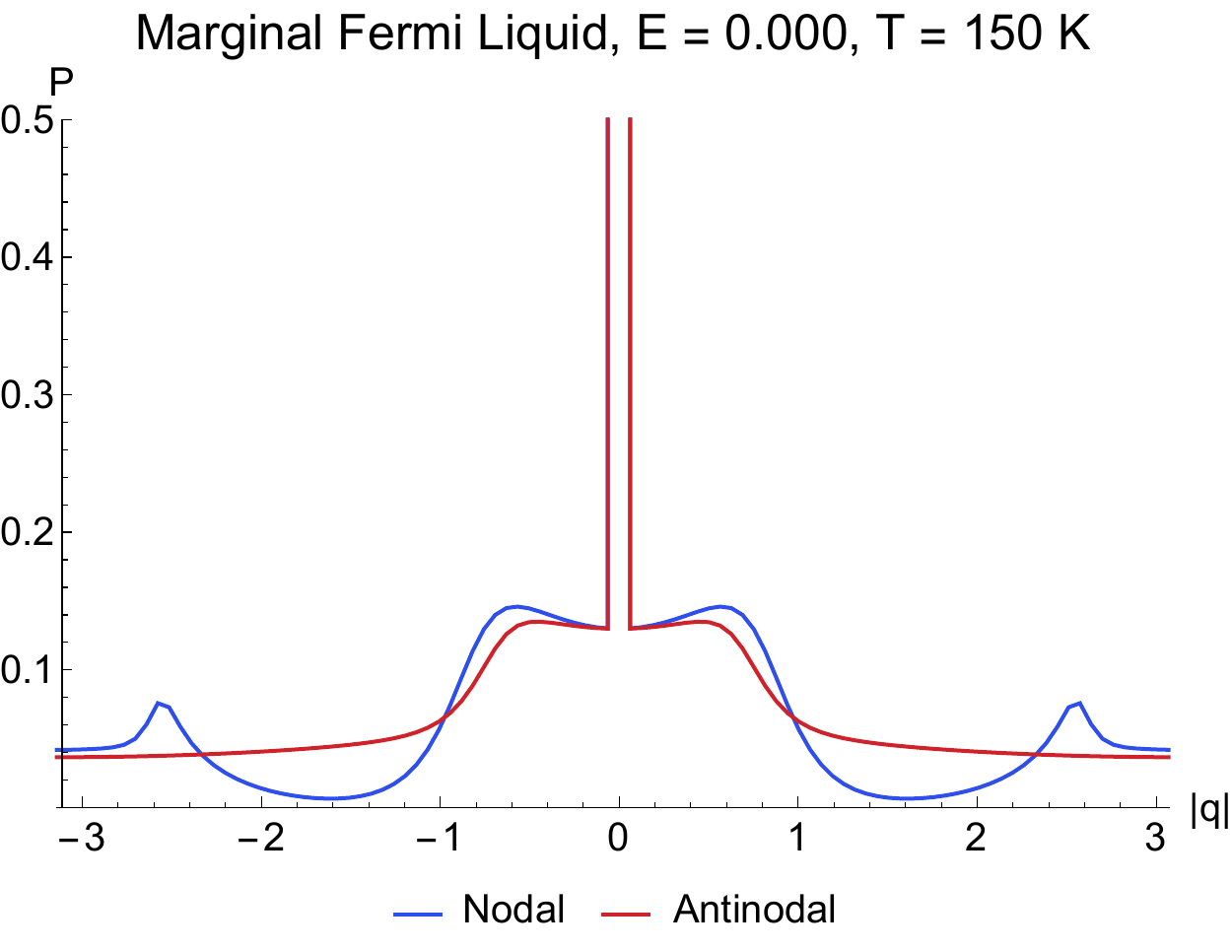}
	\includegraphics[height=0.18\textwidth]{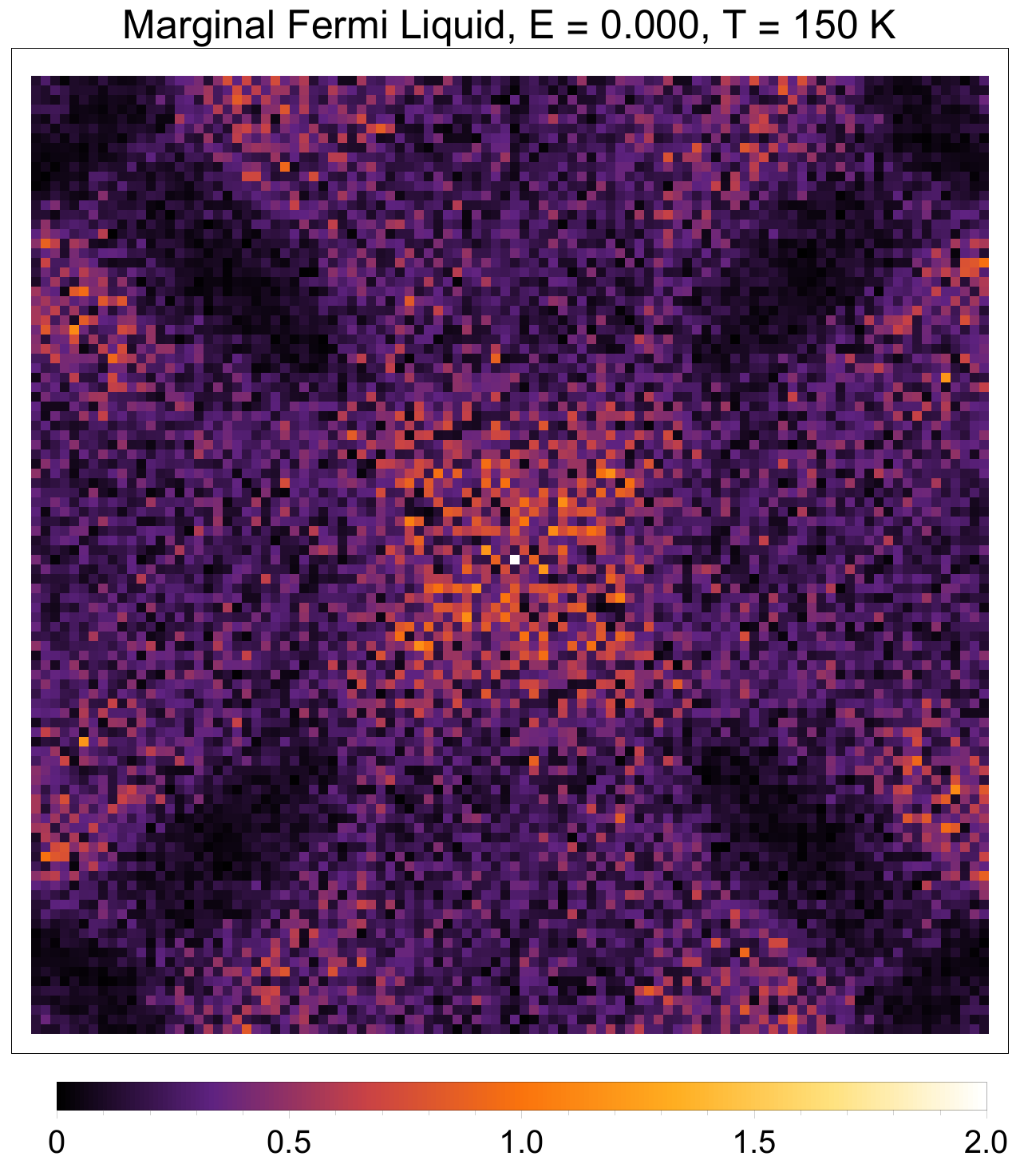}
	\includegraphics[height=0.18\textwidth]{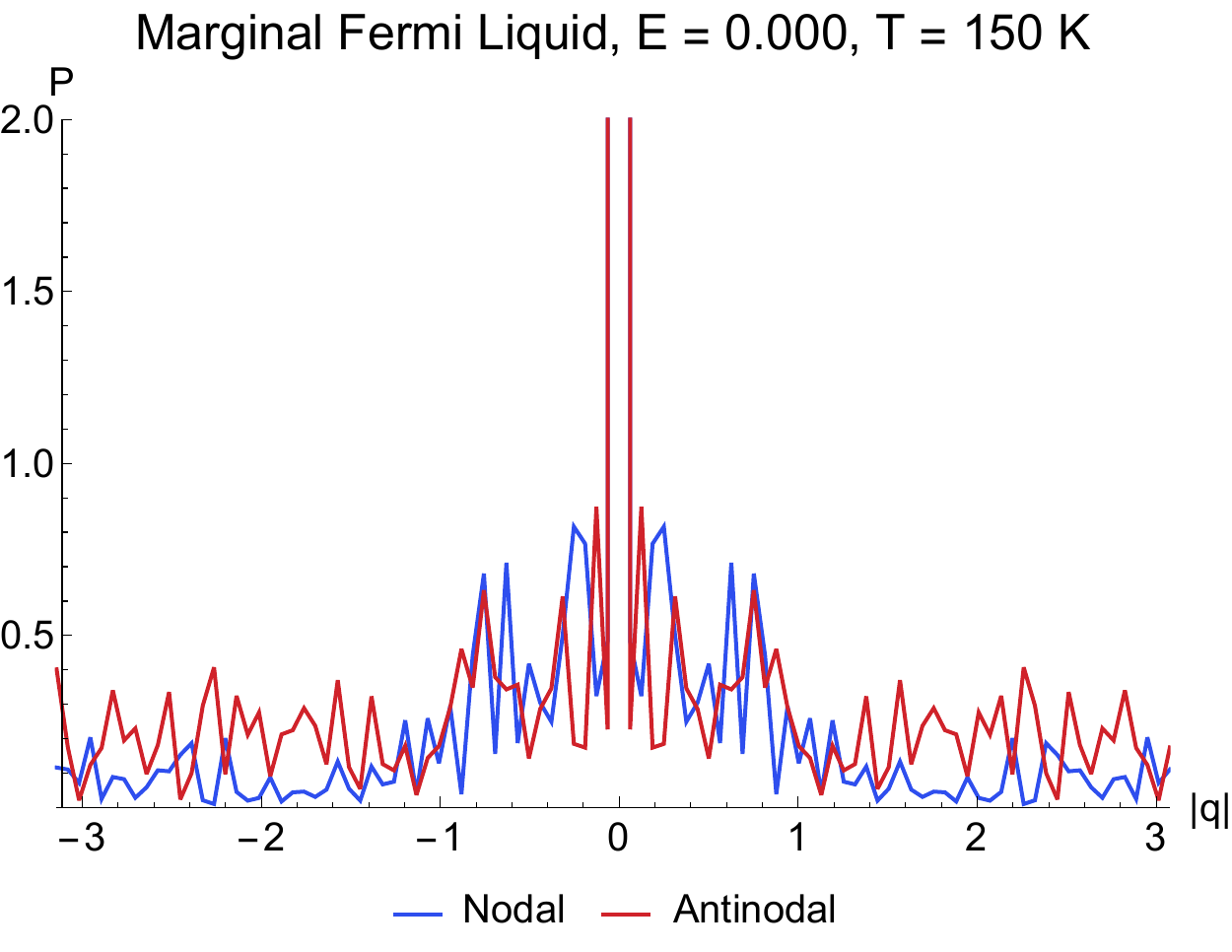} \\
	\includegraphics[height=0.18\textwidth]{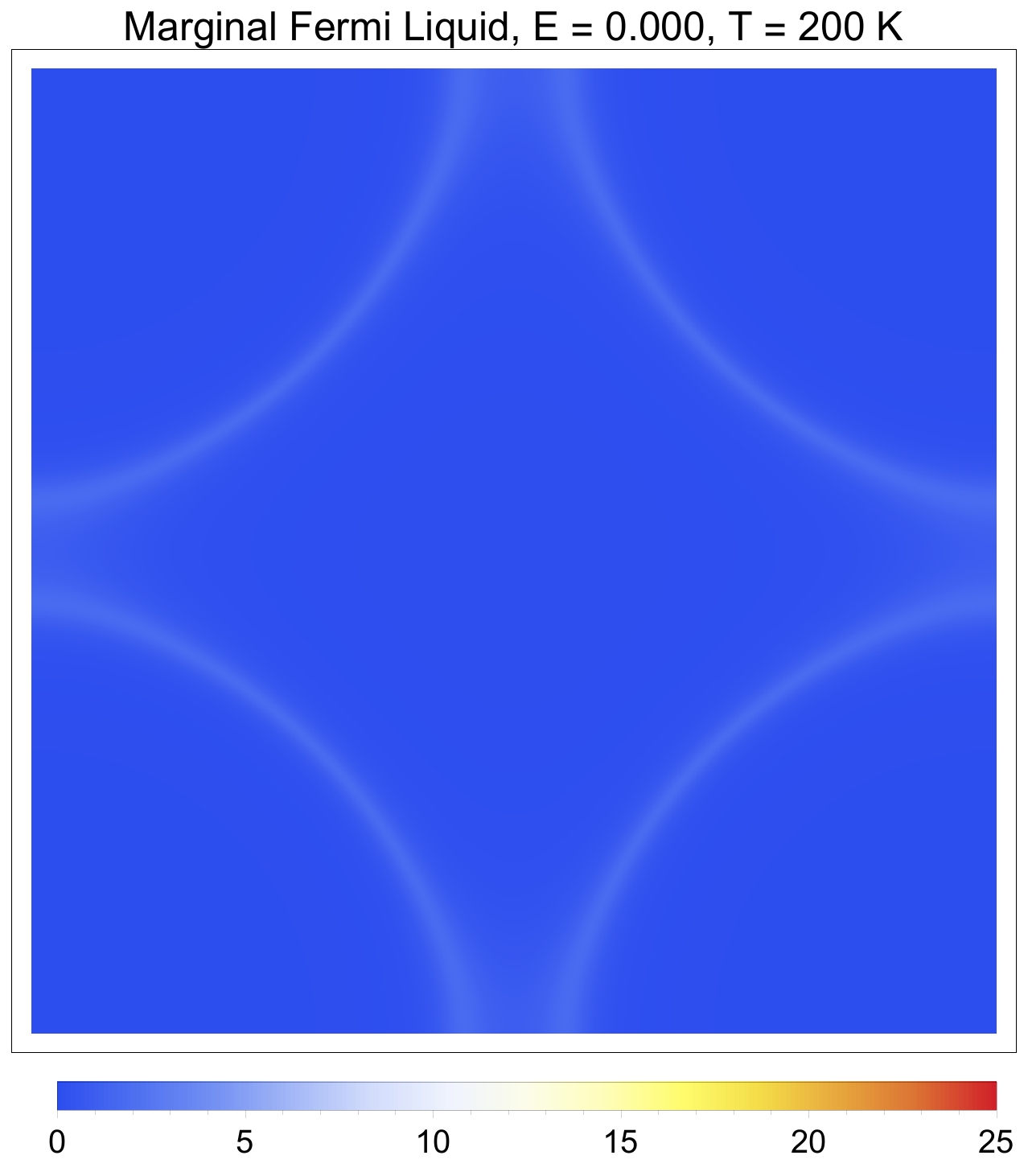}
	\includegraphics[height=0.18\textwidth]{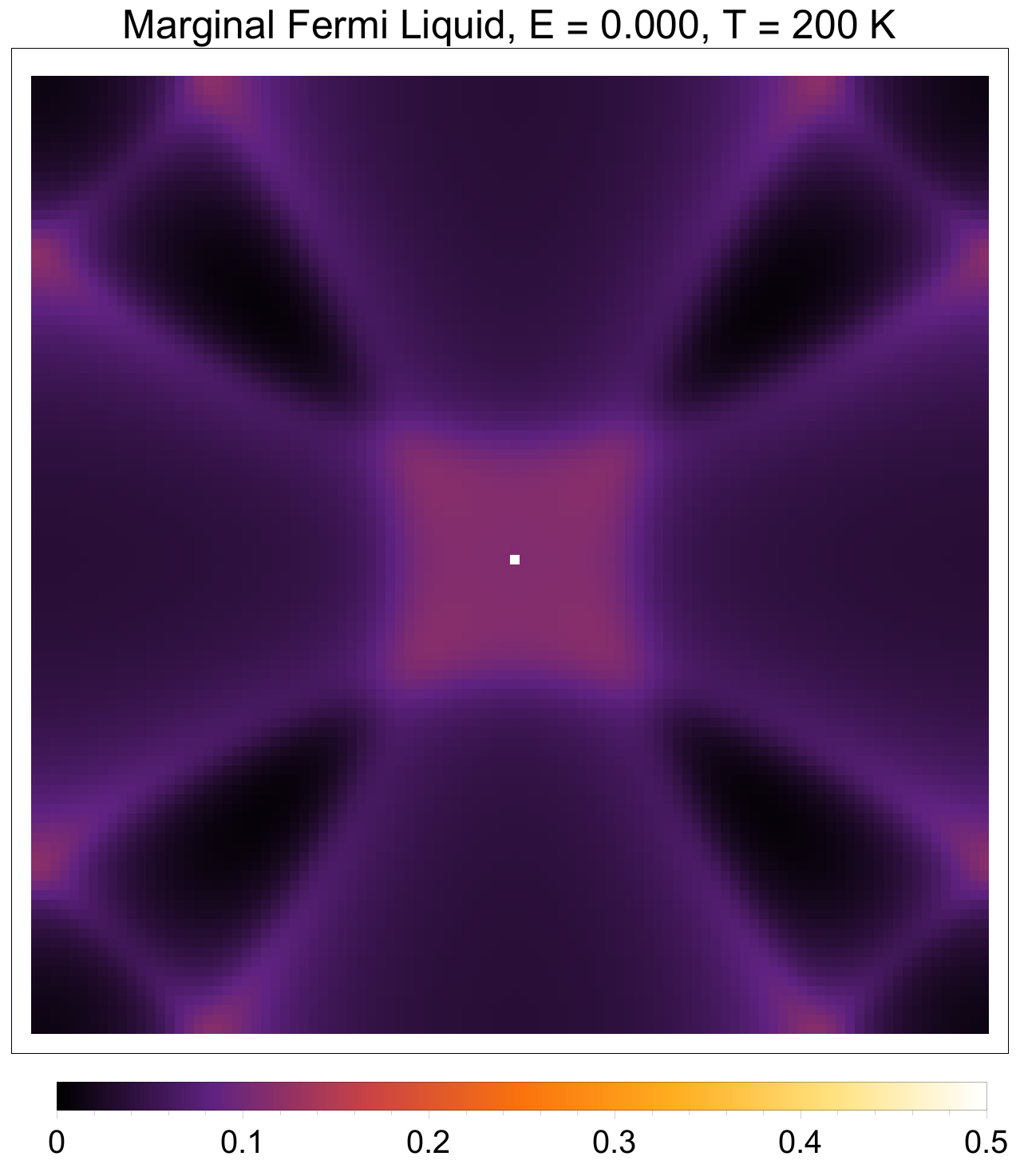}
	\includegraphics[height=0.18\textwidth]{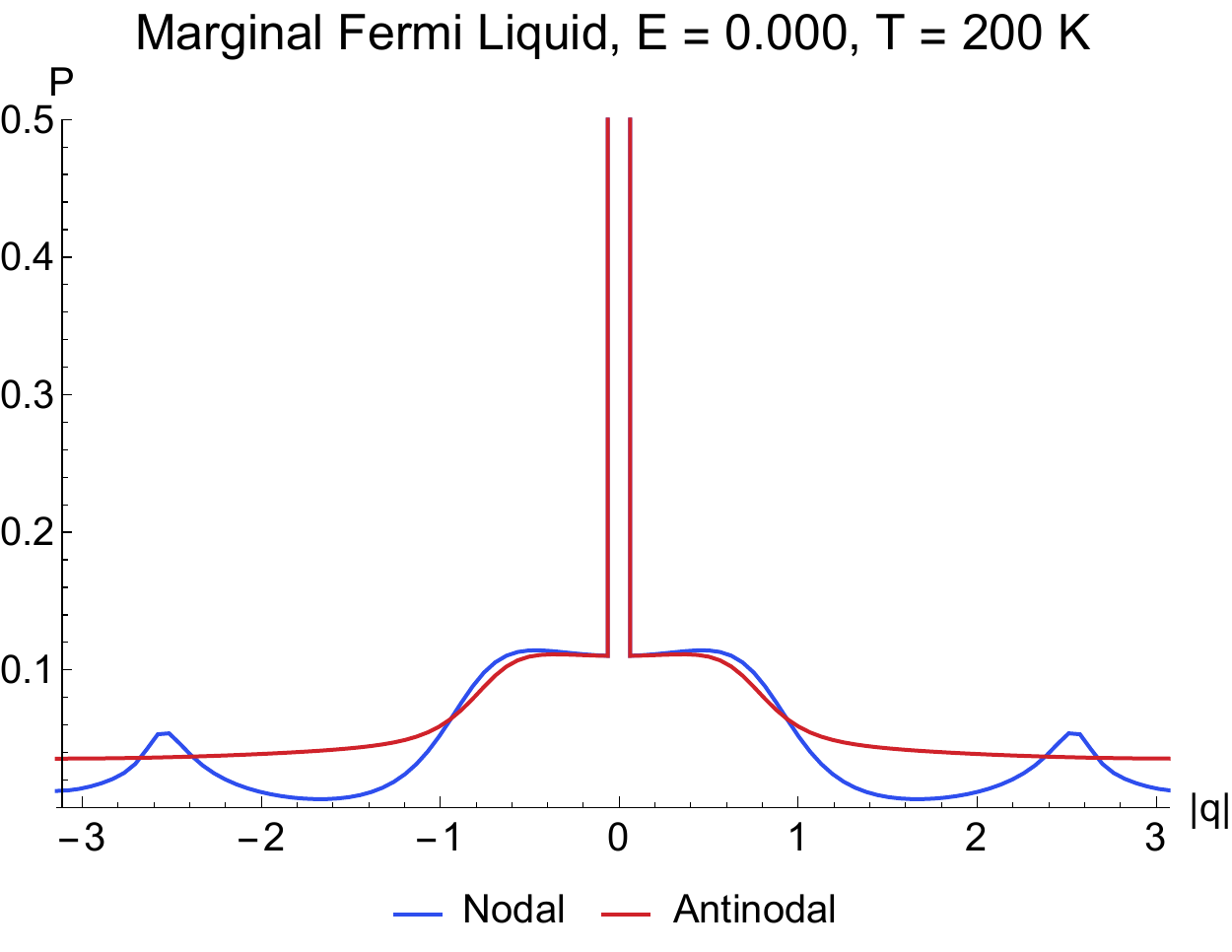}
	\includegraphics[height=0.18\textwidth]{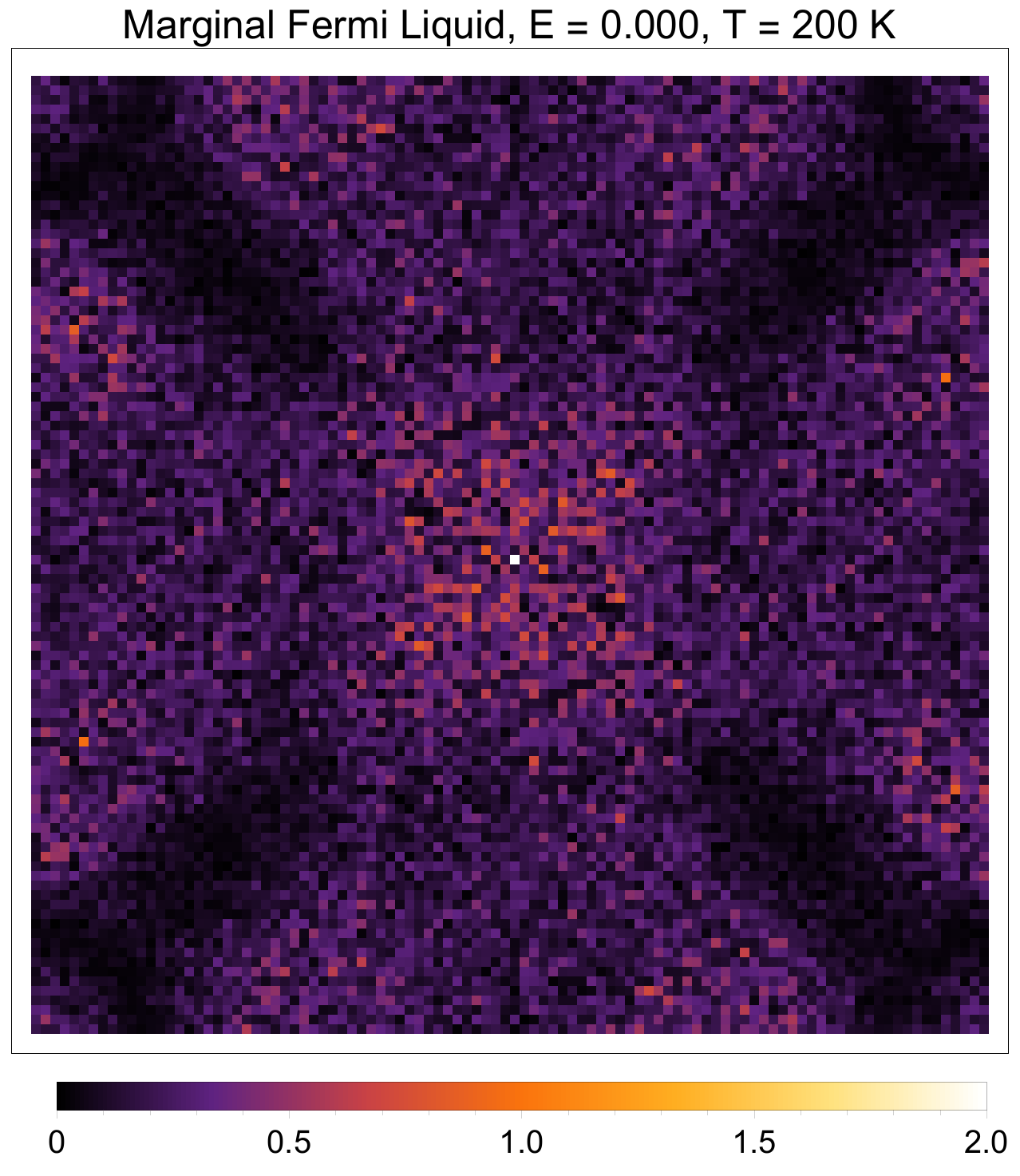}
	\includegraphics[height=0.18\textwidth]{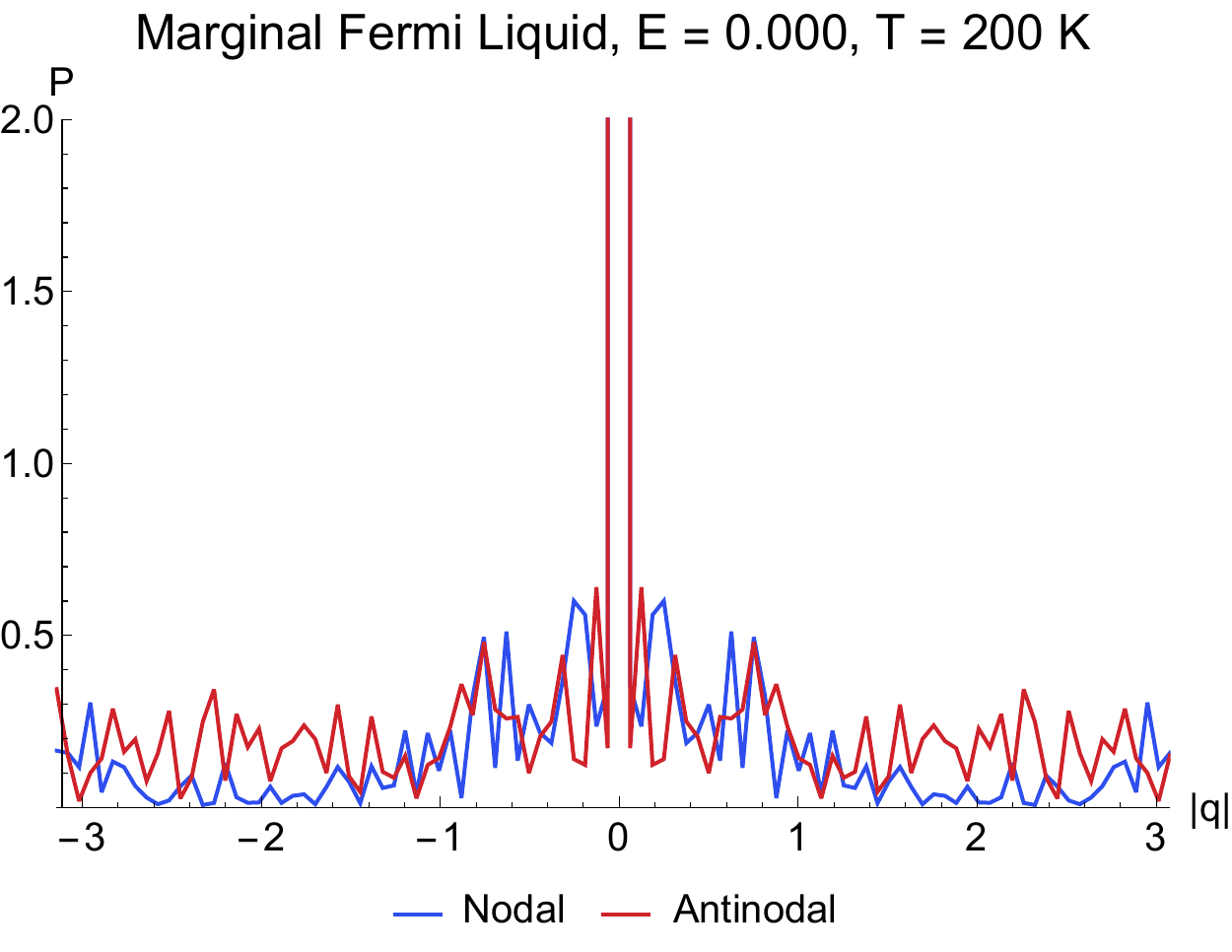} \\
	\includegraphics[height=0.18\textwidth]{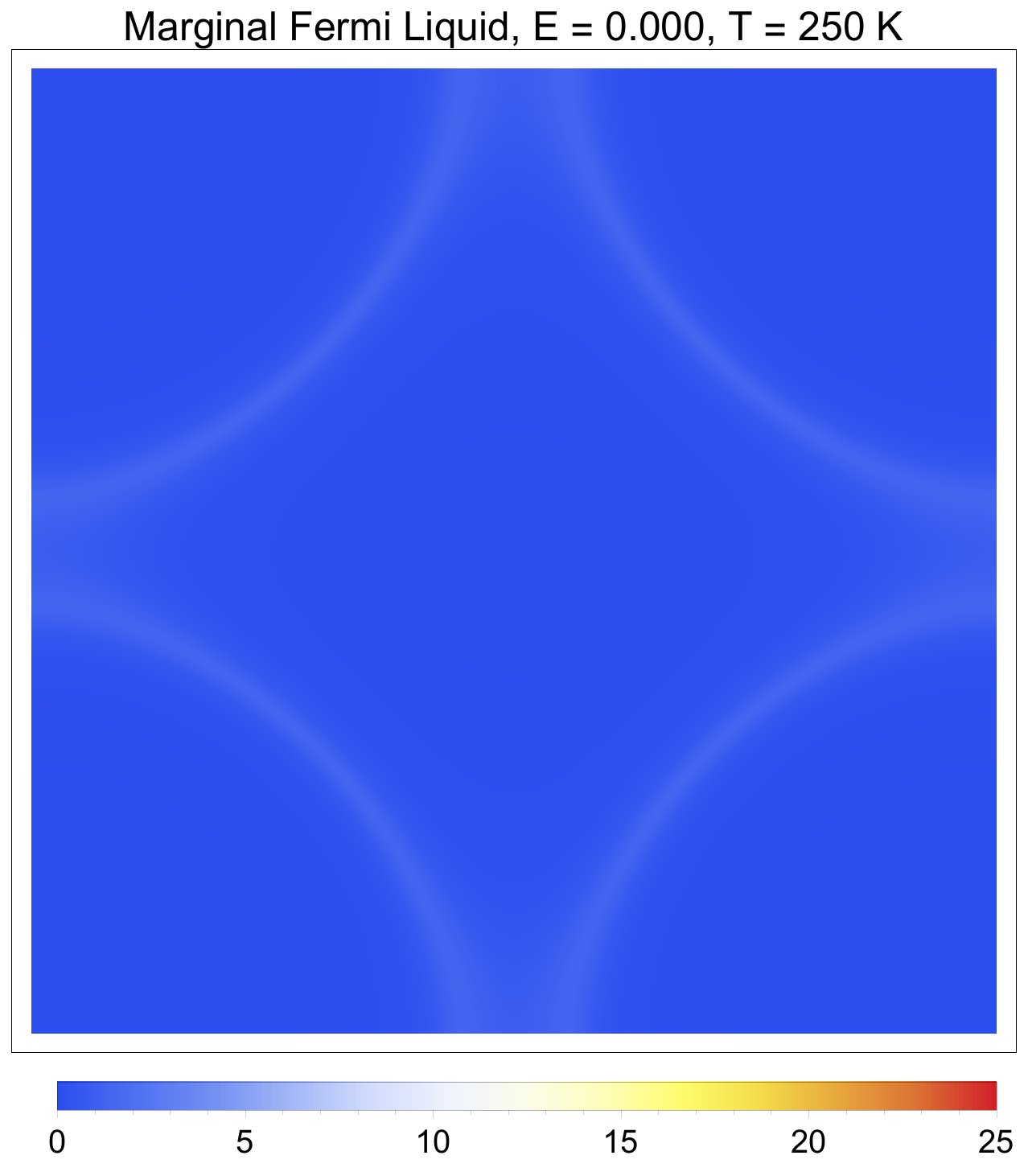}
	\includegraphics[height=0.18\textwidth]{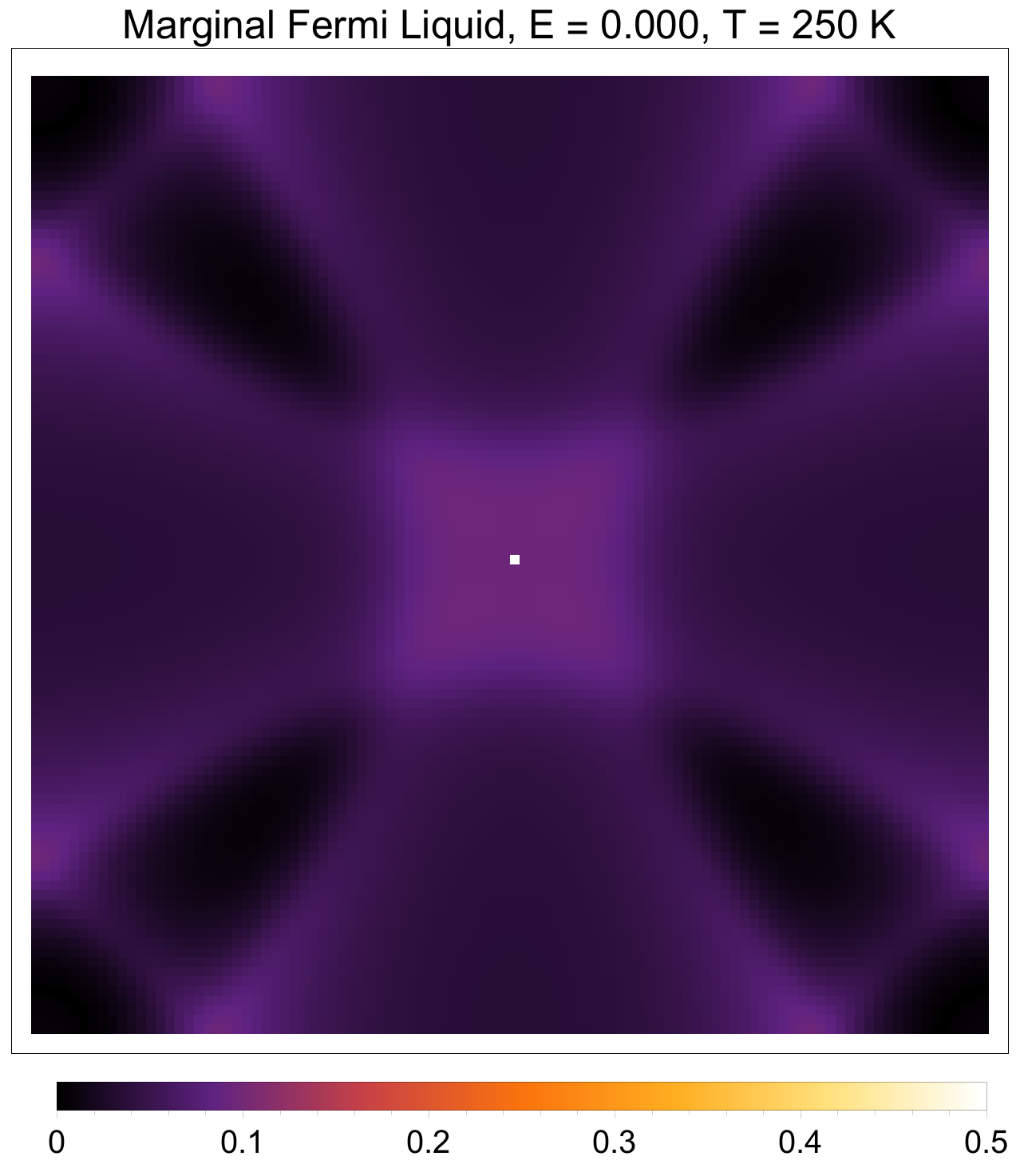}
	\includegraphics[height=0.18\textwidth]{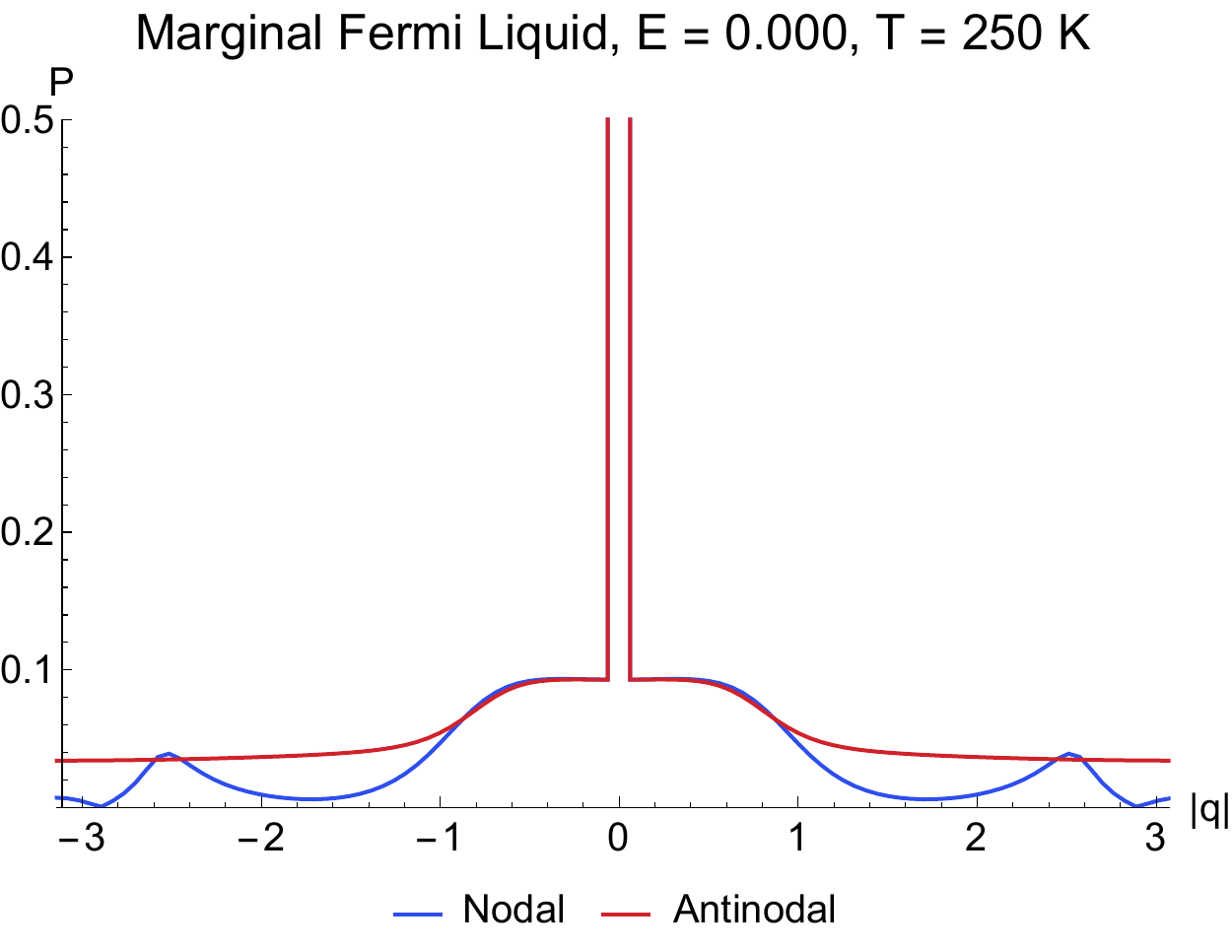}
	\includegraphics[height=0.18\textwidth]{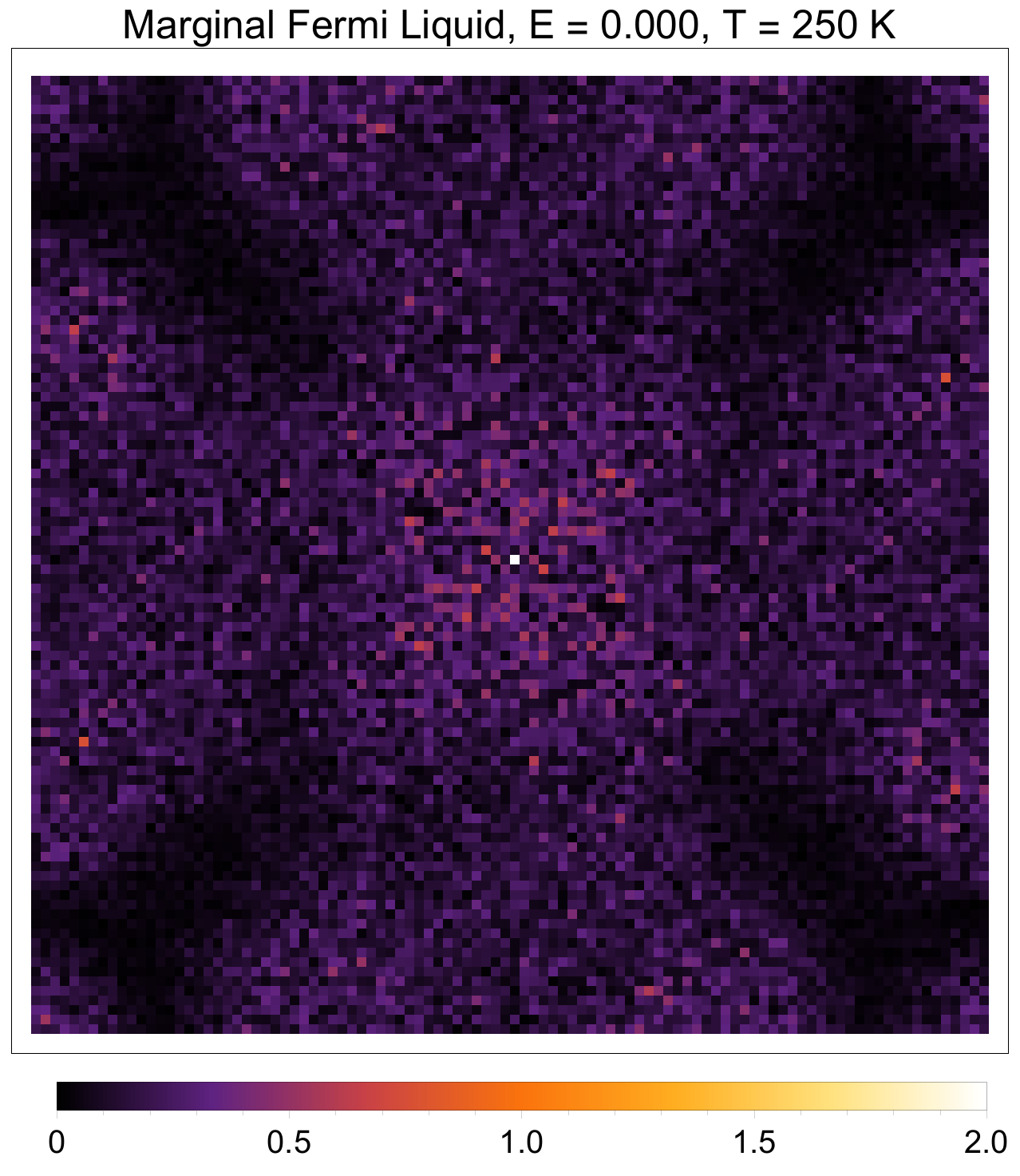}
	\includegraphics[height=0.18\textwidth]{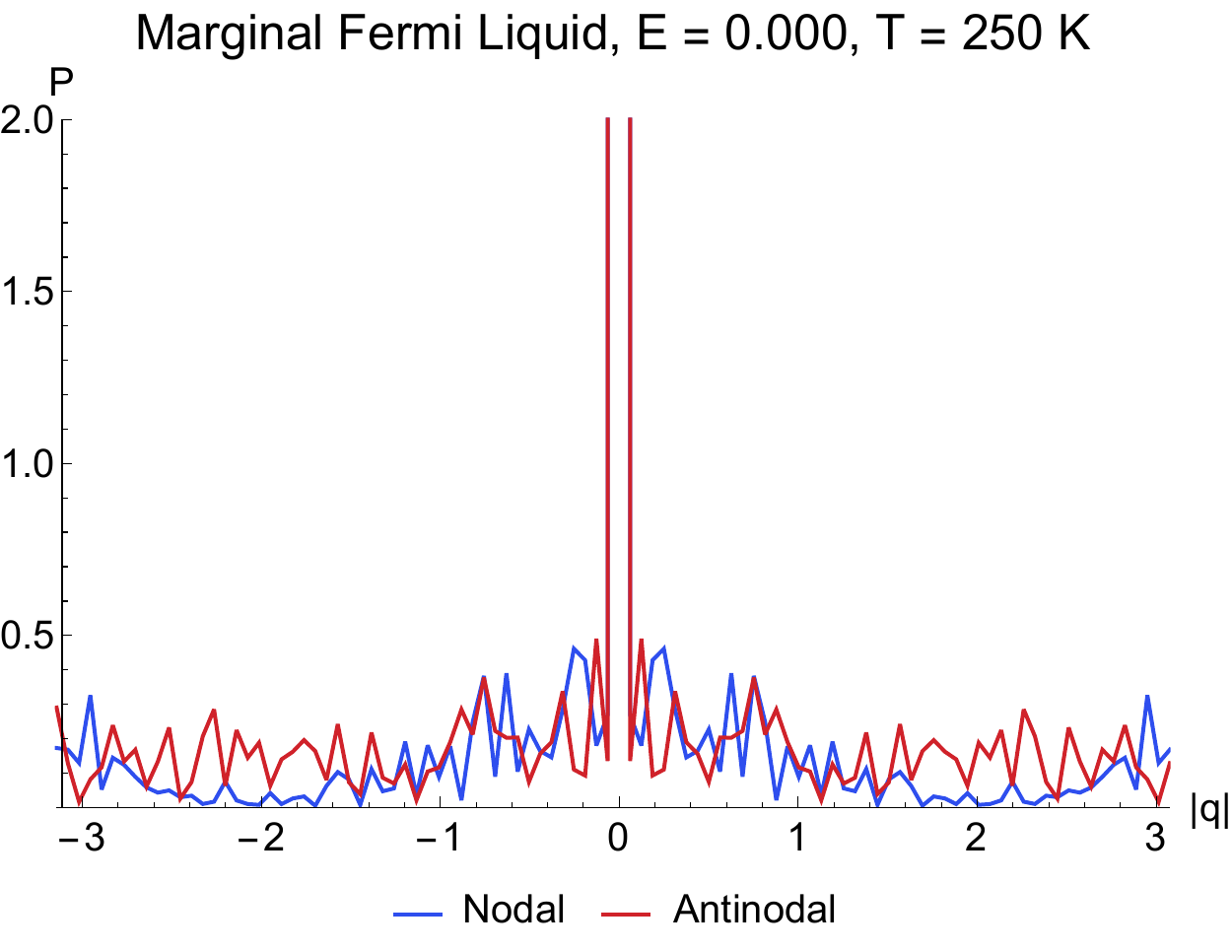} \\
	\includegraphics[height=0.18\textwidth]{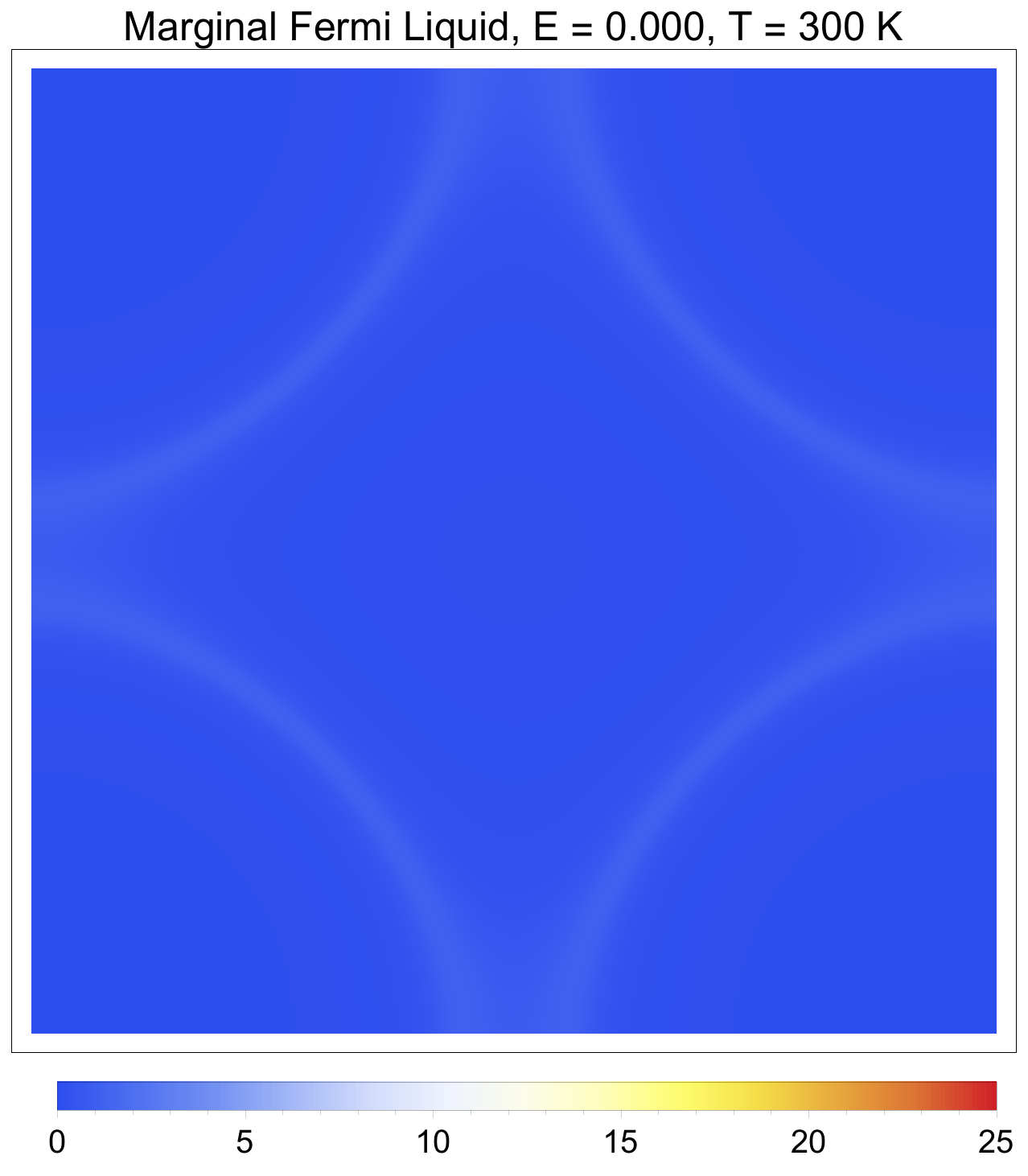}
	\includegraphics[height=0.18\textwidth]{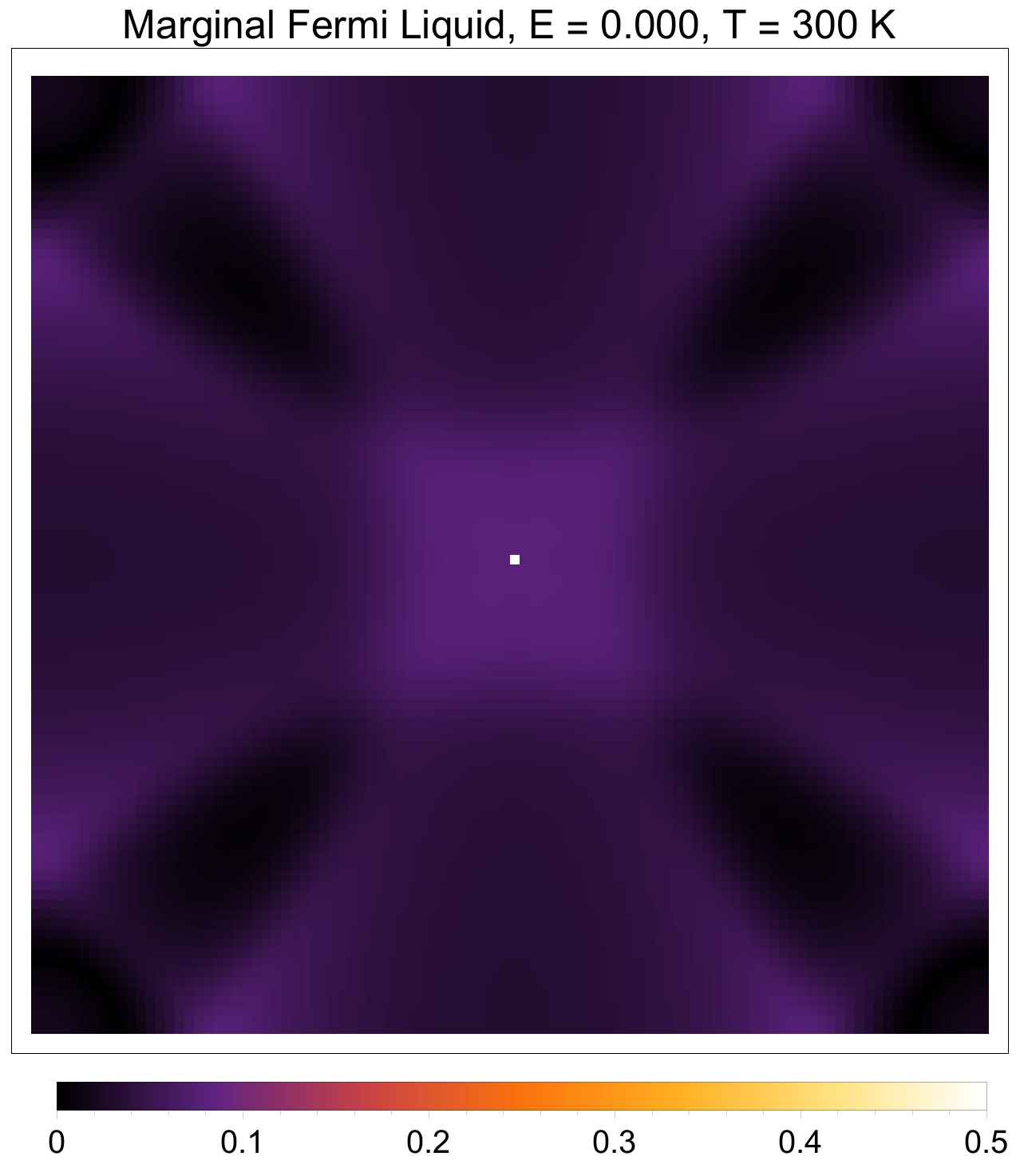}
	\includegraphics[height=0.18\textwidth]{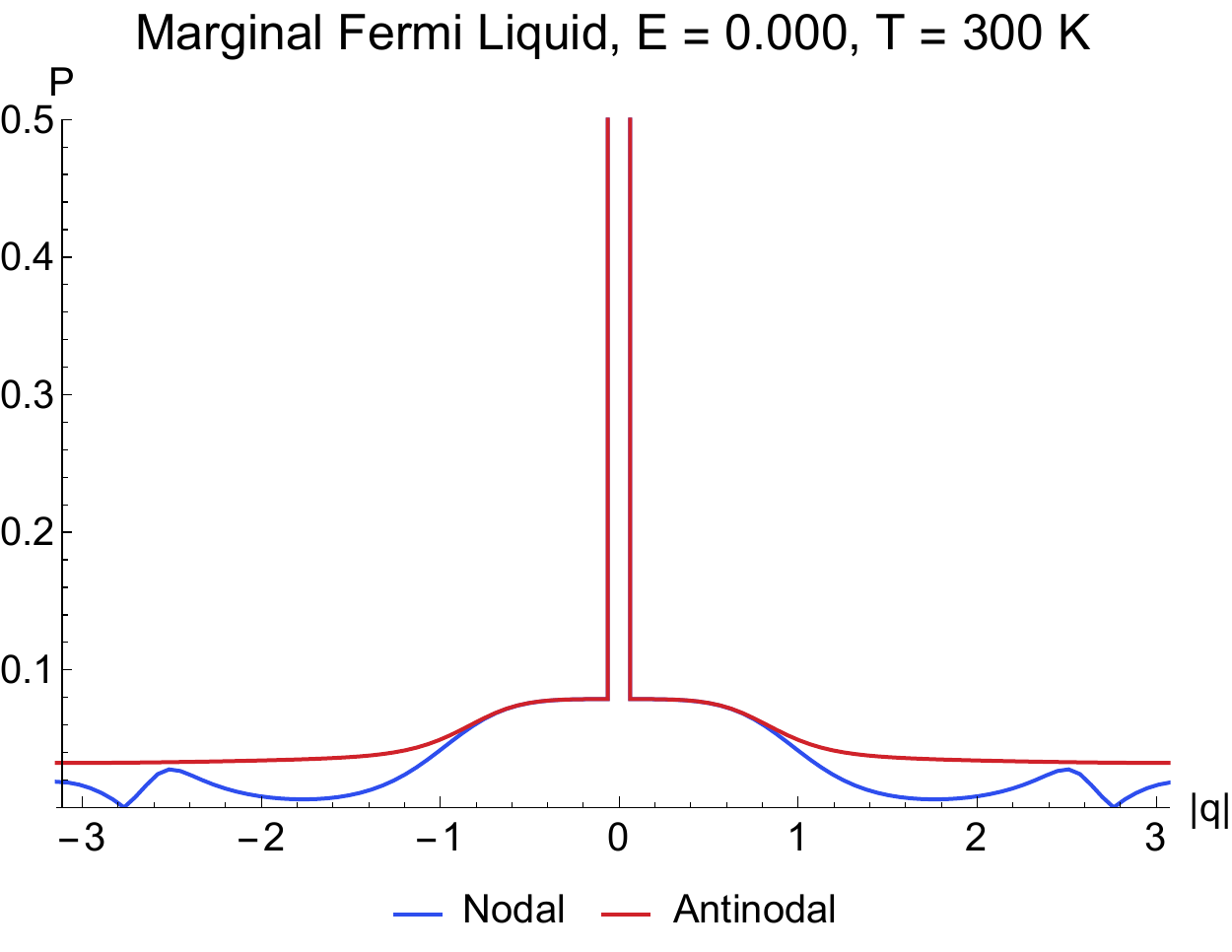}
	\includegraphics[height=0.18\textwidth]{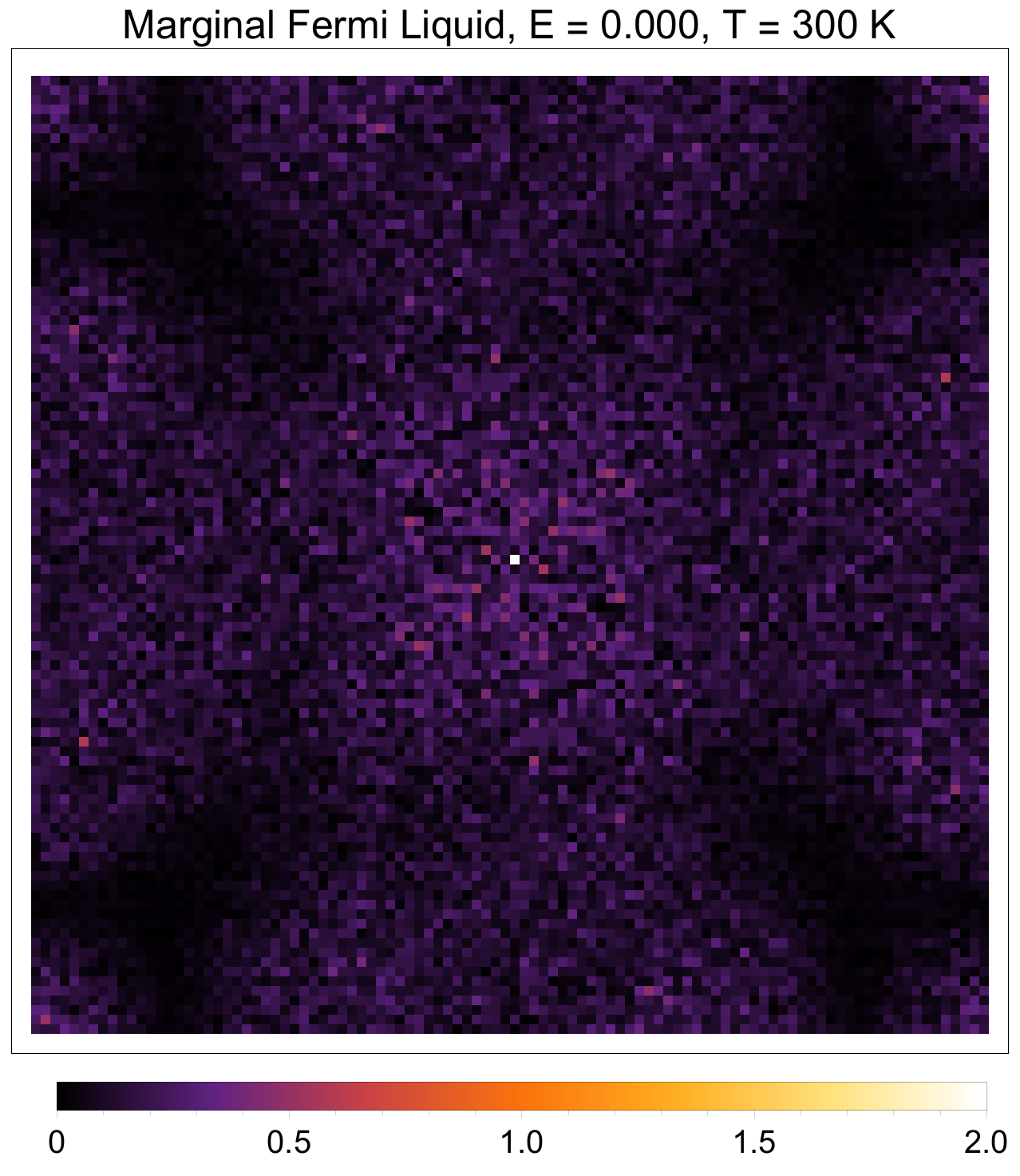}
	\includegraphics[height=0.18\textwidth]{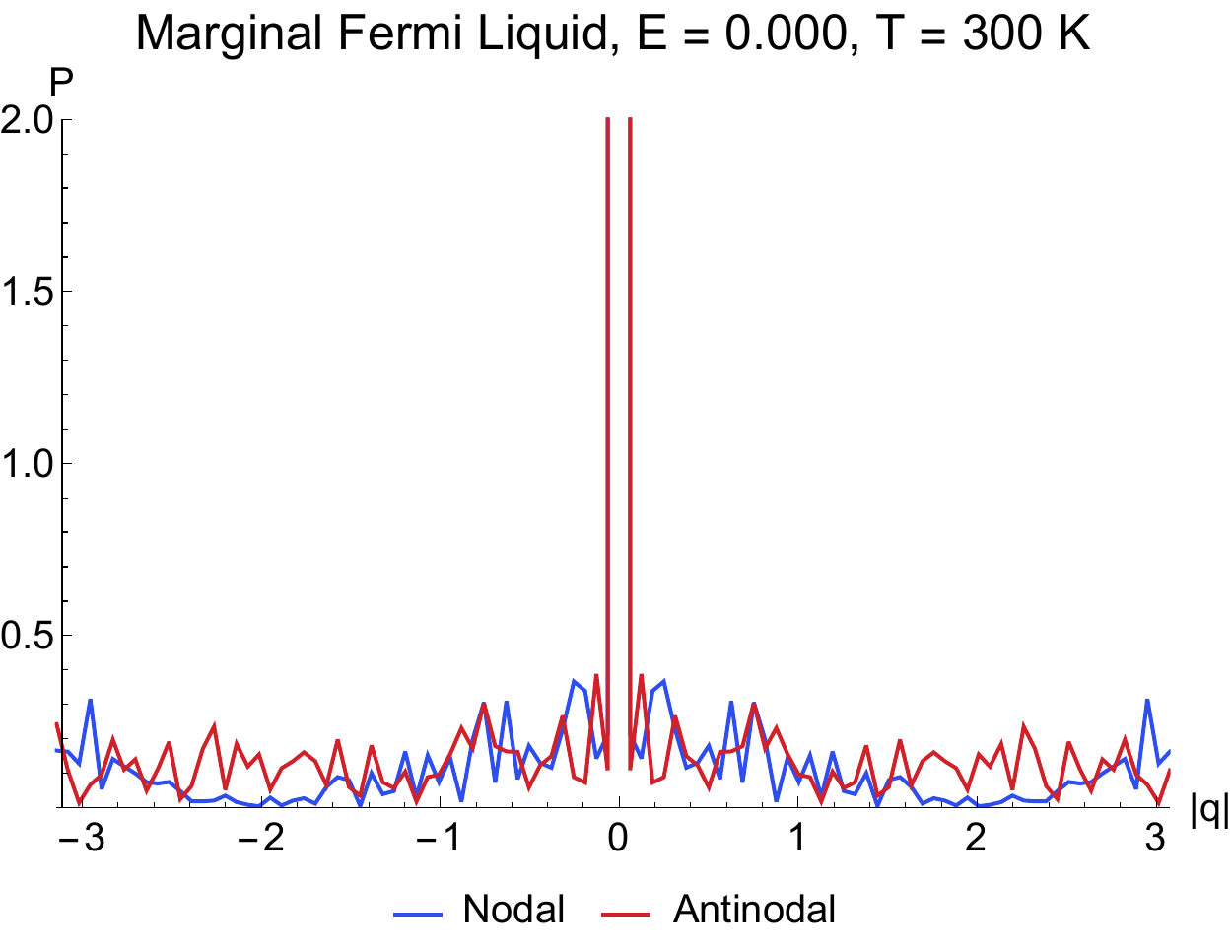} \\
		
	\caption{Marginal Fermi liquid phenomenology at various temperatures.  Left to right: The spectral function $A(\mathbf{k}, \omega)$; the Fourier transform of the LDOS $P(\mathbf{q}, \omega)$; linecuts of $P(\mathbf{q}, \omega)$ in the nodal and antinodal directions; $P(\mathbf{q}, \omega)$ in the presence of multiple weak impurities and finite-temperature smearing; and linecuts of $P(\mathbf{q}, \omega)$ in the presence of multiple weak impurities and finite-temperature smearing. All plots are taken at $E = 0.000$.}

	\label{fig:temperature_mfl}
\end{figure*}

\begin{figure*}
	\centering
	
	\includegraphics[width=0.16\textwidth]{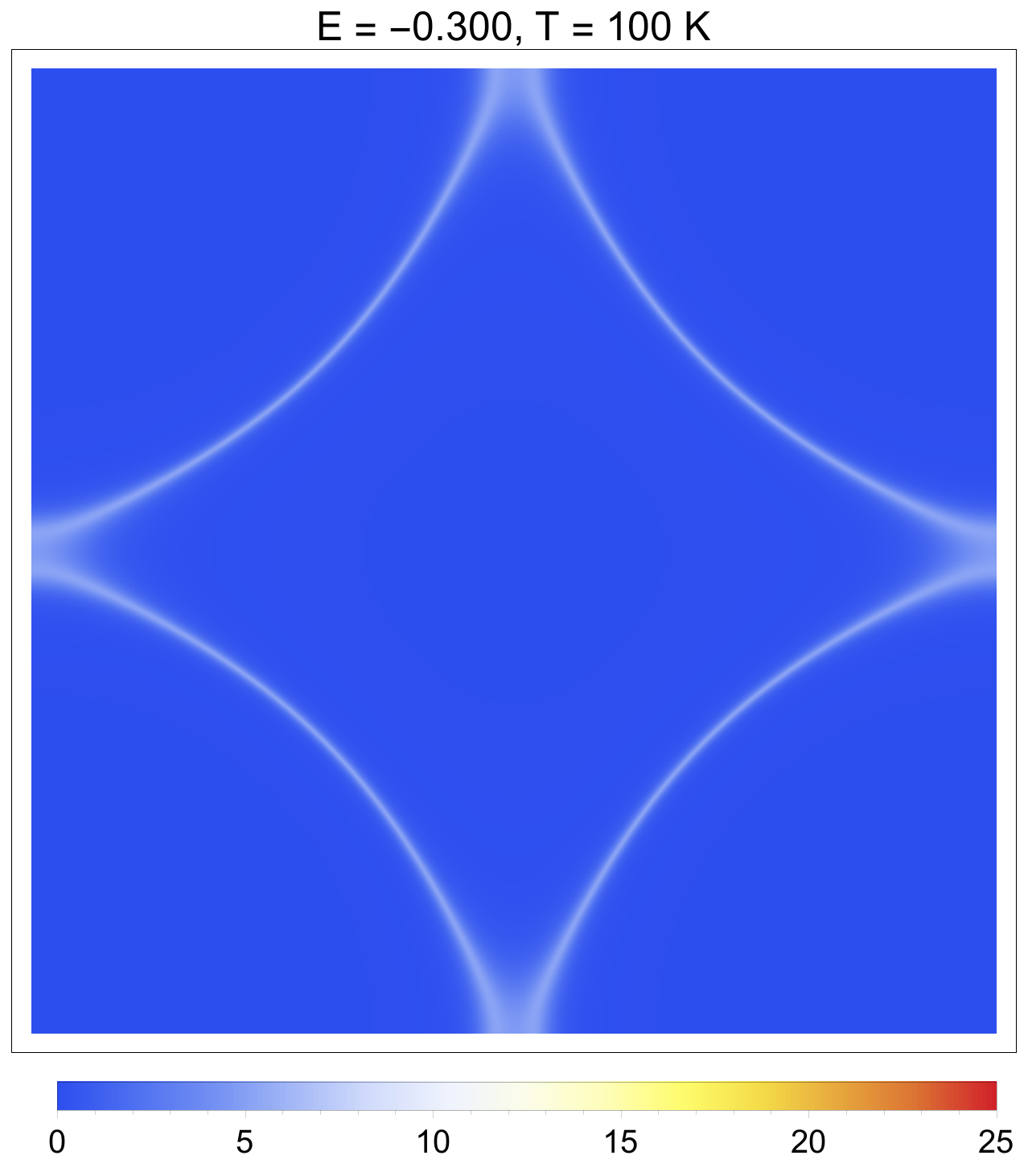}
	\includegraphics[width=0.16\textwidth]{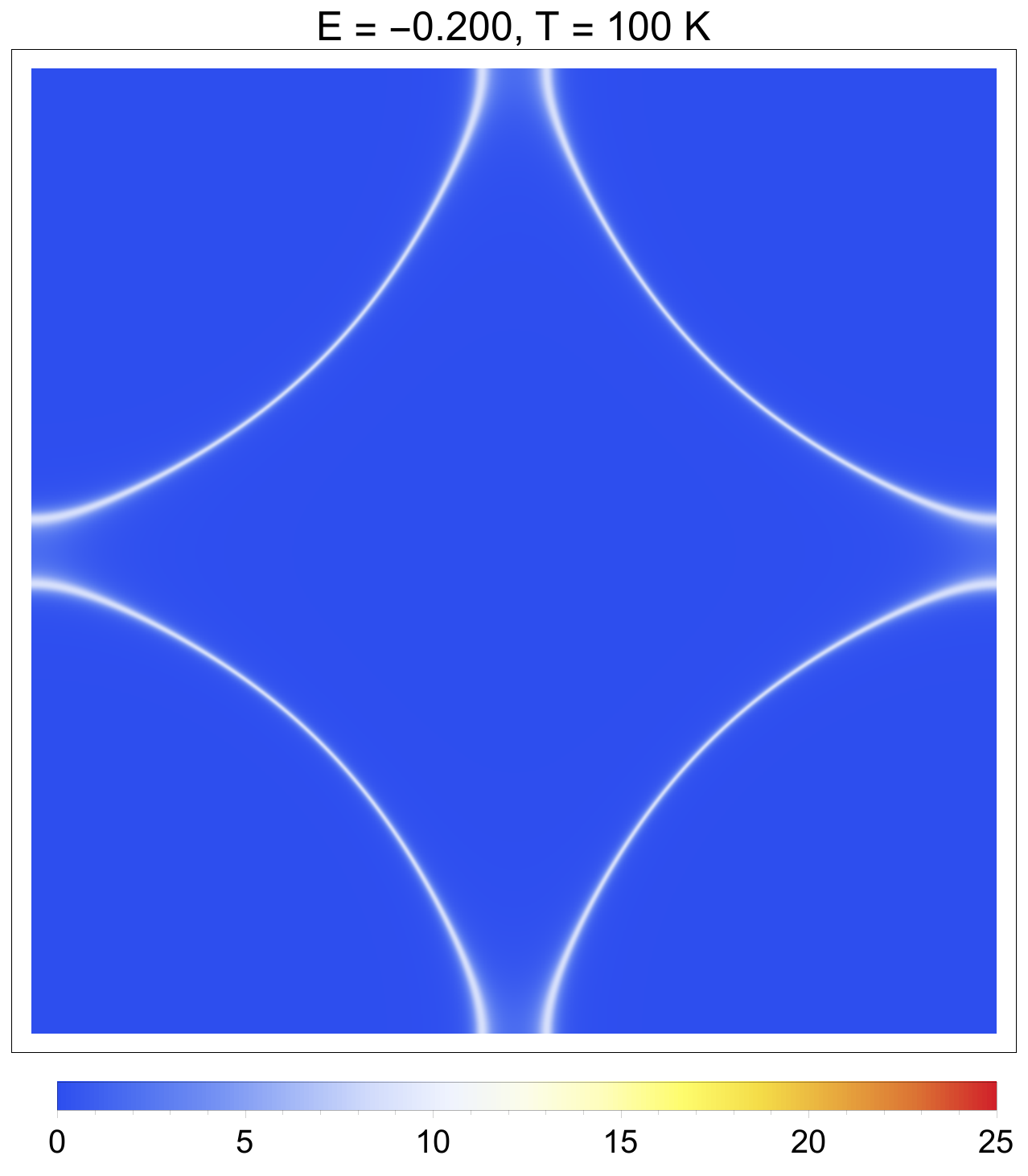}
	\includegraphics[width=0.16\textwidth]{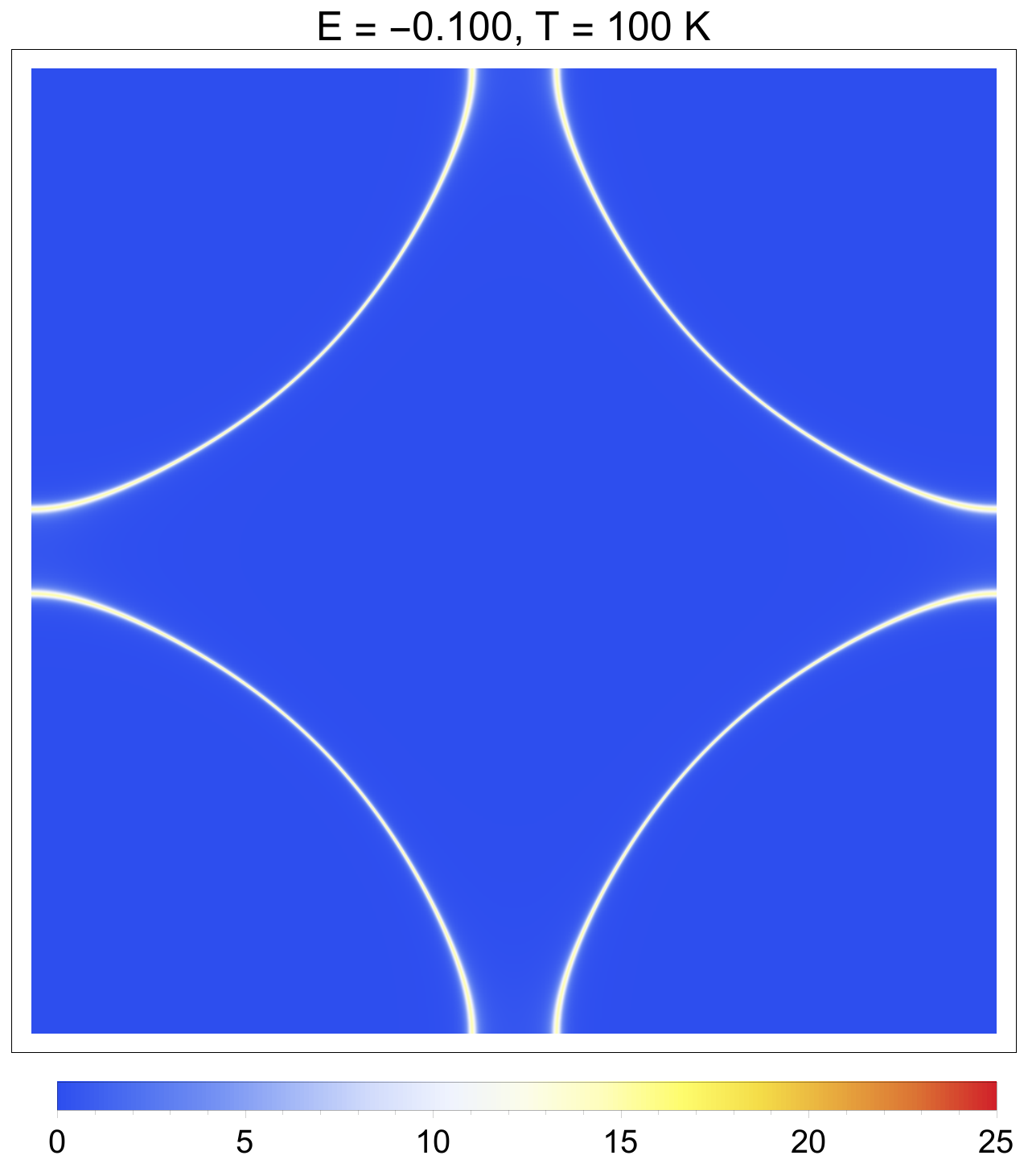}
	\includegraphics[width=0.16\textwidth]{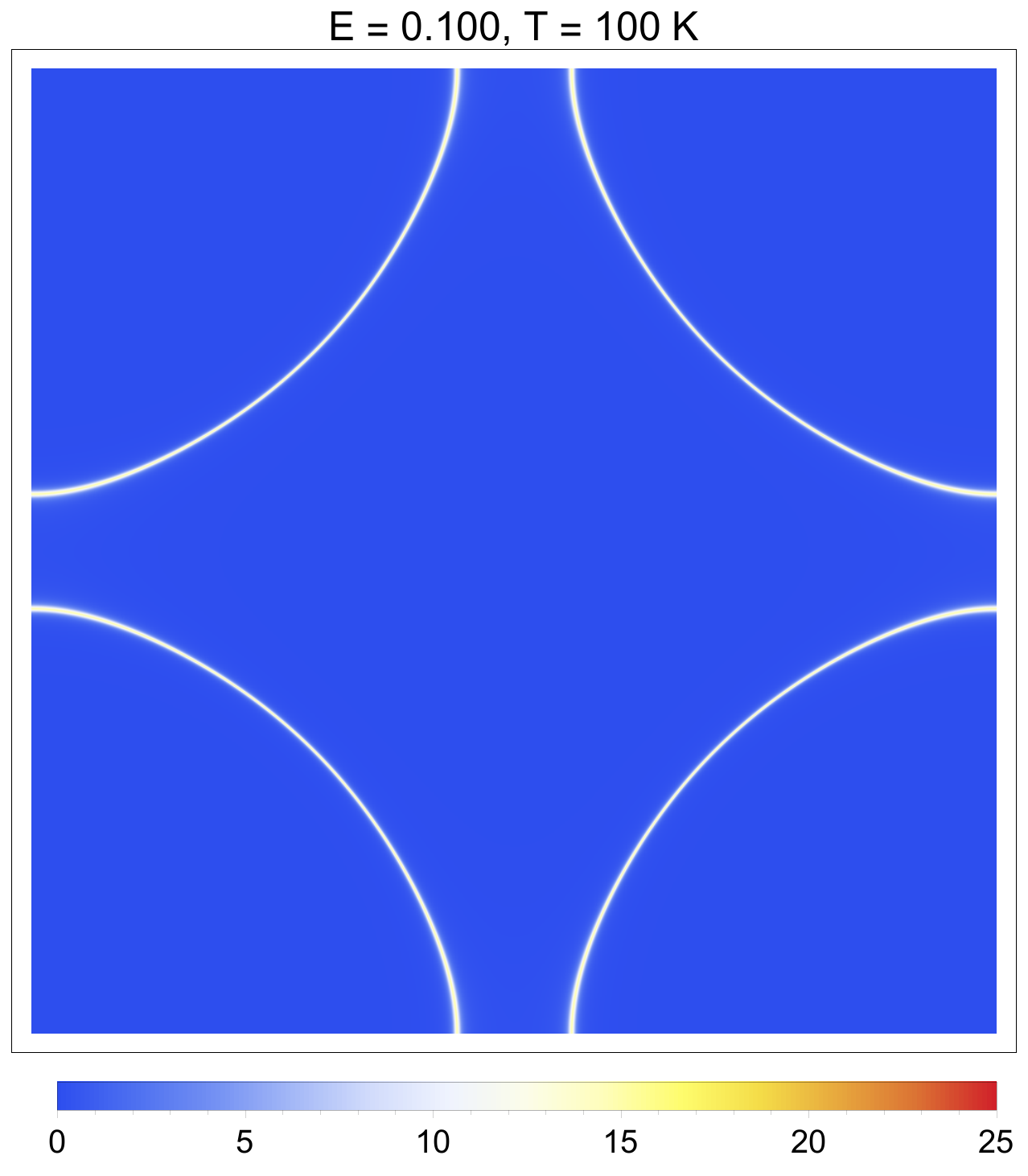}
	\includegraphics[width=0.16\textwidth]{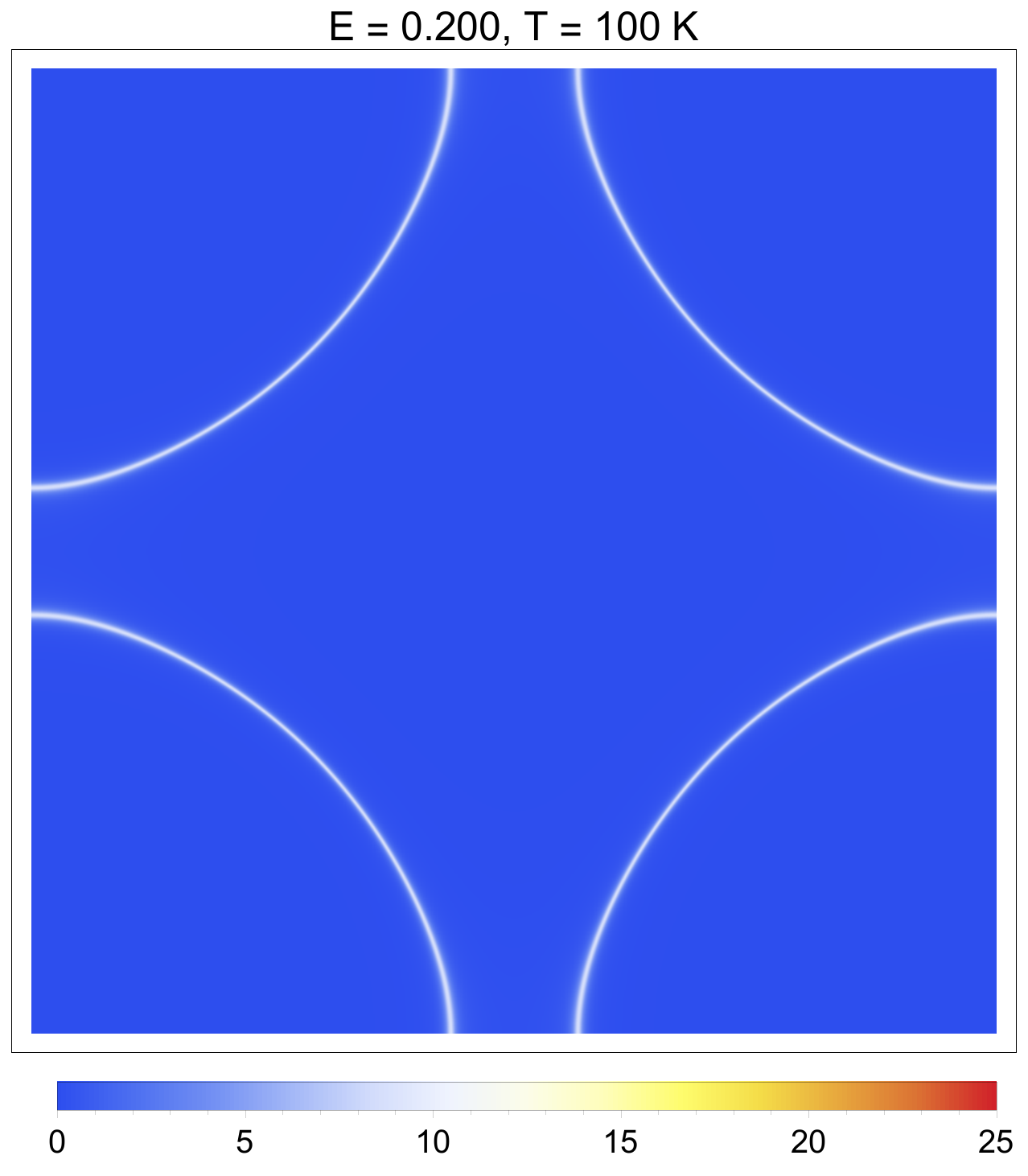}
	\includegraphics[width=0.16\textwidth]{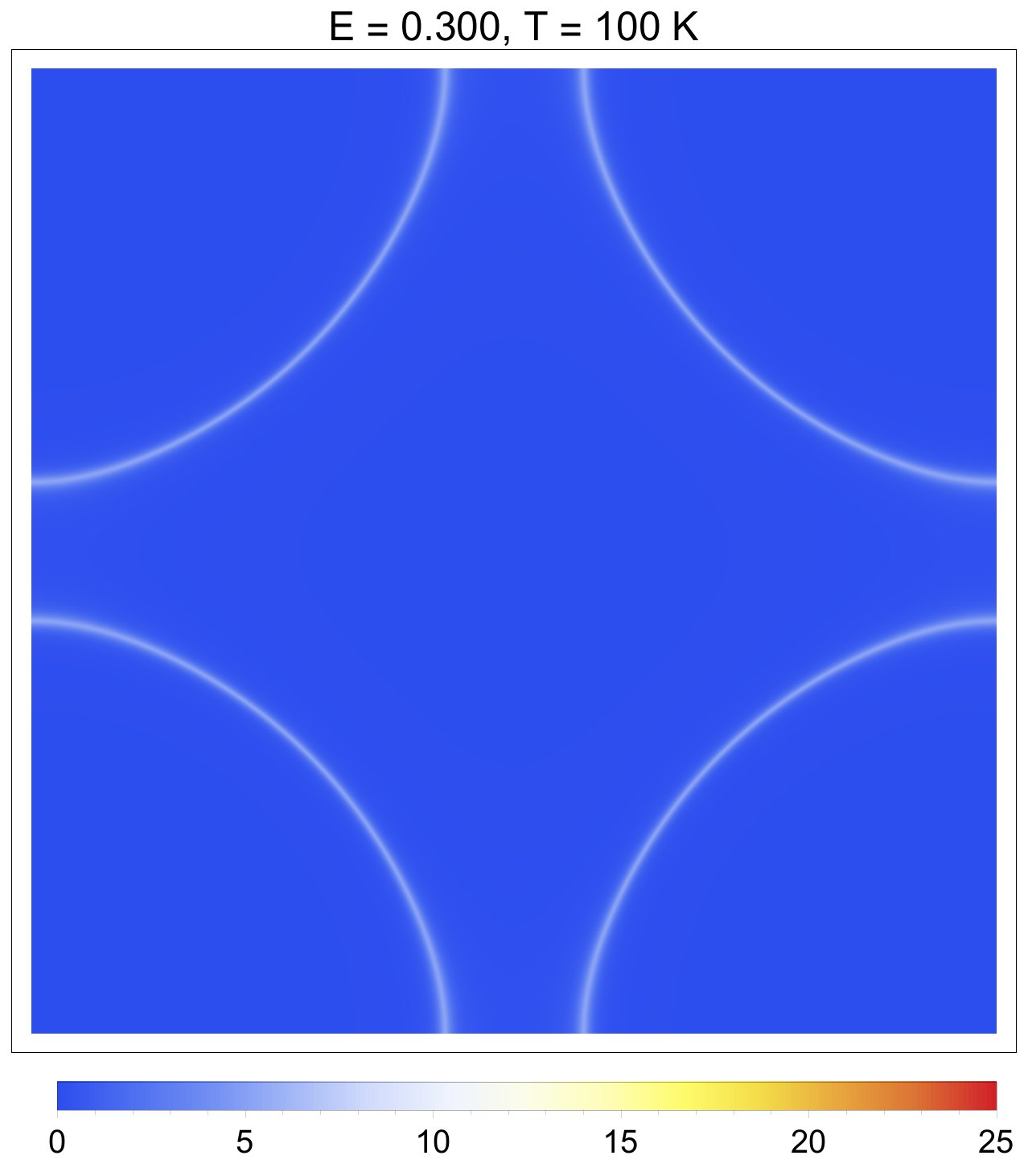} \\
	\includegraphics[width=0.16\textwidth]{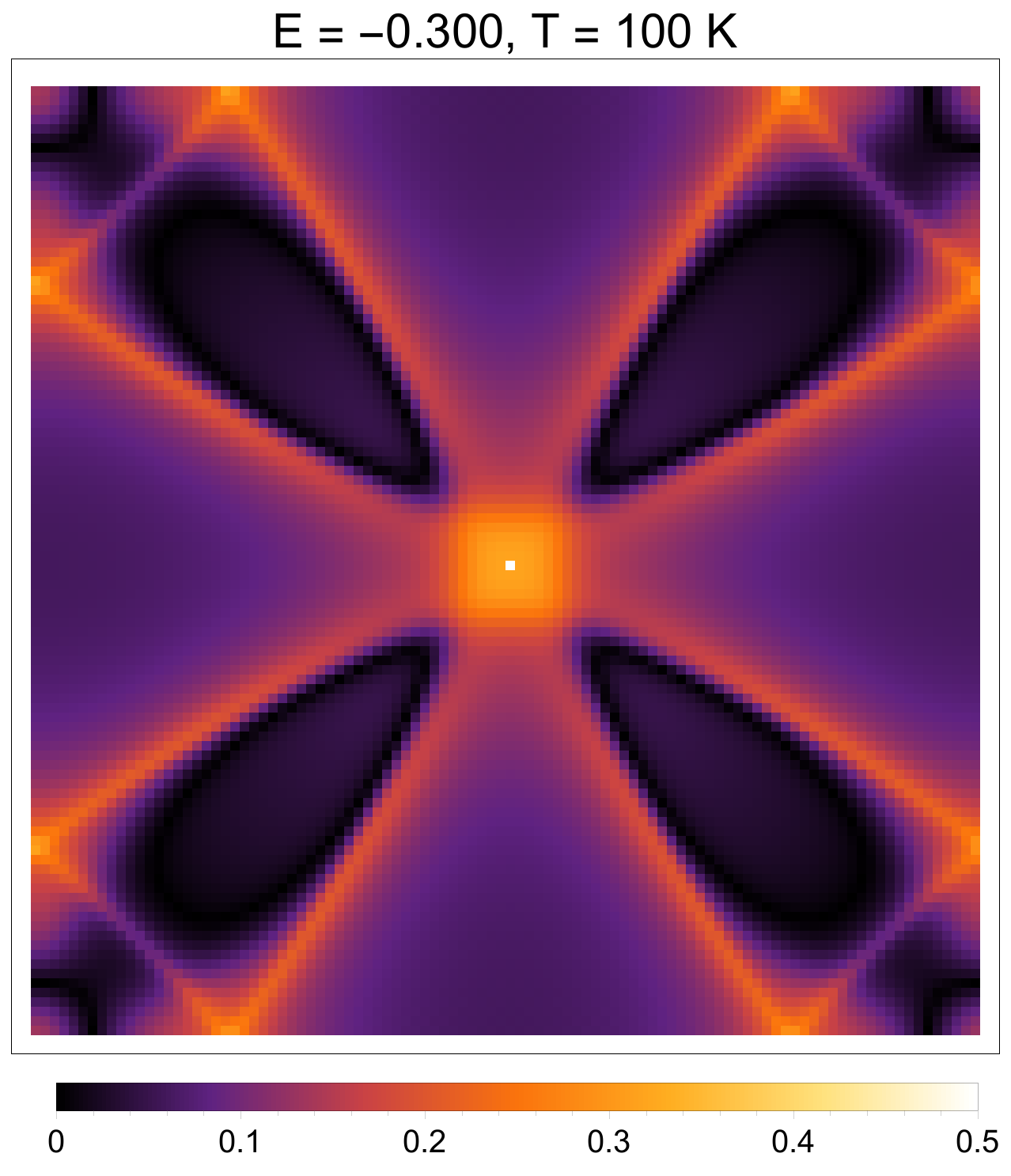}
	\includegraphics[width=0.16\textwidth]{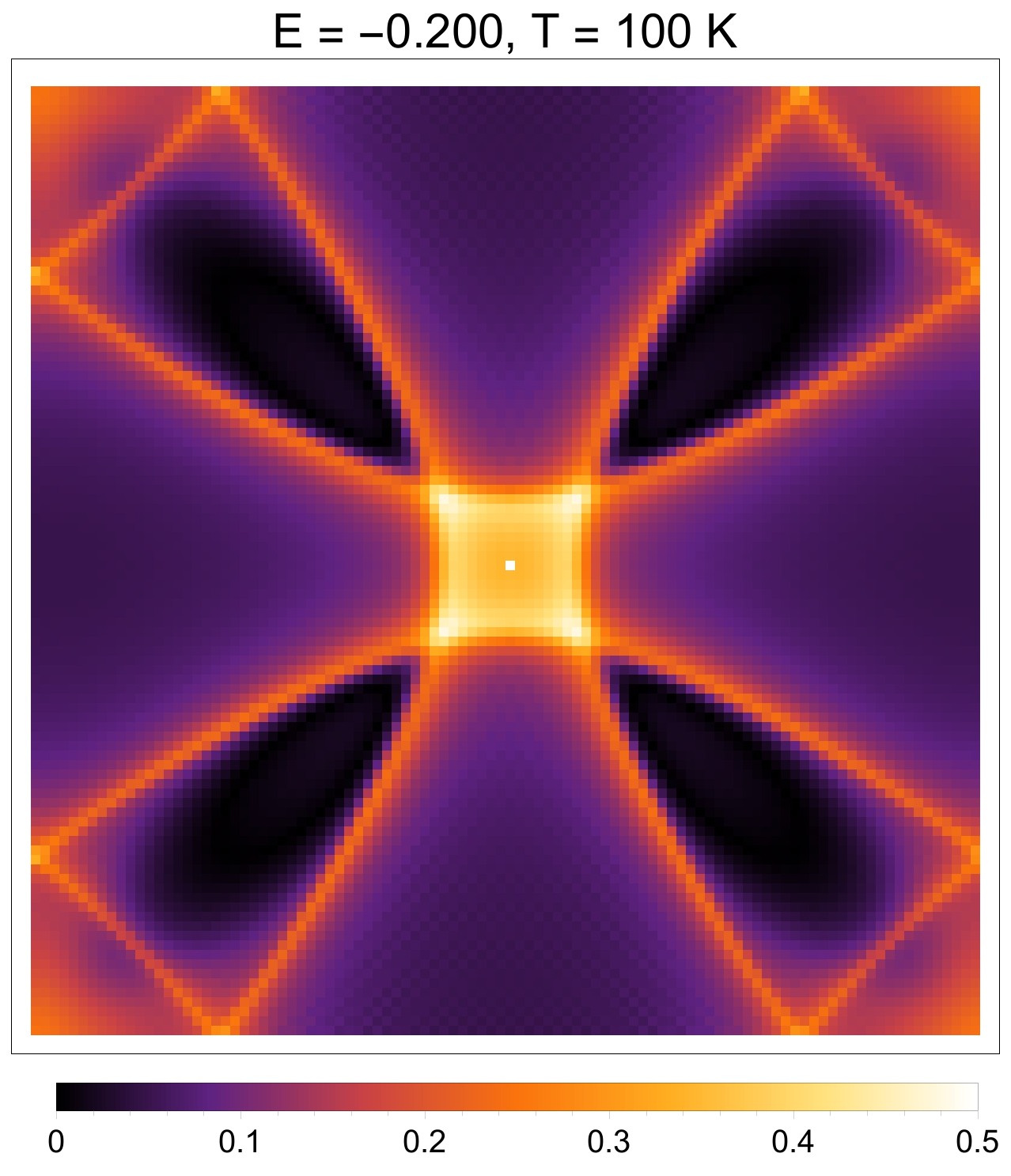}
	\includegraphics[width=0.16\textwidth]{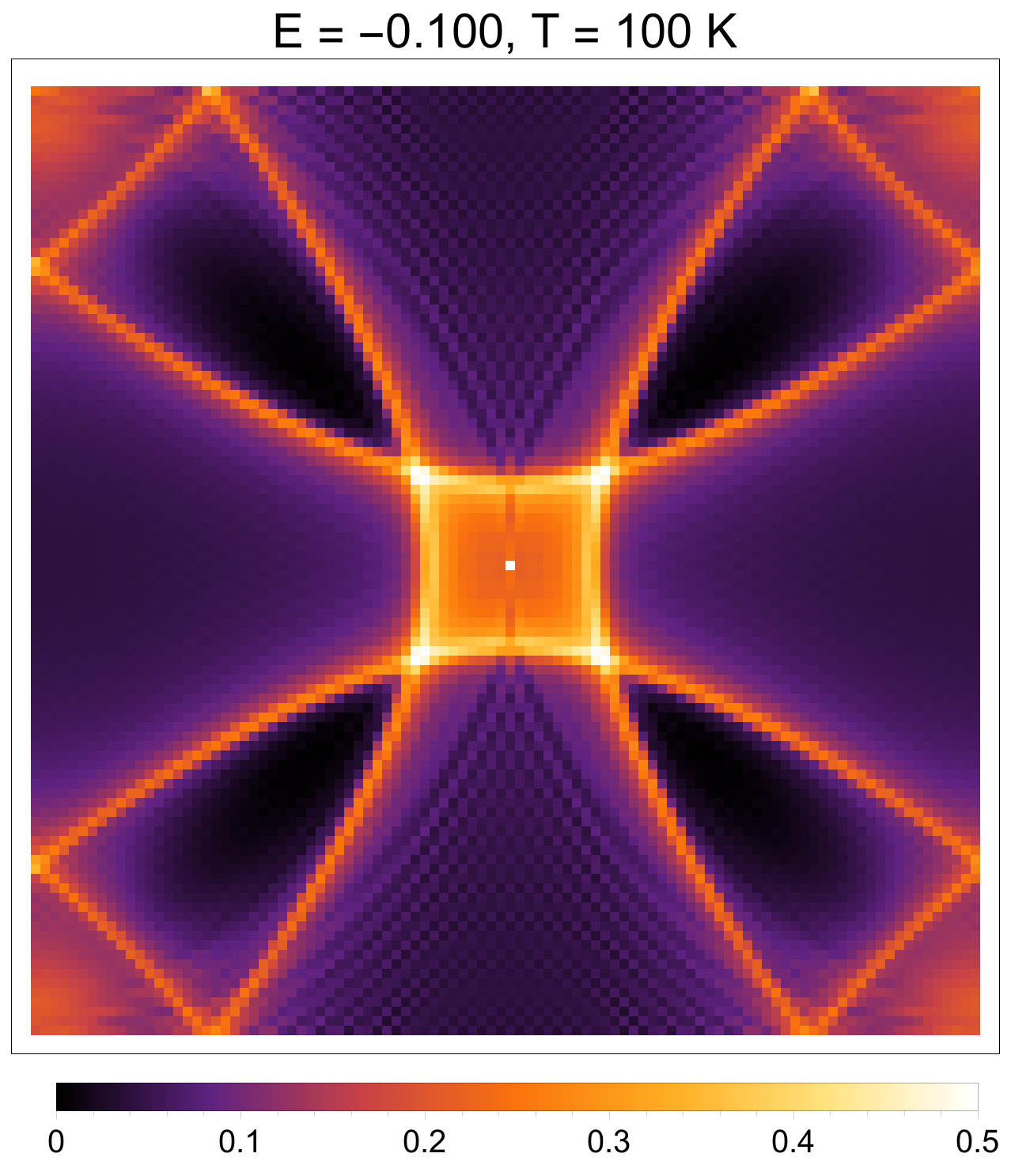}
	\includegraphics[width=0.16\textwidth]{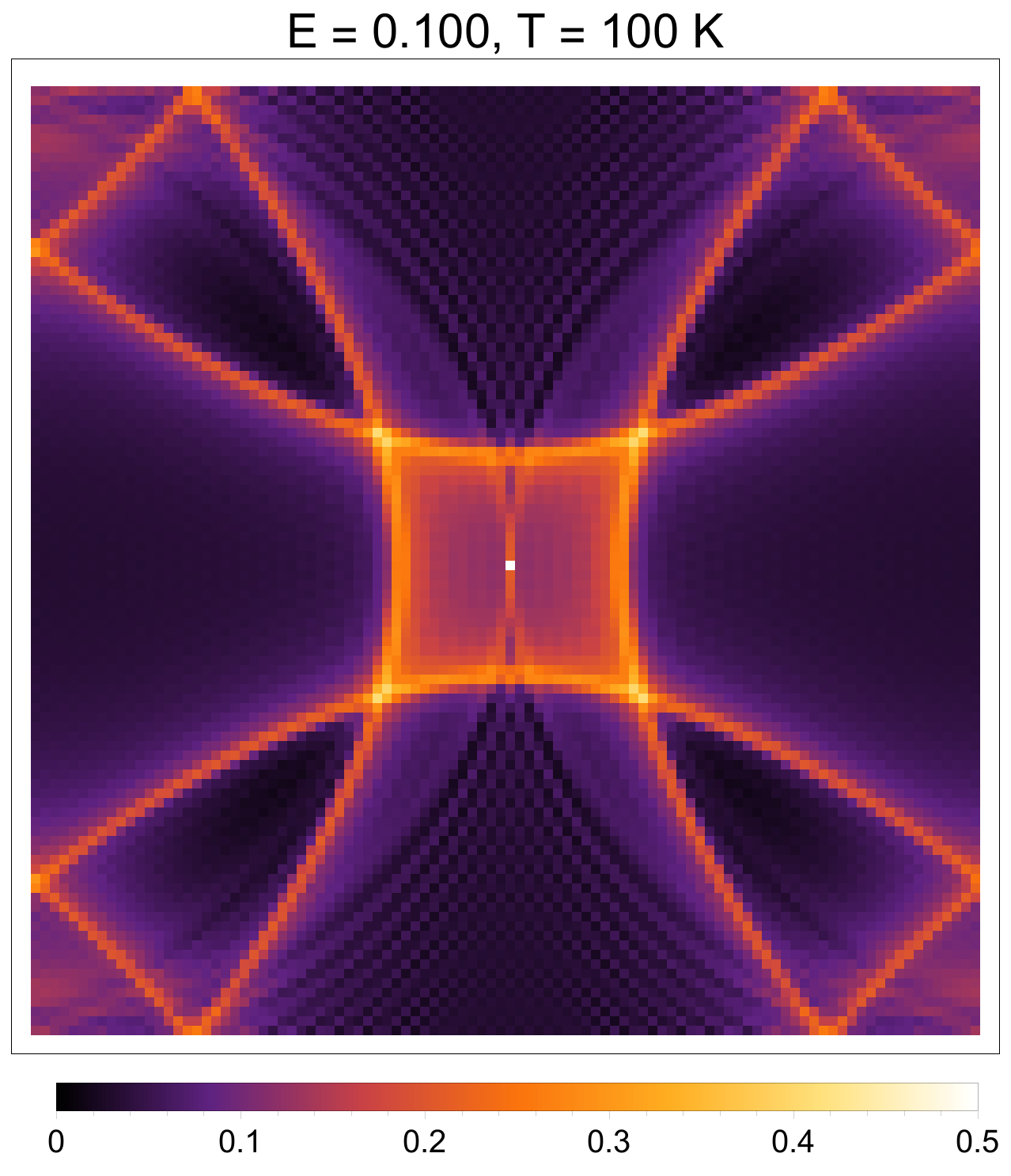}
	\includegraphics[width=0.16\textwidth]{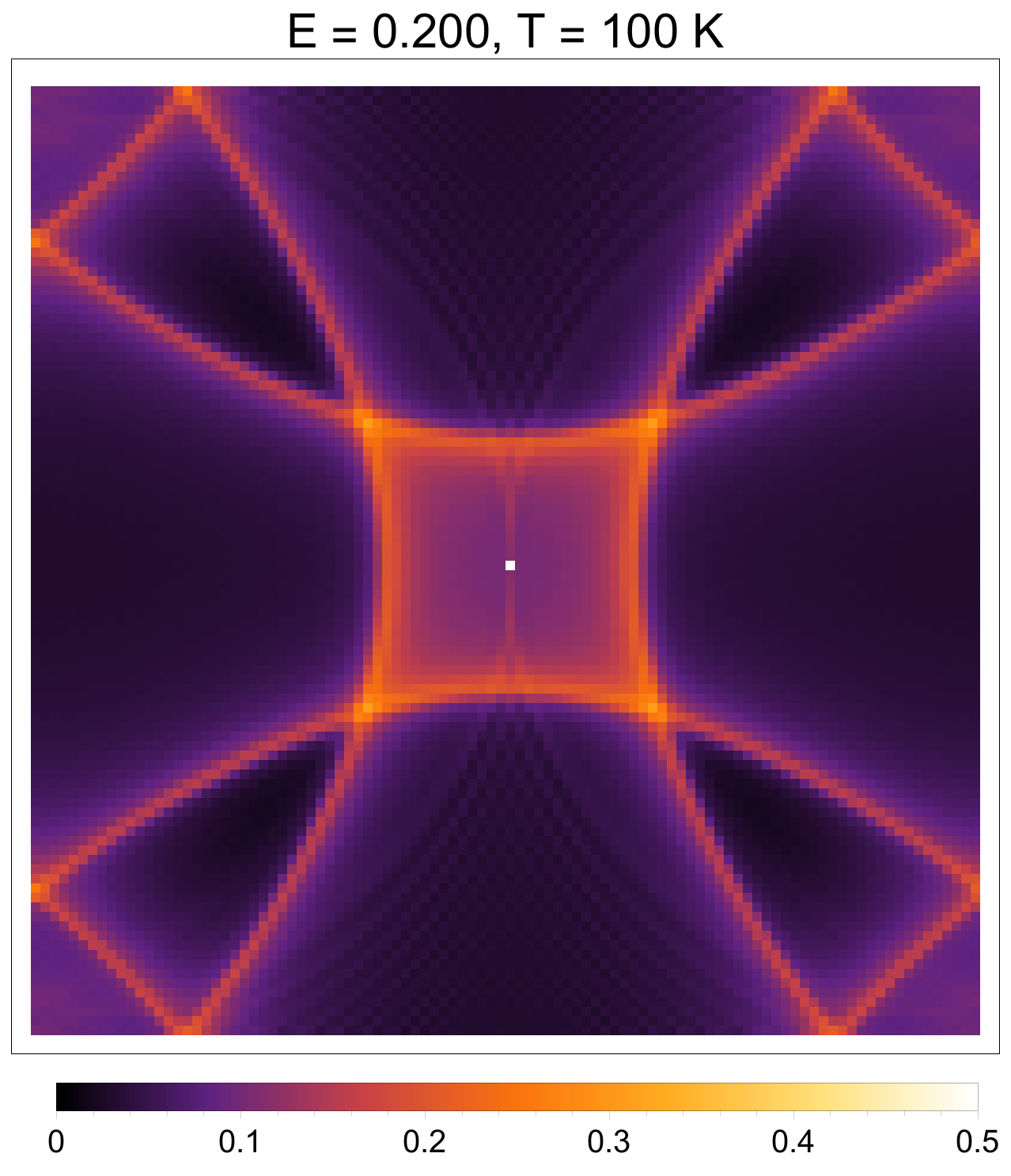}
	\includegraphics[width=0.16\textwidth]{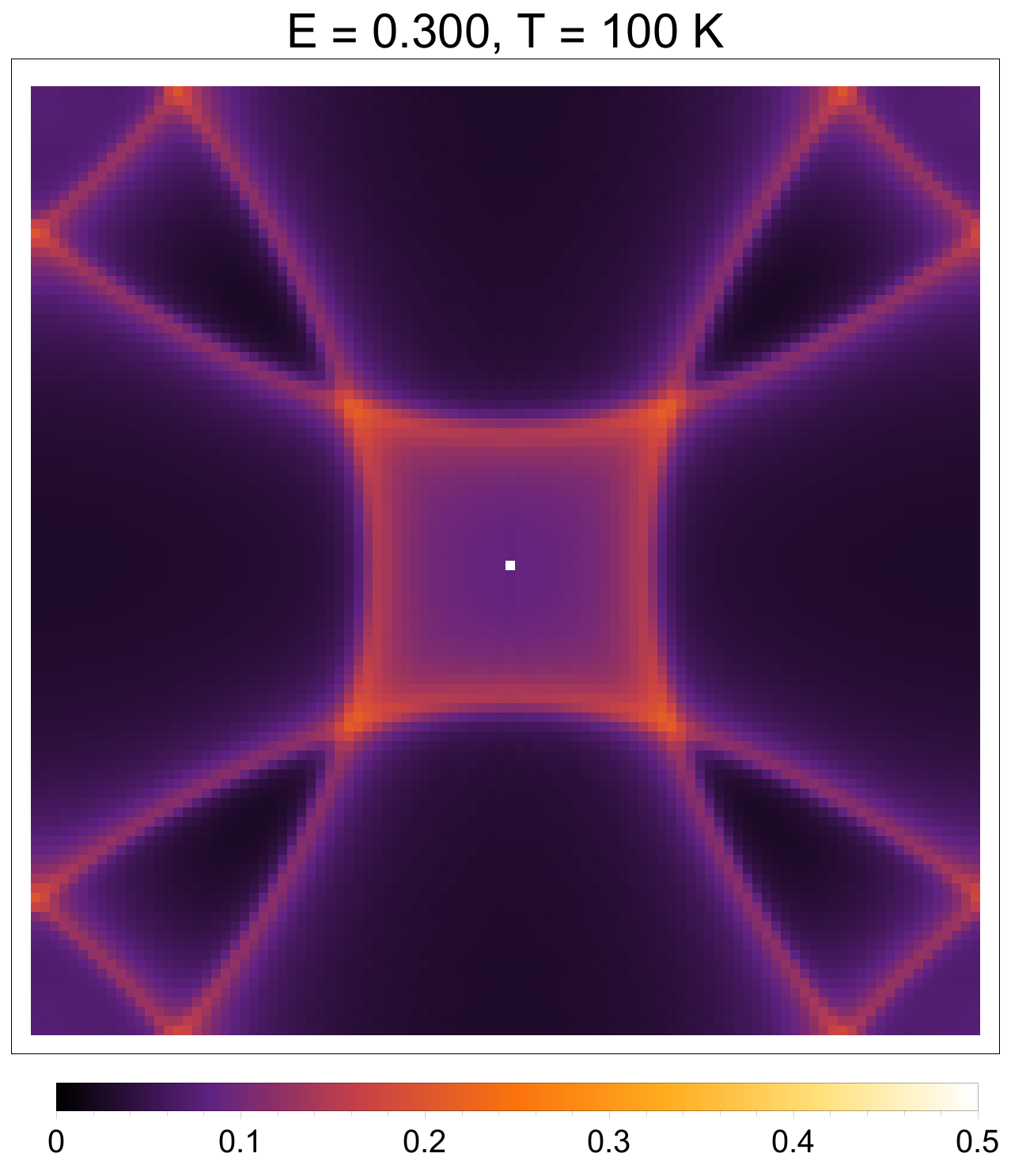}\\
	\includegraphics[width=0.16\textwidth]{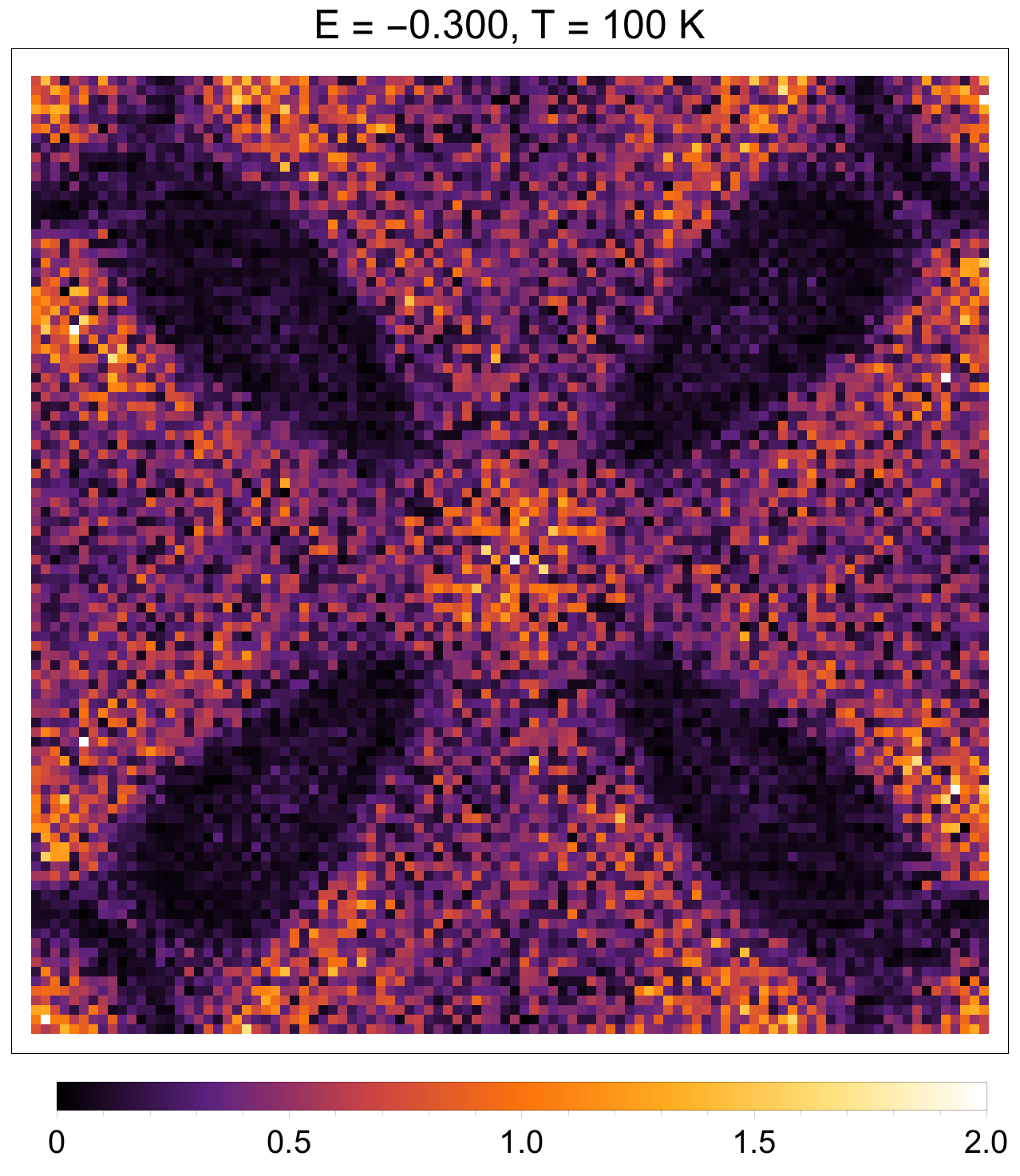}
	\includegraphics[width=0.16\textwidth]{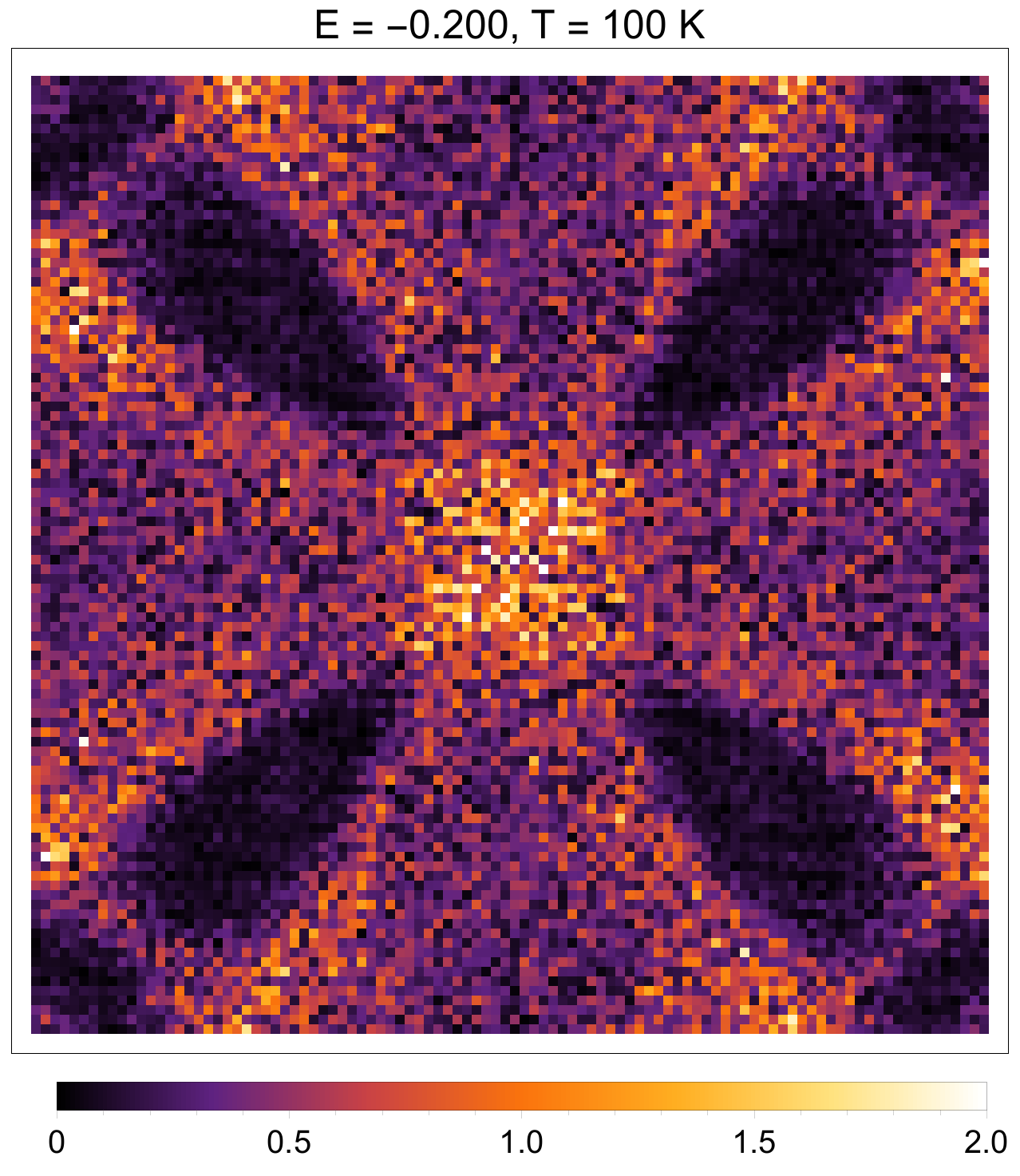}
	\includegraphics[width=0.16\textwidth]{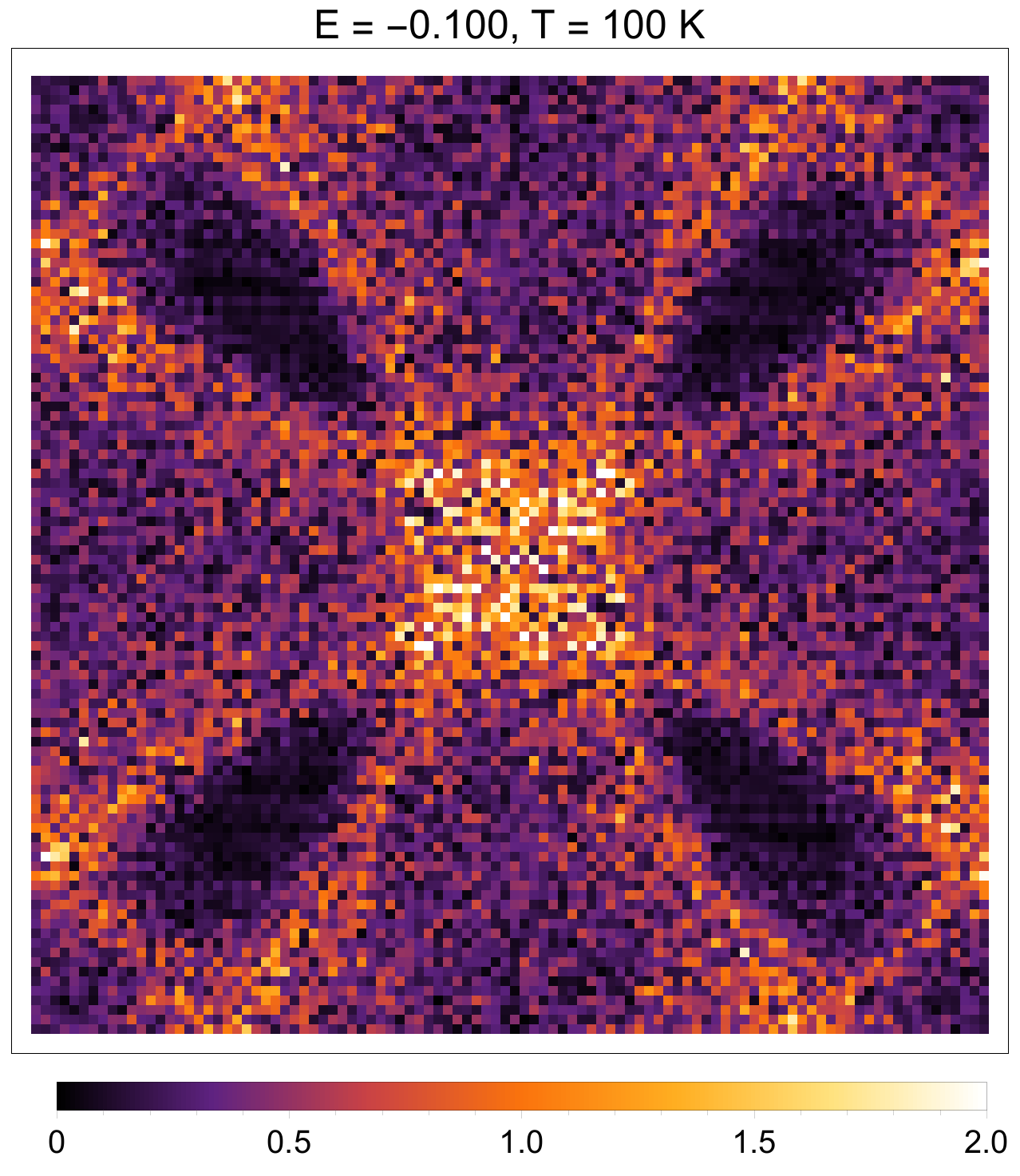}
	\includegraphics[width=0.16\textwidth]{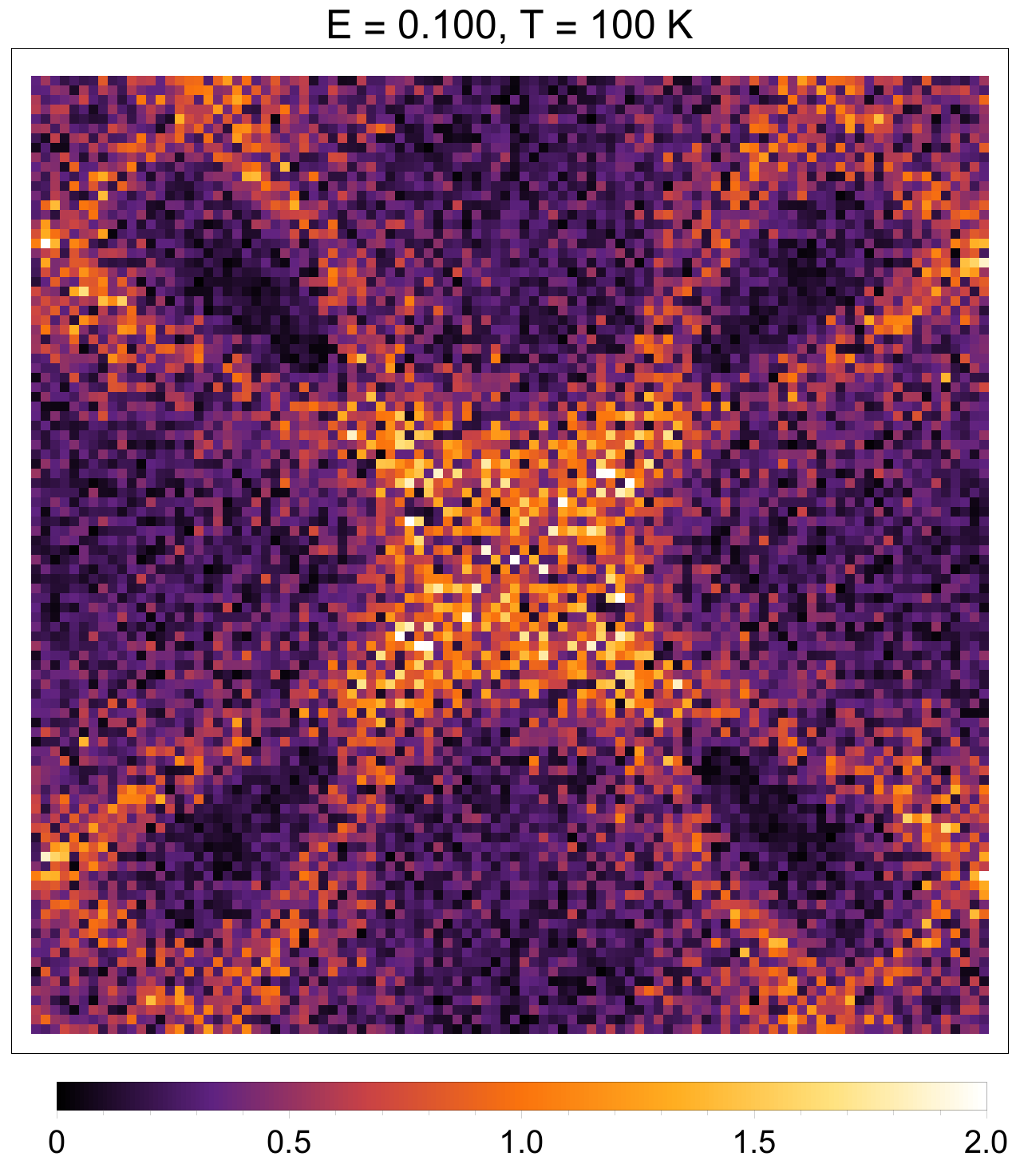}
	\includegraphics[width=0.16\textwidth]{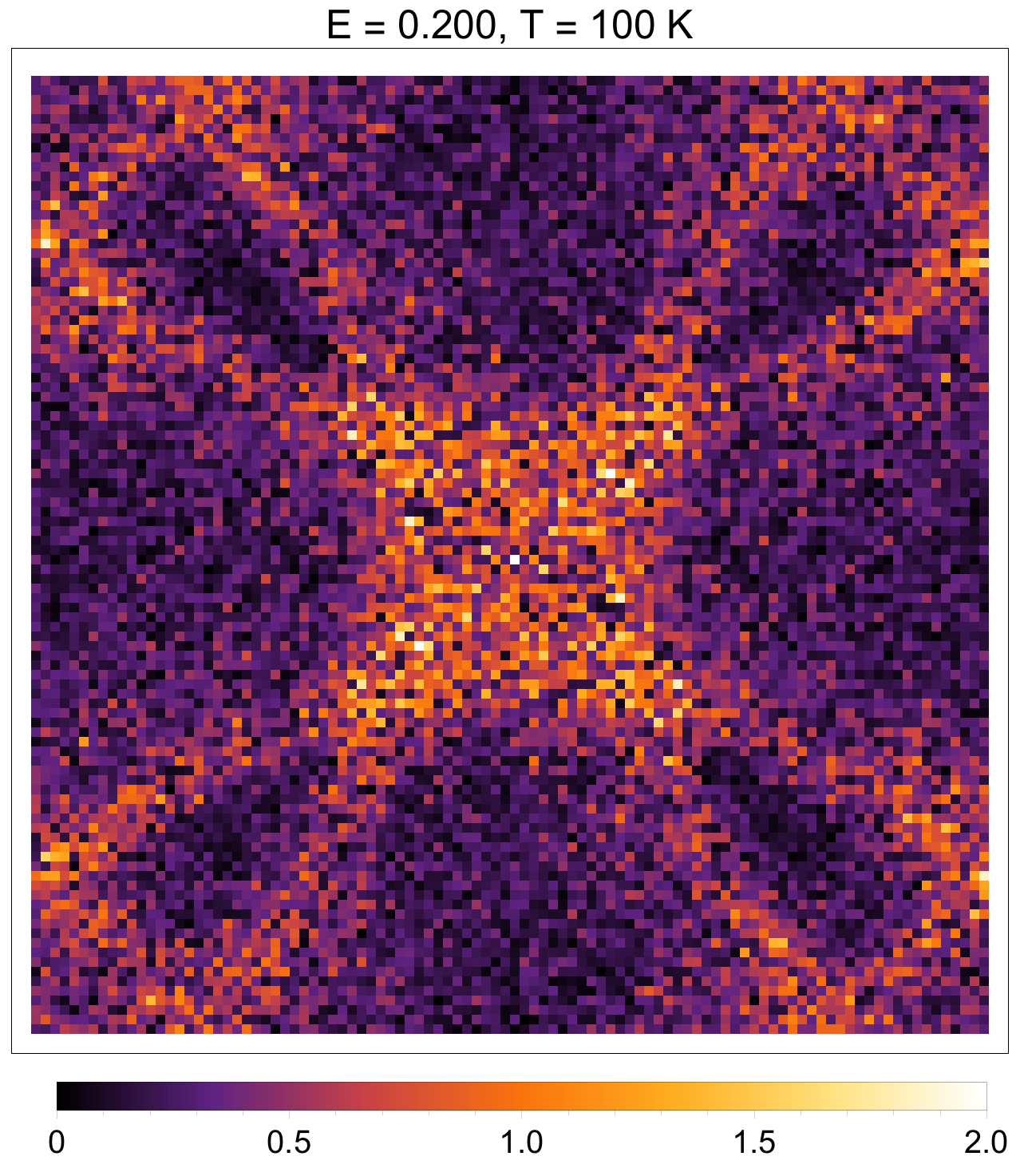}
	\includegraphics[width=0.16\textwidth]{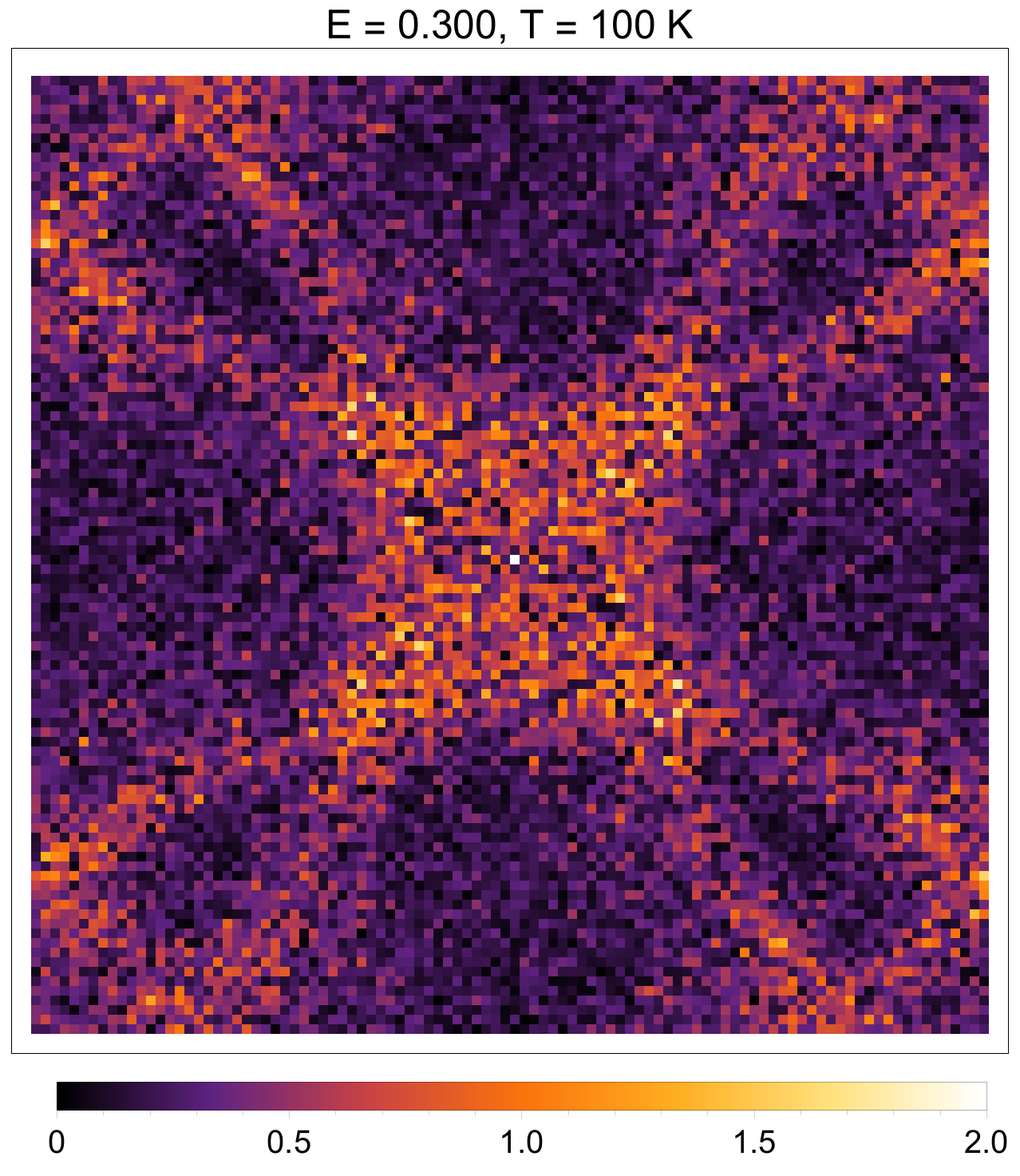}
	
	\caption{Frequency-dependence at $T = 100$ K of the spectra of an ordinary Fermi liquid. Shown are plots of the spectral function $A(\mathbf{k}, \omega)$ (upper row); the LDOS power spectrum with a single pointlike scatterer without thermal smearing (middle row); and the LDOS power spectrum with both a 0.5\% concentration of pointlike scatterers and thermal smearing (bottom row). Note that the scales used for plotting the LDOS power spectra are the same for all frequencies.}
	\label{fig:frequency_fl_100k}
\end{figure*}

\begin{figure*}
	\centering
	
	\includegraphics[width=0.16\textwidth]{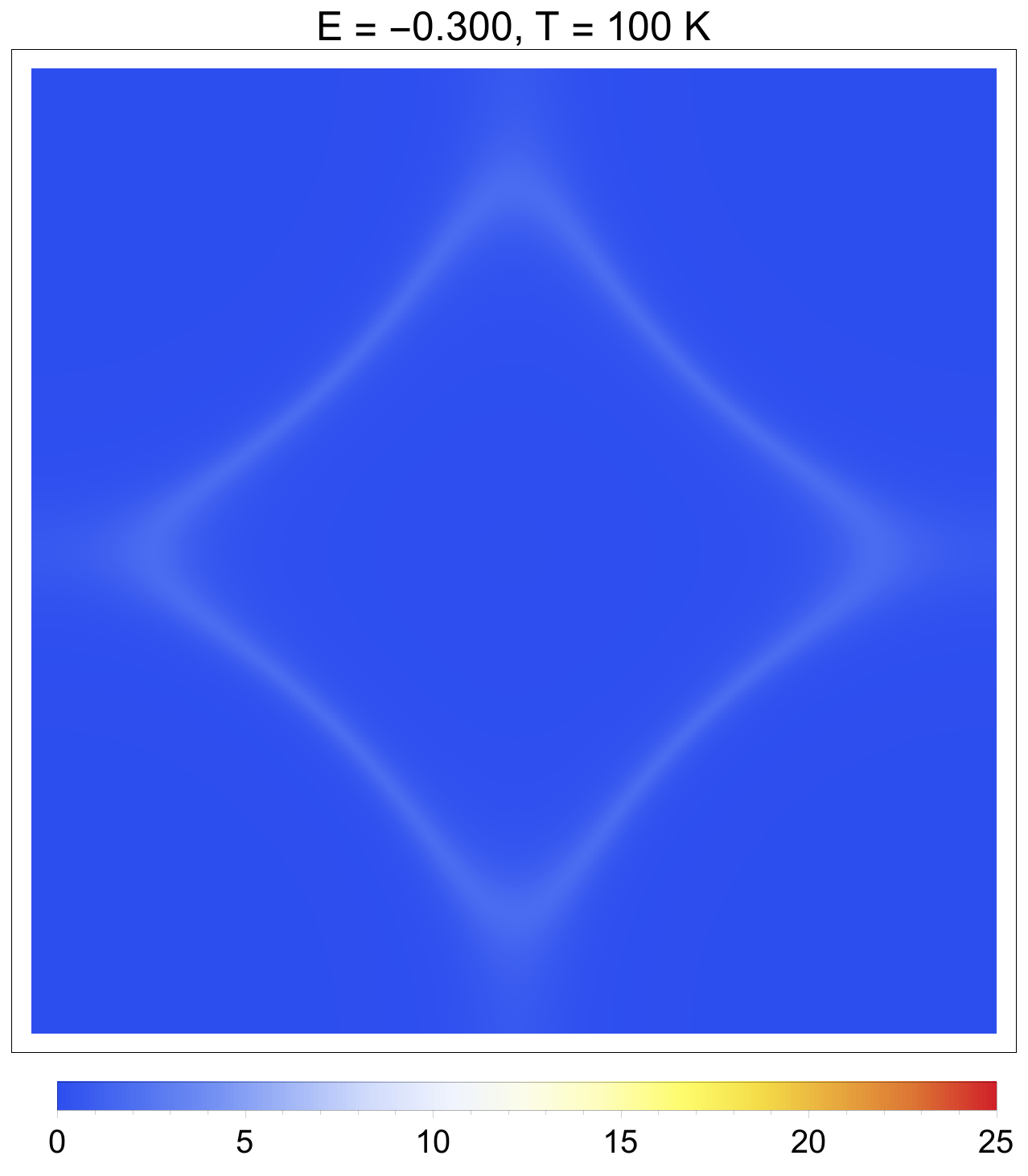}
	\includegraphics[width=0.16\textwidth]{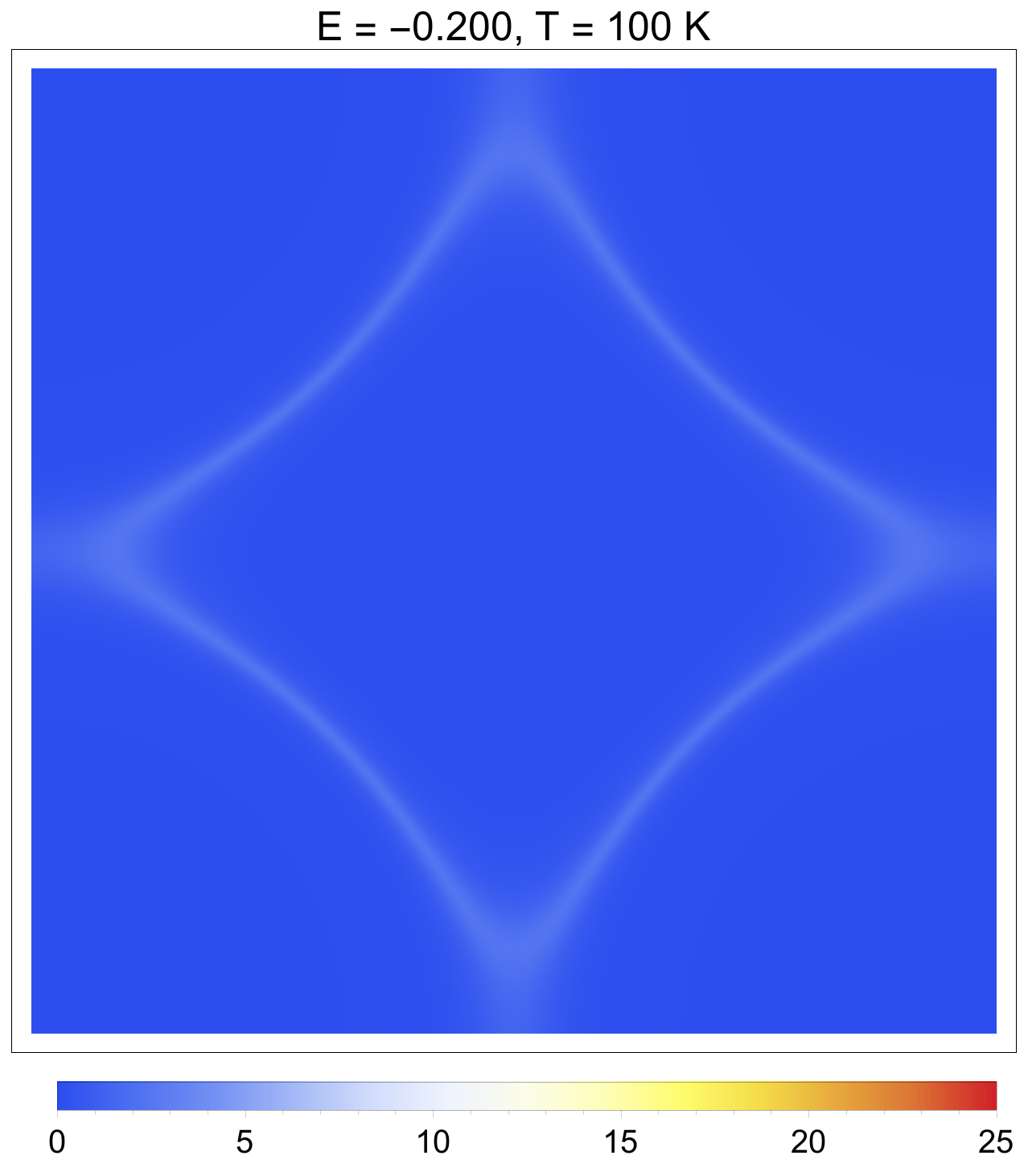}
	\includegraphics[width=0.16\textwidth]{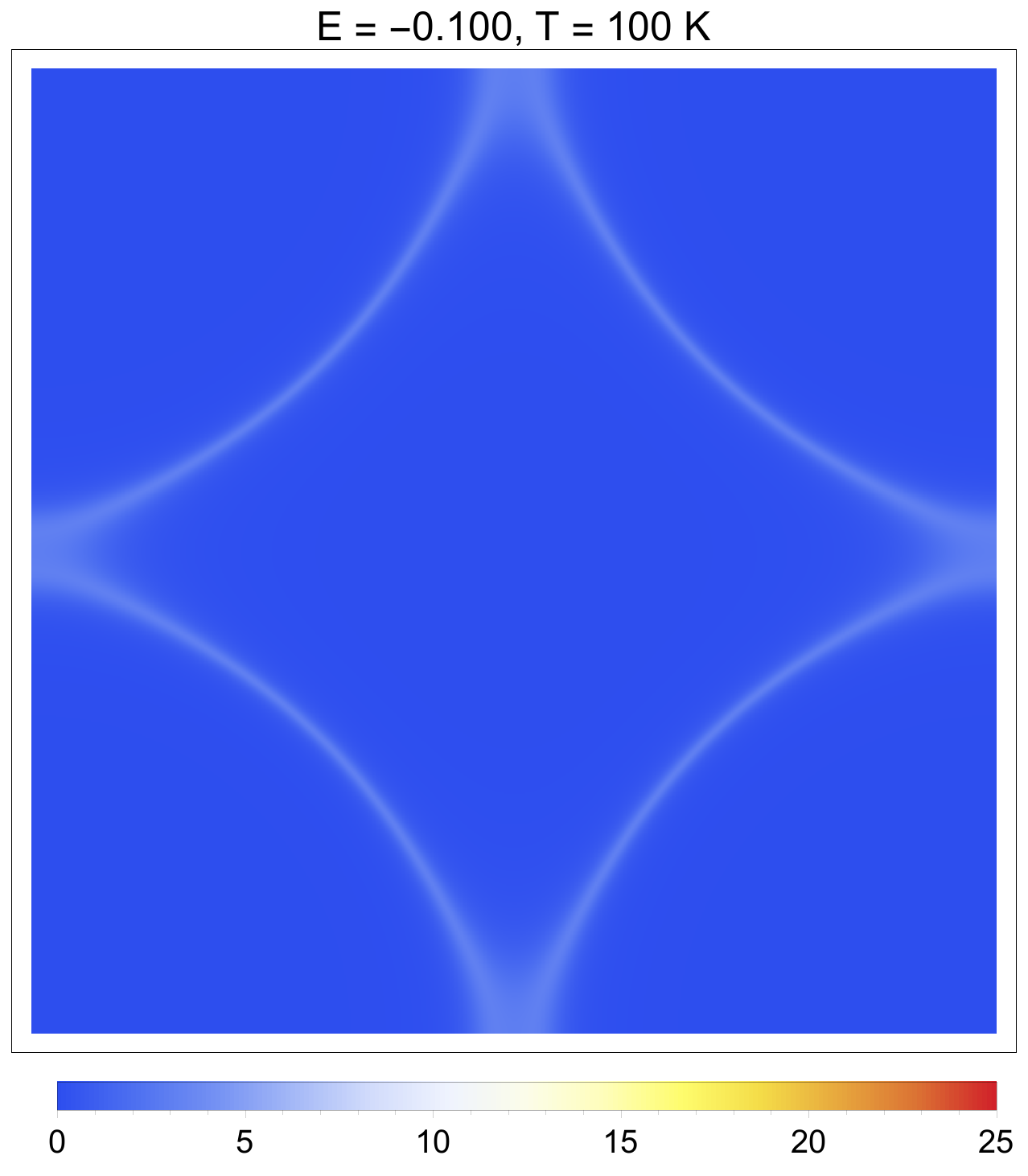}
	\includegraphics[width=0.16\textwidth]{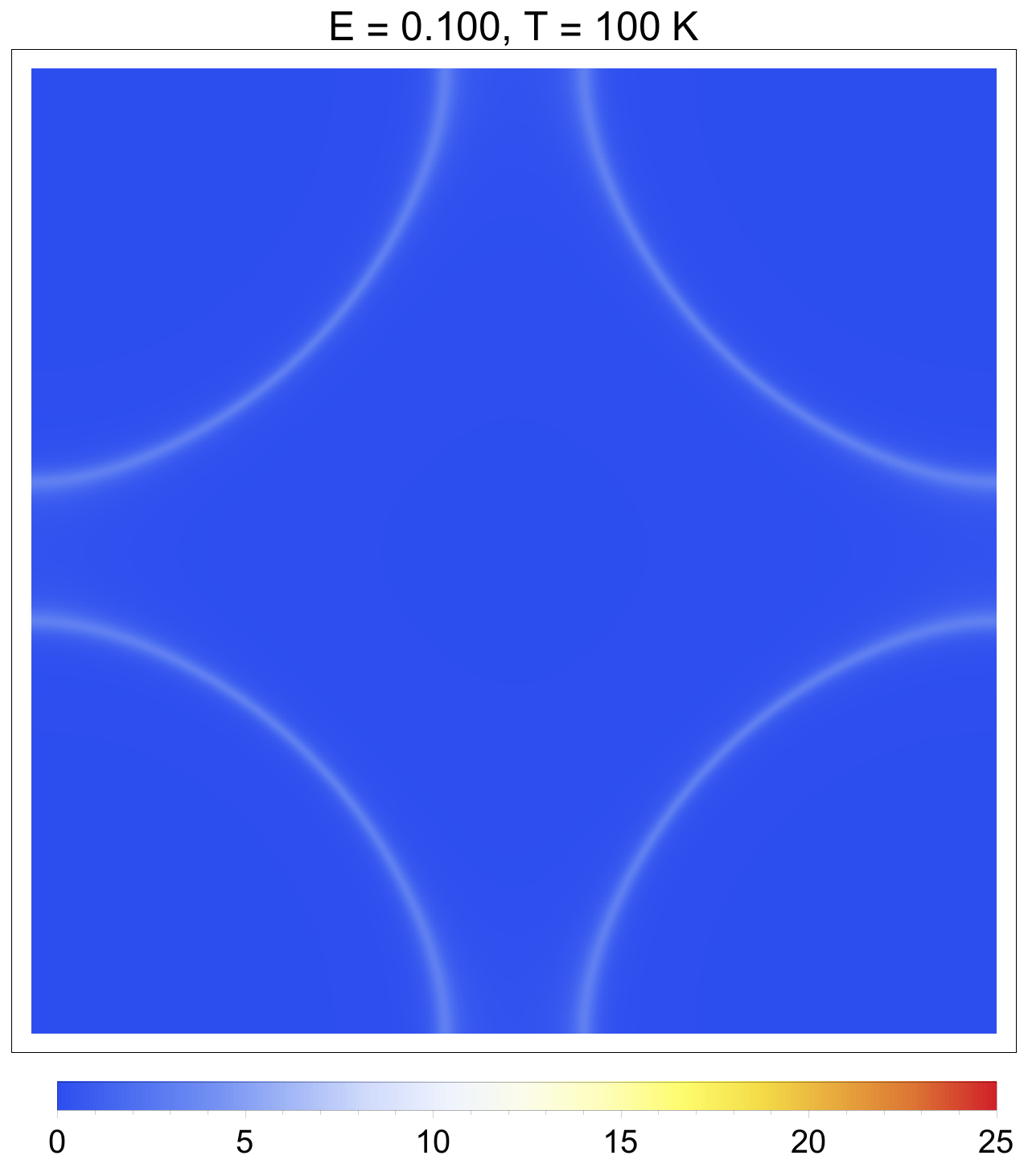}
	\includegraphics[width=0.16\textwidth]{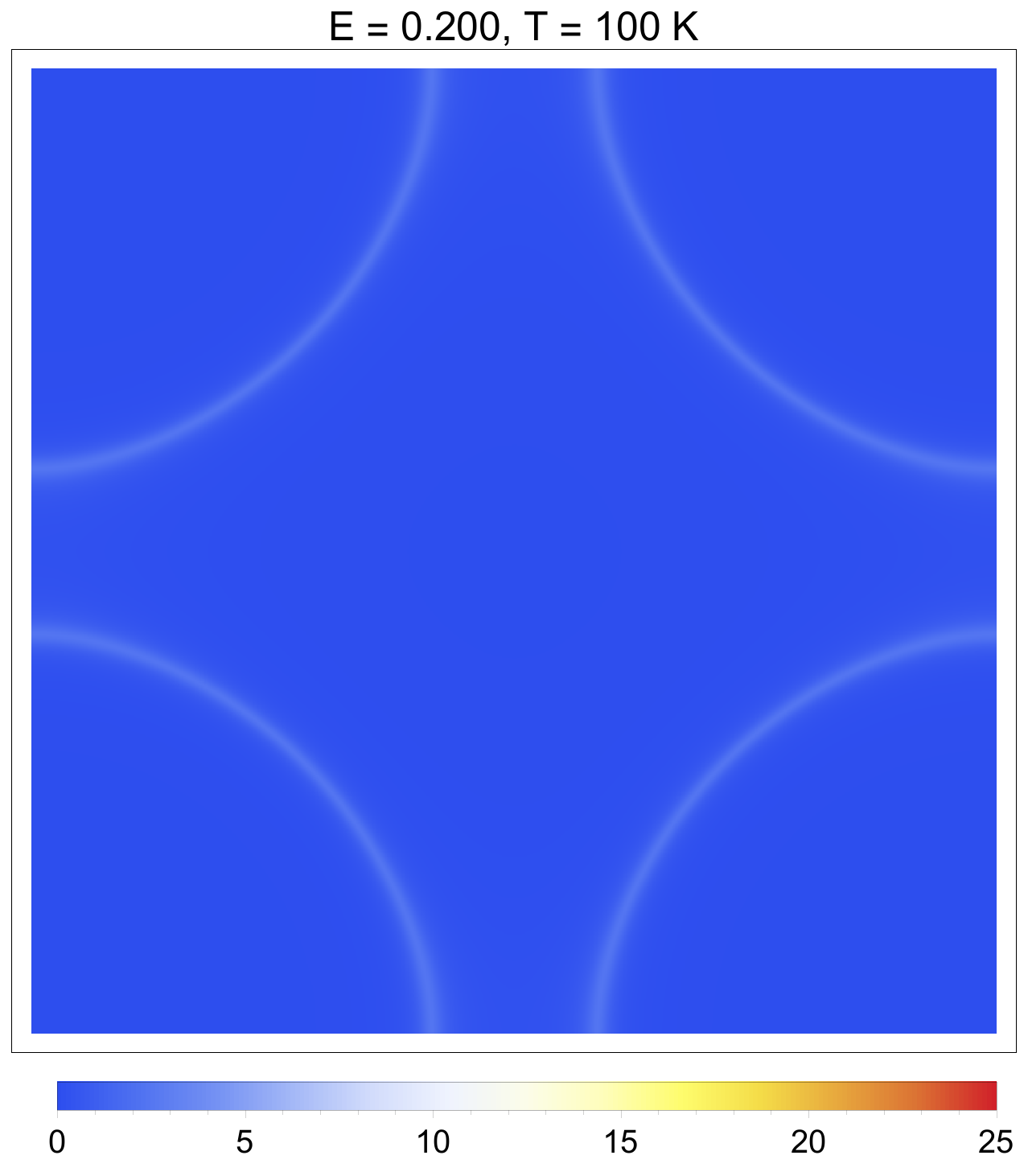}
	\includegraphics[width=0.16\textwidth]{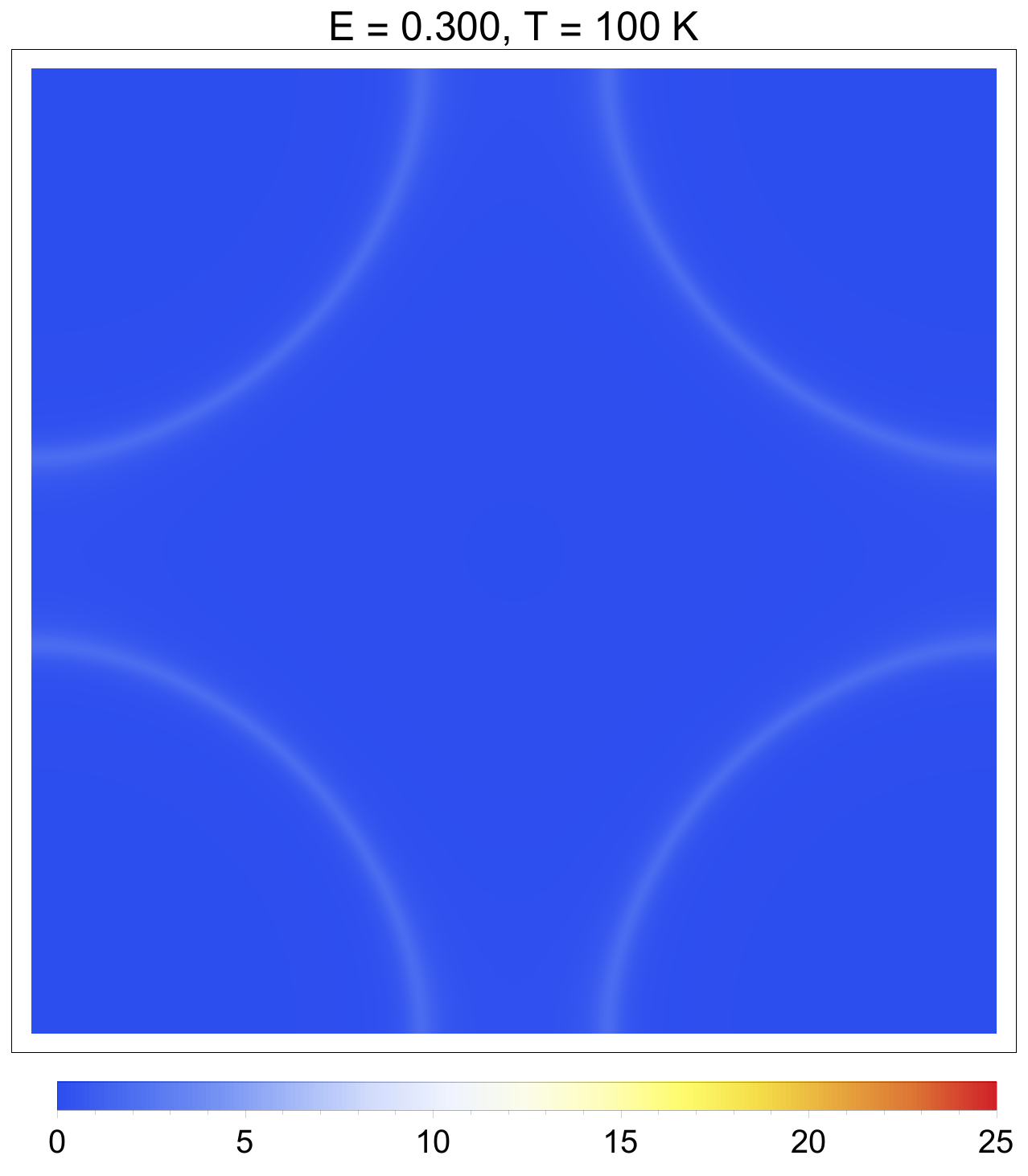} \\
	\includegraphics[width=0.16\textwidth]{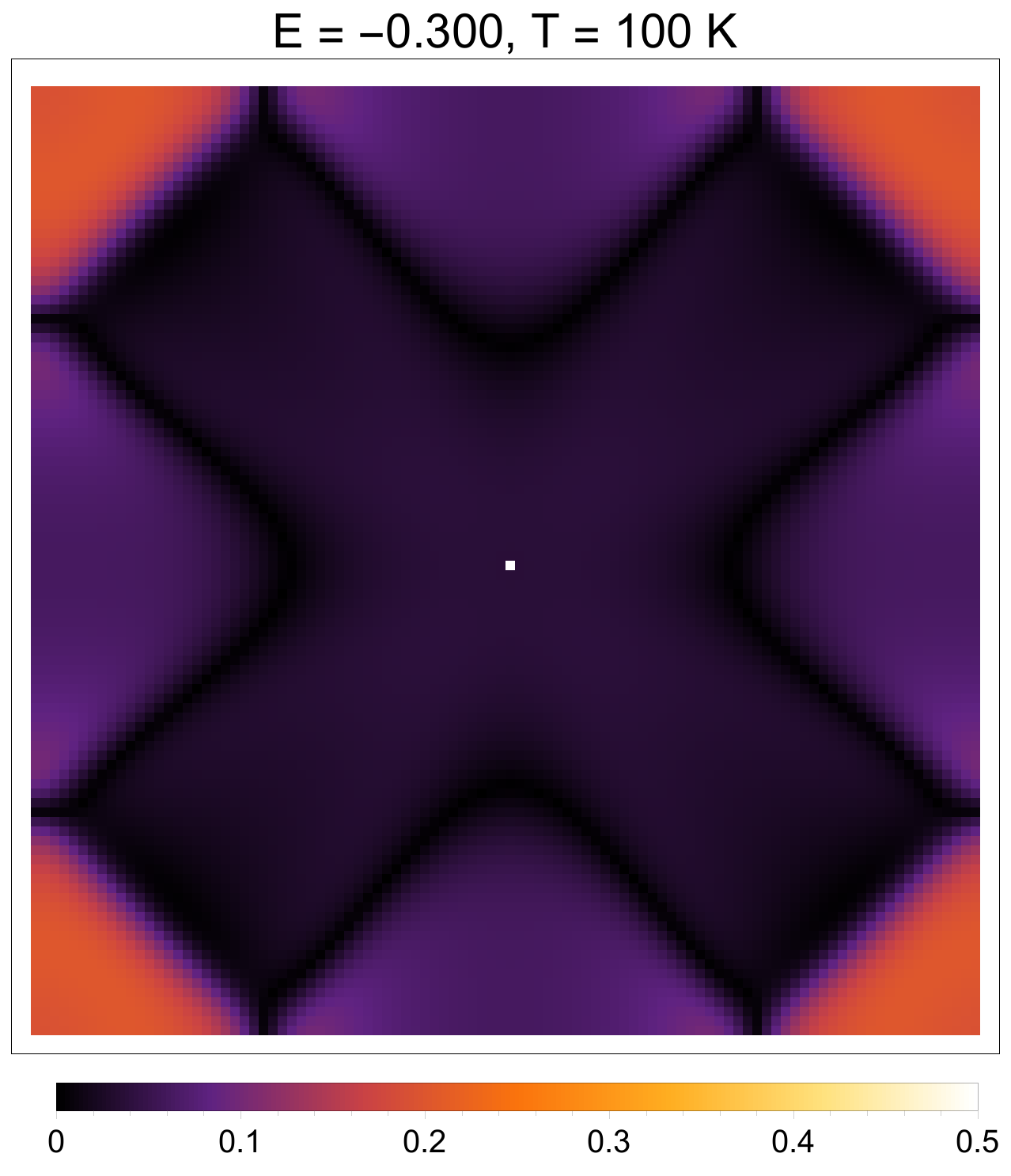}
	\includegraphics[width=0.16\textwidth]{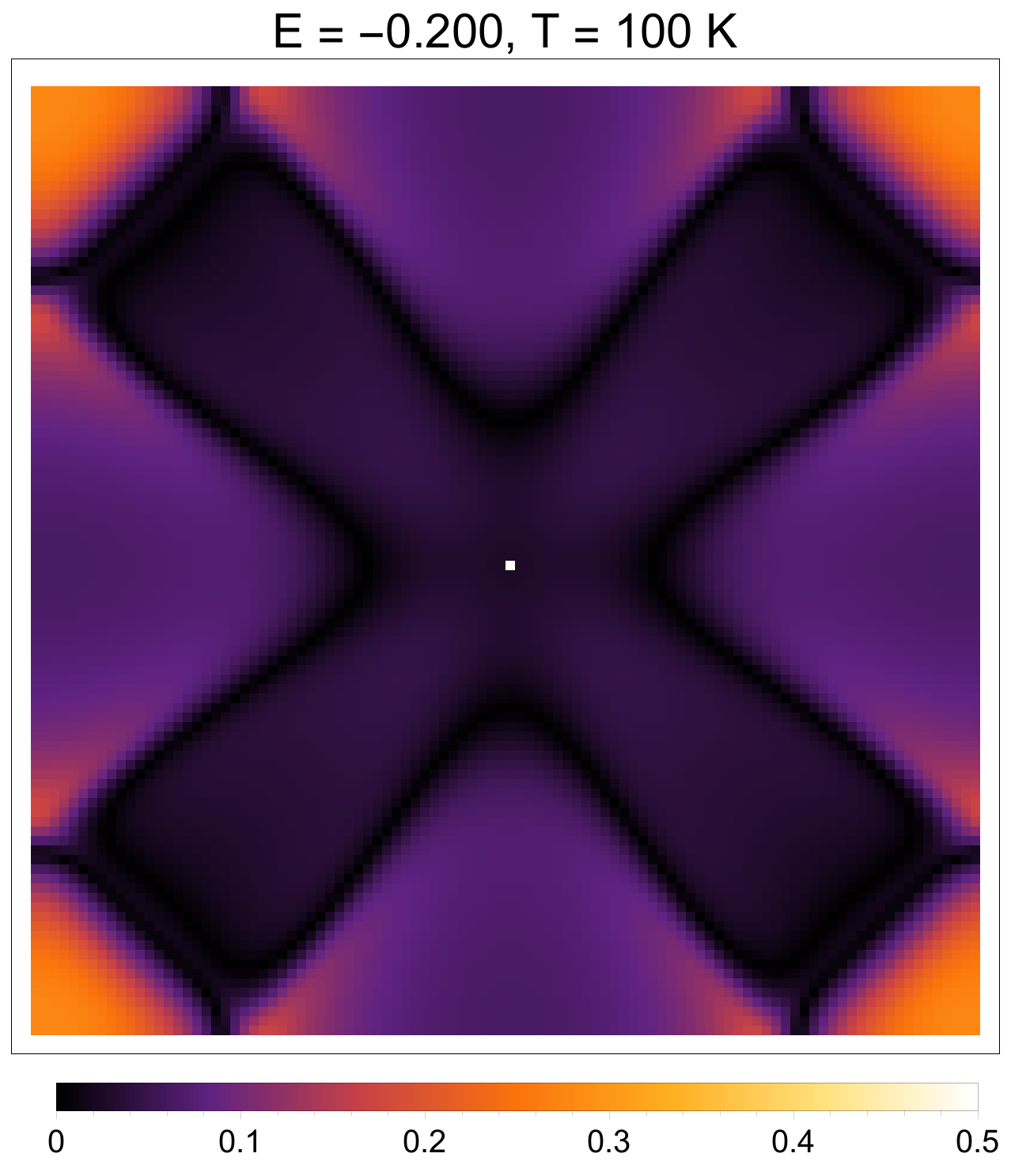}
	\includegraphics[width=0.16\textwidth]{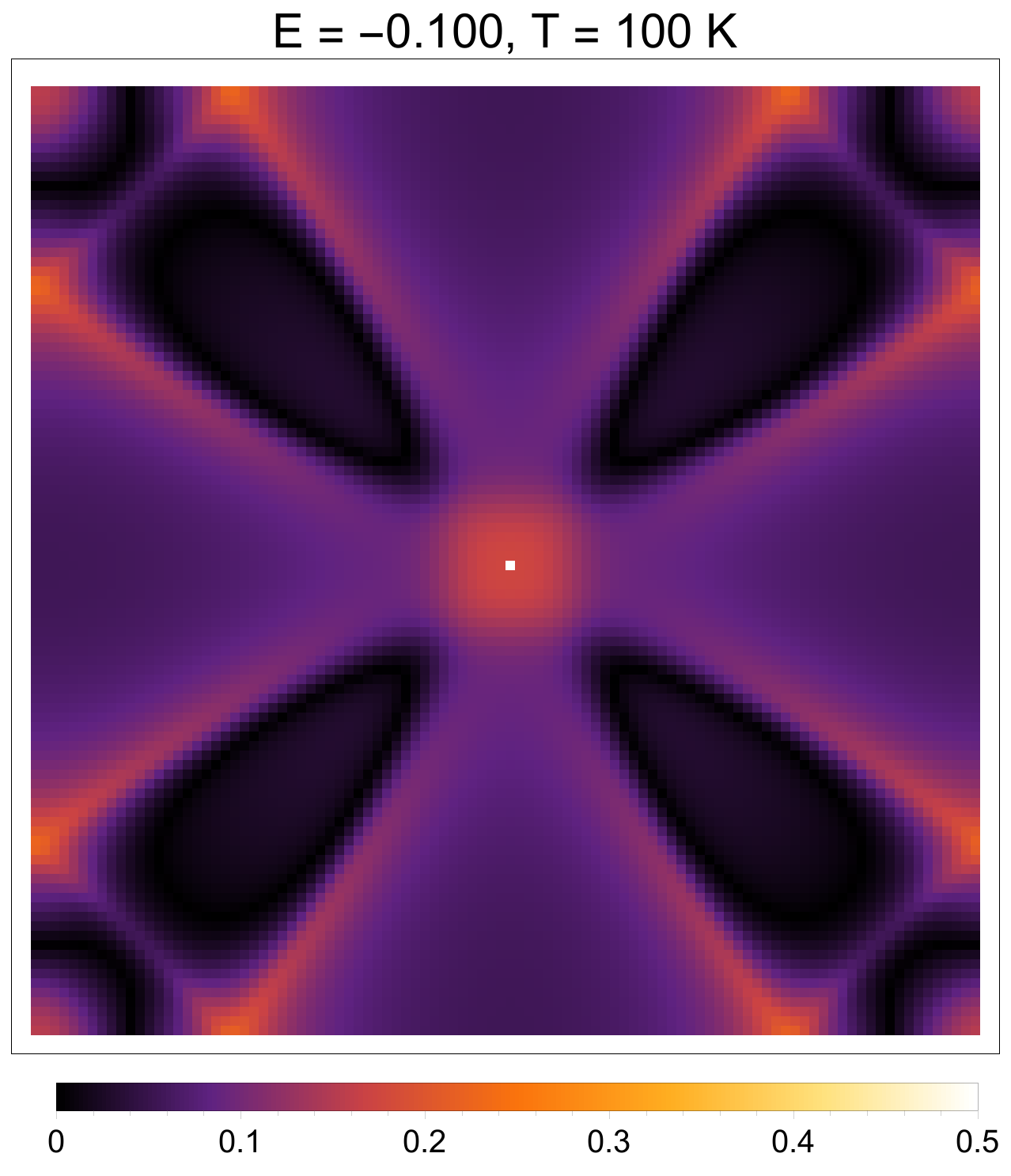}
	\includegraphics[width=0.16\textwidth]{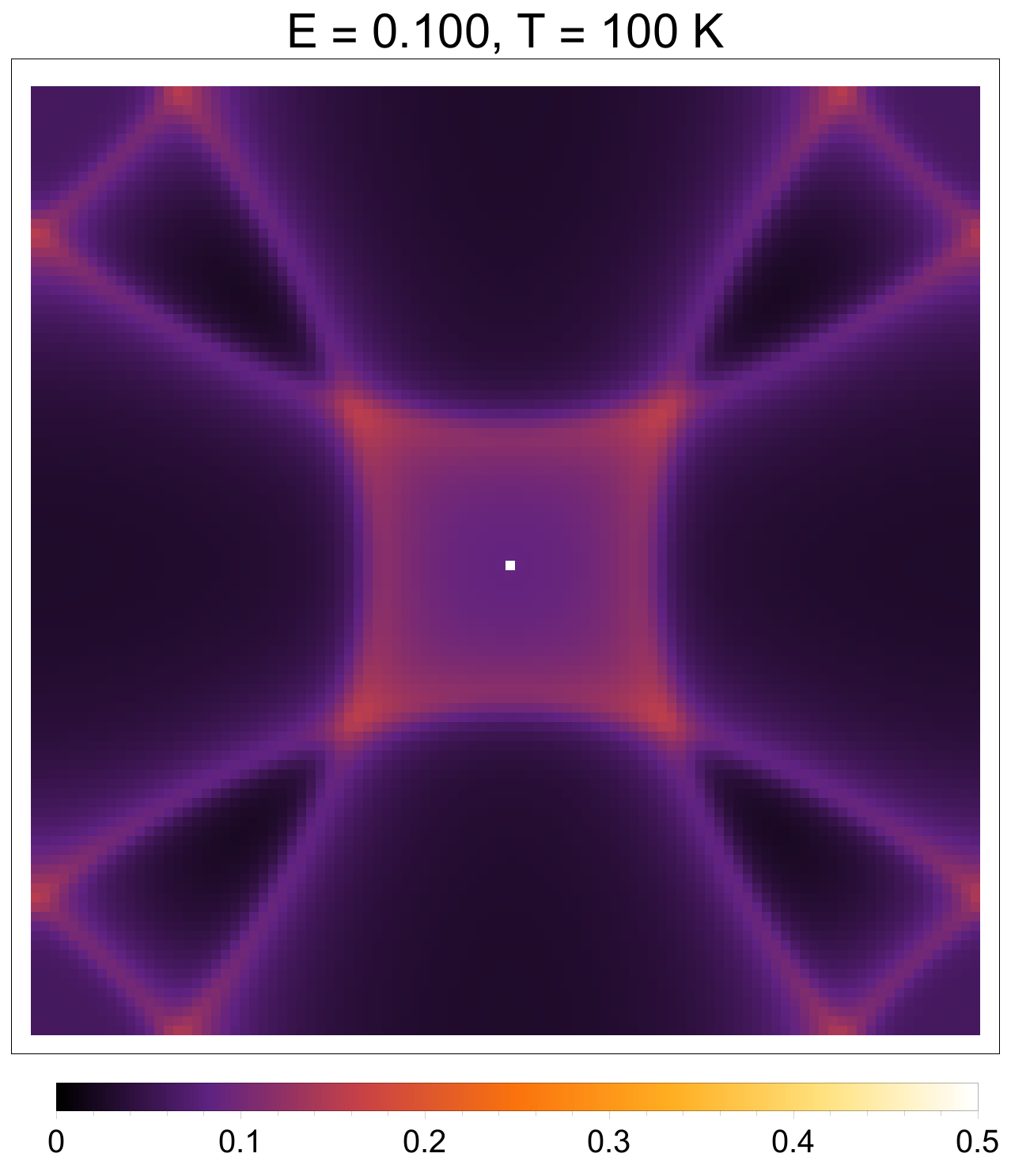}
	\includegraphics[width=0.16\textwidth]{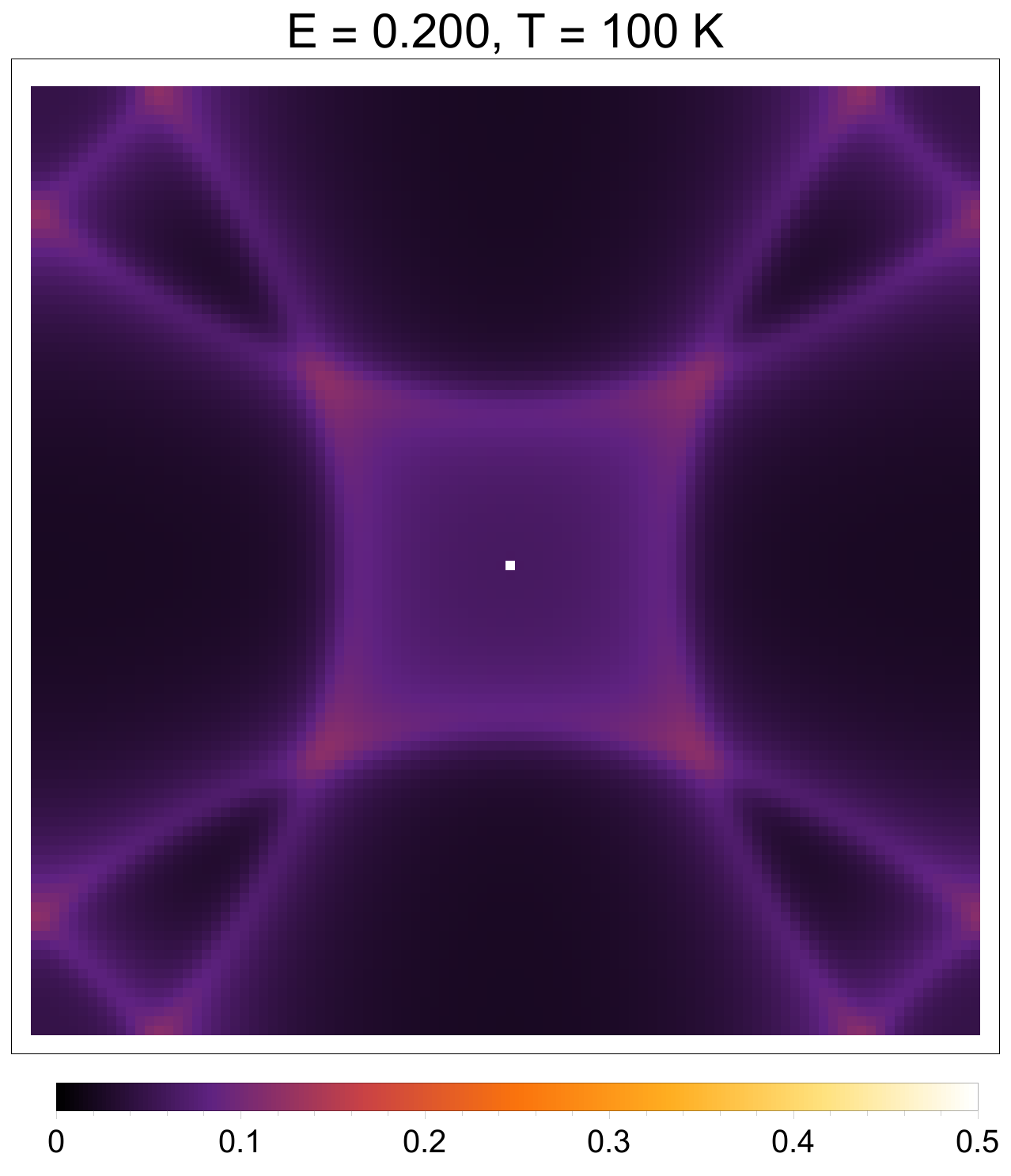}
	\includegraphics[width=0.16\textwidth]{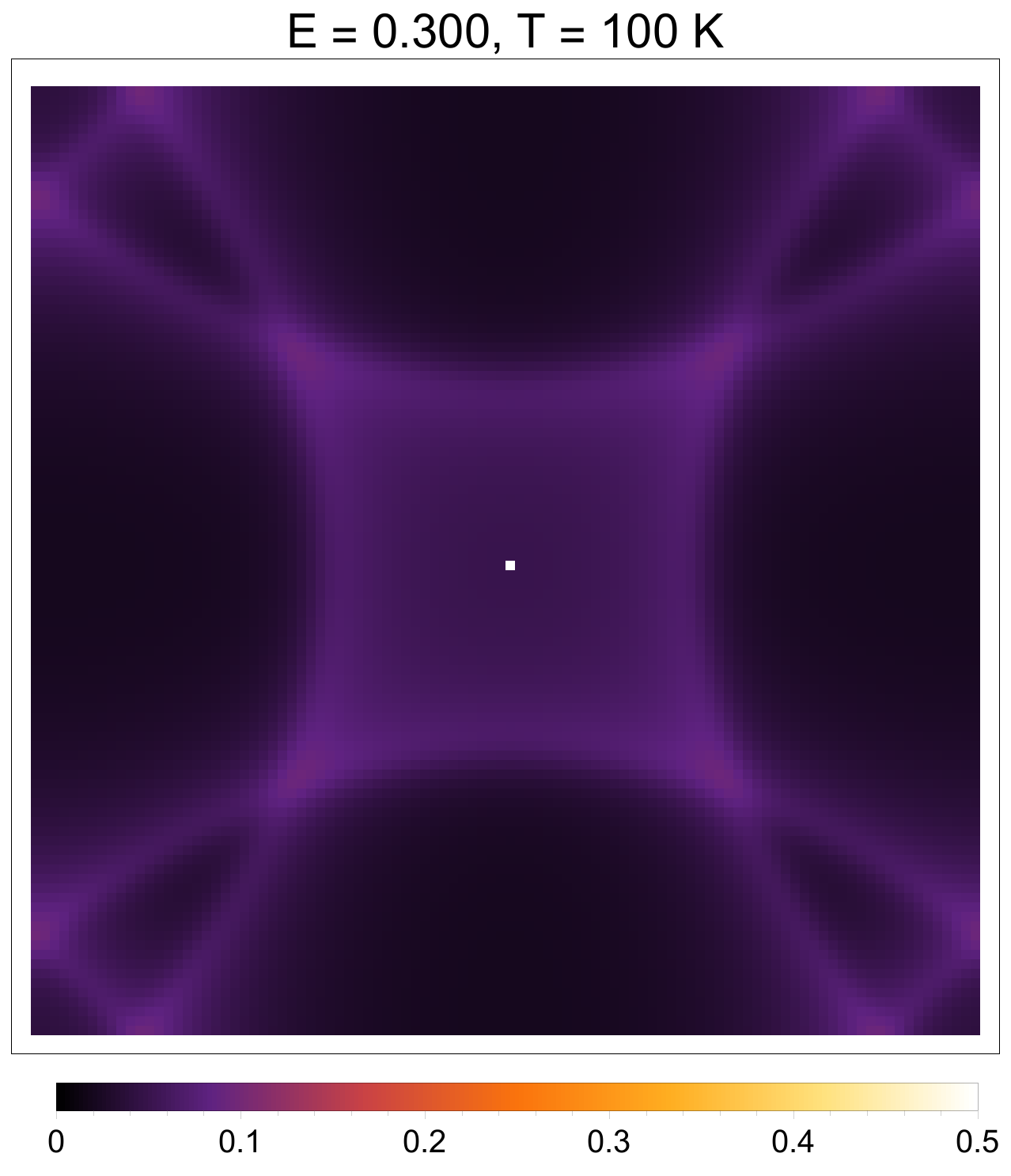}\\
	\includegraphics[width=0.16\textwidth]{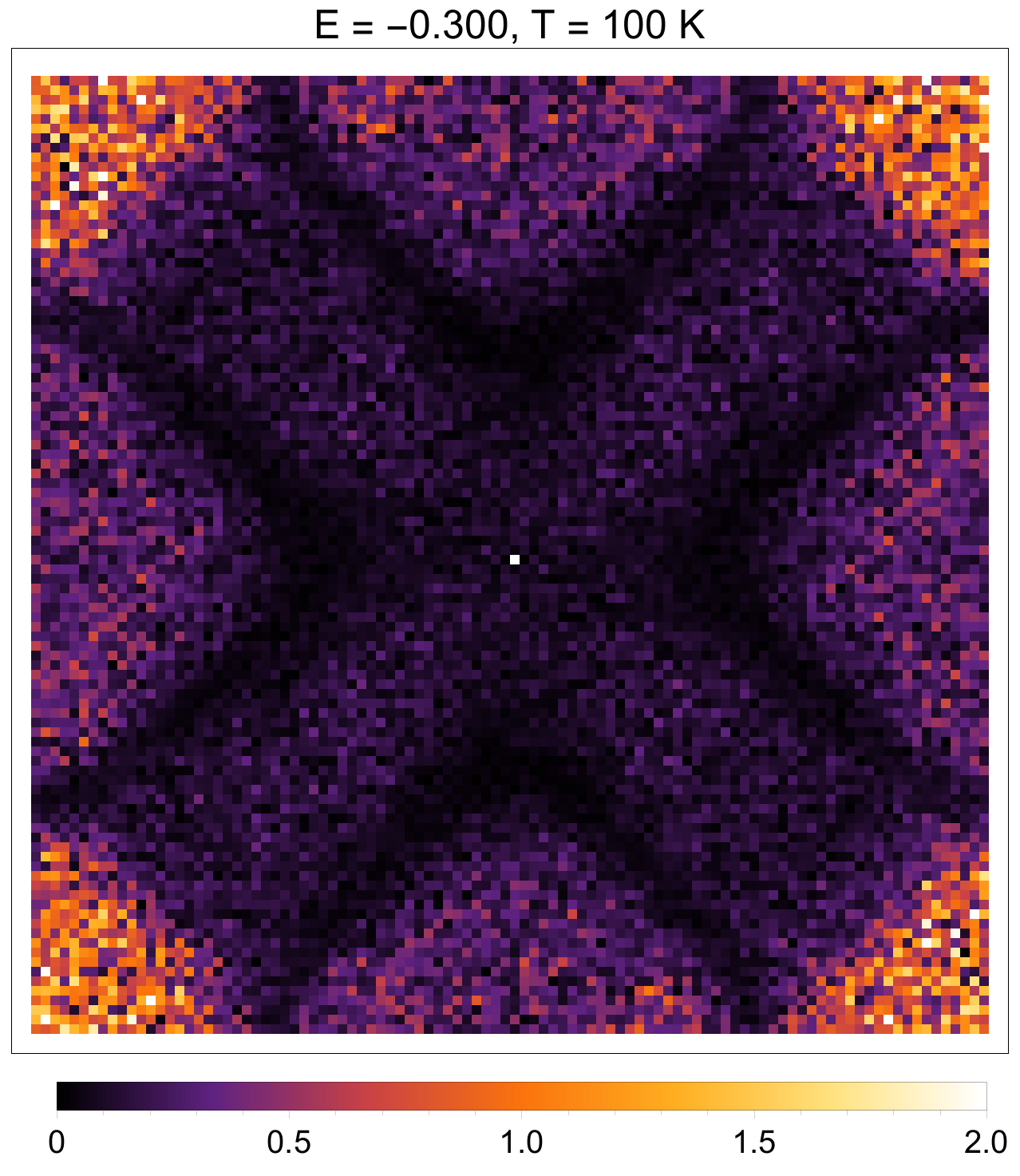}
	\includegraphics[width=0.16\textwidth]{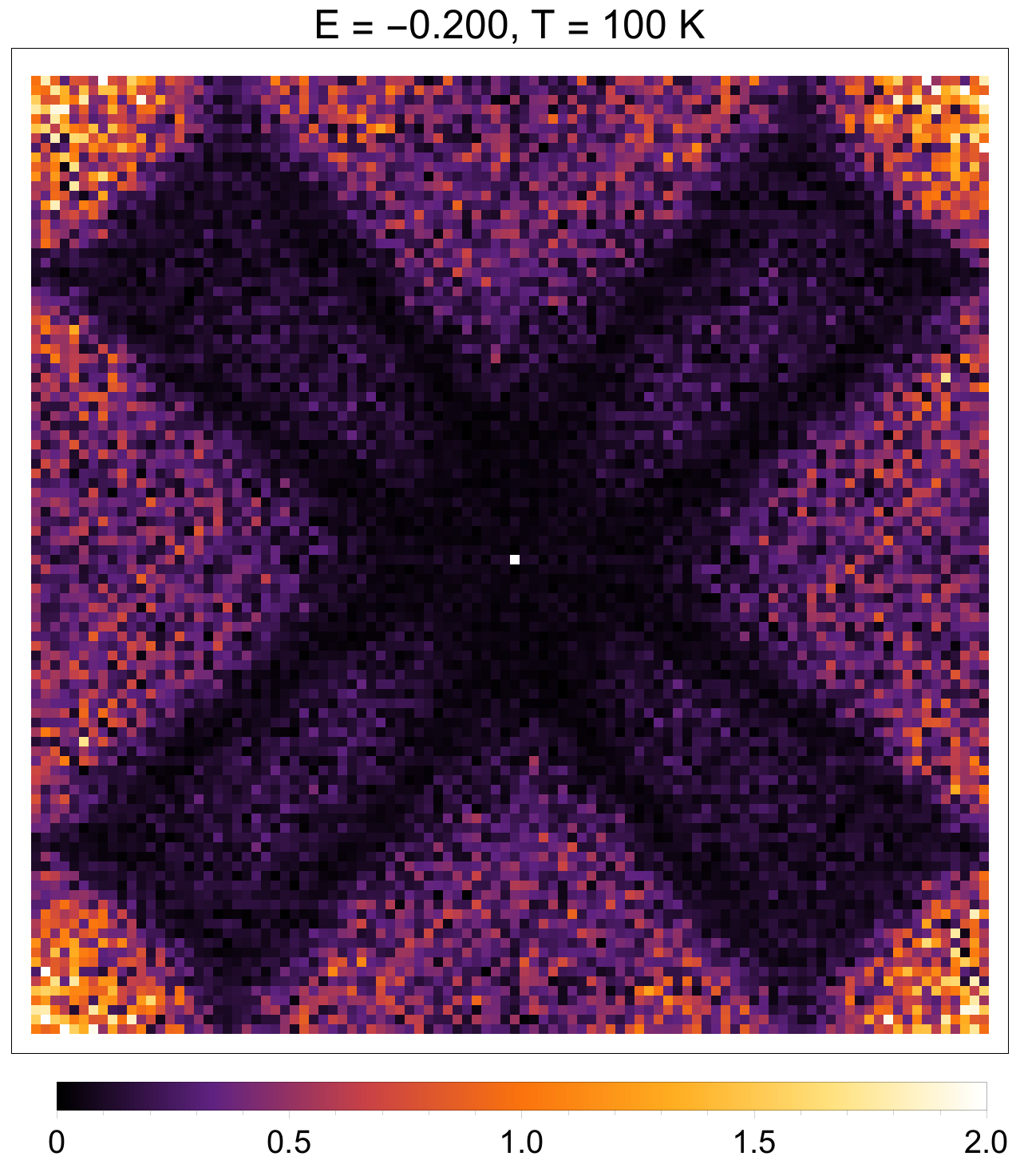}
	\includegraphics[width=0.16\textwidth]{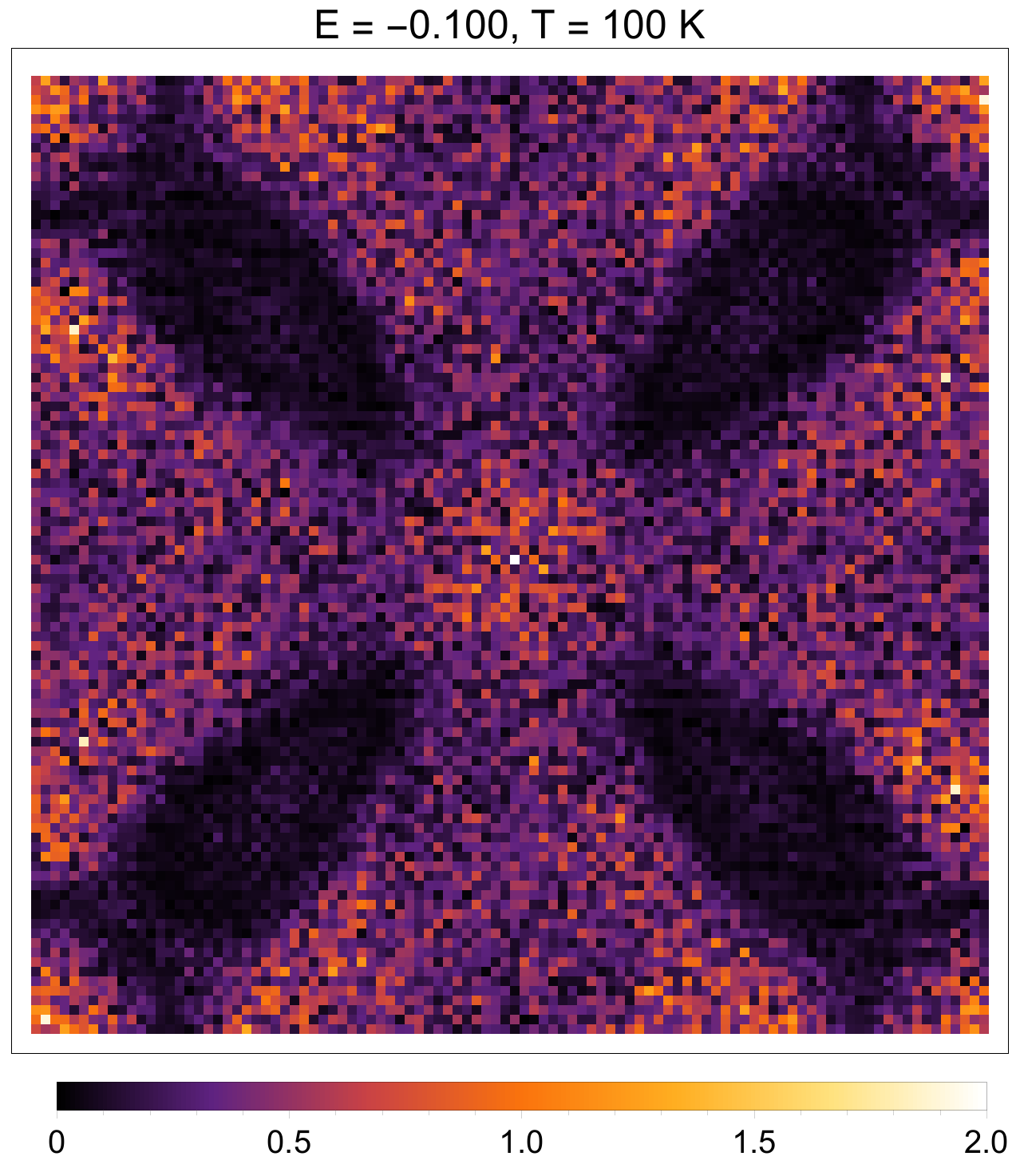}
	\includegraphics[width=0.16\textwidth]{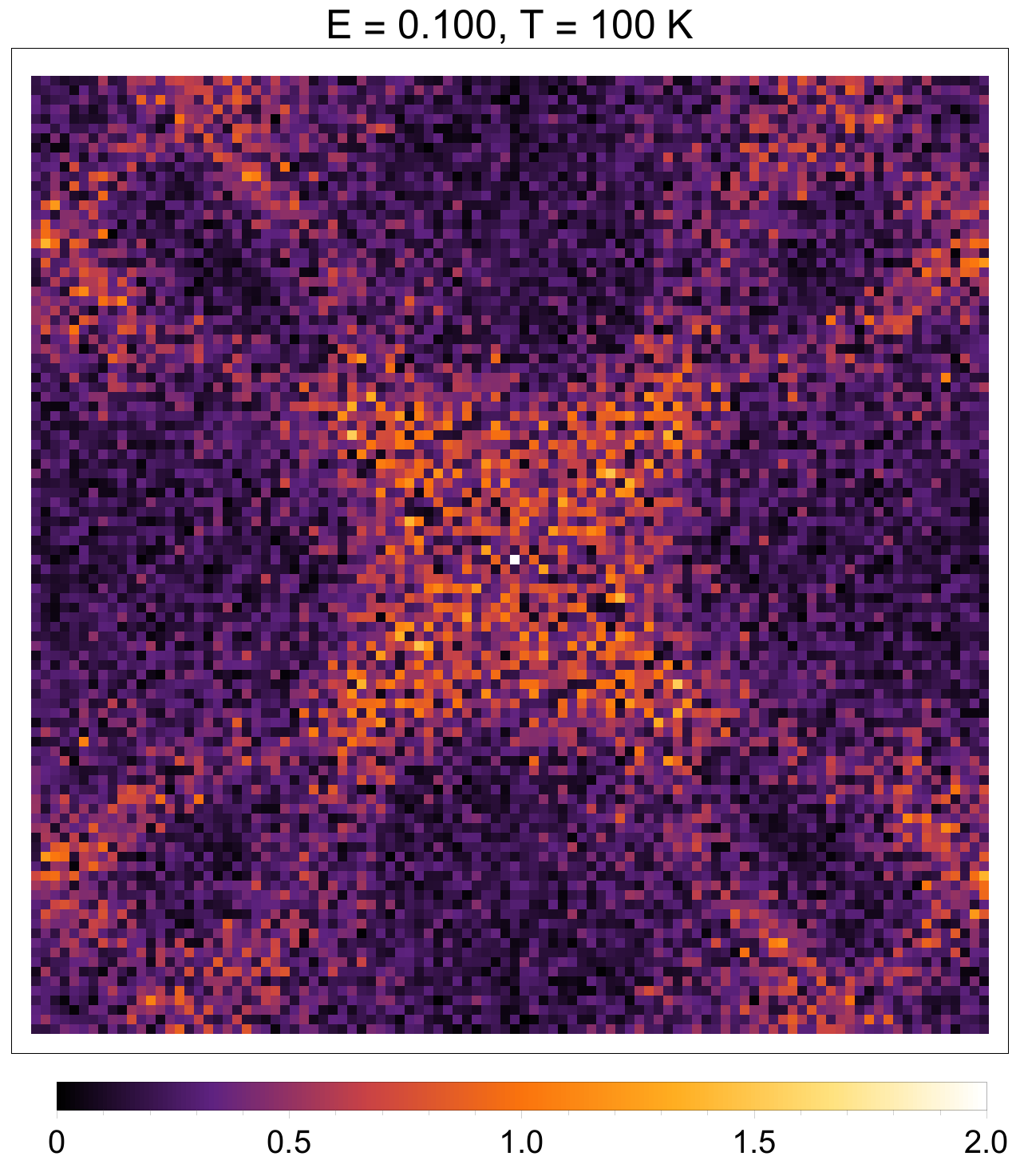}
	\includegraphics[width=0.16\textwidth]{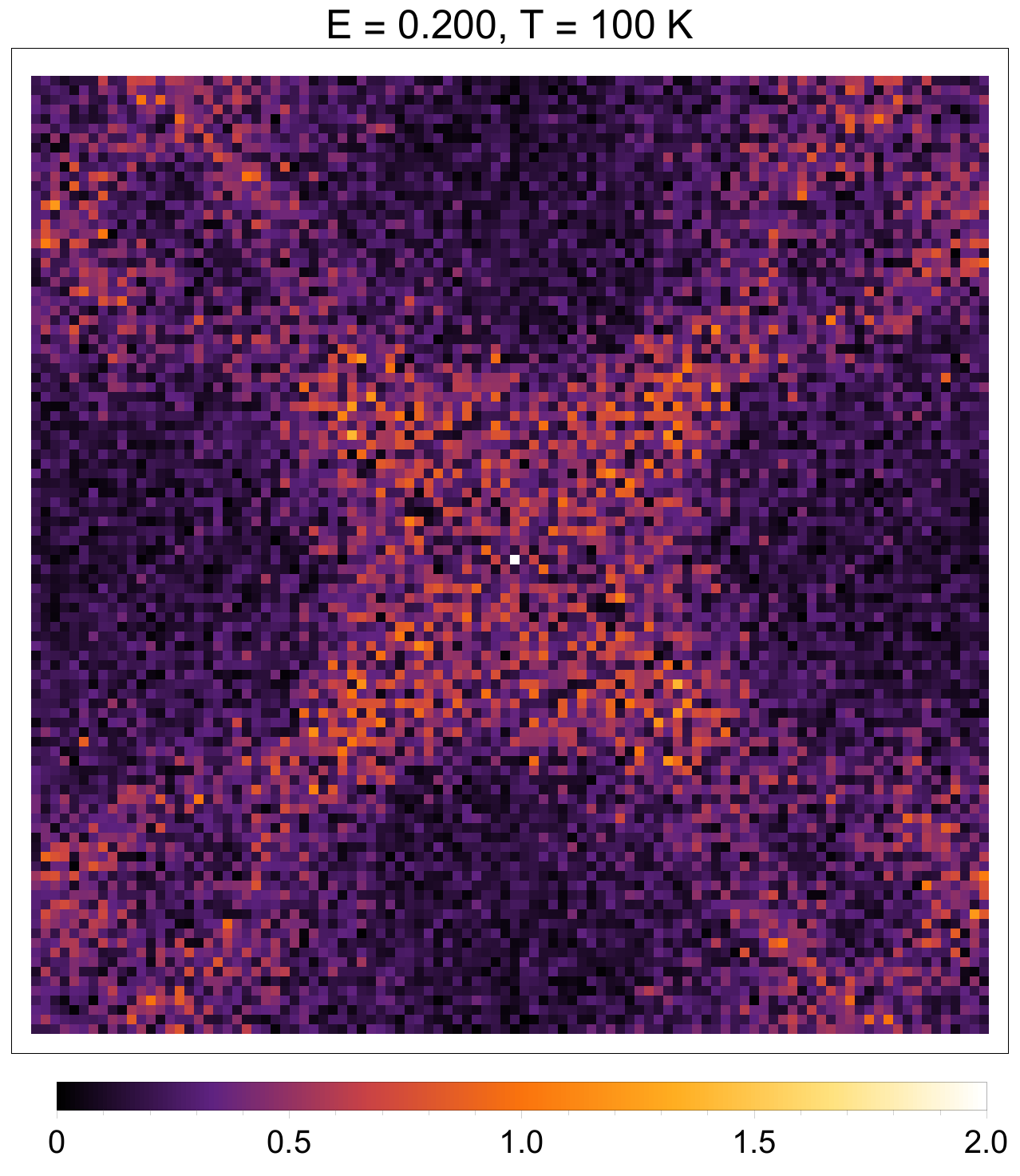}
	\includegraphics[width=0.16\textwidth]{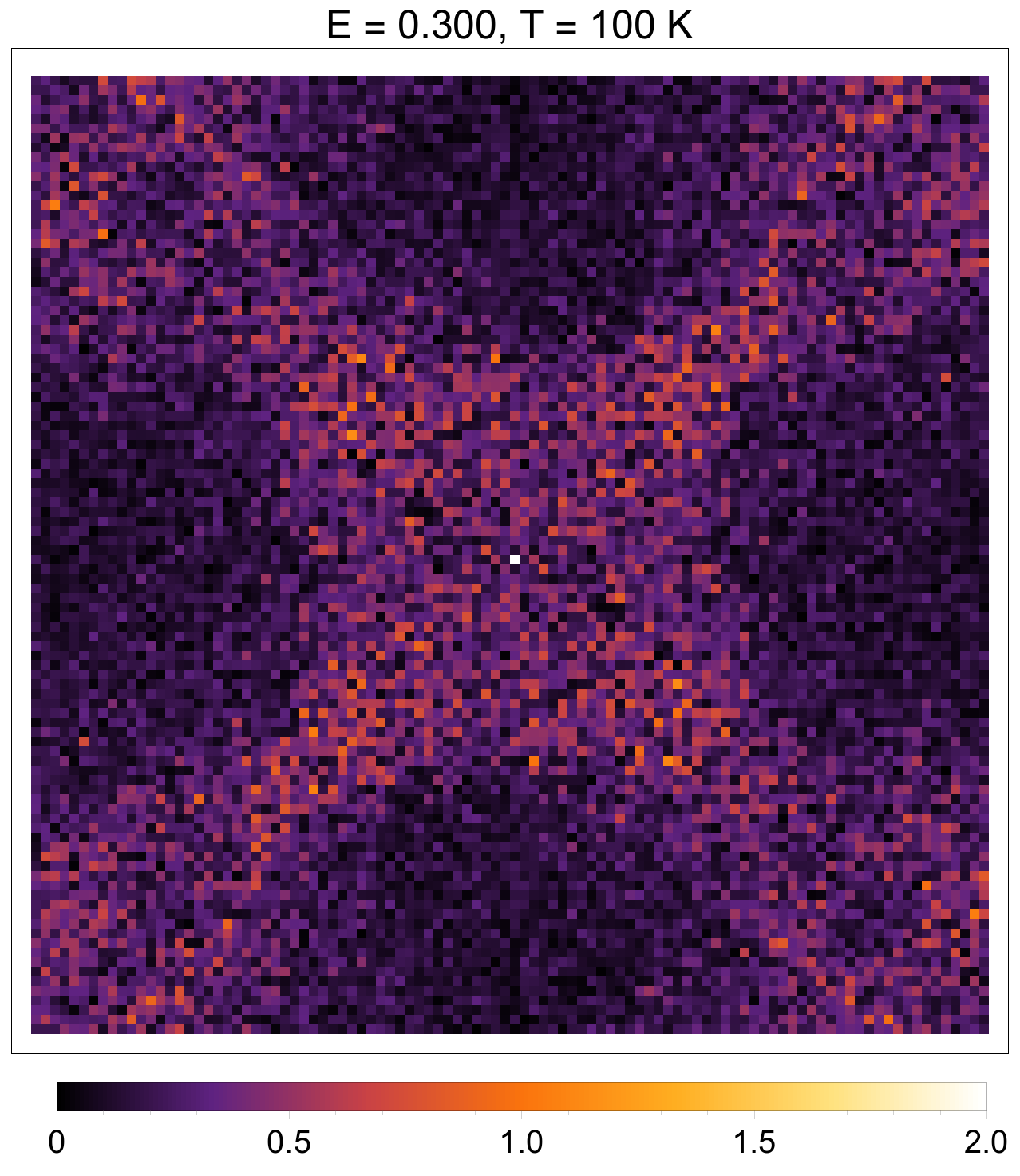}
	
	\caption{Frequency-dependence at $T = 100$ K of the spectra of a marginal Fermi liquid. Shown are plots of the spectral function $A(\mathbf{k}, \omega)$ (upper row); the LDOS power spectrum with a single pointlike scatterer without thermal smearing (middle row); and the LDOS power spectrum with both a 0.5\% concentration of pointlike scatterers and thermal smearing (bottom row). Note that the scales used for plotting the LDOS power spectra are the same for all frequencies.}
	\label{fig:frequency_mfl_100k}
\end{figure*}

\begin{figure*}
	\centering
	
	\includegraphics[height=0.24\textwidth]{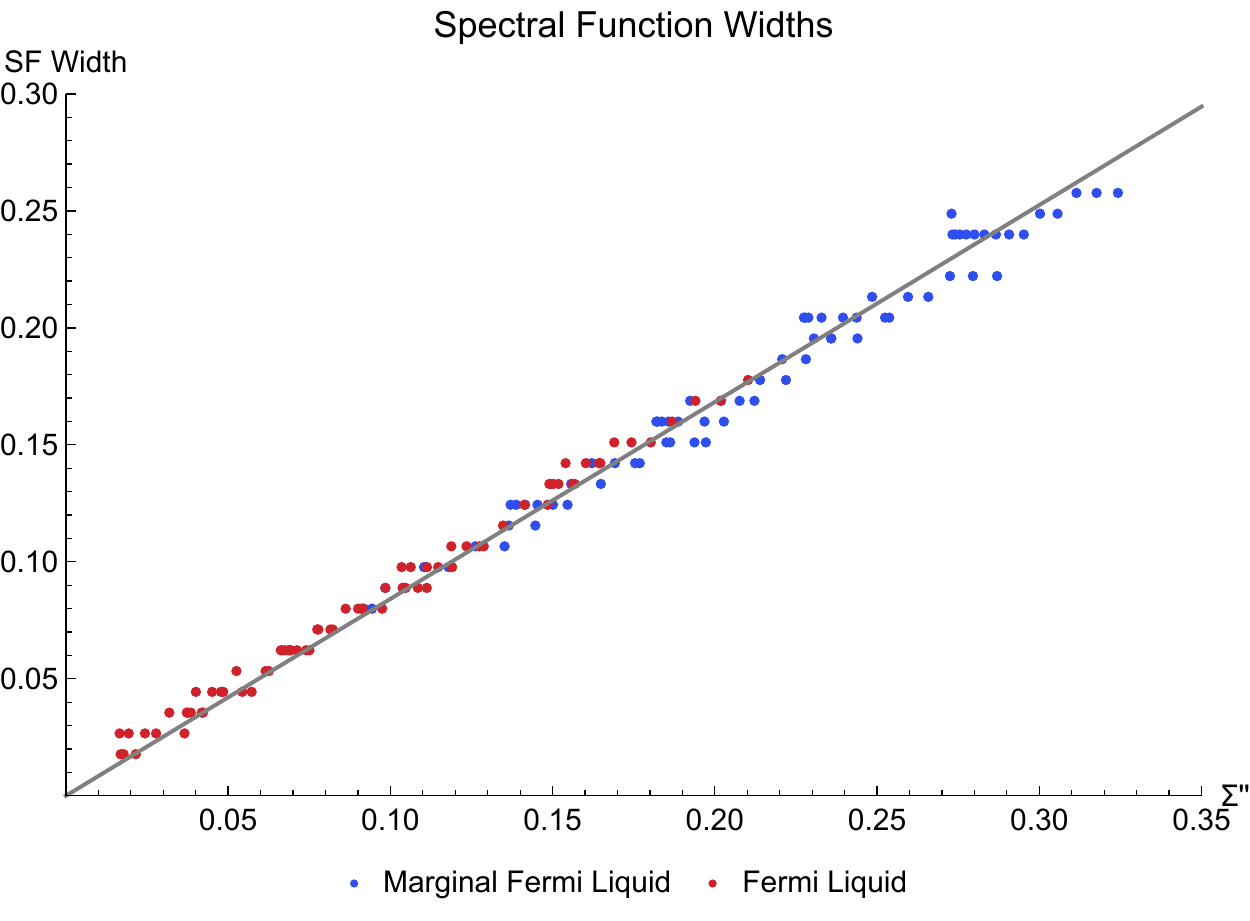}
	\includegraphics[height=0.24\textwidth]{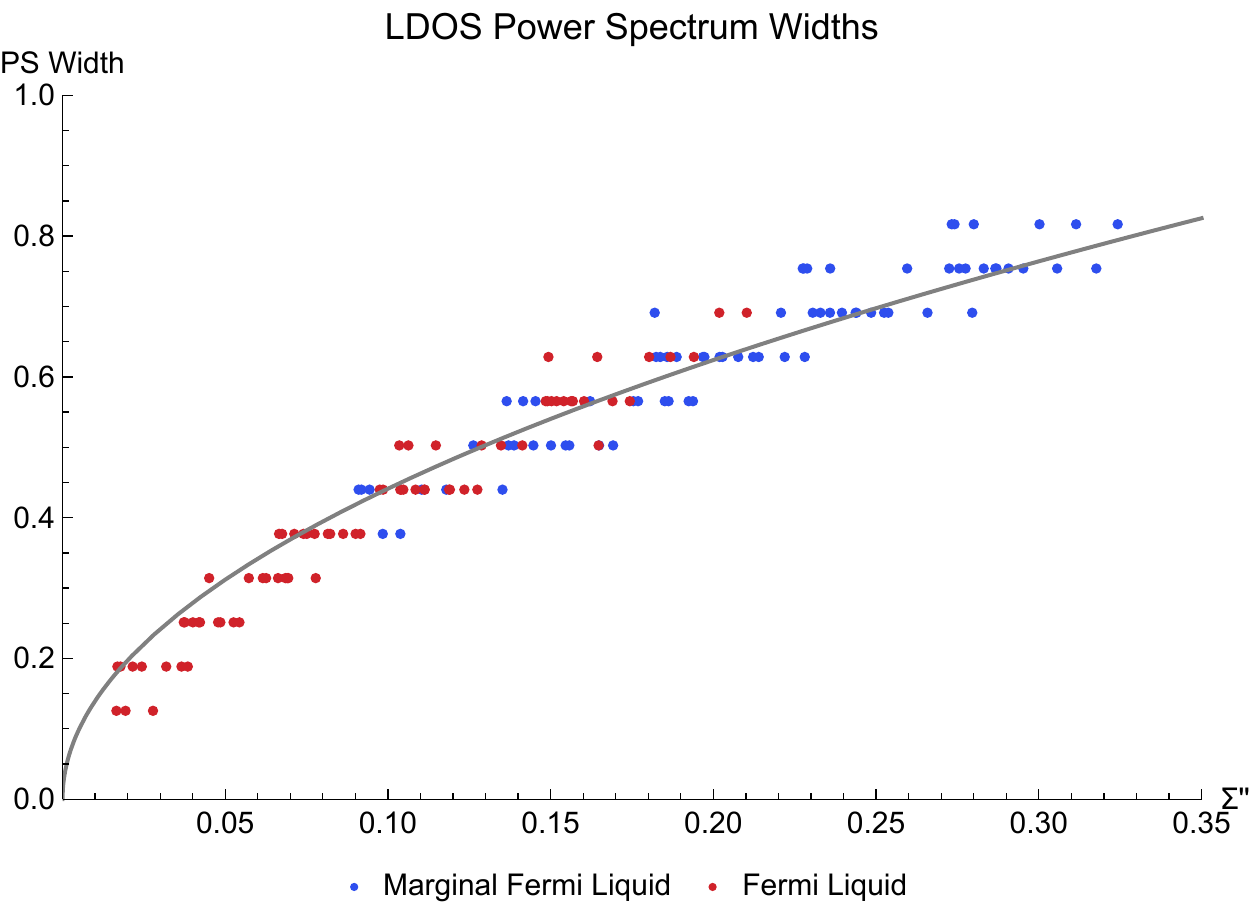}
	\includegraphics[height=0.24\textwidth]{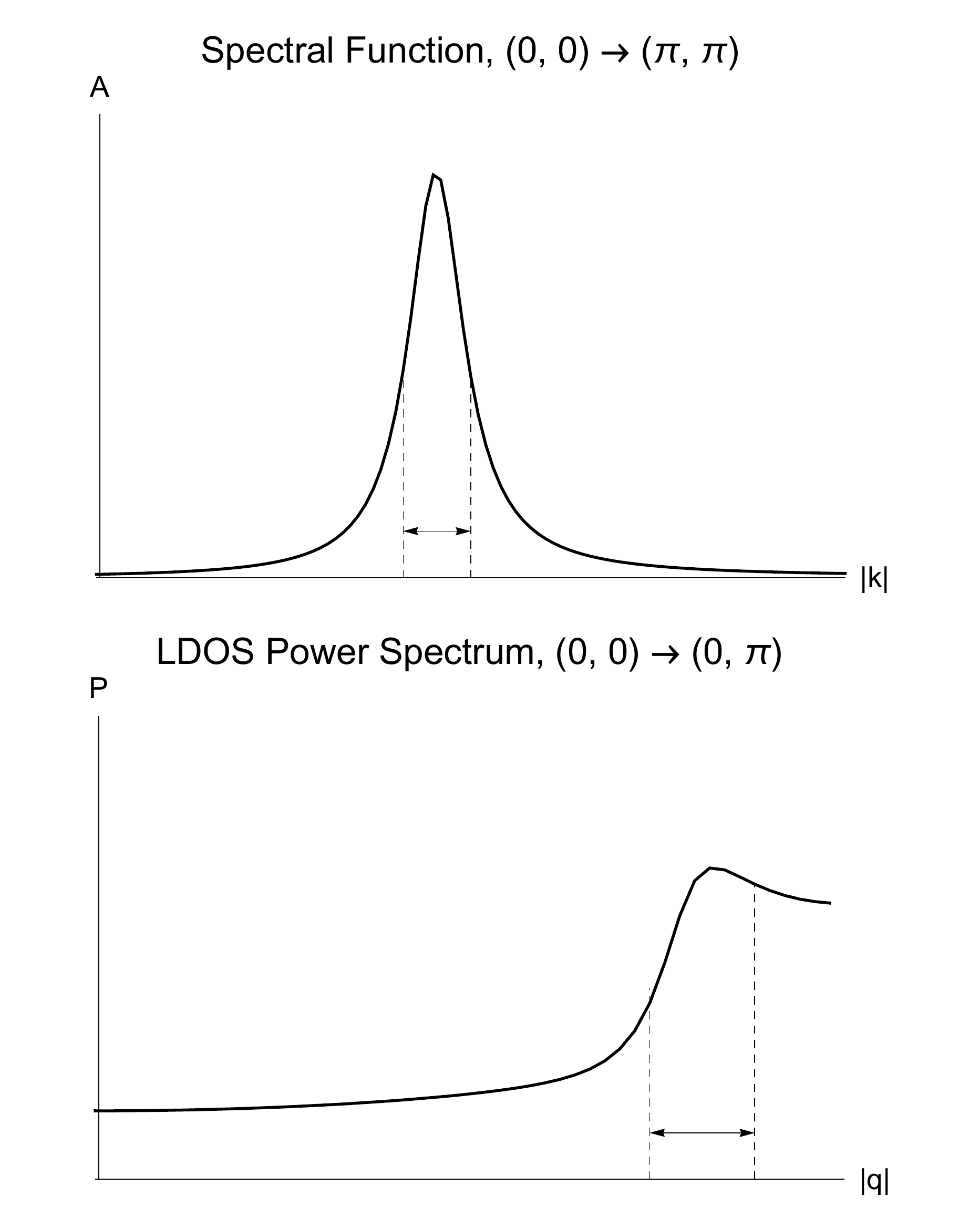}
	
	\caption{The widths of the spectral function (left) and the single-impurity LDOS power spectrum (middle) versus the imaginary part of the self-energy for the marginal Fermi liquid and the ordinary Fermi liquid, both with momentum-independent self-energies, at a variety of temperatures and frequencies. These are evaluated from the widths of the momentum-distribution curves along the nodal directions for the spectral function and from the widths of caustics along the antinodal direction for the LDOS power spectrum. The rightmost graphic illustrates how the spectral widths, as defined in the text, are extracted from linecuts of $A(\mathbf{k}, \omega)$ and $P(\mathbf{q}, \omega)$. In this example the self-energy is of marginal-Fermi-liquid form, and $T = 100$ K and $E = 0$.}
	
	\label{fig:widths}
\end{figure*}

\begin{figure*}
	\centering
	
	\includegraphics[height=0.24\textwidth]{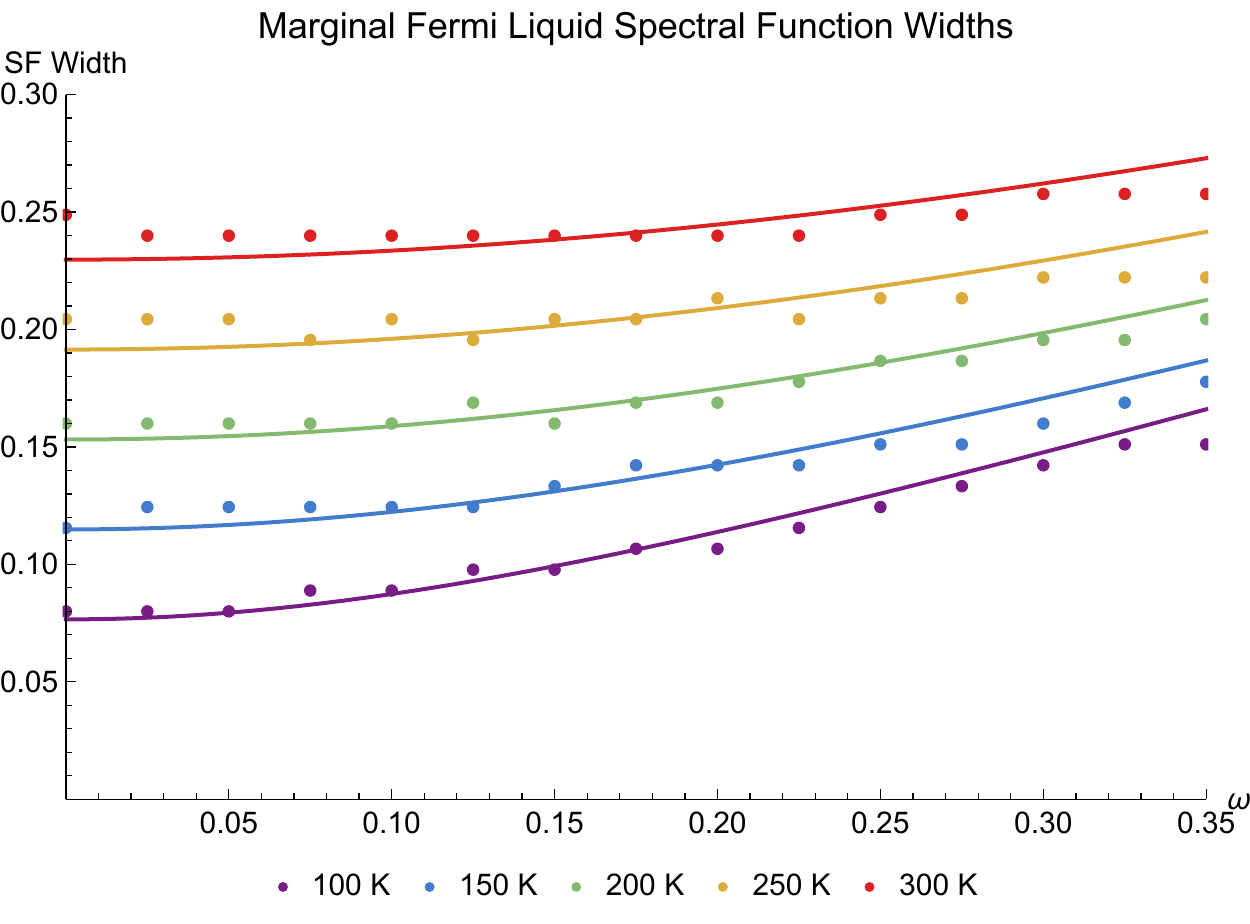}
	\includegraphics[height=0.24\textwidth]{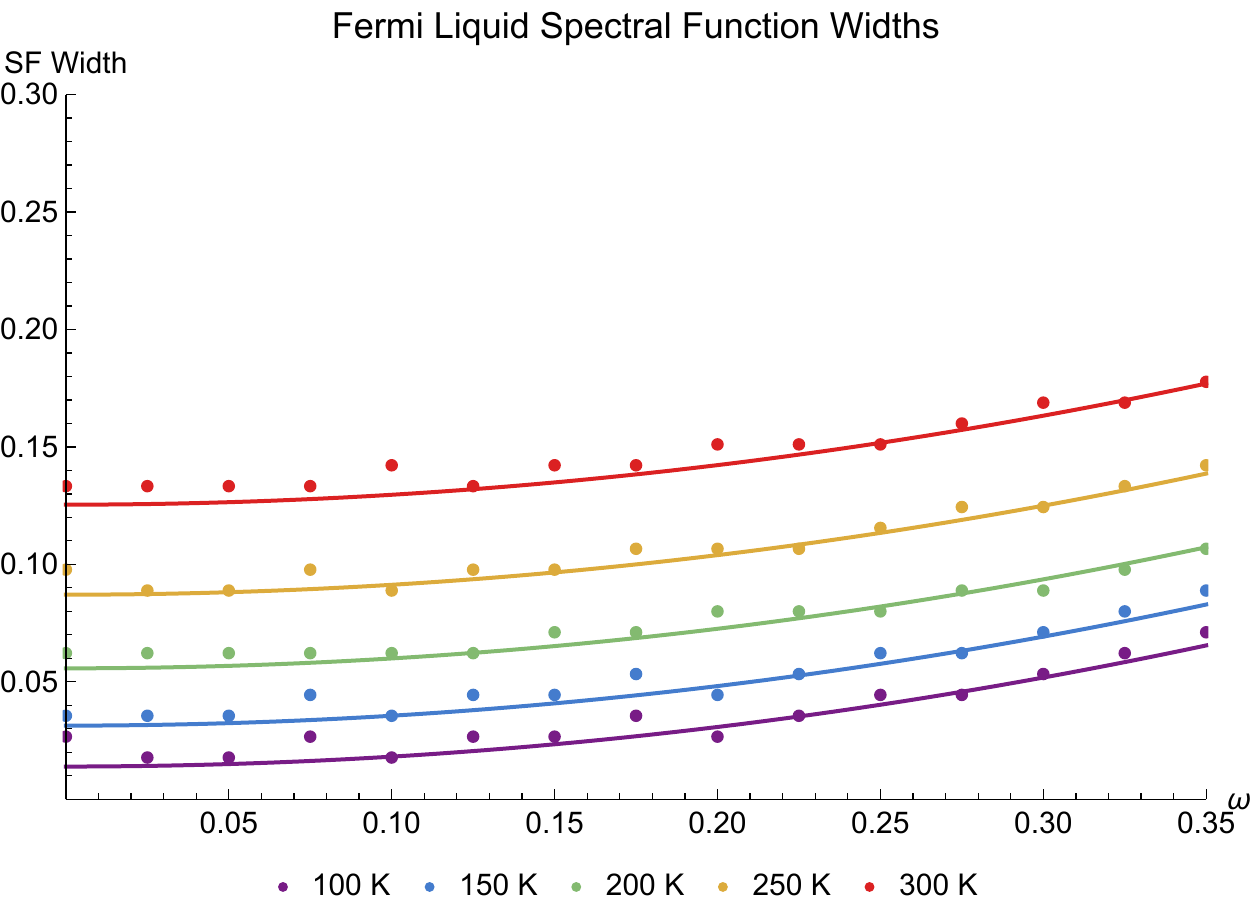} \\
	\includegraphics[height=0.24\textwidth]{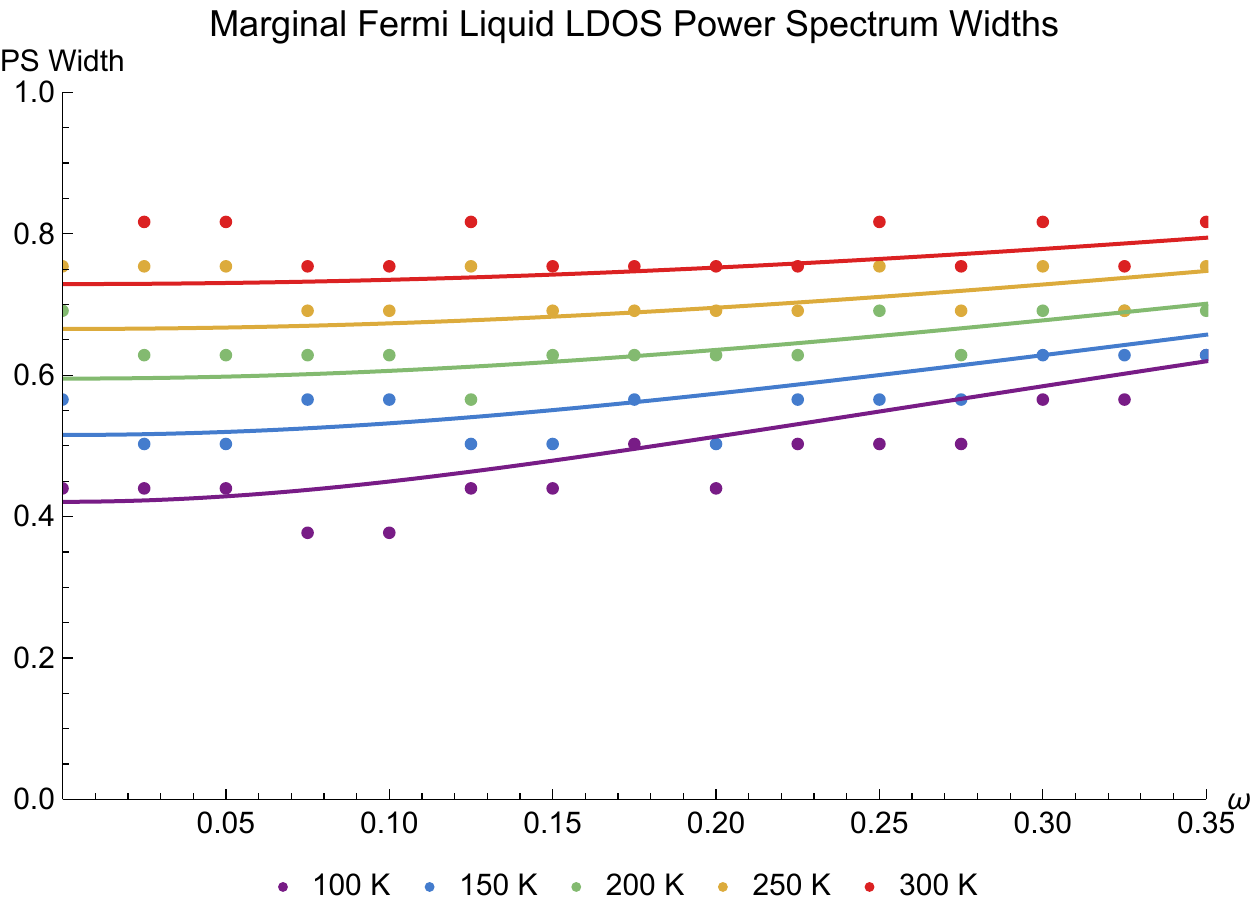}
	\includegraphics[height=0.24\textwidth]{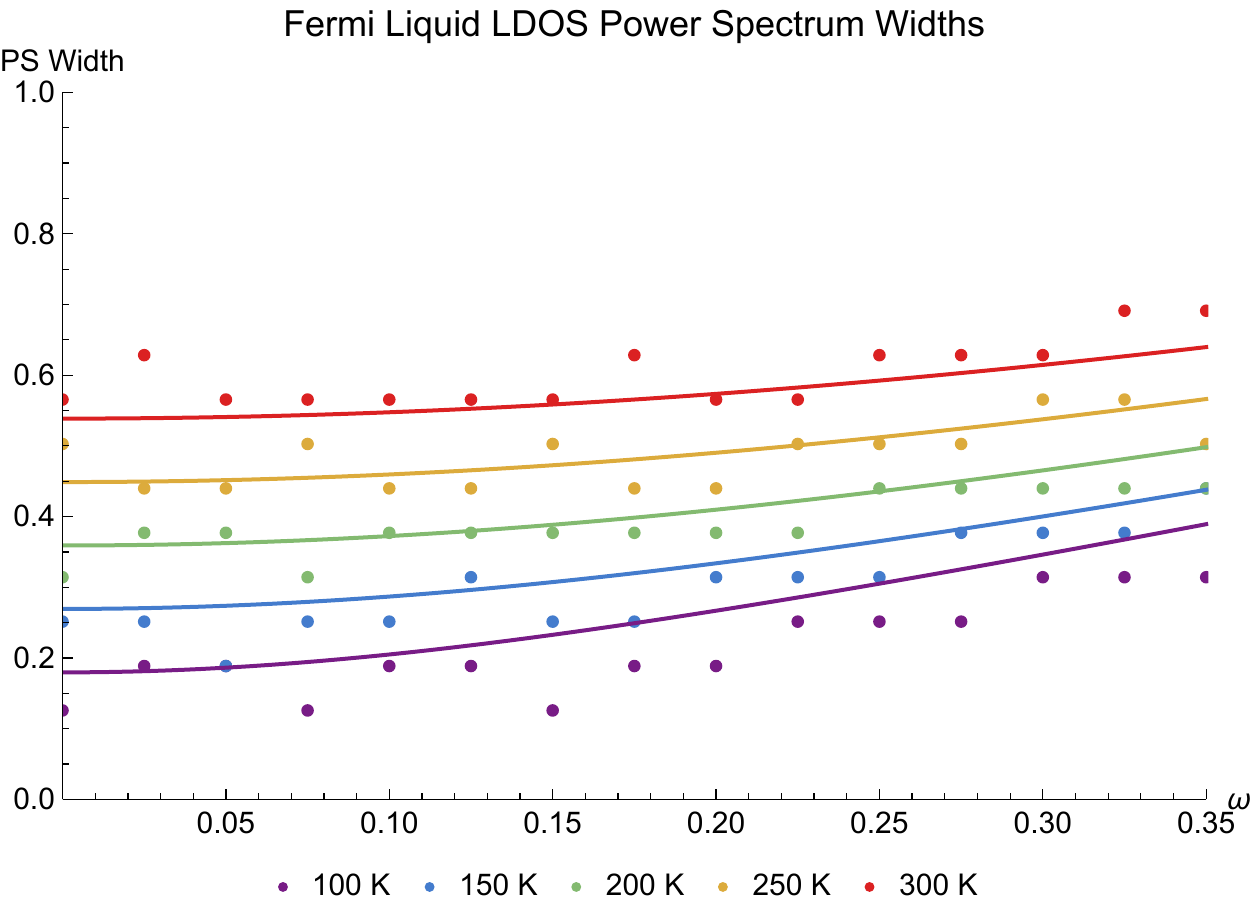} \\
	
	\caption{The widths of the spectral function (top row) and the single-impurity LDOS power spectrum (bottom row) versus frequency for the marginal Fermi liquid (left column) and the ordinary Fermi liquid (right column), evaluated at various temperatures. The fits used are taken from the complete data plotted in Fig.~\ref{fig:widths}. The limited resolution available in the LDOS power spectrum results in the relatively jagged behavior of the plots compared to that seen in plots of the spectral function.}
	
	\label{fig:widthstemperature}
\end{figure*}

\begin{figure}
	\centering
	\includegraphics[width=0.4\textwidth]{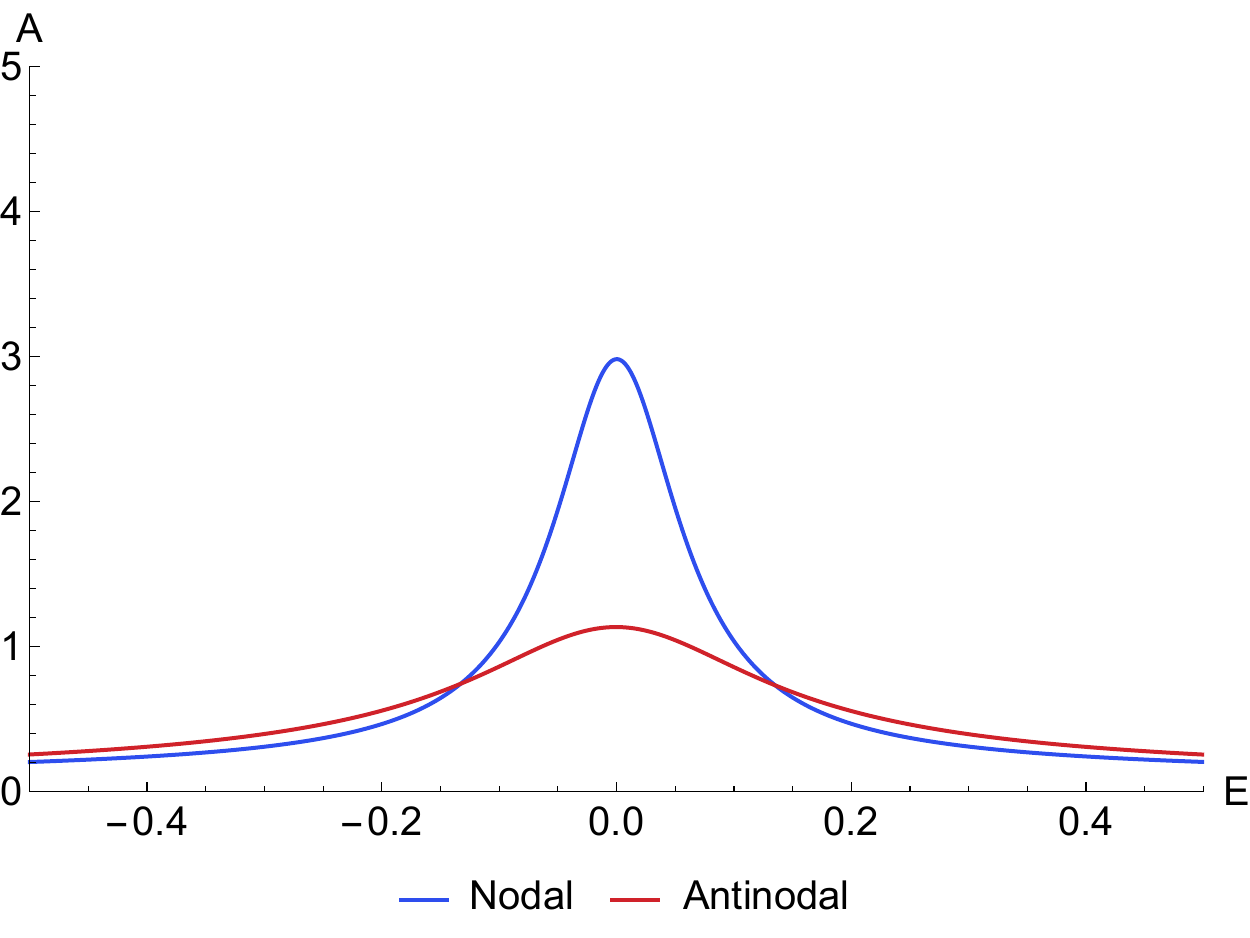}
	\caption{Energy-distribution curves taken at the nodal and antinodal points on the Fermi surface for an anisotropic marginal Fermi liquid. Here $T = 100$ K and $\beta = 0.2$ (see Eq.~\ref{eq:anisotropic_se2} for the functional form of the self-energy). } 
	\label{fig:nan_edc}
\end{figure}

\begin{figure*}
	\centering
	
	\includegraphics[height=0.18\textwidth]{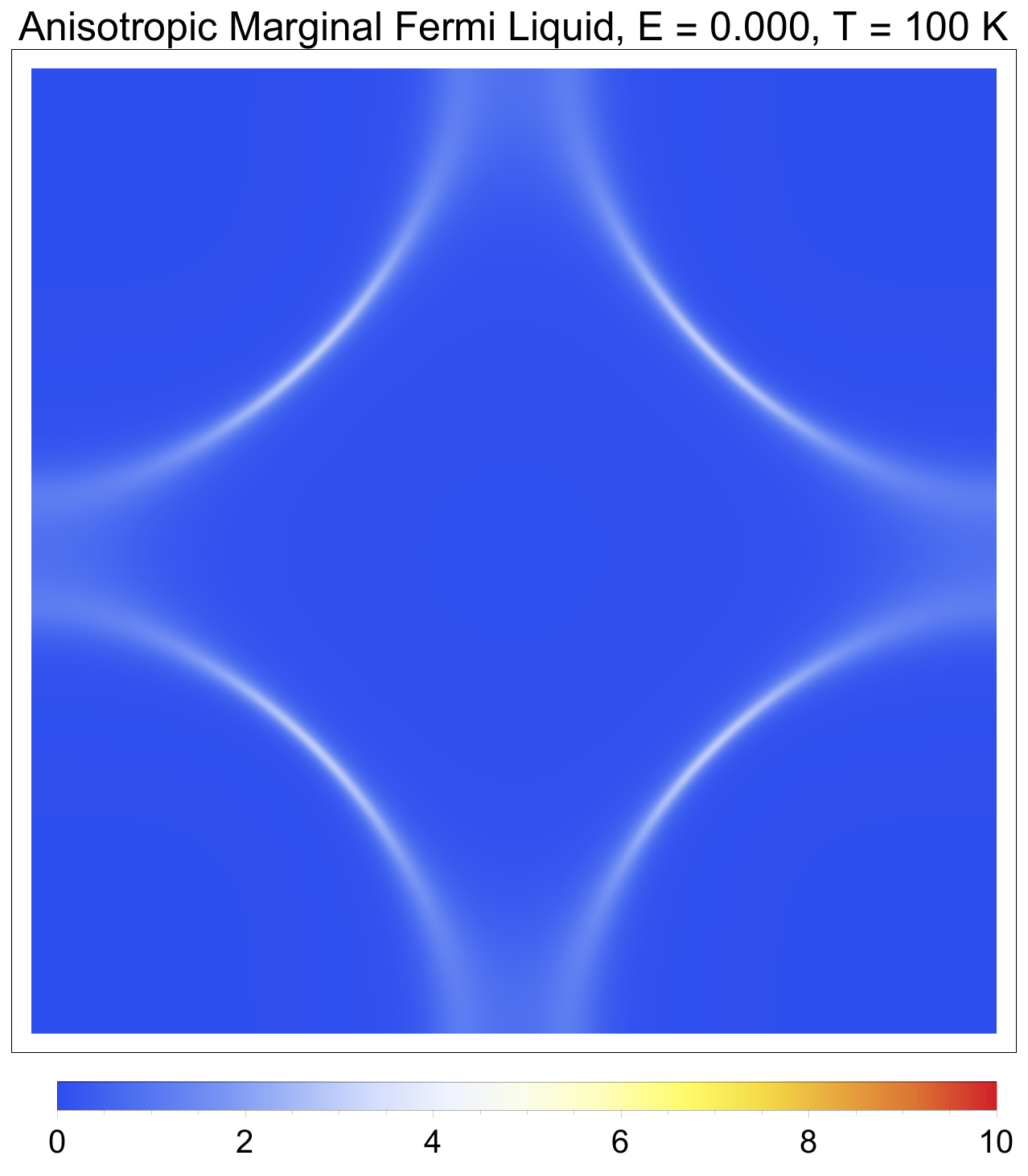}
	\includegraphics[height=0.18\textwidth]{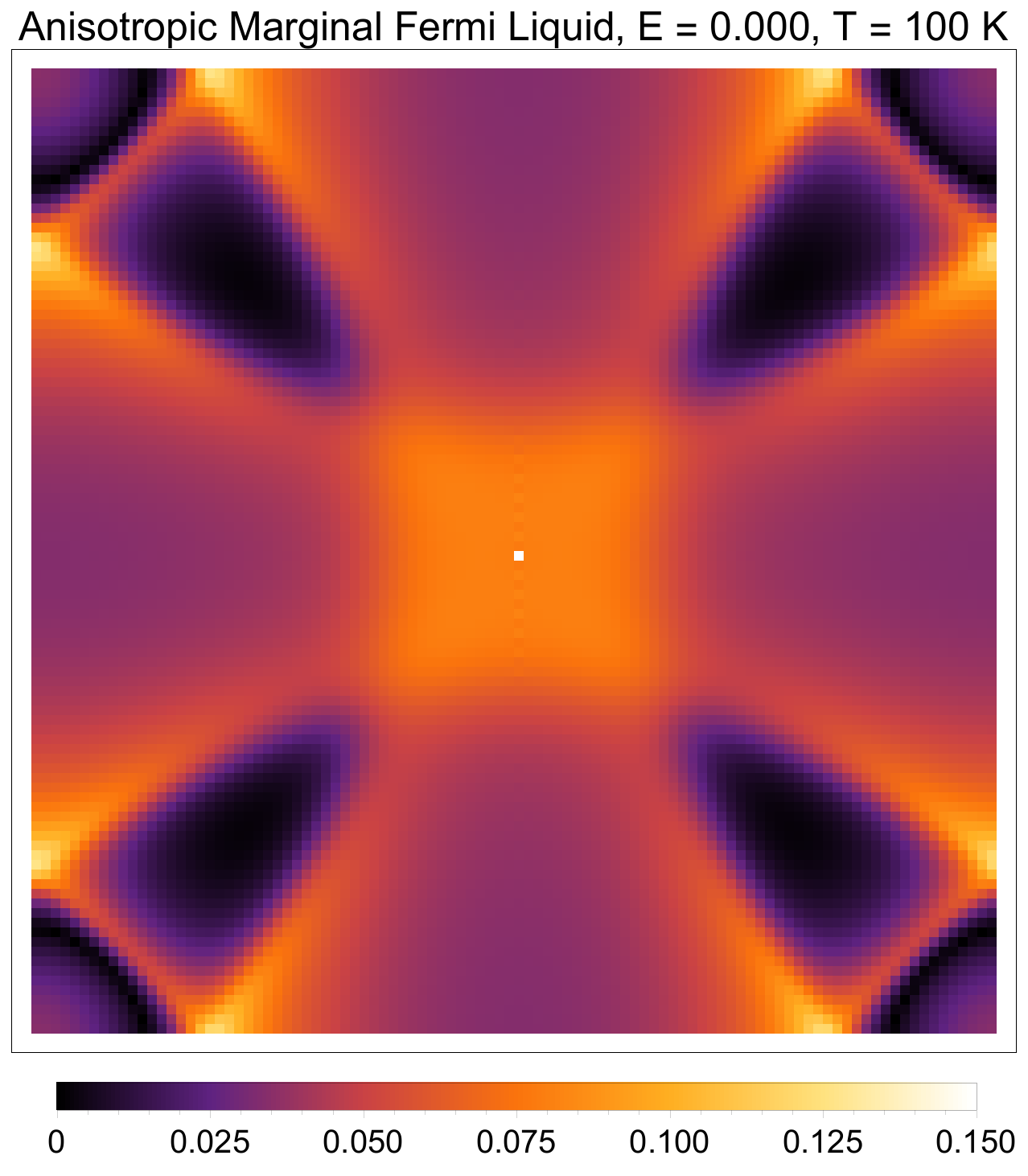}
	\includegraphics[height=0.18\textwidth]{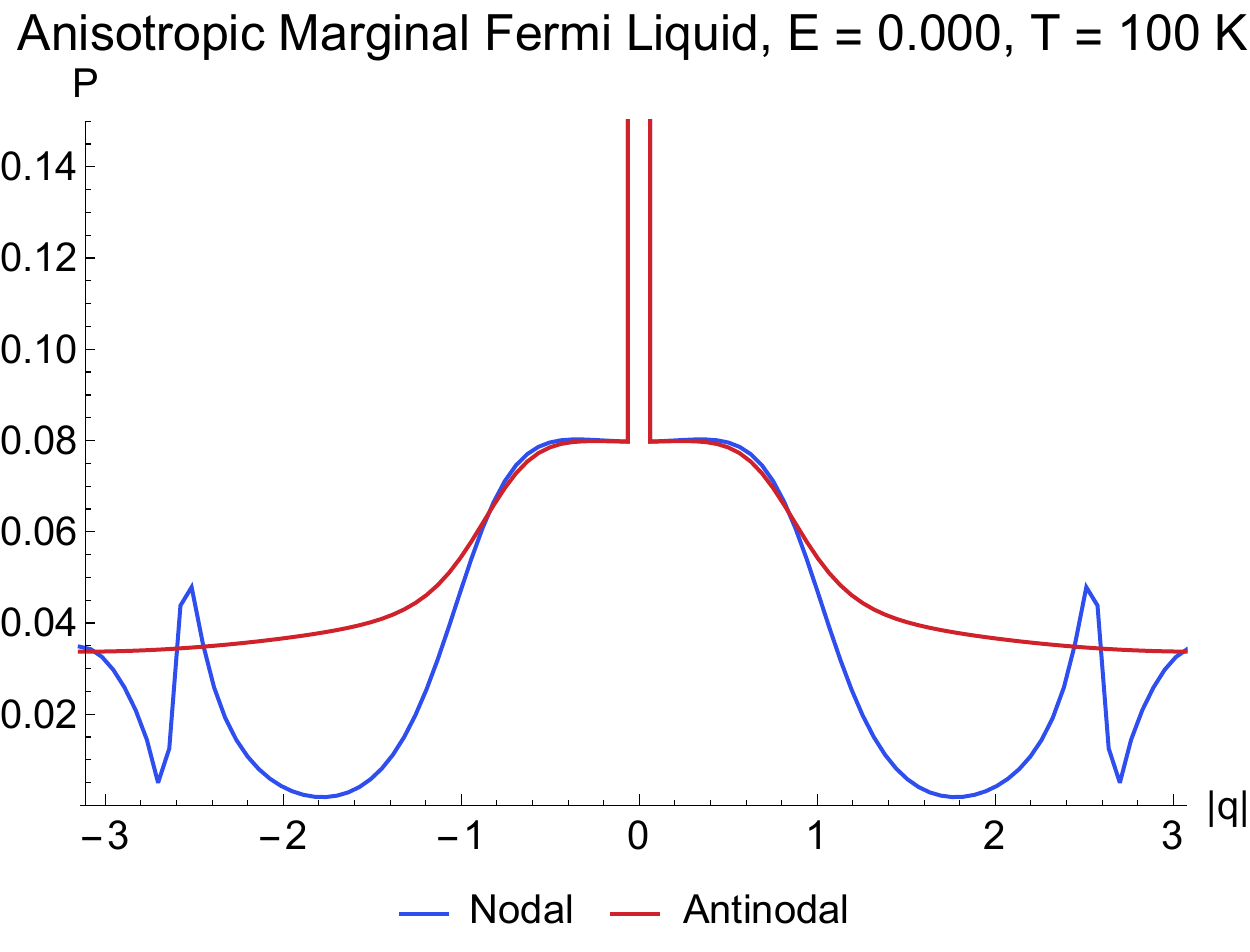}
	\includegraphics[height=0.18\textwidth]{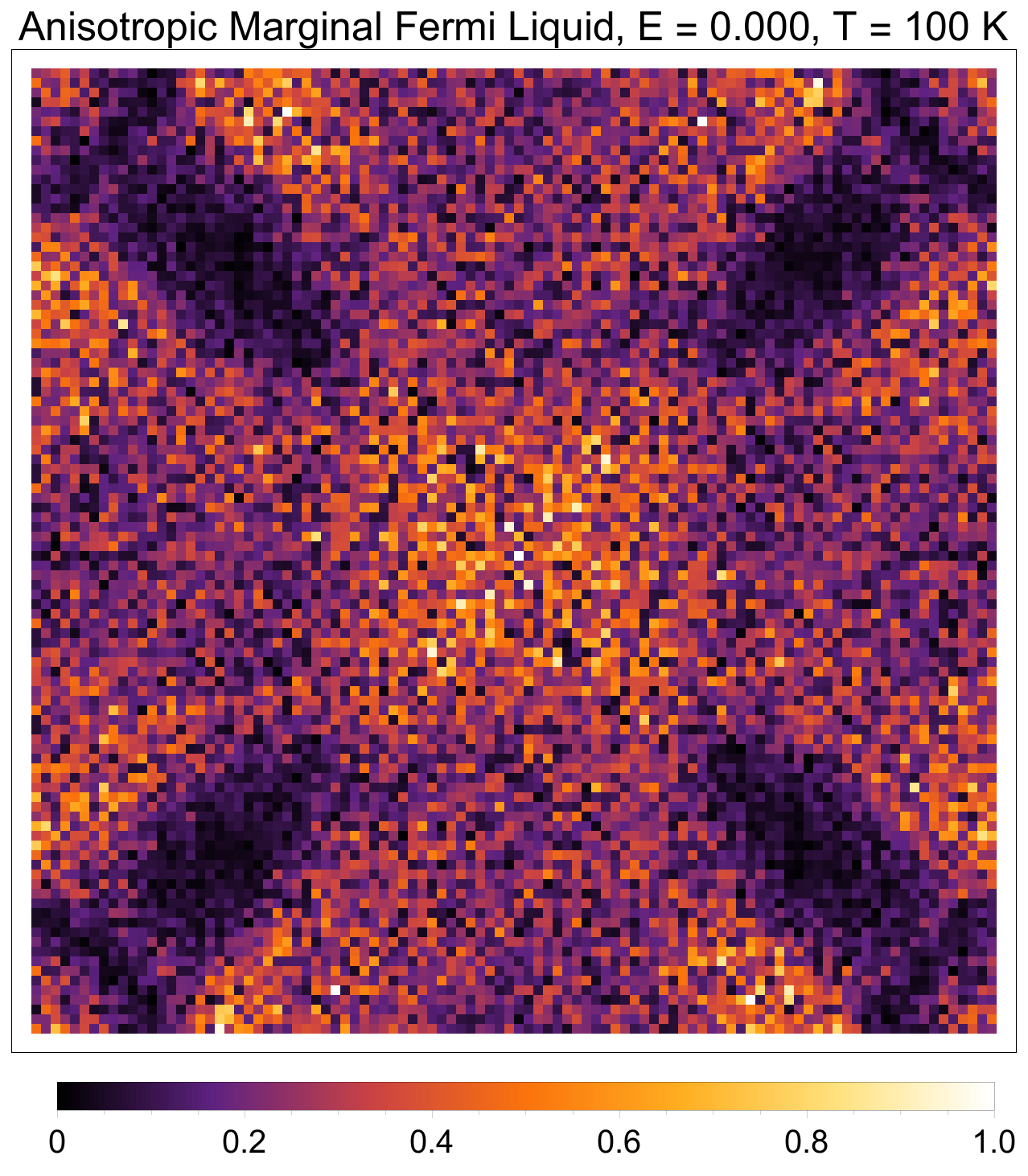}
	\includegraphics[height=0.18\textwidth]{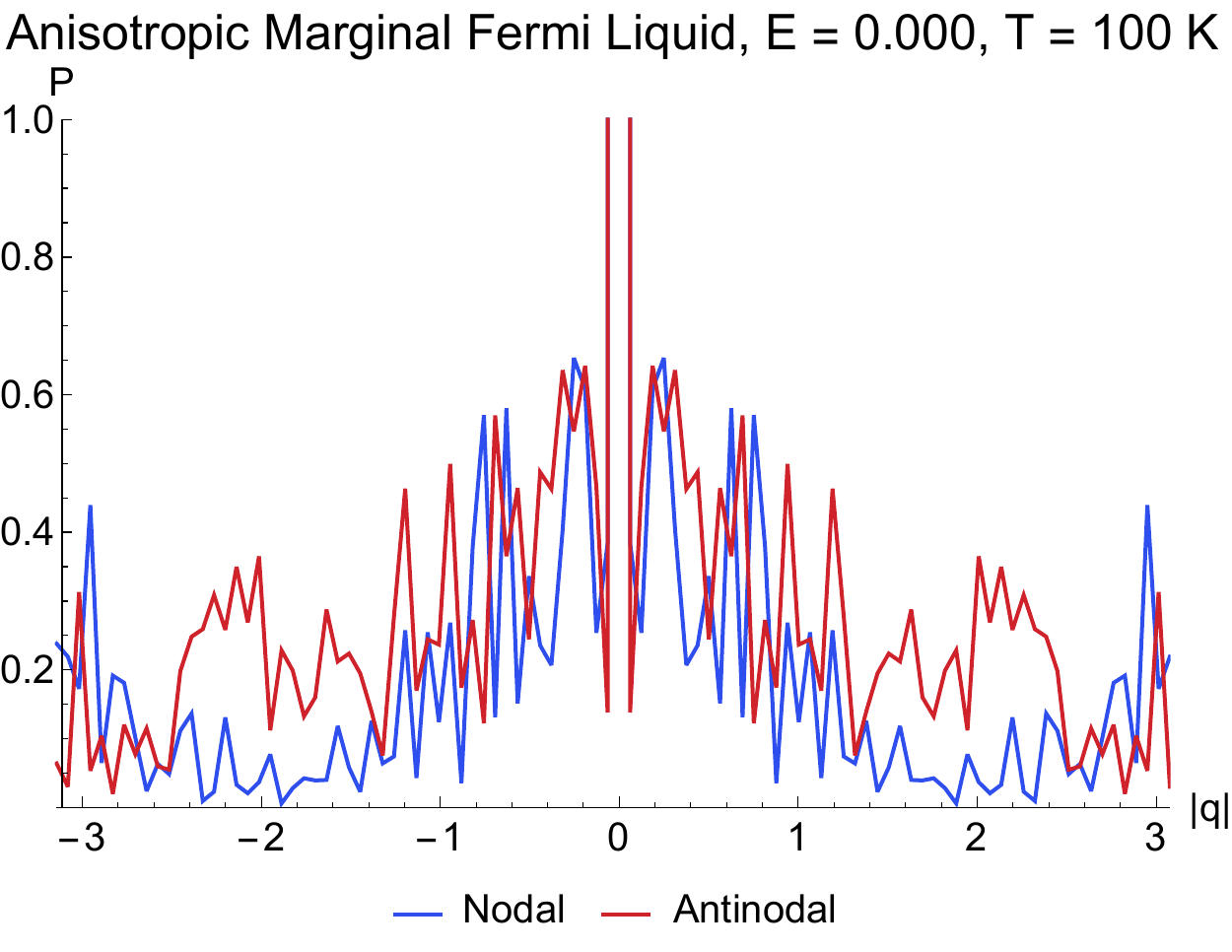} \\
	\includegraphics[height=0.18\textwidth]{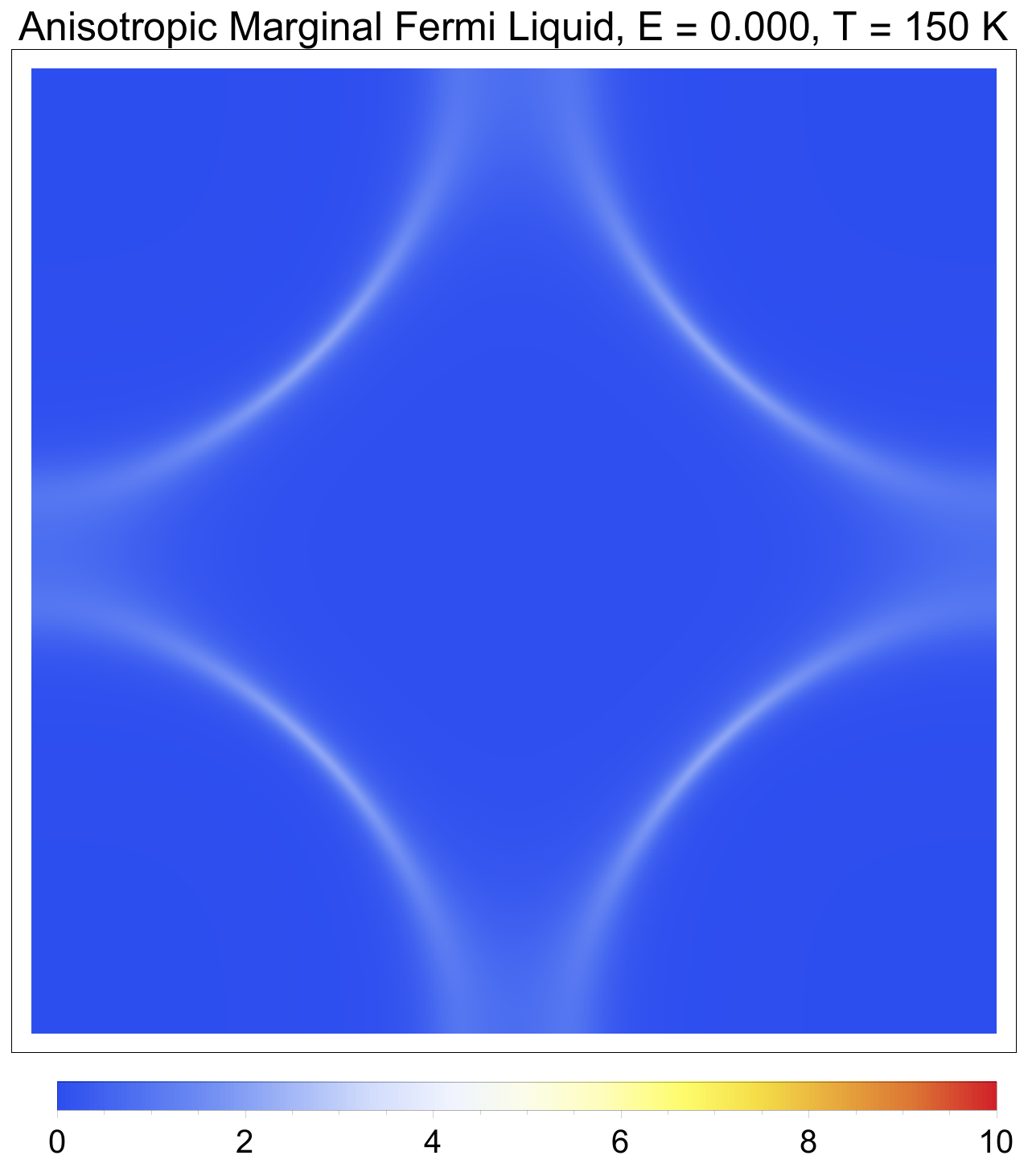}
	\includegraphics[height=0.18\textwidth]{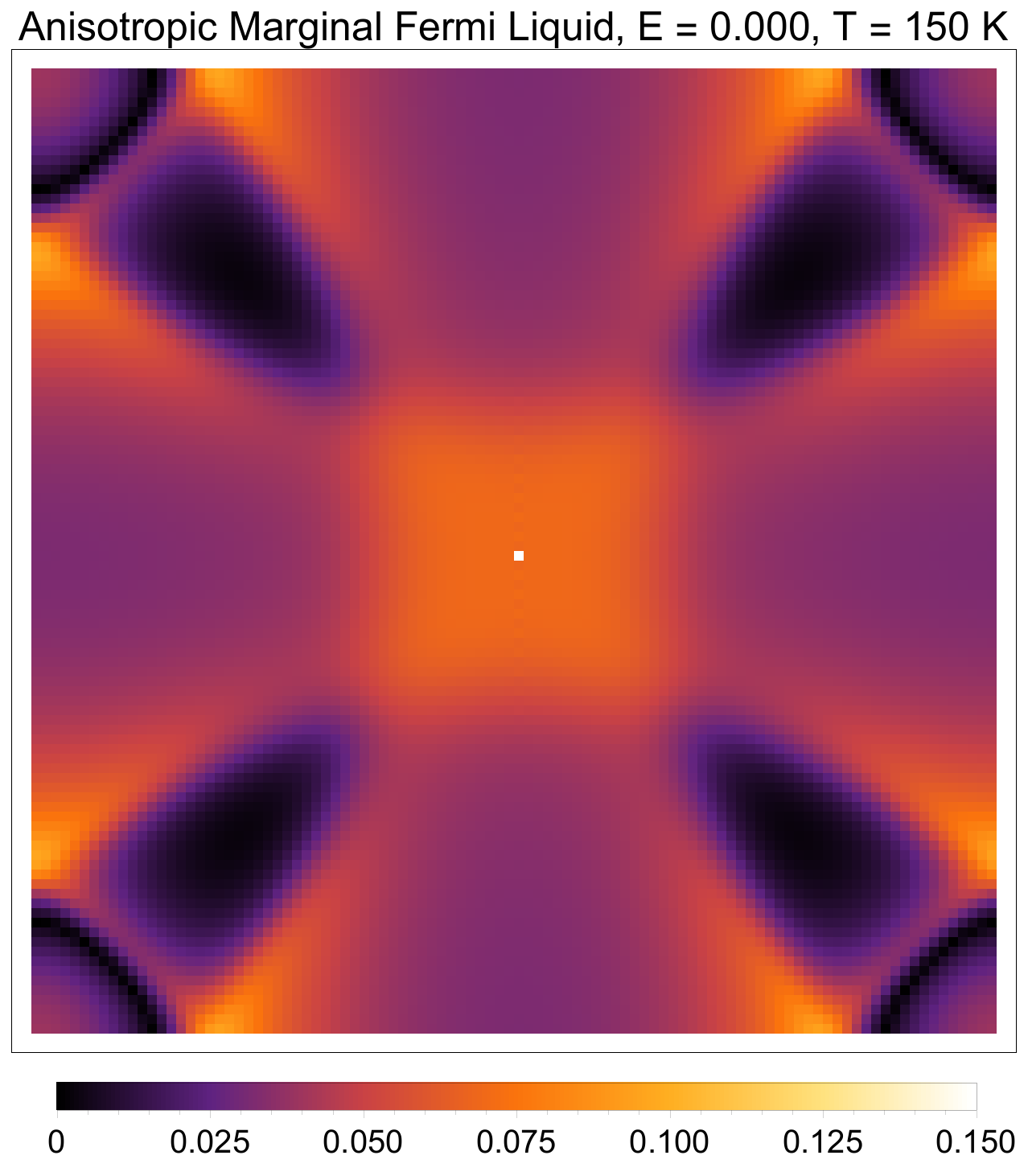}
	\includegraphics[height=0.18\textwidth]{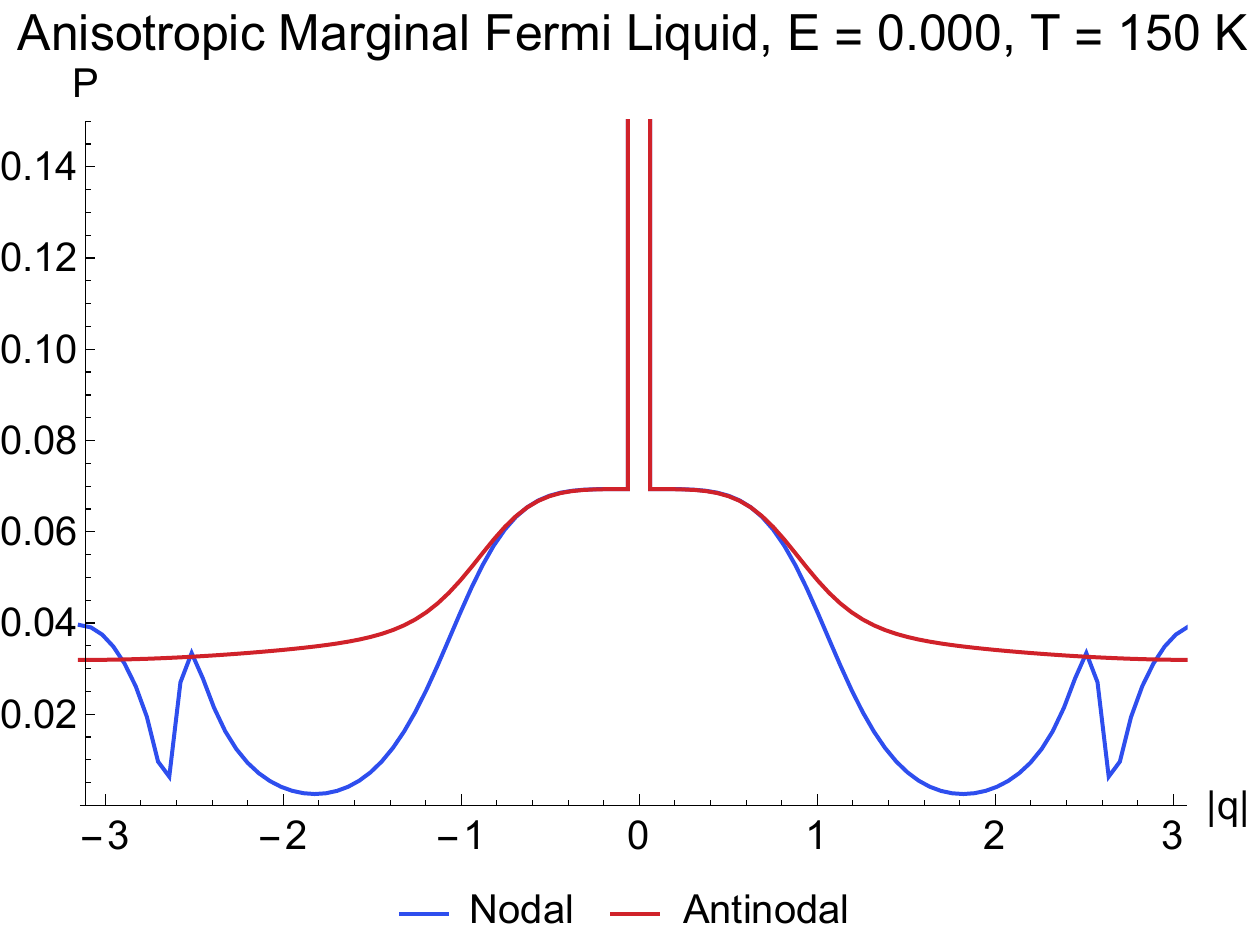}
	\includegraphics[height=0.18\textwidth]{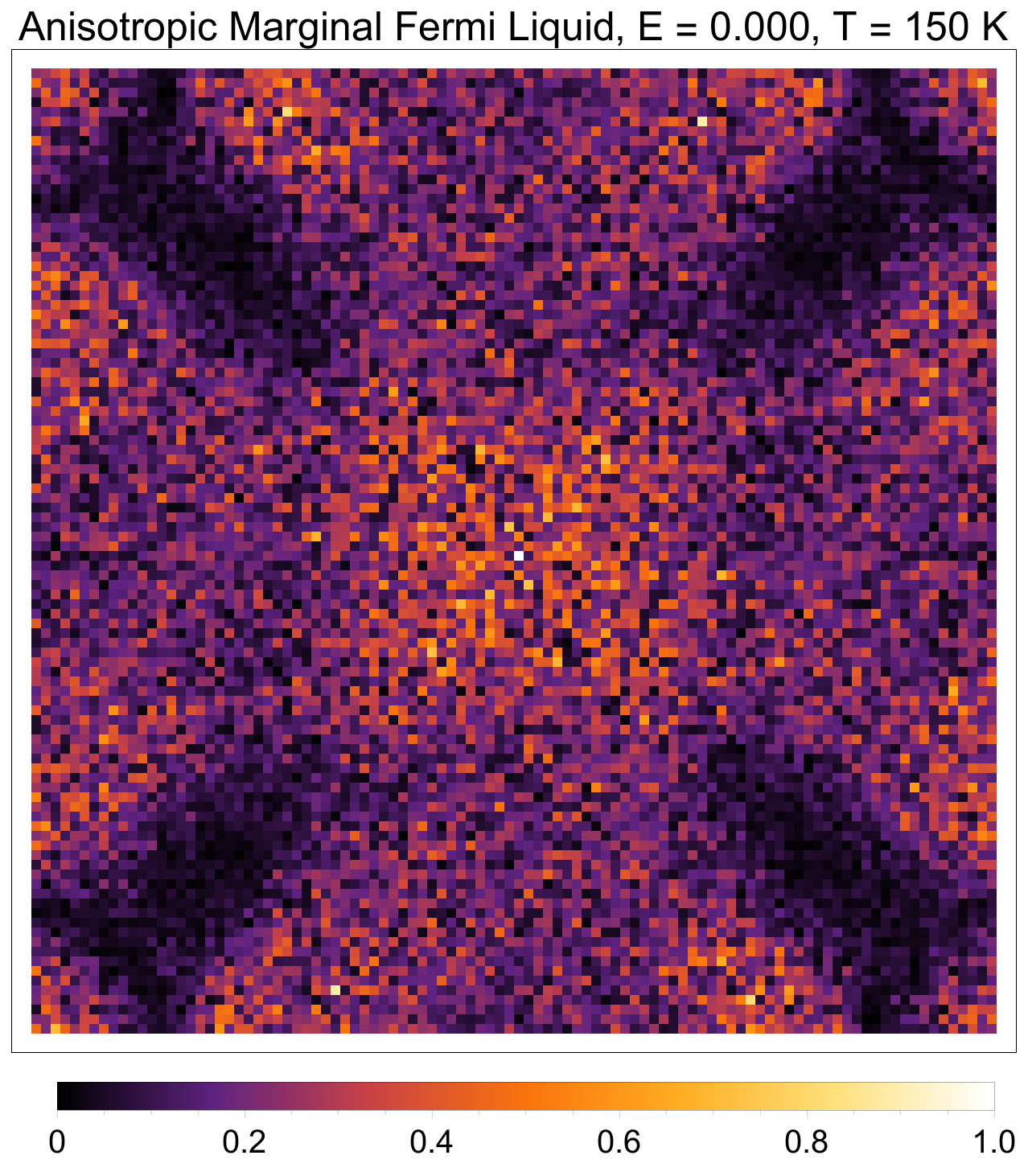}
	\includegraphics[height=0.18\textwidth]{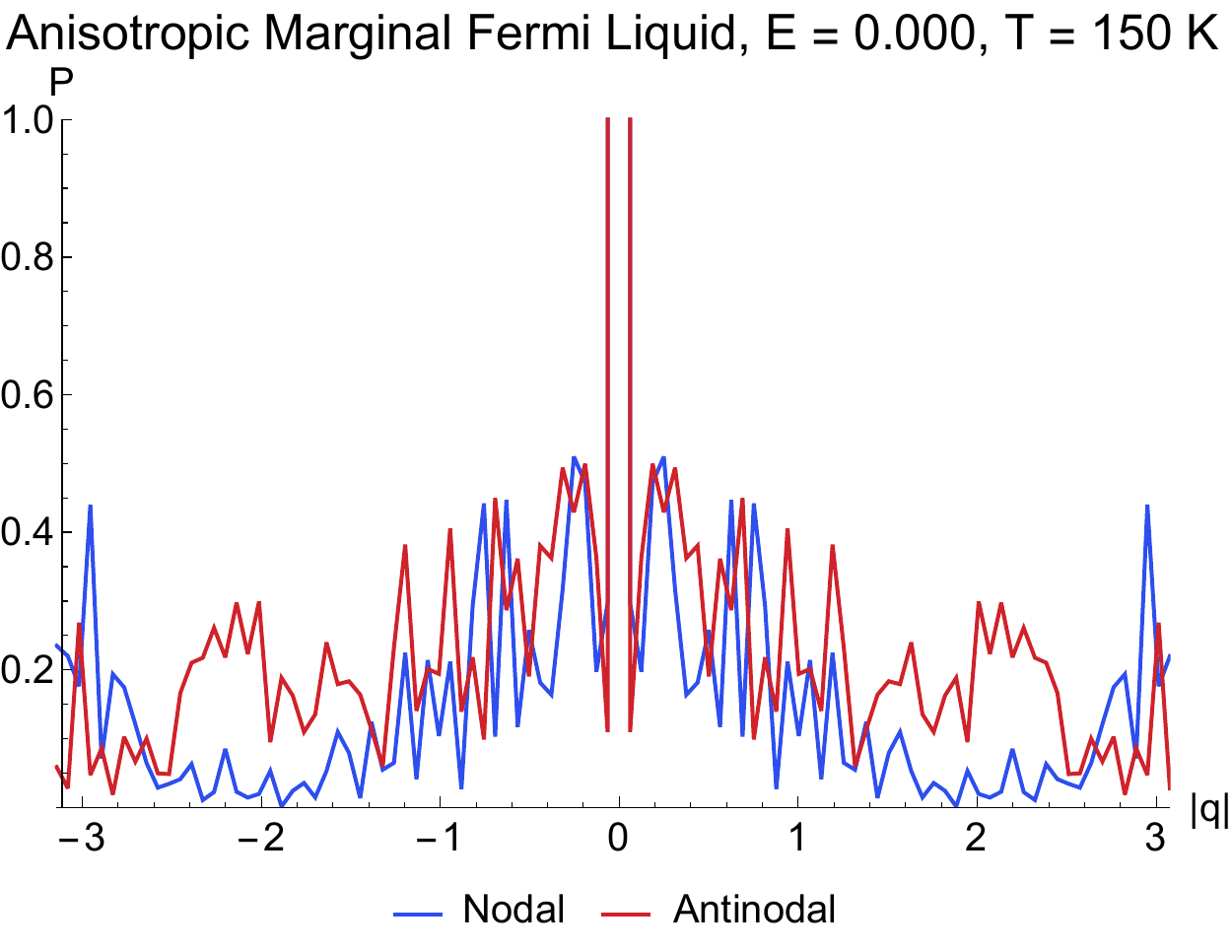} \\
	\includegraphics[height=0.18\textwidth]{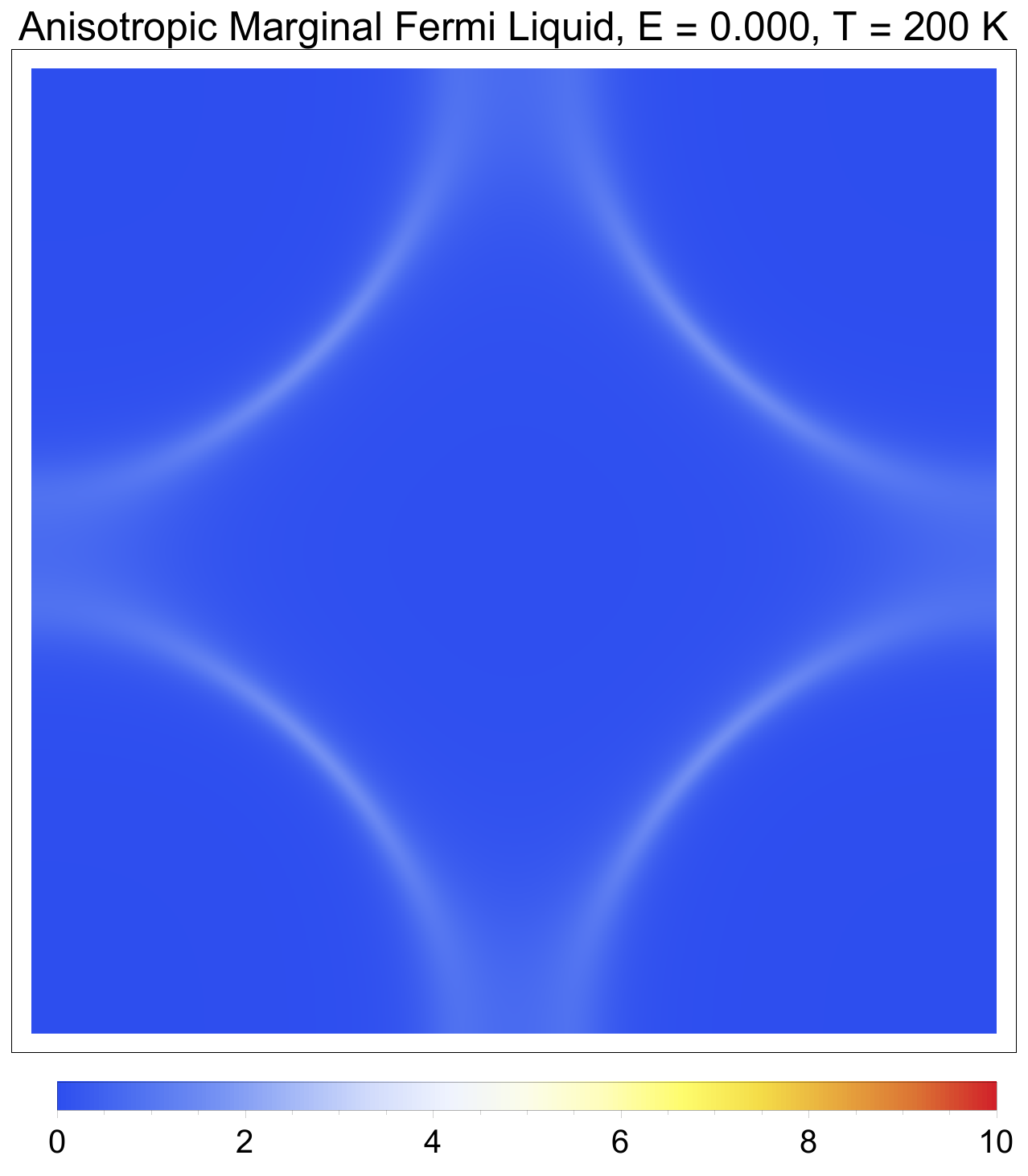}
	\includegraphics[height=0.18\textwidth]{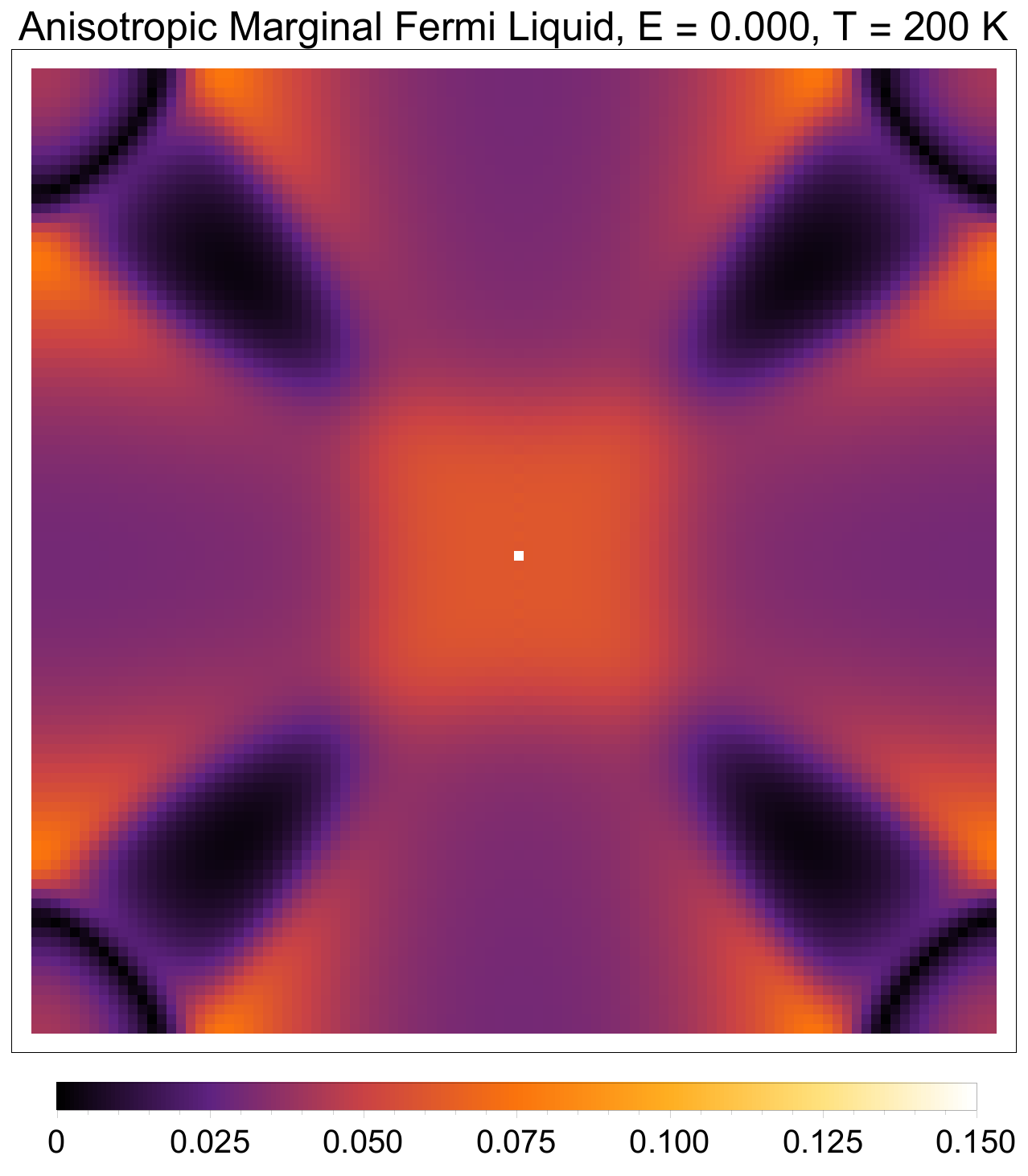}
	\includegraphics[height=0.18\textwidth]{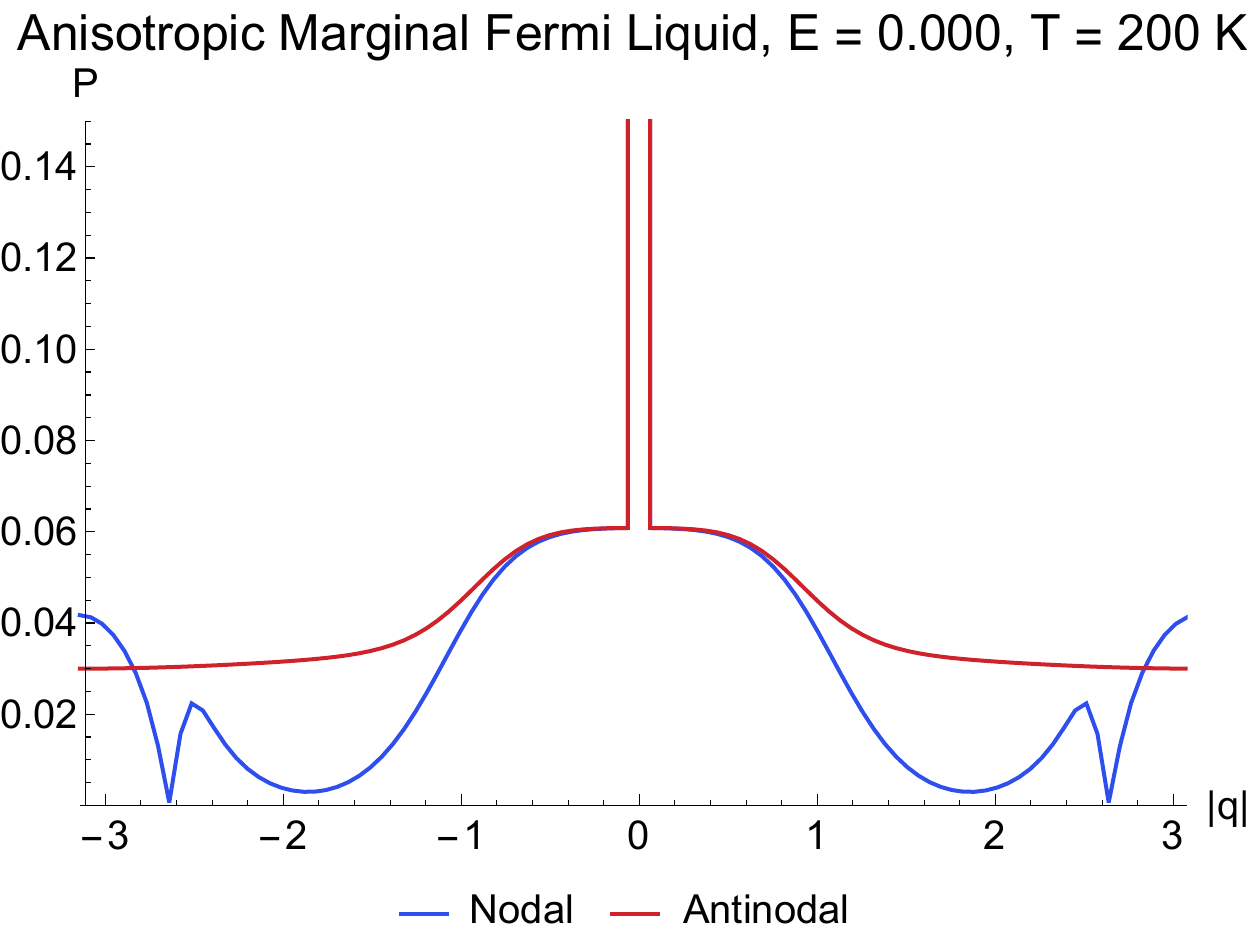}
	\includegraphics[height=0.18\textwidth]{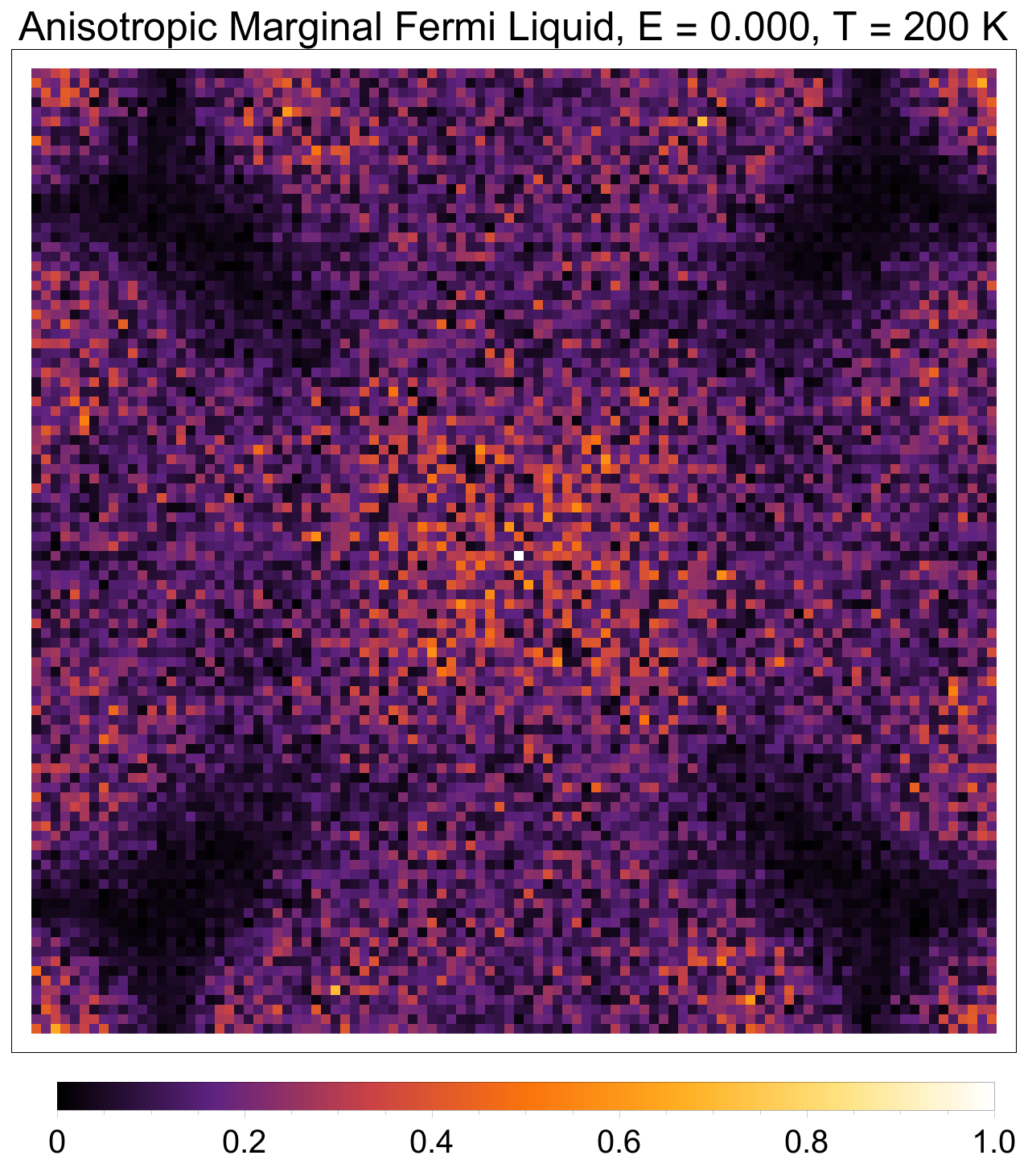}
	\includegraphics[height=0.18\textwidth]{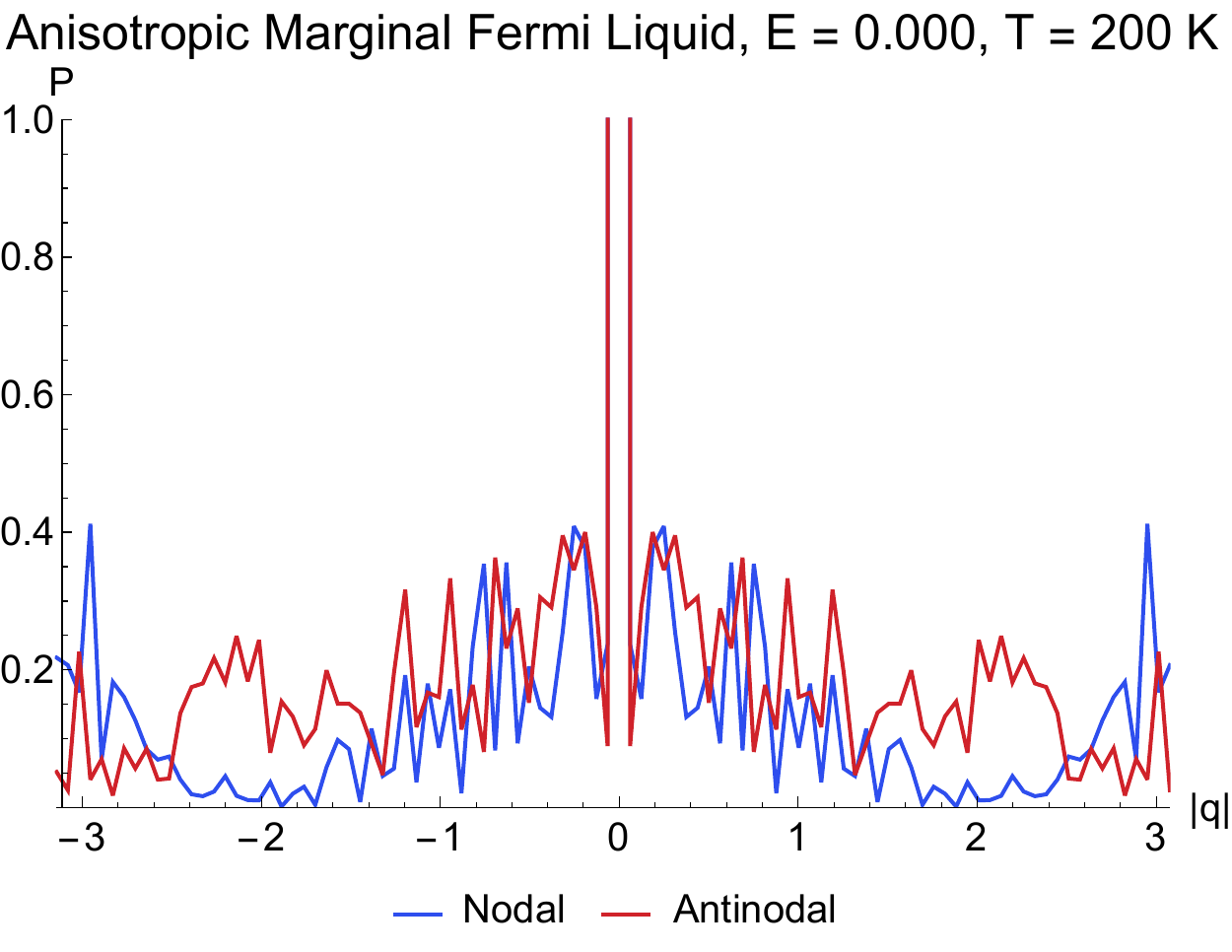} \\
	\includegraphics[height=0.18\textwidth]{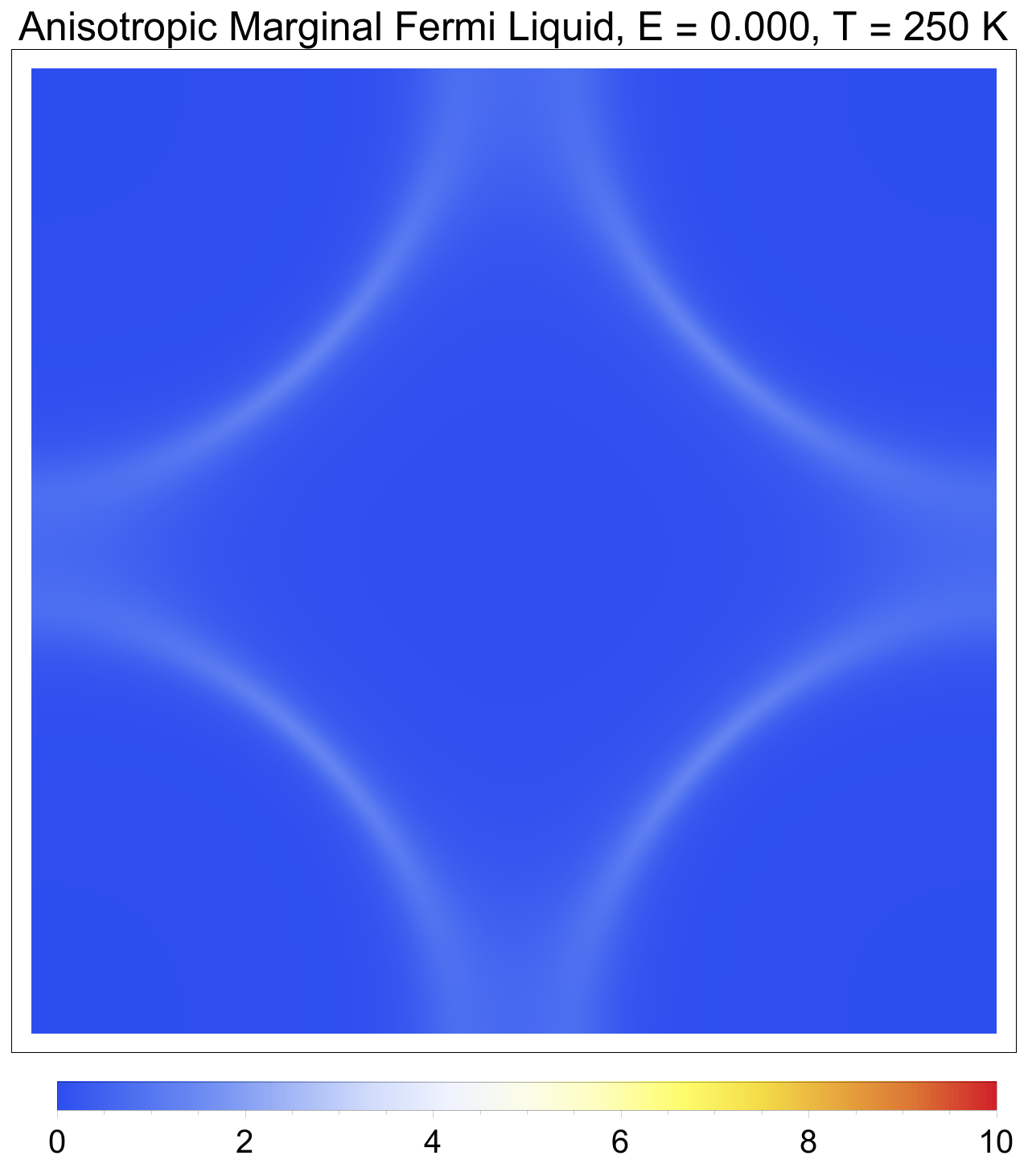}
	\includegraphics[height=0.18\textwidth]{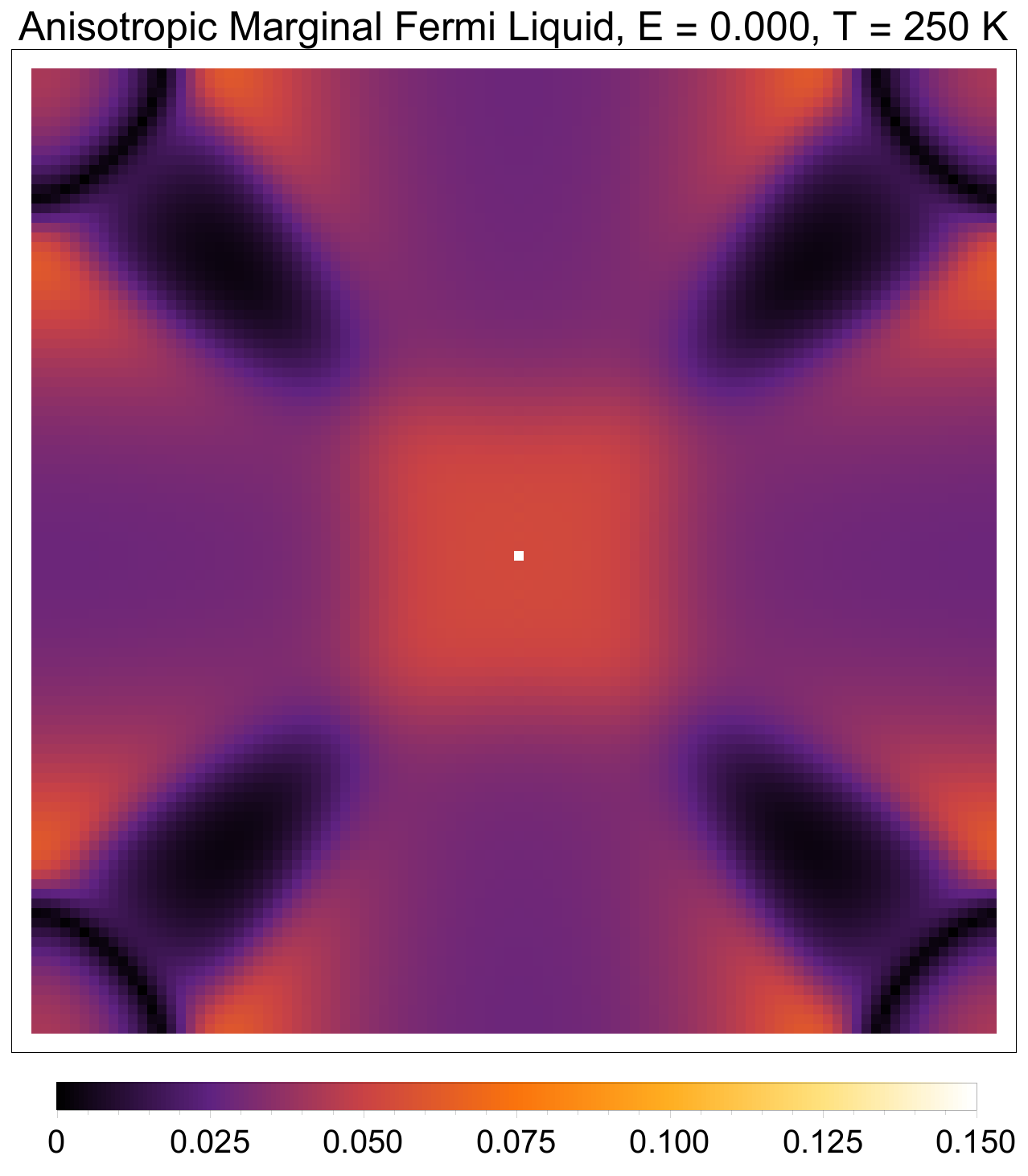}
	\includegraphics[height=0.18\textwidth]{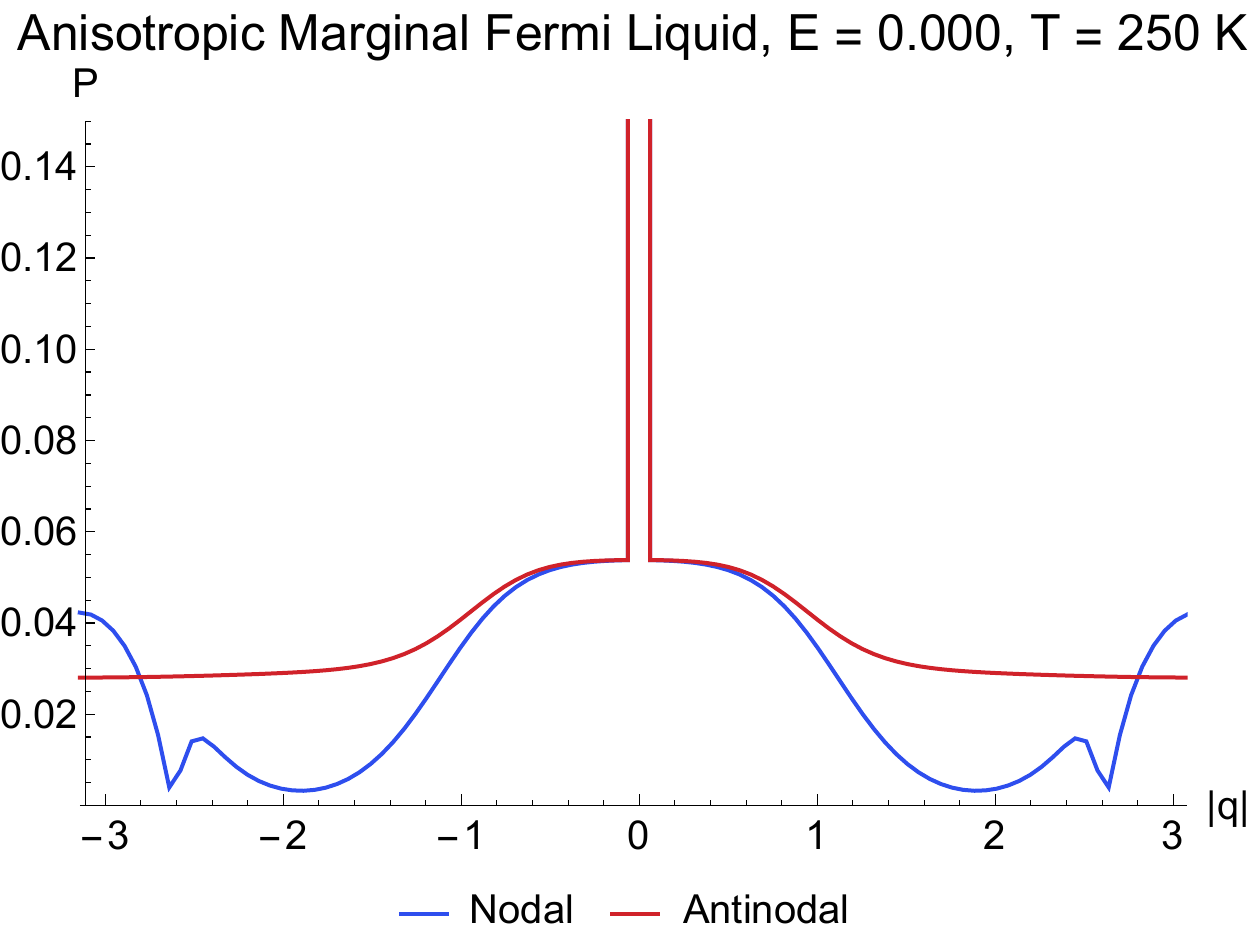}
	\includegraphics[height=0.18\textwidth]{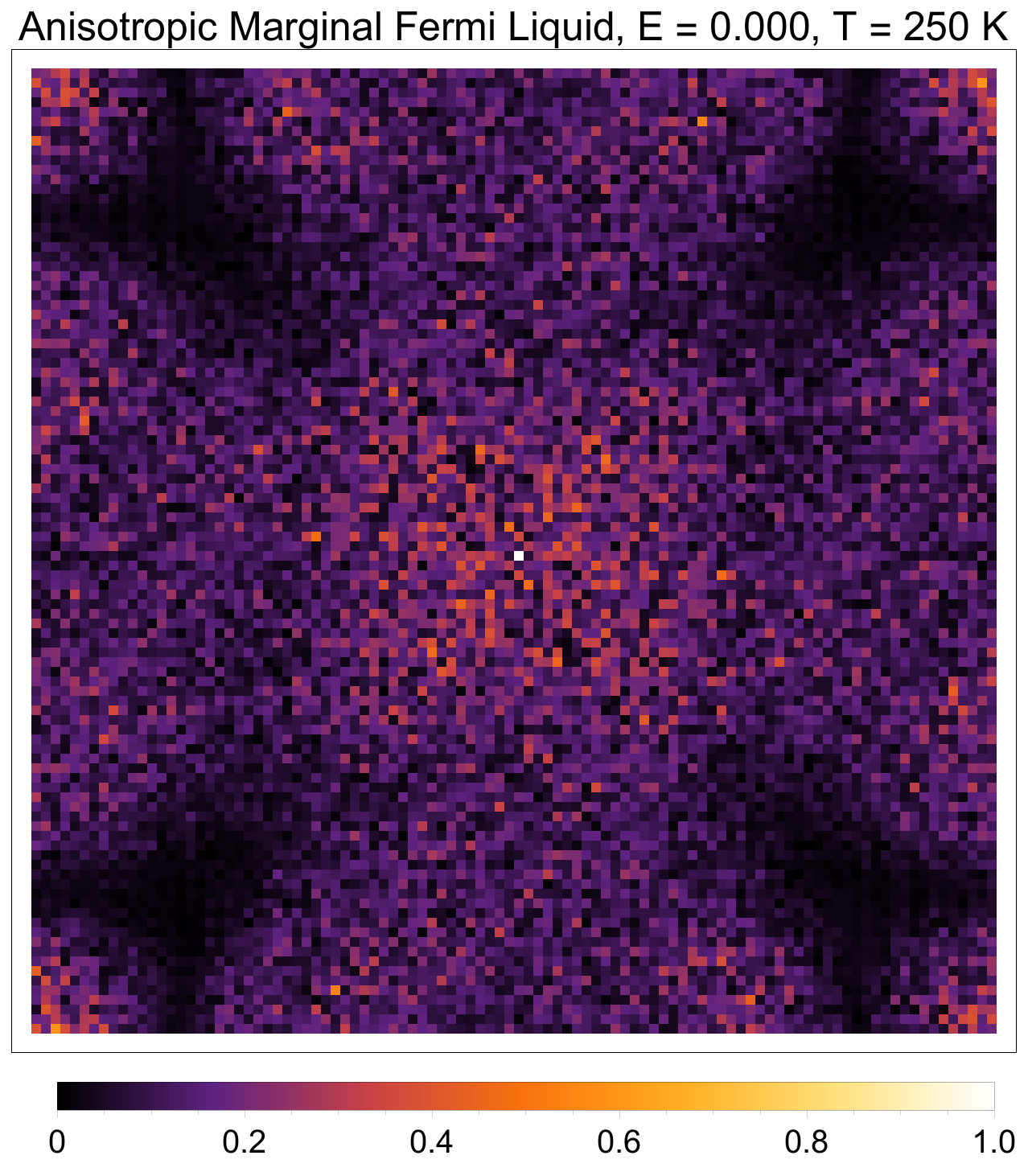}
	\includegraphics[height=0.18\textwidth]{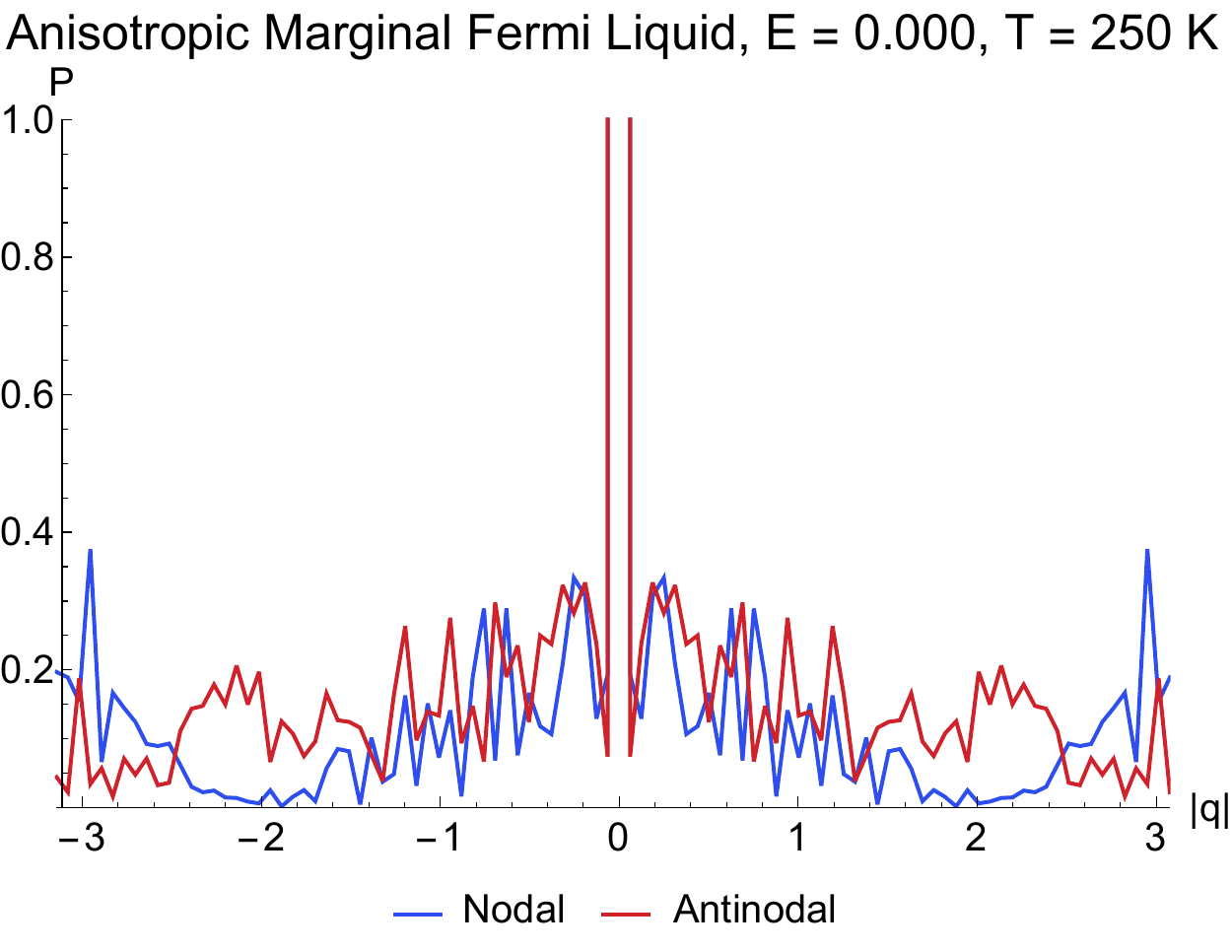} \\
	\includegraphics[height=0.18\textwidth]{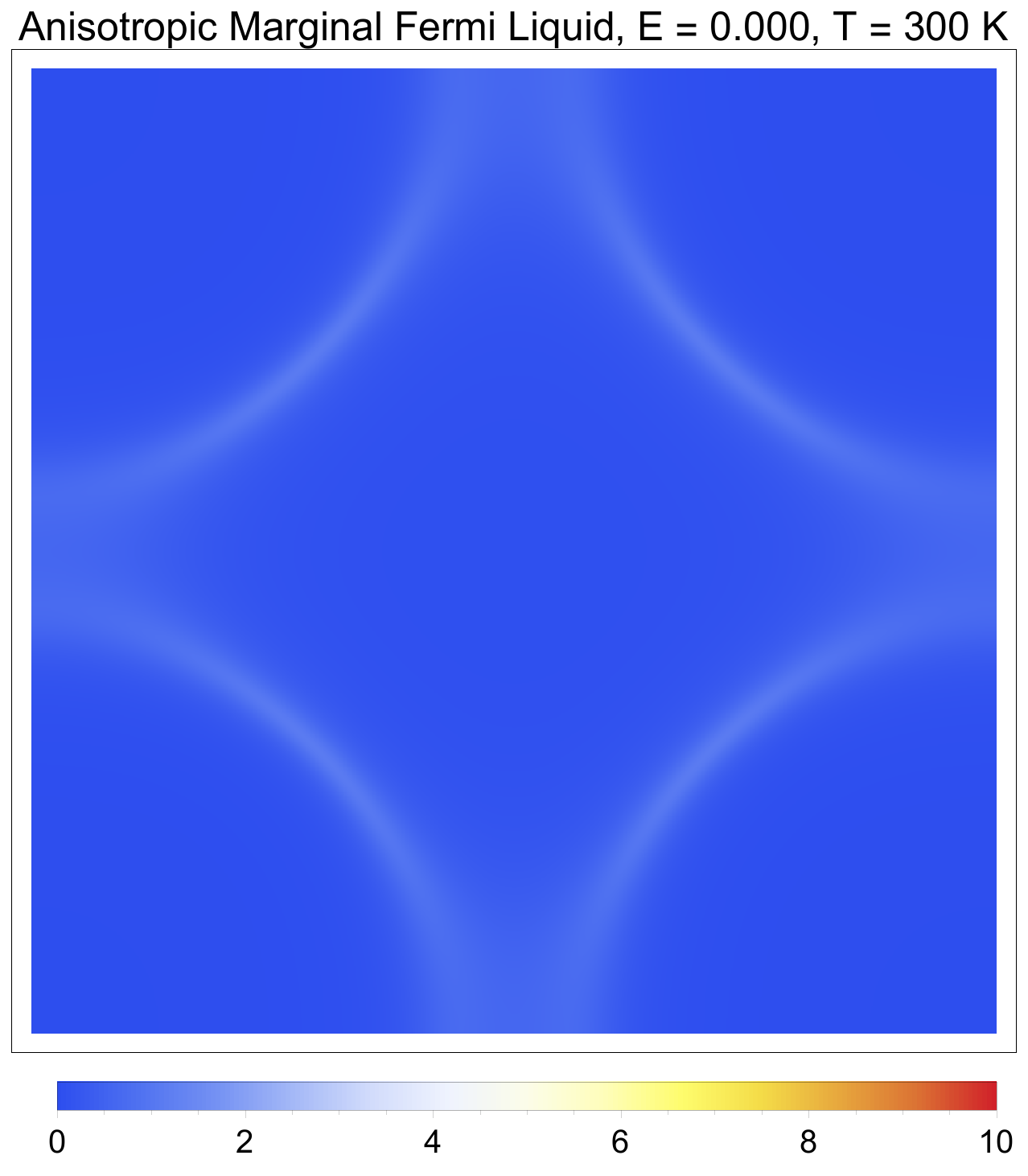}
	\includegraphics[height=0.18\textwidth]{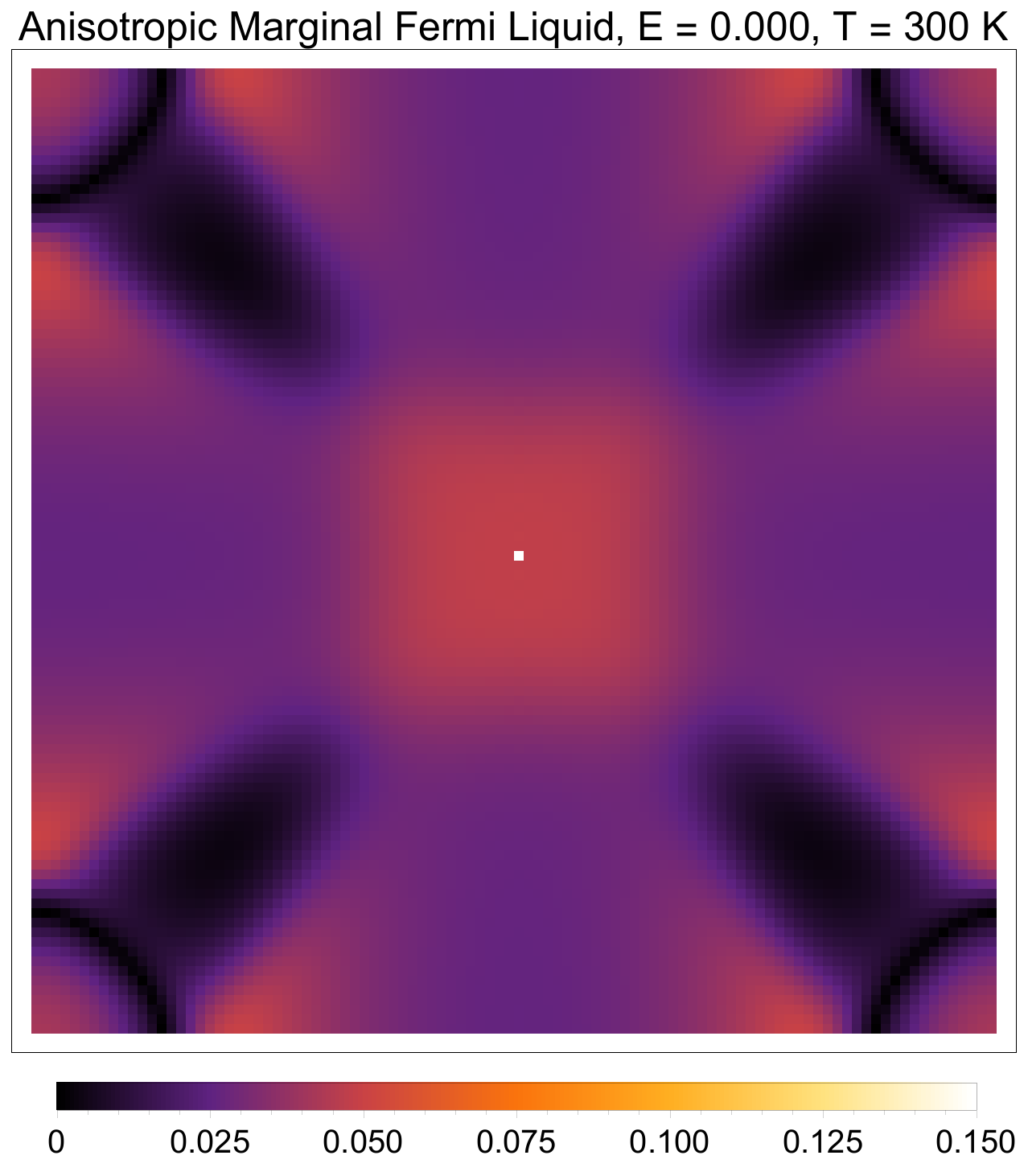}
	\includegraphics[height=0.18\textwidth]{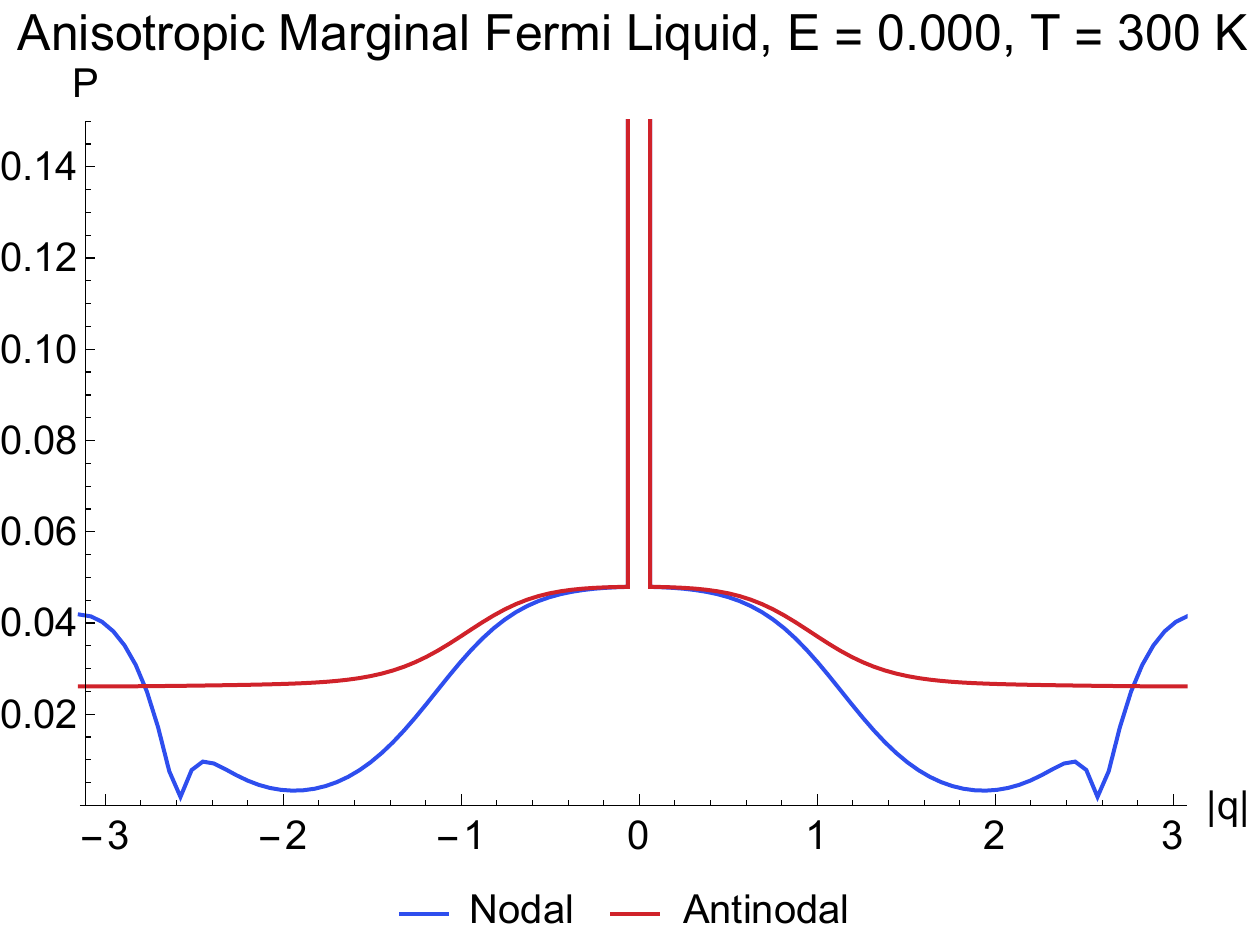}
	\includegraphics[height=0.18\textwidth]{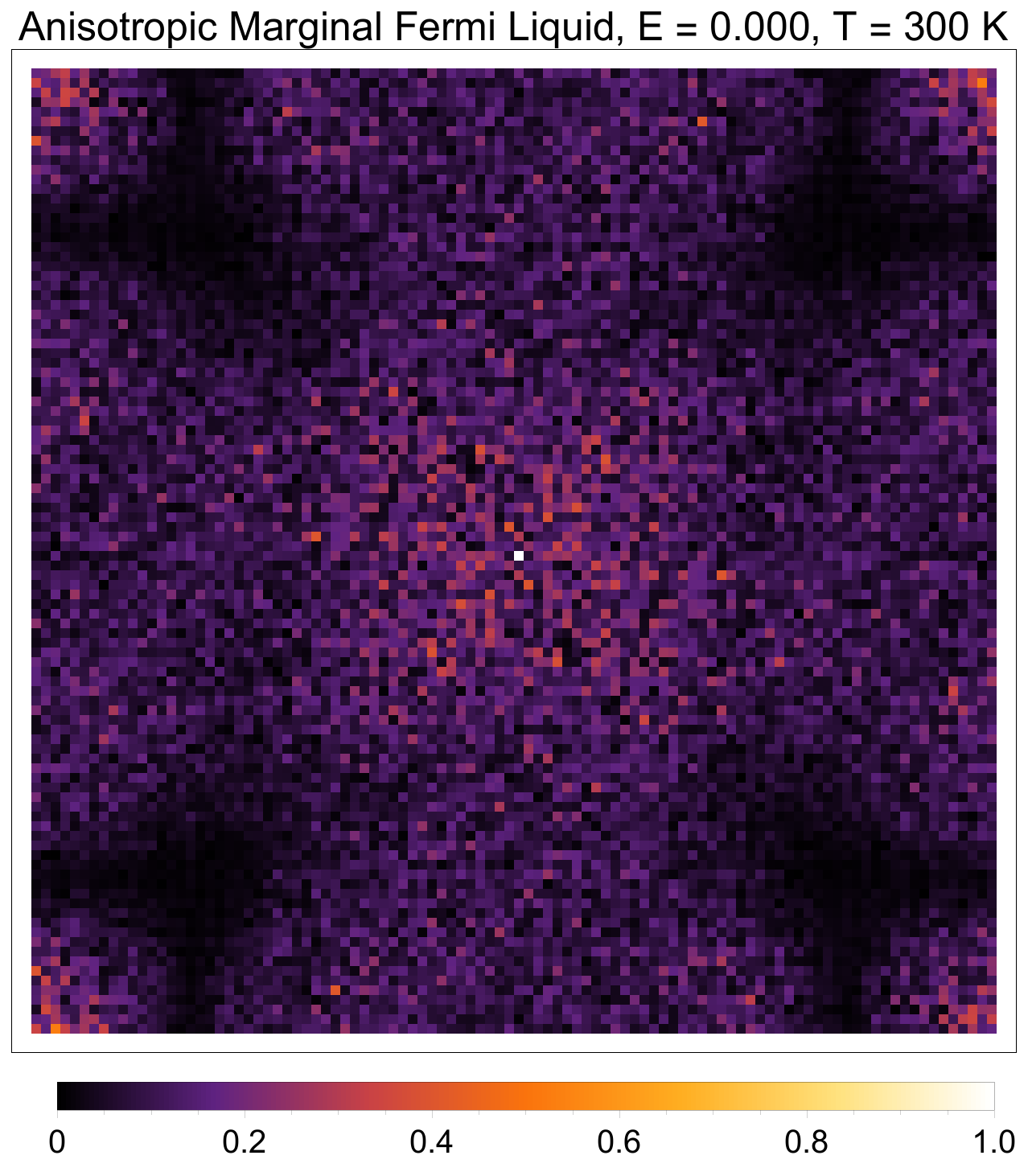}
	\includegraphics[height=0.18\textwidth]{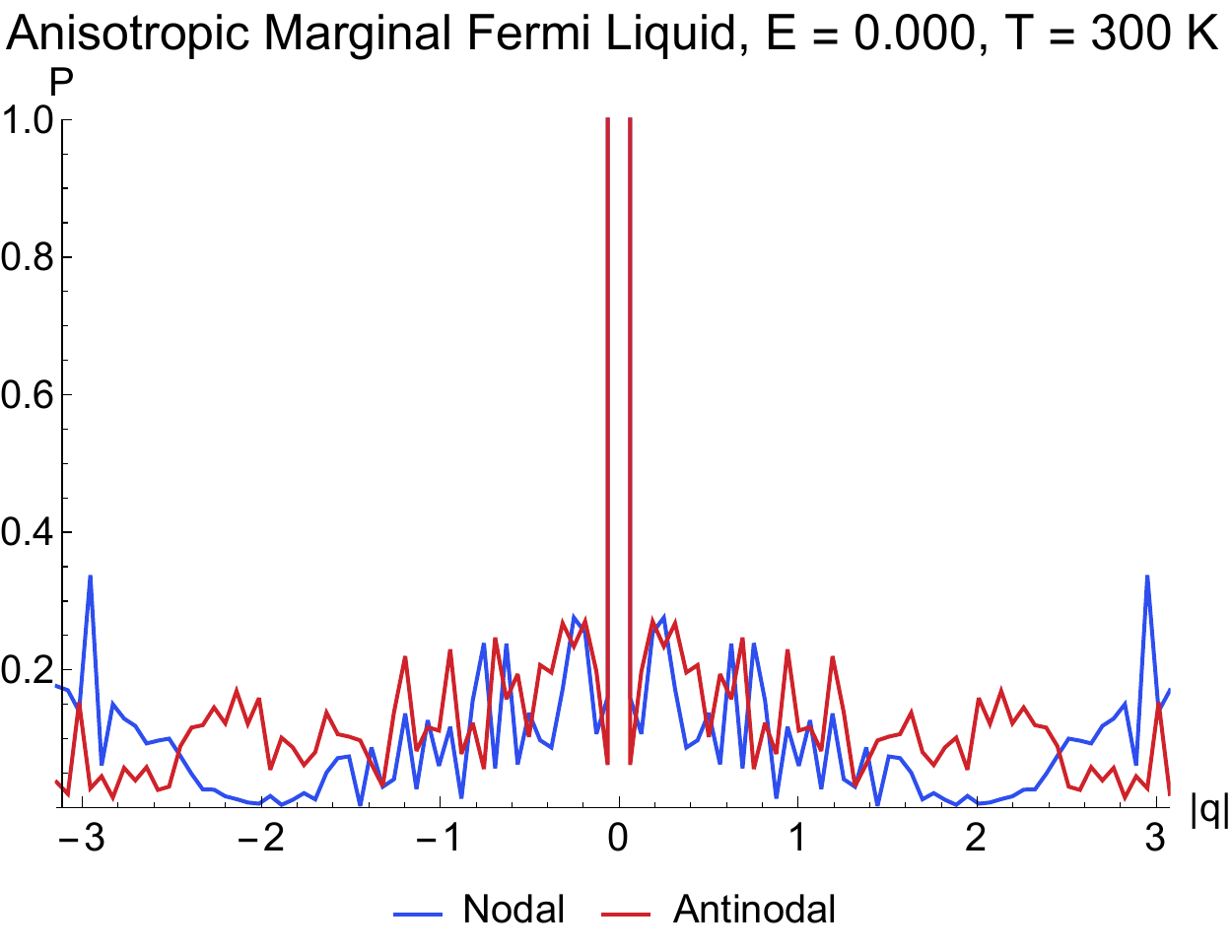} \\
	
	\caption{The spectra of an anisotropic marginal Fermi liquid at various temperatures.  Left to right: The spectral function $A(\mathbf{k}, \omega)$; the Fourier transform of the LDOS $P(\mathbf{q}, \omega)$; linecuts of $P(\mathbf{q}, \omega)$ in the nodal and antinodal directions; $P(\mathbf{q}, \omega)$ with finite-temperature smearing; and $P(\mathbf{q}, \omega)$ in the presence of many weak impurities. All plots are taken at $E = 0.000$. For ease of visualization, the scale used here is smaller than that used in the Fermi liquid and isotropic marginal Fermi liquid plots.}
	
	\label{fig:temperature_nan}
\end{figure*}

\begin{figure}
	\centering
	\includegraphics[width=0.4\textwidth]{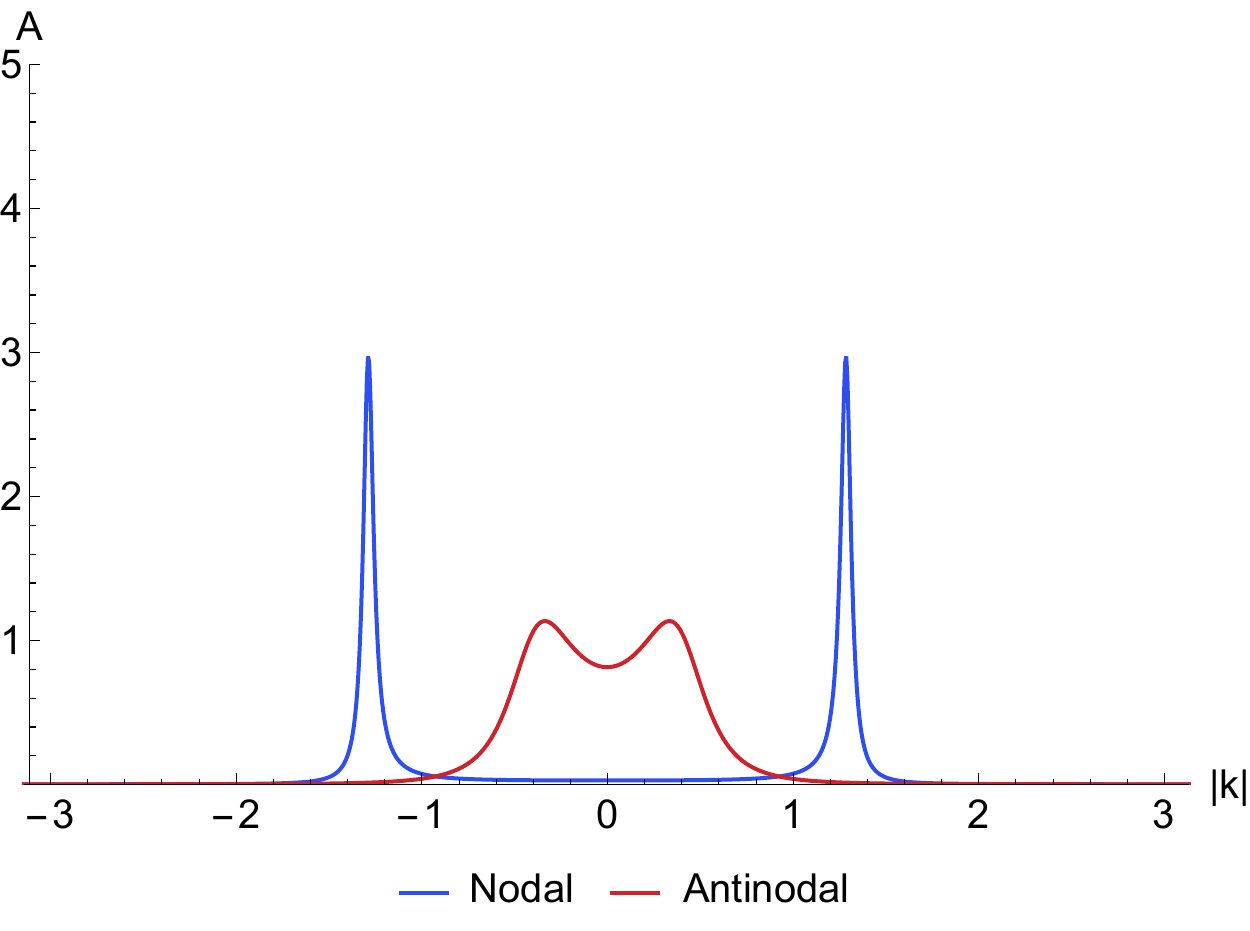}
	\caption{Momentum-distribution curves taken along nodal ($(0,0) \to (\pi,\pi)$) and antinodal ($(0,0) \to (0,\pi)$) cuts in the Brillouin zone at the Fermi energy ($E = 0$) for an anisotropic marginal Fermi liquid. Here $T = 100$ K and $\beta = 0.2$ (see Eq.~\ref{eq:anisotropic_se2} for the functional form of the self-energy). } 
	\label{fig:nan_mdc}
\end{figure}

\begin{figure*}
	\centering
	
	\includegraphics[width=0.16\textwidth]{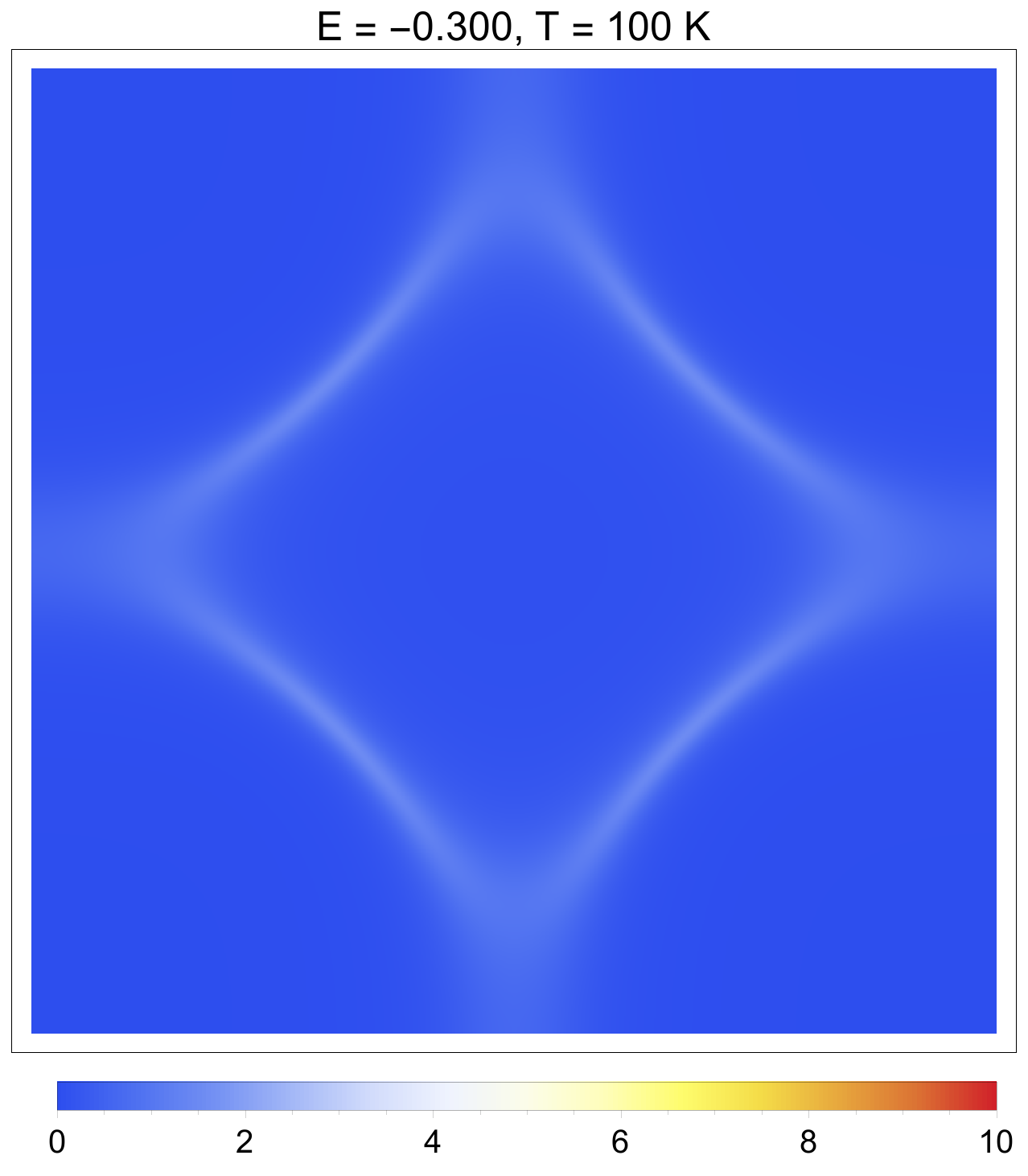}
	\includegraphics[width=0.16\textwidth]{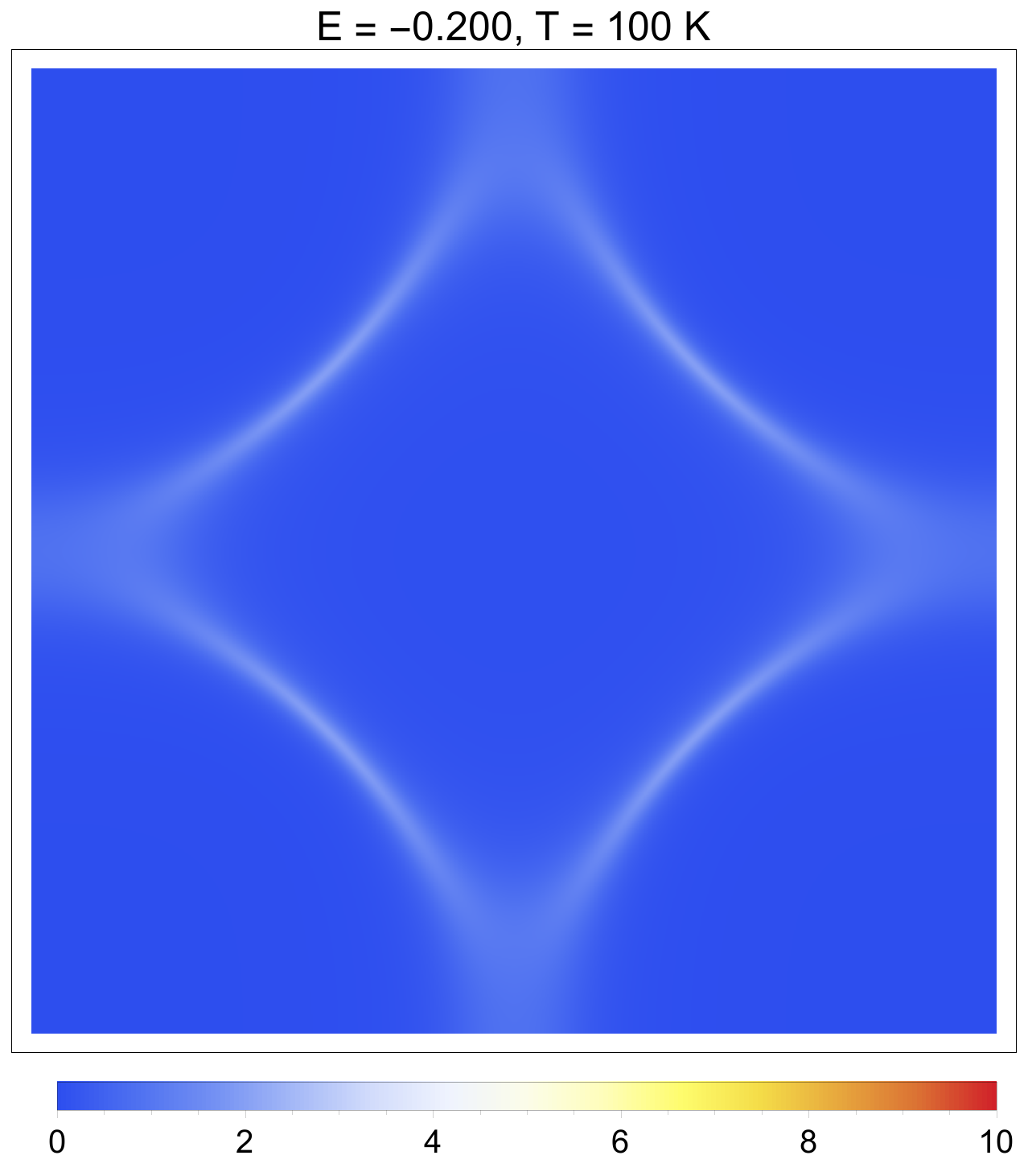}
	\includegraphics[width=0.16\textwidth]{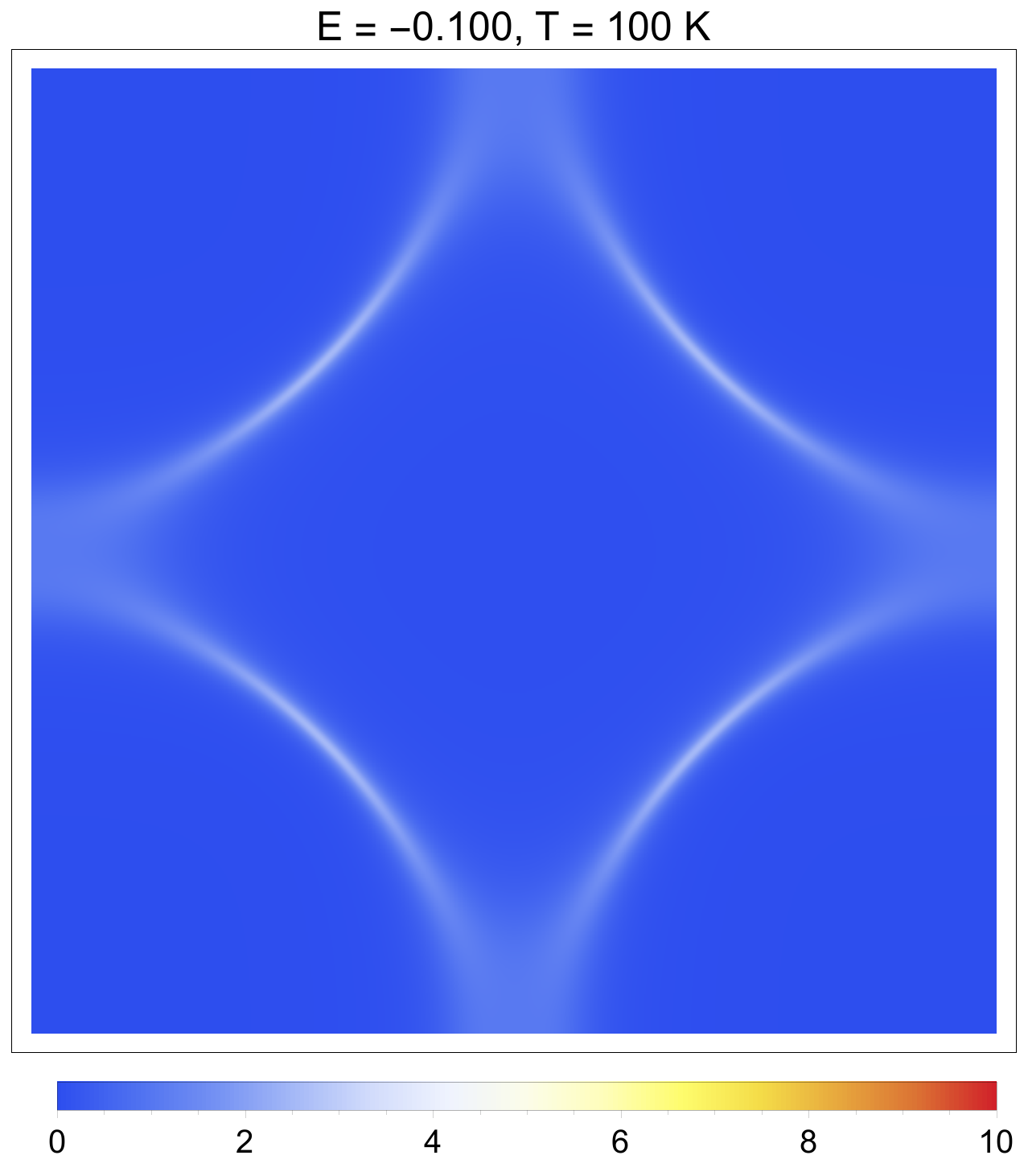}
	\includegraphics[width=0.16\textwidth]{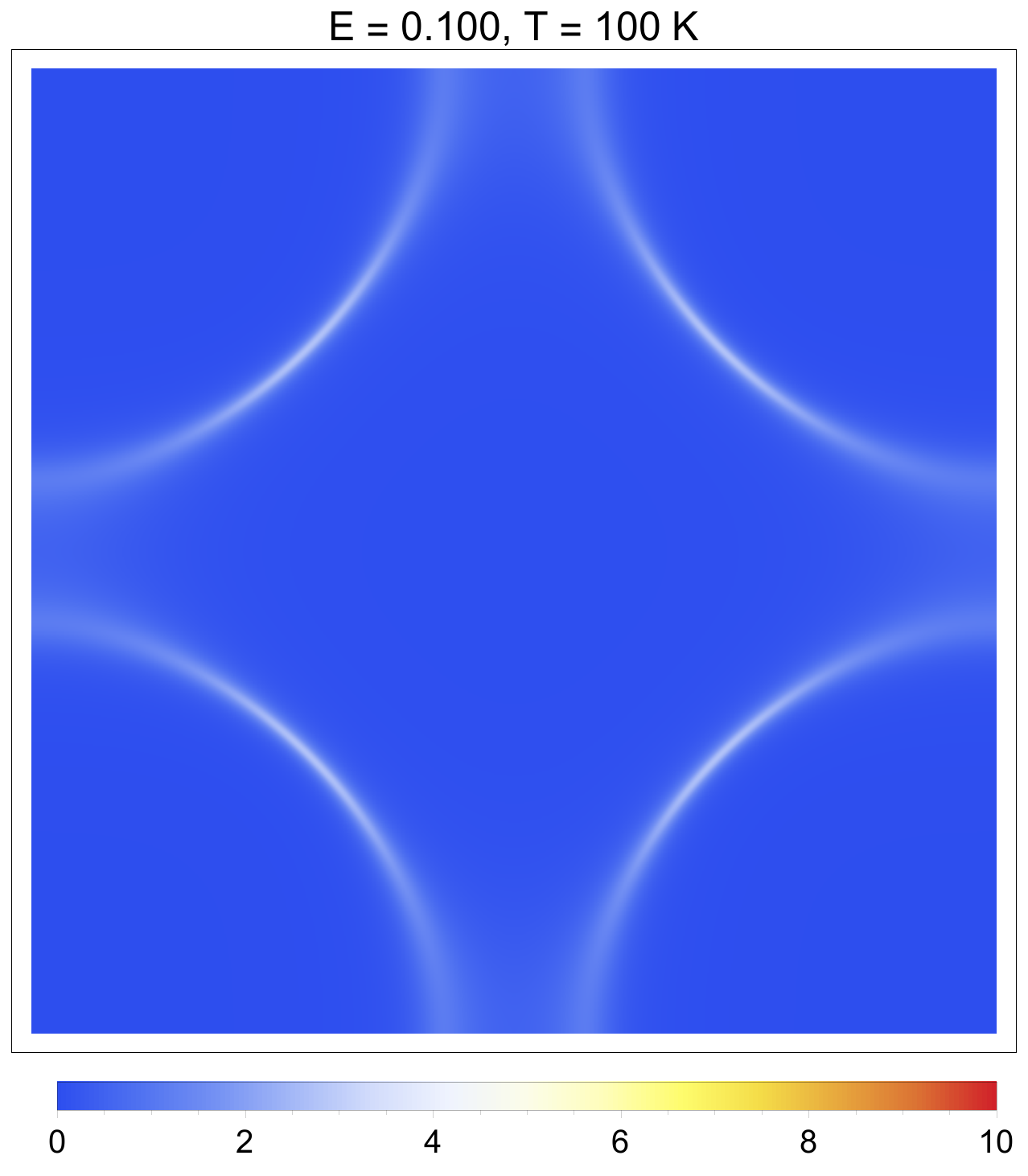}
	\includegraphics[width=0.16\textwidth]{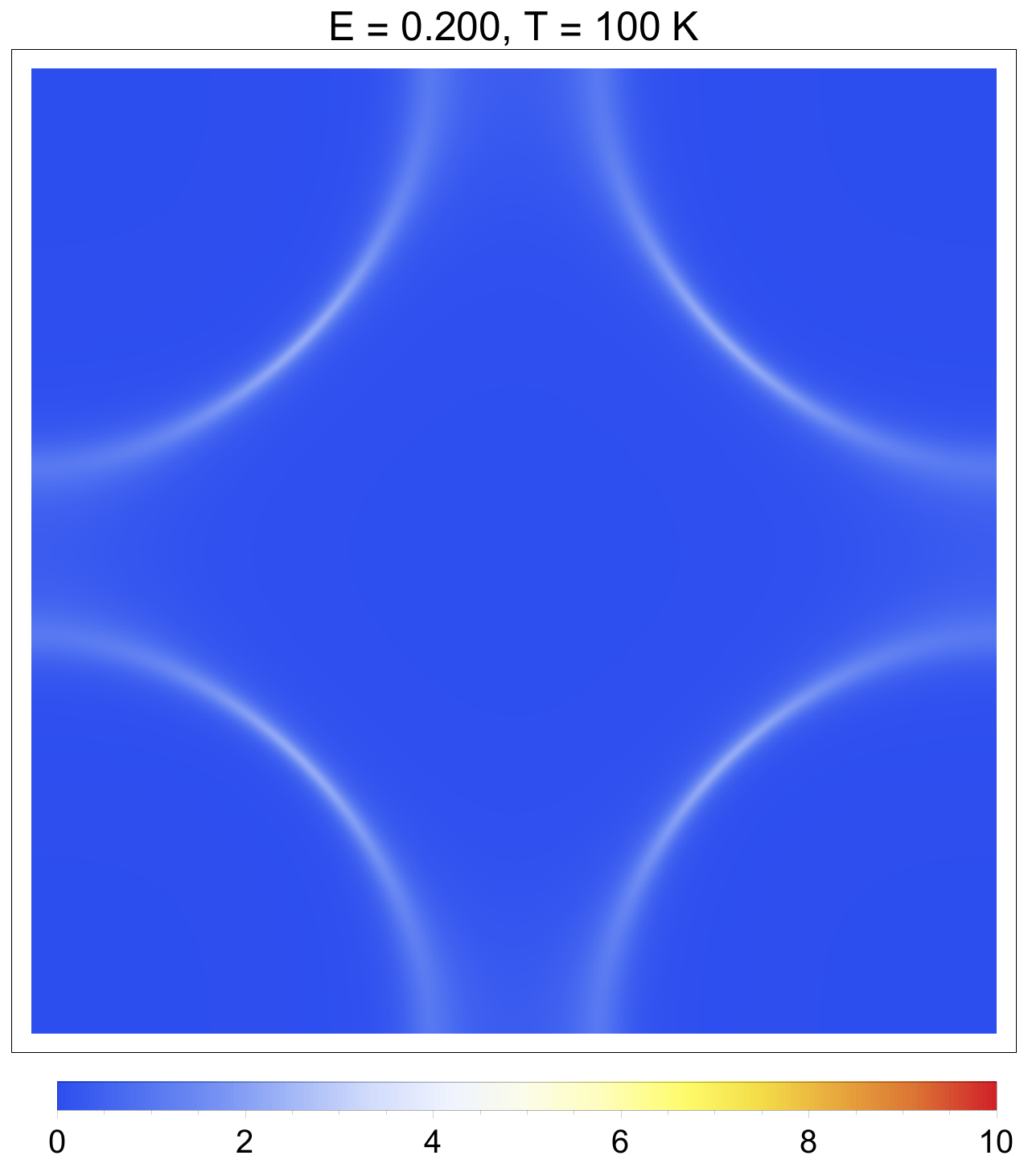}
	\includegraphics[width=0.16\textwidth]{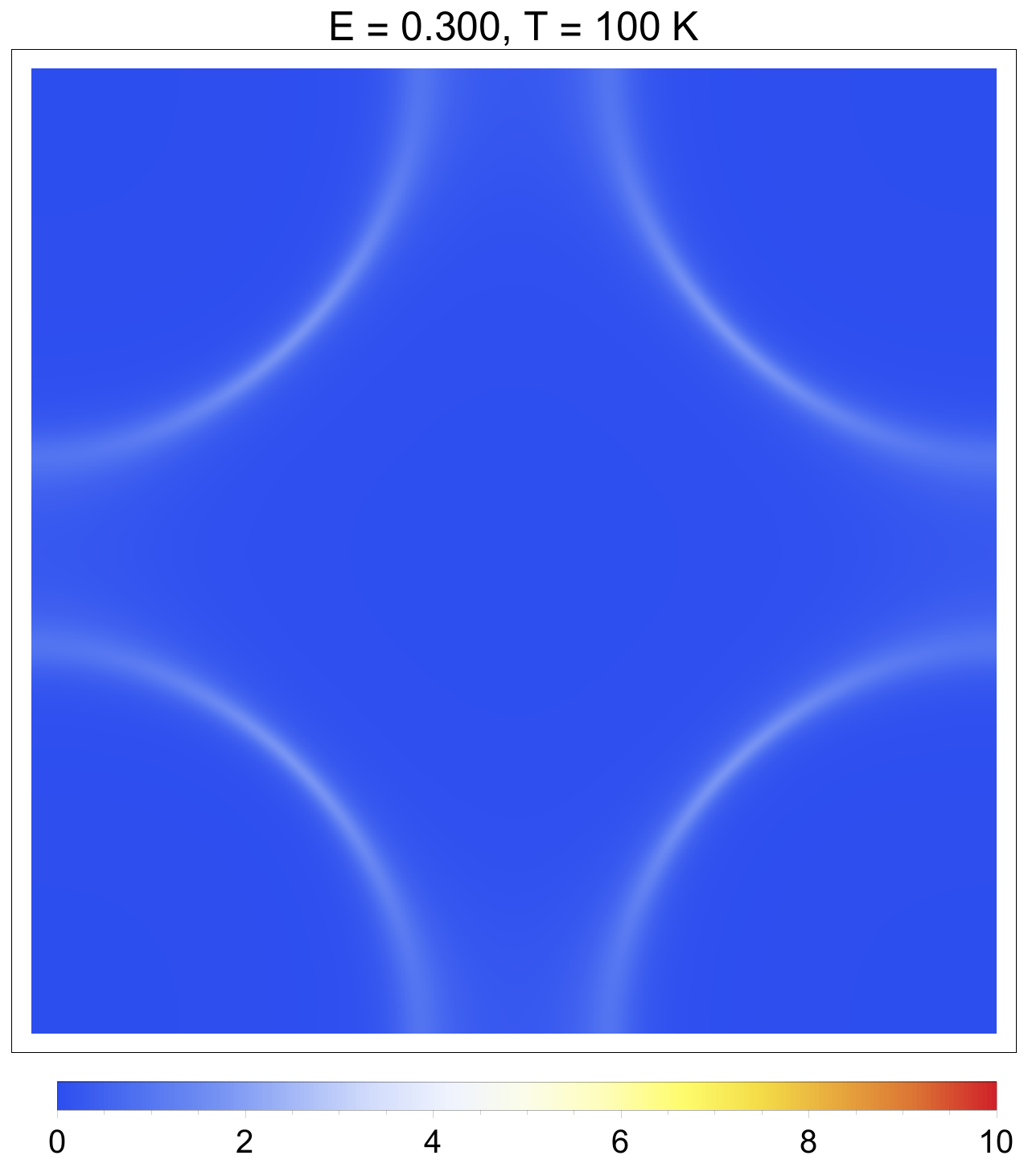} \\
	\includegraphics[width=0.16\textwidth]{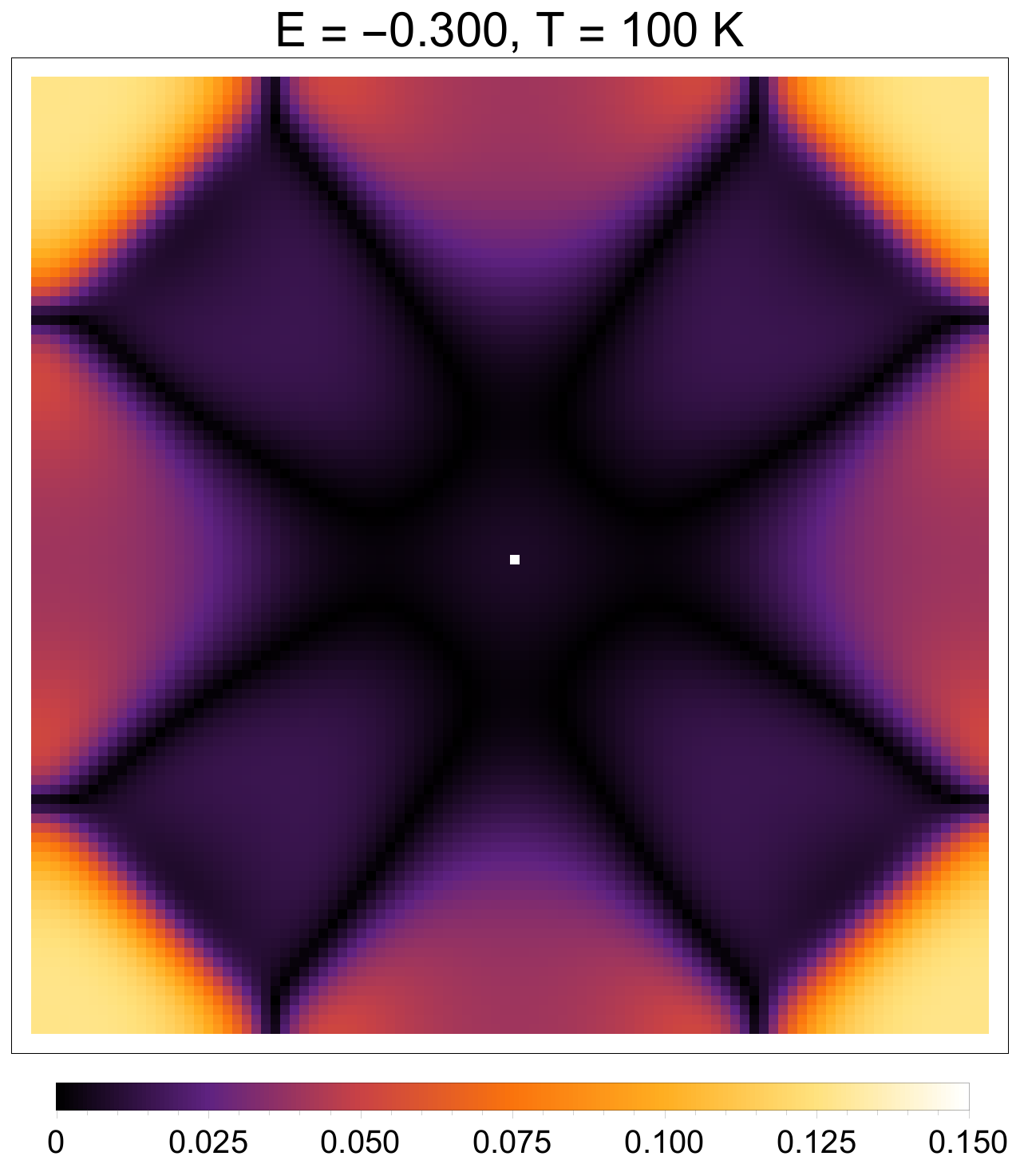}
	\includegraphics[width=0.16\textwidth]{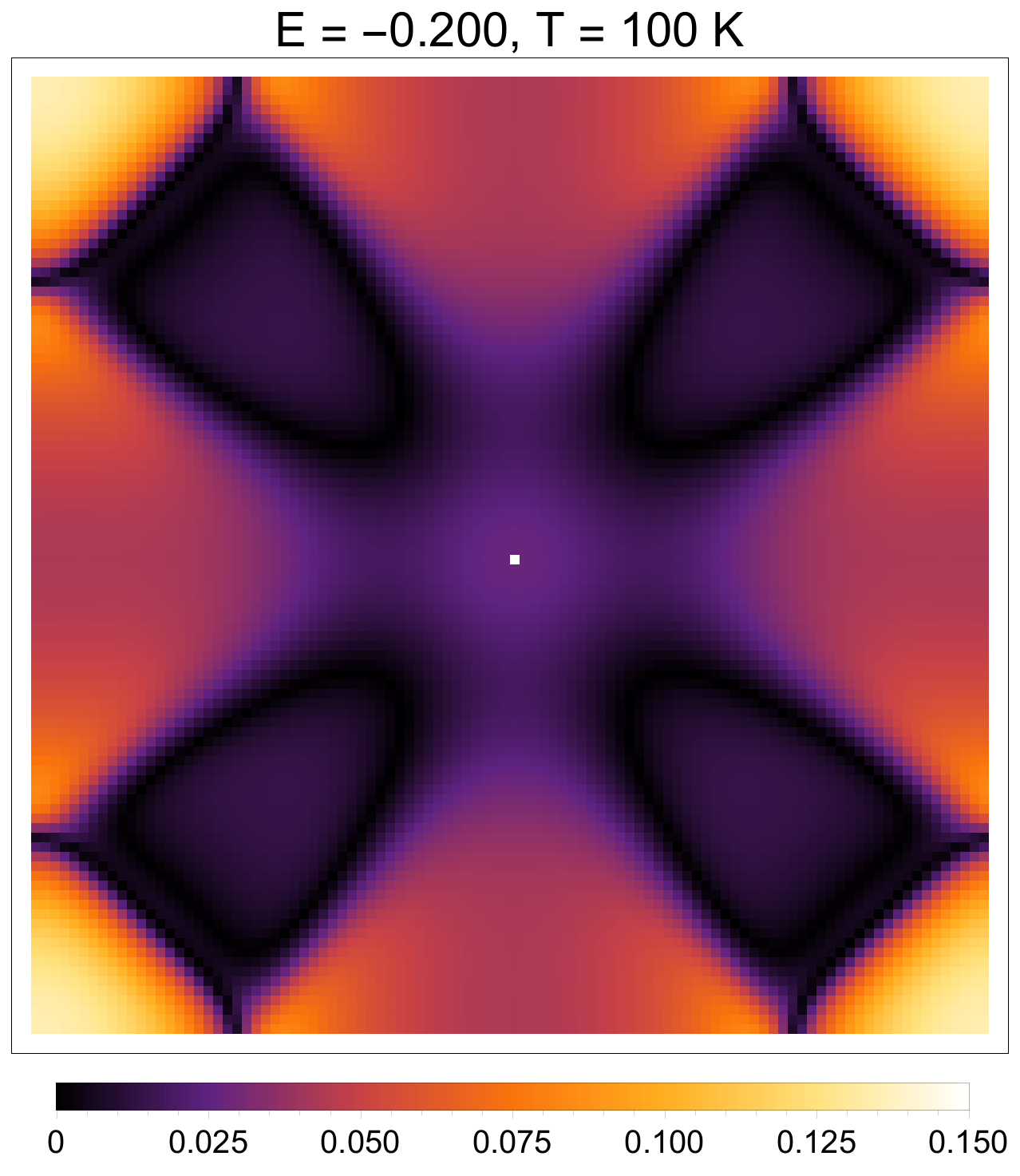}
	\includegraphics[width=0.16\textwidth]{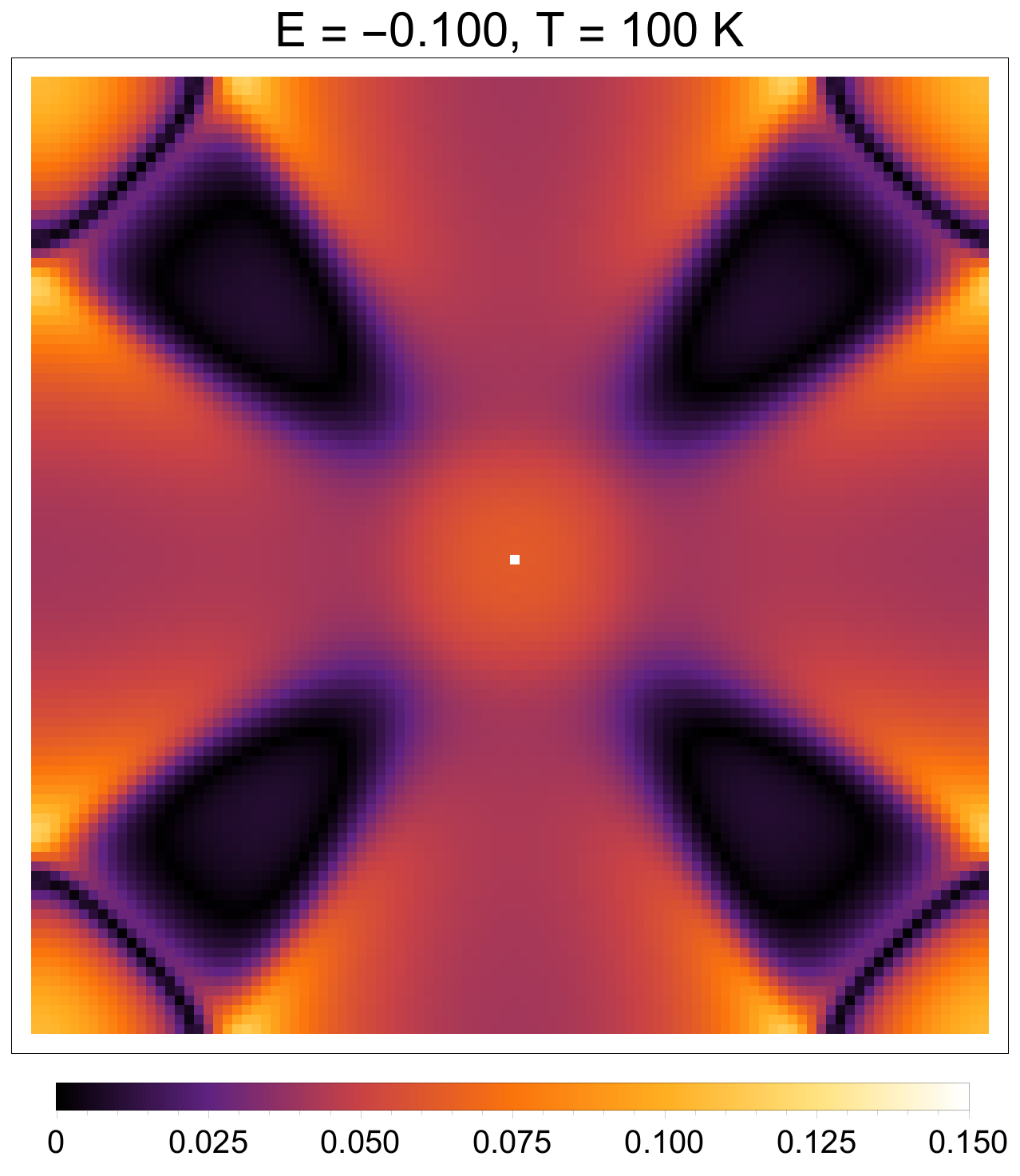}
	\includegraphics[width=0.16\textwidth]{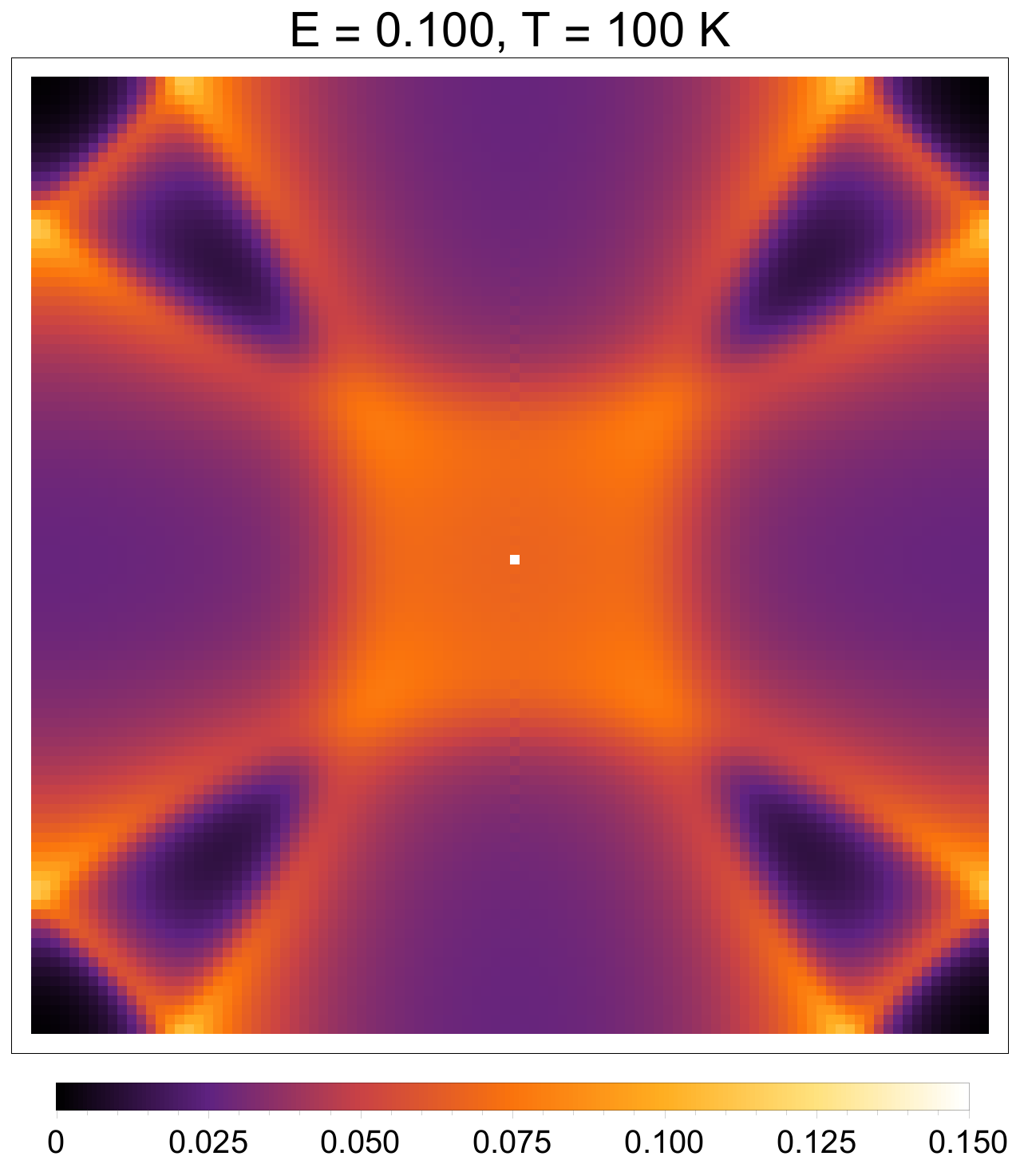}
	\includegraphics[width=0.16\textwidth]{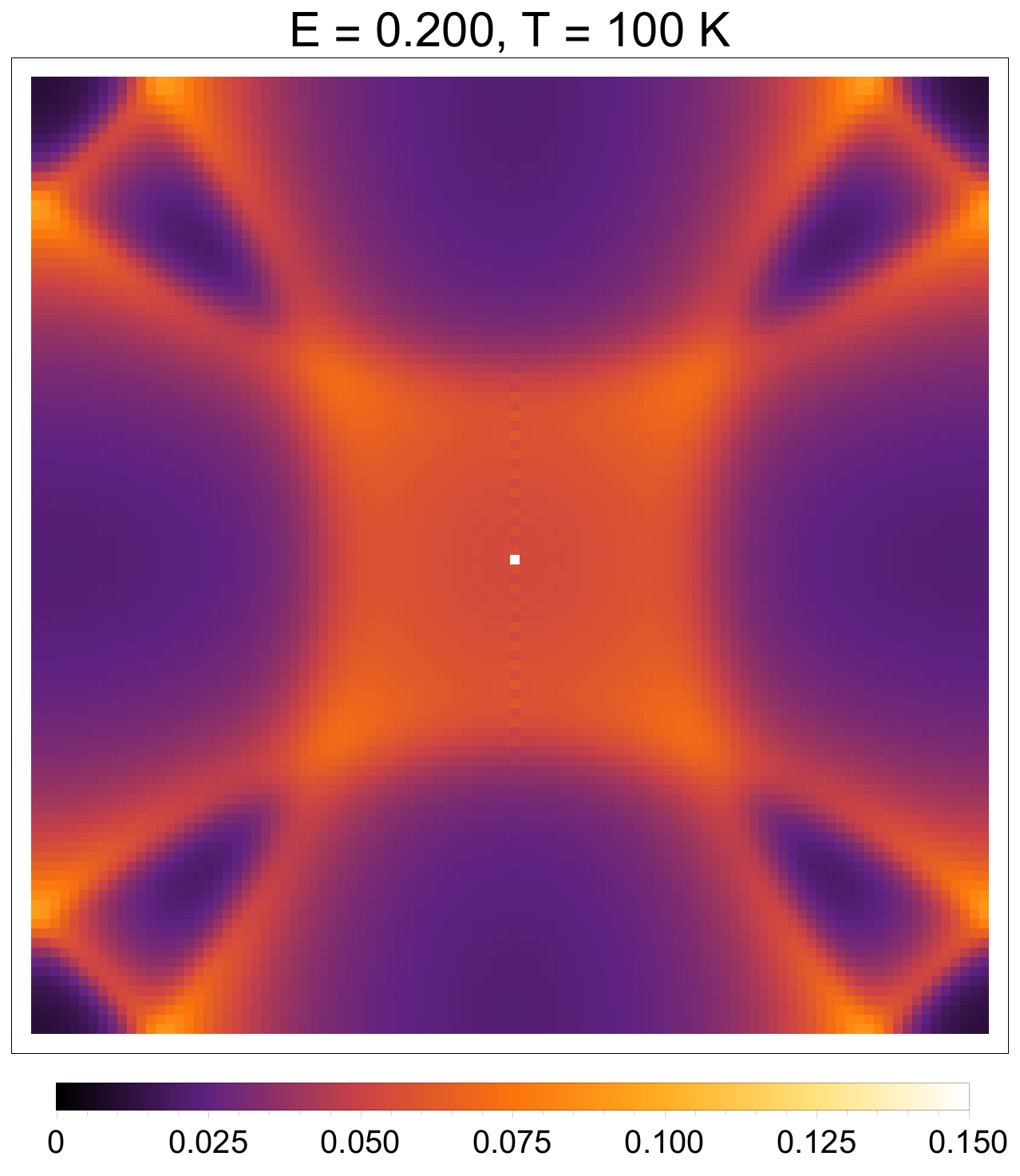}
	\includegraphics[width=0.16\textwidth]{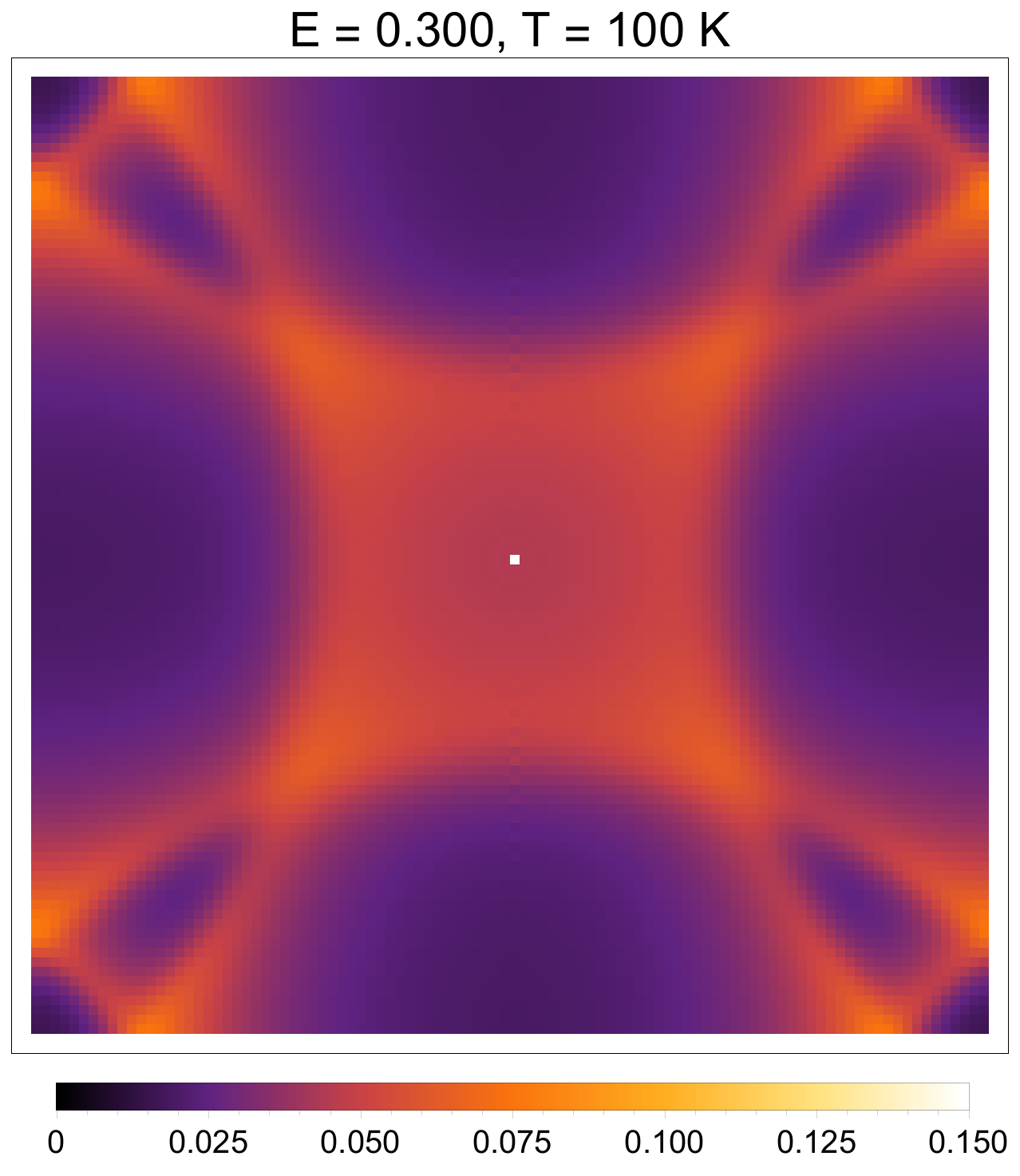}\\
	\includegraphics[width=0.16\textwidth]{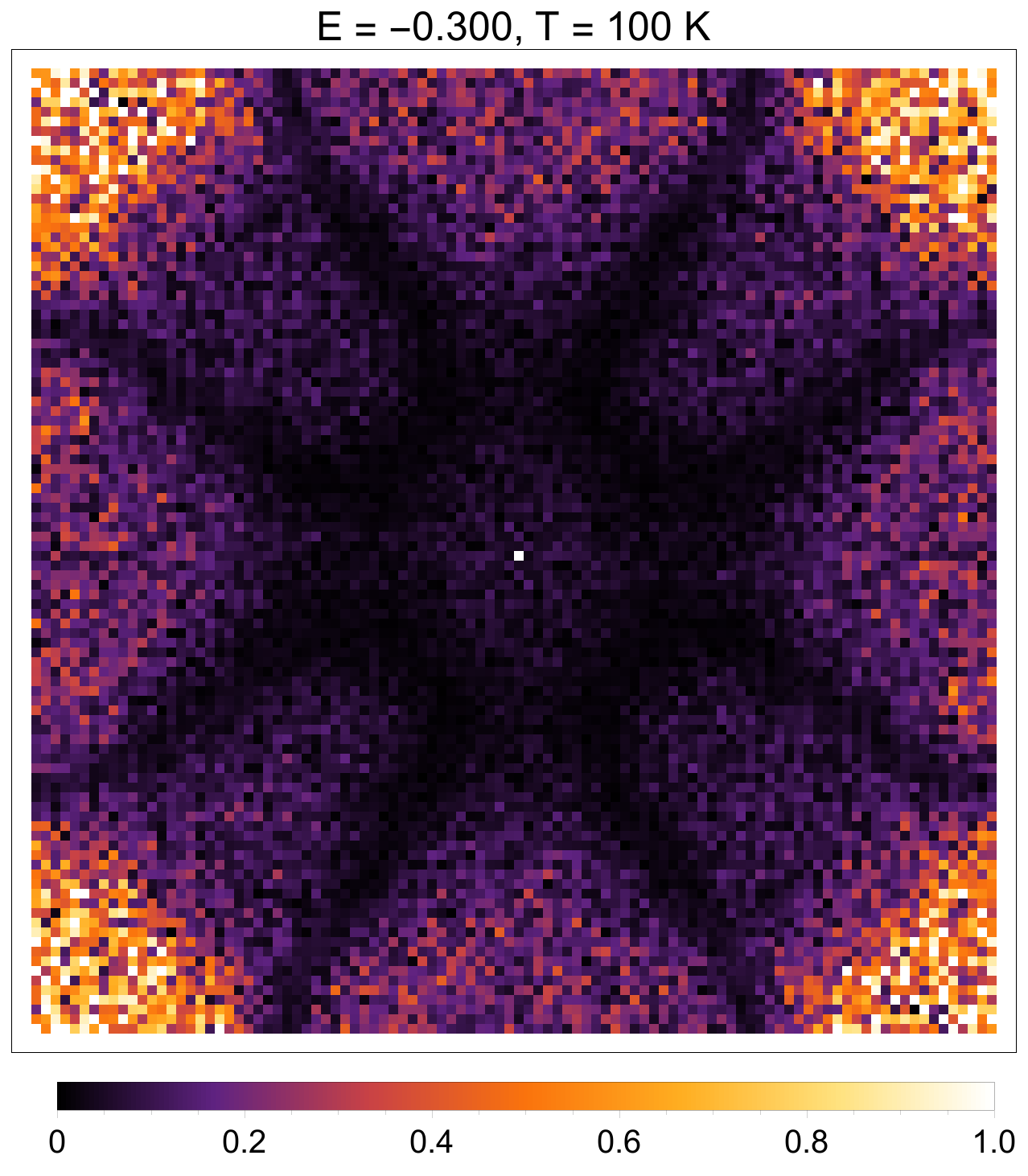}
	\includegraphics[width=0.16\textwidth]{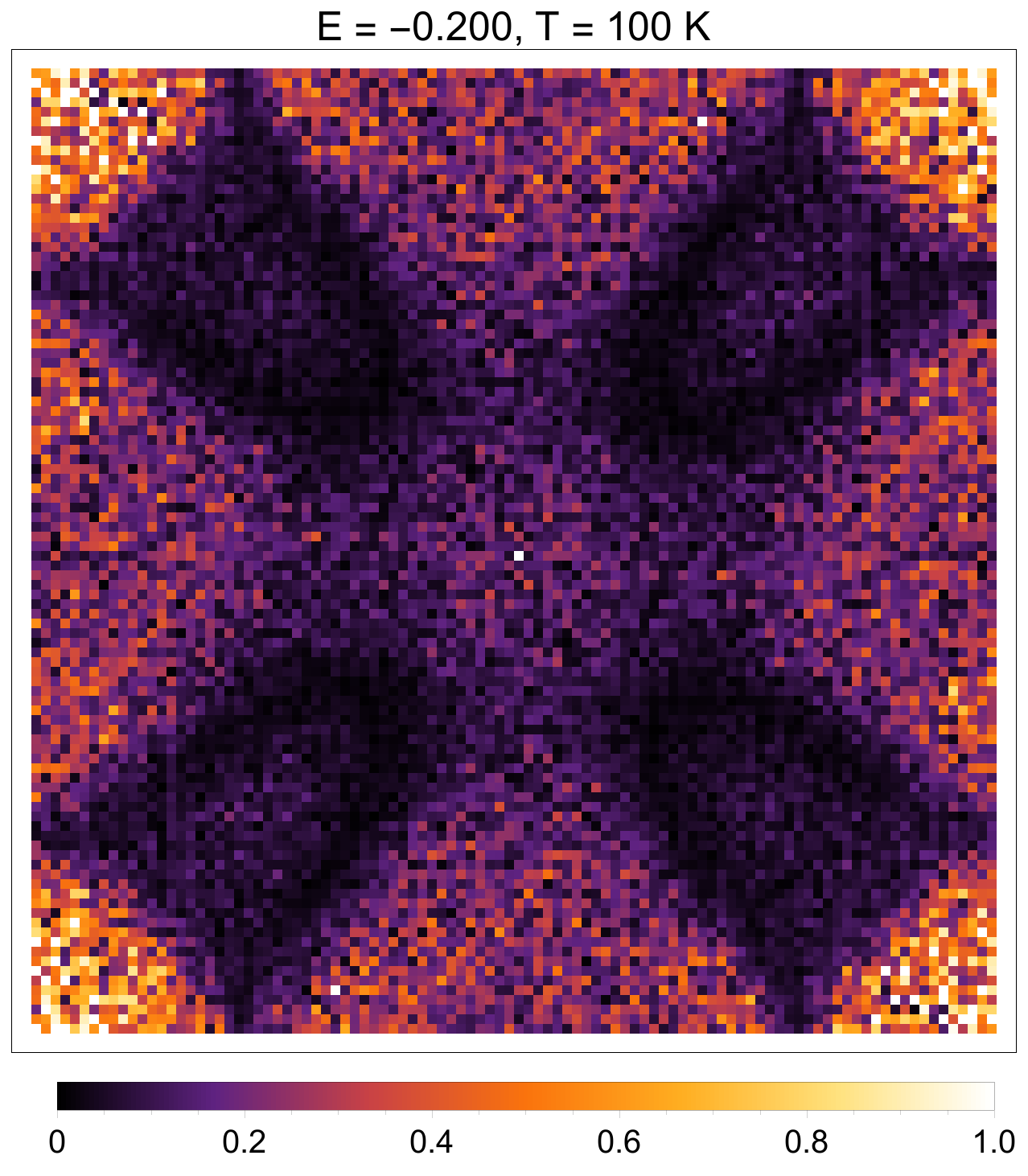}
	\includegraphics[width=0.16\textwidth]{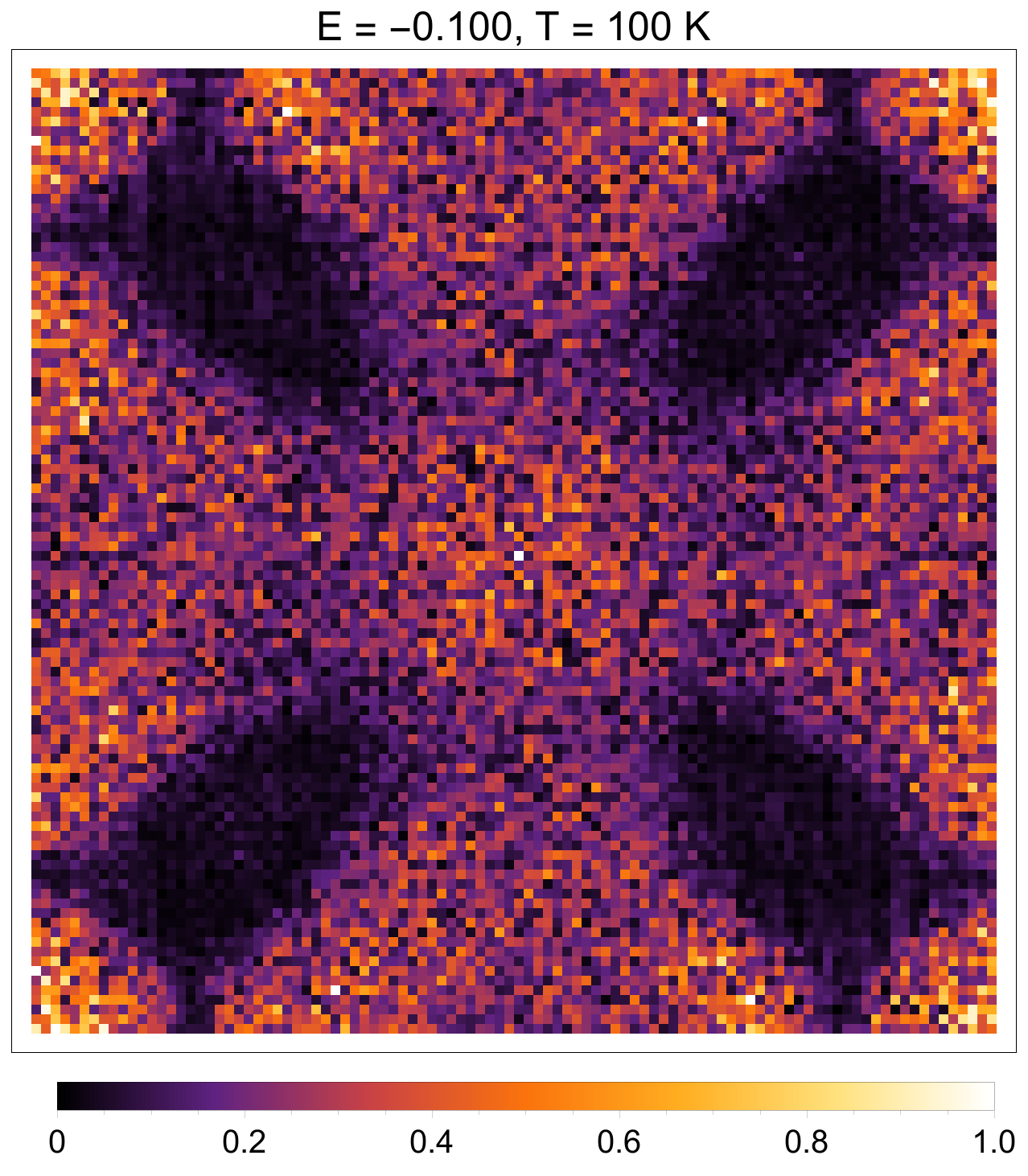}
	\includegraphics[width=0.16\textwidth]{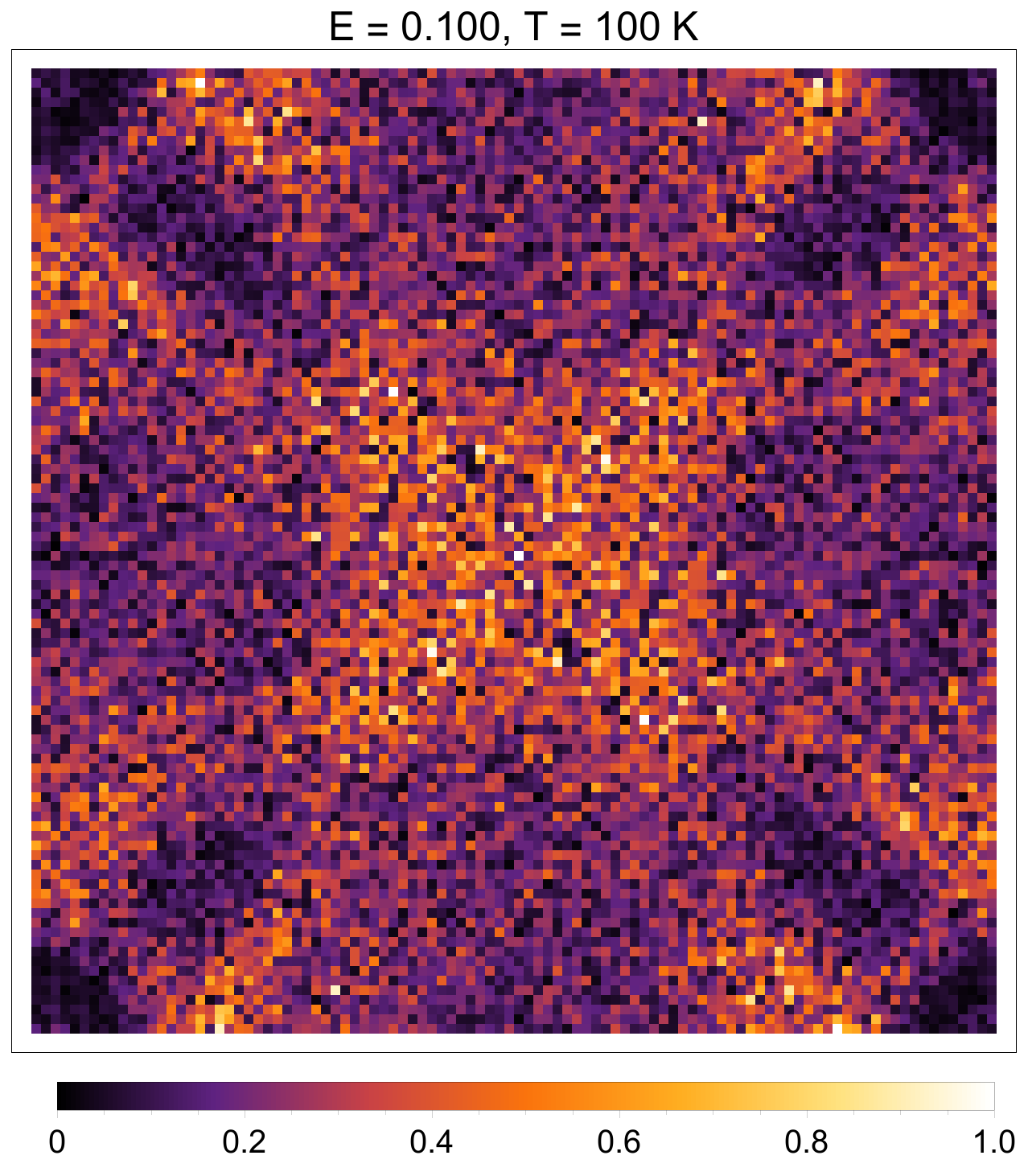}
	\includegraphics[width=0.16\textwidth]{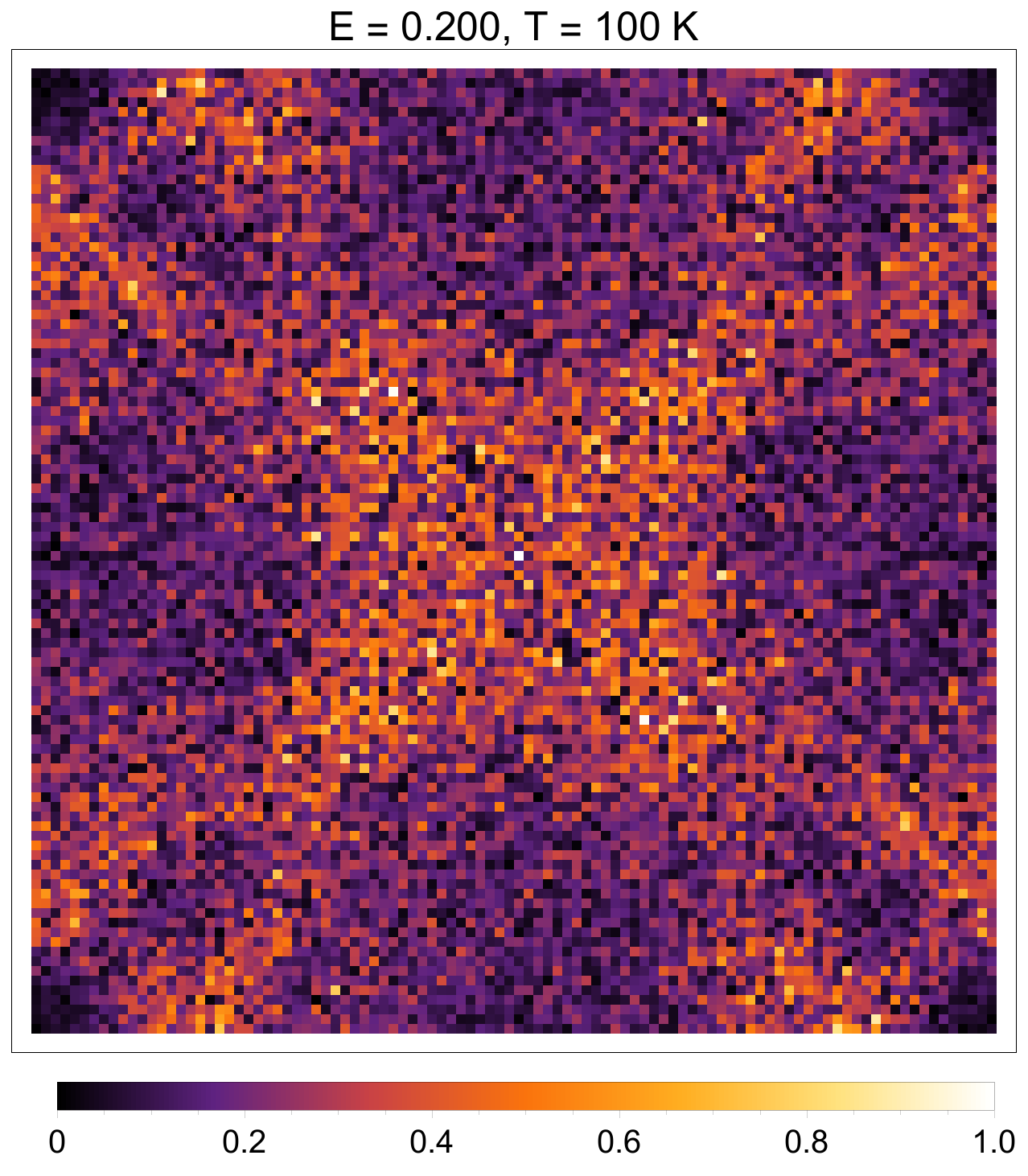}
	\includegraphics[width=0.16\textwidth]{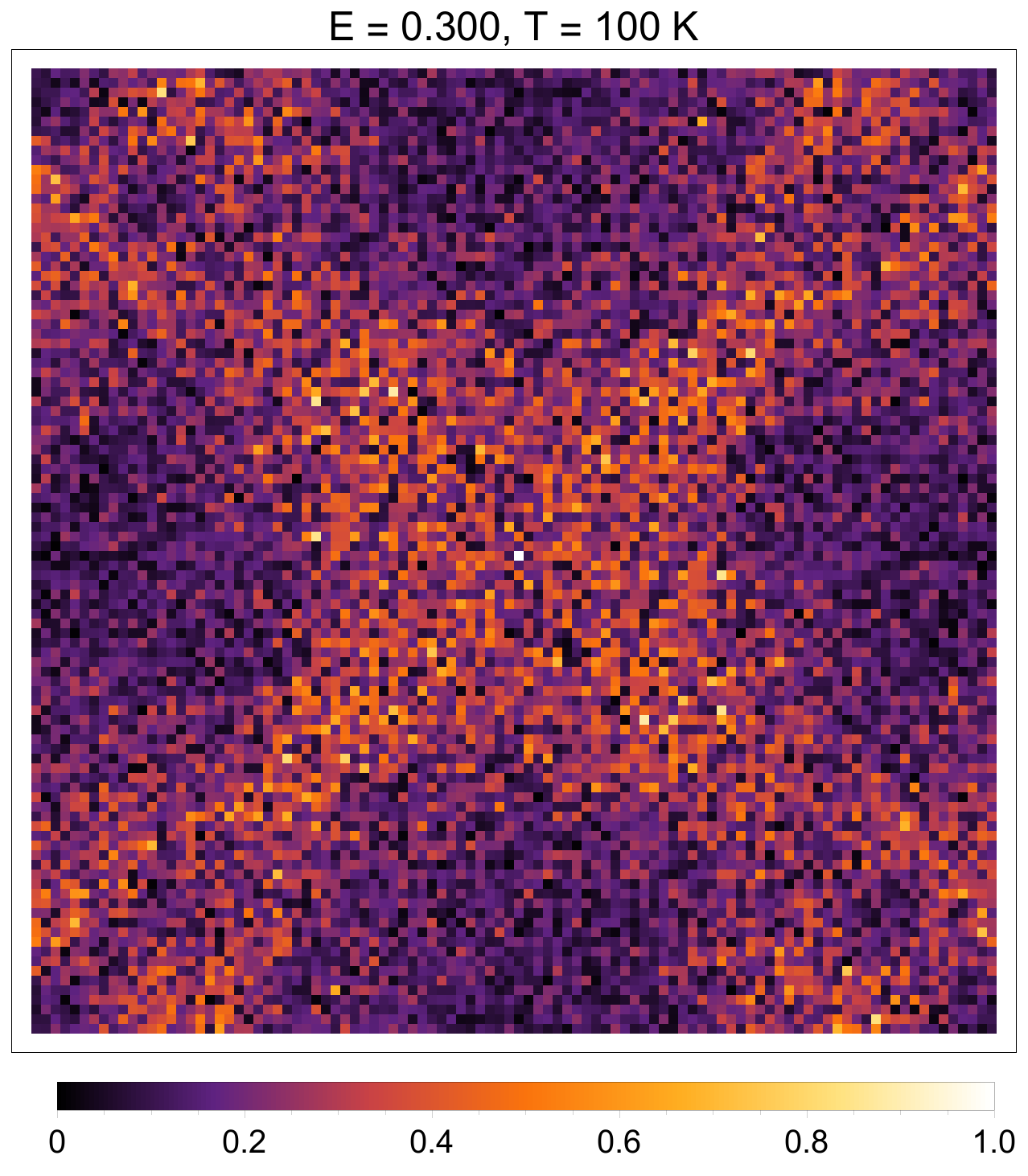}
	
	\caption{Frequency-dependence at $T = 100$ K of the spectra of an anisotropic marginal Fermi liquid. Shown are plots of the spectral function $A(\mathbf{k}, \omega)$ (upper row); the LDOS power spectrum with a single pointlike scatterer without thermal smearing (middle row); and the LDOS power spectrum with both a 0.5\% concentration of pointlike scatterers and thermal smearing (bottom row). Note that the scales used for plotting the LDOS power spectra are the same for all frequencies. For ease of visualization, the scale used here is smaller than that used in the Fermi liquid and isotropic marginal Fermi liquid plots.}
	\label{fig:frequency_nan_100k}
\end{figure*}

We next turn our attention to the effects of self-energies on the spectra in the normal state. We had briefly touched upon aspects of this in the previous section when we considered the ARPES and STS spectra at temperatures in which the gap fully closes. We will more closely examine the consequences when the self-energy in the normal state depends on frequency, temperature, and momentum. Our main focus will be on the marginal Fermi liquid phenomenology in the optimally-doped cuprates, and we will obtain concrete experimental predictions for STS which are indicative of marginal Fermi liquid behavior. We will in turn contrast the results for the marginal Fermi liquid from that of the ordinary Fermi liquid, which is argued to be the normal state of overdoped cuprates. Finally, to faithfully represent real-world ARPES data, we add at the end momentum-space anisotropy in the self-energy in order to reproduce the observation that the spectra at the antinodes are considerably more incoherent that those found in the nodal region of the Fermi surface. As with the superconducting cases considered earlier, we will evaluate the LDOS power spectrum both for a single isolated impurity without thermal smearing and for a macroscopically disordered sample with thermal smearing to incorporate effects likely to be seen in STS experiments.

We will assume that the self-energy has the ``power-law liquid'' form suggested by Reber \emph{et al.} from ARPES data on Bi-2212 across a wide range of dopings. This is simply given by
\begin{equation}
\Sigma^{''}(\omega, T) = \lambda\frac{(\omega^2 + \pi^2 T^2)^{\alpha}}{\omega^{2\alpha -1}_c} + \Gamma_0,
\label{eq:pll}
\end{equation}
where $\omega$ is the frequency, $T$ the temperature, $\Gamma_0$ a temperature- and frequency-independent impurity scattering rate, $\omega_c$ a frequency cutoff, and $\alpha$ a doping-dependent exponent which is argued from ARPES data to be equal to $0.5$ at optimal doping and near $1$ at extreme overdoping.\cite{reber2015power} This parametrization conveniently captures both the marginal Fermi liquid ($\alpha = 0.5$) at optimal doping\cite{varma1989phenomenology} and the ordinary Fermi liquid ($\alpha = 1.0$) at the overdoped side of the phase diagram. Plots of the self-energy for both the marginal Fermi liquid and the ordinary Fermi liquid at $100$ K are shown in Fig.~\ref{fig:flmflse_100K}. In our numerics the parameters are chosen to hew closely to the phenomenological fits found by Reber \emph{et al.} We will first neglect any momentum-space anisotropy in the self-energy; we will consider these effects later. We will set $\lambda = 0.5$, $\Gamma_0 = 0$, and $\omega_c = 1$ in our computations.

As an instructive case we first discuss the spectra of an ordinary Fermi liquid. Plots of $A(\mathbf{k}, \omega = 0)$ and $P(\mathbf{q}, \omega = 0)$ for this case are shown in Fig.~\ref{fig:temperature_fl}. Because of the isotropic nature of the self-energy, the spectral weight at Fermi surface is uniform at all temperatures considered. The spectral function here is narrow at the Fermi energy due to the small value of the imaginary part of the self-energy. Consequently the single-impurity LDOS power spectrum has sharp and well-defined features which broaden as temperature is increased. The main feature of $P(\mathbf{q}, \omega = 0)$ are caustics which indicate scattering wavevectors from one part of part of the Fermi surface to another, as expected from a metal. With randomly distributed impurities and thermal smearing, the LDOS spectra still manages to be visible at reasonably high temperatures, even without deconvoluting.

The situation for a marginal Fermi liquid is largely similar. In Fig.~\ref{fig:temperature_mfl} we have plotted both $A(\mathbf{k}, \omega = 0)$ and $P(\mathbf{q}, \omega = 0)$for a marginal Fermi liquid ($\alpha = 0.5$) at the Fermi energy for various temperatures. As the self-energy scales goes as $\propto T$ at the Fermi energy, the width of the spectrum at the Fermi surface also increases as $T$ increases. Like the spectral function, the LDOS power spectrum shows progressively more broadening as temperature is increased. When distributed disorder and thermal broadening are both present, the LDOS power spectra are broadened and feature speckle, but retain most of the structure present in the single-impurity case---caustics can still be observed at 100 K, but much of the spectrum becomes overwhelmed by noise at higher temperatures, rendering it difficult to extract these patterns at high temperatures without deconvoluting the data.

It has to be noted that at fixed frequency and temperature the results for the ordinary and marginal Fermi liquid cases are not drastically different from each other, except for the amount of broadening present---the marginal Fermi liquid has much more intrinsic broadening than the ordinary Fermi liquid. Thus one key signature that one may look for in ARPES and STS experiments is that, assuming that the overdoped cuprates have a Fermi-liquid normal state, the spectral widths at fixed $T$ and $\omega$ become larger as doping is decreased towards optimal doping. This is of course assuming that the normal state of the optimally-doped cuprates is in fact well-described by electrons dressed with a marginal Fermi liquid self-energy. While a marginal Fermi liquid features no quasiparticles at $T = 0$---unlike an ordinary Fermi liquid---it is clear that this description of the normal state should produce results that resemble those arising from a much more broadened version of the ordinary Fermi liquid at finite temperature. The unusual frequency- and temperature-dependence of the marginal Fermi liquid can also be measured using both ARPES and STS, and one should see a change in the scaling of the broadening of the spectra with temperature and frequency as doping is changed. If one sees these caustics in the STS spectra in the normal state of the optimally-doped cuprates, then the ``dressed Fermi liquid'' description of the normal state is valid. However, if these are not present, then a much more different theory involving exotic hidden excitations may be found to be necessary.

The frequency-dependence of the spectral function and the LDOS power spectra are plotted in Figs.~\ref{fig:frequency_fl_100k}  and~\ref{fig:frequency_mfl_100k} for the ordinary Fermi liquid and the marginal Fermi liquid, respectively, at 100 K. Note that for both these models both the spectral function and the LDOS power spectra broaden as frequency is increased at fixed temperature. The spectra do differ at high energies due to the renormalization of the band structure due to the real part of the self energy, which is different for both cases. It can be seen at negative frequencies the marginal Fermi liquid hits a van Hove singularity at a lower (negative) frequency than the ordinary Fermi liquid does owing to this renormalization. However this effect is quite troublesome to detect in practice, as disentangling this effect requires detailed knowledge of the bare band structure, and it is relatively unimportant compared to the scale set by the imaginary part of the self-energy. As such we will not direct any more focus on this phenomenon in this paper.

We next examine the precise dependence of the broadening of the spectral function and the LDOS power spectrum on the self-energy; plots of these are shown in Fig.~\ref{fig:widths}. Here, the self-energies used cover a wide range of frequencies and temperatures for both the marginal Fermi liquid and the ordinary Fermi liquid. The widths of the momentum-distribution curves along the nodal directions are proportional to the imaginary part of the self-energy. We can see this directly by obtaining the full width at half maximum of these MDCs; these widths scale \emph{linearly} with $\Sigma''$. As for the single-impurity LDOS power spectrum, the widths of the caustics broaden in a different manner from that of the spectral function. Quantifying this broadening is a bit trickier than for the spectral function, because the power spectrum features considerably more structure within the Brillouin zone due to backfolding. We define one measure of this broadening in the following manner. Along the $(0,0)\to(0,\pi)$ direction, there is a peak which corresponds to scattering between the antinodal portions of the Fermi surface. We define the width of the caustic as the distance between the midpoint between the peak and the minimum along the linecut at the central plateau near (0, 0) and the midpoint between the peak and the global minimum of this linecut. (Our analysis here differs from that of Dahm and Scalapino in that we do not assume \emph{a priori} a particular, weak-impurity form of the QPI response from which the self-energy is extracted.\cite{dahm2014quasi} Here the width of the caustics is the main measurable being considered; this analysis should hold irrespective of the type of the impurity causing the scattering.) While our resolution in $\mathbf{q}$-space is very limited, it can be seen that the widths of these caustics scale roughly as the \emph{square root} of $\Sigma''$, regardless of whether the self-energy is of marginal Fermi liquid or ordinary Fermi liquid form. Furthermore, the extracted widths of the spectra of the marginal Fermi liquid are parametrically much larger than those of the spectra of the ordinary Fermi liquid. 

We show in Fig.~\ref{fig:widthstemperature} the widths of the spectral function and the LDOS power spectrum for both the marginal Fermi liquid and the ordinary Fermi liquid as a function of frequency (for positive frequencies) for a variety of temperatures as extracted from our simulations, in addition to one-parameter fits of the form $W_{sc} = A\Sigma''(\omega, T)$ and $W_{ps} = B\sqrt{\Sigma''(\omega, T)}$ for the widths of the spectral function and the LDOS power spectrum, respectively. Here $\Sigma''(\omega, T)$ is of either marginal Fermi liquid or ordinary Fermi liquid form, and the parameters $A$ and $B$ are obtained from the data shown in Fig.~\ref{fig:widths}. It should be noted that at the energy ranges we have considered, the widths of the caustics in the LDOS spectra grow more slowly with frequency compared to the widths of the MDCs; this reflects the rough square-root dependence of the caustic widths on the imaginary part of the self-energy.

Finally, we end this section by considering a marginal Fermi liquid with a realistic amount of momentum-space anisotropy in the self-energy. A variety of ARPES measurements on optimally-doped Bi-2212 have shown that the spectral function at the antinodal region of the Brillouin zone is much less coherent than at the near-nodal region.\cite{valla1999evidence, abrahams2000angle, valla2000temperature, orgad2001evidence, kaminski2005momentum} The degree to which the spectral function is incoherent is most visible in energy-distribution curves taken at nodal and antinodal points along the Fermi surface; the nodal EDCs show a more prominent peak at the Fermi energy compared to the nodal ones. This suggests that the full self-energy is anisotropic in momentum space. Abrahams and Varma\cite{abrahams2000angle} argue that a good form of the self-energy is given by the following expression:
\begin{equation}
\Sigma^{''}(\mathbf{k}, \omega, T) = \Gamma_a(\mathbf{k}) + \lambda\sqrt{\omega^2 + \pi^2 T^2}.
\label{eq:anisotropic_se1}
\end{equation}
In this equation, the first term is the scattering rate due to disorder and is momentum-dependent and temperature- and frequency-independent. The second term contains the marginal Fermi liquid self-energy and is momentum-independent. The anisotropic elastic scattering rate is argued to arise from impurities located away from the copper-oxide planes, which induce only small-momentum scattering. To model this anisotropic scattering rate, we take it to have the following functional form:
\begin{equation}
\Gamma_a(\mathbf{k}) = \beta\left(\frac{2 + \cos 2k_x + \cos 2k_y}{4}\right).
\label{eq:anisotropic_se2}
\end{equation}
This form of the scattering rate ensures that it is small near the nodes---it is zero at ($\pm\frac{\pi}{2}, \pm\frac{\pi}{2}$), in fact---and that it has maxima at ($0, \pm\pi$) and ($\pm\pi, 0$). Importantly, this form preserves all the symmetries of the square lattice. The choice $\beta = 0.2$ gives rise to EDCs which show large anisotropy between the nodal and antinodal points on the Fermi surface, as seen in Fig.~\ref{fig:nan_edc}.

Plots of  $A(\mathbf{k}, \omega = 0)$ and $P(\mathbf{q}, \omega = 0)$ for this anisotropic marginal Fermi liquid at a variety of temperatures are shown in Fig.~\ref{fig:temperature_nan}. Note first that the spectral function at the near-nodal region is fairly sharp, while moving towards the antinodes we see that much more broadening becomes present, with considerable spectral weight being present in the regions between the Fermi surface at the antinodal regions. In the isotropic cases we considered earlier, there is zero spectral weight in these regions, as these parts of the Brillouin zone lie far beyond the bare Fermi surface, but with considerable nodal-antinodal anisotropy the antinodal regions become blurred and nonzero spectral weight results. This is even more apparent if we take momentum-distribution curves along the nodal and antinodal directions, as plotted in Fig.~\ref{fig:nan_mdc}: the MDCs along the nodal direction are quite sharp, while those along the antinodal directions are far more incoherent, although traces of peaks remain---in good agreement with ARPES experiments, which still find these antinodal peaks present in MDCs, albeit in a far weaker state compared to those at the nodes.

The LDOS power spectrum in Fig.~\ref{fig:temperature_nan} has a number of interesting features worth commenting upon. First, there is a very fuzzy square-shaped central plateau which is formed from small-momenta scattering processes between antinodal portions of the Fermi surface. Because the broadening is very large at the antinodal points, the scattering wavevectors appearing in $P(\mathbf{q}, \omega)$ consequently are severely broadened as well. Second, there is a set of very sharp features near $(\pm \pi, \pm \pi)$ which arise from internodal scattering. Recall that the spectral function remains sharp and well-defined near the nodes. As such, scattering wavevectors between near-nodal regions remain sharp in $P(\mathbf{q}, \omega)$, unlike those from antinodal-antinodal scattering. If one traces the caustics extending beyond the central plateau carefully, accounting for the backfolding of the spectra, one can make out that they decrease in width as one moves from the antinodal scattering wavevectors to the nodal ones. Linecuts along the nodal and antinodal directions are perhaps even more illuminating. The linecuts along the antinodal directions are featureless, save for the aforementioned plateau region, while the nodal linecuts show a sharp peak near the Brillouin zone boundary corresponding to nodal-nodal scattering. The contrast with the isotropic marginal Fermi liquid is quite striking, as the isotropic case (Fig.~\ref{fig:temperature_mfl}) features a central plateau which is still fairly sharply defined, while the caustics which appear beyond the plateau are of uniform width. With random disorder and thermal smearing, the resulting spectra appear very noisy---owing in part to the large intrinsic broadening at the antinodes. The central plateau visible in the single-impurity results is no longer easily seen, but there do remain sharp peaks near the zone diagonals corresponding to nodal-nodal scattering wavevectors, visible even when finite-temperature smearing is included.

Finally we note that because the frequency-dependence of the self-energy in this case is similar to the isotropic marginal Fermi liquid case considered earlier, the widths of the LDOS power spectra here should behave in the same way. This can be seen in plots of the spectral function and the LDOS power spectra at 100 K as frequency is varied, as seen in Fig.~\ref{fig:frequency_nan_100k}. As frequency is increased, the spectral function broadens throughout momentum space, and the resulting caustics in the LDOS power spectrum similarly broaden as well. This increased broadening at large frequencies contributes to the loss of signal in the disordered and thermally broadened data.

We caution the reader that because the anisotropic part of the self-energy here is presumably due to elastic scattering off of off-plane impurities, there is the danger that unless disorder is carefully taken into account, ``double-counting'' may ensue. As QPI is an intrinsically disorder-driven effect, one has to take care in these simulations that the same disorder producing QPI does not contribute additionally to the anisotropic elastic self-energy. We have taken care to use only pointlike impurities in our simulations of QPI, and the  effect of the off-plane impurities is incorporated in the anisotropic elastic scattering rate. A single weak pointlike impurity represents a very small perturbation to the system whose overall effect is negligible, while  a dilute ensemble of pointlike scatterers would presumably contribute to an \emph{isotropic} scattering rate, adding only a momentum-independent constant into the full self-energy upon disorder averaging.

\section{Discussion and Conclusion}

We have provided in this paper a comprehensive overview of the effects of self-energies on quasiparticle scattering interference, and have applied much of this insight to situations of relevance to the copper-oxide superconductors. While self-energies have been well-understood from the perspective of ARPES experiments, their effects on STS experiments have not been as similarly understood and are largely unexplored. A consistent result seen in the many scenarios we considered in this paper is the destruction of the QPI signal as broadening is increased, even when thermal smearing is ignored. In many ways, this is not an unexpected result. The physics underlying the phenomenological octet model of QPI in the superconducting state of the cuprates relies on the existence of coherent quantum-mechanical waves, which scatter elastically against impurities present in these materials. If these quasiparticles have a short lifetime---as seen in ARPES experiments in the strange metal, or even at the superconducting state near $T_c$---then the rather simple picture suggested by the octet model becomes complicated by ``off-shell'' contributions to the full, disordered Green's function. That is, in the presence of large broadening, states living away from the contours of constant energy \emph{do} contribute towards the scattering processes which determine the structure of the LDOS and its power spectrum. These effects are in fact already visible in the spectral function itself. We have seen that the contours of constant energy in both the normal and superconducting states turn from sharp, well-defined structures in momentum space into broad, incoherent entities. The effects of this broadening are particularly dramatic in the $d$-wave superconducting state, where we see that the sharp banana-shaped contours seen in the spectral function turn into incoherent arc-like streaks once the quasiparticle scattering rate is of the same order of magnitude as the superconducting gap. The loss of the sharpness in the contours of constant energy translates directly into the smearing and progressive destruction of the octet-model peaks as the scattering rate is increased.

The normal-state LDOS spectra feature no such peak-like structures, and instead what appears is a set of caustics which are continuous and whose broadening as a function of position \emph{on} the caustics directly reflects the degree of coherence of the quasiparticles of the underlying Fermi surface. As such, in the normal state the LDOS power spectrum is far less sensitive to broadening than in the superconducting state. One can see that the main feature differentiating the marginal Fermi liquid from the ordinary Fermi liquid is the amount of broadening present in both the spectral function and the LDOS power spectrum---the marginal Fermi liquid, by virtue of the fact that the imaginary part of its self-energy is much larger than that of the ordinary Fermi liquid at the same temperature and frequency, shows much more intrinsic smearing in its spectra---and how this broadening depends on temperature and frequency. Another measurable effect is the renormalization of the dispersions, due to the Kramers-Kronig relations, which can in principle be measured directly. Nevertheless this is a rather subtle effect---the bare band structure needs to be known in order for this renormalization to be detected--- and given the difficulty experimentalists are sure to face in attempting to observe this effect in STS experiments, the main signal of interest is the width of the measured power spectra.

It is worth explaining further in this section the limitations of our explicitly phenomenological approach. Our starting point consists of mean-field models of the normal and superconducting states, which are then ``dressed'' by self-energies which have a nontrivial dependence on temperature, frequency, momentum, or some combination of these. The predictions we make in this paper for STS---and, for that matter, ARPES as well---are sensible only if the actual strongly correlated phases seen in the cuprates can be adequately described by these dressed mean-field models. Much work on the two-dimensional Hubbard model, using dynamical mean-field theory, has shown that this ``dressed'' picture, involving a single-particle propagator augmented by a nontrivial self-energy, provides a reasonably accurate picture of the physics in some phases of relevance to the cuprates.\cite{civelli2008nodal, gull2015quasiparticle, sakai2016hidden}  If such a picture were to hold, then QPI will exist in some form or another. For instance, a model of the pseudogap involving a broadened $d$-wave superconductor will show QPI with the octet-model peaks decohering; nevertheless, despite the absence of sharp peaks, the power spectrum should still consist of wavevectors describing the relevant scattering processes. As another example, the marginal Fermi liquid is an exotic phase of matter without any low-temperature quasiparticle-like excitations---we remind the reader that its quasiparticle weight vanishes at the Fermi surface at $T = 0$---but still features ARPES and STS spectra that, at face value, are similar to those of an ordinary Fermi liquid. 

Having said all of this, if the phase of matter is \emph{not} describable at all by this dressed mean-field picture, there is no sense in which any of our predictions should hold. In particular, if STS were to show no evidence of these caustics in the strange-metal phase of the cuprates, then that would be one extremely convincing piece of evidence to suggest that the strange metal phase is beyond even the marginal-Fermi-liquid description. Hints of this have in fact been seen in STS studies deep inside the superconducting state: at energies larger than the superconducting gap, no well-defined caustics are seen, and instead the most dominant features are peaks corresponding to charge ordering.\cite{kohsaka2008cooper, lee2009spectroscopic, fujita2014simultaneous} In such a scenario, the appropriate theory is a strongly interacting phase of matter whose low-energy excitations are very unlike the Landau quasiparticles of the Fermi liquid. A paradigmatic example of this is the Luttinger liquid in one spatial dimension,\cite{haldane1981luttinger} whose decidedly non-quasiparticle-like excitations result in ARPES and STS spectra considerably different from those of an ordinary Fermi liquid.\cite{orgad2001evidence, kivelson2003detect} In addition, numerous examples of these phases have been constructed using holographic methods, and are known to result in physics very different from that of the ordinary Fermi liquid.\cite{iqbal2012semi, davison2014holographic} We end by noting that what high-temperature STS experiments can eventually find in the strange metal phase and in the transition to the superconducting state at optimal doping will undoubtedly be very interesting. The insights that can be gleaned from such future experiments will no doubt go a long way in illuminating the strange physics of the cuprate superconductors.

\begin{acknowledgments}
We thank M. P. Allan, L. W. Bawden, M. S. Golden, P. J. Hirschfeld, and N. Poovuttikul for useful discussions. This work was supported by the Netherlands Organisation for Scientific Research (NWO/OCW) as part of the Frontiers of Nanoscience (NanoFront) program.
\end{acknowledgments}

\nocite{*}

\bibliography{paper_selfenergies}

\end{document}